\documentclass[twoside,12pt]{Latex/Classes/PhDthesisPSnPDF}
\usepackage{amsmath,amssymb}
\usepackage{longtable}
\usepackage[export]{adjustbox}
\usepackage{threeparttablex,booktabs,threeparttable}
\usepackage{graphics,graphicx}
\usepackage{pdfpages}
\usepackage{wrapfig}
\usepackage{lscape}
\usepackage{array}
\usepackage{epsfig}
\usepackage{hyperref}
\usepackage{ulem}
\usepackage{ragged2e}
\hypersetup{colorlinks,linkcolor={blue},citecolor={blue},urlcolor={blue}} 
\makeatletter
\providecommand*{\approxident}{%
  \mathrel{%
    \mathpalette\@approxident\sim
  }%
}   
\newcommand*{\@approxident}[2]{%
  \sbox0{$#1\vcenter{}$}%
  \sbox2{$\m@th#1\equiv$}%
  \dimen2=\dimexpr\ht2 - \ht0\relax
  \sbox4{$\m@th#1\sim$}%
  \dimen4=\dimexpr\ht4 - \ht0\relax
  \dimen0=\dimexpr
    -\ht4 - \dp4 %
    + \dimen2 %
  \relax
  \vcenter{\offinterlineskip
    \copy4 %
    \kern\dimen0 %
    \copy4 %
    \kern\dimen0 %
    \copy4 %
    \ifdim\dp4=\z@
      \kern\dimexpr -\ht0 + \dimen4\relax
    \fi
  }%
}      
\usepackage{isomath}
\usepackage{scalerel,stackengine}

\usepackage{bm}
\let\vec\bm
\makeatother

\begin{document}

\renewcommand\baselinestretch{1.5}
\baselineskip=18pt plus1pt
\renewcommand{\arraystretch}{1.2}

\frontmatter
\begin{titlepage}
\begin{center}

\vspace{1cm}

\begin{Large}
\textbf{Deciphering Radio Emission from Solar Coronal Mass Ejections using High-fidelity Spectropolarimetric Radio Imaging} \\
\end{Large}

\vspace{1.0cm}


A Thesis \\ 

\vspace{0.5cm}

Submitted to the \\
Tata Institute of Fundamental Research, Mumbai \\
for the degree of Doctor of Philosophy \\
in Physics \\

\vspace{0.8cm}

by \\

\vspace{0.4cm}


\begin{large}
\textbf{Devojyoti Kansabanik}
\end{large}

\vspace{0.5cm}
Supervised by \\
\vspace{0.4cm}
\textbf{Divya Oberoi} \\


\vspace{0.5cm}

\begin{figure}[htbp]
\centering
\includegraphics[scale=0.4]{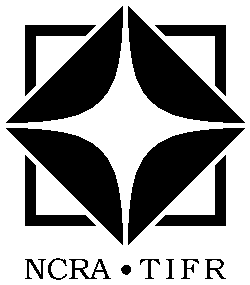}
\end{figure}

{\large National Centre for Radio Astrophysics} \\
\begin{large}
Tata Institute of Fundamental Research \\
Mumbai
 \end{large} \\
\vspace{1.0cm}
September, 2023 \\

\end{center}
\end{titlepage}

\includepdf[pages={1}]{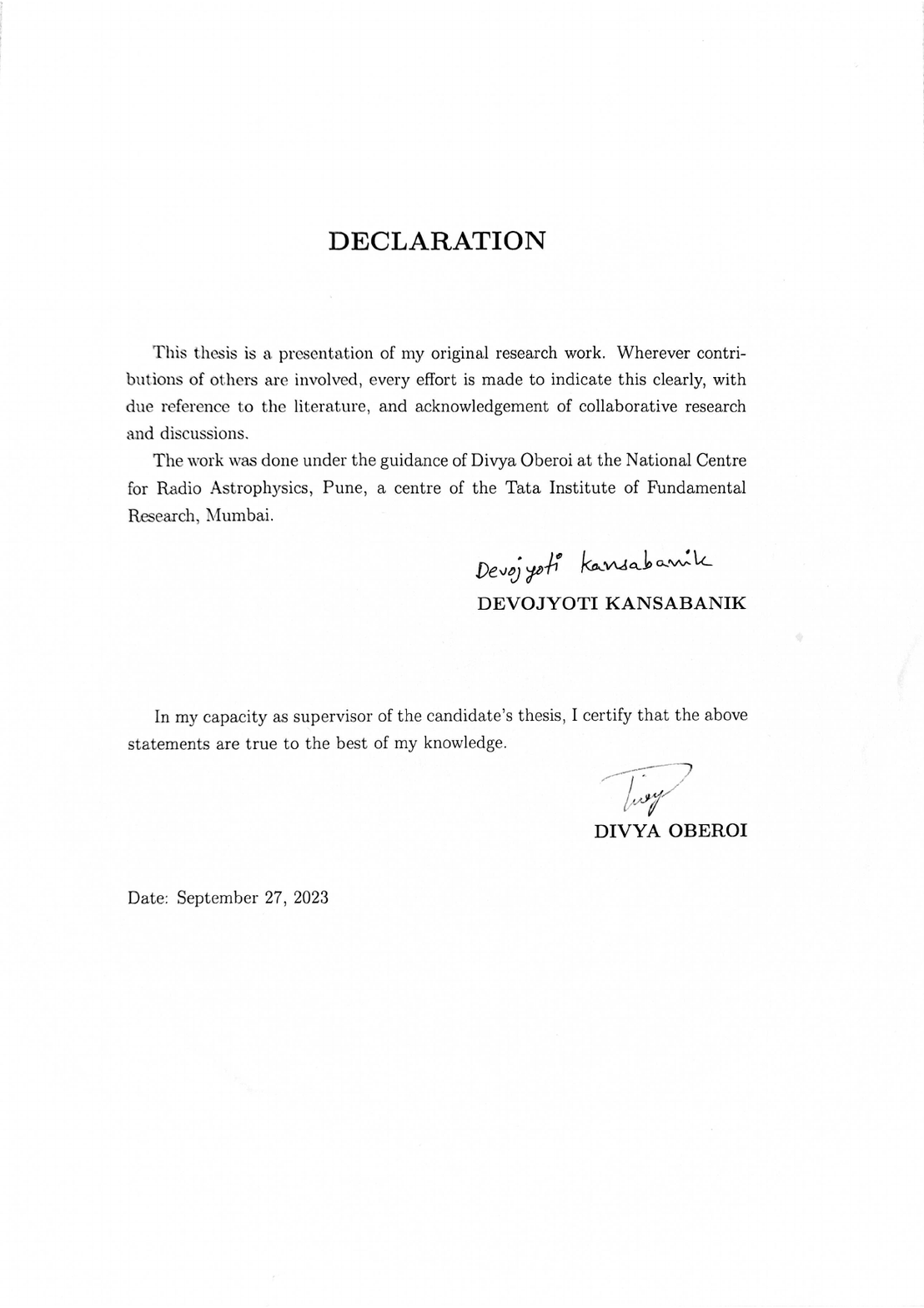}
\thispagestyle{empty}


\begin{center}
\begin{Large}
\textbf{Acknowledgements} \\
\end{Large}
\end{center}

\vspace{0.5cm}
In this humble corner of my thesis, I sincerely acknowledge my gratitude to those whose presence has woven threads of inspiration, knowledge, and support into the fabric of my academic journey. To my guiding star, my revered supervisor Divya Oberoi, your wisdom and firm belief in my potential have illuminated every corner of my academic pursuit. With every word of guidance and encouragement, you have propelled me toward intellectual heights. I am deeply grateful to you for guiding me throughout my academic journey. 

I extend my sincere appreciation to the members of my thesis committee and all faculty members of NCRA-TIFR for their insightful feedback, critical evaluation, and valuable suggestions, which have significantly contributed to the development and refinement of my research work.

I would like to offer my special thanks to my fellow travelers on this scholarly odyssey, my colleagues, and my research collaborators. Specifically, I want to thank Surajit Mondal for involving in discussions on different aspects of my thesis and providing valuable ideas and criticism. I would also like to thank Barnali Das for providing a beautiful name for the software pipeline developed as a part of this thesis. Nonetheless, I also thank Apurba Bera, Barun Maity, and Divesh Jain for their help at various stages of the thesis. 

In the tapestry of life, where parents are the intricate threads that bind us, I extend my heartfelt gratitude to them for standing beside me throughout this journey. Their love, patience, and firm belief always supported me both in times of doubt and moments of triumph. I would also want to express my heartfelt gratitude to the ideal of my life, Sri Sri Thakur Anukulchandra and present Acharyadev of Satsang Deoghar, whose guidance and continuous encouragement allow me to learn and think critically and logically about several aspects of life.

Through the valleys and peaks of research, the financial support provided by TIFR and the Department of Atomic Energy, Government of India, is precious and provides the necessary resources to breathe life into my ideas and realize the aspirations of this scholarly odyssey. I am grateful to NCRA for providing the infrastructure, facilities, and resources necessary for carrying out this research. The academic and administrative staff of NCRA have been instrumental in creating a conducive environment for learning and research. I would like to thank them all, especially Hemant Lokhande, Reena Shrikumar, J. K. Solanki, and D. V. Pawar. I would also specially thank all canteen and hostel staff members, without whom a pleasant stay was not possible. I also thank the center director of NCRA, Prof. Yashwant Gupta, for being extremely supportive of all of my foreign visits to conferences and other research institutes during my thesis tenure. 

This thesis work makes use of the Murchison Radio-astronomy Observatory (MRO), operated by the Commonwealth Scientific and Industrial Research Organisation (CSIRO), Australia. I acknowledge the Wajarri Yamatji people as the traditional owners of the Observatory site. Support for the operation of the MWA is provided by the Australian Government's National Collaborative Research Infrastructure Strategy (NCRIS) under a contract to Curtin University administered by Astronomy Australia Limited. I acknowledge the Pawsey Supercomputing Centre, which is supported by the Western Australian and Australian Governments. 

Lastly, I would like to acknowledge all those individuals, too numerous to mention individually, who have offered their assistance, provided feedback, or shown support in any form during my doctoral studies. Your contributions have played an integral part in the completion of this thesis, and I am sincerely grateful for your involvement.

As I end my modest acknowledgment, I pay respect to the eternal flame that burns bright within the hearts of seekers: the thirst for knowledge. May the light of this endeavor continue to enlighten the minds of future generations as they begin their academic journeys.

\vspace{0.5cm}
\begin{flushright}
\textbf{Devojyoti Kansabanik} \\
\vspace{0.4cm}
Pune, June 26, 2023 \\
\end{flushright}

\listoffigures	
\listoftables  
\tableofcontents            

\thispagestyle{plain}

\begin{center}
\begin{Large}
\textbf{Abstract} \\
\end{Large}
\end{center}

Solar coronal mass ejections (CMEs) are large-scale expulsions of plasma and magnetic field from the Sun into the heliosphere and are the most important driver of space weather. The geo-effectiveness of a CME is primarily determined by its magnetic field strength and topology. The evolution of CMEs while propagating through the corona and the heliosphere complicates the prediction/extrapolation of the vector magnetic field of the CMEs near the Earth based essentially on photospheric measurements. Hence, the measurement of CME magnetic fields, both in the corona and heliosphere, is essential for improving space weather forecasting. Although CMEs are routinely observed by white-light coronagraphs, these observations cannot provide a direct measure of the CME-entrained vector magnetic fields. Observations at radio wavelengths, however, can provide several remote measurement tools for estimating both strength and topology of the CME magnetic fields. Among them, gyrosynchrotron (GS) emission produced by mildly-relativistic electrons trapped in CME magnetic fields is one of the promising methods to estimate magnetic field strength and other plasma parameters of CMEs at lower and middle coronal heights. However, GS emissions from some parts of the CME are much fainter than the quiet Sun emission and require high dynamic range (DR) imaging for their detection. This thesis presents a state-of-the-art calibration and imaging algorithm capable of routinely producing high DR spectropolarimetric snapshot solar radio images using data from a new technology radio telescope, the Murchison Widefield Array (MWA). This allows us to detect much fainter GS emissions from CME plasma at much higher coronal heights than before. For the first time, robust circular polarization measurements have been jointly used with total intensity measurements to constrain the GS model parameters, which has significantly improved the robustness of the estimated GS model parameters. A piece of observational evidence is also found that the routinely used homogeneous and isotropic GS models may not always be sufficient to model the observations. In the future, with more sensitive measurements from the upcoming new generation telescopes and physics-based forward models, it should be possible to relax some of these assumptions and make this method more robust for estimating CME plasma parameters at coronal heights.

\mainmatter

\chapter {Introduction}
\label{chapter_intro}

The Sun is the star at the center of our solar system. Not only the motions of the planets are largely determined by the gravity of the Sun, but the Sun also determines the overall environment of the solar system. For us on the Earth, the Sun is the source of all energy (except nuclear energy, geothermal energy, etc.). All living creatures on the Earth essentially rely on solar energy to sustain themselves. Along with all these blessings, occasionally, the Sun also gives rise to some threats to Earth's environment due to some violent eruptions taking place on it. The large-scale expulsions of plasma from the Sun are known as coronal mass ejections (CMEs) and can affect planetary environments in the solar system. Over the past several decades, a significant amount of effort has been devoted to building an understanding of the physics behind the eruption of CMEs, their evolution, propagation from the Sun to the Earth, and their effectiveness on Earth, both from observational and modeling perspectives. These efforts have led to an improved understanding of CMEs. However, there are still several challenges, both from observational and modeling perspectives, which need to be solved to understand the CMEs and accurately determine their effectiveness on Earth. 

\section{Solar Atmopshere and Heliosphere}\label{sec:corona_heliosphere_intro}
The Sun comprises different layers starting from the core at its very center to the outermost optically thick layer, the photosphere. When we look at the Sun with our naked eye, we see the photosphere of the Sun. The photosphere is the base of the solar atmosphere. In the early days, like other stellar atmospheres, the solar atmosphere was also modeled as plane-parallel or spherically-symmetric layers of plasma at different temperatures and densities \citep{Heinzel2000}. As observations improved it became clear that the solar atmosphere is far from being plane-parallel and static. Instead, it is an inhomogeneous and very dynamic medium. Hence, instead of describing different layers of the solar atmosphere geometrically, it is more appropriate to use definitions based on the physics of the layer under consideration \citep{Carlsson2019}.

The region above the photosphere where radiative equilibrium breaks down and hydrogen is predominantly neutral is called the chromosphere \citep{Carlsson2019}. Plasma parameter, {\it plasma beta ($\beta$)}, is the ratio between gas pressure and magnetic pressure of the plasma and determines which of these pressures is the dominant one. In the solar atmosphere, both the density and the magnetic field decrease with height. However, the magnetic field does not decrease with height as rapidly as the density. This causes the lower chromosphere to be dominated by gas pressure ($\beta>1$), while the upper chromosphere is dominated by the magnetic field ($\beta<1$). In quiet Sun regions, the $\beta=1$ surface lies about 0.9 Mm above the photosphere, while in strong magnetic field regions, it lies at much lower heights \citep{Gary2001}. 

The next region of the solar atmosphere is called the transition region. Solar transition region commonly refers to a thin layer above the chromosphere where the predominantly neutral gas character of the chromosphere changes to fully ionized plasma. This change in the state of the medium is accompanied by sharp changes in the temperature and density profiles \citep{Avrett2008,russell2016space,Song_2023}. Identifying the physical mechanisms determining the height where the transition region starts or the thickness of the transition region may be the most challenging problem concerning the transition region. The transition region has an average thickness of $\sim100$ km but shows different starting heights and thicknesses over quiet solar regions \citep{Song_2023} and over sunspot regions \citep{Tian2018} along with temporal variations.

The upper solar atmosphere is known as the solar corona which is at a temperature of about a million K \citep{aschwanden2004,Cranmer2019}. Temperature rises very quickly in the transition region to more than a million K in the corona. Corona is a hot and ionized medium. In the solar corona, plasma $\beta$ is less than one \citep{Gary2001} and hence much of the coronal plasma is confined by the solar magnetic field in the form of closed loops and twisted arcade-like structures \citep{Cranmer2019}. Some coronal plasma expands into interplanetary space as a supersonic outflow known as the solar wind \citep{Parker1958,Neugebauer1962}.

The solar wind is a stream of charged particles, mostly consisting of electrons and protons arising from the coronal plasma, which is continuously expanding into the interplanetary space \citep{Parker1958,Neugebauer1962}. This solar wind fills the heliosphereic region between the solar corona and the local interstellar medium. The sphere of influence of the Sun, which is essentially filled with the plasma of solar origin, is known as the Heliosphere \citep{Parker1961,Axford1963,Miralles2011}. All objects of the solar system, planets, moons, asteroids, etc., reside within the heliosphere. In the corona, the solar wind slower than the Alfvén speed, the speed of local magnetohydrodynamic waves of interest in the present context. In the inner heliosphere, the solar wind, exceeds the super-Alfvénic speeds. The notional location, where large-scale solar wind speed is equal to the speed of local Alfvén waves, is called the Alfvén surface \citep{DeForest_2014,Adhikari2019}. 

Both the coronal and the solar wind plasma are highly conducting. This causes the magnetic fields to be frozen in it \citep{Syrovatskii1978,Manoharan2003,Lui2018}. Due to this frozen-in condition, the coronal magnetic field is dragged out and fills the heliosphere. This is known as the interplanetary magnetic field \citep[IMF;][]{Gosling1997,Owens2013}. Not only does the IMF play an important role in determining solar wind flow, but also the geo-effectiveness and structures in the solar wind by influencing the propagation of CMEs.  

\section{Coronal Mass Ejections}\label{sec:cme_intro}
CMEs are large-scale expulsions of plasma and magnetic fields from solar corona into the heliosphere \citep{Chen2011,Webb2012}. CMEs are of interest for both scientific and societal reasons. Scientifically they are of interest to understand the removal of magnetic field,  helicity and plasma in the solar corona \citep{Low1996,Low2001}, triggering mechanisms of CMEs \citep{atmos10080468,Foullon_2013,Zhelyazkov2015,Vashalomidze2022}, shocks and particle accelerations \citep{Mikić2006,Vourlidas2012,Xia_2020,Frassati_2022}. From a societal perspective, CMEs are of special interest because they produce the most intense space-weather events \citep{Owens2021,Cliver2022} which can have a significant impact on our technology-reliant society and space-based assets.
\begin{figure*}[!ht]
    \centering
    \includegraphics[scale=0.45]{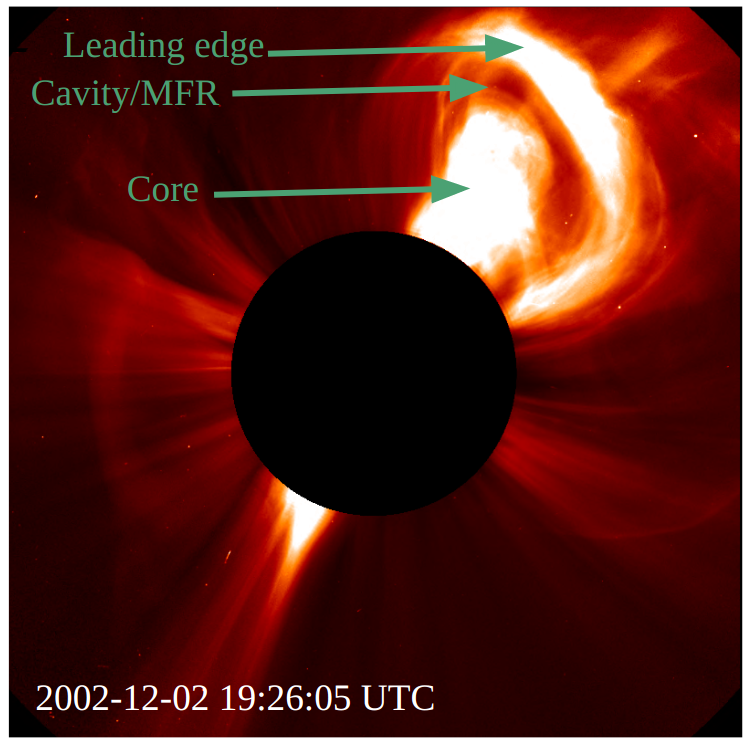}
    \caption[A classical three-part coronal mass ejection.]{A classical three-part coronal mass ejection. The image is from C2 coronagraph of LASCO onboard SOHO produced using {\it JHelioviewer} \citep{Helioviewer2017}.}
    \label{fig:cme_three_part}
\end{figure*}

Since the first space-based coronagraph observations \citep{Tousey1973,Koomen1975}, lots of progress has been made in understanding the different aspects of CMEs from remote sensing and in-situ measurements \citep[e.g.,][etc.]{rodriguez2006,Rodriguez2011,Mishra2021,Davies2021}, including the precursor of a CME, trigger mechanism, energy source, morphology, velocity, mass, propagation, and geo-effectiveness. However, there is still much more to understand about CMEs and their associated phenomena \citep{Chen2008}. For instance, it is well recognized that the coronal field changes to a more complex magnetic structure in a quasi-steady way and once a threshold is reached, this magnetic structure cannot sustain its equilibrium and begins to erupt. The details of what triggers the progenitor to transition from an equilibrium state to an eruptive state are yet to be understood \citep{MITTAL2010,Georgoulis2019}. Another open question is about the acceleration of CMEs during their early stages \citep{JamesChen2003,SURYANARAYANA20191} and the role of magnetic reconnection \citep{Wu2005} during this process. There are other open issues about the relationships between CMEs and accompanying phenomena -- solar flares \citep{YOUSSEF2012172,Kawabata_2018}, coronal dimmings \citep{Dissauer_2019}, transient coronal holes, etc. The most important open issue, from a space-weather perspective, is that of understanding CME evolution in the heliosphere and how it affects the prediction of their geomagnetic effectiveness \citep{Vourlidas2019,Ionescu2021}.

CMEs often show a three-part structure in white-light coronagraph images -- a bright leading edge, dark cavity, and a bright core \citep{Cremades2004,Vourlidas2013,Song_2022}. A classical three-part CME observed using the Large Angle Spectroscopic Coronagraph \citep[LASCO;][]{Brueckner1995} onboard the Solar and Heliospheric Observatory \citep[SOHO;][]{Domingo1995} is shown in Figure \ref{fig:cme_three_part} and the different parts are marked. Although the three-part structure is considered to be the standard morphology of CMEs, only $\sim$30\% of CMEs show all of these three-part structures \citep{Webb1987}. But, it has been shown that even CMEs that do not show a three-part structure in white-light coronagraph images have a three-part appearance in extreme-ultraviolet (EUV) images at lower coronal heights \citep{Song_2019}. Since K-corona \citep{Calbert1972} originates from the Thomson scattering of photospheric light from free electrons, the brightness of K-corona depends on the electron column density \citep{Hayes2001}. Hence, the appearance of a three-part structure of CMEs in coronagraphs implies a high-low-high electron density structure. These structures are explained as the observational manifestations of a high-density region due to background plasma pileup (leading edge), a low-density region of the twisted magnetic fields known as magnetic flux rope (MFR), and eruption of prominence (core) which again has a higher density. Efforts have been made to make a unified picture of the three-part structure of CME to explain different varieties of CMEs \citep{Vourlidas2013,Vourlidas2014,Vourlidas2017}.

CME speed, acceleration, and mass have been measured routinely using white-light coronagraph observations. Speeds of CMEs can vary from a few tens to a few thousand km/s. Depending on their speed, CMEs can take as little as a few tens of hours to a few days to reach the Earth. Measurement of CME mass requires the inversion of Thompson-scattered brightness to estimate the electron density \citep{Billings1966}. Mass of CMEs typically fall in the range of $\sim10^{10}-10^{13}$ kg \citep[e.g.,][]{Jackson1985,Vourlidas2002,Kahler2006,Vourlidas2010}. CME masses have also been estimated using radio \citep{Gopalswamy1992,Ramesh_2003}, X-ray \citep{Rust1976} and EUV observations \citep{Harrison2003,Aschwanden2009}. Besides all of these geometrical, dynamical, and thermodynamical parameters, another important parameter of CMEs is their magnetic fields.

\section{Importance of CME Magnetic Field Measurements}\label{sec:cme_mag}
The magnetic field is a crucial physical parameter of a CME, right from its eruption stage to its later stages. Although the exact mechanism of CME eruption is still not understood in detail, it is well-understood that magnetic field strength and topology is the key driver determining eruptions and characteristics of CMEs. The geo-effectiveness of a CME is determined primarily by its magnetic field strength and topology \citep{Vourlidas2019,Ionescu2021}. CME magnetic fields may get modified as it propagates through the corona and the heliosphere due to interactions with the coronal magnetic fields, IMF, and other heliospheric structures.

The geo-effectiveness of a CME is crucially determined by its southward component ($B_\mathrm{z}$) of the magnetic field, with respect to Earth's magnetic field. It is this $B_\mathrm{z}$ which reconnects with the northward component of the Earth's magnetic field causing major geomagnetic storms and injections of energetic charged particles into the Earth's atmosphere. Hence, to predict the geo-effectiveness of a CME, it is necessary to predict the $B_\mathrm{z}$ component of CME magnetic fields along with its other properties and the arrival time. However, there are several challenges which limit the accuracy of the prediction of vector magnetic field of the CMEs at 1 AU. These range from the impacts of interaction between CMEs and the background solar wind and other heliospheric structures to observations from very few vantage points, especially given the immense size of the CMEs at 1 AU with respect to the Earth \citep{Vourlidas2019,Ionescu2021}.

The size of a typical CME at 1 AU is  orders of magnitude larger than the Earth-Moon system. \cite{Mishra2021_radial_size} performed a statistical study of the CMEs in solar cycles 23 and 24, and found the radial size of CMEs at 1 AU span a range between 0.03 to 1.34 AU. This immense size of the CMEs at 1 AU is the ultimate physical challenge in the prediction of the geo-effectiveness of a CME \citep{Vourlidas2019}. For accurate predictions of the CME arrival time and/or magnetic field, one requires high-precision modeling of the 3D structure of the CME. Prediction of CME-$B_\mathrm{z}$ broadly consists of two steps -- i) estimating  3D CME magnetic field structures near the Sun and ii) propagating the near Sun measurements through the corona and the heliosphere to 1 AU. As stated earlier, the magnetic fields of CMEs evolve considerably during their propagation through the corona and the heliosphere due to interactions with other magnetized structures such as background solar wind, co-rotating interaction regions, stream interaction regions, etc. For a precise prediction of the $B_\mathrm{z}$ component of the CME magnetic field, it is necessary to improve the physics-based coronal and heliospheric models as well as constrain these models from observations of CME magnetic fields both at coronal and heliospheric heights.    

\section{Observations of CMEs at Multiple Wavelengths}\label{subsec:cme_observations}
CMEs and their associated phenomena can be observed at different wavelengths across the electromagnetic spectrum, which probe different properties of the CMEs. 

\subsection{Routine White-light Coronagraphic and EUV Observations}\label{subsec:whtie_light}
CMEs are routinely observed using both ground- and space-based coronagraphs. These coronagraphs measure the brightness of Thompson-scattered photospheric lights from the free electrons in the CME plasma \citep{Billings1966,Howard2009}. Hence, white-light images are mainly images of the density structure of the CMEs. Several observational techniques have been developed to estimate different physical properties of CMEs based on white-light coronagraph observations \citep{Webb2012}. Routine high-quality white-light coronagraph observations became available after the launch of LASCO in 1997. Multi-vantage point observations from LASCO along with coronagraph observations from other two spacecraft -- Solar Terrestrial Relations Observatory - Ahead (STEREO-A) and Behind (STEREO-B) \citep{Kaiser2008}, allow us to reconstruct the 3-dimensional (3D) geometry of CMEs. These three space-based coronagraphs improved our understanding of the geometry and dynamics of the CMEs using observations of  thousands of CMEs over the last two decades. 

However, white-light coronagraph observations cannot provide direct measurements of the magnetic fields of CMEs. Some in-direct techniques have been developed to measure the CME magnetic fields using EUV and white-light observations \citep[e.g.,][etc.]{Savani_etal_2015,Gopalswamy2017,Kilpua2021}. These methods use certain coronal features during or after the CME eruptions such as filament details, flare ribbons, post-eruption arcades, etc. at EUV bands and 3D reconstruction of MFR using multi-vantage point white-light observations to estimate magnetic fields at different parts of CMEs. The accuracy and applicability of these methods depend on the identification of the coronal features and the validity of the underlying assumptions. 

\subsection{Observations of CMEs at Radio Wavelengths}\label{subsec:radio_observation_cme}
Different parts of CMEs emit radio emissions via different emission mechanisms at different stages of their evolution. These include thermal free-free emission from CME plasma \citep[e.g.,][etc.]{Gopalswamy1992,Gopalswamy1993,Ramesh_2003,Ramesh2021}, coherent plasma emissions from CME shocks (type-II radio bursts) \citep[e.g.,][etc.]{Nelson1985,Gopalswamy_2000_typeII,Cairns2003,Gopalswamy2019_typeII,Jebaraj2021}, coherent plasma emissions from CME cores (type-IV radio bursts) \citep[e.g.,][etc.]{KRISHNAN1961,Ramaty1969_typeIV,Kumari2017_typeIV,morosan2019} and gyrosynchrotron (GS) emission from CME plasma \citep[e.g.,][etc.]{bastian2001,Tun2013,Bain2014,Carley2017,Mondal2020a}.

All of these emissions carry imprints of either the source region magnetic field or the magnetic field of the medium of propagation, or both. Hence, they provide some information of CME magnetic fields. One of the biggest advantages of observations at radio wavelengths is that most of these radio observables can provide an estimation of magnetic fields at different parts of the CMEs, which are otherwise unavailable \citep{Vourlidas2020,Carley2020}. There are some other indirect observing techniques at radio wavelengths  using background galactic/extra-galactic radio sources that can also be used to study CMEs at heliospheric heights. In the following Sections \ref{subsec:direct_radio_obs} and \ref{subsec:indirect_radio_obs}, I discuss these observing techniques at radio wavelengths which can be used to study different properties of CMEs. 

\subsubsection{Direct Observables}\label{subsec:direct_radio_obs}
\begin{figure*}[!ht]
    \centering
    \includegraphics[scale=0.5]{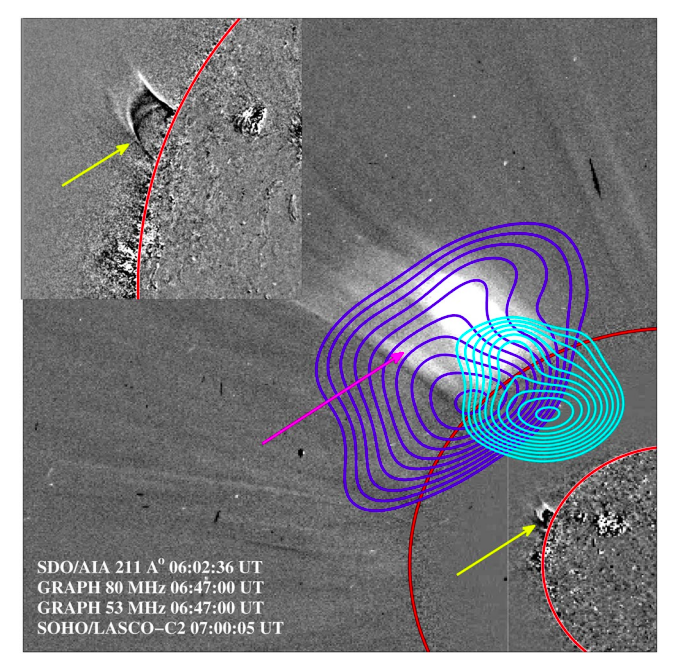}
    \caption[Thermal radio emission from a CME.]{Thermal radio emission from a CME. A composite difference image using observations at EUV, white light and radio wavelengths is shown. The cyan and blue contours correspond to radio emission at 80 and 53 MHz, respectively (Reproduced from \cite{Ramesh2021} with the permission from publisher).}
    \label{fig:cme_thermal}
\end{figure*}

\begin{figure*}[!ht]
    \centering
    \includegraphics[scale=0.6]{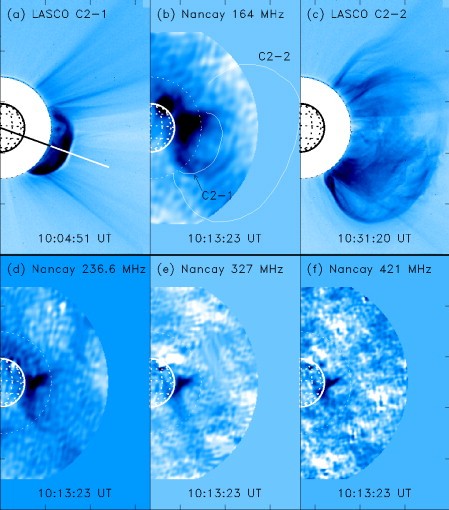}
    \caption[Gyrosynchrotron radio emission from CME loops.]{Gyrosynchrotron radio emission from CME loops. First and third columns of the first row show LASCO C2 white-light coronagraph images and other panels show radio images at different frequencies using data from from the Nan\c{c}ay Radio Heliograph (Reproduced from \cite{bastian2001} with permission from the author).}
    \label{fig:radio_cme_bastian}
\end{figure*}

\begin{figure*}[!ht]
    \centering
    \includegraphics[scale=0.8]{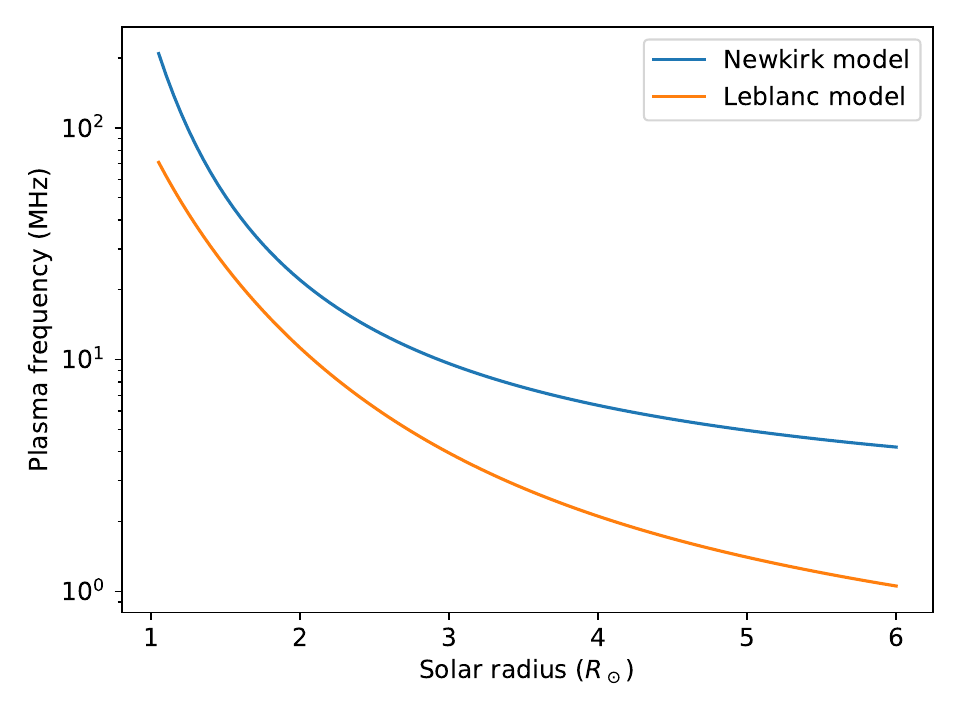}
    \caption[Variation of plasma frequency with coronal heights.]{Variation of plasma frequency with coronal heights. The blue line shows the variation of plasma frequency considering Newkirk coronal density model \citep{Newkirk1961,Newkirk1967} and the orange line shows for Leblanc coronal density model \citep{Leblanc1998}.}
    \label{fig:height_vs_vp}
\end{figure*}

There are multiple types of radio emissions produced from CMEs -- 
\begin{enumerate}
        \item {\bf Radio bursts: } Multiple types of solar radio bursts are related to CMEs, such as type-I and type-III radio bursts, which are associated with post-eruption or pre-eruption particle acceleration phenomena. 
        \begin{itemize}
        \begin{figure*}[!ht]
            \centering
            \includegraphics[trim={0cm 0cm 5cm 0cm},clip,scale=0.27]{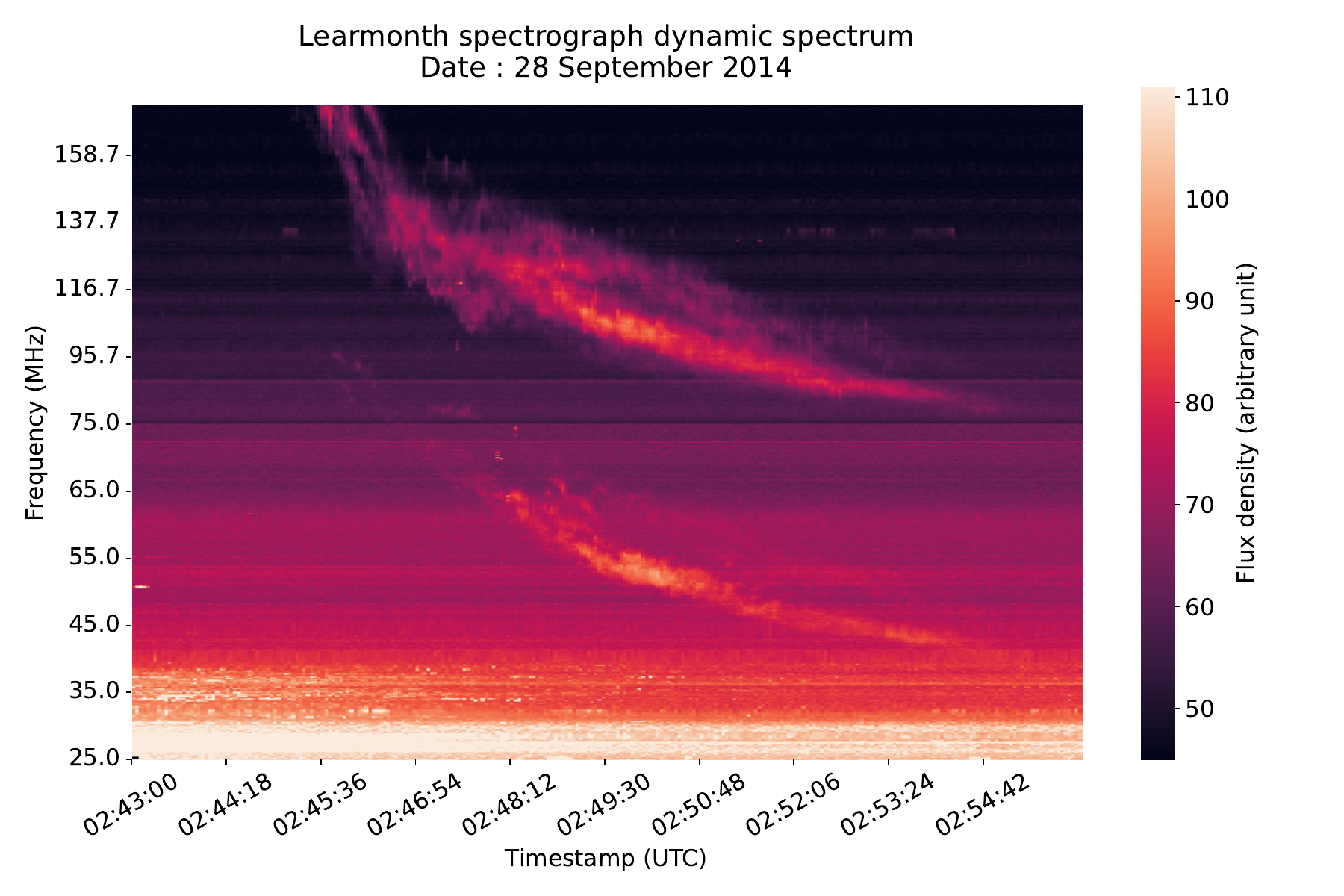}\includegraphics[trim={0cm 0cm 5cm 0cm},clip,scale=0.27]{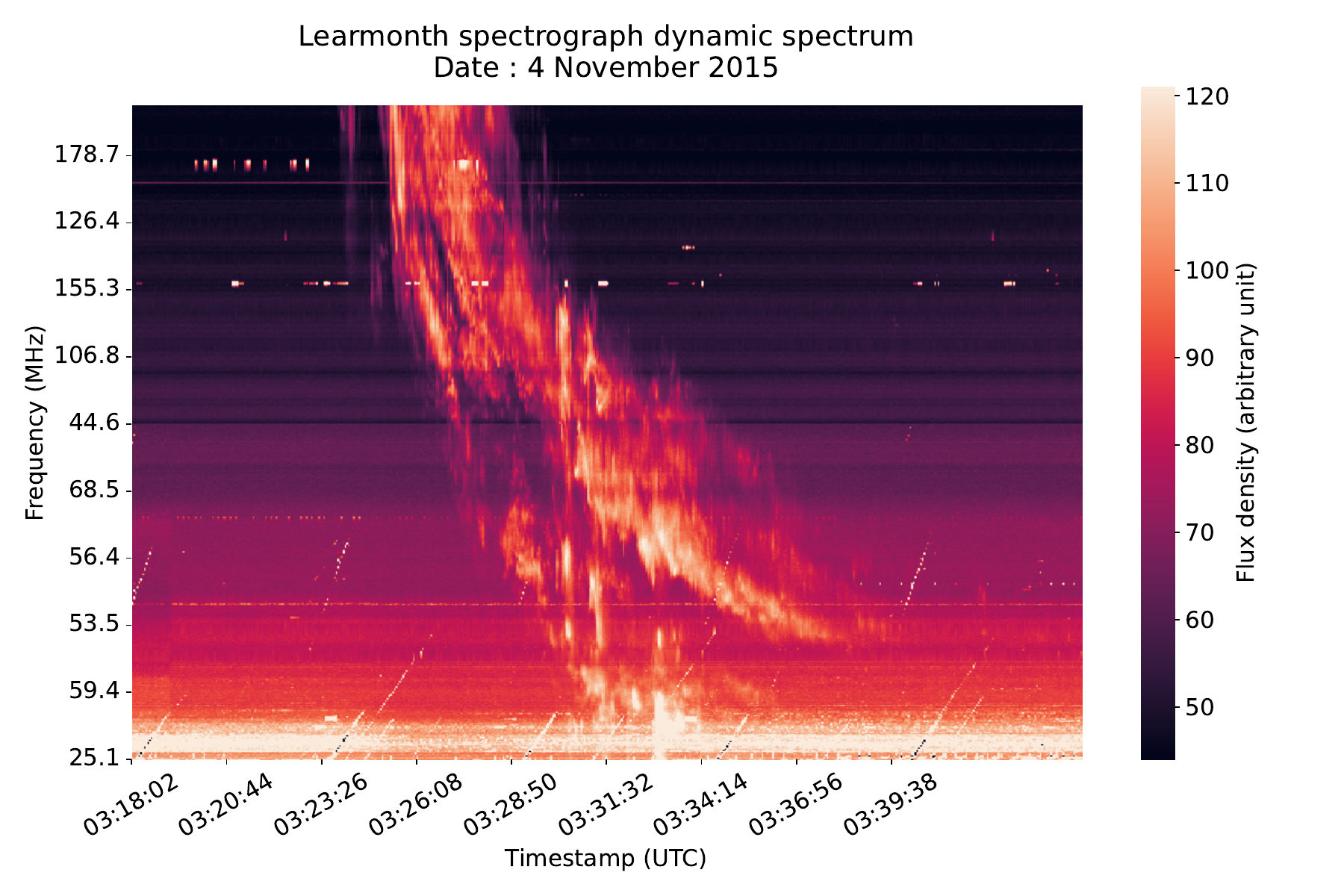}
            \caption[Sample dynamic spectra of two type-II radio bursts.]{Two sample dynamic spectra of type-II radio bursts observed using Learmonth solar radio spectrograph. {\it Left panel:}  A type-II radio burst observed on 2014 September 28 shows two lanes at the fundamental and harmonic of the local plasma frequency. {\it Right panel: } A type-II radio burst observed on 2015 November 4 shows multiple lanes.}
            \label{fig:type-II}
        \end{figure*}
        \item \textbf{\textit{Type-II radio bursts: }} Solar type-II radio bursts are produced by a coherent plasma emission mechanism when accelerated electrons at the shocks of CMEs propagate through the background plasma. Type-II bursts are identified by their slowly drifting features in the dynamic spectrum. Two sample dynamic spectra of type-II radio bursts are shown in Figure \ref{fig:type-II}. Band-splitting of type-II radio bursts \citep[e.g.,][]{Smerd1975,Vasanth2014,Hariharan2015,Kumari2017typeII_band,MAHROUS201875} and circular polarization measurements \citep[e.g.,][]{Ramesh_2022,Ramesh_2023} can provide estimates of magnetic field strength at the shock front of CMEs. 
        \begin{figure*}[!ht]
            \centering
            \includegraphics[scale=0.45]{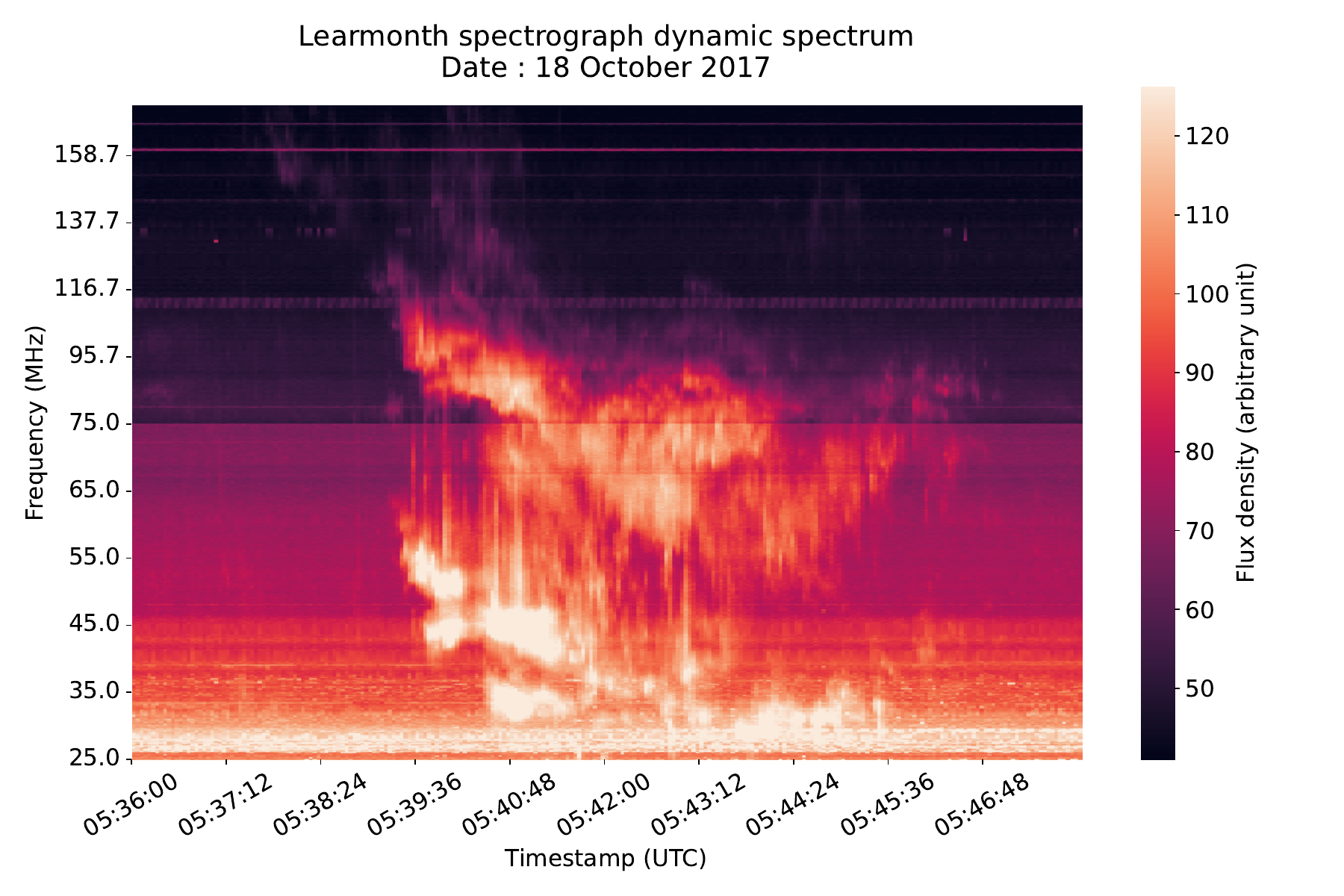}
            \caption[Sample dynamic spectrum of a type-IV radio burst.]{Sample dynamic spectrum of a type-IV radio burst occurred on 2017 October 18. The dynamic spectrum is obtained from the Learmonth solar radio spectrograph.}
            \label{fig:type-IV}
        \end{figure*}
        \item \textbf{\textit{Type-IV radio bursts: }}Type-IV radio bursts are classified as a broadband long duration and comparatively bright radio source in the dynamic spectrum \citep{Boischot1957}. Type-IV radio bursts can be divided into two sub-classes -- moving type-IV bursts (IVm), and stationary type-IV bursts (IVs). Type-IVm bursts show frequency drifts, while type-IVs do not. This implies that the radio source of type-IVm moves outward from the Sun and is associated with eruptive phenomena like CMEs, while type-IVs are stationary. A sample dynamic spectrum of a type-IVm radio burst is shown in Figure \ref{fig:type-IV}. Although it is easy to identify them in the dynamic spectrum, their emission mechanism shows variability \citep{morosan2019}. Some studies identified them as GS emissions by the mildly-relativistic electrons trapped in CME magnetic fields \citep[e.g.,][]{Boischot1968,Dulk1973}, while some others identified them as plasma emissions \citep[e.g.,][etc.]{Weiss1963,Gary1985}. \cite{morosan2019} showed that the emission mechanism can change throughout the type-IV burst as well. Spectropolarimetric modeling of type-IV radio bursts can be used to measure the magnetic field strength of CME cores \citep[e.g.,][]{Raja2014,Kumari2017_typeIV,Carley2017,Kumari2022}.
        \end{itemize}
        \item {\bf Thermal free-free emission: }Thermal emission is produced by free electrons present in the CME plasma. This emission has a much lower brightness temperature. Hence there are only a handful of studies that  have claimed to detect thermal radio emission from CMEs \citep{Gopalswamy1992,Gopalswamy1993,Ramesh2021}. A sample dual frequency difference image of thermal emission from a CME is shown in Figure \ref{fig:cme_thermal}. Thermal emission has been used to measure the mass of the CME \citep{Gopalswamy1992} and induced circular polarization has been used to measure the magnetic field of the CME plasma \citep{Ramesh2021}. 
        \item {\bf Gyrosynchrotron (GS) emission: }GS emission is produced by mildly-relativistic electrons trapped in CME magnetic fields. While in some instances type-IV radio bursts are found to be produced by GS emission, there is another type of potentially GS emission having morphology similar to the CME structures seen in white-light images. These emissions are generally faint and were first detected by \cite{bastian2001} (Figure \ref{fig:radio_cme_bastian}) who gave them the name ``radio CME". This was followed by only a handful of other successful imaging detections \citep{Maia2007,Mondal2020a}, where radio emissions trace the white-light morphology. As already mentioned, GS emission from CME cores has also been detected and modeled. While there is no consensus yet on what types of GS emissions associated with CMEs should be referred to as ``radio CME" \citep{Vourlidas2020}, GS emissions from different parts of a CME do provide a unique observational tool to estimate CME magnetic field and other plasma parameters.   
\end{enumerate}
It is expected that both the bright coherent emissions from CME shock or core and the much fainter incoherent emissions from CME plasma will be present simultaneously. Hence, to use these observables for estimating magnetic fields and other plasma parameters of the CME and the surrounding corona, one needs to detect both of these emissions simultaneously in the image plane. This requires high dynamic range (DR) spectropolarimetric imaging at radio wavelengths, which has now become possible using the new generation ground-based radio telescopes. Due to the ionospheric cutoff, ground-based radio telescopes can only observe at frequencies above $\sim$10 MHz, which corresponds to a coronal height of $\sim2\ R_\odot$ (Figure \ref{fig:height_vs_vp}). Hence, coherent plasma emissions can only be used to study the CME-shock and core  below $\sim 2\ R_\odot$, while incoherent emissions can be detected at much larger coronal heights, $\sim10\ R_\odot$ using the high DR images from the new-generation ground-based radio interferometers. 

Type-II bursts are also observed in interplanetary space at frequencies down to a few kHz using radio spectrometers onboard different space-based observatories. These non-imaging devices can provide spectro-temporal information, but no spatial information. Hence, it is difficult to localize a radio source and estimate its size using such instruments. Currently, no space-based radio imaging instrument is available. However, a mission comprising six CubeSats aimed at space-based interferometric imaging of solar type-II radio bursts is currently in an advanced stage of development. This mission is named the Sun Radio Interferometer Space Experiment \citep[SunRISE;][]{Kasper2019}, and it will perform radio interferometric imaging in the frequency range 0.1 -- 25 MHz. Combining observations from the SunRISE mission with ground-based observations will provide crucial information on CME shocks from $2\ R_\odot$ to $\sim20\ R_\odot$.

\subsubsection{Indirect Observables}\label{subsec:indirect_radio_obs}
It has already been argued that to use direct radio observations to estimate different plasma parameters of CMEs, including their magnetic field, one needs high DR spectropolarimetric imaging. At present, this is only possible from ground-based radio observations, which limits the use of direct methods up to $\sim10\ R_\odot$. Beyond 10 $R_\odot$, in the outer coronal regions and heliosphere, several CME plasma parameters can be measured using the following two in-direct methods:
\begin{enumerate}
    \item {\bf Interplanetary scintillation: }The plasma density fluctuations ($\Delta N$) in the turbulent solar wind and CMEs can be measured through a phenomenon known as ``Interplanetary scintillation (IPS)", which was first reported by \cite{HSW64}. IPS is the radio analog of the optical twinkling of stars which is caused by turbulent density fluctuations in Earth's atmosphere. In the past few decades, increasingly sophisticated algorithms have been developed for 3D reconstruction of global heliospheric parameters; density, and velocity, using observations of multiple IPS radio sources covering the entire heliosphere \citep[e.g.,][]{Jackson1997, Jackson1998}. These reconstructions have been extended to include data from IPS stations across the world, observations from heliospheric imagers, and also to incorporate MHD models, which can now also provide a reconstruction of the heliospheric magnetic fields \citep{Jackson2020}. This method has recently been used for detailed modeling of a CME \citep{Iwai2022}.
    \item {\bf Faraday-rotation measurements: }When linearly polarized radiation passes through a magnetized plasma, its plane of polarization rotates. This phenomenon is known as Faraday rotation (FR). The amount of FR is proportional to the line-of-sight (LoS) integral of the product of electron density and LoS component of the magnetic field and also to the squared of the wavelength of observation. The wavelength-independent part of FR is expressed in terms of a quantity referred to as the Rotation Measure \citep[RM;][]{Brentjens2005}. FR measurements of linearly polarized emission through CMEs \citep[e.g.,][]{Bird-CME-FR-1985,kooi2017,Kooi2021} offer a promising remote-sensing tool to measure the vector magnetic field in CMEs both at the corona and the inner heliosphere \citep[e.g.,][]{Vourlidas2020}. Recent developments about FR measurements due to heliospheric plasma are discussed in \cite{Kooi2022}.
\end{enumerate}

A limitation shared by both these indirect observables is that they rely on the presence of suitable background sources. To constrain the CME properties one requires a large number of pierce points through the CME. This, however, depends upon a variety of factors ranging from the sensitivity and FoV of the instrument being used, to the relevant background sources available in the patch of the sky where these measurements need to be made. In addition, like any other remote sensing observable, these measurements are also integrals along the LoS.

Combining both the direct and the indirect methods using ground-based radio telescopes along with observations using coronagraphs and heliospheric imagers can provide remote measurements of CME plasma parameters including its magnetic field. For precise prediction of CME evolution, it is important to understand both coronal and heliospheric models of CMEs and constrain them well using observations. Although all of these observing methods seem promising, there are several challenges to overcome, in terms of observation and in estimating the final plasma parameters from the observation, for each of them. Of these possible methods, I will focus on developing the required observing capabilities for routine observations of GS emission from CME plasma and providing a robust estimation of CME plasma parameters using spectropolarimetric modeling of the observed spatially resolved GS spectra. 

\section{Current Status: Observation and Modeling of Gyrosynchrotron Emission from CMEs}\label{sec:current_status}
GS emission from  the then unknown CMEs was first reported in the 1960s \citep{Boischot1957,Boischot1968,Dulk1973}, and was regarded as type-IVm bursts. Faint GS emission from CME loops was first imaged by \cite{bastian2001} using spectroscopic imaging observations with the NRH. While the bright type-IV bursts could be detected using non-imaging instruments, detecting faint GS emissions from CME loops needs imaging observations. Since the first detection and spatially resolved modeling, there have been only a handful of imaging studies that have imaged GS emission from CME plasma \citep{Maia2007,Mondal2020a}. This scenario started to change with the availability of high DR spectroscopic imaging using data from the new generation instruments. These data allowed one to detect fainter GS emissions even from slow and unremarkable CMEs and estimate CME plasma parameters using spatially resolved spectroscopy \citep{Mondal2020a}.

Over the last decade, several developments have drastically reduced the computational effort required to produce GS spectra, even though the level of sophistication of the models has grown, in terms of using more realistic electron distribution and physical geometry. These fast GS codes \citep{Fleishman_2010,Kuznetsov_2021} allow one to perform more realistic modeling of GS emission. On the observational front, the limitations of the earlier studies came from limited spectral coverage, limited availability of imaging observations, and unavailability of reliable polarimetric measurements. Hence, earlier studies had no choice but to rely on several assumptions to estimate CME plasma parameters from observations. While those studies provided useful ballpark estimates of magnetic fields and other parameters, the uncertainties associated with these estimates naturally depended on the validity of the assumptions made. To use these estimates to constrain CME models at coronal heights and to serve as inputs for the heliospheric models, one needs to verify these assumptions and/or improve the robustness of estimated parameters by including more observational constraints (spectropolarimetric imaging, wider and denser spectral sampling, complementary information from other wavelengths, etc.) and removing the underlying assumptions of the models.  

\section{Challenges}\label{sec:challenges}
GS emission from CME plasma is a promising, but challenging, observational technique that can provide spatially resolved estimates of CME magnetic fields and other plasma parameters at coronal heights. However, to use GS emission to estimate spatially resolved CME plasma parameters, one has to overcome several challenges. These can be classified into two categories -- observational and modeling. 

\subsection{Observational Challenges}\label{subsec:observational_challenge}
GS emission produced from CME plasma is generally faint, when compared to quiet Sun thermal mission, except when it is the emission mechanism behind the associated type-IV bursts. The brightness temperature of these emissions lies in the range $\sim10^3-10^4$ K, while quiet Sun brightness temperature is $\sim10^6$ K. Very often, these emissions are accompanied by much brighter active emissions from different types of radio bursts. Hence, their presence cannot be detected in the dynamic spectrum which only provides flux density integrated over the entire Sun. Since the solar emissions at meter-wavelength are usually highly time variable, one can not average over long temporal spans. To improve the robustness of estimated model parameters, polarization measurements are necessary. Hence, to detect these faint GS emissions even in the presence of bright active emissions and use them for robust estimation of CME plasma parameters, high DR spectropolarimetric snapshot solar radio imaging is required.

Solar radio imaging observations at meter-wavelengths have been done for decades using several radio interferometers across the globe -- some of them are dedicated to solar observations, e.g., Culgoora Radio Heliograph \citep{wild_1967}, Clark Lake Multifrequency Radioheliograph \citep{Kundu1983}, Nan\c{c}ay Radio Heliograph \citep[NRH;][]{bonmartin1983,avignon1989}, the Gauribidanur Radio Heliograph \citep[GRH;][]{sundaram2004,sundaram2005}, etc. and some of them are general purposes radio interferometers, e.g., the Very Large Array \citep[VLA;][]{Thompson1980,VLA2009}, the Giant Metrewave Radio Telescope \citep[GMRT;][]{Swarup_1991,Gupta_2017}, etc. Some new generation instruments which have started observing the Sun in the last decade are -- the Murchison Widefield Array \citep[MWA;][]{lonsdale2009,Tingay2013,Wayth2018}, the LOw Frequency ARray \citep[LOFAR;][]{lofar2013}, NenuFAR \citep{Zarka2018,Briand2022}, the Owens Valley Long Wavelength Array \citep[OVRO-LWA;][]{Hallinan2023}, etc. Among all of these instruments, the MWA is exceptionally well-suited for high DR spectroscopic snapshot imaging of the Sun with good spatial resolution. 

Although MWA is intrinsically capable of producing high DR spectroscopic snapshot solar images, several challenges need to be overcome for solar observations. The MWA is a wide field-of-view (FoV) \citep{Tingay2013,neben2015} aperture array \citep{Farhat2014} instrument. It does not have any moving parts to enable it to point at and track a specific source in the sky. Instead, it uses electronic delays to point to a particular direction in the sky. These two characteristics, coupled with the extremely high flux density of the Sun make the standard calibration methods for spectropolarimetry not applicable for solar observation with the MWA. The challenges and differences are as follows:
\begin{enumerate}
    \item Based on the full-width half-maximum (FWHM) of the primary beam, at 150 {$\mathrm{MHz}$} the FoV of the MWA is $\sim$610 $\mathrm{degree^2}$ \citep{Tingay2013}. Being an aperture array instrument, the primary beam sidelobes of the MWA are high; $\sim10\%$ of the peak \citep{Sokolwski2017,Line2018}. Hence, given the very high solar flux density, any astronomical calibrator observations during the daytime are contaminated by solar emission.
    \item Hence, at the MWA, calibrators are routinely observed either before sunrise or after sunset and used to determine antenna gains of the array. 
    \item The large time gap between the calibrator and the solar observations, in addition to the difference in the pointing direction, reduces the ability to constrain the true state of the instrument and the ionosphere using nighttime calibrator observations. 
    \item Polarization response of the MWA depends on the pointing direction in the sky. Hence, polarization calibration obtained using calibrator sources does not apply to the solar observation made using a different pointing. 
\end{enumerate}
Moreover, during the initial phase of its operation (2013 to 2015), most of the solar observations with the MWA did not have a dedicated nighttime calibrator observation with the same spectral configuration as used for solar observation. Hence, calibrating these observations involves several non-trivial steps. By its very nature, snapshot spectropolarimetric imaging with the modern high frequency and time resolution instruments leads to tens of thousands to millions of images even for minutes of observation. Generating these images in the traditional human effort-intensive manner is simply infeasible and error-prone. Additionally, the large information content of the MWA data is accompanied by a large increase in raw data volumes and the process of interferometric imaging is inherently iterative and compute-intensive. All of these aspects bring their own set of challenges that need to be met to enable routine spectropolarimetric detection of GS emission from CME plasma.

\subsection{Modeling Challenges}\label{subsec:modeling_challenges}
The next set of challenges lies in using GS emission for the estimation of CME plasma parameters at coronal heights via robust modeling of the observed GS spectrum. Fast GS codes \citep{Fleishman_2010,Kuznetsov_2021} allow one to consider either analytical or numerically derived electron energy and pitch-angle distributions to produce GS emission. But, the GS emission model for even the simplest electron energy distribution has ten free parameters \citep{Mondal2020a}, and some of them are degenerate. Hence, it is not possible to estimate all of these GS model parameters unambiguously just by using the total intensity (Stokes I) spectrum. The inclusion of circular polarization (Stokes V) measurements can significantly improve the robustness of GS model parameter estimation. It is well-known that CMEs comprise large-scale magnetized structures showing inhomogeneity in terms of density and magnetic field \citep{Mishra2015,Owens2017,Song2021}. Due to limited observational constraints available, however, all earlier studies have assumed homogeneous distributions of the plasma parameters along the LoS. In fact, none of the earlier studies have attempted to explore and quantify the error introduced due to the assumption of homogeneity on the estimated GS model parameters. Clearly, for a realistic and robust estimation of CME plasma parameters, it is important to study the impacts of these widely used assumptions on GS models.

\section{Objective of the Thesis}\label{sec:thesis_goals}
The objective of this thesis is to make use of the recent developments to push the boundaries for both the observational and modeling challenges and overcome them to the extent possible. On the observational front, I use data from the MWA, a new generation instrument that offers the densest monochromatic snapshot sampling in the Fourier domain. The intrinsic capability of these data had already been demonstrated \citep{Oberoi2023}, primarily using the output from a robust unsupervised interferometric imaging pipeline \citep{Mondal2019}, though this pipeline was limited to Stokes I imaging.

The first objective of this thesis is to develop a robust and unsupervised polarimetric calibration and imaging algorithm to tackle the challenges mentioned in Section \ref{subsec:observational_challenge}. This algorithm has been implemented as a robust software pipeline that can produce high DR and high-fidelity spectropolarimetric snapshot solar radio images using the MWA observations. Along with high DR imaging, one also needs precise calibration of flux density in physical units. To achieve that I have developed a unique and robust solar flux density calibration technique. 

The second objective of this thesis is to use this pipeline to perform spectropolarimetric imaging of GS emissions from CMEs and undertake a detailed study to understand the nature of GS spectra and their dependence on physical parameters of the CMEs, the degeneracies between different parameters, and examine the limitations arising from assumptions routinely relied upon in the simple GS models which have been used for estimating CME plasma parameters from the observation. 

This thesis is divided into several chapters, each discussing a specific aspect of the work done toward achieving the goals of the thesis. Chapter \ref{mwa} describes the MWA and its architecture in some detail. The working principle behind the calibration algorithm for MWA solar imaging is discussed in Chapter \ref{paircars_principle}. A detailed description of the polarization calibration algorithm is presented in Chapter \ref{paircars_algorithm} followed by a description of the robust solar flux density calibration algorithm in Chapter \ref{fluxcal}. These calibration and imaging algorithms have been implemented seamlessly in a robust and unsupervised software pipeline. Implementation details and architecture of this pipeline are presented in Chapter \ref{paircars_implementation}. GS emissions associated with two CMEs have been studied as a part of this thesis. Chapter \ref{cme_gs1} describes a detailed spectropolarimetric modeling study of GS emission from one of the CME. Chapter \ref{cme_gs2} describes a detailed spectropolarimetric modeling study of the GS emission from a CME-streamer interaction and possible evidence of insufficiency of homogeneous GS models. I end the thesis with a brief discussion of the conclusions,  current limitations and future works in Chapter \ref{conclusion}.  
\chapter {Murchison Widefield Array}
\label{mwa}
The Murchison Widefield Array \citep[MWA,][]{lonsdale2009} is a radio interferometer and a precursor of the Square Kilometre Array Observatory \citep[SKAO,][]{SKAO2021}. An instrument is designated to be a SKA-precursor if it demonstrates (some aspects of) the technology and the science of the future SKAO and is located at one of the sites chosen for the SKAO telescopes. MWA is located at the Murchison Radio-astronomy Observatory (MRO) in Western Australia, where the low-frequency array of the SKAO is currently being built. The MRO is a protected radio-quiet zone, to ensure an extremely low level of human-made radio frequency interference (RFI). MWA started its journey with an engineering prototype comprised of 32 elements, referred to as the ``MWA-32T" system. The full MWA, comprising 128 tiles, was commissioned in mid-2013. The array has gone through a few phases of development and upgrades. The array deployed in 2012/2013 is now called the ``MWA Phase-I" \citep{Tingay2013}. The array underwent a significant upgrade in 2018, which is referred to as the ``MWA Phase-II" \citep{Wayth2018}. The MWA is the workhorse instrument for this thesis. This chapter discusses the details of the MWA architecture relevant to this work.
\begin{figure}
    \centering
    \includegraphics[scale=0.75]{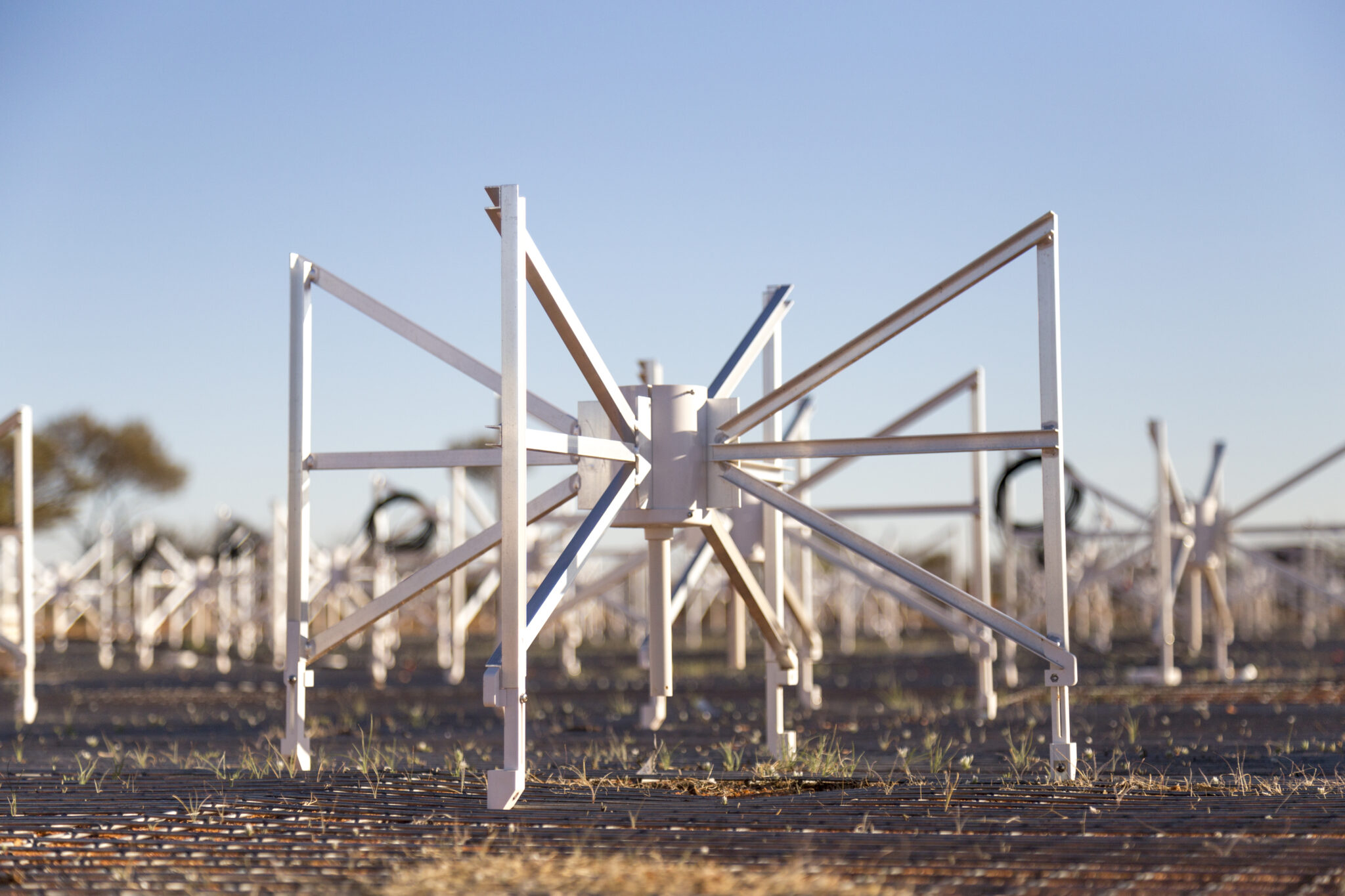}
    \caption[MWA bowtie dipole.]{One of the MWA bowtie dipoles (Image credit: ICRAR/Curtin).}
    \label{fig:mwa_dipole}
\end{figure}

\section{MWA Architecture}\label{sec:mwa_architechture}
\begin{figure*}[!ht]
    \centering
    \includegraphics[scale=0.6]{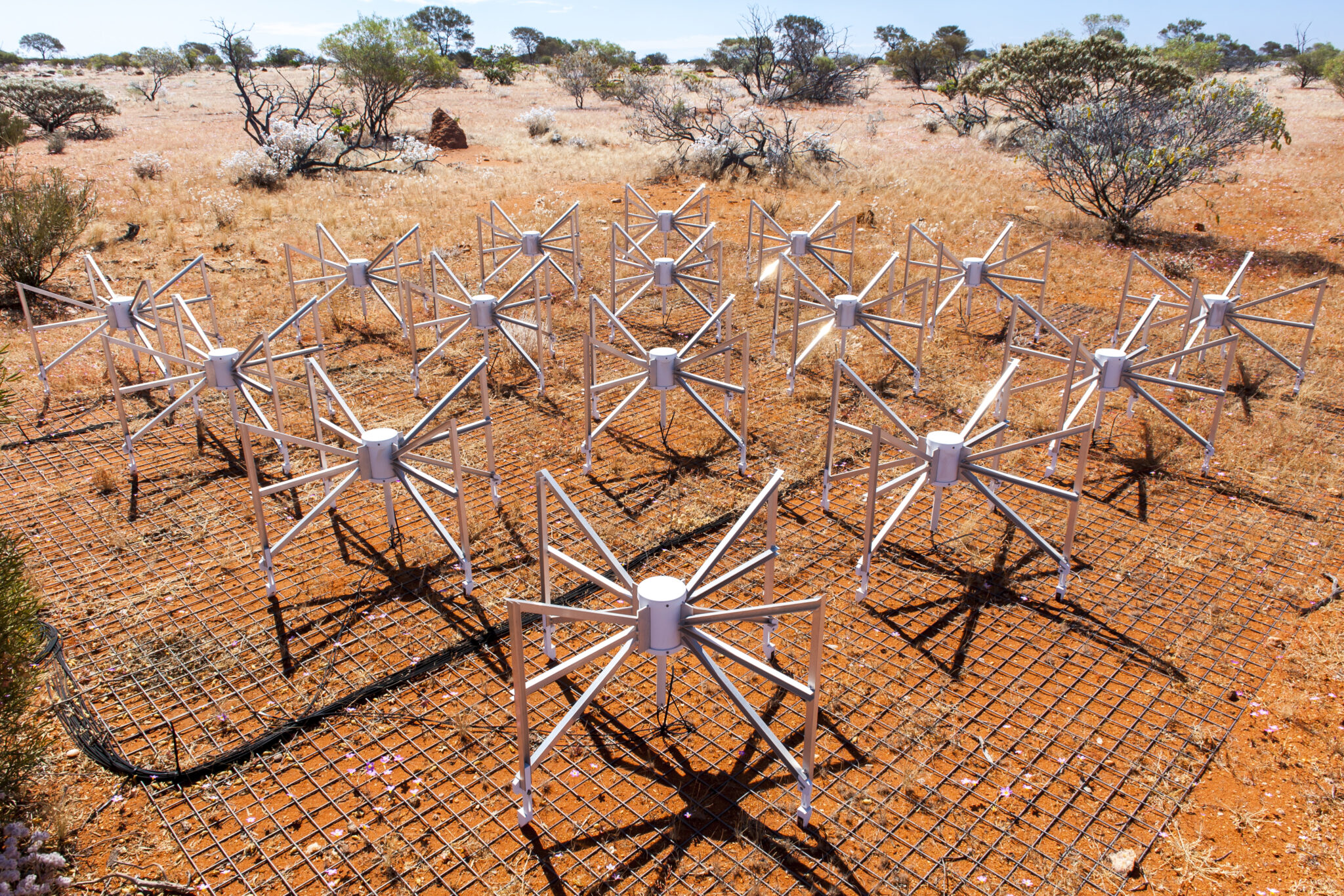}
    \caption[MWA antenna tile.]{One of the MWA tiles consists of sixteen bowtie dipoles in a 4$\times$4 grid (Image credit: Marianne Annereau, 2015).}
    \label{fig:mwa_tile}
\end{figure*}

The MWA is an aperture-array instrument \citep{Farhat2014} and does not have any moving parts, quite unlike the usual arrays with parabolic dish elements. Instead, it consists of large numbers of dual-polarization bowtie dipoles. One such dual polarization dipole is shown in Figure \ref{fig:mwa_dipole}. Each of these dipoles receives sky signals from a very wide field of view over a broad frequency range for each of the polarizations. The MWA signal path starts with these bowtie dipoles. Sixteen of them are arranged in a 4$\times$4 grid with 1.1 meter spacing. Each of these 4$\times$4 grids is called a ``tile". One such tile is shown in Figure \ref{fig:mwa_tile}.  Each of these tiles can be pointed to different parts of the sky by adding suitable delays to the signals from its sixteen dipole elements, which is accomplished independently for each of the two polarizations via an {\it analog beamformer}. Thus, despite having no moving parts, the analog beamformer allows the MWA tiles to steer their beam to different parts of the sky. 

\subsection{Array Configuration of the MWA}\label{subsec:array_config}
\begin{figure}
    \centering
    \includegraphics[trim={0cm 0cm 0cm 0cm},clip,scale=0.6]{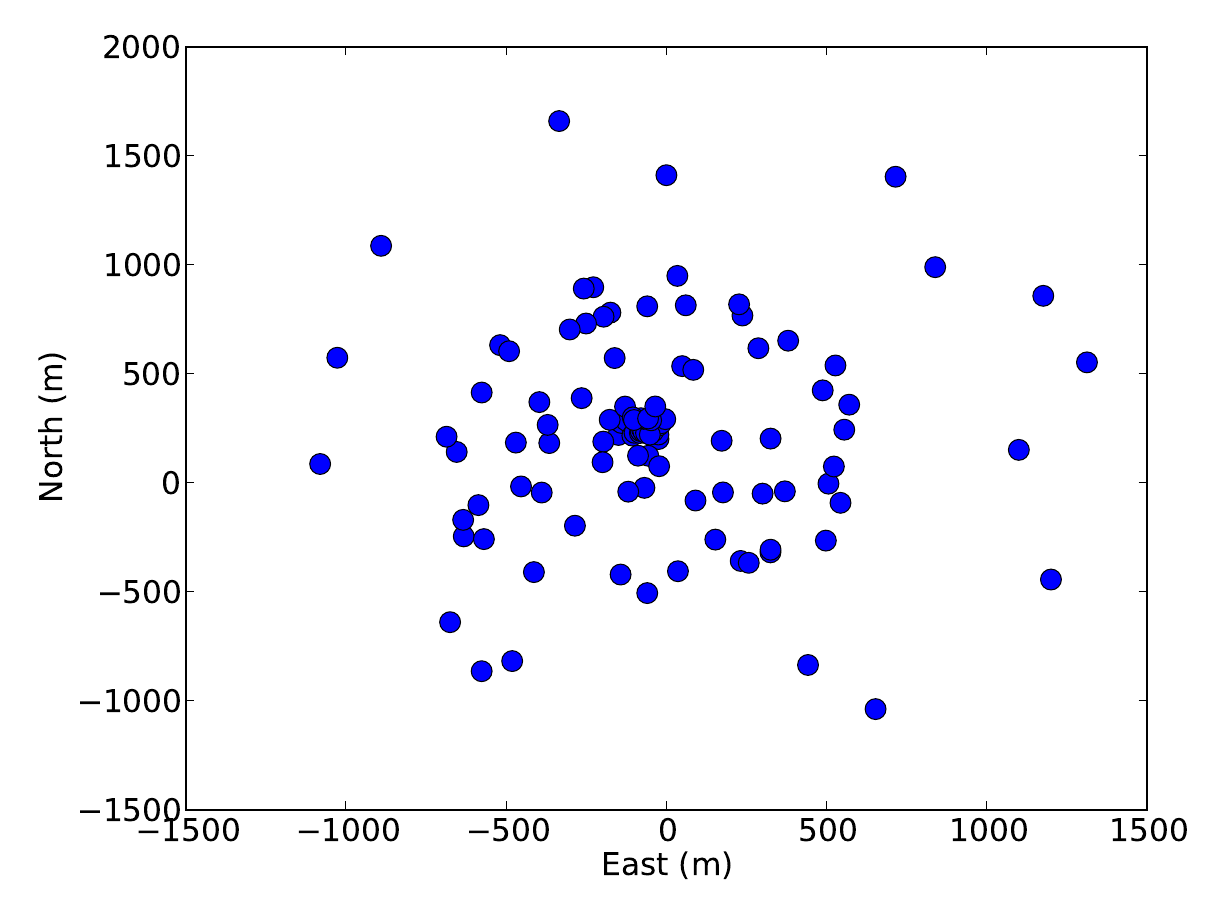}
    \caption[MWA Phase-I array configuration.]{MWA Phase-I array configuration.}
    \label{fig:phaseI_array}
\end{figure}

MWA is a 128-elements (currently 144-elements and growing) radio interferometric array, where each antenna tile serves as a single antenna element. The array configuration of the MWA Phase-I has a dense core of 100 meters in diameter which has 25\% of 128 tiles. The remaining tiles are distributed in a smooth distribution out to 1.5-km diameter, and 16 tiles are placed in an outer region of 3 km diameter. The Phase-I array configuration is shown in Figure \ref{fig:phaseI_array}. 

\begin{figure}
    \centering
    \includegraphics[trim={0cm 0cm 0cm 0.6cm},clip,scale=0.9]{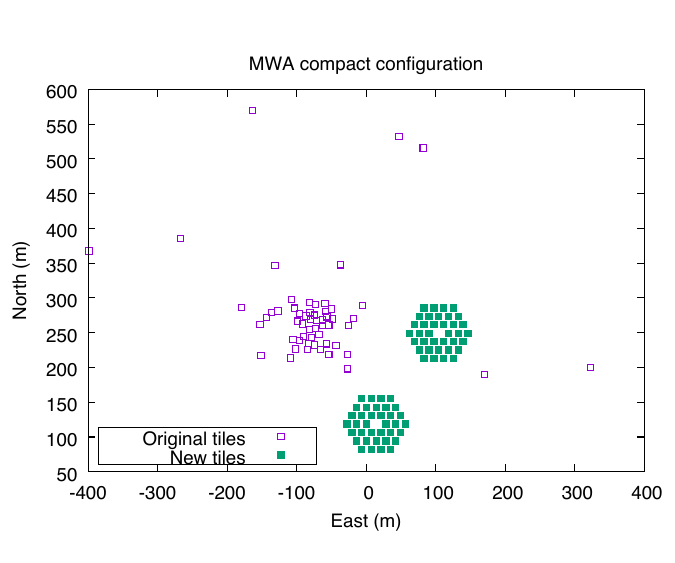} \includegraphics[trim={0cm 0cm 0cm 0.5cm},clip,scale=0.9]{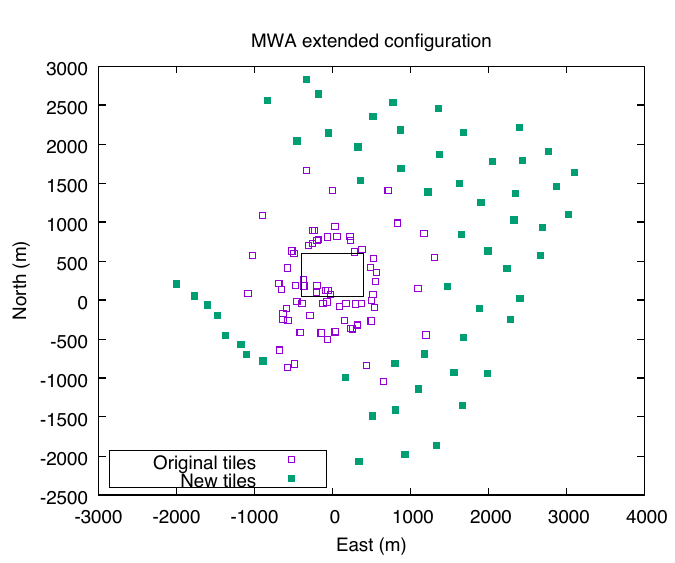}
    \caption[MWA Phase-II array configuration.]{MWA Phase-II array configurations. {\it Top panel: }Distribution of antenna tiles of Phase-II compact configuration. {\it Bottom panel: }Distribution of antenna tiles of Phase-II extended configuration. Magenta boxes represent the tiles from Phase-I and green filled boxes represent the new tiles added during Phase-II upgrade (Reproduced from \citet{Wayth2018} with permission from journal).}
    \label{fig:phaseII_array}
\end{figure}

Another 128 tiles were added to the array during the MWA Phase-II upgrade. Although the total number of tiles increased to 256, limitations of the backend instrumentation permitted only 128 of them to be used at any given time. 
Hence, the MWA Phase-II array provides two different observing configurations -- compact and extended. The compact configuration uses the core of the Phase-I array along with 72 new tiles arranged in two hexagons. The distribution of tiles for phase-II compact configuration is shown in the top panel of Figure \ref{fig:phaseII_array} and the maximum spread of the array is $\sim700$ m. Phase-II extended configuration uses Phase-I tiles outside the dense core and newly added antenna tiles and extends the longest baseline of the array to $\sim5$ km. The distribution of antenna tiles of the Phase-II extended configuration is shown in the bottom panel of Figure \ref{fig:phaseII_array}.


MWA Phase-I array has both -- a dense core and long baselines up to 3 km, which provide an excellent sampling of the Fourier plane of the sky brightness distribution (including at large angular scales) as well as good spatial resolution. This is well-matched with the need for high dynamic range spectroscopic snapshot solar imaging. Among the two Phase-II configurations, owing to its short footprint, the compact configuration offers too coarse a resolution to be of interest for solar imaging. Although the Phase-II extended configuration does not include the dense core, it has a sufficient number of short baselines as well as long baselines needed for solar observations. Hence, MWA Phase-I and Phase-II extended configurations are the best array configurations for solar imaging with the MWA. In this thesis, I have used observation done using the MWA Phase-I.

\section{Signal Chain of the MWA}\label{sec:signal_chain}
The signal chain of the MWA starts from a single bowtie dipole and ends at the correlator which combines signals from all antenna tiles of the array and allows to use of the MWA as a single radio telescope.
In the following sections, each component of the MWA signal chain and its role are briefly discussed. 

\subsection{Low-noise Amplifier}\label{subsec:lna}
\begin{figure}
    \centering
    \includegraphics[trim={0cm 0cm 0cm 0cm},clip,scale=0.75]{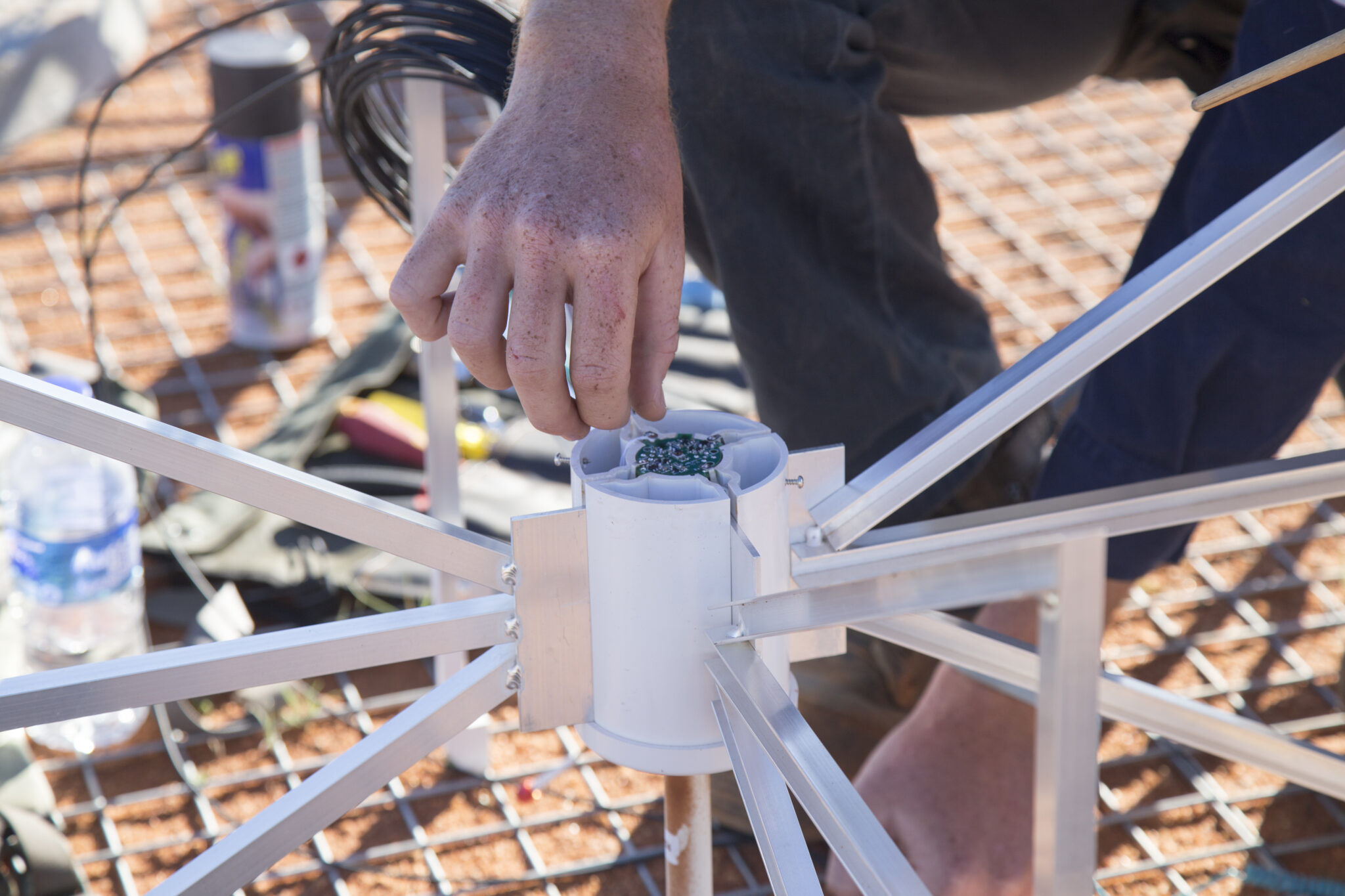}
    \caption[Low-noise amplifier of MWA.]{Low-noise amplifier (LNA) of the MWA is attached to the bowtie dipoles (Image credit: ICRAR/Curtin).}
    \label{fig:mwa_lna}
\end{figure}

Each bowtie dipole antenna is fitted with a custom-designed low noise amplifier (LNA), placed between each pair of bowtie arms as shown in Figure \ref{fig:mwa_lna}. The LNA amplifies incoming signals and is designed such that its output is sky noise dominated till about 300 MHz. The LNA for each antenna element is placed inside a protective, UV-resistant hub, where the two orthogonal bowtie arms are attached.

\subsection{Analog Beamformer}\label{subsec:analog_beamformer}
The beamformer is a device that allows the MWA to point and track objects in the sky electronically even though it does not have any moving parts. MWA uses an analog beamformer for each antenna tile. Signals from all 16 dipoles in a tile are received by the beamformer and independent delays are added to each of them to phase them up toward a specified direction, thus pointing the tile beam towards that direction. Depending on the observing frequency, the tile beam is 15 -- 50 degrees wide at full-width Half-maximum (FWHM). The delayed signals are combined and amplified by the beamformer and then sent over a coaxial cable to the MWA receiver.

The analog beamformer uses a set of five ``delay-lines" which can be switched in or out of the signal path to provide the necessary geometric delays to the signal from each of the dipoles before they are combined. These delay lines are used to provide a 32-step delay in steps of 435 picoseconds allowing a range of delays of 0.0 to 13.5 nanoseconds. As the beamformer can only provide a discrete set of delays, it is only for a limited set of pointings that the available delay value matches the required delay values precisely. These discrete pointing directions are referred to as sweet spots. All MWA solar observations are done by keeping the tile pointed at the nearest sweet spot from the Sun. Due to this reason, the Sun may not remain at the pointing center. Hence, while calibrating one has to take care of the response of the primary beam at the location of the Sun.

\subsection{MWA Receiver and Solar Attenuator}\label{subsec:receiver}
Each analog beamformer produces wideband outputs for two orthogonal polarizations. These wideband analog signals are fed to the MWA receiver node which is responsible for taking the analog radio frequency signals from eight antenna tiles, doing signal conditioning before performing digitization, coarse channelization and then transmitting the resulting digital streams using a fiber optic cable \citep{Prabu-MWA-Rx-2015}. A receiver filters out two analog signals from each tile to a bandpass of 80 -- 300 MHz, Nyquist samples the signals with an 8-bit analog-to-digital converter (ADC) and digitally filters the resulted data stream into 256$\times$1.28 MHz frequency channels. Each of these 1.28 MHz channels is called a ``coarse channel". Among these 256, coarse channel numbers below 55 ($<$70 MHz) and above 235 ($>$ 300 MHz) are highly attenuated during the analog signal conditioning. The MWA receiver gives the user the flexibility to choose any subset of 24 coarse channels in the 80 -- 300 MHz band. This provides a total instantaneous observing bandwidth of 30.72 MHz. 

The ADC can operate in a linear regime only over a certain range of input analog power levels. The MWA is designed to observe faint astronomical sources. Input analog power during solar observations is too high for the ADC to operate in its linear regime. The analog signal conditioning part of the MWA receiver provides independent amplification for each of the signal paths using adjustable attenuators, in steps of 1 dB and spanning a range of 60 dB.
These attenuators are employed to bring down the analog power level to lie within the linear range of the ADCs, while also keeping some headroom for strong active emissions.
The high attenuation settings of these attenuators are often referred to as ``solar attenuation". Although they enable solar observations, transferring calibration from the usual flux calibrator observations to the Sun, requires the response of these attenuators to be characterized. This is discussed in more detail in Chapter \ref{fluxcal}.

\subsection{The MWA Correlator}\label{subsec:correlator}
This 30.72 MHz digitized data from the receiver is transported to the central processing facility using optical fiber for correlation. MWA uses a hybrid FX correlator \citep{mwa_correlator_2015}, using both Field Programmable Gate Arrays (FPGA) and Graphical Processing Units (GPUs). Unlike a filled aperture instrument, an interferometric array like the MWA measures the level of correlation between the signals from all antenna pairs at different frequencies across the observing band. Each of these correlation products measures one single Fourier component of the 2-dimensional sky brightness distribution. Before performing correlation, each 1.28 MHz coarse channel is further filtered into 10 kHz ``fine" channels. The correlator performs the correlation between all antenna pairs and polarizations (X and Y). These correlation products are referred to as {\it visibilities} and the MWA correlator can provide visibilities at a time and frequency resolution of 0.5 s and 10 kHz. Due to the inverse relationship between Fourier conjugate pairs, visibilities between antenna pairs with small separations (short baselines) measure the sky signal at large angular scales, and visibilities from the long baselines measure the sky signal at smaller angular scales.

\begin{figure}
    \centering
    \includegraphics[trim={0cm 0cm 0cm 0cm},clip,scale=0.75]{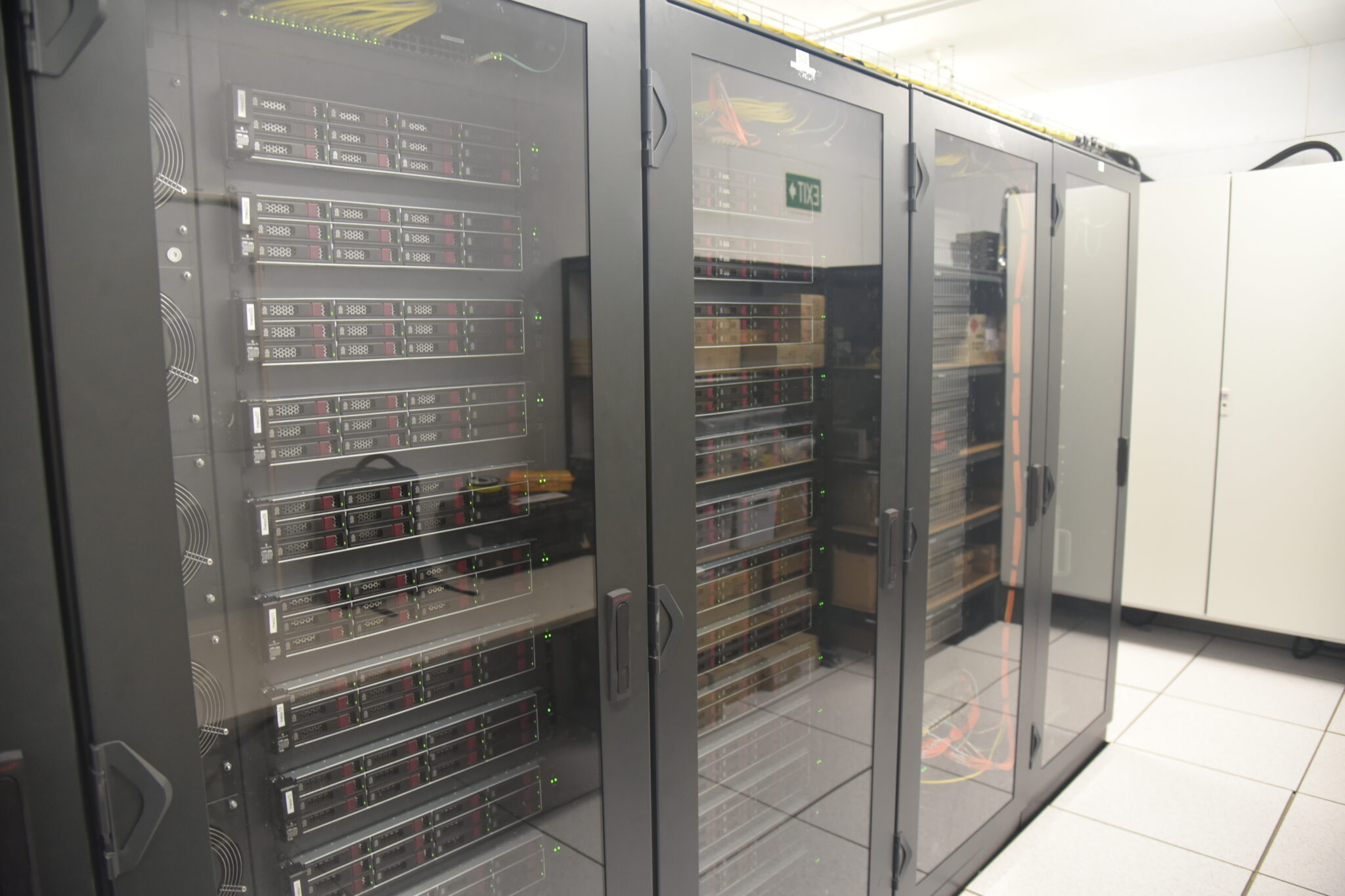}
    \caption[Correlator of the MWA]{The new MWAX correlator (Image credit: ICRAR/Curtin).}
    \label{fig:mwa_correlator}
\end{figure}

Very recently, the MWA correlator has been upgraded and the new correlator is called the MWAX Correlator \citep{MWAX2023} (Figure \ref{fig:mwa_correlator}). The MWAX provides enhanced capabilities and greater flexibility, scalability, and maintainability compared to the earlier MWA correlator. It is designed to enable future Phase-III upgrades, which will require simultaneous correlation of all 256 MWA tiles. MWAX is a fully software-programmable correlator and is highly flexible and scalable in terms of the number of antenna tiles and the number of coarse channels to be correlated. It also offers a wide range of combinations of spectro-temporal resolutions to the users. 

\section{MWA Solar Observation Configurations}\label{sec:mwa_solar_observation}
MWA solar observations have been done in multiple different spectral configurations, namely picket-fence mode, contiguous mode, or harmonic mode. Among them, the picket-fence mode is the one used most widely, where 24 coarse channels are placed at certain intervals covering the 80--300 MHz band. This mode provides sparser but wider spectral sampling. In the contiguous mode, all 24 coarse channels are placed adjacent to each other, and usually one cycles over the entire MWA band in time. For the harmonic mode, 12 coarse channels are chosen at some low-frequency part of the MWA band, and the other 12 coarse channels are placed in the first harmonic band of the low-frequency part. All MWA solar observations are done using either 10 dB or 14 dB solar attenuation.

MWA Phase-I solar observations have been done with spectral and temporal resolutions of 40 kHz and 0.5 s, respectively. MWA Phase-II extended configuration observations are done at 10 kHz and 0.5 s spectral and temporal resolutions. The MWAX correlator now provides more flexibility in terms of spectral and temporal resolution. The finest spectral and temporal resolutions with MWAX are 3.2 kHz and 0.25 s. To keep the data volumes manageable, currently, MWA solar observations in this configuration are done with resolutions of 160 kHz and 0.25 s. 

All solar observations are done in the large proposal mode under project ID G0002. Since the start of the MWA operation in mid-2013, the MWA has routinely observed the Sun for at least 100 hours per semester. All of the visibility data from the MWA solar observations are publicly available after the 18-month proprietary period at the MWA All Sky Virtual Observatory (\href{https://asvo.mwatelescope.org/}{MWA ASVO}). 

In the following chapters, I discuss the suitability of the MWA and its architecture for meter-wavelength solar observations. Ultimately to make the final solar images from the measured visibilities, one has to overcome several challenges, including some additional ones specific to solar observations. These are discussed in the following chapters. 
\chapter {Principle Behind Generating High dynamic range Spectroscopic Snapshot Solar Radio Images}
\label{paircars_principle}

The necessity for high dynamic range (DR) spectroscopic snapshot solar imaging was already discussed briefly in Section \ref{subsec:observational_challenge} of Chapter \ref{chapter_intro}. In this chapter, I discuss it in greater depth the principle behind successfully producing high DR spectroscopic solar radio images using new-generation instruments, like the MWA. The work presented in this chapter is based on \citet{Kansabanik_principle_AIRCARS}, which was published in Solar Physics.

\section{Introduction}\label{sec:intro_chapter_paircars_principle}
Several emission mechanisms, like plasma emission, thermal bremsstrahlung, and gyrosynchrotron, give rise to meter-wavelength solar emission originating in the solar corona. Low-frequency radio observations are particularly useful for measuring the coronal magnetic fields and the nonthermal electron populations, which are rather hard to do using observations at other wavelengths. Despite their well-appreciated importance, low-frequency imaging observation of the Sun is one of the least explored areas of solar physics.
\begin{figure*}[!ht]
    \centering
    \includegraphics[trim={0cm 2.5cm 0.5cm 0cm},clip,scale=0.28]{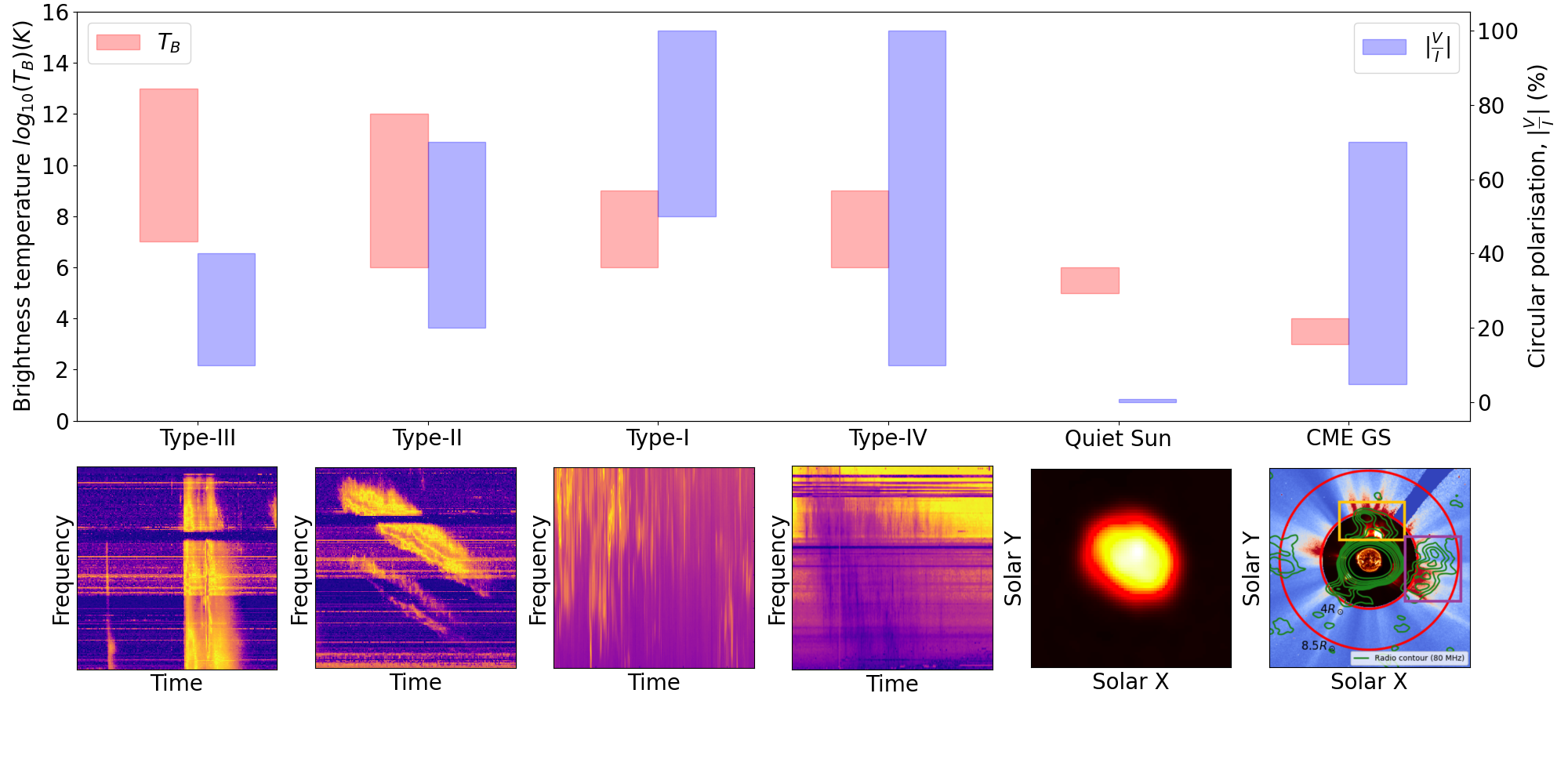}
    \caption[Spectro-temporal and spatial distribution of brightness temperature and circular polarization of meter-wavelength solar emissions.]{{\it Top panel :} The expected range of brightness temperature ($T_\mathrm{B}$) and circular polarization fraction for different kinds of low-frequency solar radio emissions are shown by the blue and red bars, respectively. {\it Bottom panel :} Sample dynamic spectra for type-I, -II, -III, and -IV radio bursts. Dynamic spectra of type-II, -III, and -IV solar radio bursts have been obtained from the Learmonth solar radio spectrograph. The dynamic spectrum of type-I solar radio burst is obtained from the MWA. Dynamic spectra have different spectro-temporal structures spanning a large range of spectral and temporal widths. The images of the last two panels show the Quiet Sun emission and gyrosynchrotron emission from a CME. These emissions have spatial structures spanning a large range of angular scales. The images of the Quiet Sun and CME are from MWA.}
    \label{fig:different_emissions}
\end{figure*}

The brightness temperature ($T_\mathrm{B})$ of the low-frequency solar emissions can vary from $\sim10^3-10^4\ \mathrm{K}$ for gyrosynchrotron emission from coronal mass ejection (CME) plasma \citep{bastian2001,Mondal2020a} to $\sim10^{13}\  \mathrm{K}$ for bright type-III radio bursts \citep{McLeanBook,Reid2014} (shown by red bars in Figure \ref{fig:different_emissions}) over a background quiescent $T_\mathrm{B}$ of $\sim10^6\  \mathrm{K}$. Depending upon the emission mechanism at play, the polarization fraction of the meter-wavelength solar emission can also vary from $\leq 1\%$ to $\sim100\%$ \citep{McLeanBook,Nindos2020} (shown by the blue bars in Figure \ref{fig:different_emissions}). 

Imaging the Sun at low radio frequencies with high DR and high-fidelity is a challenging problem. The Sun is an extended source having morphology spanning a large range of angular scales, from a few degrees  down to a few arcminutes at meter-wavelengths. The meter-wavelength solar emission varies over  short temporal and spectral scales, which imposes a requirement for snapshot spectroscopic imaging. The need to be able to see features varying vastly in $T_\mathrm{B}$, highlights the need for a high imaging DR. Only recently it has become possible to meet these exacting requirements for solar radio imaging using instruments like the MWA. The MWA has a large number of antenna elements distributed over a small array footprint. The array coverage of the MWA (Figure \ref{fig:phaseI_array} and \ref{fig:phaseII_array} in Chapter \ref{mwa}) is especially well-suited for snapshot spectroscopic imaging. Although MWA data is intrinsically capable of producing high-fidelity solar images, one first needs to precisely correct all instrumental and atmospheric (mostly of ionospheric origin at low radio frequencies) effects.
 
To perform the precise calibration of the instrumental and ionospheric effects for MWA solar observations, \citet{Mondal2019} developed a novel calibration and imaging pipeline called ``Automated Imaging Routine for Compact Arrays for the Radio Sun" (AIRCARS). AIRCARS has been used on a set of MWA solar observations at different solar conditions and successfully produced the best spectroscopic snapshot images of the Sun at low frequencies to date. The DR of the images produced by AIRCARS varies between $>300$ to about $10^5$. It has led to many discoveries over the last few years \citep{Mohan2019a,Mohan2019b,Mondal2020a,Mondal2020b,Mondal2021a,Mohan2021a,Mohan2021b}. As a part of this thesis, AIRCARS has been improved and extended to produce high-fidelity spectropolarimetric imaging. It has been given the name ``Polarimetry using Automated Imaging Routine for Compact Arrays for the Radio Sun" \citep[P-AIRCARS,][]{Kansabanik_paircars_2,Kansabanik2022_paircarsI} and is discussed in Chapters \ref{paircars_algorithm} and \ref{paircars_implementation}. In this chapter, the working principle behind AIRCARS/P-AIRCARS is discussed and demonstrated using statistical and quantitative analysis.

\section{Suitability of the Array Configuration of the MWA for High-fidelity Spectroscopic Snapshot Solar Imaging}\label{sec:suitability_of_MWA}
Radio interferometric imaging is a Fourier imaging technique \citep{McCready1947,thompson2017}. A radio interferometer is made up of multiple antenna elements (or dishes) distributed on the ground. Each antenna element of the array receives radio emission from the sky and convert them into electronic voltages. The cross-correlation of the measured voltages between the antenna pairs ($\mathrm{i}$ and $\mathrm{j}$) is known as {\it visibilities}, $V_\mathrm{ij}$. Each of these visibilities corresponds to a single Fourier component of the sky brightness distribution in a 2-dimensional Fourier plane, which is commonly known as the {\it uv-}plane. The inverse Fourier transform of the measured visibilities on the {\it uv-}plane gives the true sky brightness distribution. Ideally, the {\it uv-}plane has to be sampled at the spatial Nyquist resolution for the reconstruction of the sky brightness distribution accurately. Most of the conventional radio interferometers, like the Very Large Array \citep[VLA,][]{VLA2009}, Giant Metrewave Radio Telescope  \citep[GMRT,][]{Swarup_1991,Gupta_2017}, Westerbork Synthesis Radio Telescope \citep[WSRT,][]{WSRT2021}, LOw-Frequency ARray \citep[LOFAR,][]{lofar2013} etc, have a limited number of antennas distributed sparsely over a large area on the ground.  Hence, the instantaneous sparse sampling of the {\it uv-}plane does not meet the Nyquist criteria. These instruments use the rotation of the Earth to get the same physical baseline to sample the different Fourier components in the {\it uv}-plane. For sources whose spectra can be modeled, frequency synthesis is also routinely employed to further improve the sampling in the {\it uv}-plane.

\begin{figure*}[!ht]
    \centering
    \includegraphics[trim={2.3cm 1.5cm 3cm 1.5cm},clip,scale=0.55]{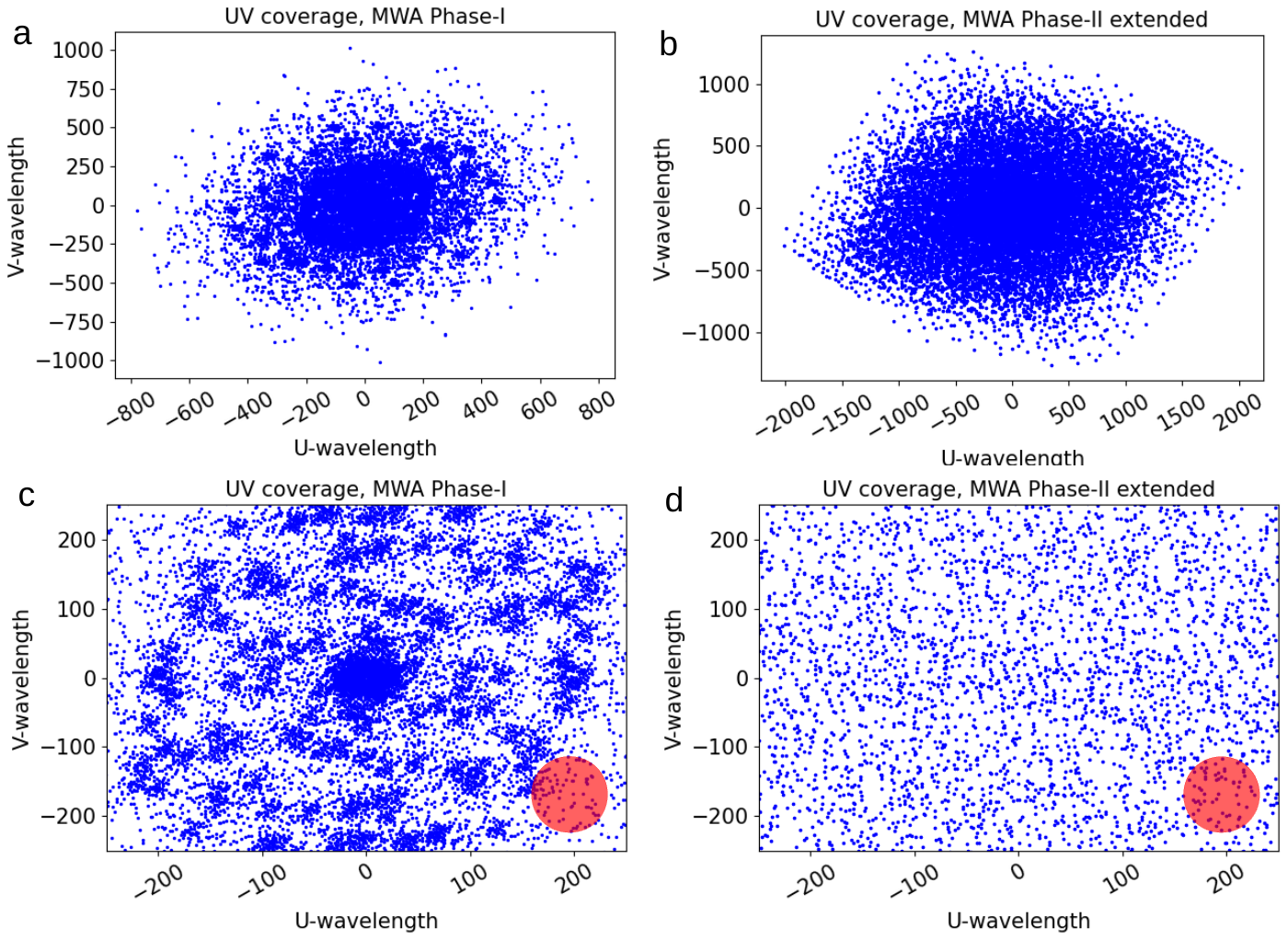}
    \includegraphics[trim={2.5cm 14.5cm 4.5cm 1.8cm},clip,scale=0.4]{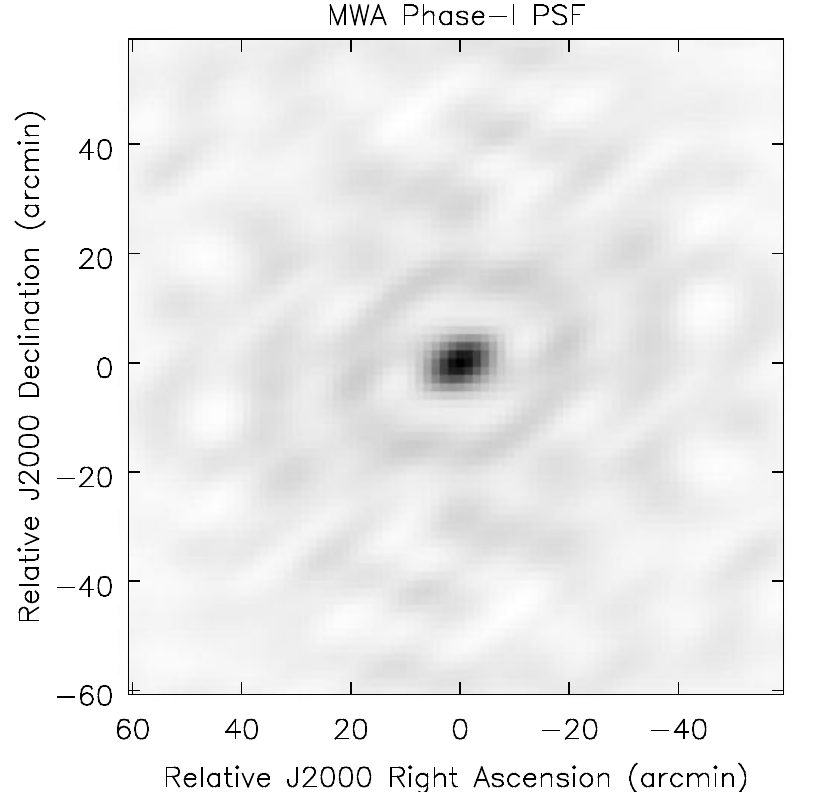}\includegraphics[trim={2.5cm 14.5cm 2.2cm 1.8cm},clip,scale=0.4]{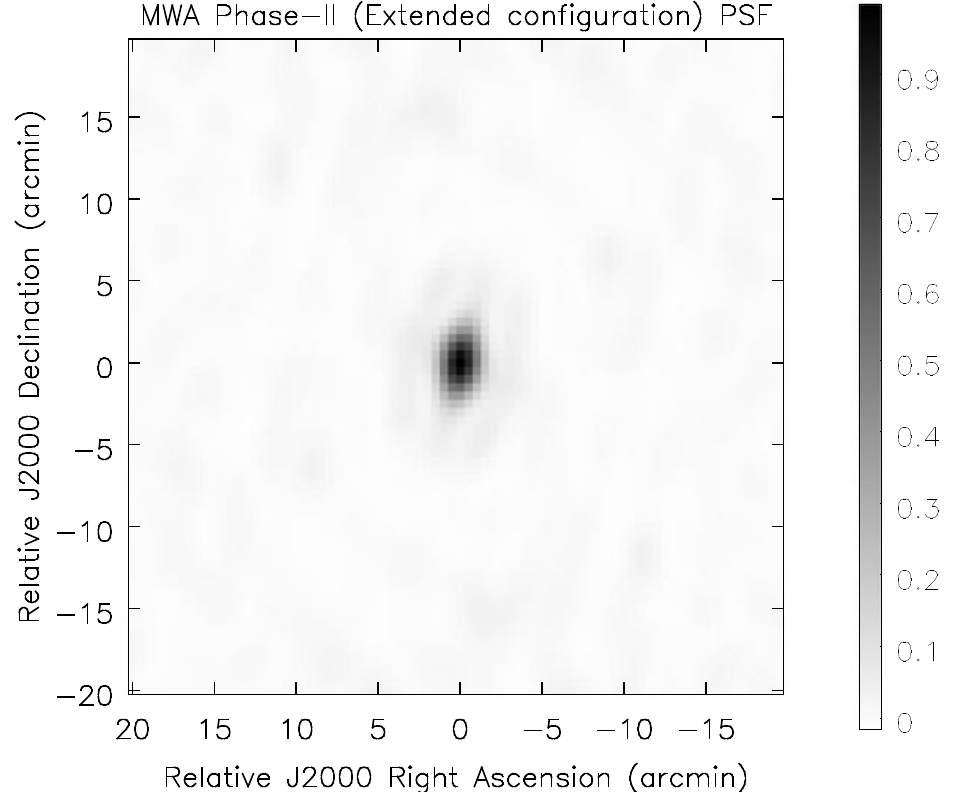}
    \caption[Snapshot {\it uv-}coverage and point spread function (PSF) of the MWA at 150 $\mathrm{MHz}$.]{Snapshot {\it uv-}coverage and point spread function (PSF) of the MWA  at 150 $\mathrm{MHz}$. {\it Top panel: a) } Snapshot {\it uv-}coverage of MWA Phase-I, and {\it b)} Phase-II extended configuration. {\it c,d)} Zoomed-in version of the {\it uv-}coverage over a region of size $250\ \lambda$. Red circles correspond to the {\it uv-}cell for a source with 1$^\circ$ angular scale. {\it Bottom panel: Left.} Un-tapered and naturally weighted PSF of MWA Phase-I, and {\it Right.} Phase-II extended configuration at 150 MHz.}
    \label{fig:uvcoverage}
\end{figure*}

One way to sample the {\it uv-}plane densely is using the so-called ``large-N” array configuration. The MWA array design follows the ``large-N" array configuration. It has 128 antenna tiles distributed over a small array footprint and provides dense spectroscopic snapshot {\it uv-}coverage. The MWA has two phases of operation and their array configurations are described in Section \ref{subsec:array_config} and shown in Figure \ref{fig:phaseI_array} and \ref{fig:phaseII_array} in Chapter \ref{mwa}. Among them, phase-I and phase-II extended configurations are the most favorable configurations for solar observations. Currently, the MWA is upgrading toward its Phase-III and can perform observations simultaneously with 144 antenna tiles. Once completed, the MWA Phase-III will be able to observe using 256 antenna tiles simultaneously and further improve its sensitivity and {\it uv}-coverage. The snapshot {\it uv-}coverage of MWA phase-I and phase-II extended configurations are shown in Figure \ref{fig:uvcoverage}a and b respectively. The zoomed-in versions over a square region of $250 \lambda$ (where $\lambda$ refers to wavelength) are shown in Figure \ref{fig:uvcoverage}c and d respectively. The red circle shows the {\it uv-}cell required for Nyquist sampling for a source with $1^\circ$ angular scale, which is the approximate angular size of the Sun at the meter-wavelengths. It is evident from Figure \ref{fig:uvcoverage}c and d that, the density of {\it uv-}sampling approaches or even exceeds the Nyquist criterion over a significant part of the uv-plane. The bottom left panel of Figure \ref{fig:uvcoverage} shows the naturally weighted and un-tapered spectroscopic snapshot point spread function (PSF) for the phase-I and the right panel is for the phase-II extended configuration. These spectroscopic snapshot PSFs are  exceptionally well-behaved, which reduces the deconvolution artifacts in the final images. These properties of the MWA array configuration make it well-suited for high DR spectroscopic snapshot solar imaging. 

\section{A Brief Overview of Radio Interferometric Calibration}\label{sec:calibration}
The true source visibility, $V_{\mathrm{ij}}$, between an antenna pair, $\mathrm{i}$ and $\mathrm{j}$, is corrupted by the complex instrumental gains and due to the atmospheric propagation effects. At low radio frequencies, the ionospheric propagation effect is the major atmospheric propagation effect. In practice, both the instrumental gain and ionospheric propagation effect are merged into a single complex gain term. The measured visibility, $V_{\mathrm{ij}}^{\prime}$, can be written in terms of the $V_\mathrm{ij}$ as
\begin{equation}\label{eq:measurement_eq}
\begin{split}
     V_{\mathrm{ij}}^\prime(\nu,\ t,\ \vec{l}) &= J_\mathrm{i}(\nu,\ t,\ \vec{l})\ V_{\mathrm{ij}(\nu,\ t,}\ \vec{l})\ J_\mathrm{j}^\dagger(\nu,\ t,\ \vec{l}) + N_\mathrm{ij}\\
      &= |J_\mathrm{i}(\nu,\ t,\ \vec{l})|\ V_{\mathrm{ij}}(\nu,\ t,\ \vec{l})\ |J_\mathrm{j}^\dagger(\nu,\ t,\ \vec{l})| e^{i\left[\phi_\mathrm{i}(t)-\phi_\mathrm{j}(t)\right]} \\&+ N_\mathrm{ij}
\end{split}
\end{equation}
where, $J_\mathrm{i}(\nu,\ t,\ \vec{l})$ and $J_\mathrm{j}(\nu,\ t,\ \vec{l})$ are the complex gain terms incorporating both the instrumental and the ionospheric effects, $N_{\mathrm{ij}}$ is the additive correlated noise. $\nu,\ t,\ \vec{l}$ represent the observing frequency, time, and direction in the sky plane respectively. $|J_\mathrm{i}|$, $|J_\mathrm{j}|$ represent the amplitude and $\phi_\mathrm{i}$, $\phi_\mathrm{j}$ are the phase parts of $J_\mathrm{i}$ and $J_\mathrm{j}$, respectively. Equation \ref{eq:measurement_eq} is popularly known in the literature as the {\it measurement equation} \citep{Hamaker1996_1} for a radio interferometer.  One has to estimate $J_\mathrm{i}(t,\ \nu,\ \vec{l})\ =\ G_\mathrm{i}(t)\ B_\mathrm{i}(\nu)\ E_\mathrm{i}(\vec{l})$ for all the antenna elements and correct for them to obtain $V_\mathrm{ij}$ from the $V_{\mathrm{ij}}^\prime$. $J_\mathrm{i}(\nu,\ t,\ \vec{l})$ can be decomposed into two major parts --
\begin{enumerate}
    \item {\bf Direction independent terms :} $G_\mathrm{i}(t)$ and $B_\mathrm{i}(\nu)$ are the two direction independent components of $J_\mathrm{i}$. $G_\mathrm{i}(t)$ represents the time variable instrumental and ionospheric gain and $B_\mathrm{i}(\nu)$ is the instrumental bandpass.
    \item {\bf Direction dependent terms : } Direction-dependent effects arise for the array with a large field of view (FoV) \citep{Lonsdale2005}. Propagation of radio emission from different parts of the sky through different parts of the ionosphere introduces direction-dependent complex gain, $E_\mathrm{i}(\vec{l})$.
\end{enumerate}

The standard practice in radio interferometric calibration is to observe a calibrator source with known flux density, spatial structure, and spectral properties, and use it to estimate the $G_\mathrm{i}(t)$ and $B_\mathrm{i}(\nu)$. For wide FoV instruments, instead of a single calibrator source, a global sky model is also used to estimate the direction-dependent gain term, $E_\mathrm{i}(\vec{l})$. Details of estimating these terms and correcting them are discussed in Chapter \ref{paircars_algorithm}. Among these terms, the time-varying component, $G_\mathrm{i}(t)$ has the largest effect on the imaging DR and hence needs to be corrected with high precision before correcting for other terms. In the rest of this chapter, I  describe the algorithm for estimating $G_\mathrm{i}(t)$ and demonstrate the working principle of this algorithm.

\subsection{Requirement of direction dependent calibration}\label{sec:direction_dependent_calibration}
For the wide FoV instruments like the MWA, direction-dependent calibration is necessary, and it is implemented in the standard calibration and image processing pipeline for the MWA \citep[RTS;][]{Mitchell2008}. But, the Sun is the source with the highest flux density in the low-frequency radio sky. The flux density of even the quiet Sun is more than $10^4\ \mathrm{Jy}$ \footnote{Jy (Jansky) is the unit of flux density used in radio astronomy. 1 Jy = $10^{-26}\ \mathrm{W\ m^{-2} Hz^{-1}}$}, which can increase by a few orders of magnitudes during periods of active emission. On the other hand, the flux densities of only a handful of sources lie in the range of hundreds of Jy, and for the vast majority of sources, flux densities lie in the range of a few Jy and below. This effectively reduces the solar observation to a small FoV problem, with a single bright source at the center of the FoV, that dominates the overall visibility. Hence, for MWA solar observation, direction-dependent calibration is not required and is not implemented in AIRCARS/P-AIRCARS.

\section{A Brief Description of the Calibration \\Algorithm}\label{sec:aircars_description}
\begin{figure}
    \centering
    \includegraphics[trim={0cm 0cm 0cm 0cm},clip,scale=1]{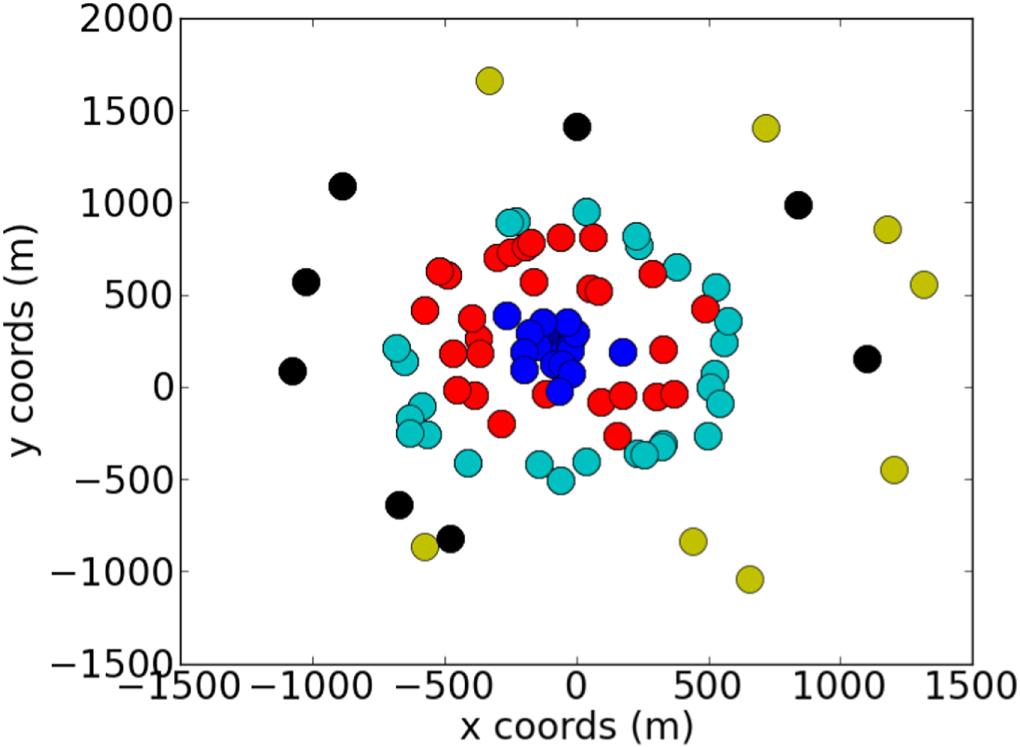}
    \caption[Antenna addition scheme implemented in AIRCARS/P-AIRCARS for MWA phase-I array.]{Antenna addition scheme implemented in AIRCARS/P-AIRCARS for MWA phase-I array. The antennas marked by blue are used in the first step, and antennas at increasingly large distances are added in subsequent steps marked by red,  magenta, black, and yellow (Reproduced from \cite{Mondal2019} with the permission from author).}
    \label{fig:antenna_addition}
\end{figure}
\begin{figure*}[!ht]
    \centering
    \includegraphics[trim={0cm 0cm 0cm 0cm},clip,scale=0.12]{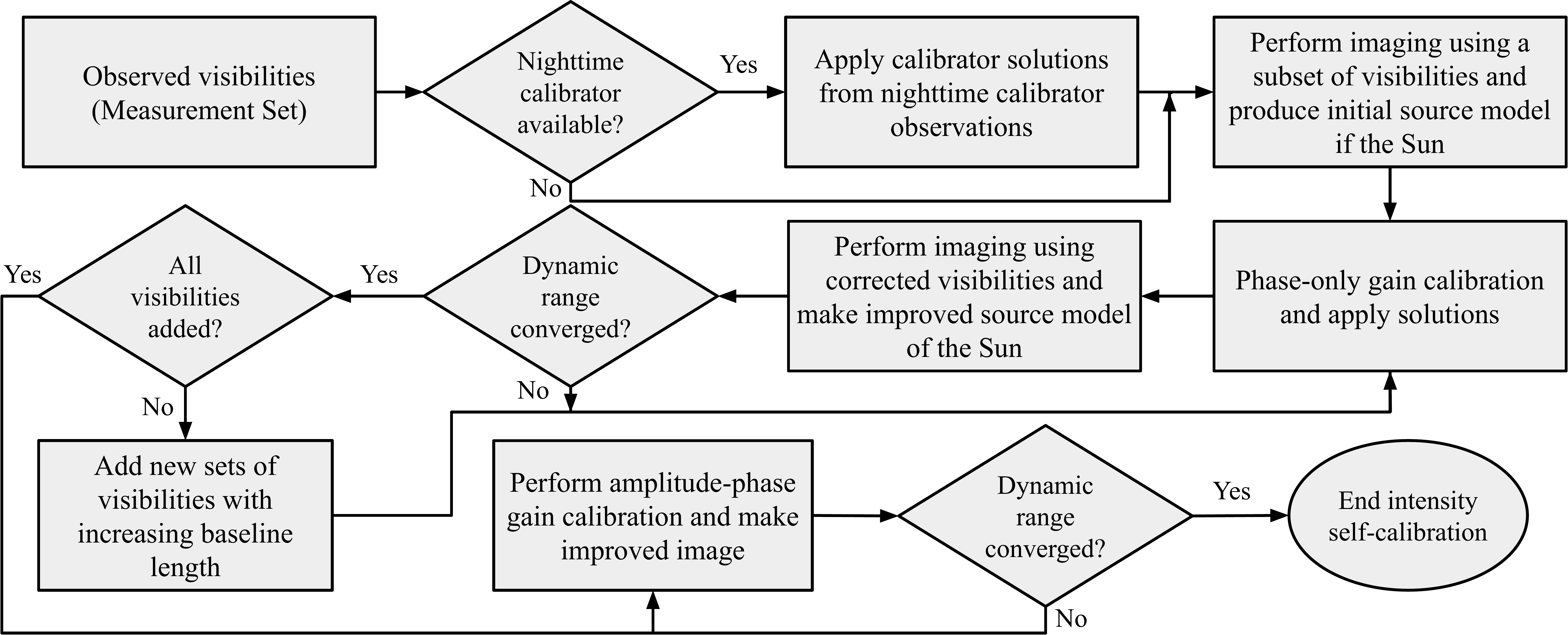}
    \caption[Flowchart of the intensity self-calibration algorithm.] Flowchart describing intensity self-calibration algorithm implemented in AIRCARS \citep{Mondal2019} and P-AIRCARS \citep{Kansabanik2022_paircarsI,Kansabanik_paircars_2}.
    \label{fig:aircars_flowchart}
\end{figure*}

In this section, I describe briefly the calibration steps followed in AIRCARS/P-AIRCARS to correct $G_\mathrm{i}(t)$s. These steps are referred to as ``intensity self-calibration" and a flowchart describing it is shown in Figure \ref{fig:aircars_flowchart}.
\begin{enumerate}
    \item When a dedicated nighttime calibrator observation is available with the same spectral settings as used for solar observation, calibration solutions obtained from nighttime calibrators are applied first. If this is unavailable, AIRCARS/P-AIRCARS starts the calibration using a simple initial source model  of the Sun.
    \item To produce the initial source model of the Sun, a subset of visibilities, $V_\mathrm{ij}$ are chosen, where $\mathrm{i}$ is from the antennas marked by blue in Figure \ref{fig:antenna_addition} and $i<j$.   
    \item Phase-only gain calibration is performed using the initial source model. This corrects the phase part of the $G_\mathrm{i}(t)$s. 
    \item Phase-only gain solutions are applied, and an improved source model is arrived at.
    \item The last two steps are continued until the changes in DR have converged. DR is deemed to have converge when the  DR changes over three consecutive rounds are smaller than a predefined value. 
    \item Once the DR has converged, antennas with increasing distance from the core are added in small steps to the self-calibration process. These additional groups of antenna tiles, which are added literately in the self-calibration process, are shown by different colors in Figure \ref{fig:antenna_addition} for MWA phase-I. 
    \item These additional antenna tiles also have an initial gain solution from the previous self-calibration rounds, because baselines were taken with the set of all antenna, ${\mathrm{j}}$. 
    \item When all antennas are added in the self-calibration, AIRCARS/P-AIRCARS starts the calibration of both amplitude and phase part of the $G_\mathrm{i}(t)$ considering all antennas. This process continues until the DR of the image has converged.
    \item AIRCARS/P-AIRCARS uses well-defined convergence criteria for DR which is determined by some user inputs and is detailed in Chapter \ref{paircars_implementation}. There is a minimum number ($\sim$5) of fixed iterations after the start of amplitude and phase self-calibrations, only after which the convergence criteria of the DR are evaluated. This is done to avoid some local convergence in the self-calibration process.
\end{enumerate}
One of the unique features of the AIRCARS/P-AIRCARS is that it can start the self-calibration process even without any {\it a priori} calibration solutions obtained from the nighttime calibrators. In the later sections, the explanation behind this unique feature of AIRCARS/P-AIRCARS is discussed in detail.

\begin{figure*}[!ht]
    \centering
    \includegraphics[trim={3cm 14cm 3cm 0cm},clip,scale=0.475]{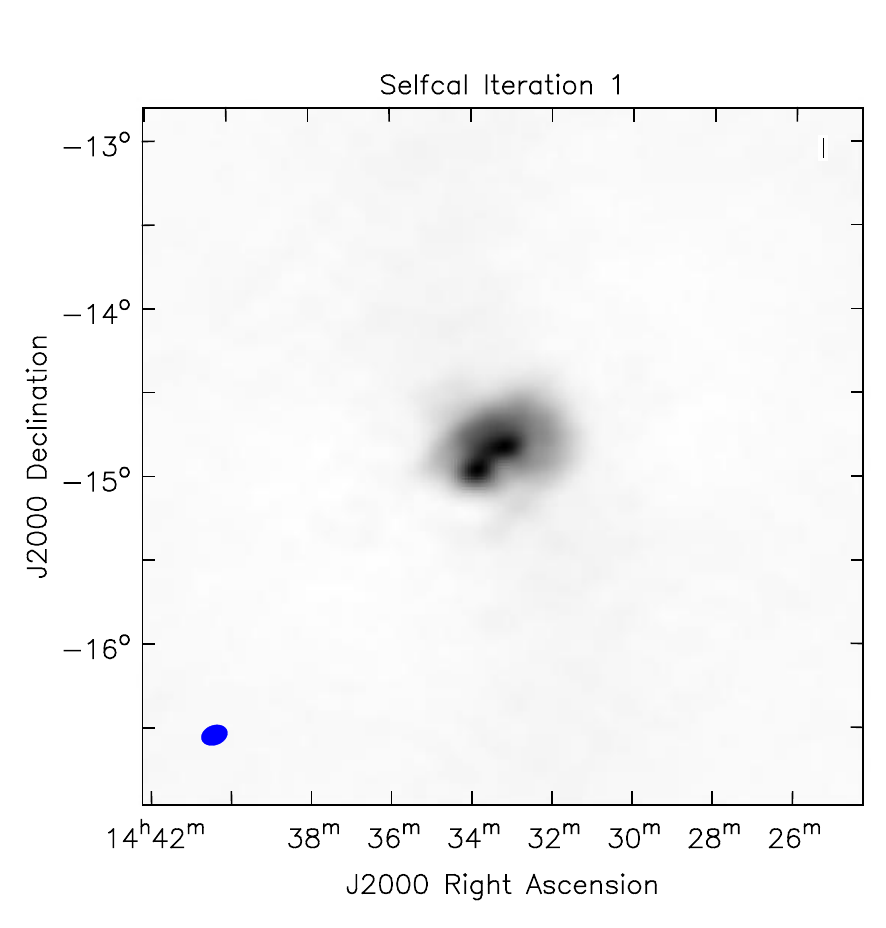}\includegraphics[trim={3cm 14cm 3cm 0cm},clip,scale=0.475]{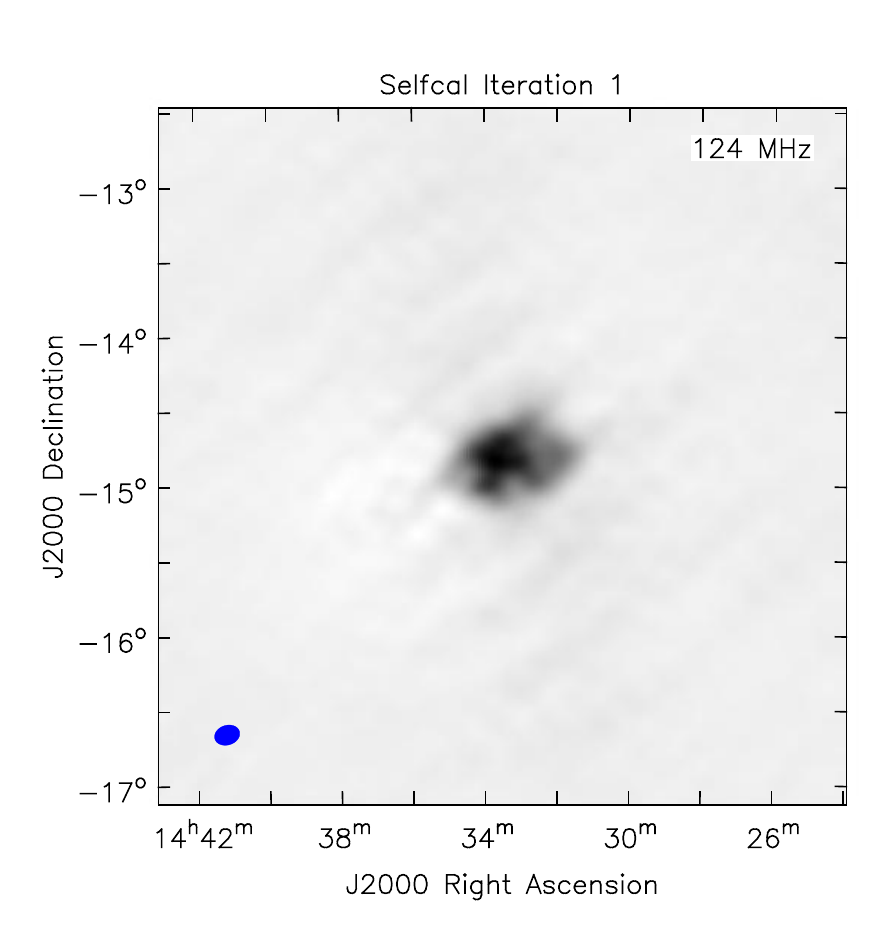}
    \caption[Initial image of the Sun to start the self-calibration.]{Initial image of the Sun to start the self-calibration. {\it Left panel: }Initial image made after applying the calibration solutions from the nighttime calibrator observation. {\it Right panel: }Initial image made directly from the first subset of uncalibrated visibilities.}
    \label{fig:ini_model}
\end{figure*}

\section{Initial Source Model of AIRCARS}
\begin{figure*}[!ht]
    \centering
    \includegraphics[trim={0cm 0cm 0cm 0cm},clip,scale=0.8]{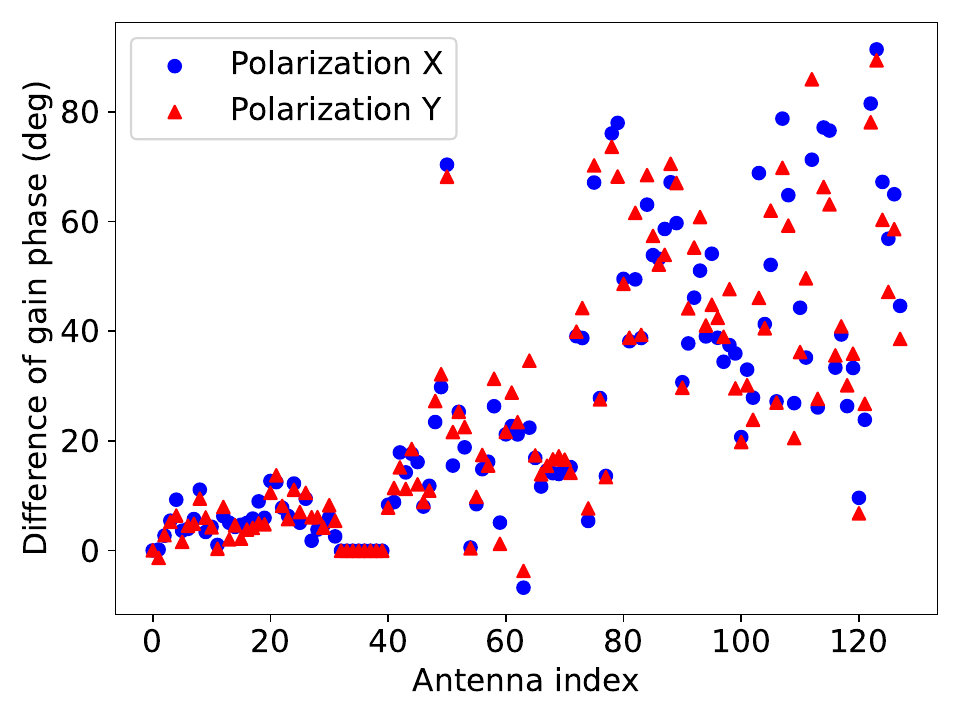}
    \caption[Difference between phases of the nighttime and daytime complex antenna gains.]{Difference between phases of the nighttime and daytime complex antenna gains. Blue circle and red triangle represent the X and Y polarization, respectively.}
    \label{fig:cal_aircars_gain_diff}
\end{figure*}

When nighttime calibration is available, the calibration solutions are applied before the initial imaging (left panel of Figure \ref{fig:ini_model}). When this is not available, the initial image is made from the uncalibrated observed visibilities (right panel of Figure \ref{fig:ini_model}) as described in Section \ref{sec:aircars_description}. There are differences between these two images. This happens because the phases of the complex gains during daytime are different from the phases during nighttime (Figure \ref{fig:cal_aircars_gain_diff}). If these differences are large, nighttime calibration solutions may not provide a significant improvement in the initial image compared to the image made from uncalibrated visibilities. In both cases, there is a significant amount of source flux concentrated near the phase center, because the phase distribution of the antenna gains is not uniformly random and the array has some level of coherency, which is demonstrated later in Sections \ref{sec:expected_gain_char} and \ref{sec:stats}. 

\section{Expected Characteristics of the Complex \\Gains}\label{sec:expected_gain_char}
As described in Section \ref{sec:calibration}, $J_\mathrm{i}(\nu,\ t,\ \vec{l})$ can be decomposed into $G_\mathrm{i}(t),\ B_\mathrm{i}(\nu)$ and $E_\mathrm{i}(\vec{l})$. As discussed in Section \ref{sec:direction_dependent_calibration}, $E_\mathrm{i}(\vec{l})$ can be neglected for solar observations. The algorithm of determining $B_\mathrm{i}(\nu)$ is discussed in Chapter \ref{fluxcal}. The only remaining term is time-dependent complex gain, $G_\mathrm{i}(t)$. $G_\mathrm{i}(t)$ has the contribution from both instrument ($g_\mathrm{i}^{\mathrm{instrumental}}(t)$) and the ionospheric ($g_\mathrm{i}^{\mathrm{ion}}(t)$). In practice, it is not necessary to separate them and this has not been done in AIRCARS/P-AIRCARS. 

\subsection{Expected Characteristics of the Instrumental Gains}\label{subsec:instrumental_gain}
The contributions from $g_\mathrm{i}^{\mathrm{instrumental}}(t)$ are not expected to originate from a uniformly random distribution. There are several reasons behind this --
\begin{enumerate}
    \item Except for the active dipoles and the low-noise amplifiers (LNA), other components of the electronic chain of the MWA are passive elements \citep{Tingay2013}, and the characteristics of the passive components are extremely stable.
    \item The characteristics of the LNA can also be well modeled \citep{Sokolwski2017} for the MWA and are similar for all the antenna elements.
    \item Temperature variation of the environment changes the effective cable length and introduces an additional phase in the complex gain. These are small for the core antennas, which are connected using shorter cables, and, grow larger for the antennas at long baselines connected using longer cables.
    \item Despite the well-modeled LNA and passive elements, there are some manufacturing tolerances, which could introduce a spread in the distribution of the instrumental gains.
\end{enumerate}

\subsection{Expected Characteristics of the Ionospheric Phases}\label{subsec:ionopsheric_phase_expectations}
\begin{figure}
    \centering
    \includegraphics[scale=0.22]{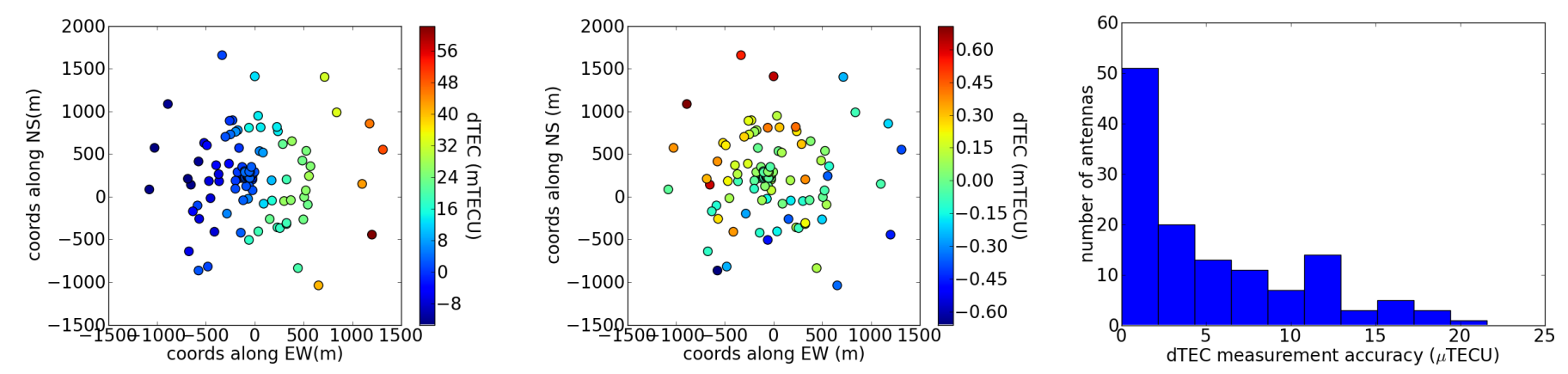}
    \caption[Daytime variation of differential ionospheric total electron content over the MWA.]{Daytime variation of differential ionospheric total electron content (dTEC) over the MWA. {\it Left panel:} The mean dTEC of the antenna tiles of the MWA phase-I with respect to the line of sight of a reference antenna, over 1 minute. The dTEC has changed by 72 mTECU over the array. {\it Middle panel: } The mean subtracted dTEC at a 0.5s snapshot, illustrating the small and fast phase variations over the array. The maximum dTEC variation in this figure is about 0.89 mTECU. {\it Right panel:} The accuracy of the dTEC measurements for a typical 0.5s snapshot (Reproduced from \cite{mondal_ionosphere} with the permission from author).}
    \label{fig:day_TEC}
\end{figure}

At low radio frequencies, another major contribution to the complex gain comes from the ionosphere. \citet{mondal_ionosphere} determined the total electron content (TEC) of the ionosphere using the daytime observation of the Sun along a single line-of-sight (LoS). They demonstrated that the daytime ionospheric differential TEC (dTEC) can vary over the MWA array, even over the core. The dTEC value varies by $\sim10\ \mathrm{mTECU}$ over the core antennas as shown in the left panel of Figure \ref{fig:day_TEC}. This variation corresponds to $\sim50\ \mathrm{degrees}$ variations in ionospheric phases \citep{Mevius2015}. \citet{mondal_ionosphere} also showed that the variation is smooth across the array (middle panel of Figure \ref{fig:day_TEC}), and the mean subtracted small scale random dTEC fluctuations over the array is $\lesssim1\ \mathrm{mTECU}$, which corresponds to a few degrees \citep{Mevius2015} of ionospheric phase variations. This demonstrates that although there are variations of the ionosphere across the MWA array,  even across the core antennas, these variations are smooth and the random fluctuations are small. During periods of solar activity or solar maxima, the level of ionospheric disturbances, in terms of electron density and its temporal variability, increase, but due to the small array footprint of the MWA, spatial variations across the array remain smooth and the level of spatial variability does not change much.

\subsection{Expected Statistical Properties of $G_\mathrm{i}$}\label{subsec:expected_stats}
As described in Section \ref{subsec:instrumental_gain} and \ref{subsec:ionopsheric_phase_expectations}, the core antenna tiles are expected to have a similar phase with a spread around a mean value due to instrumental (temperature variation across the array, manufacturing tolerances) and ionospheric effects. These effects become larger for the antenna tiles farther away from the core. Hence one can expect the following distribution of the phases of $G_\mathrm{i}$:
\begin{enumerate}
    \item {\bf Only core antenna tiles: }Distribution will be quasi-Gaussian with a small standard deviation.
    \item {\bf Only non-core antenna tiles: }Distribution will not be a peaked distribution and the standard deviation will be very large.
    \item {\bf All antenna tiles: }Since the core antenna tiles ($\sim60$) dominate the total number of antenna tiles (128), the distribution will be quasi-Gaussian with a slightly larger standard deviation compared to the distribution of only core antenna tiles.
\end{enumerate}

\begin{figure}
    \centering
    \includegraphics[scale=0.23]{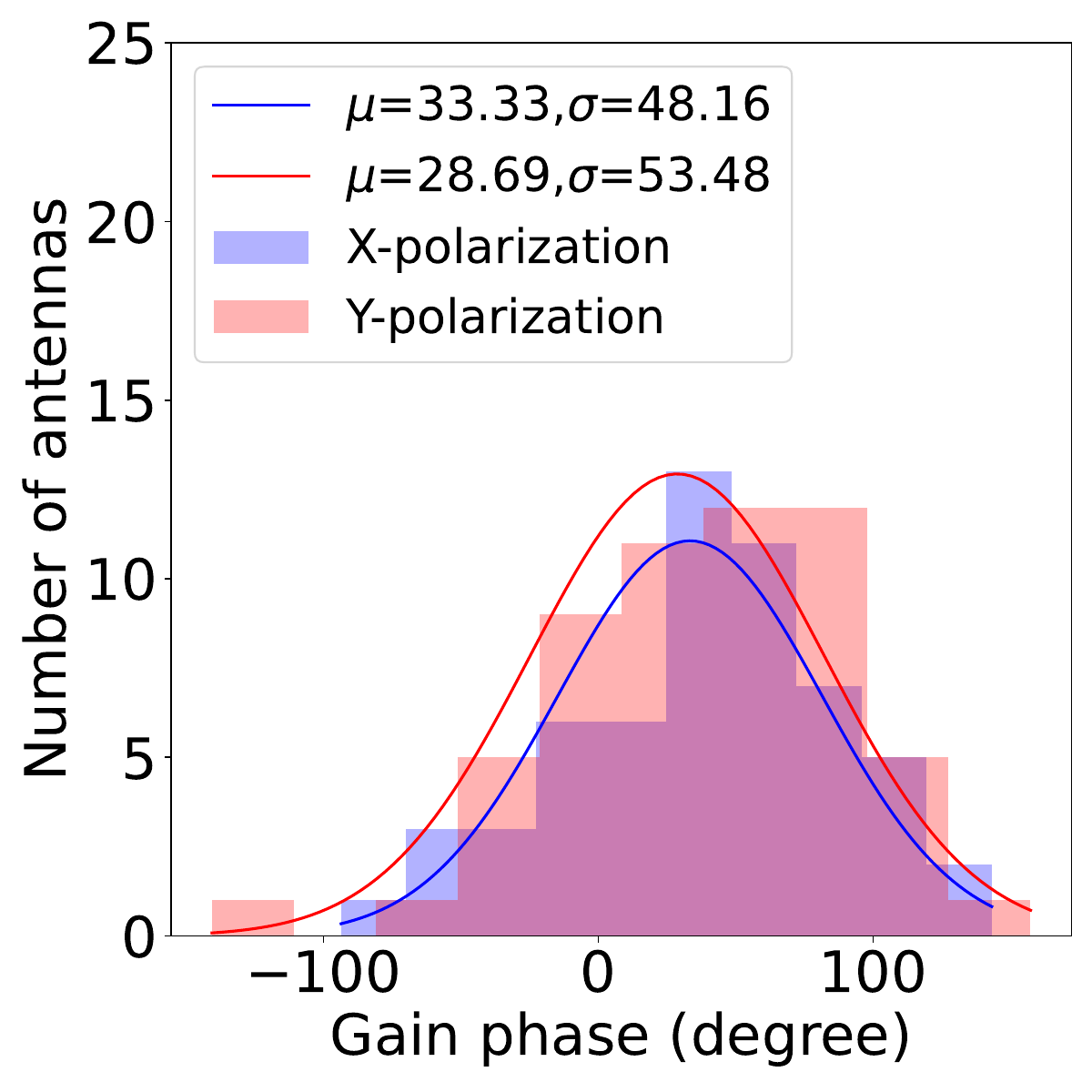}
    \includegraphics[trim={1.5cm 0cm 0cm 0cm},clip,scale=0.23]{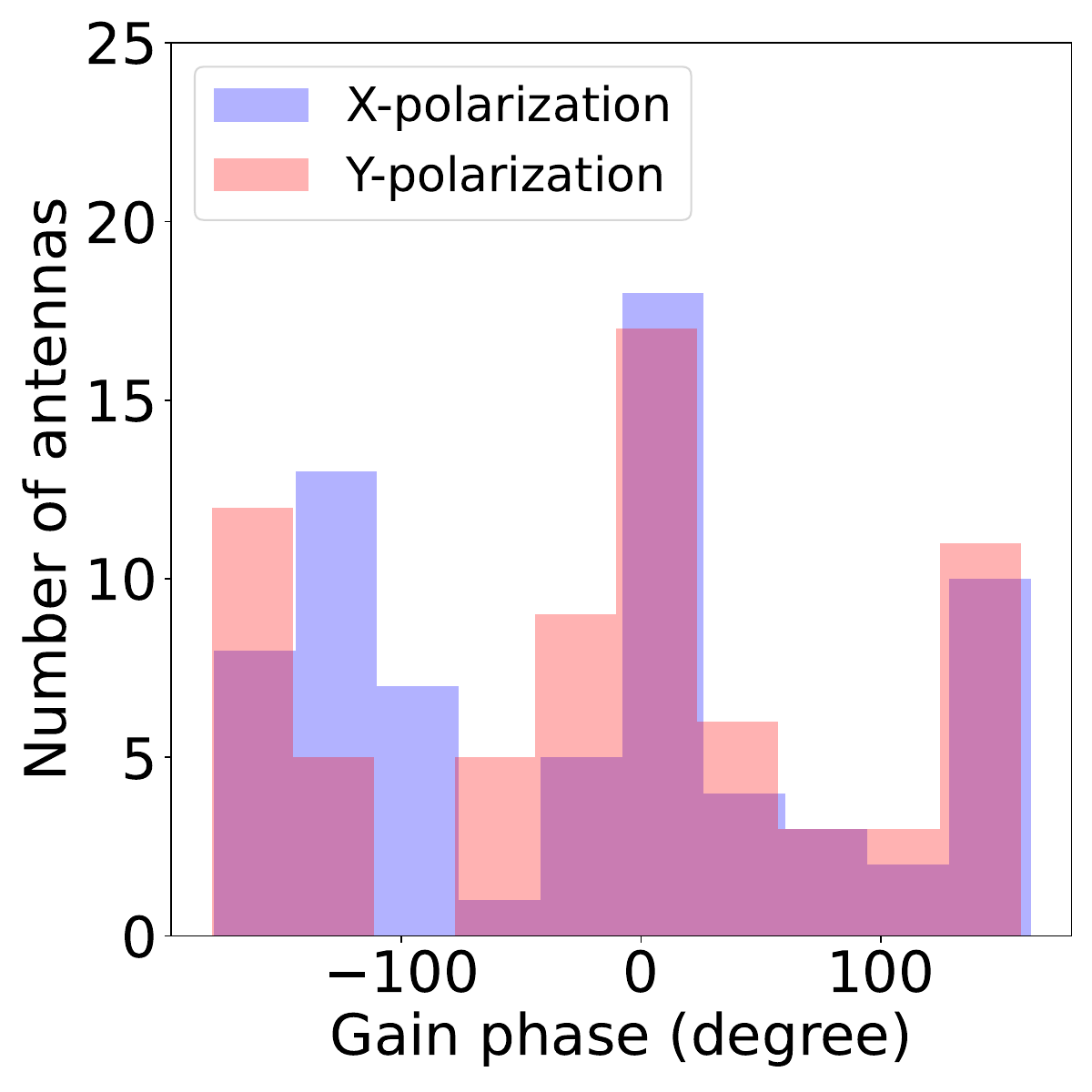}
    \includegraphics[trim={1.5cm 0cm 0cm 0cm},clip,scale=0.23]{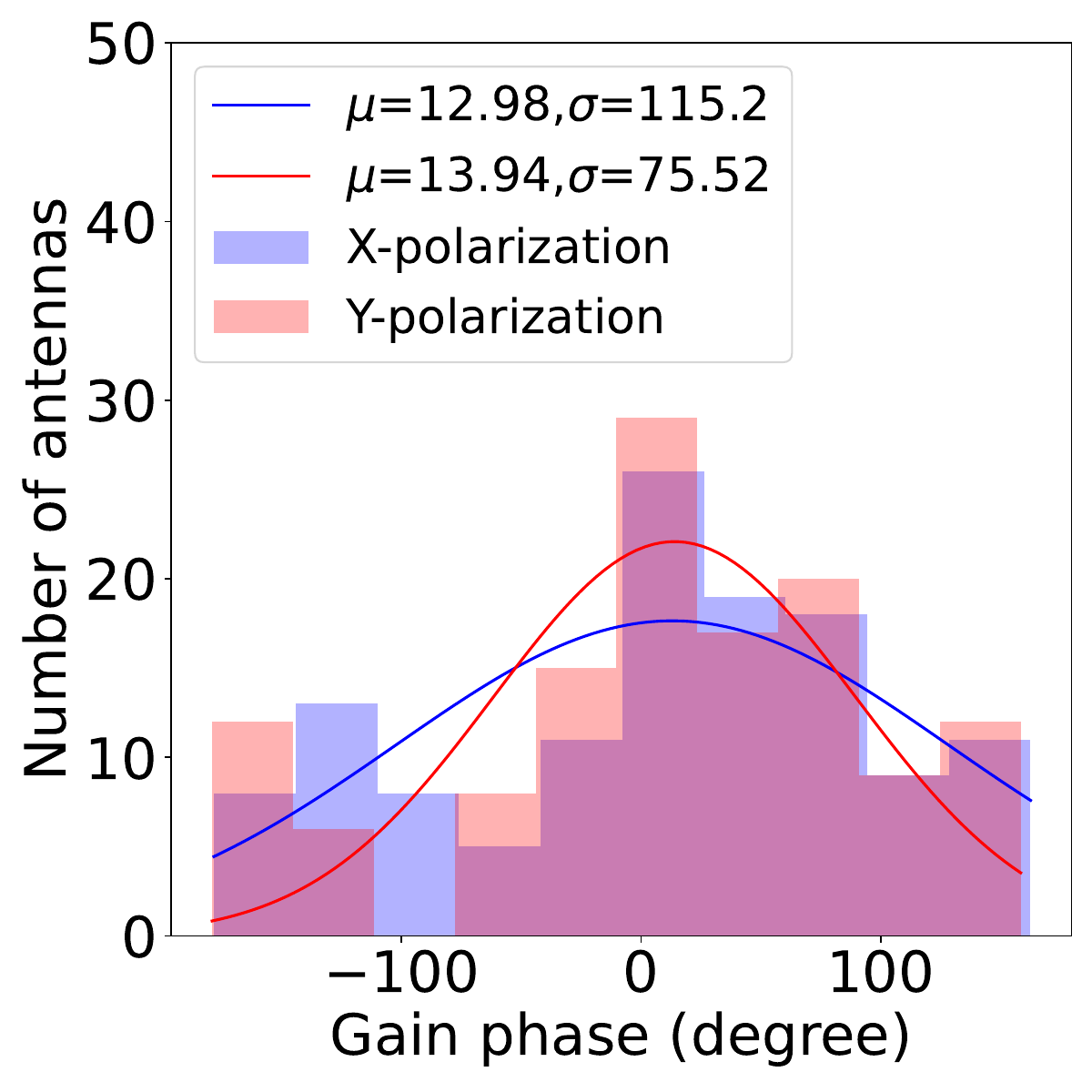}
    \caption[Distribution of phases of complex antenna gains.]{Distribution of phases of the antenna gains at 80 MHz for the observation on 2014 May 05. X polarization is shown in blue and Y polarization is shown in red. Distribution of phases of $G_\mathrm{i}$ is shown for {\it Left panel: } only the core antennas, {\it Middle panel: }for the non-core antennas, and {\it Right panel: }for all antennas.}
    \label{fig:gain_stats}
\end{figure}

\section{Comparison Between the Expected and Observed Statistical Properties of the Antenna Gains}\label{sec:stats}
A comparison between the expected and observed properties is done for the three sub-groups of antennas as mentioned in Section \ref{subsec:expected_stats}. As evident from Equation \ref{eq:measurement_eq}, the coherency of the visibilities are affected by $\phi_\mathrm{i}-\phi_\mathrm{j}$. Hence, the statistical properties of the $\phi_\mathrm{i}-\phi_\mathrm{j}$ are also discussed.

\subsection{Observed Properties of Antenna Gains}\label{subsubsec:antenna_gain}
The histograms of the phases of the complex gains are shown in Figure \ref{fig:gain_stats}. The distribution of phases only for the ``core antennas" is shown in the left panel and is well fitted with a Gaussian distribution with a standard deviation of $\sim50\ \mathrm{degrees}$. The distribution of the phases of only for the ``non-core antennas" is shown in the middle panel, and could not be fitted with a Gaussian distribution. The distribution of phase for ``all antennas" shown in the right panel can be fitted with a Gaussian distribution, but the standard deviation is larger ($\sim 100\ \mathrm{degrees}$) compared to ``core antennas". These observed properties match the expected properties as mentioned in Section \ref{subsec:expected_stats}.

\begin{figure}
    \centering
    \includegraphics[scale=0.27]{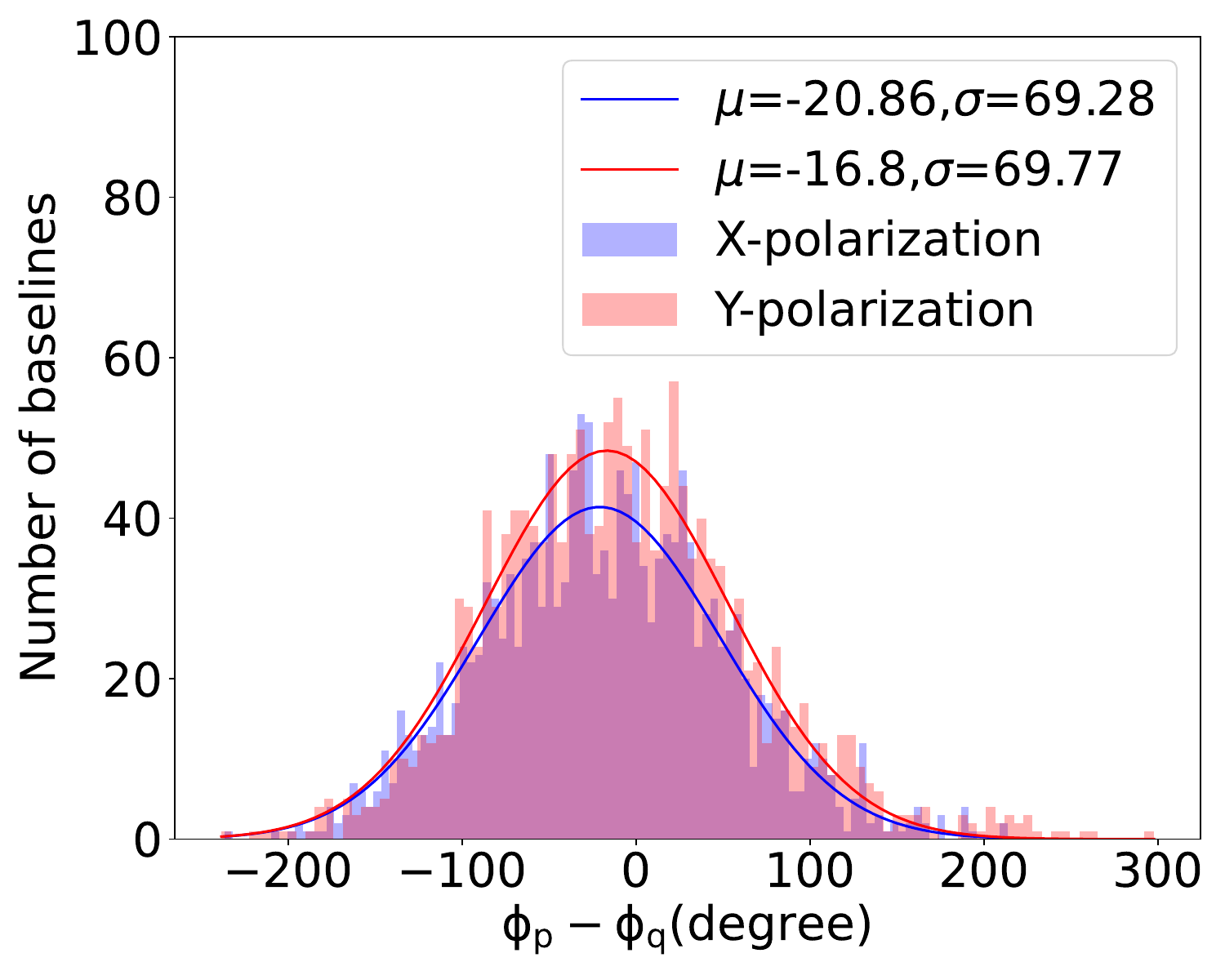}\includegraphics[scale=0.27]{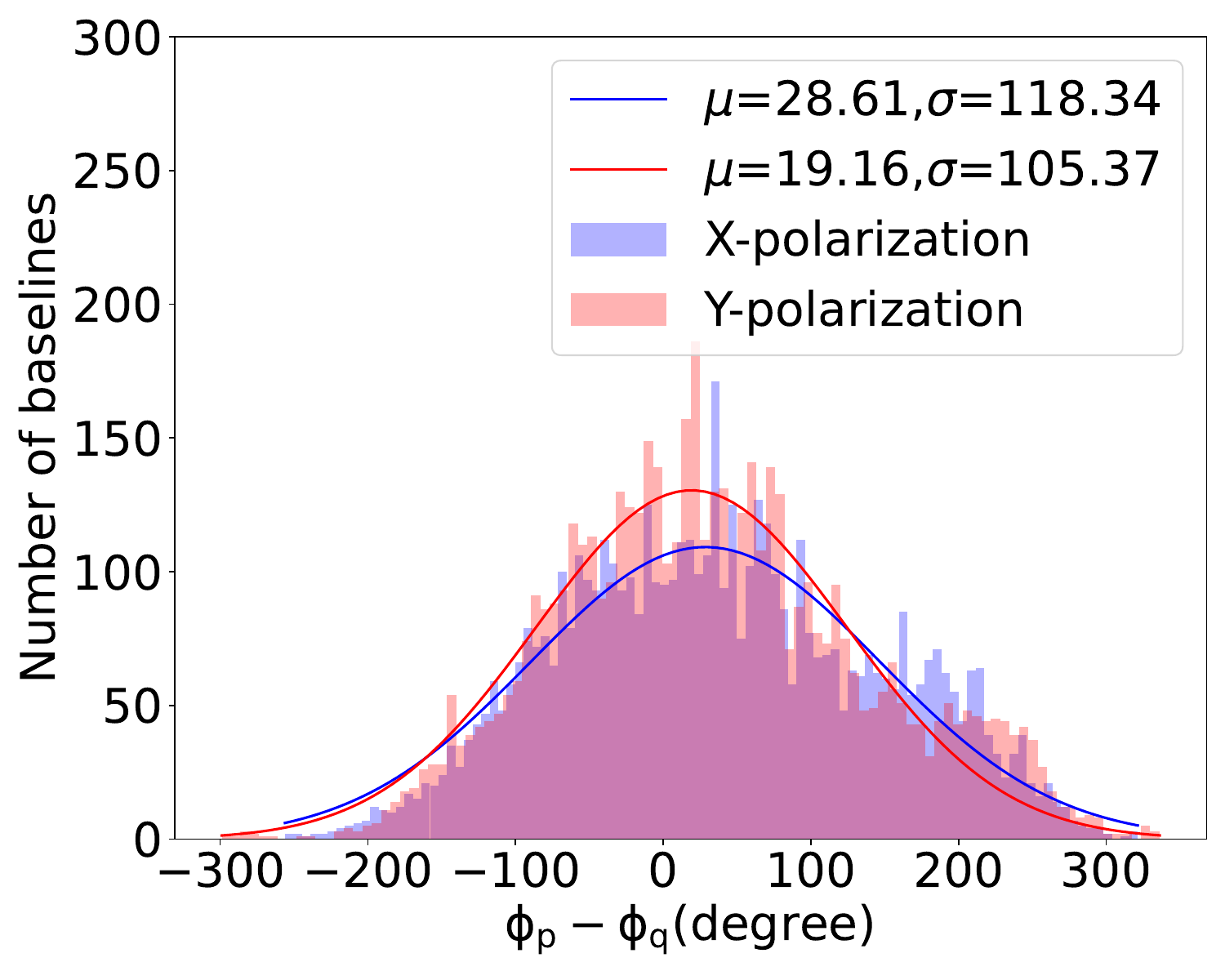}
    \caption[Distribution of the phase difference between antenna all pairs of antenna tiles.]{Distribution of $\phi_\mathrm{i}-\phi_\mathrm{j}$ at 80 MHz for the observation on 2014 May 05. X polarization is shown in blue and Y polarization is shown in red. {\it Left panel: }Distribution for core-all baselines. {\it Right panel: }Distribution of all baselines.}
    \label{fig:baseline_stats}
\end{figure}

\subsection{Observed Properties of $\phi_\mathrm{i}-\phi_\mathrm{j}$}
The histogram of $\phi_\mathrm{i}-\phi_\mathrm{j}$ for all the baselines originating from the core (core-all) is shown at the left panel and for all baselines are shown at the right panel of Figure \ref{fig:baseline_stats}. The standard deviation of the fitted Gaussian for the core-all histogram is much smaller ($\sim70\ \mathrm{degrees}$) compared to all baselines ($\sim 120\ \mathrm{degrees}$). Both these distributions follow a Gaussian distribution but, there are still slight deviations from the true Gaussian distribution at the edges, which is more prominent for the histogram including all baselines shown in the right panel.  

The observed statistical properties of both the phases and the difference between the phases of the antenna gains follow a quasi-Gaussian distribution. It is readily evident that they do not follow a ``uniformly random" distribution. On including all baselines, the standard deviation of the Gaussian becomes larger and also starts to deviate from the true Gaussian distribution, and the array loses coherency. But, the standard deviation is much smaller for core-all baselines, which provides better coherency for even the uncalibrated observed visibilities. The ability of the MWA to arrive at an initial source model (right panel of Figure \ref{fig:ini_model}) without any calibration applied from nighttime calibrator observations is due to the inherent coherency of the array just described. This is the primary reason why AIRCARS/P-AIRCARS can produce high DR images through a self-calibration approach alone even without a dedicated calibrator observation.

\section{Simulation}\label{sec:simulation}
In section \ref{sec:stats}, it is stated that AIRCARS/P-AIRCARS can proceed with the self-calibration from the uncalibrated observed visibilities because the distribution of phase and phase difference between the pairs of antenna tiles is not uniformly random. This phenomenological explanation is verified through simulation in this section.

\subsection{Description of the Simulation}
\begin{figure*}[!ht]
    \centering
    \includegraphics[trim={0cm 14cm 0cm 0cm},clip,scale=0.5]{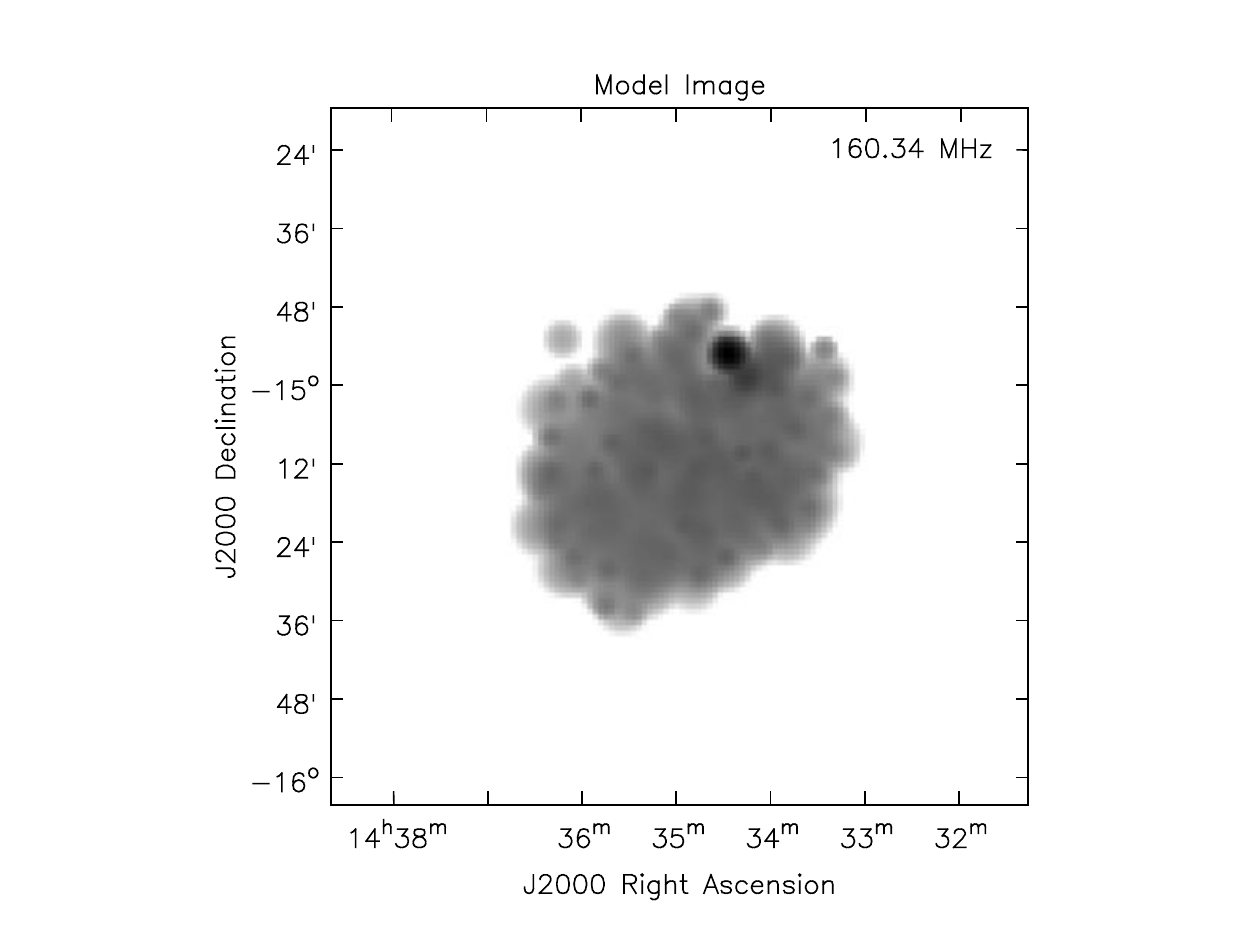}
    \caption[Model radio map of the Sun used for the simulation.]{Model image of the Sun from the observation on 2015 November 11 used for simulation.}
    \label{fig:model}
\end{figure*}
The simulation is done as follows --
\begin{enumerate}
    \item A model image of the Sun is obtained from the observation on 2015 November 11 (Figure \ref{fig:model}). This model is obtained using the imaging and deconvolution task {\it tclean}\footnote{It is based on the \textsf{CLEAN} algorithm \citep{Hogbom1974}. This algorithm assumes that the radio sky can be represented by a superposition of an adequate number of point sources. To produce the deconvolved images, it uses a simple iterative procedure to find the positions and strengths of these compact sources, convolves them with an idealized elliptical Gaussian point spread function (synthesized beam), and adds the residual noise to it.} of the commonly used software for radio interferometric data analysis, Common Astronomy Software Applications \citep[CASA,][]{mcmullin2007,CASA2022}. 
    \item The model image is then Fourier transformed to obtain the model {\it visibilities}; $V_\mathrm{{ij,model}}$.
    \item Antenna gains ($G_\mathrm{i}$) are simulated from a underlying distribution. Amplitudes are chosen to be unity. 
    \item Simulated visibilities are obtained as, $V_\mathrm{ij}^\prime=G_\mathrm{i}\ V_{\mathrm{ij,model}}\ G^\dagger_\mathrm{j}$. 
\end{enumerate}
\begin{figure*}[!ht]
    \centering
    \includegraphics[trim={0cm 0cm 0cm 0cm},clip,scale=0.32]{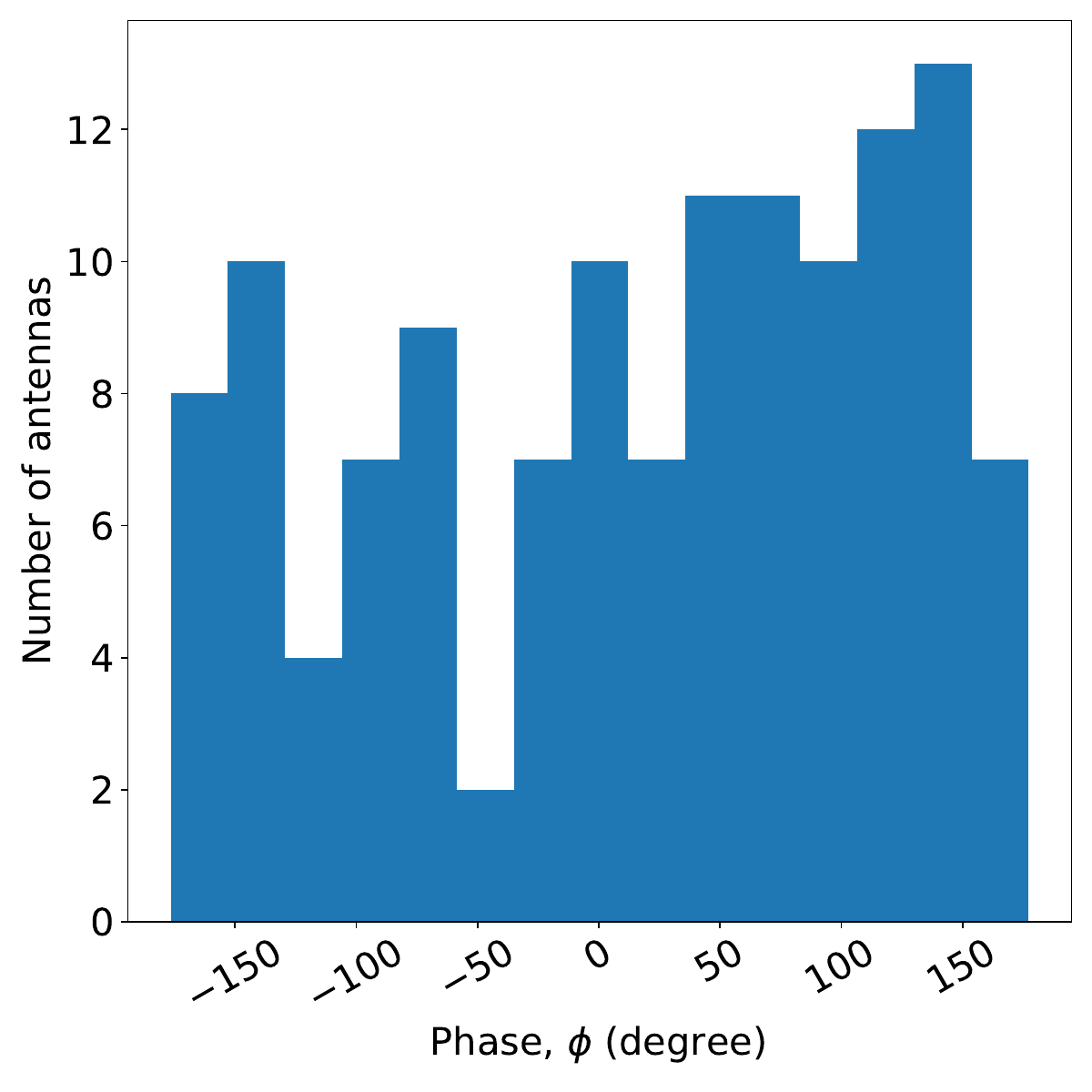}\includegraphics[trim={3cm 14cm 3cm 1cm},clip,scale=0.45]{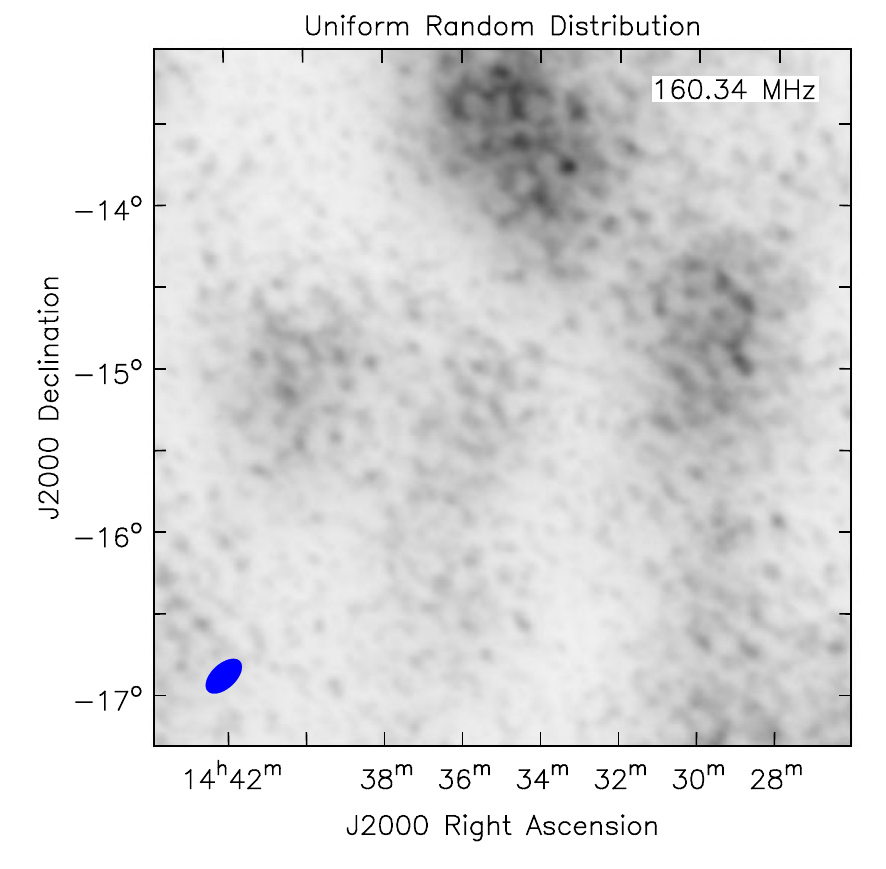}
    \caption[Dirty images from simulated visibilities from uniformly random phase distribution of antenna gains.]{{\it Left panel: }Uniformly random} distribution of the phase of the antenna gains. {\it Right panel: } Dirty image made from simulated visibilities.
    \label{fig:uniform}
\end{figure*}
The phases of the gains of antenna tiles are drawn from the two types of distributions between $-180$ to $+180\ \mathrm{degrees}$: 
\begin{enumerate}
    \item {\bf Uniformly random distribution: }The probability density function of the uniformly random distribution is given as,
    \begin{equation}
        p(x; \mathrm{a, b})=\frac{1}{\mathrm{a-b}}
    \end{equation}
     within the interval $\mathrm{[a, b)}$, and zero outside this range. 
     \item {\bf Truncated Gaussian random distribution: }The probability distribution function is given as,
     \begin{equation}
     \begin{split}
        p(x;\mu,\sigma,\mathrm{a,b)}=\frac{\phi({\frac{x-\mu}{\sigma}})}{\Phi({\frac{b-\mu}{\sigma}})-\Phi({\frac{a-\mu}{\sigma}})}
     \end{split}
     \end{equation}
     for $\mathrm{a\leq x\leq b}$ and $p=0$ otherwise. Here, $\phi(\zeta)$ is the probability distribution function of standard Gaussian distribution:
     \begin{equation}
     \begin{split}
         \phi(\zeta)=\frac{1}{\sqrt{2\pi}}\mathrm{exp}(-\frac{1}{2}\zeta^2)
     \end{split}
     \end{equation}
     and, $\Phi(\epsilon)=\frac{1}{2}[1+erf(\frac{\epsilon}{\sqrt{2}})]$ is the cumulative distribution function, where, $erf$ is the error function.
\end{enumerate}

\begin{figure}
    \centering
     \includegraphics[trim={0cm 0cm 0cm 0cm},clip,scale=0.25]{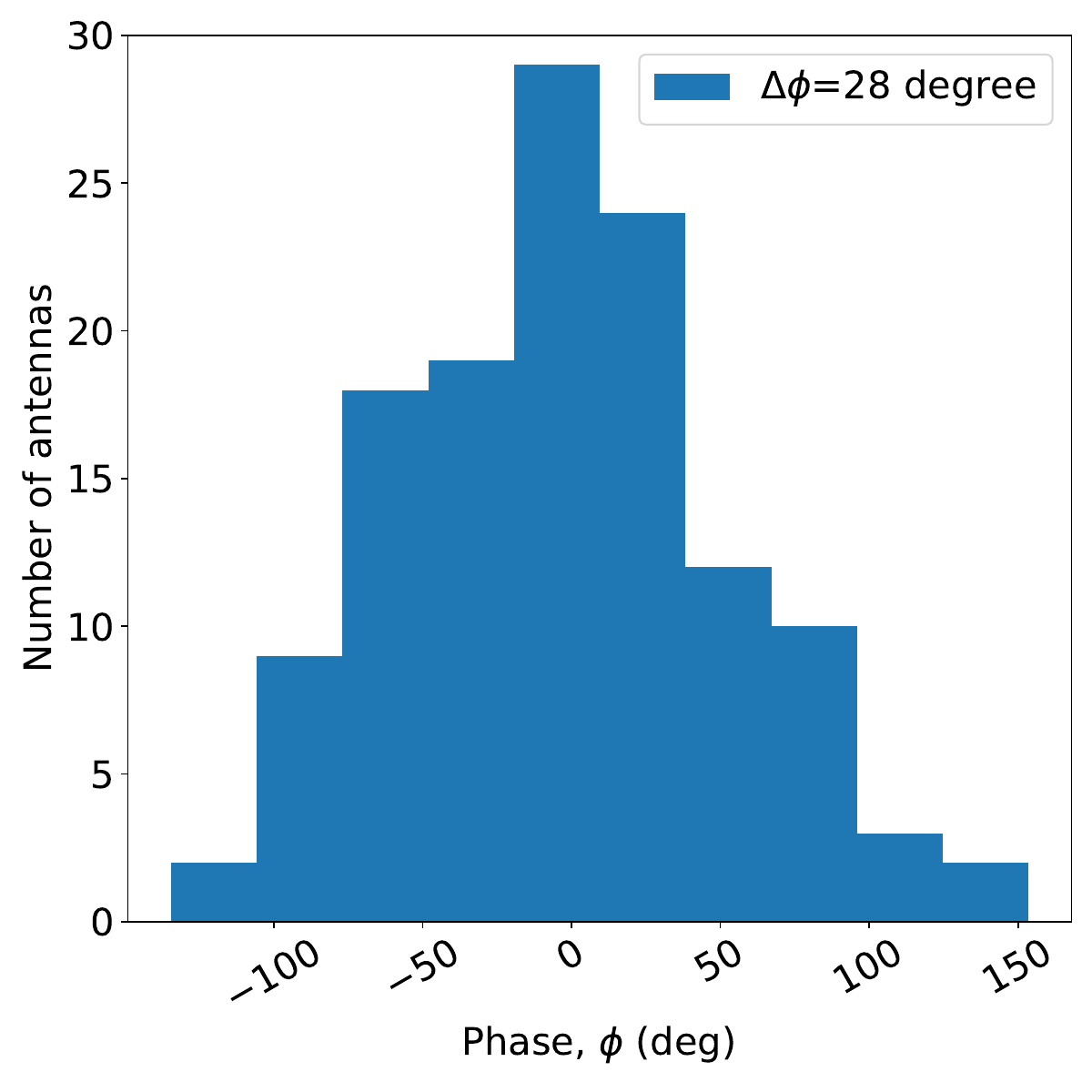}\includegraphics[trim={3cm 14cm 3cm 1cm},clip,scale=0.35]{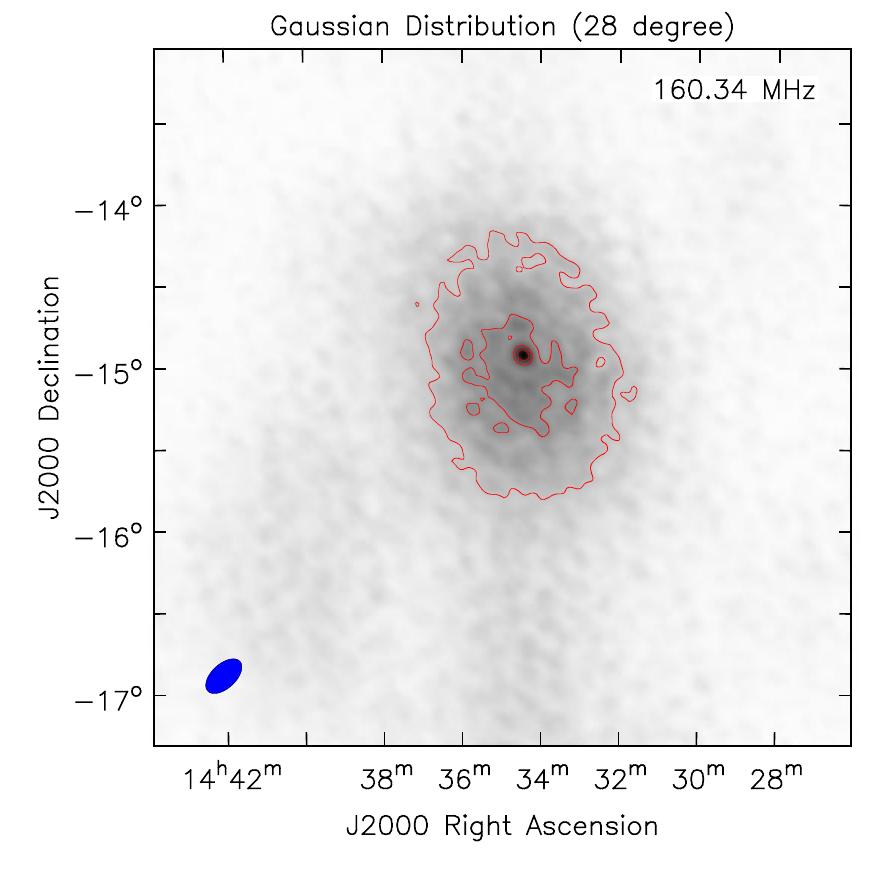}\\
    \includegraphics[trim={0cm 0cm 0cm 0cm},clip,scale=0.25]{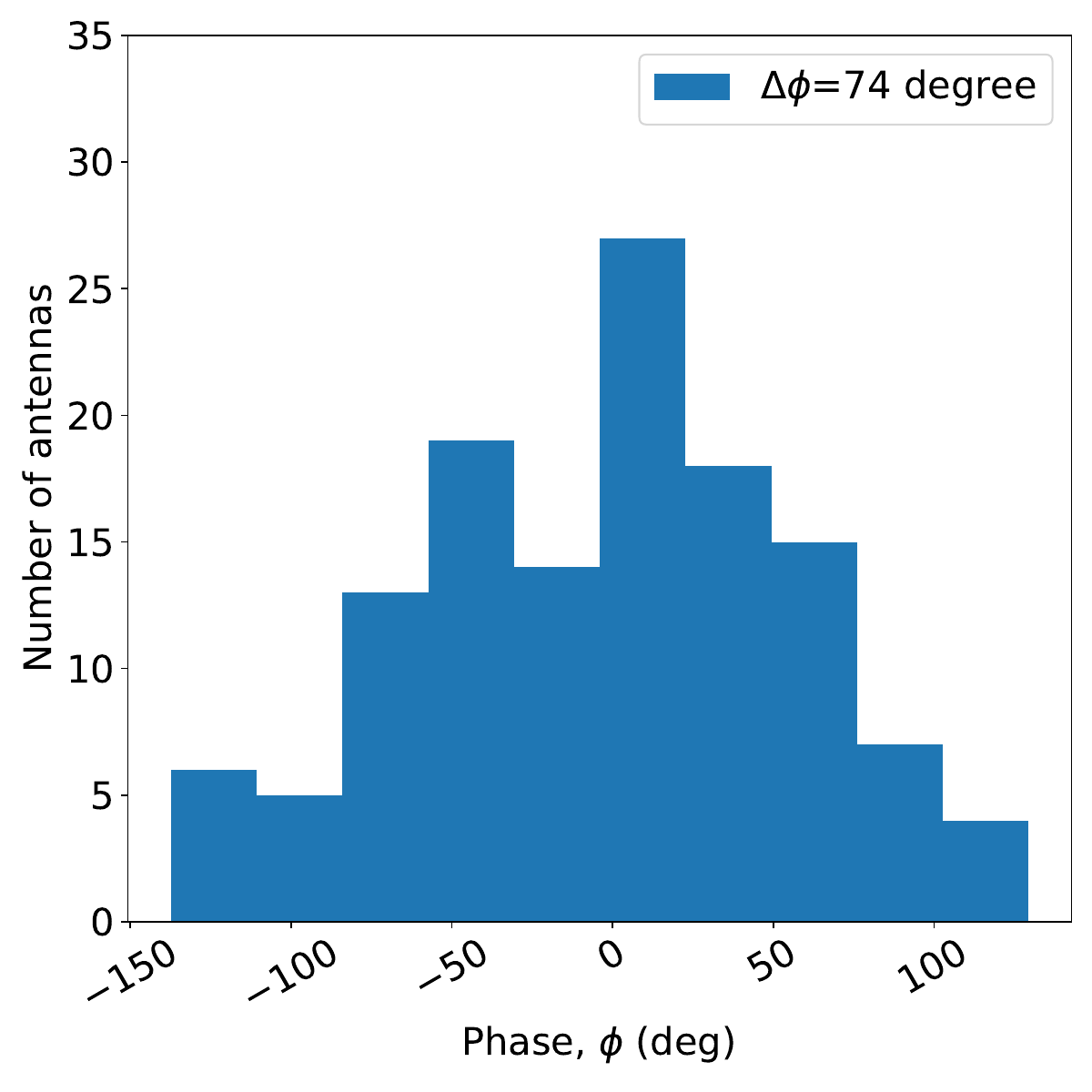}\includegraphics[trim={3cm 14cm 3cm 1cm},clip,scale=0.35]{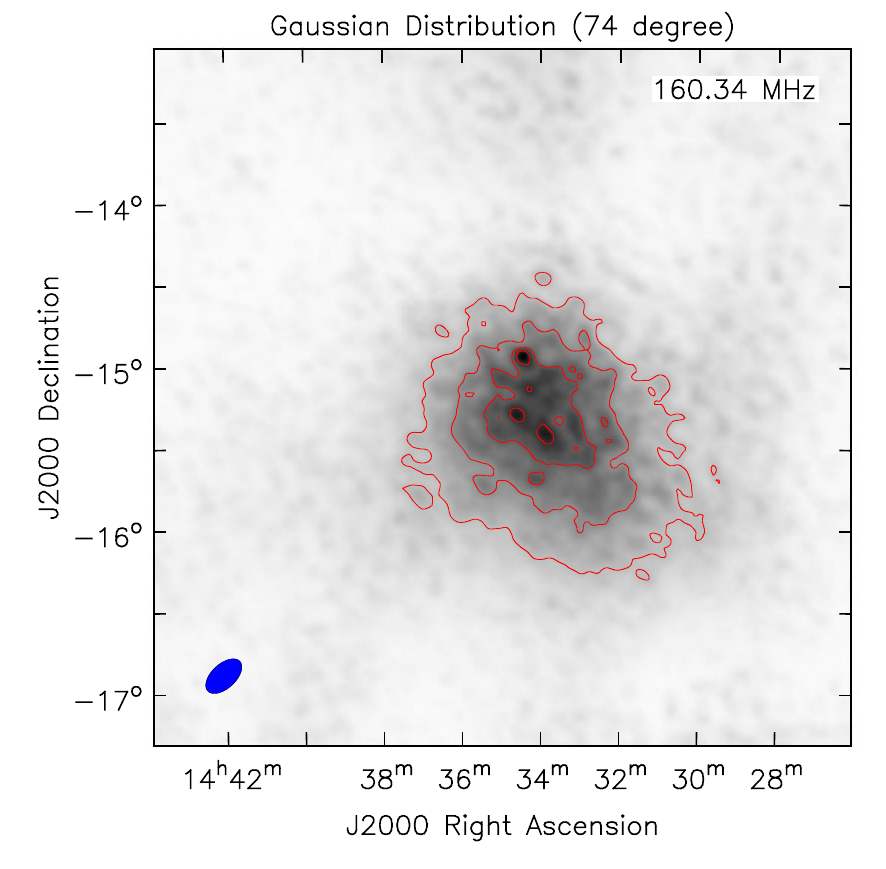}\\
    \includegraphics[trim={0cm 0cm 0cm 0cm},clip,scale=0.25]{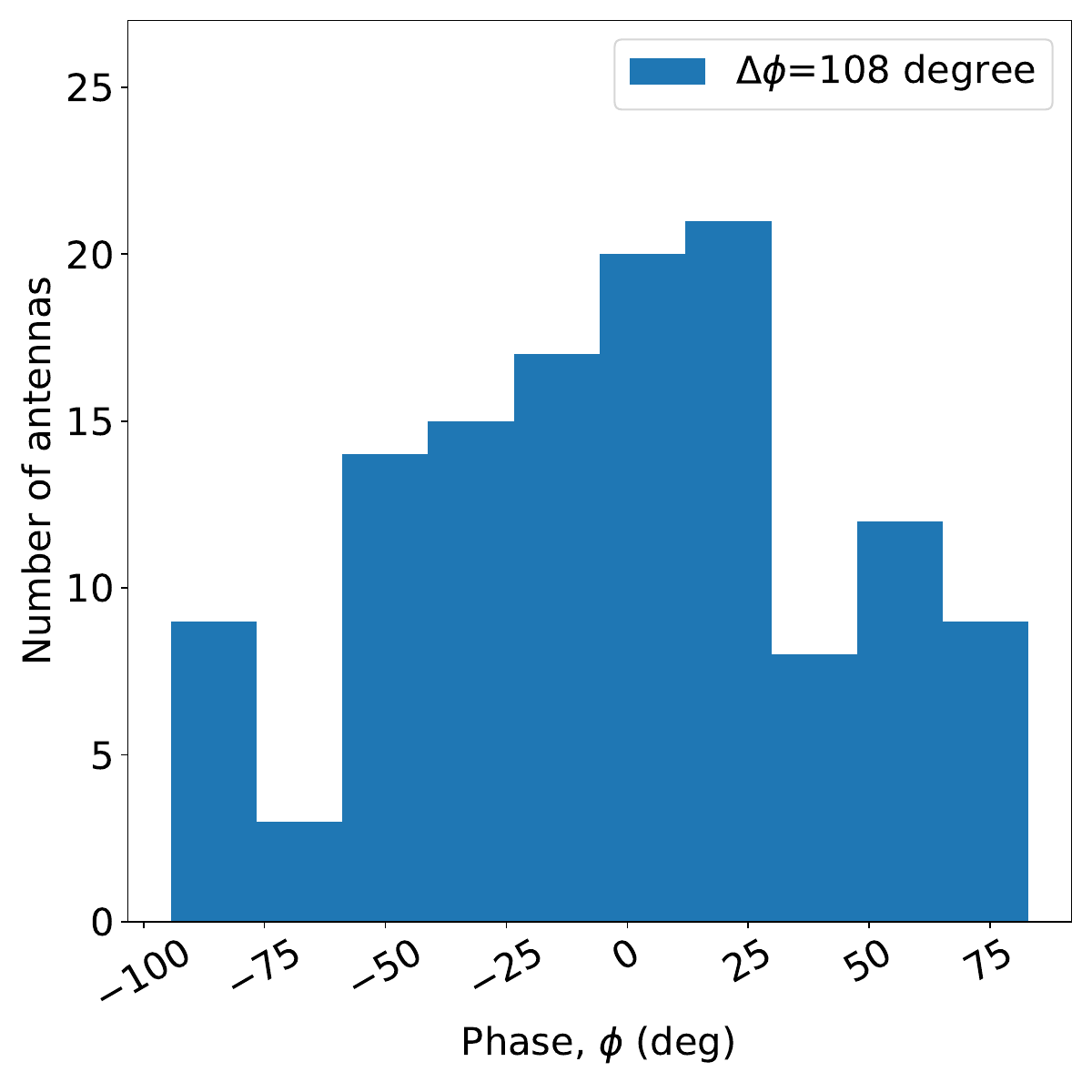}\includegraphics[trim={3cm 14cm 3cm 1cm},clip,scale=0.35]{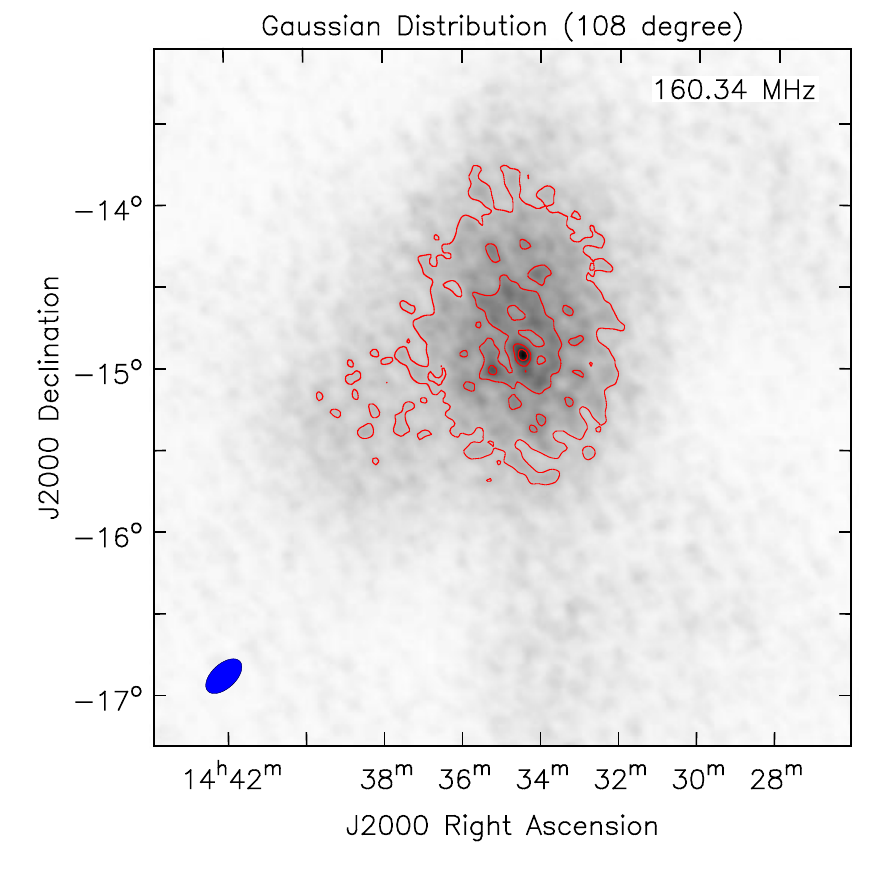}
    \caption[Dirty images from simulated visibilities from truncated Gaussian phase distribution of antenna gains.]{Dirty images from simulated visibilities from truncated Gaussian phase distribution of antenna gains. Left panels show the distribution of the simulated phases and the right panels show the corresponding dirty images. Results are shown for three truncated Gaussian distributions with standard deviations; {\it top panel: }28 degrees. {\it middle panel: } 74 degree, and {\it bottom panel: }108 degree.}
    \label{fig:gaussian}
\end{figure}

\subsection{Properties of the Initial Images Made from \\Simulated Visibilities}
The  main goal of  the simulation is to quantify the statistical parameters of a suitable distribution of the antenna phases, such that uncalibrated visibilities have some coherency and AIRCARS/P-AIRCARS can start the calibration without any dedicated calibrator observation.

The dirty image made from the simulated visibilities for a uniformly random distribution of the phase of the antenna gains is shown in Figure \ref{fig:uniform}. There is no source detected with more than 10-sigma (this is the default value used in AIRCARS/P-AIRCARS to pickup emission in the source model) significance near the phase center and the image looks noise-like. This demonstrates if the phases of the antenna gains follow a uniform random distribution, the array does not have any coherency. Hence, it is not possible to start the self-calibration without any dedicated calibrator observation.
\begin{figure}
    \centering
    \includegraphics[trim={0cm 0.2cm 0cm 0cm},clip,scale=0.6]{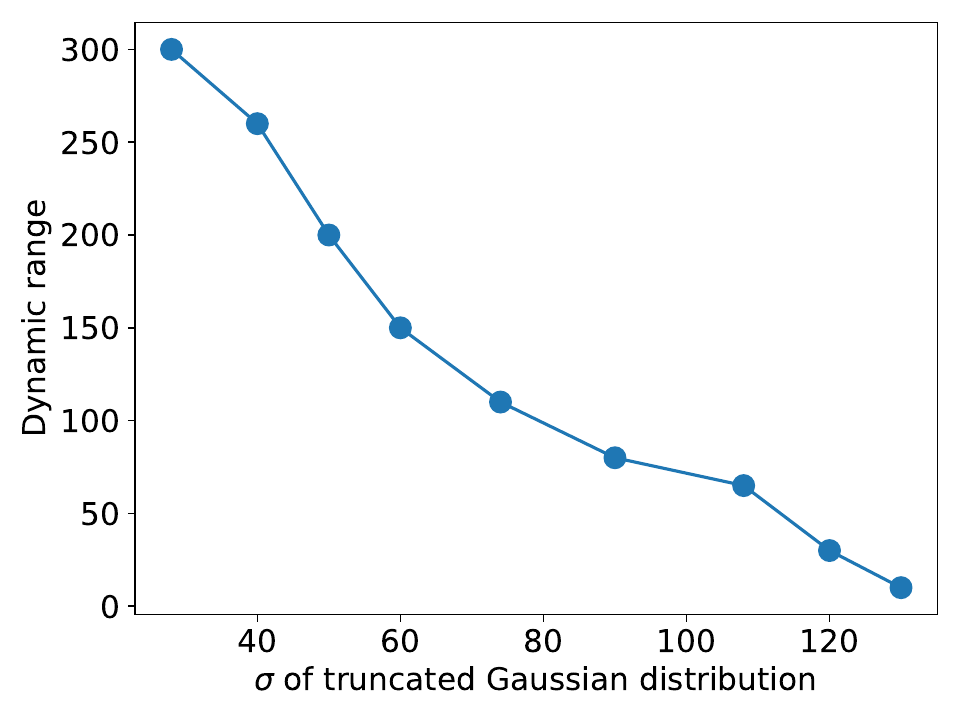}
    \caption[Variation of dynamic range of simulated images with the spread of the truncated Gaussian distributions.]{Variation of DR with the $\sigma$ of the truncated Gaussian distribution.}
    \label{fig:snr_with_phase}
\end{figure}

These simulations are also done for a number of truncated Gaussian random distributions with a wide range of standard deviations. Here the results from 3 sample standard deviations of 28, 74, and 108 degrees are shown. The distribution of the simulated phases is shown in the left panels of Figure \ref{fig:gaussian}. The dirty images made from the simulated visibilities are shown in the first column of Figure \ref{fig:gaussian}. In all three situations, there is a source emission detected with more than 10-sigma detection near the phase center. The DR of the images decreases with the increase in the standard deviation of the truncated Gaussian distribution. The DR of the images is 300, 110, and 55, respectively, for the truncated Gaussian distributions with $\sigma$ of 28, 74, and 108 $\mathrm{degrees}$. DR is plotted against $\sigma$ in Figure \ref{fig:snr_with_phase}, which monotonically decreases with the increase in $\sigma$. It has been found that for $\sigma\geq\ 120\ \mathrm{degrees}$, DR becomes lower than 20.

\section{Discussion and Summary}\label{sec:discussion_paircars_principle}
AIRCARS/P-AIRCARS is a self-calibration based algorithm. Any self-calibration based algorithm has some intrinsic limitations like the loss of absolute flux density scale and astrometric accuracy. Both the flux density calibration and astrometric accuracy are important for cross-comparisons with observations at other wavelengths or with synthetic radio maps from forward models. In the past, the solar flux density calibration was done using an instrumental gain-independent method described by \citet{oberoi2017} and \citet{Mohan2017spreads}. Recently a new technique has been developed which utilizes the instrumental characterization and very stable instrumental bandpass of the MWA \citep{Kansabanik2022}, which is discussed in detail in Chapter \ref{fluxcal}. This flux density calibration method is more general and is now the default technique for this purpose. The astrometric correction in AIRCARS/P-AIRCARS is done based on an image-based approach, which provides astrometric accuracy better than the PSF size (a few arcmins). The detailed description of that method is discussed in Chapter \ref{paircars_algorithm}. 

One of the novel features of AIRCARS/P-AIRCARS is that it can perform the calibration of the solar observation with the MWA even without any dedicated calibrator observations exploiting the partial coherency of the MWA array. The partial phase stability is demonstrated using simulation for Gaussian distribution (Figure \ref{fig:gaussian}), but in practice, the distribution of phase may not follow a true Gaussian distribution. However, the fact that the true phase distribution (Figure \ref{fig:gain_stats}) has a strong peak, and the baselines comprising the antennas whose phase lies close to the strong peak will have always some coherency between them and hence can be used to produce a reasonably accurate source model to start the self-calibration procedure even in the absence of calibrator observation.

During solar maxima, ionospheric activities are expected to be larger. Hence it is instructive to test the AIRCARS/P-AIRCARS on the datasets from both solar maxima and solar minima. All the examples shown in this chapter are from 2014 and 2015, which are close to the maxima of the solar cycle 24. AIRCARS/P-AIRCARS has been tested on a dataset covering both solar maxima and solar minima and has worked successfully. This demonstrates the robustness of the AIRCARS algorithm, which is independent of solar and ionospheric conditions. In this chapter, I demonstrate this statistically and quantitatively.

It is anticipated that AIRCARS/P-AIRCARS will serve the purpose of calibration and imaging for future radio interferometers if certain conditions are satisfied by the array:
\begin{enumerate}
   \item Instrumental gains of all antennas should be similar. This demands precision in the manufacturing of the antenna elements.
    \item The delays introduced by the electronic cables needs to be measured properly at regular intervals and corrected before performing the cross-correlation (as is usually the case with most radio interferometers). This will reduce the loss of coherency.
    \item A large number of antennas needs to be distributed over a small array footprint, such that all the baselines originating from the core are dominated by the core-core baselines independent of the array footprint.
\end{enumerate}

Among these three criteria, the third one depends on the array configuration. This is expected to be satisfied by the future Square Kilometre Array Observatory \citep[SKAO,][]{SKAO2021} and, some other next-generation radio interferometers; like the Next Generation Very Large Array \citep[ngVLA,][]{ngVLA2019}, and the Frequency Agile Solar Radiotelescope \citep[FASR,][]{Gary2003,Bastian2005,Bastian2019,Gary2022_FASR}. ngVLA is planned to observe from $1-115\ \mathrm{GHz}$ and has three separate array configurations that will operate in parallel. Among these three array configurations, the Short Baseline Array (SBA) consisting of $19\times6\ \mathrm{m}$ antennas located at the current VLA site is highly suited for high-fidelity spectroscopic snapshot imaging of the Sun. FASR will be a solar dedicated radio interferometer operating in the range $0.2-20\ \mathrm{GHz}$. Two separate array configurations have been proposed \citep{Bastian2019} for FASR, which will provide dense {\it uv-}coverage over a large bandwidth. The array footprint of FASR is similar to the MWA phase-I, hence AIRCARS is expected to work efficiently on future FASR observation.

Based on the working principle discussed in this chapter, I have developed a state-of-the-art polarization calibration and imaging algorithm, P-AIRCARS, which is a successor of AIRCARS. Since this algorithm is based on self-calibration, it can provide high DR solar images. Along with high DR, one needs precise calibration of the absolute flux density and polarization of the source. In the following chapters, I describe the polarization and flux density calibration algorithms in detail to obtain high-fidelity spectropolarimetric snapshot solar radio images with the MWA.
\chapter {State-of-the-art Polarimetric Calibration Algorithm for Solar Observations}
\label{paircars_algorithm}
The quality of a radio interferometric image can be captured in two key metrics -- dynamic-range (DR) and fidelity. DR defines the contrast in the image and determines how faint an emission can one detect reliably in the presence of much brighter emissions. On the other hand, the fidelity of an image can be regarded as a metric of the reliability of the quantities estimated from the image like source structure, flux density, and polarization. The relevance of measuring precise flux density is discussed in next Chapter \ref{fluxcal}. In this chapter, I describe a state-of-the-art calibration algorithm for obtaining high-fidelity spectropolarimetric solar radio images using the observations from the MWA and other low radio frequency interferometers with a centrally condensed configuration, including the Square Kilometre Array Observatory (SKAO)-Low. The work presented in this chapter is based on \citet{Kansabanik2022_paircarsI}, which was published in the Astrophysical Journal.

\section{Introduction}
Spectropolarimetric radio observations can provide several remote sensing observing techniques for measuring the coronal magnetic fields and other coronal plasma parameters as described in Section \ref{subsec:radio_observation_cme} of Chapter \ref{chapter_intro}. Under favorable circumstances, radio observations have been used to estimate coronal magnetic fields associated with active regions and/or CMEs \citep[e.g.][]{Vourlidas2020,Carley2020,Alissandrakis2021}. The coronal optical depth at these heights ($>1.3\ R_\odot$) becomes too low for visible and extreme ultraviolet (EUV) bands. Although the radio observables, in principle, are sensitive to coronal magnetic fields, it has been technically too challenging to extract this information regularly. Most of the radio studies have focused on active emissions, and the large-scale quiescent coronal magnetic fields at higher coronal heights have remained beyond reach.
    
The polarization properties of solar radio emission, in addition to being a direct probe of the coronal magnetic fields, can also provide strong constraints on the emission mechanisms. Despite its well-appreciated importance, low-frequency polarimetric observations of the Sun are one of the least explored areas of solar physics. The radio Sun is a complicated source. It has structures spanning a large range of angular scales. The spectral, temporal, and morphological characteristics of radio emissions are also very dynamic. The brightness temperature ($T_\mathrm{B})$ of the low-frequency solar emissions can vary from $\sim10^3-10^4\ \mathrm{K}$ for gyrosynchrotron (GS) emission from CME plasma \citep[e.g.][]{bastian2001,Mondal2020a} to $\sim10^{13}\ \mathrm{K}$ for bright type-III radio bursts \citep[e.g.][]{McLeanBook,Reid2014} over a background quiescent $T_\mathrm{B}$ of $\sim10^6\ \mathrm{K}$ (shown by red bars in the top panel of Figure \ref{fig:different_emissions} in Chapter \ref{paircars_principle}). Depending upon the emission mechanism at play, the polarization fraction can vary from $\lesssim 1\%$ to $\sim100\%$ \citep[e.g.][]{McLeanBook,Nindos2020} (shown by blue bars in the top panel of Figure \ref{fig:different_emissions} in Chapter \ref{paircars_principle}).

To date, most of the polarimetric studies of the Sun at low frequencies are based on non-imaging dynamic spectra measuring the circular polarization \citep[e.g.][]{Reid2014,Kaneda2017}. These observations cannot provide any information about the source structure or location. Some innovative instruments use simultaneous Stokes I imaging and Stokes V dynamic spectra to help in the localization of the source of active emission \citep{Raja2014}. These studies implicitly assume the locations of the peaks in the Stokes I and Stokes V emission to be the same. This assumption usually holds when there is a single dominant source of emission. This approach is not useful when multiple sources of active emission are simultaneously present on the Sun \citep{Mohan2017spreads} or for weaker and/or extended emission like GS emission from CME plasma \citep[e.g.][]{bastian2001, Mondal2020a} and the free-free emission from the quiet Sun \citep{Sastry_2009}.

The variation in the solar emission over small temporal and spectral scales imposes a requirement for snapshot spectroscopic imaging. The need to be able to see features varying vastly in $T_\mathrm{B}$ highlights the need for a high imaging DR. Only recently it has become possible to meet these exacting requirements for solar radio imaging with the MWA, as discussed in Chapter \ref{paircars_principle}. Having established the ability of MWA solar observations to deliver high DR Stokes I images \citep{Mondal2019}, the next logical step is to produce high-fidelity polarimetric images.

Full-Stokes calibration is significantly more challenging than working with Stokes I alone. These challenges are even greater for the case of low radio frequency solar imaging. On the one hand, solar emission can have a very large range of intrinsic polarizations, which can also vary rapidly across time and frequency, on the other the MWA has a large field of view (FoV). Based on the FWHM of the primary beam, at 150 {$\mathrm{MHz}$} the FoV of the MWA is $\sim$610 $\mathrm{degree^2}$, which reduces to $\sim$375 $\mathrm{degree^2}$ by 200 $\mathrm{MHz}$ \citep{Tingay2013}. The wide FoV aperture arrays tend to have large instrumental polarization imposing a strong requirement for precise calibration. In fact, some of the assumptions made for routine polarimetric calibration at higher frequencies for small FoV instruments no longer hold in this regime \citep{lenc2017}. Hence, I have developed a general algorithm for polarimetric calibration suitable for our application. This algorithm is implemented and demonstrated its efficacy on the MWA solar data. The algorithm will be well-suited for solar imaging with the future SKAO-Low and other interferometers with centrally condensed array configurations.

The chapter is organized as follows. First, I briefly discuss some basics of polarization calibration in Section \ref{sec : basic_polarimetry} to build up the base for the calibration algorithm. Section \ref{sec : challenges} describes the challenges of the polarization calibration of the Sun at low frequencies and the limitations of the conventional methods of polarization calibration. I then describe the new algorithm in Section \ref{Overview of the Algorithm}. Section \ref{sec : result} demonstrates the outcomes of the algorithm with a discussion, and Section \ref{Conclusion_paircars} provides the conclusion.

\section{Polarization Calibration Framework of A Radio Interferometer}\label{sec : basic_polarimetry}
A radio interferometer is made up of several radio antennas or antenna elements. These antennas measure the voltages corresponding to the two orthogonal polarizations of the electric field, $\vec{\mathrm{E}}$, incident on the antenna. $\vec{\mathrm{E}}$ could be measured in either of the linear or circular bases -- ($\mathrm{E_X},\ \mathrm{E_Y}$) or ($\mathrm{E_R},\ \mathrm{E_L}$), respectively. Incident $\vec{\mathrm{E}}$ induces a voltage in the antenna, and the primary observable of an interferometer is the cross-correlation between the components of the induced voltages for every antenna pair, referred to as {\it visibilities}. To capture the complete information about the state of polarization of $\vec{\mathrm{E}}$, for any given antenna pair described by indices $\mathrm{i}$ and $\mathrm{j}$, an interferometer needs to measure a set of four visibilities, $\mathrm{X_iX_j^\dagger},\ \mathrm{X_iY_j^\dagger},\ \mathrm{Y_iX_j^\dagger},\ \mathrm{Y_iY_j^\dagger}$ (or equivalently $\mathrm{R_iR_j^\dagger},\ \mathrm{R_iL_j^\dagger},\ \mathrm{L_iR_j^\dagger},\ \mathrm{L_iL_j^\dagger}$){\footnote{$\dagger$ represents conjugate transpose.}}. This complete set of visibilities is often referred to as {\it full-polar} visibilities.

The measured visibilities include corruption due to atmospheric propagation effects and instrumental effects. To arrive at the true visibilities corresponding to the astronomical sources, these corruptions need to be removed. This process is known as calibration. \cite{Hamaker1996_1} proposed a general mathematical framework for polarimetric calibration, which is commonly known as {\it measurement equation} framework. Briefly, the measured complex voltage vector, $\vec{\mathrm{V}}$, per antenna can be expressed in terms of $\vec{\mathrm{E}}$, and the antenna-based Jones matrix, $\mathrm{J}$, \citep{Jones1941} as:
\begin{equation}\label{eq:jones_matrix}
\begin{split}
    \vec{\mathrm{V}} &=\mathrm{J}\ \vec{\mathrm{E}}\\
    \begin{pmatrix}\mathrm{X}\\\mathrm{Y}\end{pmatrix} &=\mathrm{J} \begin{pmatrix}\mathrm{E_X}\\\mathrm{E_Y}\end{pmatrix},
\end{split}
\end{equation}
where $\mathrm{J}$ is a $2\times2$ matrix representing the instrumental and atmospheric effects. The measured correlation products for two antennas represented by indices $\mathrm{i}$ and $\mathrm{j}$ can be written as a $2\times2$ matrix, also known as the {\it Visibility matrix}, as follows \citep{Hamaker2000,Smirnov2011}:
\begin{equation}\label{eq:visibility_matrix}
    \begin{split}
      { \mathrm{V_{ij}}^\prime}&= 2\begin{pmatrix} \mathrm{X_iX_j^\dagger} & \mathrm{X_iY_j^\dagger}\\\mathrm{Y_iX_j^\dagger} & \mathrm{Y_iY_j^\dagger}
        \end{pmatrix}\\
       \mathrm{V_{ij}^\prime}&=2\ \mathrm{J_i}
        \begin{pmatrix}
        \mathrm{E_{X,i}E_{X,j}}^\dagger & \mathrm{E_{X,i}E_{Y,j}^\dagger}\\\mathrm{E_{Y,i}E_{X,j}^\dagger} & \mathrm{E_{Y,i}E_{Y,j}^\dagger}
        \end{pmatrix}
       \mathrm{J_j^\dagger}\\
       \mathrm{V_{ij}^\prime}&=\mathrm{J_i}\begin{pmatrix}
       \mathrm{V_I+V_Q} & \mathrm{V_U+iV_V} \\ \mathrm{V_U-iV_V} & \mathrm{V_I-V_Q}
       \end{pmatrix}_\mathrm{ij}\mathrm{J_j^\dagger}\\
      \mathrm{V_{ij}^\prime}&= \mathrm{J_i\ V_{ij}\ J_j^\dagger},
    \end{split}
\end{equation}
where $\mathrm{V_{ij}}$ is the true source visibility matrix; $\mathrm{V_{ij}^\prime}$ is the observed visibility matrix; and  $\mathrm{V_I},\ \mathrm{V_Q},\ \mathrm{V_U}$ and $\mathrm{V_V}$ are the Stokes visibilities of the incident radiation. 
Here we have followed the IAU/IEEE definition of the Stokes parameters \citep{IAU_1973,Hamaker1996_3}. The four Stokes parameters, $\mathrm{I,\ Q,\ U,\ V}$ were originally defined by \citet{stokes1851}. Stokes I represents the total intensity; Stokes Q and U represent the linear polarization and circular polarization is denoted by Stokes V. All $\mathrm{V_{ij}^\prime}$s have independent additive noise, $\mathrm{N_{ij}}$, associated with them, and the Equation \ref{eq:visibility_matrix} can be written as:
\begin{equation}\label{eq:measurement_equation}
    \begin{split}
         \mathrm{V_{ij}^\prime}&=\mathrm{J_i\ V_{ij}\ J_j^\dagger + N_{ij}}.
    \end{split}
\end{equation}
Equation \ref{eq:measurement_equation} is referred to in the literature as the {\it measurement equation} of a radio interferometer \citep{Hamaker1996_1,Hamaker2000,Smirnov2011}.

The objective of polarization calibration is to estimate $\mathrm{J_i}$s for all antennas and obtain $\mathrm{V_{ij}}$ from the $\mathrm{V_{ij}^\prime}$. 
This requires correcting four different aspects.
These aspects and their impacts are enumerated below.
\begin{enumerate}
\item {\bf Time-variable complex gain:} While they are independent in origin, in practice, it is not feasible to disentangle the atmospheric propagation effects from time-variable instrument gains. So these effects are clubbed with instrumental gains in the {\it measurement equation} formalism. The impact of these gain variations is to make the interferometer incoherent.
\item {\bf Frequency-dependent instrumental bandpass:} This can modify the true spectral signature of the source and introduce incoherence across the frequency axis.
\item {\bf Polconversion:} This can lead to leakage from Stokes I to other Stokes parameters and thus modifies the magnitude of the observed polarization vector, $\mathrm{p=(Q,\ U,\ V)}$.
\item {\bf Polrotation:} This is the mixing between Stokes Q, U, and V and leads to a rotation of $\mathrm{p}$.
\end{enumerate}

The time and frequency dependence of the instrumental gains arises due to the nature of the signal chain and the atmospheric propagation effects. These are routinely corrected for in standard interferometric calibration. Ideally, a set of orthogonal receptors are expected to receive only the matched orthogonal component of the incident $\vec{\mathrm{E}}$. In practice, reasons ranging from proximity to other receptors, and manufacturing tolerances to cross-talk between closely placed cables, imply that a given orthogonal receptor also picks up some amount of signal of the other component of $\vec{\mathrm{E}}$. This mixing of the orthogonal components of $\vec{\mathrm{E}}$ gives rise to {\it polconversion}. Things like misalignment of the dipoles with respect to the sky coordinates and the phase differences between the two orthogonal receptors give rise to {\it polrotation}. 

\section{Challenges and Limitations}
\label{sec : challenges}
Although a general mathematical foundation of polarization calibration has been provided by \citet{Hamaker2000} and a more recent review is available in \citet{Smirnov2011}, the complexity of the problem and its computation-heavy nature have restricted most commonly available implementations to make some simplifying assumptions. This is an active area of research, especially because of the upcoming ambitious facilities like the SKAO, and new algorithms and implementations are being developed by multiple groups across the world \citep[e.g.][etc.]{Smirnov2011,Mitchell2008,stefcal2014,cubical2018}. This section lists the challenges of low-frequency solar polarimetry and the limitation of conventional algorithms and earlier attempts.

\subsection{Challenges of Low-frequency Solar Polarimetry}
These challenges are related to the large FoVs and the nature of low-frequency aperture array instruments, which have been discussed in detail in Section \ref{sec:challenges} of Chapter \ref{chapter_intro}. Conventional methods correct for polconversion using observations of a strong unpolarized calibrator source. Correcting for polrotation requires observations of a single(multiple) strong source(s) with known polarization properties \citep{Hamaker2000,Hales2017}.

The large FoV of low radio frequency aperture array instruments implies that:
\begin{enumerate}
    \item As there are no moving parts and the beam is steered electronically, the primary beam of the instrument can vary dramatically with pointing direction.
    \item Given the large FoV and nature of low radio frequency sky, there is no single polarized source strong enough to dominate the observed $\mathrm{V_{ij}^{\prime}}$.
\end{enumerate}
This requires that, for good polconversion calibration, the target field and the calibrator should be observed with the same pointing, which is rarely the case. The lack of a dominant polarized source makes it hard to do polrotation calibration. In addition, the observations at low radio frequencies require one to contend with the direction-dependent ionospheric distortion and Faraday rotation (FR). A comprehensive discussion of a successful approach to deal with these challenges has been provided by \citet{lenc2017}.

\subsection{Limitations of Conventional Algorithms}
\label{sec:conventional-algos}
Most of the standard interferometric calibration and imaging packages like CASA \citep{mcmullin2007,CASA2022} and AIPS \citep{Wells1985} implemented a linearized form of the {\it measurement equation}. While they have been spectacularly successful in delivering high-quality polarimetric images, the following two assumptions must be satisfied for this formalism to be valid:
\begin{enumerate}
    \item The instrumental polarization must be small ($\lesssim 10$\%).
    \item The fractional polarization of the sources used for calibration should be low \newline($\lesssim 10$\%).
\end{enumerate}
In the usual case of steerable antennas where the FoVs are small enough that one is never too far from the optical axis of the dish, the instrumental polarization usually meets this threshold. Also, the fractional polarization of standard polarization calibrators (e.g. 3C 286, 3C 138, 3C 48) is $\leq10\%$\footnote{\href{https://science.nrao.edu/facilities/vla/docs/manuals/obsguide/modes/pol\#section-4}{Properties of standard polarization calibrators.}}. A linearized formalism is, hence, quite adequate for most of the applications. For our particular application of looking at the Sun with an aperture array, however, neither of these assumptions holds. The instrumental polarization is a strong function of the pointing direction. It increases as one goes farther from the zenith and/or cardinal directions and is often much larger than 10\%. The solar emission can vary dramatically in its intrinsic polarization from being unpolarized to being nearly 100\% polarized (Figure \ref{fig:different_emissions} in Chapter \ref{paircars_principle}). As calibrator observations are not possible during the solar observation with the MWA, one has to rely on the self-calibration using the Sun itself. The potentially very large fractional polarization of solar emission implies that one cannot rely on using them for polarization self-calibration with the linearized algorithms. This forces us to develop a more general formalism for polarimetric calibration.

The linear approximation also restricts the DR of the Stokes images \citep{Smirnov2011}. In addition, in the linearized formulation, the instrumental leakages are calibrated only using the cross-hand visibilities ($\mathrm{XY^\dagger,\ YX^\dagger}$ or $\mathrm{RL^\dagger,\ LR^\dagger}$) \citep{Hales2017}, and the parallel-hand visibilities ($\mathrm{XX^\dagger,\ YY^\dagger}$ or $\mathrm{RR^\dagger,\ LL^\dagger}$) are simply ignored. Hence, while the cross-hand visibilities are updated during polarization self-calibration, the parallel-hand correlations remain unchanged. As a consequence, this algorithm is unsuitable for iterative self-calibration-based implementation.

\subsection{Previous Attempts and Their Limitations}
\label{sec:previous-attempts}
While \citet{Mondal2019} have used self-calibration-based methods with remarkable success to obtain high DR Stokes I solar images, their algorithm does not include polarimetric imaging. It also does not include absolute flux density calibration, which needs to be done independently \citep{Kansabanik2022} and discussed in Chapter \ref{fluxcal}. \citet{Patrick2019} have demonstrated polarimetric solar radio imaging with the MWA. They used nighttime calibrator observations to estimate the instrumental and ionospheric gains. They used an ad-hoc approach to mitigate instrumental polarization, which we refer to as Method-I in the following text.
The assumptions and requirements of Method-I and the limitations they impose are listed below:
\begin{enumerate}
    \item The leakage from Stokes I to other Stokes components remains essentially constant across the angular span of the solar disk. While this assumption is valid for some pointing, it is not true in general and this is demonstrated in detail in Section \ref{sec:ideal_beam}.
    \item $\mathrm{S_Q=\ S_U=\ 0}$, where $\mathrm{S_Q}$ and $\mathrm{S_U}$ represent the Stokes Q and U components of the solar flux density, i.e. the linearly polarized emission from the Sun is assumed to be exactly zero. While no robust detection of linearly polarized emission has been reported from the Sun at low radio frequencies yet, the new generation of instruments can now provide spectroscopic snapshot images over spectral spans as small as a few $\mathrm{kHz}$, as opposed to order $\mathrm{MHz}$ available earlier. Assuming the linearly polarized flux density to be zero precludes their discovery and locks us out of an interesting discovery phase space.
    \item Method-I relies on the fact that the fractional circular polarization from the quiet Sun is expected to be small. It attempts to estimate an epoch-dependent instrumental leakage by minimizing the total number of pixels that show a fractional circular polarization larger than a chosen threshold, $\mathrm{r_c}$. It inherently assumes that this emission is coming from quiet-Sun regions. At low radio frequencies, the presence of multiple simultaneous active sources on the Sun can, however, limit the regions of the solar disk with quiet Sun emission \citep{Mohan2017spreads}. Even though the area occupied by the active regions is a small fraction of the solar disk at optical and higher frequencies, at low radio frequencies even the smallest active region gets broadened due to coronal scattering to a few arcmin \citep[e.g.,][]{Kontar2017,Mohan2021b}. In addition, the intrinsic $T_\mathrm{B}$ of various emissions often found to appear simultaneously on the Sun vary from $10^4\ \mathrm{K}$ for gyrosynchrotron emission to up to  $10^{13}\ \mathrm{K}$ for type-III solar radio bursts. The presence of very bright nonthermal emission has two consequences -- one, they lead to an increase in the system temperature and consequently the thermal noise in the image; and two, they impose a larger imaging DR requirement to be able to image the $\lesssim 10^6\ \mathrm{K}$ quiet-Sun regions in the presence of much brighter nonthermal emission.
\end{enumerate}

The primary merit of Method-I is that it enabled the authors to get to interesting science \citep{Patrick2019, Rahman2020} using an approximate and quick correction of instrumental polarization and circumventing the effort and complexities of developing and implementing a formally correct polarimetric calibration algorithm. The ionospheric phases during the solar observations are expected to be significantly different compared to those determined from nighttime calibrator observation. Hence applying the gain solutions from the nighttime calibrator observations limits the images too much poorer fidelity than the intrinsic capability of the data. These images cannot be used for reliable measurements of low levels of circular polarization. Method-I is also known to give rise to some spurious polarization for very bright solar radio bursts \citep{Rahman2020}.

\section{Details of the Current Algorithm}\label{Overview of the Algorithm}
This section describes a robust formal polarization calibration algorithm that overcomes the shortcomings mentioned in Section \ref{sec:previous-attempts} and enables high-fidelity polarimetric imaging. It is built on three pillars; i) self-calibration, ii) availability of a reliable instrumental beam model, and iii) some well-established properties of low-frequency solar radio emission. For low-radio-frequency solar observations with aperture arrays, it is not feasible to obtain calibrator observations at nearby times with the same primary beam pointing. This algorithm is, hence, designed to not require any calibrator observations, which has been discussed in Chapter \ref{paircars_principle}. We refer to this algorithm as Polarimetry using Automated Imaging Routine for Compact Arrays for the Radio Sun \citep[P-AIRCARS,][]{Kansabanik2022_paircarsI}. A detailed description of the algorithm is presented here, and that of its implementation as a robust unsupervised pipeline is discussed in Chapter \ref{paircars_implementation}.

\subsection{Full Jones Calibration Algorithm}
In conventional polarization calibration tools like CASA or AIPS, all four observed visibilities between an antenna pair are written as separate equations in terms of instrumental gains and leakages. These equations are approximated up to first-order terms in leakages and solved separately to obtain the instrumental parameters \citep[e.g.][]{Hales2017}. In full Jones calibration, the {\it Measurement Equation} is solved as a $2\times2$ matrix equation. From the Equation \ref{eq:visibility_matrix} we can write the coherence noise, $\mathrm{S}$, as,
\begin{equation}\label{eq:coherency_noise}
\begin{split}
     \mathrm{S}&=\mathrm{\sum_{ij} ||J_i^{-1}V_{ij}^\prime J_j^{\dagger{-1}} - V_{ij}||^2_F}\\
     \mathrm{S}&=\mathrm{\sum_{ij} Tr\left[ \left( J_i^{-1}V_{ij}^\prime J_i^{\dagger-1}-V_{ij}\right) \left(J_i^{-1}V_{ij}^\prime J_i^{\dagger-1}-V_{ij}\right)^\dagger \right]}
\end{split}
\end{equation}
where $\mathrm{||.||_F}$ represents the Frobenius norm \citep{horn_johnson_1985} of a matrix, $\mathrm{V_{ij}}$ is the model visibility, $\mathrm{V_{ij}}^\prime$ the observed visibility, and $\mathrm{J_i}$ and $\mathrm{J_j}$ represent Jones matrices for antennas $\mathrm{i}$ and $\mathrm{j}$, respectively. The instrumental Jones matrices are estimated by minimizing $\mathrm{S}$. Minimization of $\mathrm{S}$ leads to the matrix generalization of the conventional scalar calibration, which has been used in several standard interferometric software packages like CASA \citep{mcmullin2007,CASA2022}, AIPS \citep{Wells1985}, flagcal \citep{Flagcal2012}, and classical {\it antsol} \citep{Bhatnagar2001}, where Jones matrices were replaced by a single complex number. The full Jones calibration was introduced by \citet{Hamaker2000} and \citet{Mitchell2008} and was later optimized as {\it StefCal} \citep{stefcal2014}. We have used the recently developed full Jones calibration software package, {\it CubiCal} \citep{cubical2018,Cubical_robust2019} and its latest implementation {\it QuartiCal} \citep{Quartical2022}, which uses complex optimization and the Wirtinger derivative \citep{Wirtinger1927}. In brief, minimization of $\mathrm{S}$ reduces to an analytical update rule of $\mathrm{J_i}$ in terms of $\mathrm{V_{ij},V_{ij}^\prime}$ and the Jones matrices of other antennas as,
\begin{equation}\label{eq:jones_update}
    \mathrm{J_i^{\dagger{-1}}}=\mathrm{\left[ \sum_j V_{ij}J_j^{-1}V_{ij}^{\prime\dagger}\right] \left[ \sum_j V_{ij}J_j^{\dagger{-1}}J_j^{-1}V_{ij}^\dagger\right] ^{-1}}
\end{equation}
The Jones matrices of all the antenna elements are initialized as the identity matrix. $\mathrm{J_i}$s are estimated using Equation \ref{eq:jones_update} in subsequent iterations until the absolute value of the changes in the Jones terms and $\mathrm{S}$ fall below a small positive number ($\epsilon \sim 10^{-6}$) between two consecutive iterations. We find that solutions generally converge within $\sim$ 20 -- 30 iterations. 

\subsection{Self-calibration Algorithm of P-AIRCARS}\label{subsec:p_algorithm}
In radio interferometry, it is standard practice to write the instrumental Jones matrices as a chain of independent 2$\times$2 matrices, each with its distinct physical origin and referred to as the Jones chain \citep{Smirnov2011}. In P-AIRCARS, the net Jones matrix for the $i$th antenna is given by,
\begin{equation}\label{eq:jones_terms}
J_\mathrm{i} (\nu,\ t,\ \vec{l}\ )=G_\mathrm{i}(t,\ \vec{l}
)\ B_\mathrm{i}(\nu)\ K_{\mathrm{cross}}(\nu,\ t)\ D_\mathrm{i}(\nu,\ t,\ \vec{l})\ E_\mathrm{i}(\nu,\ t,\ \vec{l}\ )
\end{equation}
where $\nu$, $t$ and $\vec{l}$ refer to the frequency, time and direction ($\theta, \phi$) of the source, respectively.

These individual terms in Equation \ref{eq:jones_terms} for antenna $\mathrm{i}$ are:
\begin{enumerate}
\item $G_\mathrm{i}(t,\ \vec{l}) = \begin{pmatrix}
g_\mathrm{i,X}(t,\ \vec{l}) & 0\\ 0 & g_\mathrm{i,Y}(t,\ \vec{l})
\end{pmatrix}$ represents the frequency-independent time-variable instrumental gain.
This term also includes the direction-dependent gains arising due to propagation effects. 
At low radio frequencies, they correspond primarily to the ionospheric phase. Its characteristics and procedure of estimation and corrections are described in Chapter \ref{paircars_principle}.

\item $B_\mathrm{i}(\nu) = \begin{pmatrix}
b_\mathrm{i,X}(\nu) & 0\\ 0 & b_\mathrm{i,Y}(\nu)
\end{pmatrix}$ represents the time-independent instrumental bandpass response. The procedure to estimate this term is discussed in Chapter \ref{fluxcal}.

\item $K_\mathrm{cross}(\nu,\ t) = \begin{pmatrix}
e^{i\frac{\psi(\nu,\ t)}{2}} & 0\\ 0 & e^{-i\frac{\psi(\nu,\ t)}{2}}
\end{pmatrix}$ is the Jones matrix representing the phase difference between the two orthogonal receptors for the reference antenna,  given by $\psi(\nu, t)$.

\item $E_\mathrm{i}(\nu,\ t,\ \vec{l}) = E_\mathrm{i}(\nu,\ t,\ \theta,\ \phi) \\= \begin{pmatrix}
E_{\mathrm{i,X}\theta}(\nu,\ t, \theta) & E_{\mathrm{i,X}\phi}(\nu,\ t,\ \phi)\\E_{\mathrm{i,Y}\theta}(\nu,\ t, \theta) & E_{\mathrm{i,Y}\phi}(\nu,\ t,\ \phi)
\end{pmatrix}$ is the direction-dependent modeled instrumental primary beam response.

\item $D_\mathrm{i}(\nu,\ t,\ \vec{l}) = \begin{pmatrix}
d_\mathrm{i,XX}(\nu,\ t,\ \vec{l}) & d_\mathrm{i,XY}(\nu,\ t,\ \vec{l})\\ d_\mathrm{i,YX}(\nu,\ t,\ \vec{l}) & d_\mathrm{i,YY}(\nu,\ t,\ \vec{l})
\end{pmatrix}$ is the error on the ideal instrumental primary beam model.
\end{enumerate}

The Sun is by far the dominant source in the sky implying that we are in essentially a small FoV regime. This offers the advantage that the direction dependence due to the ionospheric effects included on $G_\mathrm{i}(t,\, \ \vec{l})$ can be ignored. We have verified that the direction dependence of $E_\mathrm{i}(\nu,\ t,\ \vec{l})$ cannot be ignored over the angular span of the Sun, but that of the $D_\mathrm{i}(\nu,\ t,\ \vec{l})$ can be, as discussed in detail in Section \ref{subsec:imaged_based_cor}. Henceforth in this work, their direction dependence is ignored and regard $G_\mathrm{i}(t,\, \ \vec{l}) \approxident G_\mathrm{i}(t)$ and $D_\mathrm{i}(\nu,\ t,\ \vec{l}) \approxident D_\mathrm{i}(\nu,\ t)$.

\begin{figure*}[!t]
    \centering
    \includegraphics[trim={0.8cm 1.5cm 0.0cm 1cm},clip,scale=0.46]{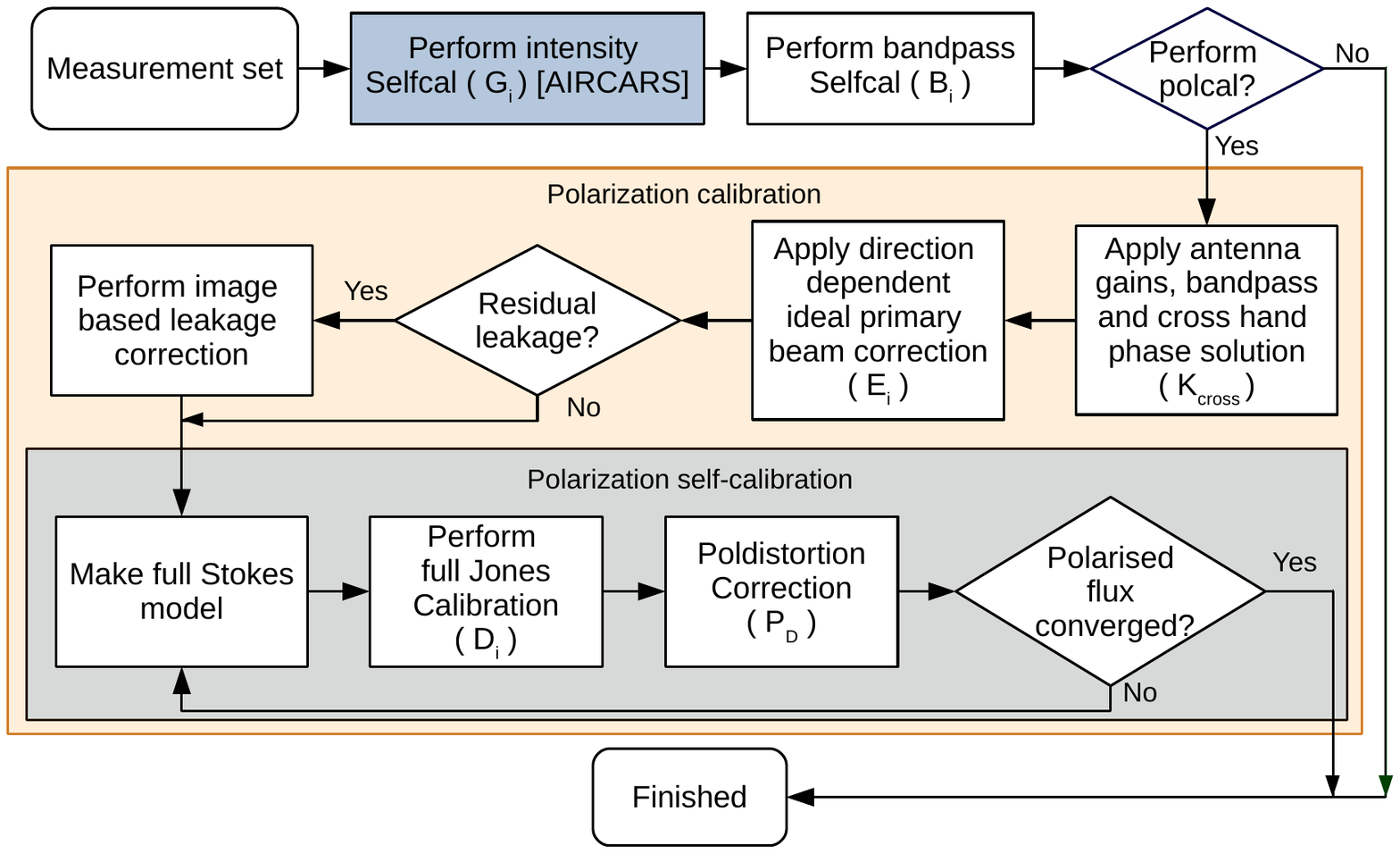}
    \caption[Flowchart of polarization calibration algorithm]{Flowchart describing the self-calibration algorithm of the P-AIRCARS. The first major calibration block is intensity self-calibration, which is similar to that implemented in AIRCARS (marked in blue) and described in Section \ref{sec:aircars_description} of Chapter \ref{paircars_principle}. The orange-shaded box represents the polarization calibration block. The gray shaded box is a subset of the polarization calibration and shows the steps for polarization self-calibration. The Jones terms in Equation \ref{eq:jones_terms} are being solved at each step and are denoted inside brackets.}
    \label{fig:polcal_flowchart}
\end{figure*}

If we write the $\mathrm{V_{ij}}$ in terms the of sky brightness matrix, $\mathbf{B}(\vec{l})$ using the van Cittert-Zernike theorem \citep{Thomson2017}, and neglecting the noise term, Equation \ref{eq:measurement_equation} can be written as ,
\begin{equation}\label{eq:vcz}
\begin{split}
 \mathrm{V_{ij}^\prime}=&J_\mathrm{i}\ \left[ \iint \mathbf{B}(\nu,\ t,\ \vec{l})\ e^{-2\pi i(u_\mathrm{ij}l+v_\mathrm{ij}m+w_\mathrm{ij}(n-1))}\frac{dl\ dm}{n} \right]\ J_\mathrm{j}^\dagger
\end{split}
\end{equation}
where $l,\ m$ and $n$ are the direction cosines of $\vec{l}$; and $u,\ v$, and $w$ are the components of the baseline vector in units of the wavelength in the Fourier plane of the sky brightness distribution. Using Equations \ref{eq:jones_terms} and \ref{eq:vcz}, we can write $\mathrm{V_{ij}}$ as,
\begin{equation}\label{eq:decomposed_me}
\begin{split}
\mathrm{V_{ij}}^\prime\ (\nu,\ t)=&\ G_\mathrm{i}(t)\ B_\mathrm{i}(\nu)\ K_{\mathrm{cross}}(\nu,\ t)\ D_\mathrm{i}(\nu,\ t)\\
 &\times \left[ \iint E_\mathrm{i}(\nu,\ t,\ \vec{l})\ \mathbf{B}(\nu,\ t,\ \vec{l})\ E_\mathrm{j}^\dagger(\nu,\ t,\ \vec{l})\right. \\
 &\times \left. e^{-2\pi i(u_\mathrm{ij}l+v_\mathrm{ij}m+w_\mathrm{ij}(n-1))}\frac{dl\ dm}{n} \right] \\
 &\times D_\mathrm{j}^\dagger(\nu,\ t)\ K_{\mathrm{cross}}^\dagger(\nu,\ t)\ B_\mathrm{j}^\dagger(\nu)\ G_\mathrm{j}^\dagger(t)\\
\mathrm{V_{ij}}^\prime\ (\nu,\ t)=&\ G_\mathrm{i}(t)\ B_\mathrm{i}(\nu)\ K_{\mathrm{cross}}(\nu,\ t)\ D_\mathrm{i}(\nu,\ t)\ V_\mathrm{ij,app}(\nu,\ t)\\
&\times D_\mathrm{j}^\dagger(\nu,\ t)\ K_{\mathrm{cross}}^\dagger(\nu,\ t)\ B_\mathrm{j}^\dagger(\nu)\ G_\mathrm{j}^\dagger(t)
\end{split}
\end{equation}
Here the quantities outside the square bracket are the direction-independent Jones terms and the quantities inside the square brackets are the direction-dependent terms. We estimate each of these Jones terms step-by-step. The flowchart of the P-AIRCARS algorithm for estimating these Jones terms is shown in Figure \ref{fig:polcal_flowchart}.

\subsection{Intensity Self-calibration}
\begin{figure*}
    \centering
    \includegraphics[trim={1.5cm 11cm 2cm 1cm},clip,scale=0.7]{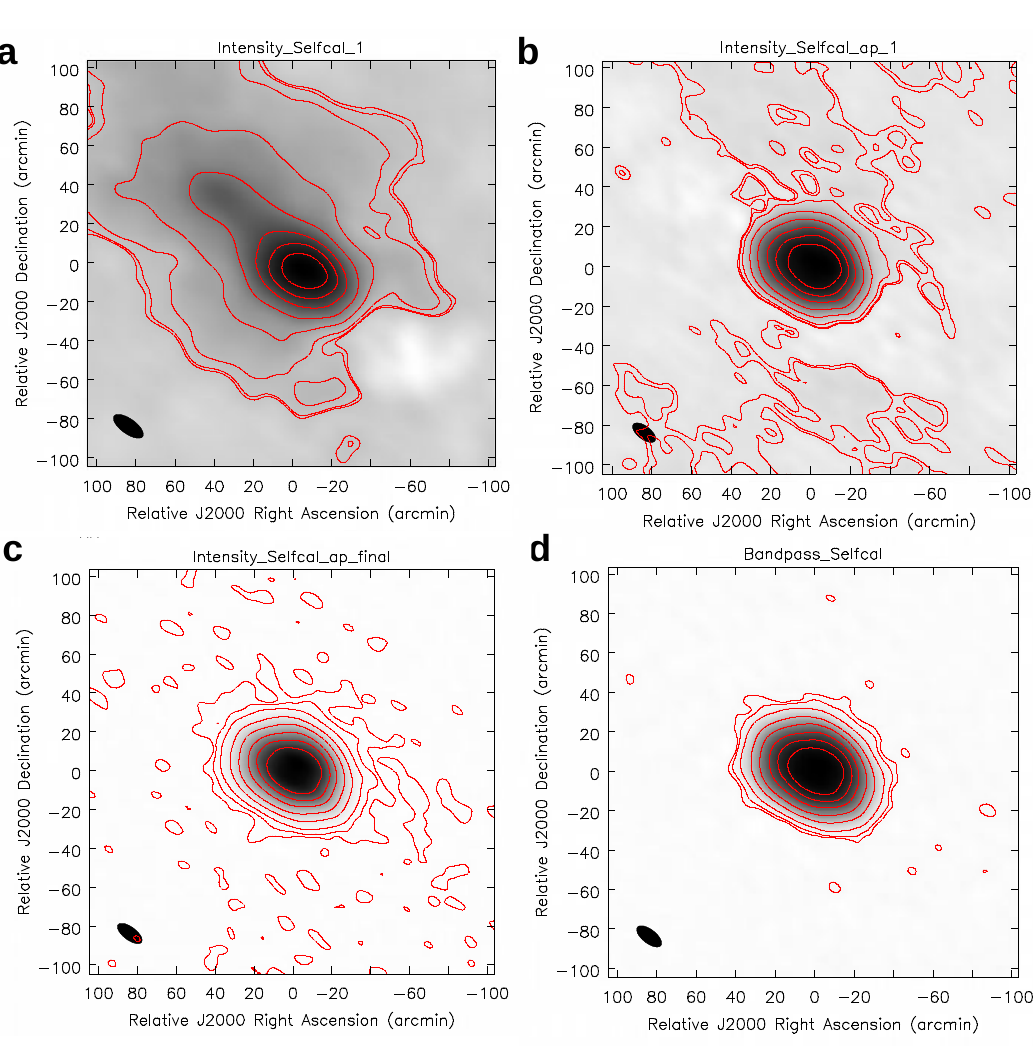}
    \caption[Demonstration of improvements in image quality with self-calibration]{Improvements in image quality during the intensity and bandpass self-calibration. Images shown here are at 88 $\mathrm{MHz}$ with a frequency resolution of 160 $\mathrm{kHz}$ and temporal resolution of 2 $\mathrm{s}$. The red contours levels are at 0.3, 0.6, 2, 8, 20, 40, 60, and 80 \% of the peak flux density. The black ellipse at the bottom left is the point spread function. a. Image after the first round of intensity self-calibration. Bright emission is present near the phase center, but the solar disk is distorted. b. Image of the first amplitude-phase intensity self-calibration. c. Image after the end of the amplitude-phase intensity self-calibration. The noise level decreases significantly in comparison. d. Image showing the effect of bandpass self-calibration. The image noise is reduced further after the bandpass self-calibration. The DRs of the images are 24, 385, 491, and 793, respectively.}
    \label{fig:intensity_bp_selfcal}
\end{figure*}

We first estimate the time-variable instrumental and ionospheric gain, $G_\mathrm{i}(t)$, normalized over all antenna elements of the array. We choose a single frequency ($\nu=\nu_0$) channel for intensity self-calibration and set $B_\mathrm{i}(\nu_0)=1$. We write Equation \ref{eq:decomposed_me} as,
\begin{equation}\label{eq:gaincal}
 \begin{split}
 \mathrm{V_{ij}}^\prime(\nu_0,\ t)=&G_\mathrm{i}(t)\ \mathrm{V_{ij,Gcor}}(\nu_0,\ t)\ G_\mathrm{j}^\dagger(t)
 \end{split}
\end{equation}
where
\begin{equation}\label{eq:gaincal_1}
\begin{split}
        \mathrm{V_{ij,Gcor}}(\nu_0,\ t)=&B_\mathrm{i}(\nu_0)\ K_{\mathrm{cross}}(\nu_0,\ t)\ D_\mathrm{i}(\nu_0,\ t)\ \mathrm{V_{ij,app}}(\nu_0,\ t)\\
         &\times D_\mathrm{j}^\dagger(\nu_0,\ t)\ K_{\mathrm{cross}}^\dagger(\nu_0,\ t)\ B_\mathrm{j}^\dagger(\nu_0),
\end{split}
\end{equation}
is the apparent model visibility for a single spectral channel after intensity self-calibration.

The intensity self-calibration algorithm in P-AIRCARS follows the same philosophy as AIRCARS \citep{Mondal2019}, which is described in Section \ref{sec:aircars_description} of Chapter \ref{paircars_principle} and the implementation in P-AIRCARS incorporates additional improvements and optimizations. Making use of the compact and centrally condensed array configuration of the MWA and the very high flux density of the Sun, we estimate both antenna gains, $G_\mathrm{i}(t)$ and $\mathrm{V_{ij,Gcor}}(t)$ iteratively. AIRCARS assumes $\mathrm{V_{ij,Gcor}}(\nu_0,\ t)$s to be unpolarized and effectively uses the same source model for the $\mathrm{XX}$ and $\mathrm{YY}$ polarizations. Unlike AIRCARS, no assumptions are made about the polarimetric properties of $\mathrm{V_{ij,Gcor}}(\nu_0, t)$. A $2\times2$ matrix calibration is performed without any constraints on $\mathrm{V_{ij,Gcor}}(\nu_0, t)$ except that $G_\mathrm{i}(t)$ is assumed to be diagonal.

Figure \ref{fig:intensity_bp_selfcal}a shows the image after the first round of phase-only self-calibration, where the solar disk is rather distorted. The DR of this image is only 24. The image after the first round of amplitude-phase self-calibration is shown in Figure \ref{fig:intensity_bp_selfcal}b. The coherence of the array has improved remarkably, the solar disc is well formed, and the DR has increased to 385. Figure \ref{fig:intensity_bp_selfcal}c shows the final output of the intensity self-calibration process. The improvement in image quality is self-evident, and the DR has reached 491. We note that before AIRCARS, the highest imaging DR for spectroscopic snapshot solar imaging at meter wavelengths was a few hundred ($\lesssim$ 300) and the imaging fidelity was too poor to be able to reliably detect features of strength few percent of the peak \citep{Mercier2009}. We have chosen a quiet featureless Sun for this illustration, the high DR and high-fidelity imaging of which continue to remain challenging at low radio frequencies even for the new generation instruments \citep{Vocks2020}. 

\subsection{Bandpass self-calibration}\label{subsec:bandpass_cal}
As the instrumental bandpass amplitude and phase vary across frequencies, bandpass calibration is required before combining multiple spectral channels to make an image. As AIRCARS was designed for spectroscopic imaging and the flux density calibration was done using an independent non-imaging technique \citep{oberoi2017}, it did not need to include bandpass calibration. Conventionally, instrumental bandpass is determined using standard flux density calibrator sources with known spectra. The lack of availability of suitable calibrator observations during the daytime pushes us to rely on bandpass self-calibration. This in turn required us to find a way to deal with the degeneracy between the instrumental bandpass shape and the intrinsic spectral structure of the source (Sun). To avoid intrinsic spectral structure, we carefully choose sufficiently quiet times for the bandpass self-calibration from the initial flux-density calibrated dynamic spectrum. We obtain the initial flux density calibrated dynamic spectrum using the non-imaging technique mentioned earlier, which is independent of instrumental gains and is computationally much faster. Though precise flux density calibration of MWA solar observations can be done using the method \citep{Kansabanik2022}, which is described later in Chapter \ref{fluxcal}, it is only applicable to bandpass calibrated visibilities or images, which are not yet available at this stage in the calibration process. Bandpass self-calibration is performed for narrow bandwidths of 1.28 $\mathrm{MHz}$ at a time, referred to as a ``picket". The spectrum of the quiet Sun can justifiably be assumed to be flat across a picket and apparent source visibility after bandpass calibration, $\mathrm{V_{ij,Bcor}}(\nu,\ t)=\mathrm{V_{ij,Bcor}}(\nu_0,\ t)$.

A single time slice ($\mathrm{t=t_0}$) is chosen for performing bandpass self-calibration. Bandpass self-calibration starts with $\mathrm{V_{ij,Gcor}}(\nu_0,\ t_0)$ as the initial model. Equation \ref{eq:gaincal_1} over the band can then be rewritten as,
\begin{equation}\label{eq:bandpass}
\begin{split}
 \mathrm{V_{ij,Gcor}}(\nu,\ t_0)=& B_\mathrm{i}(\nu)\ \mathrm{V_{ij,Bcor}}(\nu_0,\ t_0)\ B_\mathrm{j}^\dagger(\nu)
 \end{split}
\end{equation}
where $\mathrm{V_{ij}}(\nu,\ t_0)=V_{ij,Bcor}(\nu_0, t_0)$ is the apparent model visibility after bandpass self-calibration given as,
\begin{equation}\label{eq:bandpass_1}
\begin{split}
        \mathrm{V_{ij,Bcor}}(\nu,\ t_0)=&\mathrm{V_{ij,Bcor}}(\nu_0, t_0)=K_{\mathrm{cross}}(\nu,\ t_0)\ D_\mathrm{i}(\nu,\ t_0)\\
        &\times \mathrm{V_{ij,app}}(\nu_0,\ t_0)\ D_\mathrm{j}^\dagger(\nu,\ t_0)\ K_{\mathrm{cross}}^\dagger(\nu,\ t_0)
\end{split}
\end{equation}
We find that inter-picket bandpass phases for MWA can be modeled well by a straight line \citep{Sokolwski2020}. Inter-picket bandpass amplitudes show a more complicated variation and are corrected using an independent method \citep{Kansabanik2022} described in a later Chapter \ref{fluxcal}. As expected, bandpass self-calibration improves the image quality, and the DR increases from 431 (Figure \ref{fig:intensity_bp_selfcal}c) to 793 (Figure \ref{fig:intensity_bp_selfcal}d).
 
\subsection{Polarization self-calibration}\label{subsec:polselfcal}
This section describes the different parts of the polarization self-calibration algorithm marked by the orange shaded region in Figure \ref{fig:polcal_flowchart}.

\subsubsection{Cross-hand Phase Calibration}
\label{subsec:cross-hand-phase}
During the intensity and bandpass self-calibration, the phase of the reference antenna is set to zero for each of the orthogonal receptors. There is, however, an arbitrary phase difference between the two orthogonal receptors of the reference antenna. This cross-hand phase, $\psi$, is thus a single number and is applied to all antennas by a single Jones matrix $K_\mathrm{cross}(\nu,\ t)$. In the case of linearly polarized receptors, $K_\mathrm{cross}(\nu,\ t)$ causes leakage from Stokes U to Stokes V and vice versa. 

There are, however, only a few bright polarized sources available for cross-hand phase calibration at low frequencies \citep{lenc2017,lenc2018}. In addition to the reasons given in Section \ref{sec:conventional-algos}, the requirement of a strong polarized point source at the phase center dominating the measured visibility, implies that the standard cross-hand phase calibration method available in CASA cannot be used for aperture arrays with large FoVs. Cross-hand phase cannot be determined using the self-calibration-based approaches as well. Hence, we use an image-based method for calibration of $\psi$, tailored for low-frequency aperture array instruments with large FoV \citep{bernardi2013}. An observation of a linearly polarized source with sufficiently high rotation measure (RM) is chosen and the direction-independent instrumental gain and bandpass calibrations obtained from suitable unpolarized calibrator observation are applied. It is important to distinguish the true source polarization from the Stokes I due to the leakages caused by the instrumental primary beam. The ideal primary beam corrections are hence applied to account for instrumental leakage. Even after the ideal primary beam correction, the residual leakages from Stokes I to other Stokes parameters remain. For well-designed and well-modeled instruments, departures from non-orthogonality of the dipoles are small, which implies that the Stokes Q to Stokes V leakage is small. For the MWA it is small enough to be ignored \citep{bernardi2013,lenc2017}.

\begin{figure*}[!ht]
 \centering
 \includegraphics[trim={1.3cm 3cm 1.5cm 0cm},clip,scale=0.5]{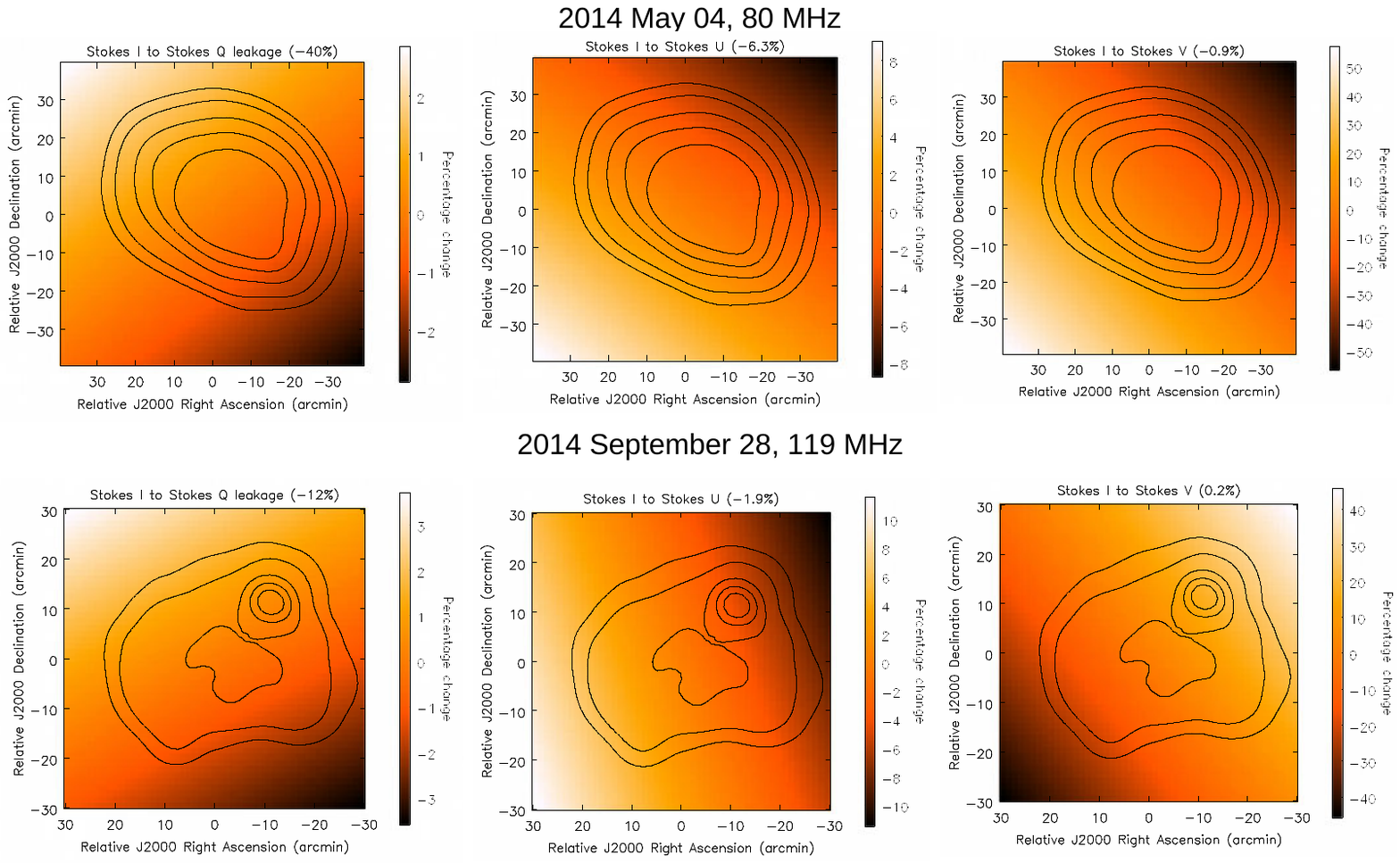}
 \caption[Fractional change in polarization leakages due to ideal MWA beam response.]{Fractional change of leakages from Stokes I to other Stokes due to ideal primary beam response, $E_\mathrm{i}(\vec{l})$, over the angular extent of the Sun. Black contour represents the Stokes I solar emission. Contour levels are at 10\%, 20\%, 40\%, 60\%, 80 \% of the peak flux density. Polarization leakage from Stokes at the center of the Sun is mentioned in the title of each panel. {\it Top row:} Variation of leakages for the observation of 2014 May 04 at 80 MHz are shown. {\it Bottom row:} Variation of leakages for the observation of 2014 September 28 at 119 MHz are shown.}
 \label{fig:stokes_leakage}
\end{figure*}
When a linearly polarized emission passes through magnetized plasma, the polarization angle ($\chi$) rotates. The observed $\chi \propto \lambda^2$, where $\lambda$ is the wavelength of observation. The proportionality constant is referred to as Rotation Measure (RM) and, like all propagation effects, is integral along the entire LoS. It depends on the distributions of the LoS component of magnetic field strength and electron density. Multiple mediums (e.g., interstellar medium, interplanetary medium, and ionosphere) with different RMs contribute to the total rotation. RM synthesis \citep{Brentjens2005} is a Fourier synthesis technique to separate these RM components in the Fourier domain. The leakage flux from Stokes I to Stokes V can also be thought of as yet another medium contributing to the observed RM. This leakage flux must appear at the instrumental RM in the Fourier domain, which is typically a few $\mathrm{rad\ m^{-2}}$. As the linearly polarized source is chosen to be at an RM significantly higher than the instrumental RM, any Stokes V emission detected at source RM in the Fourier space can only arise due to leakages from components that rotate with the source RM. Hence the observed Stokes V emission at source RM must arise due to the leakages from Stokes U. Thus the Stokes U and Stokes V flux density are estimated from the image using RM synthesis. For linearly polarized receptors, $K_\mathrm{cross}(\nu,\ t)$ causes the leakage from Stokes U to Stokes V. We vary $\psi$ between $-180^{\circ}$ and $+180^{\circ}$ and determine the value of $\psi$ for which the spurious Stokes V emission is minimized. 

It has been found that the $\psi$ is extremely stable across both time and frequency for the MWA. \citet{lenc2017} found that $\psi$ essentially remains constant across the MWA frequency band from 80 $-$ 300 $\mathrm{MHz}$. The GaLactic and Extragalactic All-sky MWA (GLEAM) survey team (private communication with Xiang Zhang, ICRAR) have recently determined that $\psi$ is stable over time scales of years. P-AIRCARS generally uses the values of $\psi$ available {\it a priori} but does provide the flexibility to estimate it from the nearest observations of a linearly polarized source and apply the corresponding correction. The extreme stability across time and frequency allows the use of nighttime observations from even months away to estimate $\psi$.

After the correction of the cross-hand phase, $\mathrm{V_{ij,Xcor}}$ at any time, $t$ and any frequency, $\nu$, is obtained from Equation \ref{eq:bandpass_1} as, 
\begin{equation}\label{eq:crossphase_corrrection}
\begin{split}
\mathrm{V_{ij,Xcor}}(\nu,\ t) &=K_{\mathrm{cross}}(\nu,\ t)^{-1}\ \mathrm{V_{ij,Bcor}}(\nu,\ t)\ K_{\mathrm{cross}}(\nu,\ t)^{-1\dagger}\\
&= D_\mathrm{i}(\nu,\ t)\ \mathrm{V_{ij,app}}(\nu,\ t)\ D_\mathrm{j}^\dagger(\nu,\ t)\\
&=D_\mathrm{i}(\nu,\ t)\ \left[ \iint E_\mathrm{i}(\nu,\ t,\ \vec{l})\ \mathbf{B}(\nu,\ t,\ \vec{l})\ E_\mathrm{j}^\dagger(\nu,\ t,\ \vec{l})\right.\\
&\times \left. e^{-2\pi i(u_\mathrm{ij}l+v_\mathrm{ij}m+w_\mathrm{ij}(n-1))}\frac{dl\ dm}{n} \right]\ D_\mathrm{j}^\dagger(\nu,\ t)
\end{split}
\end{equation}

\subsubsection{Ideal Primary Beam Correction}\label{sec:ideal_beam}
For aperture array instruments, a significant part of the instrumental leakage comes from the pointing-dependent instrumental primary beam. In reality, it is rarely feasible to measure the full-Stokes primary beam for aperture arrays. Hence, an ideal model of the primary beam, obtained using electromagnetic simulations is generally used. There are a few primary beam models available for the MWA -- an analytical beam model \citep{Ord_2010}; an average embedded element (AEE) beam model by \citet{Sutinjo2015}; and full embedded element (FEE) beam models by \citet{oberoi2017} and \citep{Sokolwski2017}. These models have steadily improved in their sophistication and performance. We have used the most recent ideal primary beam model for the MWA \citep{Sokolwski2017} for which the FEE beams have been simulated using the electromagnetic simulation tool, FEKO {\footnote{\url{www.feko.info}}}. We note that P-AIRCARS allows the flexibility to use any primary beam model and will enable us to benefit from improved models as and when they become available. Naturally, the true primary beam response can deviate from the ideal beam model. 

Method-I assumes that the leakage from Stokes I to other Stokes components remains essentially constant across the angular span of the solar disc. While this assumption is valid for some pointings, it is not true in general. The percentage variation of this leakage across the solar disc is estimated to be as large as 50\%. We define the percentage variation as,
\begin{equation}\label{eq:frac_leakage_change}
 \begin{split}
 \Delta l(\theta,\ \phi) = \frac{l(\theta,\ \phi)-l_\mathrm{centre}}{l_\mathrm{center}}\times100 \%,
 \end{split}
\end{equation}
where $l(\theta,\ \phi)$ is the leakage from Stokes I at the sky coordinate $(\theta,\ \phi)$, $l_\mathrm{center}$ is the leakage at the center of the solar disc, and $\Delta l(\theta,\ \phi)$ is the percentage change of $l(\theta,\ \phi)$ with respect to $l_\mathrm{center}$. The variation of $\Delta l(\theta,\ \phi)$ over the Sun for two observing epochs is shown in Figure \ref{fig:stokes_leakage}. The Sun was close to half power point of the primary beam during the first epoch, 2014 May 04. During the second epoch, 2014 September 28, the Sun was close to the peak of the primary beam. We have found that $l_\mathrm{center}$ is smaller for the second epoch as compared to that for the first epoch. This is expected as the primary beam model is more accurate near its peak. Nonetheless, the fractional variation of the leakages over the Sun for both epochs is not similar and not negligible. Hence, for precise polarization calibration, it is essential to correct the direction-dependent primary beam response.

For a homogeneous array comprising identical antenna elements, the modeled primary beam response can be assumed to be identical for all antenna elements. Thus we can substitute $E_\mathrm{i}(\nu,t,\vec{l})=E_\mathrm{j}(\nu,t,\vec{l})=E(\nu,t,\vec{l})$ in Equation \ref{eq:crossphase_corrrection},
\begin{equation}\label{eq:beamcor_1}
\begin{split}
     \mathrm{V_{ij,Xcor}}(\nu,\ t)&=D_\mathrm{i}(\nu,\ t)\ \left[\iint E(\nu,\ t,\ \vec{l})\ \mathbf{B}(\nu,\ t,\ \vec{l})\ E^\dagger(\nu,\ t,\ \vec{l})\right.\\
 &\times \left. e^{-2\pi i(u_\mathrm{ij}l+v_\mathrm{ij}m+w_\mathrm{ij}(n-1))}\frac{dl\ dm}{n}\right]\ D_\mathrm{j}^\dagger(\nu,\ t)
\end{split}
\end{equation}
When a calibrator source is available, $\mathbf{B}(\nu,\ t,\ \vec{l})$ is known and the only unknowns in Equation \ref{eq:beamcor_1} are the $D_\mathrm{i}(\nu, t)$s. As no suitable calibrator observation is available, $D_\mathrm{i}(\nu,t)$ is not known {\it a priori} there is a degeneracy between $\mathbf{B}(\nu,\ t,\ \vec{l})$ and $D_\mathrm{i}(\nu,t)$. A perturbative approach is used to break this degeneracy. As $D_\mathrm{i}(\nu,\ t)$ is small compared to $E(\nu,\ t,\ \vec{l})$s, we approximate $D_\mathrm{i}$ as identity matrix in Equation \ref{eq:beamcor_1} and obtain the source visibility, $\mathbf{B}_0(\nu,\ t,\ \vec{l})$, while incorporating the correction for $E(\vec{l})$. 
Equation \ref{eq:beamcor_1} is then takes the form,
\begin{equation}\label{eq:beamcor_2}
\begin{split}
     \mathrm{V_{ij,Xcor}}(\nu,\ t)&=\iint E(\nu,\ t,\ \vec{l})\ \mathbf{B}_0(\nu,\ t,\ \vec{l})\ E^\dagger(\nu,\ t,\ \vec{l})\\
 &\times e^{-2\pi i(u_\mathrm{ij}l+v_\mathrm{ij}m+w_\mathrm{ij}(n-1))}\frac{dl\ dm}{n}\\
    &=\iint \mathbf{B}_{0,app}(\nu,\ t,\ \vec{l})\\
    &\times e^{-2\pi i(u_\mathrm{ij}l+v_\mathrm{ij}m+w_\mathrm{ij}(n-1))}\frac{dl\ dm}{n},
\end{split}
\end{equation}
where $\mathbf{B}_{0,app}(\nu,\ t,\ \vec{l})$ is the apparent source visibility, and $\mathbf{B}_0(\nu,\ t,\ \vec{l})$ is the source visibility without the correction for $D_\mathrm{i}(\nu, t)$s and $\mathrm{E}(\nu, t, \vec{l})$. $\mathbf{B}_0(\nu,\ t,\ \vec{l})$ and $\mathbf{B}_{0,app}(\nu,\ t,\ \vec{l})$ are related as
\begin{equation}\label{eq:beamcor_3}
    \begin{split}
        \mathbf{B}_0(\nu,\ t,\ \vec{l})=E^{-1}(\nu,\ t,\ \vec{l})\ \mathbf{B}_{0,app}(\nu,\ t,\ \vec{l})\ E^{-1\dagger}(\nu,\ t,\ \vec{l}).
    \end{split}
\end{equation}
$\mathbf{B}_0(\nu,\ t,\ \vec{l})$ differs from the true brightness matrix, $\mathbf{B}(\nu,\ t,\ \vec{l})$, due to the following two reasons:
\begin{enumerate}
    \item $D_\mathrm{i}(\nu, t)$ has been ignored in Equation \ref{eq:beamcor_1}, which introduces errors in \\$\mathbf{B}_0(\nu,\ t,\ \vec{l})$.
    \item In the absence of calibrator observations or other independent astronomical constraints, the degeneracy between $G_\mathrm{i}$ and $\mathrm{V_{ij,Gcor}}$ cannot be broken unambiguously.
\end{enumerate}

Although $D_\mathrm{i}(\nu, t)$ is expected to be small, this may not hold for all pointings, especially the ones at low elevations. The top row of Figure \ref{fig:res_stokes_leakage} shows the residual leakages after the ideal primary beam correction at 80 MHz for 2014 May 04. For this epoch, the beam pointing was at the lowest permitted elevation for the MWA. The residual leakages for Stokes I to Stokes Q are $\sim34-35\%$ and Stokes I to Stokes U and Stokes V is $\sim3-4\%$. For comparison, the bottom row of Figure \ref{fig:res_stokes_leakage} shows the residual leakages for observation on 2014 September 28 at 119 MHz. The beam pointing for this epoch was close to the meridian, and the residual leakages are only a few percent. No systematic spatial variation of the residual leakages is seen in Figure \ref{fig:res_stokes_leakage}. The spatial variations over small angular scales are at 1--2\% and are discussed in detail in Section \ref{subsec:imaged_based_cor}. For precise polarization calibration, it is essential to reliably correct for $D_\mathrm{i}(\nu, t)$s. The next section describes our approach for an image-based first-order correction for $D_\mathrm{i}(\nu, t)$ s to $\mathbf{B}_0(\nu,\ t,\ \vec{l})$.

\subsection{Image-based Leakage Correction}\label{subsec:imaged_based_cor}
As discussed in Section \ref{sec : basic_polarimetry}, the objective of polarimetric calibration is to correct for instrumental {\it polconversion} and {\it polrotation}, which cannot be achieved using self-calibration-based approaches alone. The most accurate MWA primary beam model \citep{Sokolwski2017} used in P-AIRCARS successfully reduces the instrumental leakages significantly. Though it can come close, no primary beam model can reproduce the true response exactly. To remove the residual instrumental leakages, I use some well-established physical properties of the quiet-Sun thermal emission at meter wavelengths to design an image-based correction. The characteristics of the quiet-Sun thermal emission we rely on are:
\begin{enumerate}
\item The brightness temperature of the quiet-Sun thermal emission is well known to lie in the range of $10^5-10^6\ \mathrm{K}$ \citep{Mercier2015, oberoi2017, Mondal2019, Vocks2020, Sharma2020,Zhang_2022}.
\item Quiet-Sun thermal free-free emission {\it at meter-wavelength} has a very low level of circular polarization ($\lesssim1\%$), which arises due to propagation effects through the magnetized corona \citep{Sastry_2009}.
\item No linearly polarized emission is expected from the quiet-Sun thermal free-free emission \citep{Alissandrakis2021}. 
\end{enumerate}
No assumptions are made about the polarization properties of any active emissions as they depend on the emission mechanism, magnetic field strength, and topology, and are variable across time and frequency. 
\begin{figure*}[!ht]
 \centering
 \includegraphics[trim={1cm 1.5cm 1cm 2.5cm},clip,scale=0.5]{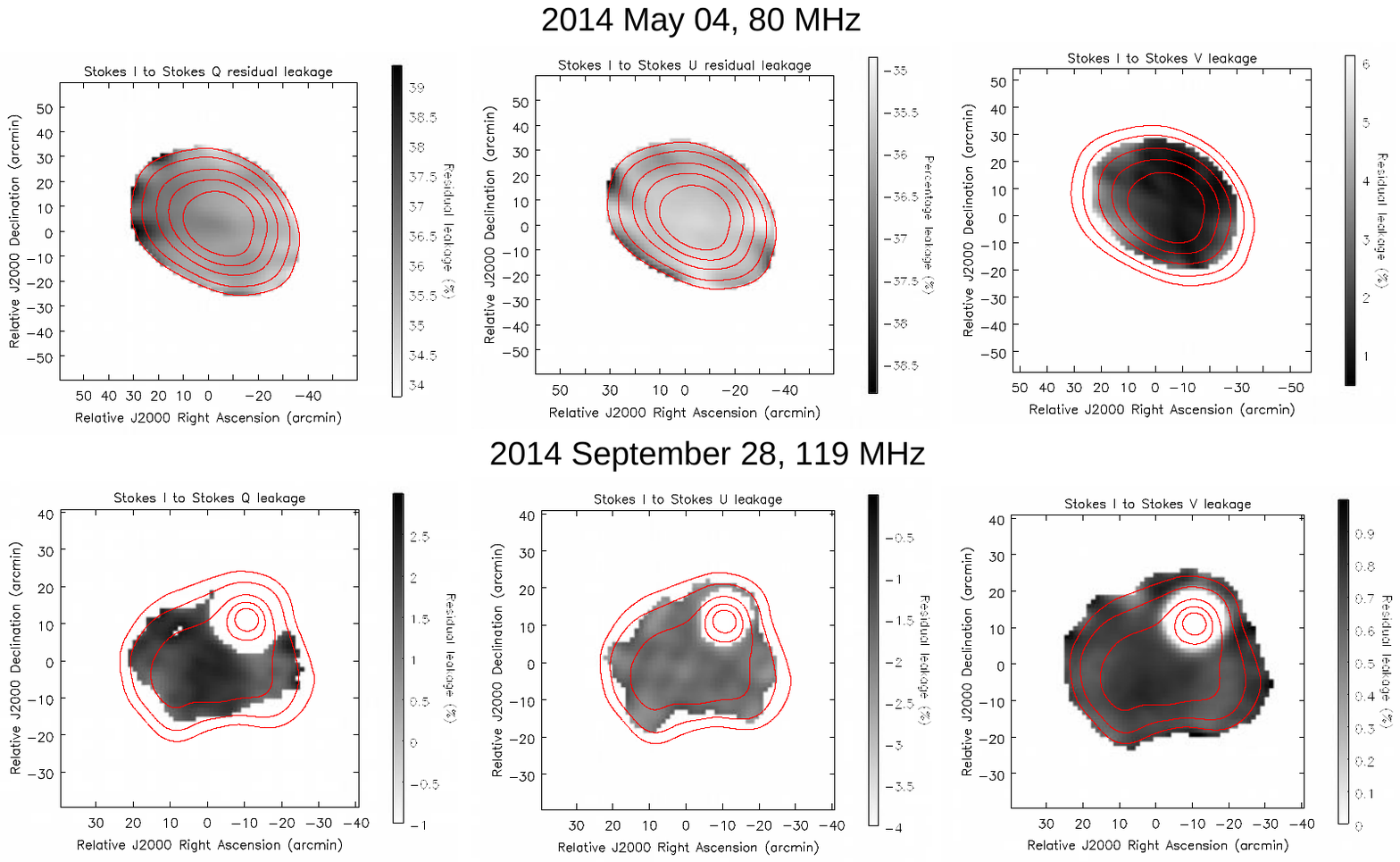}
 \caption[Residual leakages from Stokes I to other Stokes over the quiet-Sun region.]{Residual leakages from Stokes I to other Stokes over the quiet-Sun region. Red contour represents the Stokes I solar emission. Contour levels are at 10, 20, 40, 60, and 80 \% of the peak flux density. Quiet-Sun regions with more than 3$\sigma$ detection in each Stokes plane are shown. {\it Top row: } Residual leakages for the observation on 2014 May 04 at 80 MHz are shown. The beam pointing is at the lowest elevation for the MWA and residual leakages are large. {\it Bottom row: }Residual leakages for the observation on 2014 September 09 at 119 MHz are shown. The primary beam pointing is close to the meridian, and residual leakages are small. In the Stokes V image for this epoch, no quiet-Sun regions were having more than $3\sigma$ detection. Thus we have only used the $3\sigma$ threshold on Stokes I emission only.}
 \label{fig:res_stokes_leakage}
\end{figure*}

Given these properties, one can argue that any Stokes Q and U emission seen in the quiet Sun region must arise due to residual instrumental leakages. Also, as the circular polarization of the quiet-Sun is $\lesssim1\%$, the dominant contribution to the Stokes Q and U leakage must come from Stokes I. For the MWA, the misalignment of the dipoles with respect to the sky coordinates has been established to be small enough to give rise to insignificant mixing between Stokes Q and U \citep{lenc2017}. An explicit correction for $K_\mathrm{cross}(\nu,\ t)$ has also been applied to correct for the mixing between Stokes U and V. Based on these arguments, instrumental leakages are corrected as follows:
\begin{enumerate}
\item A $T_\mathrm{B}$ map is first made using the flux-density calibrated solar images \citep{Kansabanik2022}.
\item Regions where solar emission has reliably been detected are identified using an $n\sigma$ lower threshold, where $\sigma$ is the map rms noise in a region far from the Sun and $n$ is usually chosen to lie between 10 and 6. This is the same threshold as used in the deconvolution process during imaging and is determined and applied for each Stokes plane independently.
\item The regions lying between $10^5$ and $10^6\ \mathrm{K}$ are considered to correspond to the quiet Sun.
\item Median values of Stokes Q and U fractions from the quiet-Sun regions are computed and deemed to represent the leakages from Stokes I to Stokes Q and U, respectively.
\item The leakages thus determined are then subtracted from the Stokes Q and U maps of $\mathbf{B}_0(\nu,\ t,\ \vec{l})$.
\end{enumerate}

We have only considered the leakages from Stokes I to Stokes Q and U. Stokes I to V leakages have generally been found to be consistent with 0\% after the correction using the latest FEE beam model \citep{Sokolwski2017}. As discussed in Section \ref{subsec:cross-hand-phase} Stokes Q to Stokes V leakages are also small enough to be ignored for the MWA. Occasionally, however, when the beam pointing is at very low elevations, residual Stokes I to V leakage can grow to be as large as a few tens of percent. In such instances, an approach similar to what is used for estimating the first-order corrections for Stokes Q and U maps is used for obtaining a first-order estimate of Stokes I to V leakage. This is employed only when the median circular polarization in the quiet-Sun region is found to be $> 2\%$. Method-I used an older beam model \citep{Sutinjo2015} which could not correct for the Stokes I to V leakage effectively, and needed to rely exclusively on this approach to estimate Stokes I to V leakage. In contrast, we use it sparingly, only when the Stokes I to V leakage is so large that our perturbative approach breaks down. These corrections account for {\it polconversion}, which arises due to the $D_\mathrm{i}(\nu, t)$ terms. There is little mixing between any of Stokes Q and U or Stokes Q and V for the MWA \citep{bernardi2013,lenc2017}. As $K_\mathrm{cross}$ has also been corrected, there is no mixing between Stokes U and V. Hence, no additional correction for polrotation is required.

For aperture arrays, errors on the ideal beam are expected to be direction-dependent. This is true for the MWA as well \citep{lenc2017}. We find that, although the leakages due to the ideal primary beam vary significantly over the angular extent of the Sun, the residual leakages due to $D_\mathrm{i}(\nu, t)$s do not. Figure \ref{fig:res_stokes_leakage} shows the residual leakages from Stokes I to other Stokes parameters after the ideal primary beam correction. No significant systematic variations are seen across the solar disc after the subtraction of the median leakage, validating the assumption to treat the residual leakage or $D_\mathrm{i}(\nu, t)$s as direction-independent quantities, as mentioned in Section \ref{subsec:p_algorithm}. This image-based correction for leakages yields the source brightness matrix, $\mathbf{B}_1(\nu,\ t,\ \vec{l})$, which incorporates the first-order corrections for these leakages. Paralleling  Equation \ref{eq:beamcor_2}, $\mathbf{B}_{1,app}(\nu,\ t,\ \vec{l})$ can be written as:
\begin{equation}\label{eq:leakge_cor1}
    \begin{split}
        \mathbf{B}_{1,app}(\nu,\ t,\ \vec{l})&=E(\vec{\nu,\ t,\ l})\ \mathbf{B}_1(\nu,\ t,\ \vec{l})\ E^{\dagger}(\nu,\ t,\ \vec{l})
    \end{split}
\end{equation}

We can now rewrite Equation \ref{eq:beamcor_1} using $\mathbf{B}_{1,app}(\vec{l})$ and including the first-order correction of error matrices, $D_\mathrm{1,i}(\nu, t)$s, as
\begin{equation}\label{eq:leakge_cor2}
    \begin{split}
        \mathrm{V_{ij,Xcor}}(\nu,\ t)&=D_\mathrm{1,i}(\nu,\ t)\
        \left [\iint \mathbf{B}_\mathrm{1,app}(\nu,\ t,\ \vec{l})\ e^{-2\pi i(u_\mathrm{ij}l+v_\mathrm{ij}m+w_\mathrm{ij}(n-1))}\frac{dl\ dm}{n}\right]\\
        &\times D_\mathrm{1,j}^\dagger(\nu,\ t)\\
        \mathrm{V_{ij,Xcor}}(\nu,\ t)&=D_\mathrm{1,i}(\nu,\ t)\ \mathrm{V_{1,ij,app}}(\nu,\ t)\ D_\mathrm{1,j}^\dagger(\nu,\ t),\\
    \end{split}
\end{equation}
where $\mathrm{V_{1,ij,app}}(\nu, t)$ is the apparent source visibility after first-order correction. $D_\mathrm{1,i}(\nu, t)$s can now be solved for iteratively using $\mathrm{V_{1,ij,app}}$ as the initial model, as discussed next.

\subsubsection{Perturbative Correction : Residual Poldistortion Estimation and Correction}\label{sec:poldist}
The $\mathrm{V_{1,ij,app}}(\nu, t)$s can be thought of as the full-Stokes sky model but with small errors arising largely from the deficiencies of first-order polarization calibration. The situation is analogous to the missing flux density problem in intensity self-calibration \citep{Grobler2014}, where it is addressed by using normalized solutions over all antenna elements of the array. For the same reasons, a normalization factor also needs to be estimated in the case of polarization self-calibration. 

As the first-order corrections have already been applied in the process of determining $\mathrm{V_{1,ij,app}}(\nu, t)$, $D_\mathrm{1,i}(\nu, t)$ is expected to be small. $D_\mathrm{1,i}$ is estimated using full Jones matrix solver {\it QuartiCal}. We define
\begin{equation}\label{eq:pol_selfcal_1}
 \begin{split}
 D_\mathrm{1,i}(\nu, t)&=D_\mathrm{i}(\nu, t)\ P_\mathrm{D}(\nu, t),
 \end{split}
\end{equation}
where $D_\mathrm{i}(\nu, t)$ represents the true values of the errors on the ideal instrumental primary beam mentioned in Equation \ref{eq:beamcor_1}, and $P_\mathrm{D}(\nu, t)$ is the residual {\it poldistortion} left behind in the data after the first-order corrections. The mean of all $D_\mathrm{i}(\nu, t)$ is expected to lie close to the identity matrix and is the full-Stokes analog of the scalar normalization used in intensity self-calibration. For ease of notation we drop the explicit $\nu$ and $t$ dependence of $\mathrm{V_{1,ij,app}}$s, $D_\mathrm{1,i}$s, $D_\mathrm{i}$s and $P_\mathrm{D}$s in the following text.

We initiate the self-calibration process using $\mathrm{V_{1,ij,app}}$s as the initial model. While $D_\mathrm{1,i}$ is not assured to be close to the Identity matrix, the first-order calibration already applied makes them small enough for a self-calibration-like approach to converge. $P_\mathrm{D}$ is estimated assuming that all $D_\mathrm{i}$ are close to identity. 
To estimate $P_\mathrm{D}$, $S_\mathrm{D}$, the sum of the variance of $D_\mathrm{i}$ with respect to identity matrix, $I$, is minimized. $S_\mathrm{D}$ is defined as,
\begin{equation*}
\begin{split}
 S_\mathrm{D} & = \sum_\mathrm{i}\mathrm{var}(D_\mathrm{i}-I)\\
 & = \sum_\mathrm{i} \mathrm{var}(D_\mathrm{1,i}\ P_\mathrm{D}^{-1}-I)\\
 & = \sum_\mathrm{i} Tr \left[ (D_\mathrm{1,i}\ P_\mathrm{D}^{-1}-I)(D_\mathrm{1,i}\ P_\mathrm{D}^{-1}-I)^\dagger \right]\\
 & = \sum_\mathrm{i} Tr \left[D_\mathrm{1,i}\ P_\mathrm{D}^{-1}\ P_\mathrm{D}^{\dagger-1}D_\mathrm{1,i}^\dagger\right]-Tr\left[P_\mathrm{D}^{\dagger-1}\ D_\mathrm{1,i}^\dagger\right]\\
 & \qquad\qquad -Tr\left[D_\mathrm{1,i}\ P_\mathrm{D}^{-1}\right]+Tr\left[I\right]\\
\end{split}
\end{equation*}
For minimization, $\frac{\partial S_\mathrm{D}}{\partial P_\mathrm{D}} =0$ is imposed, leading to the following relation,  
\begin{equation}
\begin{split}
\left (\sum_\mathrm{i} P_\mathrm{D}^{-1} D_\mathrm{1,i} P_\mathrm{D}^{-1}\right)^\dagger & =\biggl(\sum_\mathrm{i} P_\mathrm{D}^{-1} P_\mathrm{D}^{\dagger-1} D_\mathrm{1,i}^\dagger D_\mathrm{1,i} P_\mathrm{D}^{-1} \biggr)^\dagger\\
P_\mathrm{D}^{\dagger-1}\left(\sum_\mathrm{i}D_\mathrm{1,i}^\dagger\right)P_\mathrm{D}^{\dagger-1} &=  P_\mathrm{D}^{\dagger-1} \left(\sum_\mathrm{i} D_\mathrm{1,i}^\dagger D_\mathrm{1,i} P_\mathrm{D}^{-1}\right) P_\mathrm{D}^{\dagger-1}\\
 \left(\sum_\mathrm{i} D_\mathrm{1,i}^\dagger D_\mathrm{1,i}\right) P_\mathrm{D}^{-1} & =\sum_\mathrm{i} D_\mathrm{1,i}^\dagger\\
 P_\mathrm{D}^{-1} & =\left(\sum_\mathrm{i} D_\mathrm{1,i}^\dagger D_\mathrm{1,i}\right)^{-1} \sum_\mathrm{i} D_\mathrm{1,i}^\dagger\\
 P_\mathrm{D} & =\left[ \left(\sum_\mathrm{i} D_\mathrm{1,i}^\dagger D_\mathrm{1,i}\right)^{-1} \sum_\mathrm{i} D_\mathrm{1,i}^\dagger\right]^{-1}
\end{split}
\end{equation}

We then correct each of the $D_\mathrm{1,i}$ for the $P_\mathrm{D}$ and obtain $D_\mathrm{i}$ as given by,
\begin{equation} \label{eq:model-error}
\begin{split}
 D_\mathrm{i}&=D_\mathrm{1,i}\ P_\mathrm{D}^{-1}.
\end{split}
\end{equation}
Using Equation \ref{eq:model-error}, Equation \ref{eq:leakge_cor2} can be written as,
\begin{equation}
 \begin{split}
\mathrm{V_{ij,Xcor}}=&D_\mathrm{1,i}\ \mathrm{V_{1,ij,app}}\ D_\mathrm{1,j}^\dagger\\
=&D_\mathrm{i}\ P_\mathrm{D}\ \mathrm{V_{1,ij,app}}\ P_\mathrm{D}^\dagger\ D_\mathrm{j}^\dagger\\
=& D_\mathrm{i}\ \mathrm{V_{ij,app}}\ D_\mathrm{j}^\dagger\\
\mathrm{V_{ij,app}}=& D_\mathrm{i}^{-1}\ \mathrm{V_{ij,Xcor}}\ D_\mathrm{j}^{-1\dagger}, 
 \end{split}
\end{equation}
where $\mathrm{V_{ij,app}}$ is the final apparent source visibility. 
The final true source brightness matrix, $\mathbf{B}(\nu,\ t,\ \vec{l})$, is then obtained using the ideal primary beam as,
\begin{equation}
 \mathbf{B}(\nu,\ t,\ \vec{l})=E(\nu,\ t,\ \vec{l})^{-1}\ \mathbf{B}_\mathrm{app}(\nu,\ t,\ \vec{l})\ E(\nu,\ t,\ \vec{l})^{-1\dagger}.
\end{equation}

\begin{figure*}
    \centering
    \includegraphics[trim={3.2cm 1.5cm 6.5cm 1cm},clip,scale=0.7]{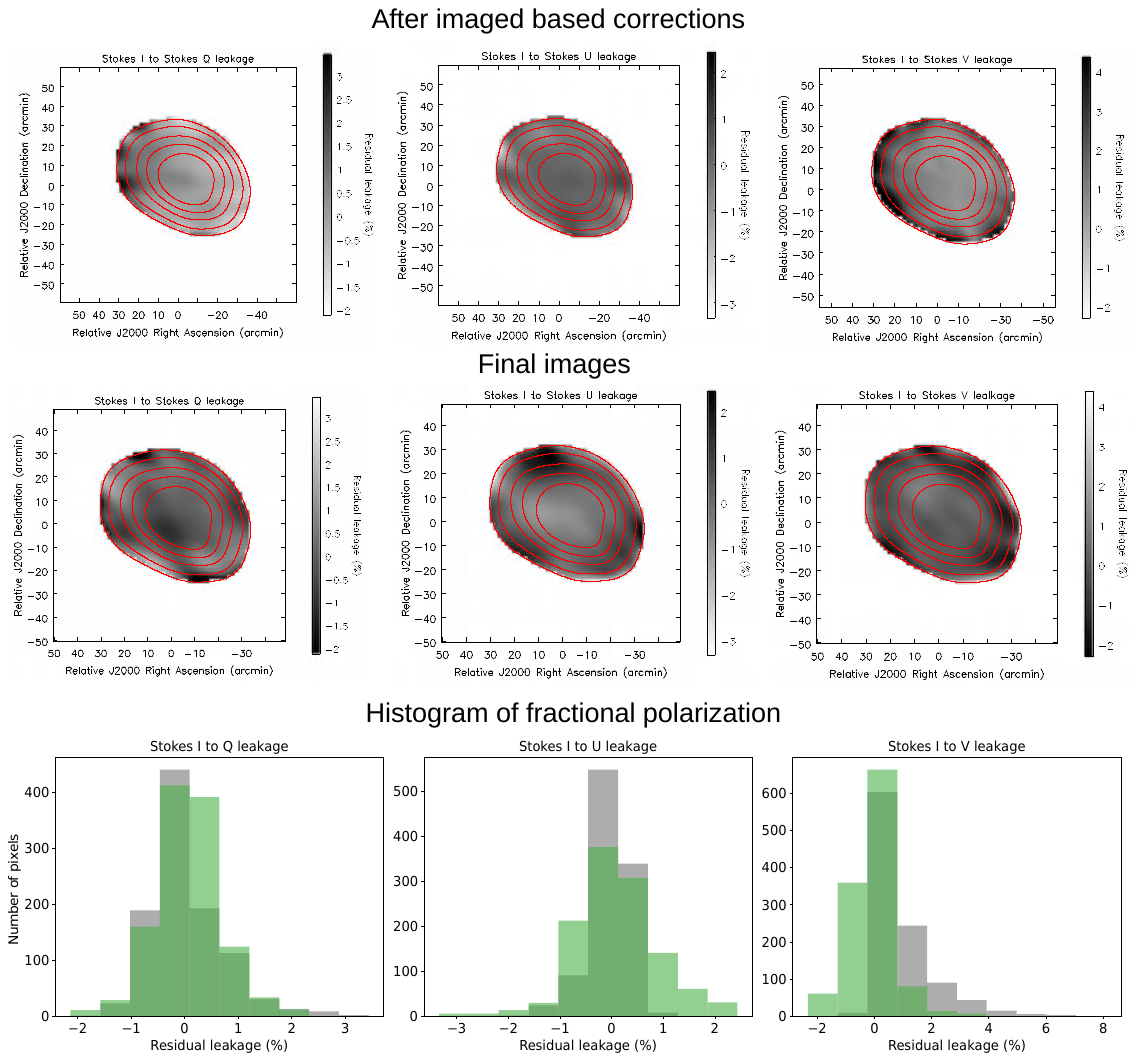}
    \caption[Residual Stokes leakages after the image-based correction and in the final images.]{Residual Stokes leakages after the image-based correction and in the final images for the observing epoch 2014 May 04 at 80 MHz. Red contours represent the Stokes I emission. Contour levels are at 10, 20, 40, 60, and 80\% of the peal Stokes I flux density. {\it Top row: }Polarization fraction over the quiet-Sun region obtained after the image-based correction is shown. No systematic variation is present. $L_\mathrm{residual}$ for Stokes Q, U, and V are respectively $\lesssim3.5,\ 2$ and 1\%. {\it Middle row: }Polarization images over the quiet-Sun region after the final self-calibration iteration is shown. $L_\mathrm{residual}$ for the Stokes Q, U, and V images are $\lesssim1.8,\ 0.8$ and 0.08\% respectively. {\it Bottom row: }Histogram of the pixel values of each Stokes image after the image-based correction is shown in gray in the background and of the final images in green in the foreground.}
    \label{fig:polcal_progress}
\end{figure*}

\subsection{Estimation of Residual Leakages}\label{subsec:res_leakage_estimation}
For astronomical observations, residual leakages in the polarization images are usually determined using unpolarized celestial sources. The polarized emission detected from such sources, after polarization calibration, provides an estimate of the residual leakages from Stokes I to other Stokes parameters. For large FoV instruments, the leakage can be a strong function of direction. Despite this, an approach based on observations of multiple unpolarized sources has been successfully used \citep{lenc2017}. They first determined the leakages toward the individual sources. Next, a 2D second-order polynomial surface was fitted to determine the leakages as a function of direction. It has recently been shown that numerous background sources can be detected in Stokes I with the MWA even in the presence of the Sun \citep{Kansabanik2022}. However, to pursue a similar approach for determining leakages, a large number of these sources need to be detected in other Stokes parameters as well, which is yet to be demonstrated.
 
Instead, I use the quiet-Sun regions in our images to estimate residual leakage fraction, $L_\mathrm{residual}$, as follows,
\begin{enumerate}
    \item When Stokes Q, U, and V emission is detected with more than $3\sigma$ significance over more than 50\% of the quiet-Sun region, we use the median values of the pixels detected in each Stokes plane, and then define the residual leakages as,
    \begin{equation}\label{eq:median_leakage}
        L_\mathrm{residual}=\frac{\mathrm{med}(L)_\mathrm{Q,U,V}}{I_\mathrm{max}},
    \end{equation}
    where $\mathrm{med}(L)_\mathrm{Q}$, $\mathrm{med}(L)_\mathrm{U}$ and $\mathrm{med}(L)_\mathrm{V}$ are the median values of the Stokes Q, U, and V pixels detected with more than $3\sigma$ significance and $I_\mathrm{max}$ is the maximum pixel value in $\mathrm{Jy}$ per beam in the quiet-Sun regions.
    \item When no polarization is detected over the quiet-Sun region, we define a residual leakage limit based on the $3\sigma$ limit. In such situations we define the $L_\mathrm{residual}$ as:  
    \begin{equation}\label{eq:res_leakage}
  L_\mathrm{residual}<\left|\frac{3\times \sigma_\mathrm{Q,U,V}}{I_\mathrm{max}}\right|,  
\end{equation}
   where $\sigma_\mathrm{Q},\ \sigma_\mathrm{U},\ \sigma_\mathrm{V}$ are the rms noise values in $\mathrm{Jy}$ per beam of the Stokes Q, U, and V images, respectively, close to the Sun and $I_\mathrm{max}$ is defined as before.
   \item Quiet-Sun emission in Stokes I may not always be detectable during the presence of very bright radio bursts (e.g. when a type-III radio burst is in progress). In such cases, the residual leakage is estimated using the closest time stamp where the Stokes I quiet-Sun emission is detected. 
\end{enumerate}

\subsubsection{A P-AIRCARS Stress Test}
\label{subsec:stress-test}
As a part of the process of development of P-AIRCARS and evaluating its efficacy, it was tested on some particularly challenging datasets. The most challenging of these observations comes from 2014 May 04 when the MWA was pointed to its lowest permissible elevation, where the sensitivity of the MWA and its polarization response is the poorest. This observation was at 80 MHz, where the flux density of the Sun is the lowest and the Galactic background the strongest. Additionally, these data also correspond to quiet-Sun conditions, where the Sun is essentially a featureless extended source and is the hardest to calibrate and image. Processing these data, hence, corresponding to a stress test of P-AIRCARS. It is quite reasonable to expect that, if P-AIRCARS can meet the challenges of calibrating and imaging these data, it will be able to successfully deal with most other MWA solar data. This section substantiates the performance of P-AIRCARS on these data.

The Stokes images after the corrections of the modeled primary beam response are shown in the top row of Figure \ref{fig:res_stokes_leakage}. $L_\mathrm{residual}$ for the Stokes Q, U, and V at this stage is $\sim34,\ -35,$ and $2.8$\% respectively. As the Stokes I to V leakage was more than $2\%$, an imaged-based leakage correction was performed for Stokes V, as discussed in \ref{subsec:imaged_based_cor}. The top row of Figure \ref{fig:polcal_progress} shows the percentage polarization over the quiet-Sun regions with more than $3\sigma$ detection in Stokes I after the image-based leakage corrections. The DR of the Stokes I image is $\sim$500, and it is evident that Stokes I to Q, U leakages have reduced by about an order of magnitude (top panel of Figure \ref{fig:polcal_progress}) beyond what was obtained by primary beam correction (top panel of Figure \ref{fig:res_stokes_leakage}). In this case, Stokes Q emission is detected at $> 3\sigma$ significance over the quiet-Sun region, but that at Stokes U and V is not detected. Hence Equation \ref{eq:res_leakage} is used to obtain the limit of $L_\mathrm{residual}$ for Stokes U and V. $L_\mathrm{residual}\approx-3.5$ for Stokes Q calculated using Equation \ref{eq:median_leakage} and $L_\mathrm{residual}<|2| $ and $|1|\%$ for Stokes U and V, respectively, which are calculated using Equation \ref{eq:res_leakage}. No systematic variation of the polarized emission is seen across the solar disc. After the image-based leakage correction, several rounds of polarization self-calibration are performed and $P_\mathrm{D}$ is corrected for at each iteration. This process is deemed converged when the absolute total polarized flux densities for Stokes Q, U, and V become stable. The rms of the Stokes images and the residual leakages are found to improve with every iteration of polarization self-calibration. The final Stokes Q, U, and V images are shown in the middle row of Figure \ref{fig:polcal_progress}. Most of the regions in Stokes Q, U, and V images are found to be consistent with noise. The $L_\mathrm{residual}$ of the final Stokes Q, U, and V images have reached the values $<|1.8|,\ |0.8|$ and $|0.08| \%$ respectively. 

The histogram of the pixel values of the Stokes Q, U, and V leakage fractions are shown in the bottom row of Figure \ref{fig:polcal_progress}. The median leakage values are already close to zero after image-based corrections, as expected, but tend to be asymmetric in some cases. In some cases, low-level artifacts are seen after image-based corrections. The regions with $>5\sigma$ detection after the image-based correction are used for subsequent rounds of polarization self-calibration. The histograms of pixel values for the final images (shown in green in the bottom panels of Figure \ref{fig:polcal_progress}) grow more symmetric and demonstrate a reduction in artifacts in the polarization images. The magnitude of this improvement can vary across different Stokes planes. For example, the histograms for Stokes Q, both after image-based correction (gray) and polarization self-calibration (green), are symmetric and very similar, demonstrating that image-based correction was already very good and did not give rise to any significant artifacts. On the other hand, after image-based correction, the Stokes V histogram shows a very skewed distribution with a positive tail extending out to 7.5\%. Polarization self-calibration leads to a much more symmetric distribution with a smaller span of $\pm$2\%. This demonstrates the efficacy of polarization self-calibration in reducing the artifacts in the Stokes V image. The situation for Stokes U is found to lie between these two regimes. While the improvements are always seen after polarization self-calibration, the trends seen here are not general and can differ from observation to observation.

\subsection{Computational Load}
P-AIRCARS uses the full Jones calibration package {\it QuartiCal} \citep{Quartical2022} for calibration and {\it WSClean} \citep{Offringa2014} for imaging.  The computational load of P-AIRCARS naturally depends on the details of the data, the choices made during analysis, and the computation hardware available. I present some numbers here to provide a general sense of the computational load and clock time taken for calibration. We use a 2 GHz CPU core with hyperthreading (two threads per core) as the benchmark device. The very first Stokes I self-calibration run on a chosen time and frequency slice (referred to as the reference time and frequency) takes about an hour. Once the gain solutions obtained from this reference slice have been applied to the dataset, the next step, bandpass self-calibration is performed in parallel on each of the 1.28 MHz wide 24 spectral chunks. This takes about 15$-$20 minutes per hyperthreaded core. The final step of polarization self-calibration takes about 45 minutes per spectral slice and is also parallelized across the frequency axis. Thus for a typical P-AIRCARS run, full-Stokes calibration takes about 2 hours for a usual MWA dataset with 30.72 MHz bandwidth when parallelized across 25 cores (50 threads). As calibration (and imaging) are both done in a spectroscopic snapshot mode, only a tiny fraction of the entire dataset is needed for any individual calibration run making the memory footprint very small. More details about the implementation of the algorithm, the optimizations used, and the computational load will be presented in Chapter \ref{paircars_implementation}.

\section{Results and Discussion}\label{sec : result}
\begin{figure*}[!ht]
    \centering
    \includegraphics[trim={2cm 9cm 3cm 1cm},clip,scale=0.6]{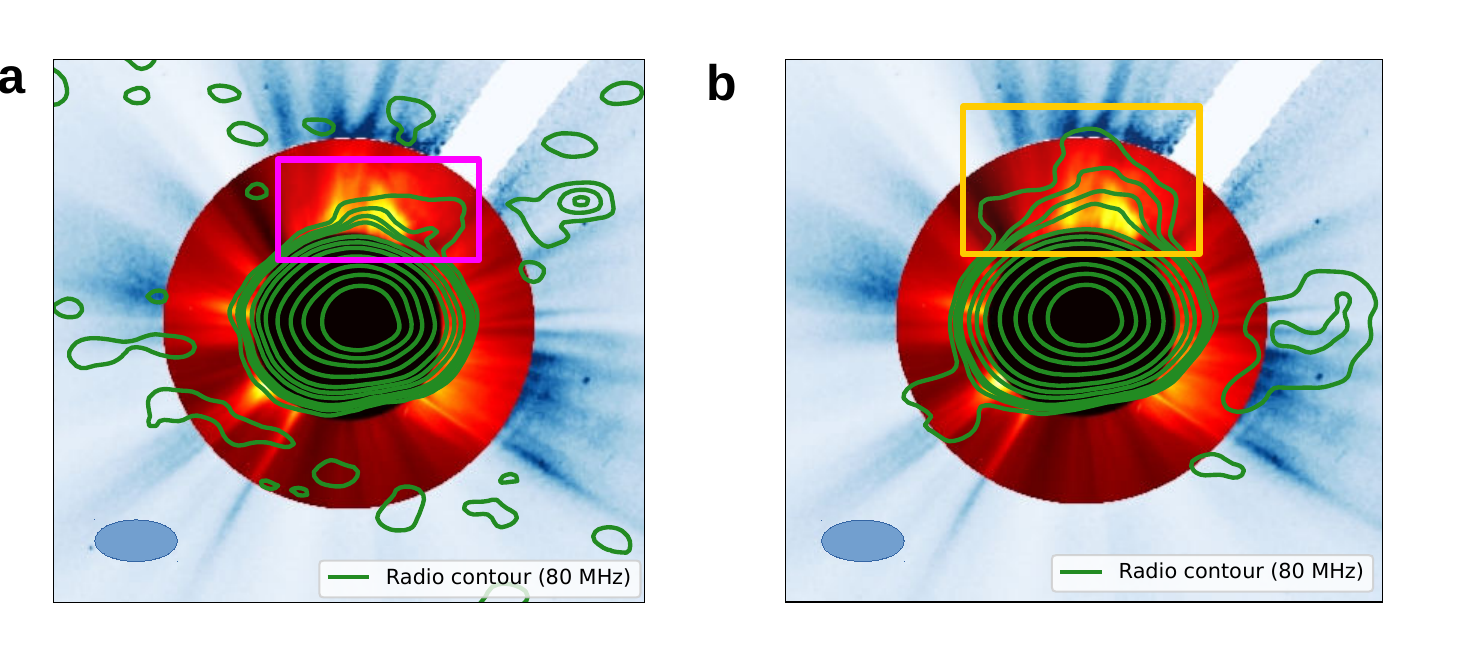}
    \caption[Comparison between images made using AIRCARS and P-AIRCARS.]{Comparison between images made using AIRCARS and P-AIRCARS. The background images are LASCO white light coronagraph images. The red color map represents the LASCO C2 images, and the blue color map represents the LASCO C3 images. The green contours represent the radio images at 80 MHz. The contour levels are 0.2\%, 0.4\%, 0.6\%, 0.8\%, 2\%, 4\%, 8\%, 20\%, 40\%, 60\%, 80 \% of the peak flux density. The filled ellipses at the lower left of the images are the PSF. {\it{a.}} The image made using AIRCARS has small extended emissions from the CME as marked by the pink box. There are noise peaks at the 0.2\% level. {\it{b.}} The image made using P-AIRCARS has extended emission over a larger region as marked by the yellow box. The radio emission covers the full white light CME.}
    \label{fig:aircars_paircars}
\end{figure*}

The performance of P-AIRCARS on a particularly challenging quiet-Sun dataset has already been substantiated in Section \ref{subsec:stress-test}. This section substantiates the various improvements that P-AIRCARS images represent and the science opportunities they enable. P-AIRCARS usually achieves $<|1|\%$ residual leakages for Stokes Q and $<|0.1|\%$ residual leakage for Stokes U and V. These values are comparable to what is generally achieved for high-quality astronomical observations \citep{lenc2017,lenc2018,Risley2018,Risley2020} with the MWA and much better than those achieved by earlier spectropolarimetric solar studies \citep{Patrick2019,Rahman2020}. P-AIRCARS not only delivers a very small residual leakage, but it also provides improved Stokes I imaging fidelity as compared to AIRCARS, as shown next.

\subsection{Improvements in Stokes I Imaging}
In addition to the operational efficiency and polarimetric imaging capability, P-AIRCARS includes multiple improvements over the earlier Stokes I state-of-the-art pipeline \citep[AIRCARS;][]{Mondal2019}, which was focused on spectroscopic snapshot imaging. It is well unknown that the lack of calibration of instrumental polarization leakages leads to a reduction in the DR \citep{Bhatnagar2001}. Besides that, while imaging over multiple frequency channels, if the instrumental bandpass is not corrected, it leads to artifacts in the images. 
P-AIRCARS includes both these capabilities, which lead to a significant improvement in Stokes I image fidelity and noise properties.

\begin{figure*}[!ht]
    \centering
    \includegraphics[trim={1.8cm 6.9cm 1.8cm 1cm},clip,scale=0.5]{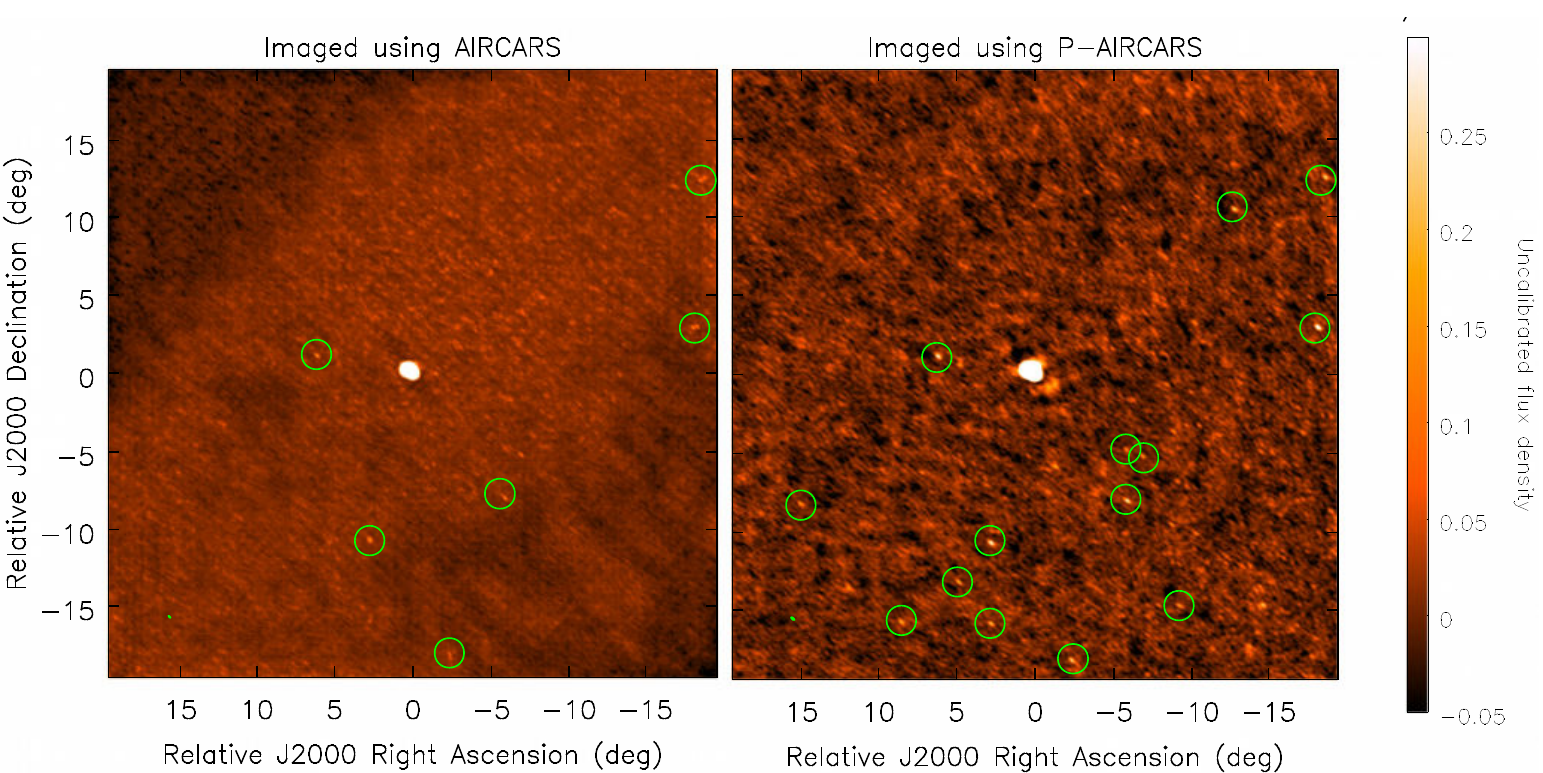}
    \caption[Comparison of the noise characteristics of the images made using AIRCARS and P-AIRCARS.]{Comparison of the noise characteristics of the images made using AIRCARS and P-AIRCARS. The noise characteristic of the same image, as shown in Figure \ref{fig:aircars_paircars} is shown for the 30$^{\circ}\times$30$^{\circ}$ FoV. The detected background galactic and extragalactic radio sources are marked by green circles. 
    {\it Left panel: } The AIRCARS image shows a non-uniform noise behavior. A bright wide strip of enhanced noise runs across the full image passing from southwest to the northeast. The DR of this image is $\sim1000$, and 6 background sources are detected. {\it Right panel: } The P-AIRCARS image shows a much more uniform noise characteristic over the full FoV. Both small and large angular scale artifacts have reduced in strength. The DR of the image is $\sim1800$. The substantially improved DR leads to the detection of 14 background sources with higher detection significance in the image.}
    \label{fig:aircars_paircars_rms}
\end{figure*}

A comparison of radio maps from AIRCARS (left panel) and P-AIRCARS (right panel) made using the same data is shown in Figure \ref{fig:aircars_paircars}. The radio maps at 80 MHz have been superposed on Large Angle and Spectrometric Coronagraph \citep[LASCO,][]{Brueckner1995} images and use data of duration of 2 minutes and a bandwidth of 2 MHz. AIRCARS image shows a small weak emission feature from a CME in 2014 May 04 (marked by the pink box in the left panel) over a region comparable to the point spread function (PSF). The P-AIRCARS image shows emission at a similar strength but extended over a much larger region covering the full extent of the white light CME (marked by the yellow box in the right panel). It is evident that the imaging artifacts near the Sun in the P-AIRCARS image are at a lower level, and another weak extended emission feature lying on the western limb overlapping with the LASCO C3 FoV has a reliable detection. The noise characteristics of the P-AIRCARS image have also improved significantly. To substantiate this, Figure \ref{fig:aircars_paircars_rms} shows the entire FoV of the image shown in Figure \ref{fig:aircars_paircars}. It is evident that the noise characteristics of the AIRCARS image (left panel Figure \ref{fig:aircars_paircars_rms}) are not uniform across the image, there is a bright and extended noise band running diagonally across the image from southwest to northeast. The DR of this image is $\sim1000$. On the other hand, the noise characteristic of the P-AIRCARS image (right panel of Figure \ref{fig:aircars_paircars_rms}) is quite uniform, and the large angular scale feature seen in the AIRCARS image is no longer evident. The DR of the P-AIRCARS image has improved by $\sim80\%$, which leads to the value $\sim1800$. 

\begin{figure*}[!ht]
    \centering
    \includegraphics[trim={2cm 9cm 1cm 1cm},clip,scale=0.5]{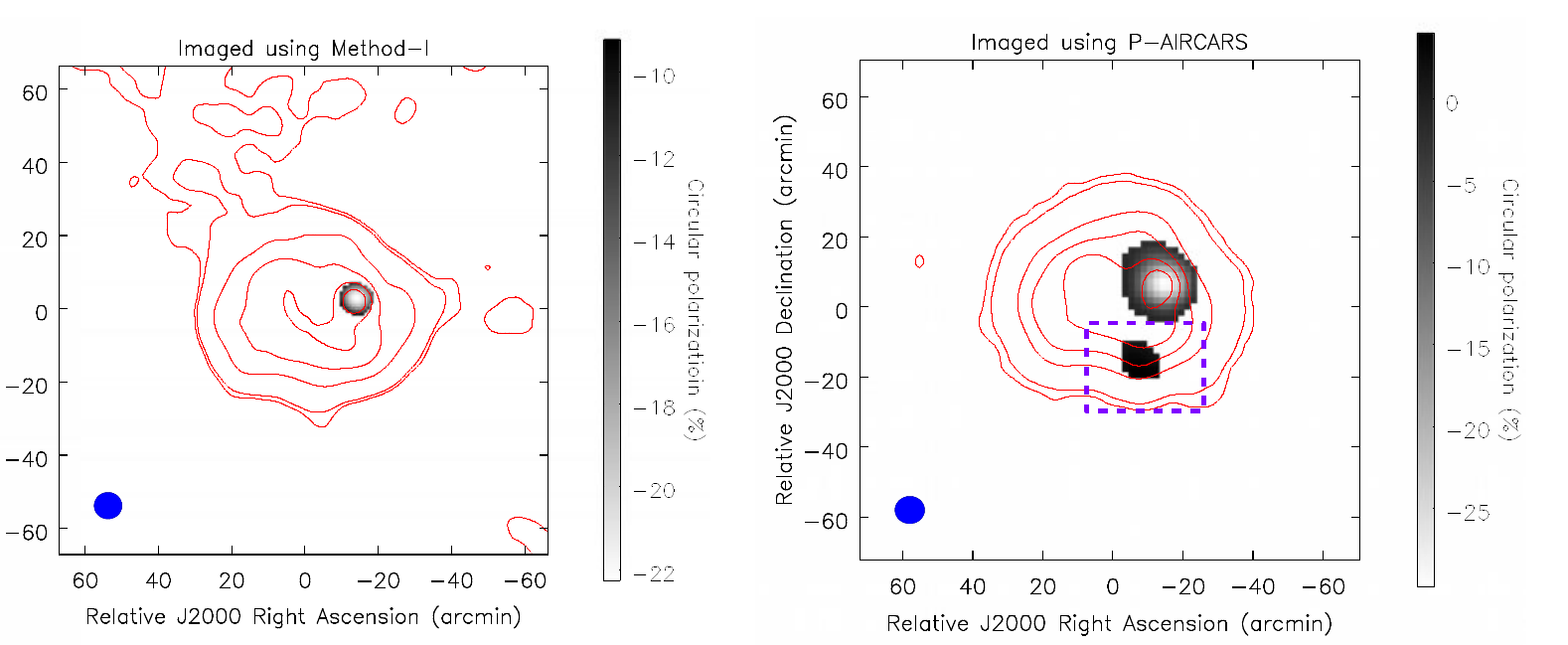}
    \caption[Comparison between Stokes I and Stokes V images made using the method-I and P-AIRCARS.]{Comparison between Stokes I and Stokes V images made using method-I and P-AIRCARS. The Stokes I and Stokes V images of an active solar emission observed at 159 MHz on 2014 October 24, 03:46:00 UTC are shown. Circular polarization fractions are shown by a gray color map. Only the Stokes V emission detected at more than 3$\sigma$ detection is shown. The red contours represent the Stokes I emission. The contour levels are at 0.8\%, 2\%, 20\%, 40\%, 60\%, and 80\% of the peak flux density. {\it Left panel: } The image is made following method-I. The DR of the Stokes I image is $\sim$35, and that of the Stokes V image is $\sim$30. There are noise peaks at the 2\% level of the peak flux density in the Stokes I image. The residual Stokes V leakage is $\lesssim$5\%. {\it Right panel: }The image is made using P-AIRCARS. The DR of the Stokes I image is $\sim$577, and that of the Stokes V image is $\sim$375, which is an order of magnitude better than that for the method-I. The first noise peak appears at the $0.8\%$ level of the peak flux density in the Stokes I image. We have detected another positive circularly polarized source (marked by the purple dashed box), which could not be detected in the image made using method-I. This source has a circular polarization fraction $\sim3\%$. The residual Stokes V leakage is $\lesssim0.5\%$, which is also an order of magnitude improvement compared to that for method-I.}
    \label{fig:paircars_methodI}
\end{figure*}

GS emission from CMEs has only been detected in a handful of cases to date, of which an even smaller fraction are at meter wavelengths \citep{bastian2001, Bain2014}. Using AIRCARS, \citet{Mondal2020a} have recently demonstrated the ability to detect GS emission from CME plasma and model spatially resolved spectra to estimate plasma parameters of the CME for a slow and unremarkable CME. With further improved imaging from P-AIRCARS, it is now feasible to detect much fainter GS emissions from CMEs and trace them out to higher coronal heights. This capability will make it possible to use this powerful tool for routinely measuring the CME plasma parameters and magnetic fields in the middle and upper corona, which are discussed and demonstrated in detail in Chapters \ref{cme_gs1} and \ref{cme_gs2}.

\subsection{Comparison With Method-I}
Section \ref{sec:previous-attempts} discussed the earlier approach to spectropolarimetric solar imaging using the MWA, which has been referred to as the Method-I in this chapter \citep{Patrick2019}, and its limitations. The robust first-principles-based polarization calibration approach of P-AIRCARS, on the other hand, ensures that all known instrumental effects are corrected. The Stokes images delivered by P-AIRCARS are limited primarily by the thermal noise of the data, and the rms noise seen in Stokes I and V images is significantly smaller than that seen in images from Method-I. 

A comparison between images delivered by Method-I (left panel) and P-AIRCARS (right panel) is shown in Figure \ref{fig:paircars_methodI}. Both images have been made from the same observation on 2014 October 24, 03:46:00 UTC at 159 MHz. The red contours represent the Stokes I emission. The Method-I image shows artifacts and noise peaks at $\sim$2\% of peak flux density. The largest noise peak in the P-AIRCARS image is at  $\sim$0.8\% of the peak flux density. The circular polarization percentage is shown by gray scales in regions where the Stokes I and V emissions are both detected with $> 3\sigma$ significance. The DRs of both the Stokes I and V images have increased by an order of magnitude -- for the Stokes I image from $\sim35$ to $\sim577$ and for the Stokes V image from $\sim30$ to $\sim375$. 

This improvement in DR enables the detection of a much weaker Stokes V emission with $\sim$3\% circular polarization in the P-AIRCARS image (purple dotted box, right panel, Figure \ref{fig:paircars_methodI}), which was not detected using Method-I, and the region over which the circular polarization is detected with confidence in the P-AIRCARS image is substantially larger than that in the method-I image. $L_\mathrm{residual}$ has been estimated to be $\sim5\%$ for the Method-I and $<|0.5|\%$ for P-AIRCARS. The peak circular polarization fraction of the image from Method-I is $\sim-22$\% and from P-AIRCARS $\sim-27\%$. Considering the $\sim5\%$ uncertainty on circular polarization fraction of Method-I \citep{Patrick2019}, both these values are consistent. The very small $L_\mathrm{residual}$ together with the high imaging DR delivered by P-AIRCARS bodes very well for the detection of the very low level of circularly polarized emission from the quiet-Sun thermal emission. This is also a very promising development for the detection of polarized emission from GS emission from the CME plasma.

\subsection{Polarization Images of the Sun Using P-AIRCARS}
\begin{figure*}
    \centering
    \includegraphics[trim={2cm 6.5cm 2.5cm 1.2cm},clip,scale=0.55]{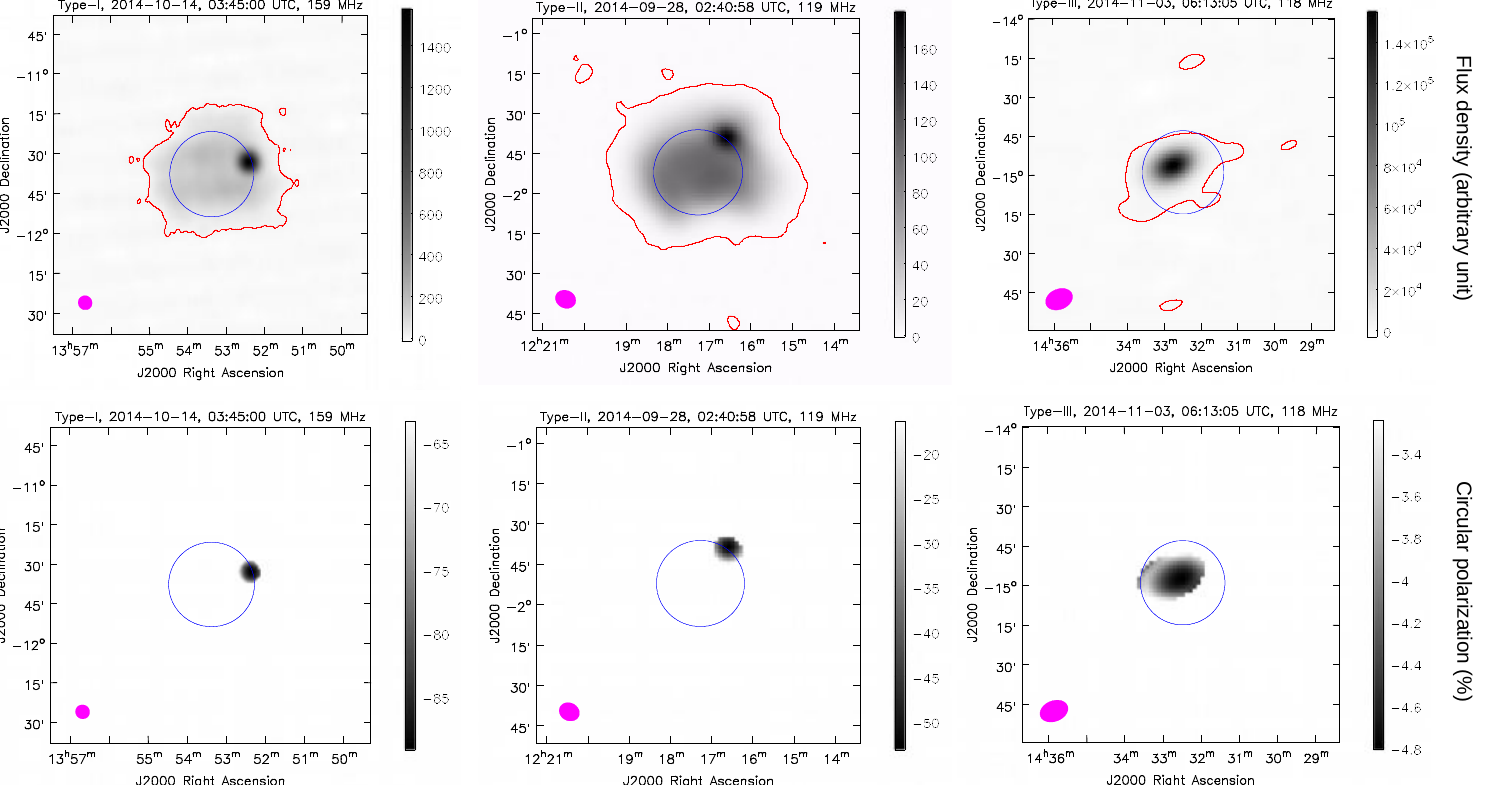}
    \caption[Example Stokes V images of type-I, -II, and -III solar radio bursts.]{Example Stokes V images of type-I, -II, and -III solar radio bursts. Stokes I images are shown in the top panels and circular polarization fractions are in the bottom panels. Only the regions with $> 5\sigma$ detection in both Stokes I and Stokes V are shown in the bottom panels. The magenta ellipses at the bottom left of each panel show the PSF. The red contours in each image represent the first contour, which picks up the noise in Stokes I images. The optical disk of the Sun is shown by the blue circles. The residual Stokes V leakage in all of these images is $\lesssim1\%$. {\it Left panel: } Circularly polarized emission from a type-I noise storm at 159 MHz. The DR of the Stokes I and Stokes V images is 788 and 518, respectively. The red contour is at 0.8\% of the peak Stokes I emission. The maximum circular polarization fraction is -89\%. {\it Middle panel: } Circular polarization from a type-II solar radio burst at 119 MHz. The DR of the Stokes I and Stokes V images is 900 and 950, respectively. The red contour is at 0.5\% of the peak Stokes I emission. The maximum circular polarization fraction is -53\%. {\it Right panel: } Stokes V image of type-III solar radio burst at 118 MHz. The DR of the Stokes I and Stokes V images are 1200 and 233, respectively. The red contour is at 0.8\% of the peak Stokes I emission. The maximum circular polarization fraction is -4.8\%.}
    \label{fig:radio_bursts}
\end{figure*}

Figure \ref{fig:radio_bursts} shows the Stokes I and V images for example type-I, type-II, and type-III solar radio bursts. These images are made at the native time and frequency resolution, 0.5 s and 40 kHz, of the observation. We have chosen different contour levels for different images to show the first noise peak of that image. Fractional Stokes V images are shown in grayscale over the regions where both Stokes I and Stokes V emissions are detected with $> 5\sigma$ significance. The peak Stokes V for these type-I, type-II and type-III radio bursts are -89\%, -53\%, and -4.8\%, respectively. The Stokes V flux densities of these bursts are $\sim$ 70, 180, and 70 SFU, respectively. In all cases, we obtain images with DR varying between $\sim500$ and $1200$. The high DR of these images enables us to detect a significant part of the quiet-Sun emission in Stokes I, even in the presence of bright active emissions. Interestingly, the peak of the Stokes V emission from the type-III burst is slightly displaced toward the southeast from the peak of the Stokes I emission, while they are coincident for the type-I and type-II bursts shown.

These, along with the quiet-Sun data showcased in \ref{subsec:stress-test} span a large range along the flux density axis. This demonstrates the capability of P-AIRCARS to produce high-DR images in a variety of different solar conditions. Additionally, the residual instrumental polarization leakages of $\lesssim 1\%$ in all the observations shown in Figure \ref{fig:radio_bursts}, represent an improvement approaching an order of magnitude over the $\sim5-10\%$ leakages obtained earlier \citep{Patrick2019,Rahman2020}. To the best of our knowledge, these are the lowest to have been achieved for any meter-wavelength solar radio images and bring the exciting science target of measuring the weak circular polarization from the quiet Sun within reach.

\subsection{Heliospheric Measurements Using \\Background Radio Sources}
The Sun is the source with the highest flux density in the low-frequency sky, with even the quiet Sun spewing flux density exceeding $10^4-10^5\ \mathrm{Jy}$ at the MWA band. Apart from a handful of sources whose flux density goes up to hundreds of $\mathrm{Jy}$, that of the bulk of other celestial sources lies in the range of a few $\mathrm{Jy}$ or weaker. Imaging a reasonably dense grid of background sources in the presence of the Sun in the FoV, hence, imposes a large DR requirement, which had not been met until recently. The high DR of the images from P-AIRCARS now routinely allows us to detect multiple background galactic and extragalactic radio sources even in the presence of the Sun. An example is shown in the right panel of Figure \ref{fig:aircars_paircars_rms}, which is made over 2.56 $\mathrm{MHz}$ and 10 $\mathrm{s}$. The P-AIRCARS image shows the detection of 14 background sources with $\gtrsim5\sigma$ detection significance. The closest of these is at $\sim$20 $R_{\odot}$ from the Sun and has a flux density of 4.9 $\mathrm{Jy}$. Another example is available in \citet{Kansabanik2022} and presented in Chapter \ref{fluxcal}, where the independently available flux densities of the background sources were used for arriving at robust solar absolute flux density calibration for MWA solar observations. The rms noise in these images approaches that achieved by GLEAM, once the excess system temperature due to the Sun is taken into account.

Measurements of interplanetary scintillation (IPS) of background radio sources have long been used to measure the electron density, properties of turbulence and velocities of CMEs, and solar wind in the heliosphere \citep[e.g.][]{Coles1978,Manoharan1990, Jackson1998} and more recently for driving the boundary conditions for magnetohydrodynamic models of the solar wind \citep{Yu2015}. At the MWA, IPS observations have generally been done over small time windows while keeping the Sun at the null of the primary beam to avoid any contamination from solar emission \citep{Morgan2018,Morgan2018a,Chettri2018}. By providing the ability to routinely make Stokes I images of background sources, P-AIRCARS brings us a step closer to removing this limitation and opens the possibility of performing IPS observations without necessarily requiring to have the Sun in a null of the MWA primary beam.

IPS is remote-sensing technique and provides information complementary to what is available from other observations. However, it is not sensitive to heliospheric magnetic fields -- the key driver of space weather phenomena. By measuring the FR due to the heliospheric and/or CME plasma along the lines of sight to background linearly polarized sources or the diffuse galactic emission, radio observations provide the only known remote sensing tool for measuring these magnetic fields. This approach has been successfully implemented by using radio beacons from satellites \citep[e.g.][etc.]{bird1990, Jensen2013, Wexler2019} as background sources, and more interestingly also using astronomical sources \citep[e.g.][etc.]{Mancuso2000, kooi2017, Kooi2021}. These observations were carried out at higher frequencies and using small FoV instruments, which can sample only a small part of the heliosphere at any given time. Wide FoV instruments like the MWA can sample large swaths of the sky at any given time and can potentially track CMEs as they make their way across the heliosphere. Measurements of FR simultaneously for large numbers of pierce points across the CME/heliosphere, open the very exciting possibility of constraining the models for CME/heliospheric magnetic fields using these data \citep{Bowman2013, Nakaraiakov2015}. P-AIRCARS delivers precise polarization calibration and produces high DR full-Stokes images and can already provide Stokes I maps of background sources. Similar efforts of demonstrating making similar full-Stokes maps of background sources and enabling these heliospheric FR measurements are in progress. 

\section{Conclusions and Future Work}\label{Conclusion_paircars} 
We have developed a robust and comprehensive state-of-the-art polarization calibration algorithm tailored to the needs of low-frequency solar observations. P-AIRCARS builds on the learnings from the earlier Stokes I imaging pipeline \citep[AIRCARS;][]{Mondal2019} and uses the advantages endowed by the MWA design features to perform full polarimetric calibration without requiring dedicated observations of calibrator sources. The key MWA design advantages in this context are the dense and compact core of the MWA array layout and its simple and well-characterized hardware.
Together these ensure that nearby antennas, the ones looking through essentially the same ionospheric patch, maintain a good degree of coherence. A detailed discussion and demonstration of these aspects have already been presented in Chapter \ref{paircars_principle}. All the P-AIRCARS images shown here were made without using any calibrator observations.

Polarization \ self-calibration \ was \ first \ demonstrated \ on \ simulated data by \\\citet{Hamaker2006}, but it had never been used for solar imaging. This work presents the first demonstration of solar polarization self-calibration and its ability to achieve high DR and high-fidelity full-Stokes images over a large range of solar conditions. The residual Stokes leakages for these images are on par with the usual astronomical images. 

Though P-AIRCARS was developed with polarization calibration of the solar observations in mind, at its core the algorithm is general and does not impose any solar-specific constraints. Its perturbative approach can be used for full Jones polarization self-calibration of the astronomical observations when a good initial sky model is available for a first-order calibration. 

The perturbative algorithm used in P-AIRCARS works well for homogeneous arrays like the MWA, where the ideal primary beam response of all antenna elements is essentially identical. MWA antenna elements are made of a total of 16 bow-tie dipoles arranged in a 4$\times$4 grid \citep{Tingay2013}. The MWA beam is modeled assuming that all of the 16 dipoles are healthy \citep{Sokolwski2017}. It has been shown using satellite measurements that even when one or two of the dipoles fail, it does not change the primary beam response close to its peak in a significant manner \citep{Line2018}. However, for precise polarization calibration being pursued here, these small changes do need to be accounted for. Presently, in P-AIRCARS we reject all antenna elements with even a single dipole failure. Though it does lead to a loss of sensitivity, it is usually tolerable as the number of such elements is usually small. However for science applications close to the edge of the sensitivity limits, e.g. detection of CME GS emission, it can become important to retain the sensitivity offered by the elements with defective dipoles. While the MWA beams can be modeled well for any subset of working dipoles \citep{Sokolwski2017}, it breaks the assumption of the array being a homogeneous one. The implication for P-AIRCARS is that an image-based approach for corrections for primary beams is no longer tenable (Equation \ref{eq:beamcor_2}). One must then use a class of algorithms referred to in the literature as {\it projection} algorithms, which can correct for image plane effects in the visibility domain. These algorithms can be used for correcting artifacts arising from a wide range of causes, ranging from the so-called {\it w-term} to the antenna-to-antenna differences in primary beams even for an array with identical elements and ionospheric phase screens. The algorithm of relevance is the one referred to as the {\it aw-projection} algorithm \citep{Jagannathan2017,Jagannathan2018,Sekhar2021}. It applies baseline-based corrections for primary beams in the visibility domain and is computationally very intensive.  As efficient implementations of such algorithms become available and the computational capacity available to us grows, it will become interesting to explore their use for scientifically interesting datasets with significant numbers of dipole failures to squeeze the most out of these data.

P-AIRCARS has been developed with the future SKAO in mind. It can be adapted straightforwardly for unsupervised generation of high-fidelity high DR full-Stokes solar images from the SKAO and other similar instruments with a dense central core. Producing high-quality solar radio interferometric images involves a steep learning curve, and its practice has remained limited to a small subset of the solar physics community. It is believed that the lack of availability of a robust tool suitable for the nonspecialist has long limited the use of radio observations in solar studies. We envisage that P-AIRCARS will prove to be a very useful tool for the solar and heliospheric physics community in times to come by filling this gap and making high-quality full Stokes solar radio imaging accessible. We aspire to make P-AIRCARS available as a stable, mature, and user-friendly software pipeline to the larger solar physics community. The present implementation of P-AIRCARS is described in Chapter \ref{paircars_implementation}.

\chapter {Robust Absolute Solar Flux Density Calibration}
\label{fluxcal}

In astronomy, as in any quantitative science, it is essential to associate a specific magnitude and units with any measurement. One of the most important measurables in astronomy is flux density, which defines the amount of energy received per unit time from an astronomical source if collected over a unit area for monochromatic emission. At radio wavelengths, the unit of this flux density is Jansky (Jy) and is defined to be $10^{-26}\ \mathrm{W\ m^{-2}\ Hz^{-1}}$. The extremely feeble radio emission from a celestial object incident on a radio telescope are picked up a tiny voltage fluctuations by the sensitive detectors and amplified by almost ten orders of magnitude before they are digitized and eventually stored after much processing in some arbitrary units. However, the quantity of interest from an astrophysical perspective is the energy incident on the telescope in units of Jy. This is achieved by a process referred to as flux density calibration. Precise measurement of the flux density in units of Jy is an essential requirement for arriving at reliable and accurate estimates of other physical properties of the object under study. In this chapter, I discuss the technique which I have developed to perform precise flux density calibration of solar observations with the MWA. This algorithm meets the requirements for the high-fidelity measurements from the high dynamic range (DR) images routinely produced by P-AIRCARS. The work presented in this chapter is based on \citet{Kansabanik2022}, which was published in the Astrophysical Journal.

\section{Introduction}\label{sec:intro}
The quiet Sun is just about the source with the highest flux density in the meter-wavelength radio sky, and its flux density can increase by multiple orders of magnitude during periods of active emissions. It is challenging to build sensitive radio instruments capable of providing a linear response spanning the entire range from the very faint astronomical sources of usual interest to the mega-Jy solar bursts. Usually the low-noise amplifiers, the very first element in the signal chain of a radio telescope (see Section \ref{subsec:lna} in Chapter \ref{mwa}), have sufficient DR, but that does not hold true for the downstream signal chain. Solar observations, therefore, typically require the use of {\textit {attenuators}} early on in the signal path to bring down the signal levels sufficiently, so that they lie in the linear regime of the downstream signal chain. Since attenuators can bring down the signal level by multiple orders of magnitude, it becomes hard to observe other astronomical sources, including most of the so-called flux density calibrators, with this attenuation in place, except for some exceptionally bright sources (e.g, Virgo-A, Crab, Cen-A, etc.). Flux density calibrators are comparatively bright radio sources whose flux density and spectra are known accurately and are typically used for absolute flux density calibration \citep[e.g][]{perley2017}. The standard flux density calibrator observations are generally done without any additional attenuators. Transferring the flux density calibration determined using observations of standard flux density calibrators to the Sun, generally observed with attenuators, is not straightforward, and it requires a detailed and accurate characterization of the attenuators. It is very effort intensive to do this characterization, and in practice, it is rarely available. This is also the case with the MWA.

Absolute solar flux density calibration usually relies on the use of a few very bright sources whose flux density is large enough for them to be observed using the same attenuation setting as solar observations. This approach is followed at the Nan\c{c}ay Radio Heliograph \citep[NRH; e.g.][]{bonmartin1983,avignon1989}, the Gauribidanur Radio Heliograph \citep[GRH; e.g.][]{sundaram2004,sundaram2005} and the Low Frequency Array \citep[LOFAR; e.g.][]{breitling2015}. Conventionally, calibrator observations are scheduled to be observed adjacent to and/or interspersed with observations of the target source(s). This is done to minimize the impacts of any drifts in instrumental gains in the time between observations of the calibrator and target source(s). The wide field of view (FoV) \citep{Tingay2013} and high primary beam sidelobes \citep{neben2015,Sutinjo2015,Sokolwski2017} of the MWA imply that observations of flux density calibrators, while the Sun is above the horizon, can have significant contamination from the solar signal. Hence, at the MWA, the practice has been to observe the flux density calibrators before sunrise and after sunset, and to broaden the pool of suitable calibrators, without the use of any additional attenuation. Additionally, during the early periods of MWA solar observations (2013 July to 2014 April), no calibrator observations in the same spectral configuration as the solar observations are available, further complicating the task of absolute flux density calibration for these observations.

Taking advantage of the fact that the MWA is a very well-characterized instrument, including a well-modeled primary beam, and the availability of reliable sky model \citep{haslam1982}, \citet{oberoi2017} implemented an innovative and computationally lean non-imaging technique for absolute solar flux density calibration. They estimated that the uncertainty in the absolute flux density estimates obtained using this technique generally lies in the range 10 $-$ 60\%. 
\citet{suresh2017} and \citet{sharma2018} used it successfully to get reliable flux density calibration for non-imaging studies. \citet{Mohan2017spreads} presented a technique to transfer this to interferometric images, which has been used in multiple works \citep[e.g.][]{Mondal2019, Mohan2019a, Mohan2019b, Mondal2020a, Sharma2020, Mohan2021a}. Though successful, this technique relied crucially on the availability of multiple very short baselines ($\lesssim10\lambda$) to obtain the total solar flux density and also to average over baseline-to-baseline fluctuations arising largely from manufacturing tolerances. In Phase-II, the MWA now has multiple configurations (see Section \ref{subsec:array_config} in Chapter \ref{mwa}). The extended configuration of MWA Phase-II, which offers the highest angular resolution and is more desirable from a solar imaging perspective, has few baselines short enough to meet the criterion imposed by \citet{oberoi2017}. An independent approach has also been used in the past to calibrate the flux density of the solar MWA data \citep[e.g.][etc.]{mccauley2017,rahman2019}. In this approach, the solar maps obtained using the MWA data are scaled such that the integrated solar flux density matches the prediction of data-driven models like FORWARD \citep{gibson2016}. Limitations of this technique have been discussed in \citet{Sharma2020} and mainly stem from the inadequacies of the model.

With improved imaging quality now available from instruments like the MWA, the need for precise solar flux density calibration has also become evident. Applications requiring precise flux density calibration include:
\begin{enumerate}
    \item Modeling of the gyrosynchrotron emission from the Coronal Mass Ejection (CME) plasma to estimate physical parameters of CMEs \citep[e.g.][]{bastian2001,Tun2013, Carley2017, Mondal2020a}.
    \item Measuring the quiet coronal plasma parameters including electron density \citep{Mercier2015,Vocks2020} and magnetic field \citep{Sastry_2009}.
    \item Attempts at a detailed comparison of the radio observable, including flux densities, obtained from elaborate models implemented by tools like FORWARD \citep{gibson2016} with the radio maps from the MWA \citep{Sharma2020}. 
\end{enumerate}

This work aims to use multiple independent approaches to develop robust absolute solar flux density calibration across the MWA frequency range which is applicable independent of MWA array configuration, observing epoch and pointing direction on the sky. The approaches explored here include:
\begin{enumerate}
    \item  Using the serendipitous presence of a strong source of known flux density in the range of a few hundreds of Jy, in the wide FoV of MWA during solar observations.
    \item Using the high DR imaging delivered by P-AIRCARS to detect numerous galactic and extragalactic background sources, the flux densities of which are known independently from the GLEAM survey \citep{Wayth2015,walker2017} to be in the range from a few to a few tens of Jy in the wide FoV of the MWA during solar observation.
    \item Observations of a strong flux density calibrator source with and without the additional attenuation.
\end{enumerate}
Using these approaches and a database of MWA calibration solutions that have recently been made available \citep{Sokolwski2020}, this flux density calibration scheme is developed.

This chapter is organized as follows: Section \ref{sec:obs_fluxcal} describes the observations and data analysis, and Section \ref{sec:results_fluxcal} presents the results, including a comparison of the solar imaging quality with the GLEAM survey. It is followed by a short Section \ref{sec:apply_fluxcal} describing the method to apply the flux scaling parameters to any solar observation, and finally, the conclusions in Section \ref{sec:conclusion_fluxcal}.

\section{Observations and Data Analysis}\label{sec:obs_fluxcal}
I have analyzed four datasets that were recorded with the MWA on 2014 May 04, July 12, September 28, and 2020 June 20. These observations are referred to by their observation dates, 20140504, 20140712, 20140928, and 20200620, in the following text. The details of these observations are given in Table \ref{tab:obs_settings}. For 20140928, data corresponding to frequencies above 132 $\mathrm{MHz}$ were bad and had to be discarded. The spectral setting of ``picket fence" refers to a spectral configuration where the 30.72 $\mathrm{MHz}$ of observing bandwidth is distributed in 12 sub-bands each of width 2.56 $\mathrm{MHz}$, centered close to 80, 88, 96, 108, 120, 132, 145, 160, 196, 210, 218, and 240 $\mathrm{MHz}$. Each of these sub-bands is referred to as a `picket'. In the contiguous mode, we observe the band from $\sim$119--151 $\mathrm{MHz}$. All of the observations were done using channel widths of 40 kHz and a time resolution of 0.5 s. The observations on 2014 May 04, 2014 September 28, and 2020 June 20 are solar observations, while Virgo-A observations on 2014 July 12 were done at night time. All these observations were calibrated using P-AIRCARS.

The MWA signal chain includes a so called ``digital gain" ($D_\mathrm{G}$), which can be applied independently every individual 1.28 MHz coarse channel of each of the polarizations from every tile. This is intended to bring the digital counts for each coarse spectral channel to the optimal level for further processing. The values of $D_\mathrm{G}$ are recorded as a part of the metadata and are corrected for before the visibility data is written out as a Measurement Set. 

\begin{table*}[!ht]
\centering
    \begin{tabular}{|p{2.5cm}|p{2.5cm}|p{2.5cm}|p{2.5cm}|p{2.5cm}|}
    \hline
    Label & 20140504 & 20140712 & 20140928 & 20200620\\ 
    \hline \hline 
    Date & 2014 May 04 & 2014 July 12 & 2014 September 28 & 2020 June 20\\
    \hline
    Attenuator settings (dB) & 10 & 10 & 14 & 10\\
    \hline
    Array \newline{configuration} & Phase-I & Phase-I & Phase-I & Phase-II \newline{extended} \\
    \hline
    Spectral mode & Picket-fence & Picket-fence & Picket-fence & Continuous \\
    \hline
    Sun in FoV & Yes & No & Yes & Yes \\
    \hline
    Source(s) used & Multiple faint sources & Virgo-A & Virgo-A & Crab \\
    \hline
    Imaging \newline{bandwidth (MHz)} & 2.56 & 2.56 & 0.16 & 0.16 \\
    \hline
    Imaging \newline{integration time (s)} & 120 & 0.5 & 0.5 & 9 \\
    \hline
    \end{tabular}
    \caption[Details of the different observations used.]{Details of the different observations used.}
    \label{tab:obs_settings}
\end{table*}

\begin{figure*}
    \centering
    \includegraphics[trim={5.5cm 1.5cm 7cm 0cm},clip,scale=0.4]{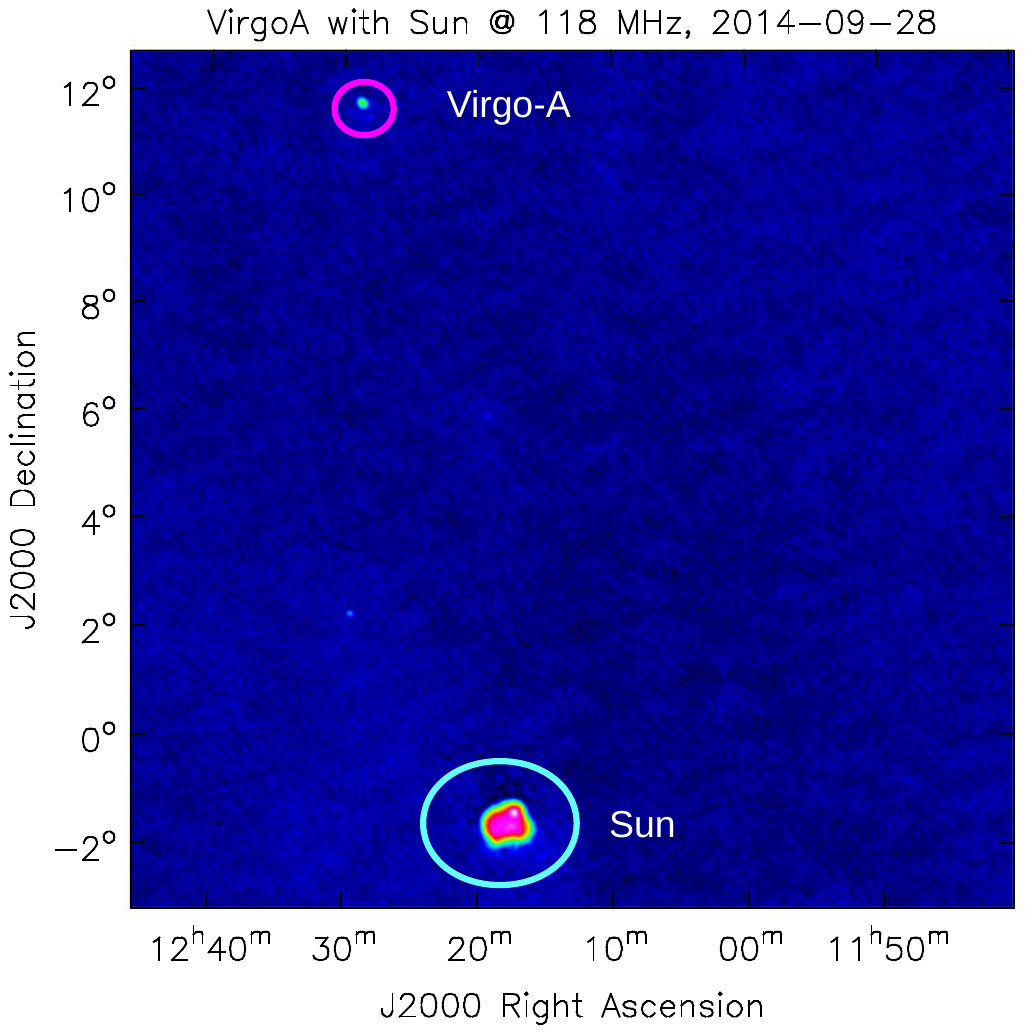}\includegraphics[trim={5.5cm 2cm 5.5cm 0cm},clip,scale=0.41]{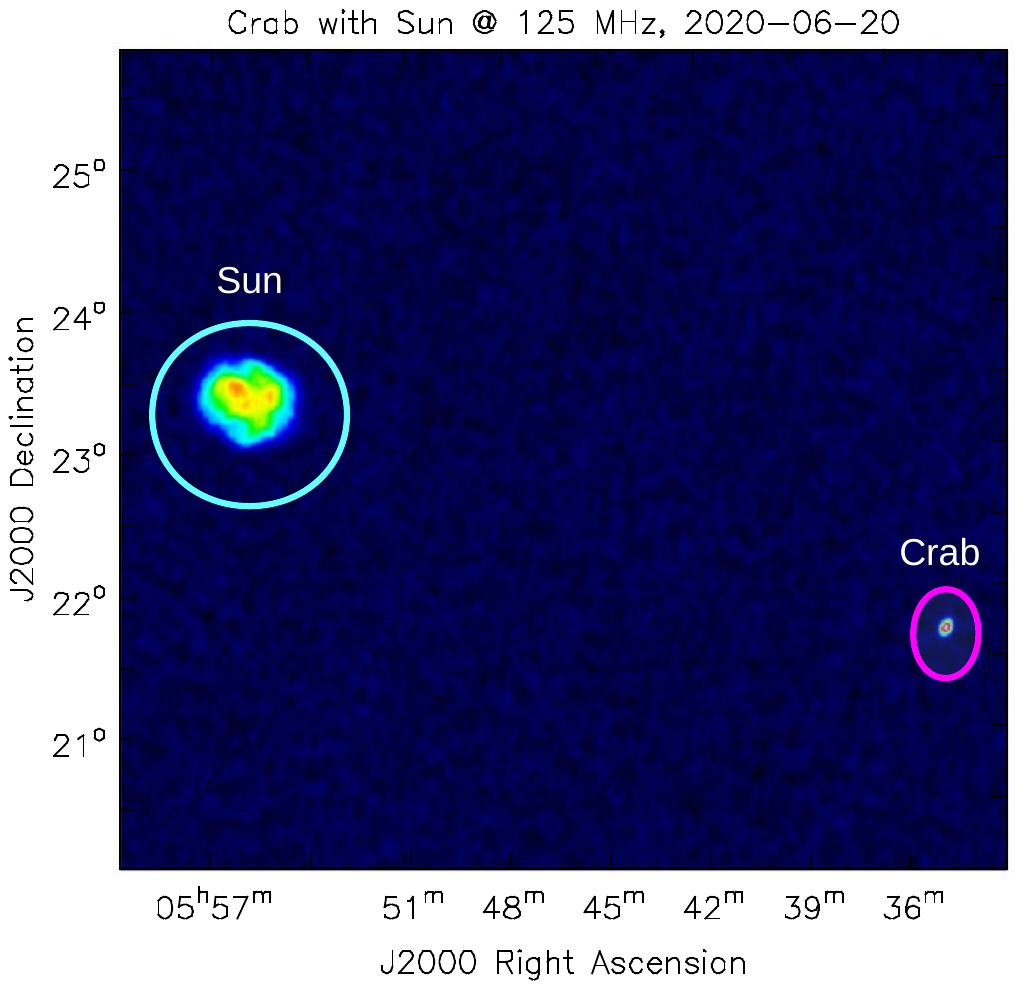}
    \caption[Crab and Virgo-A in the presence of the Sun in the field-of-view.]{{\it Left panel :} A map at 118 $\mathrm{MHz}$ showing Virgo-A and the Sun. {\it Right panel :} A map at 125 $\mathrm{MHz}$ showing Crab and the Sun.}
    \label{fig:crab_and_virgo}
\end{figure*}

\subsection{Data Analysis for 20140928 and 20200620} \label{sec:strong_source}
For these observations, the data analysis was fairly straightforward as the sources of interest, Crab and Virgo-A, are of comparable surface brightness to the quiet Sun \citep{perley2017}. The data were imaged using AIRCARS \citep{Mondal2019}. The data quality for 20140928 is bad above 145 $\mathrm{MHz}$, and hence those data are not used here. While the spectral sampling of calibration scans was matched to solar observations, the calibrator sources were observed without the additional signal attenuation used for solar observations. Hence, for these observations, no flux density scale is derived from calibrator observations. An independent relative self-calibration is performed for each of the coarse channels to determine normalized gain solutions. Appropriate masks were used to ensure that both the Sun and Crab/Virgo-A were ``cleaned'' and included in the model used for self-calibration. The images of Virgo-A and Crab with the Sun in the FoV are shown in Figure \ref{fig:crab_and_virgo}. Different temporal averaging is used for different datasets and listed in Table \ref{tab:obs_settings}. All other AIRCARS parameters were left at their default values. After the final images were obtained, the $imfit$ task available in the Common Astronomical Software and Analysis \citep[CASA,][]{mcmullin2007,CASA2022} was used to estimate the flux density of these sources. Crab could be fitted well with a single Gaussian with residual flux density $<3$\%. For these two epochs, reliable images are available for frequencies $<145$ $\mathrm{MHz}$. At these low frequencies, Virgo-A can also be modeled with a small number of Gaussian components with a residual flux density $<4\%$. No significant emission feature is seen in either of the residual images. The flux-scaling factor for reference epochs (Table \ref{tab:obs_settings}), $F_\mathrm{ref}(\nu)={F_\mathrm{cat}(\nu)}/{F_\mathrm{app,ref}(\nu)}$, was computed by comparing the obtained primary beam corrected apparent integrated flux density, $F_\mathrm{app,ref}(\nu)$, with the values, $F_\mathrm{cat}(\nu)$, available from NASA/IPAC Extragalactic Database\footnote{https://ned.ipac.caltech.edu} or the GLEAM catalog. The primary beam correction was obtained using the model by \citet{Sokolwski2017}. We calculate the uncertainty on $F_\mathrm{ref}(\nu)$, $\Delta F_\mathrm{ref}(\nu)$, by considering only the errors on the measured flux densities from the image. It is given by $\Delta F_\mathrm{ref}(\nu)= F_\mathrm{ref}(\nu) \times \Delta F_\mathrm{intg}(\nu)/F_\mathrm{intg}(\nu)$ where $F_\mathrm{intg}(\nu)$ and $\Delta F_\mathrm{intg}(\nu)$ are the integrated flux density of the source and the uncertainty on it, respectively. It is noted that the noise in the final radio image follows a Gaussian distribution before a correction for the primary beam is applied. However, quantities like $F_\mathrm{ref}(\nu)$ and $\Delta F_\mathrm{ref}(\nu)$ can only be computed after correcting for the primary beam. Hence, the distribution of $\Delta F_\mathrm{ref}(\nu)$ is not expected to follow a Gaussian as is discussed in Section \ref{sec:other_sources}.

\subsection{Data Analysis for 20140504} \label{subsec:weak_sources}
\begin{figure*}[!ht]
    \centering
    \includegraphics[scale=0.35]{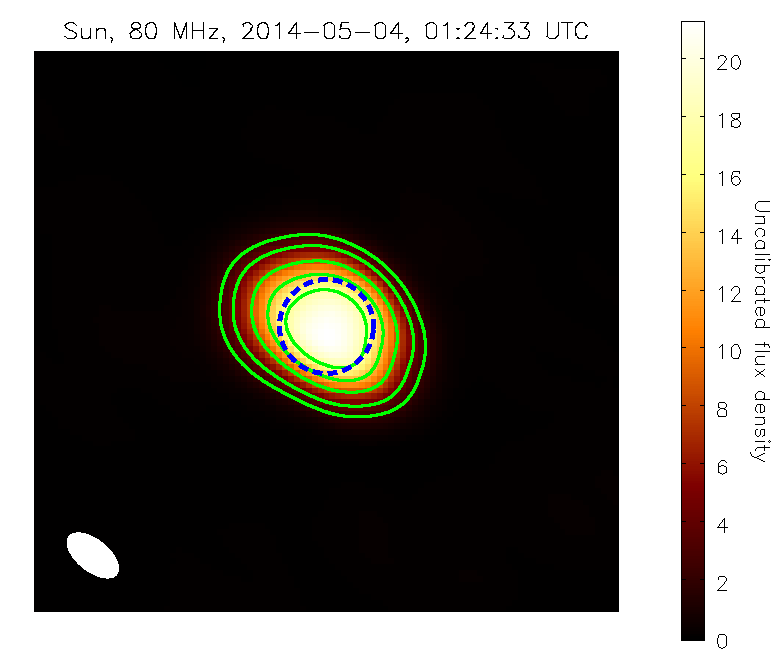}
    \caption[Image of the quiet-sun on 2014 May 04.]{A map of the radio Sun at 80 $\mathrm{MHz}$. The green contours are at 0.1, 0.2, 0.4, 0.6, and 0.8 of the peak flux density. The blue dashed circle represents the optical disk of the Sun, and the point spread function is shown by the ellipse at the bottom left.}
    \label{fig:SUN}
\end{figure*}

As none of the sources in the FoV were strong enough to be detectable in snapshot images in the presence of the Sun and the solar attenuation, this dataset required significantly more involved analysis than the other three. Although the Sun is quiet during this epoch (Figure \ref{fig:SUN}), detecting the much fainter background sources, requires averaging over the entire 2.56 $\mathrm{MHz}$ bandwidth of the spectral pickets. Since the instrumental bandpass amplitude and phase is not constant over frequency, one requires bandpass calibration before making the image over 2.56 $\mathrm{MHz}$. The large difference between the observation times of calibrator sources and the Sun can potentially lead to changes in the shapes of the bandpass amplitudes and the phases between the calibrator and target sources. Hence, an independent {\it relative} bandpass calibration for each 2.56 $\mathrm{MHz}$ picket is performed, but any {\it absolute} bandpass gain is not determined to account for the variation in spectral gain across the MWA band spanned by the pickets.

\subsubsection{Imaging the Faint Sources}
The basic approach was to image and model the solar emission, subtract it from the observed visibilities, and image the residual visibilities to look for the background sources. 
This was implemented as follows: P-AIRCARS was used to generate a solar image integrated over the full 2.56 $\mathrm{MHz}$ bandwidth of each spectral picket at 10 $\mathrm{s}$ resolution. This solar emission was modeled using the \textit{tclean} task in CASA without the \textit{w-projection} algorithm switched on. A mask that limited the cleaning only to the solar disc was used. The final solar images and deconvolved models were generated and the model visibilities for the Sun corresponding to the deconvolved model were subtracted from the calibrated visibilities using the \textit{uvsub} task in CASA. The residual visibilities thus obtained over the 2 minutes were imaged using the \textit{tclean} task in CASA with the \textit{w-projection} algorithm switched on and with a uniform weighting of the visibilities. While imaging the residual visibilities, the phase center of the image was shifted close to the direction of the peak of the primary beam.

\subsection{Data Analysis for 20140712}
These observations were designed to look at a strong astronomical source with and without the attenuation usually used for solar observations and were done at night. For this observation, since the Sun was not present in the FoV, the analysis was straightforward. A similar calibration approach is followed by what was implemented in P-AIRCARS. Each of the 12 bands of spectral width 2.56 $\mathrm{MHz}$ was calibrated using a normalized bandpass of each sub-band using the $\textit{bandpass}$ task of CASA for every 10 $\mathrm{s}$ time interval. This imaging was done for each of these 12 bands. At higher frequencies, Virgo-A is resolved by the MWA baselines. Hence, the ``Python Blob Detector and Source Finder"  \citep[PyBDSF;][]{Mohan2015} was used to fit multiple Gaussians to it, and the sum of the flux densities of all the Gaussian components was regarded as the total flux density of Virgo-A. Flux densities of Virgo-A from \cite{perley2017} are used to obtain $F_\mathrm{ref}(\nu)$ comparing with the primary beam corrected values obtained from the 2.56 $\mathrm{MHz}$ images.

\section{Results}\label{sec:results_fluxcal}
\subsection{Determining $F_\mathrm{ref}$ From Weak Sources} \label{sec:other_sources}
\begin{figure*}[!ht]
\centering
\includegraphics[trim={0cm 0cm 0cm 0cm},clip,scale=2]{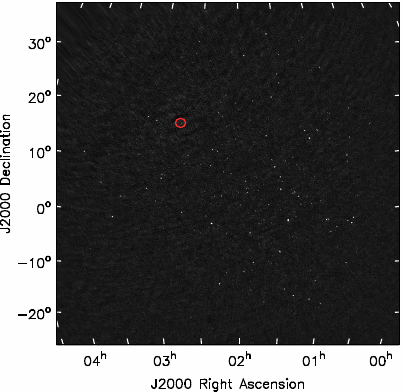}
\caption[Background radio sources within the solar field-of-view.]{Background radio sources within the solar field-of-view. Image centered at 80 $\mathrm{MHz}$, $\sim 60^{\circ}$ on each side, obtained over 2 minutes and 2.28 $\mathrm{MHz}$ after subtraction of modeled solar visibilities from the data. The image is from the observation on 2014 May 04, and the red circle with radius 2 $R_\odot$ is the region where the Sun was present.}
\label{fig:other_sources}
\end{figure*}

\begin{figure*}[!ht]
    \centering
    \includegraphics[scale=0.65]{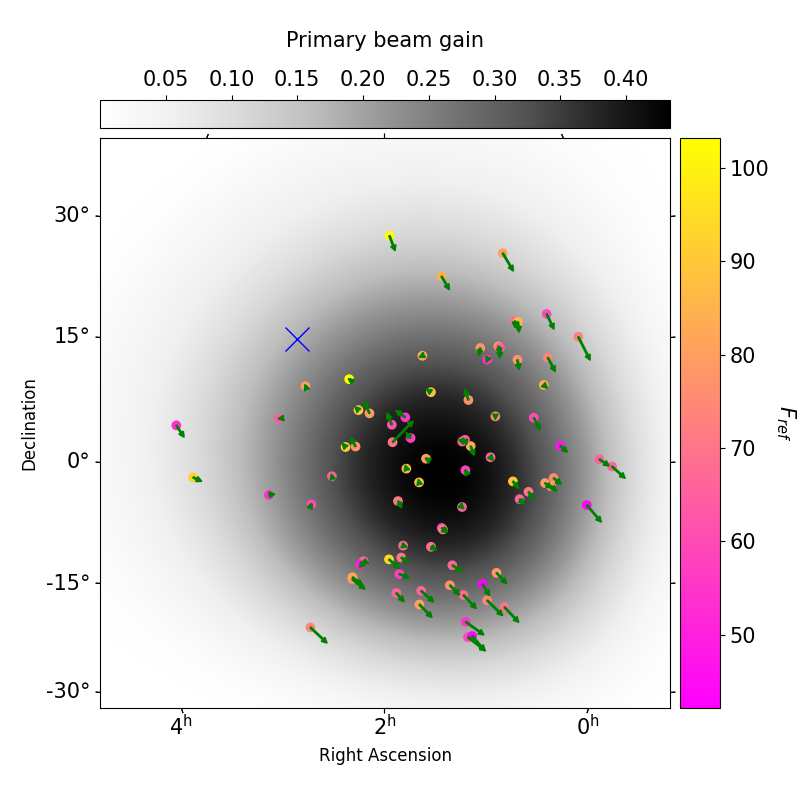}
    \caption[Locations of background radio sources with respect to the primary beam.]{The locations of the subset of detected sources from the observation on 2014 May 04 used for determining $F_\mathrm{ref}(\nu)$ are shown by colored dots. The colors of the dots denote their flux-scaling factor determined for each of the sources, and the primary beam gain at those locations is shown by the background grayscale. The arrows mark the shift of these sources from their GLEAM catalog positions, multiplied by a factor of 15 to make them visible. The location of the Sun is shown by a blue cross mark.}
    \label{fig:sources_detected_shited}
\end{figure*}

A prerequisite for determining $F_\mathrm{ref}(\nu)$ is to either have one or more strong flux density calibrator sources in the FoV or several weak sources so that the fluctuations due to uncertainty on individual estimates of $F_\mathrm{ref}(\nu)$ can get averaged out. Naturally, detecting weak sources is more challenging, and to the best of our knowledge, imaging of multiple background sources in the vicinity of the Sun is yet to be demonstrated at meter wavelengths. Figure \ref{fig:other_sources} is perhaps the first image to show the detection of numerous background sources with a high signal-to-noise ratio (S/N) at low radio frequencies. The closest source to the Sun is at $\sim$20 $R_{\odot}$ with a flux density 4.9 $\mathrm{Jy}$. The $F_\mathrm{ref}(\nu)$ obtained for different sources from 20140504 observation is shown in Figure \ref{fig:sources_detected_shited}. No systematic variation of $F_\mathrm{ref}(\nu)$ with the primary beam is apparent. All results stated here come from the observations on 20140504, and unless otherwise mentioned, they correspond to 80 $\mathrm{MHz}$. DR of the images at other frequencies below 145 $\mathrm{MHz}$ are comparable. 

Images at all frequencies were searched for sources independently before primary beam correction using the source-finding software PyBDSF. PyBDSF was tuned such that only sources with at least a 7$\sigma$ significance, where $\sigma$ is the local rms in the image as calculated by PyBDSF, were selected. The local rms is expected to drop with increasing angular distance from the Sun and was found to vary by a factor of about 2 across the FoV. It was also ensured that none of the sources included any pixels with flux density below 5$\sigma$. No sources were detected at frequencies above 145 $\mathrm{MHz}$. We attribute this to the combined effects of the flux density of the sources dropping at higher frequencies due to their typical negative spectral indices ($S_{\nu} \propto \nu^{\alpha}$) and the increasing solar flux density leading to higher system temperature (and lower sensitivity) at higher frequencies. In addition, the area of the FoV also decreases with increasing frequency. The sensitivity achieved here fell short of what was needed for the detection of sources with sufficient S/N at higher frequencies. 

All of the detected sources were carefully examined visually. To avoid possible impacts of differences in the sensitivities of the images on extended low surface brightness features on $F_\mathrm{ref}(\nu)$ estimates, we use only unresolved sources. The detected compact sources were cross-matched with the MWA GLEAM survey catalog \citep{walker2017}. Cross-matching was done using {\it Aladin} \citep{bonnarel1999}, assuming that the sources can shift at most by $20$arcmin from their reference positions. The choice of maximum allowed shift was motivated by visual inspection of the sources detected in our image and the GLEAM sources. This shift can occur due to multiple reasons, including the following:

\begin{enumerate}
    \item The calibrator observations were carried out at night, whereas these observations come from close to solar transit. The refractive shift estimated and corrected by the calibrator many hours apart and in a different direction is not applicable for solar observation. In addition, the process of self-calibration can also introduce an artificial direction-independent shift in the source locations. Both these effects contribute to giving rise to a direction-independent shift in the source locations. 
    \item 
    Nighttime direction-dependent refractive shifts due to ionospheric structures have already been convincingly demonstrated \citep[e.g.][etc.]{Loi2015a,Loi2015b,Jordan2017,Walker2018,Helmboldt2020}. The daytime ionosphere can have about an order of magnitude higher electron column density. It is hence reasonable to expect significant direction-dependent shifts due to ionospheric refraction.
\end{enumerate}

Cross-matching catalogs that have been observed with different resolutions can be tricky. The GLEAM survey data used here were obtained with the MWA Phase-I by observing the fields close to transit. In the case of 20140504, the solar elevation was $31.3^{\circ}$, and hence the angular resolution is poorer than that of GLEAM. Conventionally, under such circumstances, only the sources that are unresolved in both images are used. To avoid losing the sources unresolved in our images but resolved in GLEAM, we pursue an approach inspired by \citet{rogers2004}. Sources that produce a single match in the GLEAM catalog and remain unresolved in GLEAM pose no challenge. Instances, when a single unresolved source in our images matches more than one source in the GLEAM catalog, require some thought. In such instances, the integral of flux densities inside a region of the size of the PSF is regarded as an image centered on the brightest of the matched GLEAM sources as the effective flux density of the source. It was verified that the results obtained following this procedure and those obtained using unique unresolved GLEAM sources are consistent.

The GLEAM flux density at the frequency of interest is obtained by linear interpolation of flux densities measured by GLEAM at the frequency bands straddling the frequency of interest. Within the FWHM of the primary beam, all sources detected above a GLEAM flux density threshold of 10 $\mathrm{Jy}$ at 80 $\mathrm{MHz}$. The sensitivity levels are comparable at other frequencies. At locations where the primary beam gain is higher, we can detect much fainter sources. The weakest detected source had a GLEAM flux density of 4.6 $\mathrm{Jy}$ at 80 $\mathrm{MHz}$ near the peak of the primary beam. As a 7$\sigma$ threshold is used for choosing the sources, this implies that, close to the peak of the primary beam, the image rms is approximately 0.6 $\mathrm{Jy}$. 

Another issue to remain mindful of is the variability of flux densities \citep{Bell2013,Row2016,Lynch2017} of different sources. To avoid incurring errors in $F_\mathrm{ref}(\nu)$ estimation by using sources with time variable flux densities, only sources for which the estimated $F_\mathrm{ref}(\nu)$ lies between the 10$^\mathrm{th}$ and 90$^\mathrm{th}$ percentile are used. These criteria are met by 81 of the detected sources at 80 MHz, and their locations are shown in Figure \ref{fig:sources_detected_shited}. The $F_\mathrm{ref}(\nu)$ determined from each of these sources, the primary beam gain at their locations, and the observed shifts from their GLEAM catalog positions are also shown in Figure \ref{fig:sources_detected_shited}. The vectors showing the observed shifts of neighboring sources tend to be similar and vary smoothly across the large FoV. This systematic variation across the image is consistent with an ionospheric origin \citep{Loi2015a,Loi2015b,Jordan2017,Walker2018,Helmboldt2020}.

\begin{figure*}[!ht]
    \centering
    \includegraphics[trim={0.5cm 0cm 0cm 0cm},clip,scale=0.5]{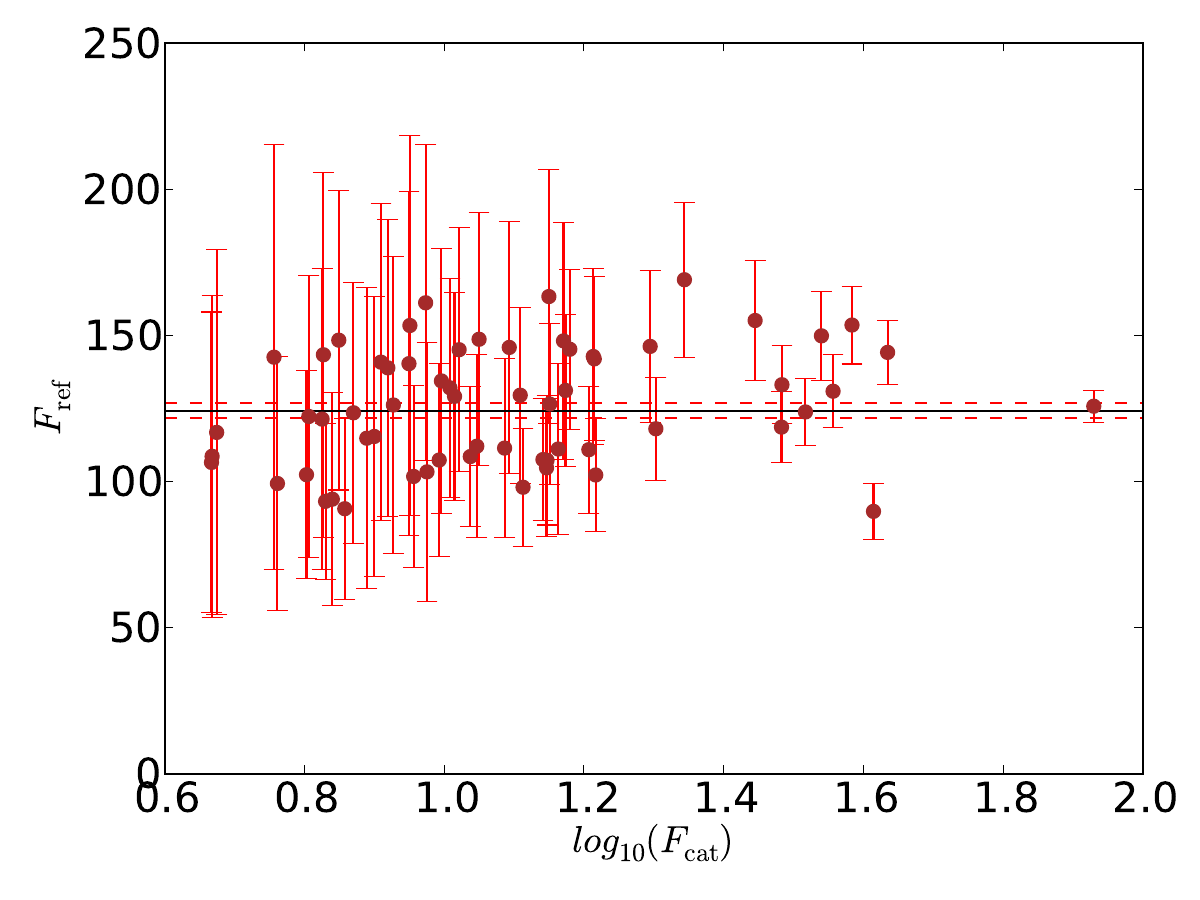}\\\includegraphics[trim={0.5cm 0cm 0cm 0cm},clip,scale=0.45]{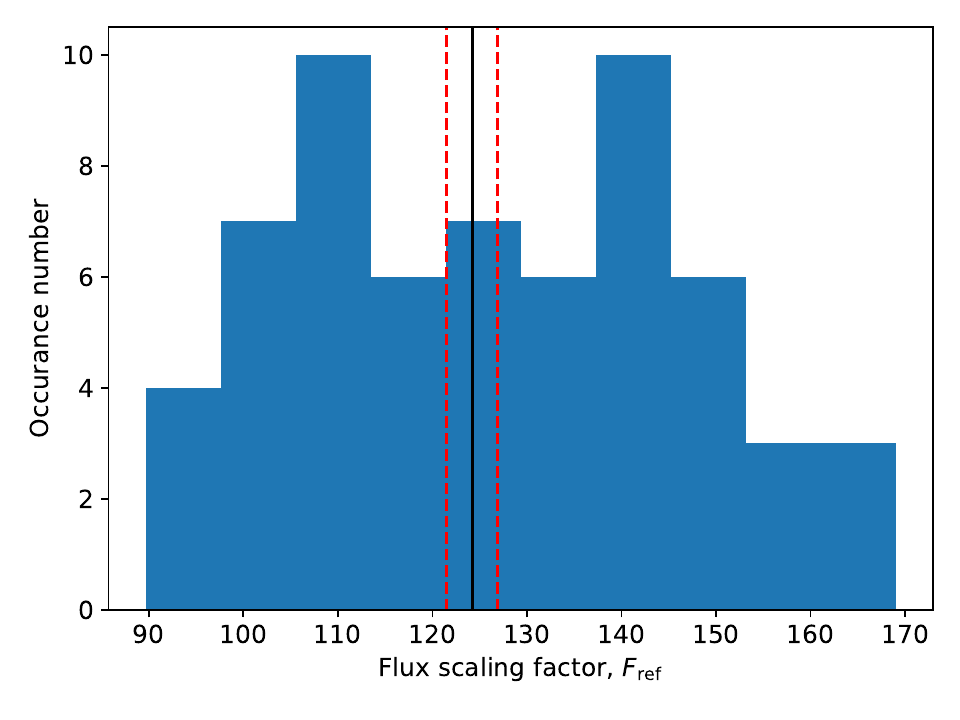}\includegraphics[trim={0.5cm 0cm 0cm 0cm},clip,scale=0.45]{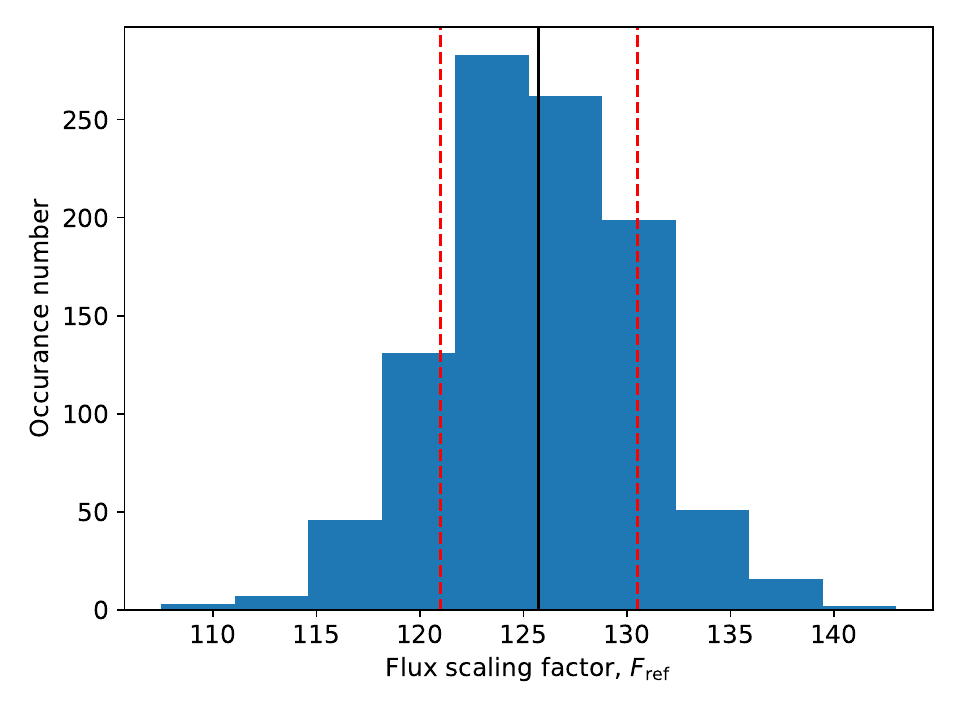}
    \caption[Statistical properties of $F_\mathrm{ref}$.]{{\it Top panel:} Estimated $F_\mathrm{ref}(\nu)$ against the logarithm of GLEAM flux density, $log_{10}(F_\mathrm{cat})$, for individual sources at 80 $\mathrm{MHz}$. {\it Bottom left panel:} Histogram of reference flux-scaling factors, $F_\mathrm{ref}(\nu)$. {\it Bottom right panel:} Histogram of $F_\mathrm{ref}(\nu)$ obtained from random sampling. The black solid line shows the inverse variance weighted averaged scaling factor, and the red dashed lines show the 1$\sigma$ uncertainty on that averaged value.}
    \label{fig:attenuator_gain_comparison}
\end{figure*}

The top panel of Figure \ref{fig:attenuator_gain_comparison} shows $F_\mathrm{ref}(\nu)$ estimated from individual sources, along with the associated uncertainty, as a function of their GLEAM flux densities. Since the levels of rms noise in the GLEAM images are much smaller than that of our image as shown later in Section \ref{subsec:compare_with_gleam}, the contribution of error in GLEAM flux densities to $\Delta F_\mathrm{ref}(\nu)$ is small and has been ignored. $\Delta F_\mathrm{ref}(\nu)$ is calculated as mentioned in Section \ref{sec:strong_source}.

The bottom left panel of Figure \ref{fig:attenuator_gain_comparison} shows the histogram of $F_\mathrm{ref}(\nu)$. The observed spread in the histogram results from the intrinsic uncertainties in the values of $F_\mathrm{ref}(\nu)$ determined for each of the sources. Using all sources at 80 $\mathrm{MHz}$, the inverse variance weighted mean of $F_\mathrm{ref}(\nu)$ ($<F_\mathrm{ref}>$) is found to be $124.2\pm2.7$, where 
\begin{equation}
    <F_\mathrm{ref}>=\frac{\sum_\mathrm{i}\frac{F_\mathrm{ref}}{\Delta F_\mathrm{ref}^2}}{\sum_\mathrm{i}\frac{1}{\Delta F_\mathrm{ref}^2}}.
    \label{eq:variance_weighted_avg}
\end{equation}

As discussed in Section \ref{sec:strong_source}, the distribution of $\Delta F_\mathrm{ref}$ is not expected to follow a Gaussian. The noise characteristics of the image itself were  Gaussian before the primary beam correction, so it is reasonable to expect the estimated uncertainty on each of the individual values of $F_\mathrm{ref}$ determined to be drawn from their own Gaussian distribution -- one with a mean and sigma corresponding to $F_\mathrm{ref}(\nu)$ and $\Delta F_\mathrm{ref}(\nu)$, respectively. As a consistency check, 1000 sets were created, each with a value drawn from a Gaussian distribution with a mean and rms equal to $F_\mathrm{intg}(\nu)$ and $\Delta F_\mathrm{intg}(\nu)$, corresponding to each of the sources used. The mean $F_\mathrm{ref}(\nu)$ was then computed for each of these 1000 sets. The bottom right panel of Figure \ref{fig:attenuator_gain_comparison} shows the histogram of the mean $F_\mathrm{ref}(\nu)$ from these 1000 sets, which shows a well-defined Gaussian. This approach yields an $F_\mathrm{ref}(\nu)$ of $125.6\pm 4.9$ at 80 $\mathrm{MHz}$. This exercise makes it evident that there are no systematic errors associated with the determination of $F_\mathrm{ref}(\nu)$, which otherwise would show as a departure from Gaussian distribution. This also verifies that the inverse variance weighted mean of the $F_\mathrm{ref}$ shown in the left panel is consistent with the estimate from this exercise.

\subsection{Flux Scaling Parameters}\label{sec:fluxscale_param}
In this section, we discuss the origin of different flux density scaling parameters and how they are derived.
\subsubsection{Formulating flux scale parameters}\label{sec:formulating_fluxscale}
It is common practice in radio interferometric calibration to decompose the {\it antenna-dependent} instrumental bandpass gain, $B_\mathrm{tot}(t,\ \nu)$ into a purely time-dependent part, $G_\mathrm{mean}(t)$, and, a purely frequency-dependent part, $B_\mathrm{full}(\nu)$. Following this approach, we decompose, $B_\mathrm{tot}(\nu, t)$ as follows, 
\begin{equation}
\begin{split}
    B_\mathrm{tot}(\nu,\ t) &= G_\mathrm{mean}(t)\ B_\mathrm{full}(\nu).
\end{split}
\label{eq:1}
\end{equation}
In the case of solar observations with the MWA, the conventional approach of flux density calibration is not followed, as has been discussed in Section \ref{sec:aircars_description} of Chapter \ref{paircars_principle}. This section provides the prescription to implement  Equation \ref{eq:1} for the MWA solar observation.

P-AIRCARS performs spectrally local bandpass calibration, normalized to unity, for each spectral picket. The MWA has a significant variation in its response across the band, which is not taken into account by the bandpass calibration done by P-AIRCARS. To take the bandpass gain variation across pickets into account, we decompose $B_\mathrm{full}(\nu)$ into the picket bandpass, $B_\mathrm{picket}(\nu)$, and the inter-picket bandpass, $B_\mathrm{inter}(\nu)$, both normalized to unity,
\begin{equation}\label{eq:bfull_vs_binter}
    B_\mathrm{full}(\nu)=B_\mathrm{picket}(\nu)\ B_\mathrm{inter}(\nu)
\end{equation}
Equation \ref{eq:1} then can then be expressed as,
\begin{equation}
    \begin{split}
         B_\mathrm{tot}(\nu, t)&= G_\mathrm{mean}(t)\ B_\mathrm{picket}(\nu)\ B_\mathrm{inter}(\nu).\\
    \end{split}
    \label{eq:2}
\end{equation}
Of these, $B_\mathrm{picket}(\nu)$ is already corrected by P-AIRCARS. P-AIRCARS computes and corrects for $B_\mathrm{picket}(\nu)$ when imaging is done over a picket of 1.28 $\mathrm{MHz}$. When imaging over a much narrower bandwidth of 160 kHz, it is simply assumed that $B_\mathrm{picket}(\nu)=1$, as the bandpass is essentially flat over this tiny bandwidth.  Then only the two remaining terms need to be computed and corrected. Thus $B_\mathrm{tot}(\nu,\ t)$ can be expressed as,
\begin{equation}\label{eq:btot_vs_gmean_and_binter}
B_\mathrm{tot}(\nu,\ t)=G_\mathrm{mean}(t)\ B_\mathrm{inter}(\nu)
\end{equation}
To determine these parameters, the catalog flux densities, $F_\mathrm{cat}(\nu)$, of the background galactic and extragalactic sources have been used. Before comparing the observed flux densities with $F_\mathrm{cat}(\nu)$, the images are corrected for the primary beam response.
The best available MWA primary beam model \citep{Sokolwski2020} normalized with respect to the direction of the brightest source (usually the Sun)in the FoV is applied using its latest implementation\footnote{\href{https://pypi.org/project/mwa-hyperbeam/}{MWA Hyperbeam}}.
The flux densities observed in these primary beam corrected images are referred to as $F_\mathrm{app}(\nu, t)$, the apparent flux density, which can differ from epoch to epoch. In addition, for solar observations, one also needs to account for the effect of the attenuators, $A(\nu,t)$ 
and digital gain, $D_\mathrm{G}(\nu,t)$, which in the most general case are a function of time and frequency. $D_\mathrm{G}(\nu,t)$ values are available in the metadata of the observations. Since, $D_\mathrm{G}(\nu,t)$ applied during the observation are known {\it a-priori}, the inverse correction is applied while obtaining $F_\mathrm{app}(\nu,t)$.
$F_\mathrm{app}(\nu,t)$ can then be expressed in terms of $F_\mathrm{cat}(\nu)$, $B_\mathrm{tot}(\nu,t)$ and $A(\nu,t)$ using Equations \ref{eq:1} and \ref{eq:btot_vs_gmean_and_binter} as,
\begin{equation}
\begin{split}
      F_\mathrm{app}(\nu,t) & =A(\nu,t)\ |B_\mathrm{tot}(\nu,t)|^2\ F_\mathrm{cat}(\nu)\\
      &= A(\nu,t)\ |G_\mathrm{mean}(t)|^2\ |B_\mathrm{full}(\nu)|^2\ F_\mathrm{cat}(\nu)\\
      &= A(\nu,t)\ |G_\mathrm{mean}(t)|^2\ |B_\mathrm{inter}(\nu)|^2\ F_\mathrm{cat}(\nu)\\
      \frac{F_\mathrm{cat}(\nu)}{F_\mathrm{app}(\nu,\ t)}&=\frac{1}{A(\nu,\ t)\ |G_\mathrm{mean}(t)|^2\ |B_\mathrm{inter}(\nu)|^2}\\
      F_\mathrm{scale}(\nu,\ t)&=\frac{1}{A(\nu,\ t)\ |G_\mathrm{mean}(t)|^2\ |B_\mathrm{inter}(\nu)|^2}\\
      F_\mathrm{scale}(\nu,\ t)&=\frac{1}{A(\nu,\ t)\ |B_\mathrm{tot}(\nu, t)|^2}\\
\end{split}
\label{eq:3}
\end{equation}

We regard observations on all four dates, 20140504,
20140712, 20140928, and 20200620, as the reference epoch,
$t_\mathrm{ref}$. The justification for this is presented in Section \ref{sec:time-variation-atten}. For $t=t_\mathrm{ref}$, $F_\mathrm{scale}(\nu,\ t)$ is considered as $F_\mathrm{ref}(\nu)$, which is similar to that defined in Section \ref{sec:strong_source} as follows,
\begin{equation}\label{eq:fref}
    \begin{split}
        F_\mathrm{ref}(\nu)&=F_\mathrm{scale}(\nu, t=t_\mathrm{ref})\\
        F_\mathrm{ref}(\nu)&=\frac{1}{A_\mathrm{ref}(\nu)\ |G_\mathrm{mean,ref}|^2\ |B_\mathrm{inter}(\nu)|^2}\\
        F_\mathrm{ref}(\nu)&=\frac{1}{A_\mathrm{ref}(\nu)\ |B_\mathrm{ref}(\nu)|^2}\\
        F_\mathrm{ref}(\nu)&=\frac{F_\mathrm{cat}(\nu)}{F_\mathrm{app,ref}(\nu)}
    \end{split}
\end{equation}
where $B_\mathrm{ref}(\nu)=G_\mathrm{mean,ref}B_\mathrm{inter}(\nu)$ is the un-normalized inter-picket bandpass gain without attenuation, $G_\mathrm{mean,ref}$ is the $G_\mathrm{mean}(t)$ for $t=t_\mathrm{ref}$, and $F_\mathrm{app,ref}(\nu)$ represents the apparent flux densities of the background sources for $t_\mathrm{ref}$. The time variations in $F_\mathrm{scale}(\nu, t)$ arise due to those in $G_\mathrm{mean}(t)$. These time variations are computed by comparing the instrumental bandpass gain amplitudes for any epoch, $t$, with the overlapping frequency of the $B_\mathrm{ref}(\nu)$. For most of the MWA solar observations, the calibrator observations were done without the use of attenuators. When applying the bandpass calibration to solar data, we also need to take the additional attenuation into account, and that is done using $F_\mathrm{scale}(\nu, t)$.

\begin{figure*}[!ht]
    \centering
    \includegraphics[trim={0cm 0.5cm 0.5cm 0cm},clip,scale=0.7]{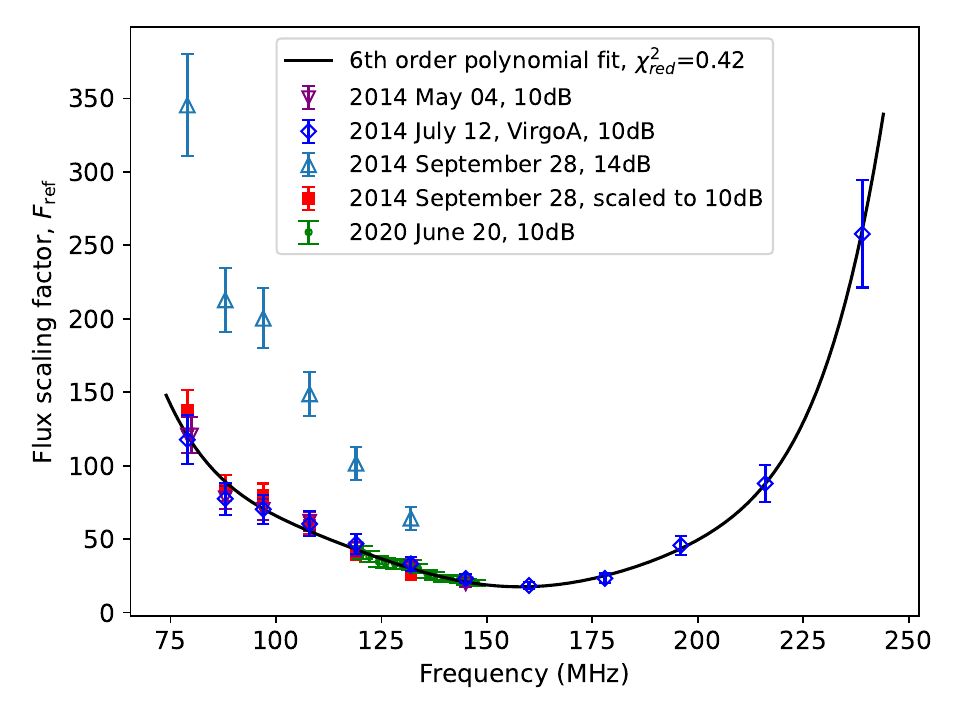}\\
    \includegraphics[trim={0.50cm 0 0cm 0cm},clip,scale=0.7]{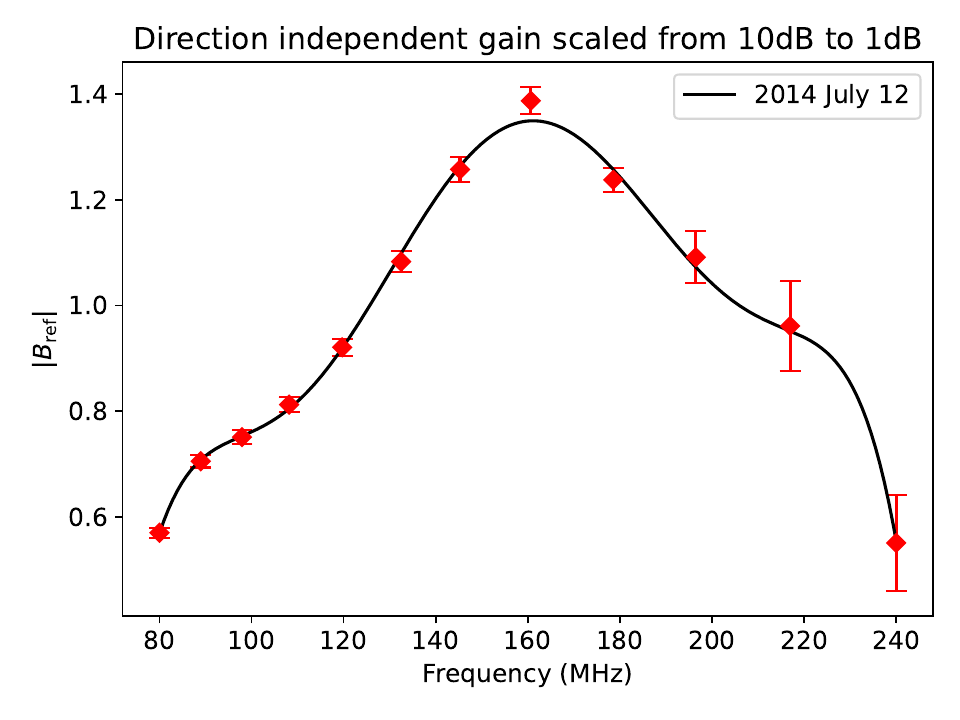} 
    \caption[Flux scaling polynomials.]{{\it Top panel:} The reference flux scaling factor, $F_\mathrm{ref}(\nu)$,  obtained for different observations. The black line shows the fitted polynomial. {\it Bottom panel:} Mean reference bandpass amplitude, $(|B_\mathrm{ref}|)$, for 2014 July 12.}
    \label{fig:fluxscale_polyfit}
\end{figure*}

\subsubsection{Computing Flux Scale Polynomials}\label{sec:fluxscale_compute}
In this section, the recipe for computing $F_\mathrm{ref}(\nu)$ and $B_\mathrm{ref}(\nu)$ is discussed. The top panel of Figure \ref{fig:fluxscale_polyfit} shows the variation of $F_\mathrm{ref}(\nu)$, obtained using different datasets and approaches, as a function of frequency, following the Equation \ref{eq:fref} and using $F_\mathrm{cat}(\nu)$ and $F_\mathrm{app,ref}(\nu)$ as described in Section \ref{sec:strong_source}. The frequency dependence of $F_\mathrm{ref}(\nu)$ comes from the overall bandpass response of the MWA. The uncertainties on each of the measurements were obtained by adding $\Delta F_\mathrm{ref}(\nu)$ and the 8\% systematic flux density uncertainty of the GLEAM survey for the sources in the declination ($\delta$) range $-72^\circ \leq \delta \leq 18^\circ.5$ \citep{walker2017} in quadrature. It is evident that $F_\mathrm{ref}(\nu)$ varies smoothly across the MWA observing band. The $F_\mathrm{ref}(\nu)$ values obtained on 20140928 are systematically higher as compared to other days. We note that the attenuator setting used on this day was higher (14 dB) than what was used on the other three days (10 dB), and we will return to a discussion of these observations later in this section. Until then we focus on the observations made with 10 dB attenuation.

Since calibrator observation of Virgo-A with and without the attenuator was only available for 2014 July 12, and two close-by epochs, 2014 July 11 and 2014 July 13, we have used these epochs to determine $B_\mathrm{ref}(\nu)$ %
for 1 dB attenuation, which is routinely used for all astronomical observations using the MWA.
These calibrator observations were done in contiguous mode at 140--170 and 170--200 $\mathrm{MHz}$. The direction-independent bandpass for these two calibrator observations was estimated following the method described in \cite{Sokolwski2017}. A mean instrumental gain amplitude has been computed for these observations by averaging over all antennas and both polarizations. An average instrumental bandpass gain, $B_\mathrm{tot}(\nu,t_\mathrm{ref})=\ G_\mathrm{mean}(t_\mathrm{ref})B_\mathrm{inter}(\nu)=\ G_\mathrm{mean,ref}B_\mathrm{inter}(\nu)$, spanning the entire MWA band was similarly computed using 20140712 Virgo-A observations, which were done using 10 dB attenuation. It is assumed that the instrumental gain amplitudes, $G_\mathrm{mean}(t)$, are similar for the calibrator observations on these close epochs. A ratio of $|G_\mathrm{mean}|^2$, with and without attenuation, has been computed using the overlapping parts of the band. The ratio is $r\sim$0.09$\pm$0.003. Though the attenuators are calibrated in power units (of the voltage squared), in the MWA signal chain, the attenuation is applied to the analog voltages. 
Hence, for 10 $\mathrm{dB}$ attenuation, a change in $|G_\mathrm{mean}|^2$ is expected by a factor of 0.1. The ratio $r$ is reassuringly close to the expected value. This suggests that, for the observations done with 14 $\mathrm{dB}$ attenuation, $r$ is likely to lie close to its expected value of $0.04$. The amplitudes of $B_\mathrm{tot}(\nu,t_\mathrm{ref})$ is obtained using observation 
20140712 and dividing by the factor of $\sqrt{r}$ just determined,
as the amplitude of $B_\mathrm{ref}(\nu)$ 
with 1 dB attenuation.
It is shown in the bottom panel of Figure \ref{fig:fluxscale_polyfit} and reflects the MWA spectral response across the band. 

To make it convenient to use $F_\mathrm{ref}(\nu)$ and amplitudes of $B_\mathrm{ref}(\nu)$ for any given frequency, we fit the observed values of $F_\mathrm{ref}(\nu)$ and the amplitudes of $B_\mathrm{ref}(\nu)$ with a polynomial. A sixth-order polynomial is found to provide a good fit,
\begin{equation}
    y = a_6\nu^6+a_5\nu^5+a_4\nu^4+a_3\nu^3+a_2\nu^2+a_1\nu+a_0
    \label{eq:attn}
\end{equation}
where $y$ is either $F_\mathrm{ref}(\nu)$ or the amplitudes of $B_\mathrm{ref}(\nu)$. The best-fit polynomial is shown in Figure \ref{fig:fluxscale_polyfit} and the polynomial coefficients are listed in Table \ref{tab:poly_coeff}.

All of the MWA solar observations thus far have used one of two attenuation settings discussed here: 10 dB and 14 dB. The observations with 14 dB attenuation used here cover only the part of the MWA band below $\sim133$ $\mathrm{MHz}$. The ratio of the $F_\mathrm{ref}(\nu)$ values obtained at 10 and 14 dB are remarkably consistent with each other at all frequencies $\lesssim 133$ $\mathrm{MHz}$, with both the mean and median of these numbers being $\sim 2.51$ and the standard deviation being 0.07. 
This value is close to the expected change in $F_\mathrm{ref}(\nu)$ due to 4 dB changes in attenuation. To extend the $F_\mathrm{ref}(\nu)$ for 14 dB to the rest of the MWA band, the best-fit polynomial for $F_\mathrm{ref}(\nu)$ arrived at for 10 dB observations is multiplied by a factor of 2.51.

Note that, due to inaccuracies in the primary beam model, the observed flux densities can show systematic declination and frequency-dependent biases \citep{Sutinjo2015, walker2017}. To arrive at the best-fit polynomial, these systematic errors have been ignored and only the random errors are taken into account.

\begin{table}
    \centering
    \begin{tabular}{|c|c|c|c|}
    \hline
    Coefficients & $F_\mathrm{ref}(\nu)$ & $|B_\mathrm{ref}(\nu)|$ \\[0.1cm]
    \hline
    $a_6$  & $+2.30\pm0.70\ (\times 10^{-6})$ & $-9.38\pm1.24\ (\times 10^{-14})$\\[0.1cm]
    $a_5$  & $-2.10\pm0.64\ (\times 10^{-3})$ & $-8.87\pm1.15\ (\times 10^{-11})$ \\[0.1cm]
    $a_4$  & $+7.84\pm2.40\ (\times 10^{-1})$ & $-3.39\pm0.43\ (\times 10^{-8})$\\[0.1cm]
    $a_3$  & $-1.54\pm0.47\ (\times 10^{2})$ & $+6.70\pm0.84\ (\times 10^{-6})$ \\[0.1cm]
    $a_2$  & $+1.67\pm0.51\ (\times 10^{4})$ & $-7.21\pm0.89\ (\times 10^{-4})$\\[0.1cm]
    $a_1$  & $-9.56\pm2.84\ (\times 10^{5})$ & $+4.01\pm0.49\ (\times 10^{-2})$ \\[0.1cm]
    $a_0$  & $+2.31\pm0.65\ (\times 10^{7})$ & $-8.96\pm0.11\ (\times 10^{-1})$\\[0.1cm]
    \hline
    \end{tabular}
    \caption[Best-fit flux scaling polynomials.]{Best-fit values of polynomial coefficients used in Figure \ref{fig:fluxscale_polyfit}. $\nu$ is in ${\mathrm{MHz}}$.}
    \label{tab:poly_coeff}
\end{table}

\begin{figure*}[!ht]
    \centering
    \includegraphics[trim={0.40cm 0 0cm 0},clip,scale=0.8]{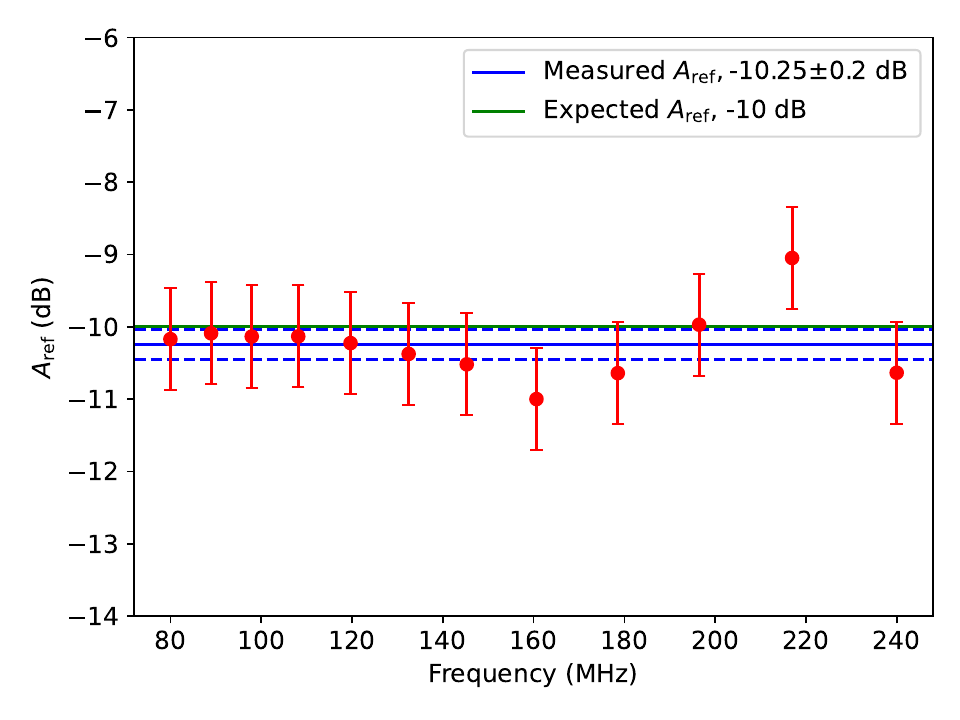} 
    \caption[Spectral variation of attenuation.]{Variation of the product of $A_\mathrm{ref}(\nu)$ with frequency. Red points show the estimated $A_\mathrm{ref}$. Solid green and blue lines represent the expected and estimated mean value of $A_\mathrm{ref}$. Dashed blue lines represent standard deviation over the mean of $A_\mathrm{ref}$.}
    \label{fig:fscale_mult_bref}
\end{figure*}

\subsection{Variation of Instrumental Gain In Time}
\label{sec:time-variation}
A database of MWA calibration solutions has recently been made available by \citet{Sokolwski2020}. This database provides robust amplitude and phase calibration solutions for individual antenna tiles at multiple epochs per day for a large fraction of MWA data available at the data archive hosted by the Pawsey Supercomputing Centre via the MWA ASVO interface\footnote{https://asvo.mwatelescope.org/}. This database is used to estimate the stability of $B_\mathrm{inter}(\nu)$ in time and the variation of $G_\mathrm{mean}(t)$ seen in the MWA data. For this exercise, calibration solutions were chosen from this database at intervals of two to three weeks spanning the period from 2013 June to 2020 June, and a normalized bandpass, $B_\mathrm{inter}(\nu)$, was computed for each epoch. The 1$\sigma$ variation of the amplitude of the $B_\mathrm{inter}(\nu)$ was found to lie in the range of 4 -- 5\% at the edges of the MWA band and 2 -- 3\% in the middle part of the MWA band. The variation in the amplitude of $B_{inter}(\nu)$ is comparable to the $\sim$3\% uncertainty in the $F_{ref}(\nu)$ and that on its best-fit polynomial description. It is also much smaller than the $\sim$8\% uncertainty associated with GLEAM absolute flux density calibration. This implies that the epoch-to-epoch variations in the spectral shape of the $B_\mathrm{inter}(\nu)$ lead to an insignificant increase in the overall uncertainty in absolute flux density calibration. On the other hand, $G_\mathrm{mean}(t)$ shows much larger variations of 10--30\% from epoch to epoch. This needs to be corrected to avoid leaving a large systematic uncertainty in the absolute flux density estimates.

\subsection{Stability of attenuator response}\label{sec:time-variation-atten}
For the value of $t_\mathrm{ref}$ Equation \ref{eq:fref} can be written as,
\begin{equation}
\begin{split}
F_\mathrm{ref}(\nu)\ |B_\mathrm{ref}(\nu)|^2&=\frac{1}{A_\mathrm{ref}(\nu)}
\end{split}
\end{equation}\label{eq:fref_mult_bref2}
The inverse of the product of $F_\mathrm{ref}(\nu)$ and $|B_\mathrm{ref}(\nu)|^2$ is shown in Figure  \ref{fig:fscale_mult_bref}. As is evident from this figure, there is no systematic trend of $A_\mathrm{ref}$ with frequency, and it has an inverse variance weighted mean and rms of -10.25 dB and 0.2 dB, respectively, for -10 dB attenuation. This mean value of $A_\mathrm{ref}$ is consistent with the expected value of -10 dB within the uncertainty.

One can use any calibrator observation to determine the un-normalized bandpass gain, $B_\mathrm{tot}(\nu,t)=G_\mathrm{mean}(t)\ B_\mathrm{inter}(\nu, t)$. The stability of the spectral shape of the $B_\mathrm{inter}(\nu)$, as discussed in Section \ref{sec:time-variation}, allows us to take the same $B_\mathrm{inter}(\nu)$ for both $t$ and $t_\mathrm{ref}$. The scaling value, $S(t)$, is then can be determined as,
\begin{equation}
\begin{split}
   S(t) &=\frac{|B_\mathrm{ref}(\nu)|^2}{|B_\mathrm{tot}(\nu,t)|^2}\\
   &= \frac{|G_\mathrm{mean,ref}|^2\ |B_\mathrm{inter}(\nu)|^2}{|G_\mathrm{mean}(t)|^2\ |B_\mathrm{inter}(\nu)|^2}\\
   &= \frac{|G_\mathrm{mean,ref}|^2}{|G_\mathrm{mean}(t)|^2} 
\end{split}\label{eq:6}
\end{equation}
$B_\mathrm{ref}(\nu)$ is needed to obtain the $S(t)$ from any calibrator observation without attenuation in any spectral configuration to scale the $F_\mathrm{ref}(\nu)$ to $F_\mathrm{scale}(\nu,\ t)$. 
\begin{figure*}[!ht]
    \centering
    \includegraphics[trim={0.5cm 0 1cm 1.4cm},clip,scale=0.65]{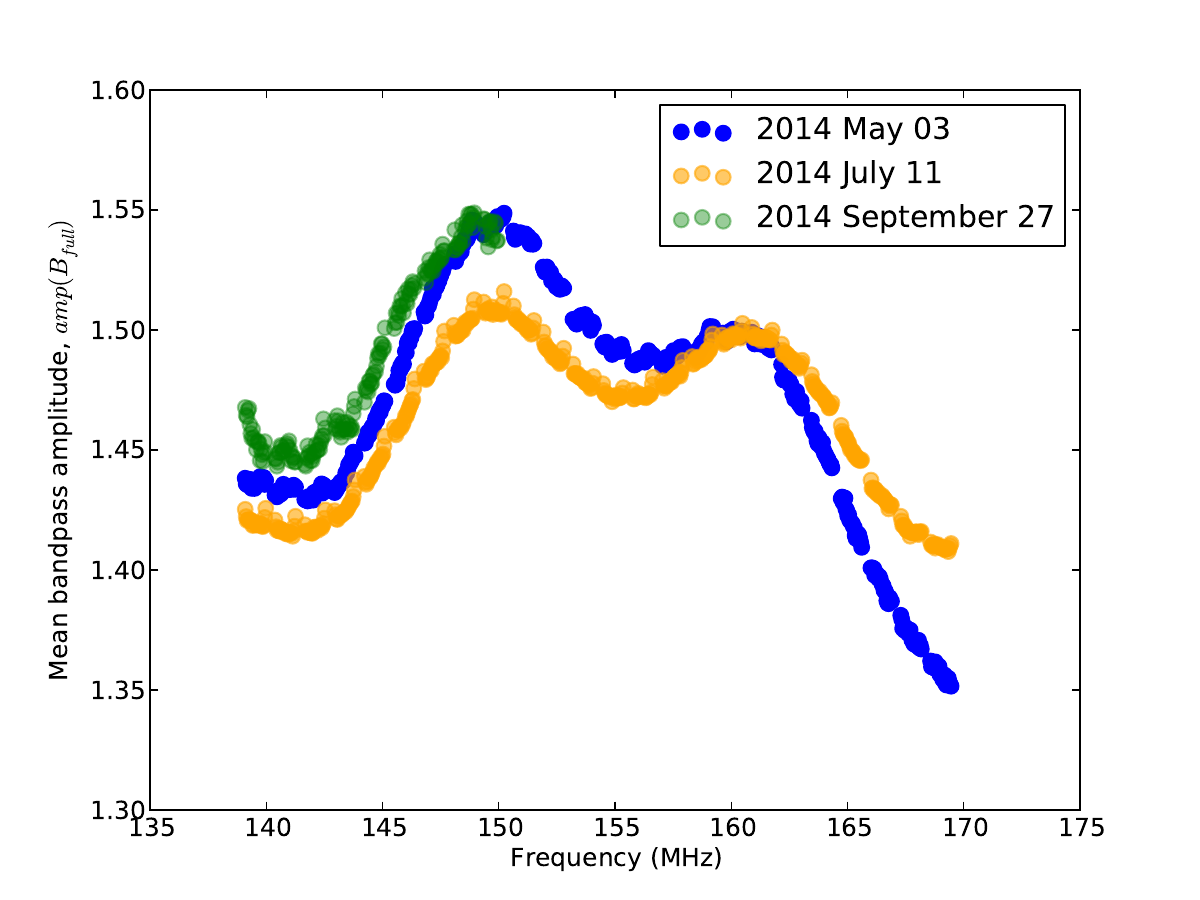} 
    \caption{Mean bandpass amplitudes, $\mathrm{amp}\ (B_\mathrm{full})$, for the calibrator observations. Three colors represent three observing epochs.}
    \label{fig:epochs_meanband}
\end{figure*}

The consistency of $F_\mathrm{ref}(\nu)=F_\mathrm{scale}(\nu,\ t=t_{ref})$ over reference epochs has been demonstrated in the top panel of Figure \ref{fig:fluxscale_polyfit}. $F_\mathrm{scale}(\nu,\ t)$ has contributions from both attenuator response and instrumental gain. However, it is likely that the consistency of $F_\mathrm{scale}(\nu,\ t)$ across epochs arises from the individual stability of the bandpass amplitude or the attenuator response. Thus, one cannot formally claim this is due to the degeneracy just mentioned. An independent estimate for the stability of the attenuator response can be arrived at by exploring the stability of the bandpass amplitude solutions for these epochs. We have shown the bandpass amplitudes of three epochs, 2014 May 03, July 11, and September 24, in Figure \ref{fig:epochs_meanband} close to the reference epochs, when calibrator observations were available. The data quality of 2014 September 27 was poorer for frequencies greater than 150 $\mathrm{MHz}$, and those data are not shown here. These mean bandpass amplitudes show that even data taken months apart are consistent within $\sim$2\%. This is very similar to the variability observed in $F_\mathrm{scale}(\nu, t)$, and hence implies that the attenuator response must have remained essentially constant across these observations. This is consistent with the expectation that, because they are passive devices, attenuators are not prone to significant evolution in their characteristics, and suggests that it is reasonable to assume that the attenuator performance, $A(t)$,  has remained steady across time. Calibrator observations with the same frequency range as shown in Figure \ref{fig:epochs_meanband} for 2020 June 20 are not available. Thus, a direct comparison of the bandpass amplitude with the other three epochs is not possible. Since the $F_\mathrm{scale}(\nu,\ t)$ for 2020 June 20 also matched well with other epochs and the time-independent nature of 
$A(\nu)$
just mentioned, 2020 June 20 is also considered as a reference epoch. Since $A(\nu)$ is independent of time 
one could use the scaling values for $F_\mathrm{ref}(\nu)$ and $B_\mathrm{ref}(\nu)$ from 10 dB to 14 dB as mentioned in Section \ref{sec:time-variation} for any other epochs. Using Equation \ref{eq:fref} and scaling values mentioned in Section \ref{sec:time-variation}, 
it is found find the value of $A(\nu)$ for 14 dB attenuation can be obtained by adding a constant 4 dB to the values obtained for 10 dB attenuation.

\subsection{A Comparison With GLEAM}\label{subsec:compare_with_gleam}
It is instructive to compare the image presented in Section \ref{sec:other_sources} with the typical imaging quality delivered by the MWA GLEAM survey \citep{walker2017}, as it provides a good benchmark for the quality of the imaging being provided by our solar imaging pipelines \citep{Mondal2019,Kansabanik2022_paircarsI,Kansabanik_paircars_2}. Here, I present the calculations of the expected rms noise for the 20140504 observation and its comparison with the GLEAM survey \citep{walker2017}. Expected rms noise per polarization, $\Delta F_\mathrm{X,Y}$, can be written as
\begin{equation}\label{eq:sesitivity}
    \begin{split}
        \Delta F_\mathrm{X,Y}=\frac{SEFD_\mathrm{X,Y}}{\sqrt{N_\mathrm{ant}(N_\mathrm{ant}-1)\Delta \nu \Delta t}}
    \end{split}
\end{equation}
where $\mathrm{X}$ and $\mathrm{Y}$ refer to the two orthogonal polarizations, $SEFD$ is the system equivalent flux density \citep{Wrobel1999,Thomson2017}, $N_\mathrm{ant}$ is the number of antennas used for imaging, $\Delta t$ is the total integration time used for imaging, and $\Delta \nu$ the total imaging bandwidth. $SEFD$ can also be expressed in terms of effective collecting area, $A_\mathrm{eff}$, and system temperature, $T_\mathrm{sys}$, as,
\begin{equation}\label{eq:aeff}
    SEFD=\frac{2\ K\ T_\mathrm{sys}}{A_\mathrm{eff}},
\end{equation}
where $K$ is the Boltzmann constant. The value of $A_\mathrm{eff}$ has been calculated using the Full Embedded Element primary beam model developed by \citet{Sokolwski2017}. $T_\mathrm{sys}$ has the contributions from beam-averaged sky temperature ($T_\mathrm{sky}$), receiver temperature ($T_\mathrm{rec}$) and ground pick-up ($T_\mathrm{pick}$). Values of $T_\mathrm{rec}$ and $T_\mathrm{pick}$ used here are provided by \citet{Daniel2020}.

\begin{table*}[!ht]
\centering
    \begin{tabular}{|p{3cm}|p{3cm}|p{3cm}|}
        \hline
        \textbf{Observation} & \textbf{GLEAM} & \textbf{20140504}\\
        \hline
        \hline
        $A_\mathrm{eff,X}$ & 17.33 & 11.61\\
        \hline
        $A_\mathrm{eff,Y}$ & 13.98 & 11.69 \\
        \hline
        $SEFD_\mathrm{X}$ (Jy) & 134098 & 230594 \\
        \hline
        $SEFD_\mathrm{Y}$ (Jy) & 160703 & 226891 \\
        \hline
        $T_\mathrm{sky,X}$ (K) & 731 & 700 \\
        \hline
        $T_\mathrm{sky,Y}$ (K) & 703 & 691 \\
        \hline
        $T_\odot$ (K) & 0 & 159 \\
        \hline
        $T_\mathrm{sys,X}$ (K) & 842 & 970 \\
        \hline
        $T_\mathrm{sys,Y}$ (K) & 814 & 961 \\
        \hline
        $\Delta \nu$ (MHz) & 7.68 & 2.28 \\
        \hline
        $\Delta t$ (s) & 120 & 120 \\
        \hline
    \end{tabular}
    \caption[Parameters used to estimate the theoretical rms noise.]{Parameters used to estimate the theoretical rms noise.}
    * We have used $T_{rec}$ = 91 K and $T_{pick}$ = 20 K.
    \label{table:sensitivity}
\end{table*}

For the 20140504 observation, images were obtained with an integration time of 2 minutes and 2.28 $\mathrm{MHz}$ bandwidth. The GLEAM survey lists typical rms values at 72 $\mathrm{MHz}$ and 240 $\mathrm{MHz}$ for its integration time of 2 minutes and bandwidth of 7.68 $\mathrm{MHz}$. In addition to the differences in time and frequency integration, an apples-to-apples comparison requires us to also take two other considerations into account. The first is the increase in system temperature due to the Sun, and this information is available in \citet{oberoi2017}. The second consideration is that because it is an aperture array, the sensitivity of the MWA is a function of the elevation of the pointing direction, which then also needs to be accounted for. GLEAM observations were done using the primary beam pointings at higher elevations. A GLEAM pointing is chosen at 108 $\mathrm{MHz}$ for the same part of the sky as the 20140504 solar observation. The theoretical thermal rms noise of the GLEAM image, without any contribution from the Sun, was calculated using Equations \ref{eq:sesitivity} and \ref{eq:aeff}, as well as the parameters listed in Table \ref{table:sensitivity}.  The theoretical rms noise for GLEAM was estimated to be $\sim$38 $\mathrm{mJy}$. The noise obtained in the GLEAM images is $\sim$150 $\mathrm{mJy}$  \citep{walker2017}, $\sim$4 times larger than the theoretical value.

To estimate the theoretical thermal rms for solar images, the primary beam-averaged contribution of the Sun ($T_{\odot}$) is added to the $T_\mathrm{sys}$. $T_\odot$ at 108 $\mathrm{MHz}$ was estimated to be 159 $\mathrm{K}$ and represents the average of the values at 103 and 117 $\mathrm{MHz}$ from the work of \citet{oberoi2017}, which provides the values for a quiet solar time. In addition, while the GLEAM survey in general used a full 128 MWA antenna elements (tiles), only 115 of them were used for observations on 20140504. Taking these differences into consideration leads to a theoretical thermal noise of 120 $\mathrm{mJy}$ for solar observations on 20140504 using the parameters mentioned in Table \ref{table:sensitivity}. Scaling up the thermal noise by the factor of 4 estimated for GLEAM leads to an expectation of 480 $\mathrm{mJy}$ for the observed rms noise. The actual value of the rms observed in the solar map is 720 $\mathrm{mJy}$, a factor 1.5 higher than the expectations based on GLEAM.

\section{Applying the Flux Scale}\label{sec:apply_fluxcal}
For the MWA solar observations, the following prescription can be used to obtain absolute flux density calibrated images in units of $\mathrm{Jy/beam}$:
\begin{enumerate}
    \item 
    Correct MWA observations with the digital gains known {\it a-priori}.
    \item Compute normalized bandpass, $B_\mathrm{picket}(\nu)$, for each 1.28 $\mathrm{MHz}$ coarse channel for the solar observations independently, and correct for it.
    \item Correct the solar images for the primary beam response using the Full Embedded Element Beam model \citep{Sokolwski2017} 
    using its latest implementation\footnote{\href{https://pypi.org/project/mwa-hyperbeam/}{MWA Hyperbeam}}
    for every 1.28 $\mathrm{MHz}$ coarse channel.
    \item Compute the value of $F_\mathrm{ref}(\nu)$ corresponding to the value of attenuation used for the observation (10 dB or 14 dB) using the polynomial coefficients in Table \ref{tab:poly_coeff} at the desired observing frequency.
    \item Obtain bandpass gains for any calibrator observation, $B_\mathrm{tot}(\nu,\ t)$, in any spectral configuration without any additional attenuation from a nearby epoch following the calibration methods described in \cite{Sokolwski2017}. Compute $S(t)$ from Equation \ref{eq:6} using the part of the band overlapping between $B_\mathrm{tot}(\nu, t)$ and $B_\mathrm{ref}(\nu)$.
    \item Since $A(\nu)$ is considered to be independent of time, $A(\nu,\ t)=A_{ref}(\nu)$. From Equations \ref{eq:3}, \ref{eq:fref} and \ref{eq:6}, $F_\mathrm{scale}(\nu,\ t)$ for any observing epoch, $t$, can be written as
    \begin{equation}
        \begin{split}
            F_\mathrm{scale}(\nu,\ t)& =\frac{1}{A_\mathrm{ref}(\nu)\ |B_\mathrm{tot}(\nu,\ t)|^2}\\
            & = \frac{1}{A_\mathrm{ref}(\nu)\ |G_\mathrm{mean}(t)|^2\ |B_\mathrm{inter}(\nu)|^2}\\
            & = \frac{S(t)}{A_\mathrm{ref}(\nu)\ |G_\mathrm{mean,ref}|^2\ |B_\mathrm{inter}(\nu)|^2}\\
            & = \frac{S(t)}{A_\mathrm{ref}(\nu)\ |B_{ref}(\nu)|^2}\\
            F_\mathrm{scale}(\nu,t)& = S(t)\ F_\mathrm{ref}(\nu)
        \end{split}
    \end{equation}
    \item 
    Divide $F_\mathrm{scale}(\nu)$ with the Stokes I primary beam towards the center of the brightest source to obtain final $F_\mathrm{scale,final}(\nu,\ t)$.
    \item Multiply the primary beam corrected solar images with $F_\mathrm{scale,final}(\nu,\ t)$ to obtain the final flux density calibrated image in $\mathrm{Jy/beam}$ units. $F_\mathrm{scale,final}(\nu,\ t)$ corrects the MWA bandpass response, temporal variation of the instrumental gain, and the response of the attenuator.
    \item An approximate way to account for the uncertainties due to multiple contributions is to add, in quadrature, an additional 10\% uncertainty to the values obtained from the best-fit polynomial.
\end{enumerate}

This method can also be employed for solar observations with no corresponding calibrator observations, as was the case during early solar observations with the MWA. It can correct for the large variation in the overall amplitude of the frequency-independent gain of the MWA bandpass, which would otherwise be the dominant source of uncertainty, using $S(t)$. The uncertainty in the $F_\mathrm{scale,final}(\nu,t)$ comes primarily from the $\sim$8\% uncertainty of GLEAM flux densities, 3--4\% uncertainty on $F_\mathrm{ref}(\nu)$ due to the thermal noise, and the 2--5\% variations in the bandshape across epochs. Together, they lead to an overall uncertainty of $\sim$10\% in the final flux density estimates.

\subsection{Validating Quiet Solar Flux Density}
\begin{figure*}[!ht]
    \centering
    \includegraphics[trim={0cm 1cm 0cm 0cm},clip,scale=0.7]{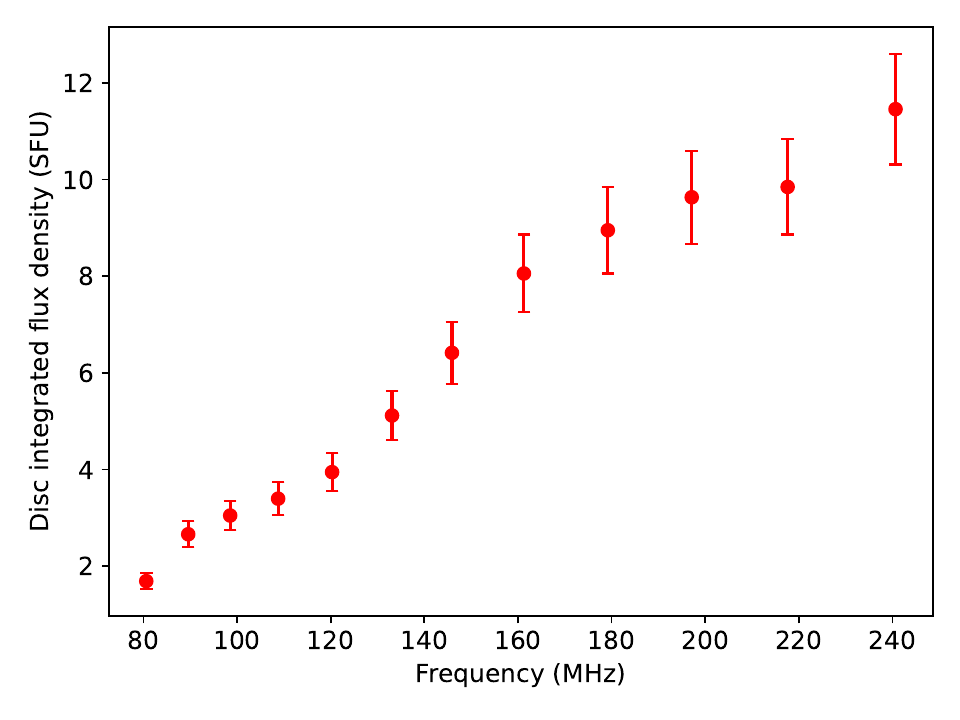}
    \caption[Disc integrated flux density spectrum of quiet Sun.]{Disc integrated flux density spectrum of quiet Sun on 2014 May 04.}
    \label{fig:QS_spectra}
\end{figure*}

Quiet Sun emission is believed to originate due to thermal bremsstrahlung emission from the hot solar atmosphere. At the MWA observing frequencies, quiet Sun flux density is expected to increase with increase in frequency \citep{Sharma2020,Zhang_2022}. At the lower part of the MWA band ($\sim100$ MHz), disc integrated quiet Sun flux density is expected to be $\sim1$ SFU (1 SFU (Solar Flux Unit) = $10^4$ Jy) and at the higher frequency ($\sim240$ MHz) flux density becomes $\sim10$ SFU. On 2014 May 04, the Sun was extremely quiet and no active emission is observed at the MWA observing bands (Figure \ref{fig:SUN}). Disc integrated flux density spectrum for 20140504 is shown in Figure \ref{fig:QS_spectra}, which matches well with the expected quiet Sun spectrum \citep{McLeanBook,oberoi2017,Sharma2020,Zhang_2022}. This verifies the consistency of the flux density calibration method presented in this chapter with expectations.

\section{Conclusion}\label{sec:conclusion_fluxcal}
In this chapter, a robust flux density calibration method for solar observations with the MWA has been described.  The earlier approach for absolute flux density calibration, though innovative, had several limitations \citep{oberoi2017}. It had an uncertainty in the range of 10--60\% which varied across the band. The uncertainty depended on the Galactic background seen in the large MWA FoV and the state of the Sun, with uncertainty increasing with an increase in solar flux density. Even more limiting fact was that it required several very short baselines, to which the Sun would appear as an unresolved source -- a requirement that was not met by the extended configuration of MWA Phase-II. The current method delivers $\sim$10\% uncertainty across the observing frequency band and is independent of the type of emission present on the Sun and the Galactic background against which the Sun is seen. It is equally applicable independent of array configuration or spectral configuration. Not only that, but the intrinsic simplicity of its application also makes it much less computationally intensive than the earlier approach. Only two different attenuation settings have been used for MWA solar observations, and the scaling between them has been determined here. The method can also provide flux density calibration for solar data even for the epochs without any matching calibrator observations. This approach has been incorporated in the P-AIRCARS \citep{Kansabanik2022_paircarsI,Kansabanik_paircars_2}, and will enable routine generation of solar radio images in absolute flux density units.

It is also noted that the MWA sensitivity is sufficient to observe some of the stronger flux density calibrators using the attenuation typically used for solar observations even at our highest observing frequency of 240 $\mathrm{MHz}$. This will be a good practice to follow for future solar observations. Being able to do so will simplify the flux density calibration process and will also allow us to take into account any variations in the spectral behavior of $F_\mathrm{ref}(\nu)$ at scales too fine to be captured by the polynomial fit employed here, if present. These small-scale variations in bandpass amplitudes are evident from Figure \ref{fig:epochs_meanband}. Understanding and modeling these small-scale variations will be important for characterizing the spectral properties of weak nonthermal emissions like the Weak Impulsive Narrow Band Quiet Sun Emissions \citep[WINQSEs;][]{Mondal2020a,Mondal2021a,Mondal2023}. The level of accuracy this method provides is quite sufficient for studying and modeling broadband emissions like GS emissions from CMEs. In the subsequent analysis in this thesis, this method is used to obtain flux density calibrated solar images.

\chapter {Implementation of P-AIRCARS}
\label{paircars_implementation}
It is already evident from Chapters \ref{paircars_principle}, \ref{paircars_algorithm} and \ref{fluxcal} that spectropolarimetric calibration and imaging of the MWA solar observations is an elaborate non-trivial multi-step process. Although the MWA data is intrinsically capable of producing high dynamic range (DR) spectroscopic snapshot polarimetric solar images, it is an impossibly tedious job to perform this calibration and imaging manually, and this also makes it even more susceptible to human errors. To eliminate this tedium, streamline the generation of science-ready spectropolarimetric solar images from the MWA observations while making efficient use of the available computing resources, a state-of-the-art software pipeline, based on the algorithm described in Chapter \ref{paircars_algorithm}, has been implemented. The pipeline shares the same name as the algorithm -- ``Polarimetry using Automated Imaging Routine for the Compact Arrays for the Radio Sun (P-AIRCARS)". This chapter describes the implementation details of P-AIRCARS and its key features. This chapter is based on \citet{Kansabanik_paircars_2}, which was published in the Astrophysical Journal Supplement Series.

\section{Introduction}
Solar phenomena span an enormous range of time scales, from the solar cycle to flares and in terms of energy from the most massive coronal mass ejections (CMEs) to the barely discernible nanoflares. It is now well understood that the solar magnetic field is the primary driver of all of these phenomena. The observed polarization properties of low-frequency coronal radio emissions can serve as excellent remote probes of coronal magnetic fields, even at middle and higher coronal heights. This is because the magnetic field affects the polarization of radio emissions arising from the coronal plasma \citep[e.g.][]{Alissandrakis2021}. Polarization observations also enable a detailed understanding of the emission mechanism of these low-frequency coronal radio emissions. Successful examples exist in the literature, though their numbers have been rather small and these studies have remained limited to comparatively brighter and highly polarized emissions. Most of these studies have relied on dynamic spectra, \citep[e.g.,][etc.] {McLeanBook,hari2014,Kumari2017_typeIV,Pulupa2020,Ramesh_2022}. In only a handful of instances, either the information of spatial location \citep[e.g.][]{Mercier1990,Morosan2022} and/or spatial structure \citep[e.g.][]{Patrick2019, Rahman2020} of the sources are also available. 

High DR spectropolarimetric solar imaging studies at low radio frequencies are very rare. Brightness temperature ($T_\mathrm{B}$) of solar radio emission varies by as much as about nine orders of magnitude, and their fractional polarization can vary by about two orders of magnitude (Figure \ref{fig:different_emissions} in Chapter \ref{paircars_principle}). These emissions can change drastically over short temporal and spectral spans. Very often, faint emissions can simultaneously be present with very bright emissions. This imposes the need for high DR polarimetric imaging over short temporal and spectral spans. These challenging requirements along with the technical and instrumental limitations at low radio frequencies have severely limited polarimetric solar radio imaging studies, despite their well-appreciated importance.

The essential requirements for high-fidelity spectroscopic snapshot solar imaging are described in Section \ref{sec:suitability_of_MWA} of Chapter \ref{paircars_principle} and are met to a large extent by one of the new technology instruments, the MWA. Though the MWA data are intrinsically capable of yielding high-fidelity solar images, doing so involves surmounting several challenges. These challenges have successfully been dealt with in the robust polarization calibration (Chapter \ref{paircars_algorithm}) and absolute flux density (Chapter \ref{fluxcal}) calibrations. But unlocking the potential of low radio frequency solar science requires the ability to perform snapshot spectroscopic polarimetric imaging over extended temporal and spectral spans, leading to tens or even hundreds of thousands of images for individual investigations. Doing this manually is an impossibly tedious job and also prone to human errors. Hence, a state-of-the-art calibration and imaging pipeline, \textit{Polarimetry using Automated Imaging Routine for the Compact Arrays of the Radio Sun} (P-AIRCARS)\footnote{Documentation available online at \url{https://p-aircars.readthedocs.io/en/latest}}, has been developed to meet this need. 

This chapter describes the implementation and architecture of P-AIRCARS. I have organized this chapter as follows. I first discuss the design principles of P-AIRCARS in Section \ref{sec:key_principles}. Section \ref{sec:pipeline_arch} describes the architecture of the pipeline. Section \ref{sec:determine_params} describes the choices of parameters for calibration and imaging followed by the salient features of P-AIRCARS in Section \ref{sec:features}. Aspects related to hardware and software requirements for P-AIRCARS and its performance are discussed in Section \ref{sec:requirement}. P-AIRCARS has led to several discoveries and scientific results, which are presented very briefly in Section \ref{sec:paircars_discoveries}. The chapter ends with the conclusions in Section \ref{sec:conclusion_paircars_implementation}.

\section{Design Principles of P-AIRCARS}\label{sec:key_principles}
The instantaneous bandwidth of the MWA is 30.72 MHz, which can be split into 24 {\it coarse channels} of 1.28 MHz each and can be distributed across the entire observing band from 80 to 300 MHz. MWA solar observations are typically done with 10 kHz and 0.5 s resolution. Making images at this temporal and spectral resolution over the useful part of the complete band leads to approximately 370,000 images for an observing duration of 4 minutes. For the MWA Phase-III, the data volume will dramatically increase. The future SKAO is expected to produce even larger volumes of data. Performing the calibration and snapshot spectropolarimetric imaging of such large volumes of data manually is infeasible. One necessarily needs a software pipeline, ideally with the following capabilities:
\begin{enumerate}
    \item The calibration and imaging algorithm and its implementation must be robust.
    \item It should be capable of unsupervised operation.
    \item The algorithms it implements should be data-driven and not rely on ad-hoc assumptions.
    \item The software implementation should provide efficient parallelization which scales well with the available hardware resources.
\end{enumerate}
A state-of-the-art software framework and the calibration approach are implemented in P-AIRCARS to meet these requirements. In addition, P-AIRCARS has also been developed to be deployable across a variety of hardware environments -- ranging from laptops and workstations to high-performance computers (HPCs). This makes it very flexible. 

Radio interferometric imaging inherently involves a steep learning curve. A consequence has been that solar radio imaging has been the domain of a comparatively small number of expert practitioners and has not found widespread adoption in solar physics, as compared to other wavebands. One of the objectives for P-AIRCARS is to overcome this barrier, as the new generation and much more capable radio instrumentation is becoming available. To achieve this, P-AIRCARS has been designed to work without requiring any radio interferometry-specific input from the user. 

As a corollary of the above requirement, P-AIRCARS is designed to be fault-tolerant, in the sense that when it encounters issues, it makes smart decisions about updating the parameters for calibration and imaging based on the nature of the issue faced. For a well-informed user, P-AIRCARS allows complete flexibility to tune the algorithms as desired. The rest of the chapter describes the software framework, calibration, and parallelization strategies adopted for P-AIRACRS following the design principles described here.

\section{Architecture of the Pipeline}\label{sec:pipeline_arch}
\begin{figure}[!ht]
    \centering
    \includegraphics[trim={1.5cm 2.5cm 5cm 1cm},clip,scale=0.6]{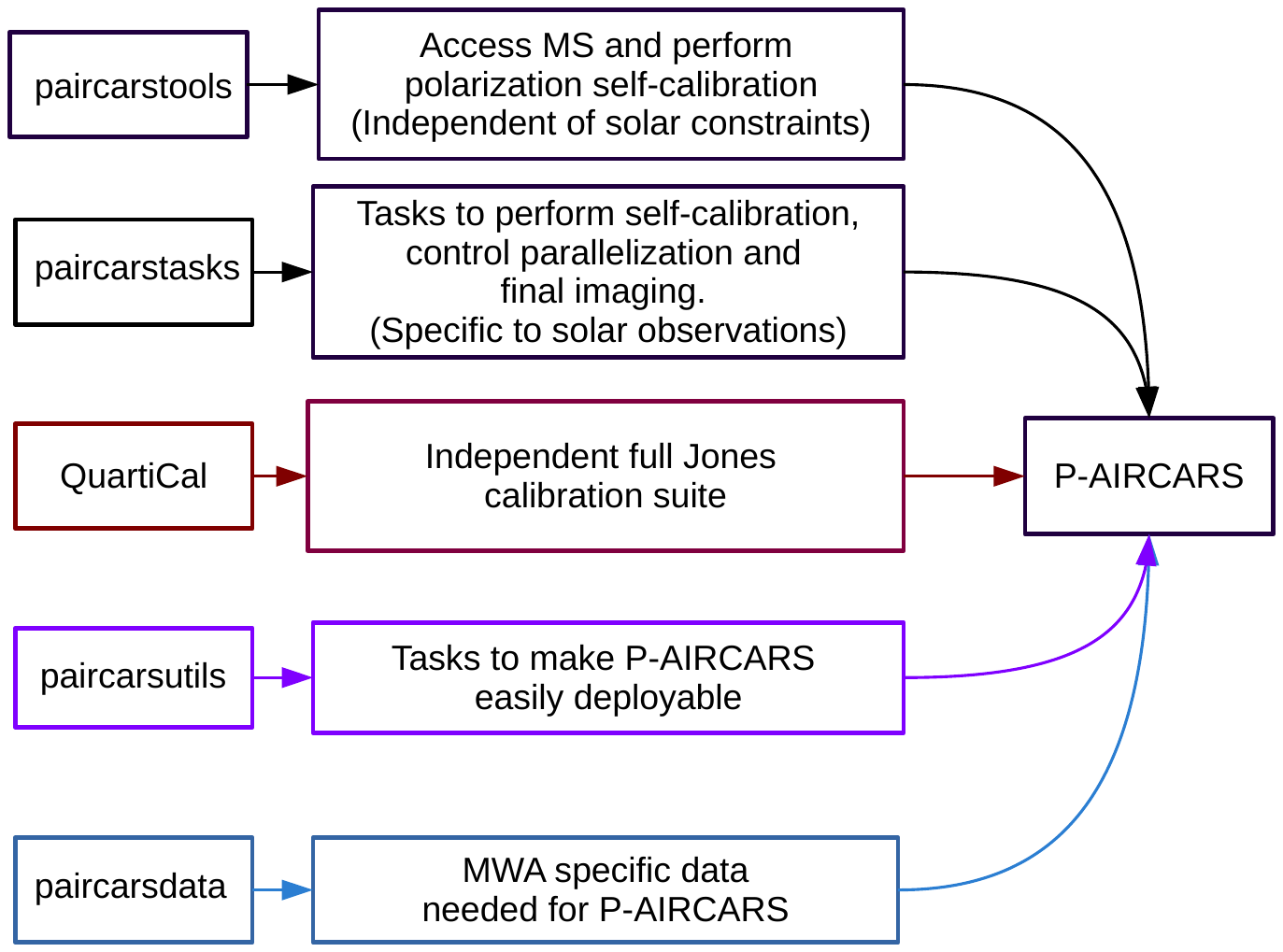}
    \caption[Schematic diagram of P-AIRCARS highlighting its main modules.]{Schematic diagram of P-AIRCARS highlighting its main modules.}
    \label{fig:module_schematic}
\end{figure}

P-AIRCARS architecture is highly modular. It has been written with ease of maintenance and adoption to other interferometers with compact core configurations in mind. A large fraction of the P-AIRCARS is written in {\it Python 3}. Some of its core modules used for calibration and flagging are written in {\it C/C++}. A schematic diagram of P-AIRCARS describing all of its modules is shown in Figure \ref{fig:module_schematic}.

The two core modules of P-AIRCARS are \textsf{paircarstools} and \textsf{paircarstasks} and are shown by black boxes in Figure \ref{fig:module_schematic}. \textsf{paircarstools} contains functionalities to perform full polarization self-calibration without imposing any constraint(s) specific to solar observation and/or the interferometer used. The optimization is specific to solar observing and is done by \textsf{paircarstasks}, which uses the functionality provided by \textsf{paircarstools}.

The third module, \textsf{QuartiCal} \citep{Quartical2022} is a successor of the full Jones calibration software suite, \textsf{CubiCal} \citep{cubical2018,Cubical_robust2019}. The \textsf{paircarsutils} module provides the utilities for the deployment of P-AIRCARS across a range of hardware and software architectures and its efficient parallelization. The \textsf{paircarsdata} module provides a collection of information specific to the MWA (e.g. the MWA beam shapes \citep{Sokolwski2017}) and MWA observations (e.g. a database of solar observations, calibration database \citep{Sokolwski2020}). 

All functions of these modules can broadly be divided into two major categories -- {\it Calibration block} and {\it Imaging block}. Instead of describing these modules function-by-function, I present the workflows of these two major blocks in Sections \ref{sec:calibration_block} and \ref{sec:image_block}, respectively. Interested users can find the details of these functions in the documentation of P-AIRCARS available online\footnote{\url{https://p-aircars.readthedocs.io/}}.  
\begin{figure*}[!ht]
    \centering
    \includegraphics[trim={1cm 7cm 1cm 1cm},clip,scale=0.5]{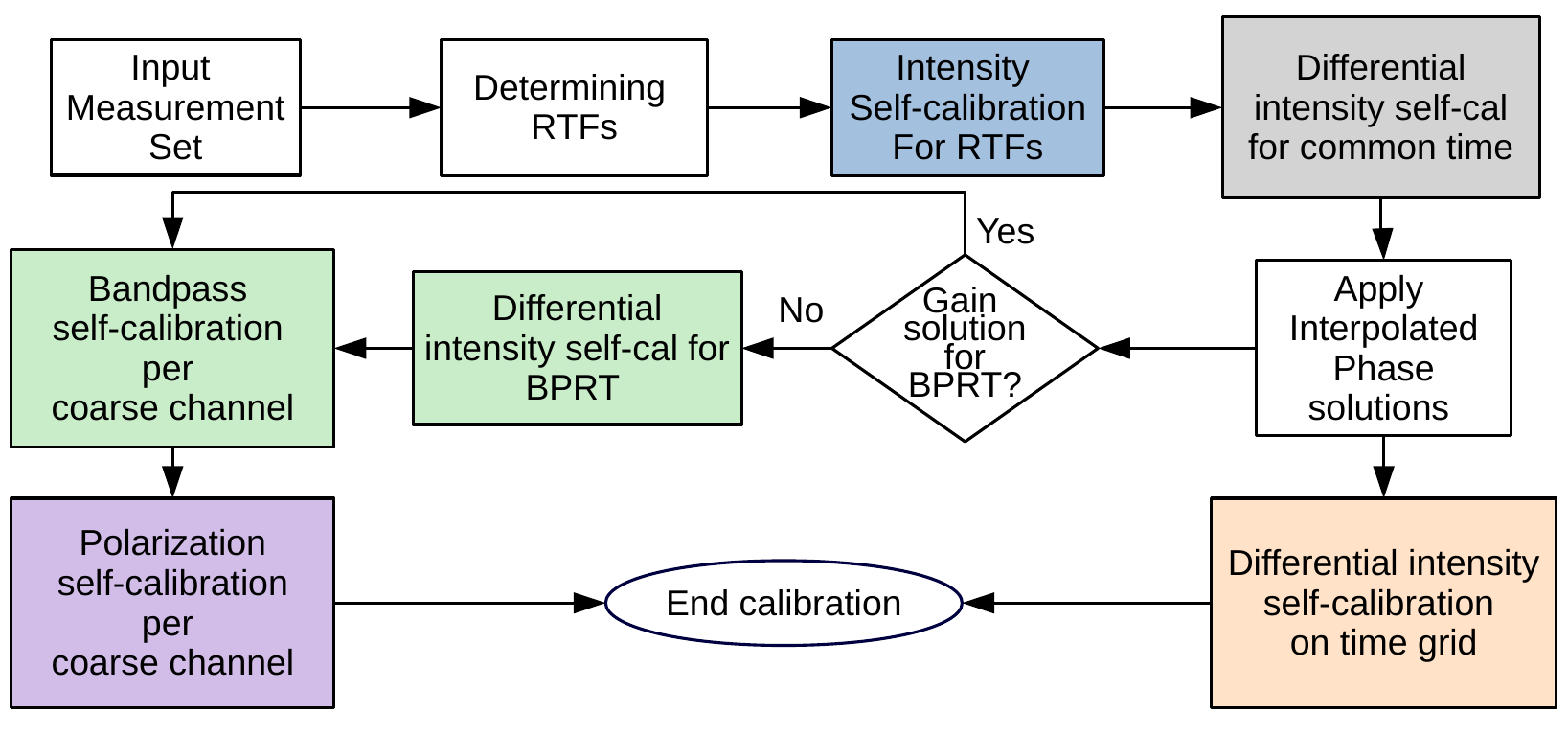}
    \caption[Flowchart describing the calibration block of P-AIRCARS.]{Flowchart describing the calibration block of P-AIRCARS. Major stages of the calibration block are shown by color boxes. {\it RTF} stands for ``reference time and frequency" and {\it BPRT} stands for ``band-pol reference time".}
    \label{fig:calibration_block}
\end{figure*}

\subsection{Implementation of Calibration Block} \label{sec:calibration_block}
The first major block of the P-AIRCARS is the calibration block. To obtain true {\it visibilities} ($V_\mathrm{ij}$) from the measured ones ($V_\mathrm{ij}^\prime$), each of the terms in Jones matrices (described in Section \ref{subsec:p_algorithm} in Chapter \ref{paircars_algorithm}) needs to be estimated precisely and corrected for. They are estimated in the following three major calibration steps (all of the symbols used here are defined in Section \ref{Overview of the Algorithm} of Chapter \ref{paircars_algorithm}):
\begin{enumerate}
    \item {\bf Intensity self-calibration: }$G_\mathrm{i}(t)$s are estimated and corrected in this step following the approach detailed in Section \ref{sec:aircars_description} of Chapter \ref{paircars_principle}.
    \item {\bf Bandpass self-calibration: }This step estimates and corrects for $B_\mathrm{i}(\nu)$s over each of the 1.28 $\mathrm{MHz}$ coarse channels. Data from quiet solar times are used for this and the integrated solar flux density is assumed to remain constant across a coarse channel.
    \item {\bf Polarization calibration: }This involves first correcting for $K_\mathrm{cross}(\nu,\ t)$, $E_\mathrm{i}(\nu,\ t,\ \vec{l})$ which are estimated independently. Next the $D_\mathrm{i}s$ are estimated and corrected using a perturbative self-calibration-based algorithm described in Section \ref{sec:poldist} of Chapter \ref{paircars_algorithm}.
\end{enumerate}
These three calibration steps form the three main pillars of the full Jones calibration algorithm of P-AIRCARS. 
\begin{figure}[!ht]
    \centering
    \includegraphics[trim={6.7cm 1.4cm 3cm 2cm},clip,scale=0.4]{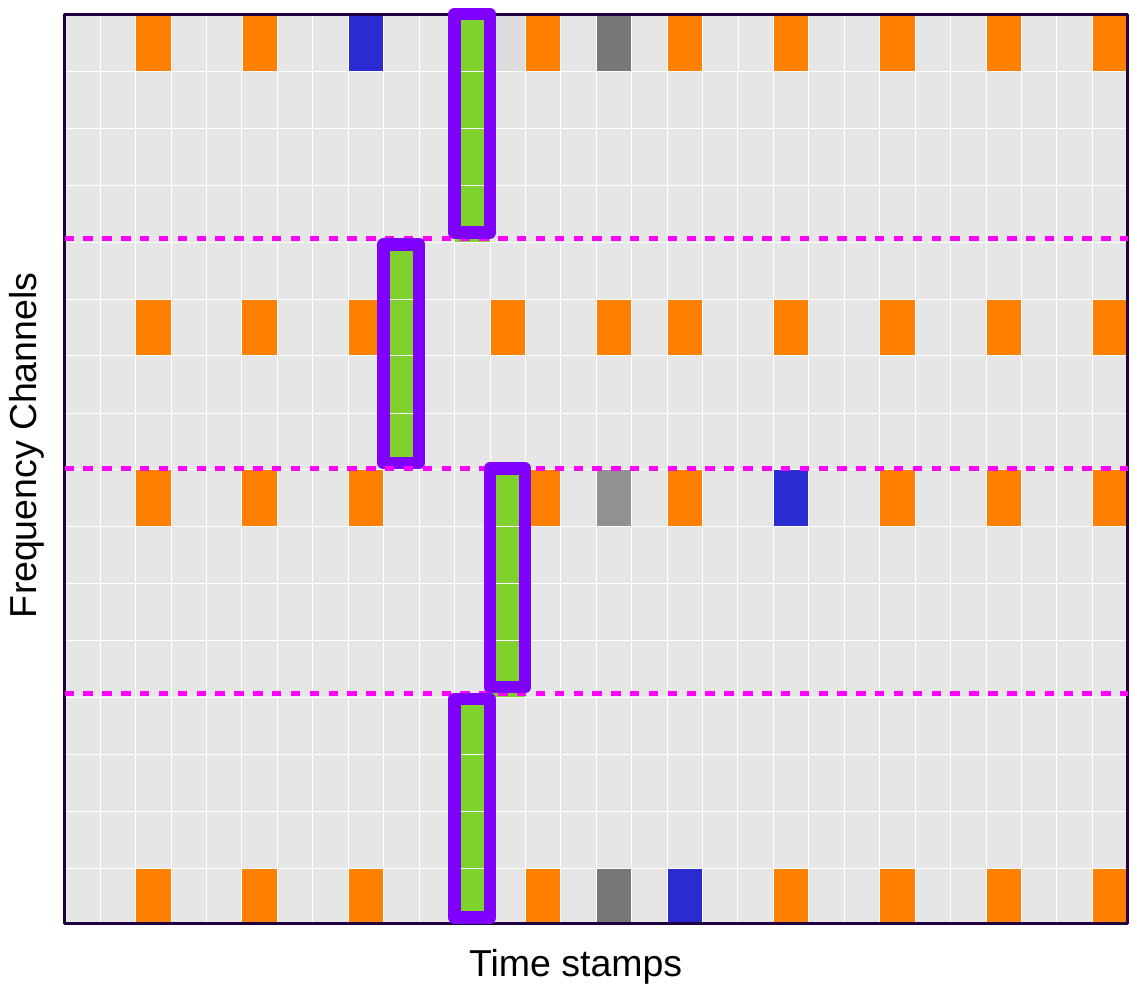}
    \caption[Time-frequency grid for performing parallel calibration.]{Time-frequency grid for parallel calibration. Blue blocks represent the RTFs. Phase part of the gain solutions is interpolated on a common timeslice shown by the dark grey cells. Orange blocks represent the time and frequency slices where differential intensity self-calibration is performed. Bandpass and polarization calibrations are performed for individual coarse channels, which are marked by green. Pink dotted lines demarcate the 1.28 $\mathrm{MHz}$ wide coarse channels.}
    \label{fig:parallel_mechanism}
\end{figure}

Since the antenna gains vary over time and frequency, in principle, one should perform self-calibration for each timestamp and frequency channel independently. Due to the intrinsic spectro-temporal variability of solar emissions, one is forced to make an independent source model for every time and frequency slice during self-calibration, which makes self-calibration for every time and frequency slice extremely compute-intensive. As the calibration for each of the time and frequency slices are independent, it can be cast in an embarrassingly parallel framework, which is implemented in P-AIRCARS. The flowchart of the calibration block is shown in Figure \ref{fig:calibration_block}. 

To start the process of calibration, a maximum of three 1.28 MHz coarse channels are chosen spanning the entire bandwidth of the data. Each of these chosen spectral channels is defined as a ``reference frequency" (RF). Next, a time slice, defined as ``reference time" (RT), is chosen for each of these RFs separately on which to perform the calibration. These are referred to as ``reference time and frequency" (RTF) slices.

Figure \ref{fig:parallel_mechanism} shows an example with four coarse channels with their boundaries marked by dashed magenta lines. RTFs are shown by blue cells in this figure. If calibrator observations are available, P-AIRCARS first applies the gain solutions obtained from them. Otherwise, intensity self-calibration is initiated from the raw data. The calibration process is initiated using the highest time resolution available in the data and if necessary the temporal span of the data used for calibration is progressively increased in an attempt to arrive at reliable gain solutions. Care is taken to not exceed the timespans over which solar emissions or ionospheric conditions are expected to evolve. The default value of this maximum timespan is set to 10$\mathrm{s}$. Once this is done on RTFs the pipeline moves to the next stage, namely bandpass self-calibration.

\citet{Sokolwski2020} demonstrated that the variation of phases across the 80--300 MHz band for the MWA antenna tiles can be modeled well by a straight line, though the amplitudes show more complex variations. Hence, it is reasonable to interpolate the phases across the MWA band using a linear model. The phase variations over a large bandwidth cause a significant frequency-dependent shift of the source from the phase center. To avoid this problem, the phase part of the gain solutions is interpolated across the entire observing band, while the amplitudes are held constant at unity. Phases are interpolated across frequency at a common timeslice marked by grey cells in Figure \ref{fig:parallel_mechanism}. A time grid is defined and marked by orange cells in Figure \ref{fig:parallel_mechanism} for each of the coarse channels, and the differential gain solutions are computed in parallel. Simultaneously, bandpass and polarization calibrations are performed at ``band-pol reference time (BPRT)" individually for each coarse channel marked by green boxes with purple borders in Figure \ref{fig:parallel_mechanism}. If the gain solutions are not available at BPRT, a differential intensity self-calibration is performed at BPRT at RF.

Once all the calibrations are complete, this information is compiled in a single calibration table spanning the entire time and frequency range. Linearly interpolated gain solutions are drawn from this final calibration table and applied during imaging. The calibration block requires one to identify RTF and BPRT. The criteria for the choice of RTF and BPRT are discussed in Sections \ref{subsec:rtf_choice} and \ref{subsec:bprt_choice}, respectively. Before this, P-AIRCARS identifies any bad data from the solar dynamic spectrum as described in Section \ref{subsec:bad_data_ds} and excludes them to avoid any problem during calibration.

\begin{figure*}[!htbp]
    \centering
    \includegraphics[trim={0.4cm 0cm 0.6cm 0cm},clip,scale=0.43]{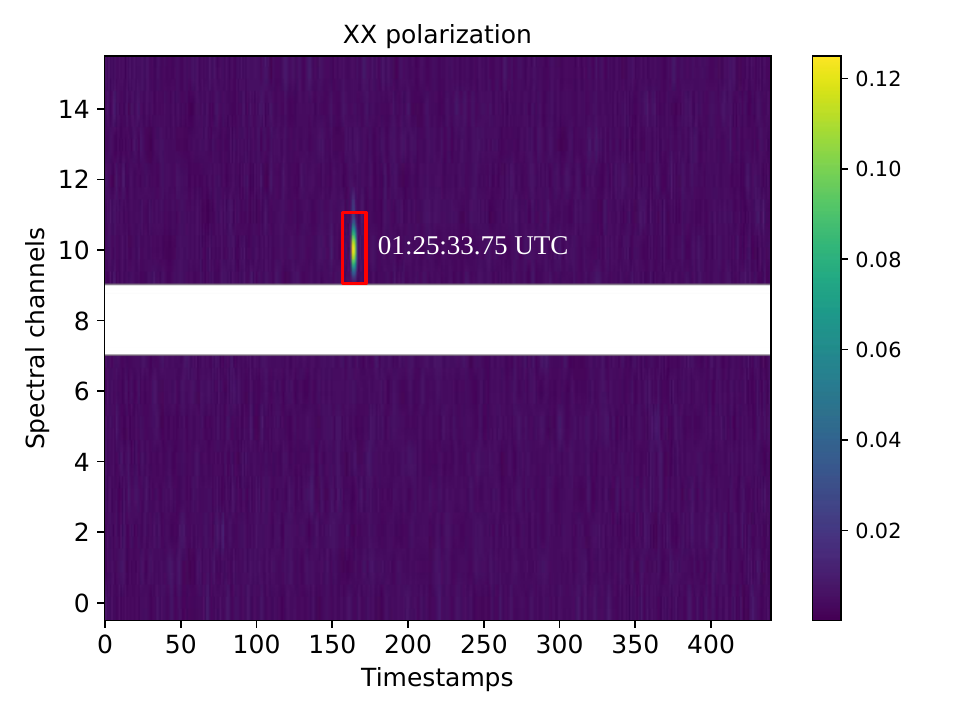}~\includegraphics[trim={0.4cm 0cm 0.5cm 0cm},clip,scale=0.43]{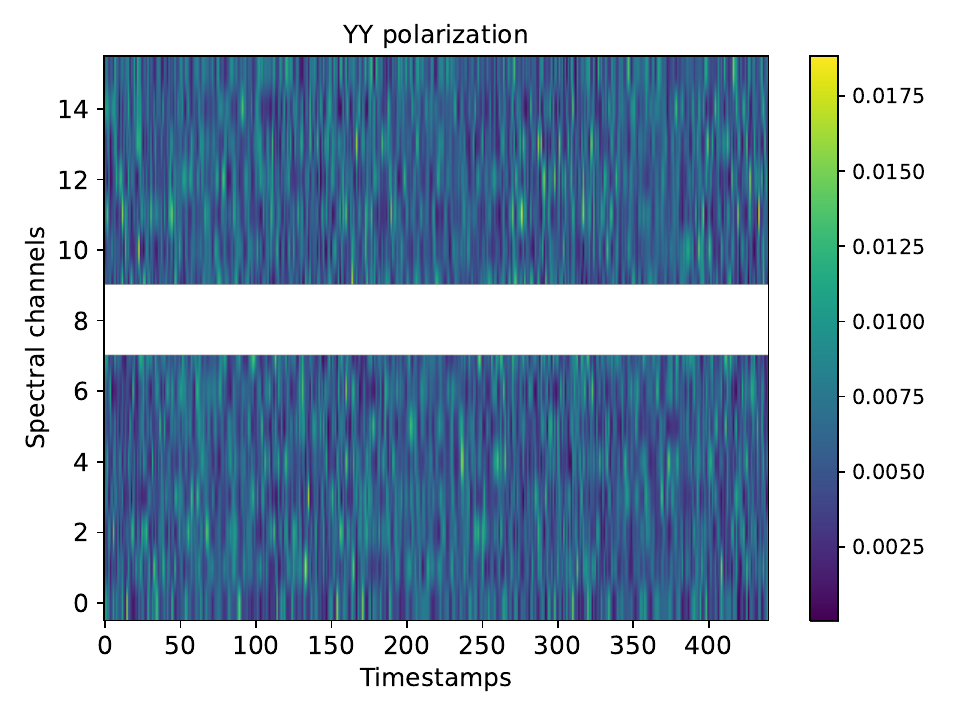}\\
     \includegraphics[trim={0.1cm 0cm 0cm 0.1cm},clip,scale=0.35]{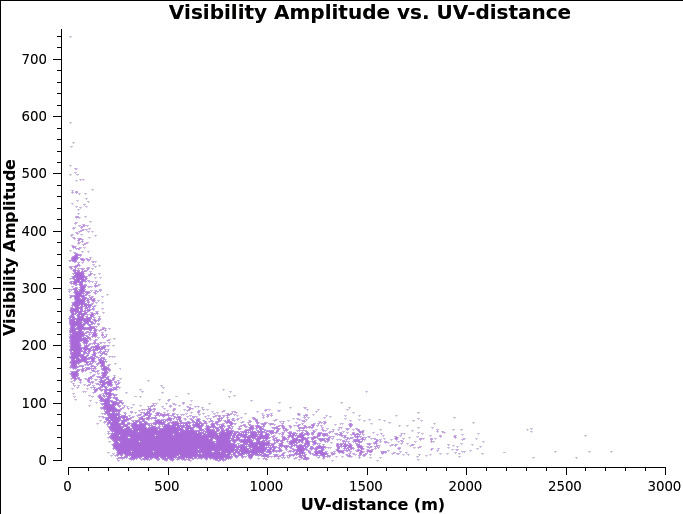}\includegraphics[trim={0.1cm 0cm 0cm 0.1cm},clip,scale=0.35]{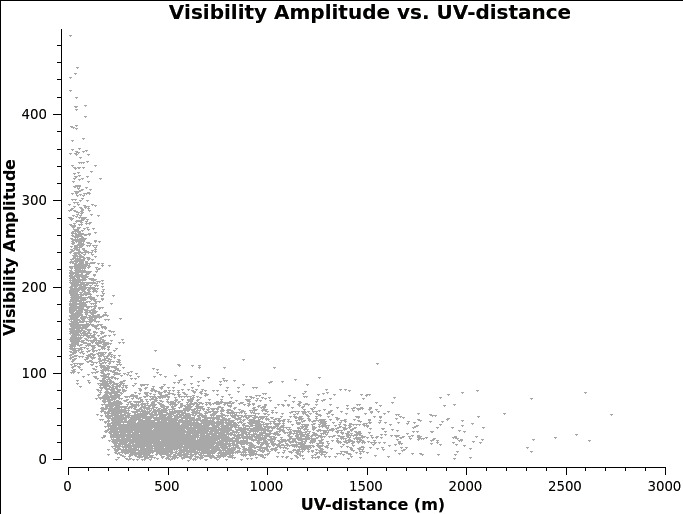}\\
     \vspace{0.5cm}
     \includegraphics[trim={0.1cm 0cm 0cm 0.1cm},clip,scale=0.35]{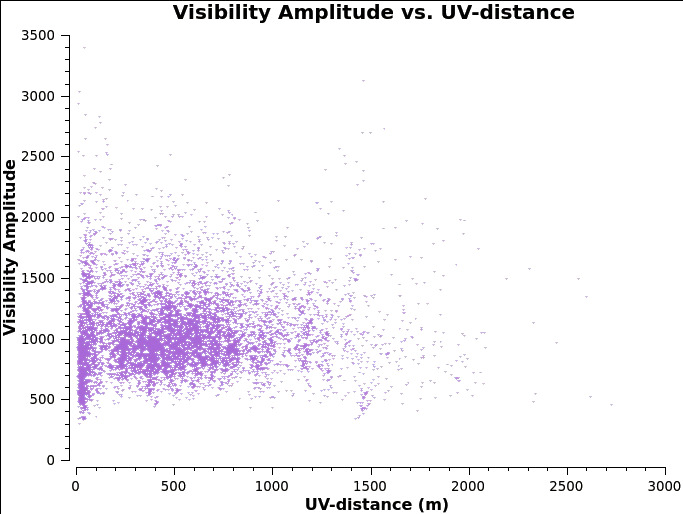}\includegraphics[trim={0.1cm 0cm 0cm 0.1cm},clip,scale=0.35]{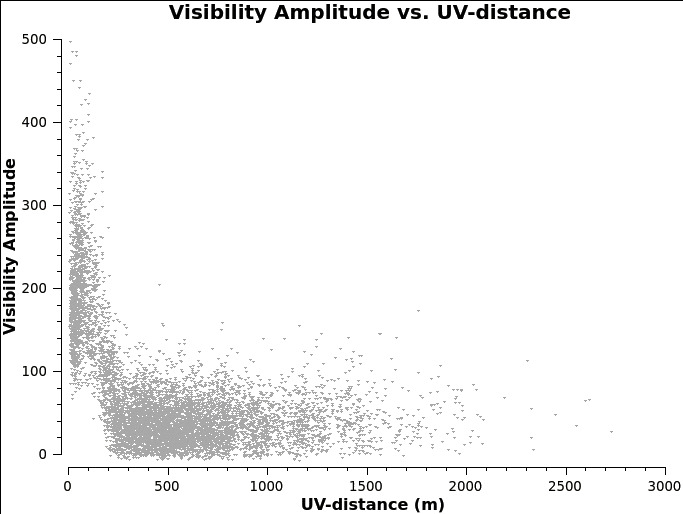}
    \caption[Demonstration of flagging of the bad data based on solar dynamic spectrum.]{ Demonstration of flagging of the bad data based on solar dynamic spectrum. {\it Top panel :} It shows the dynamic spectrum of $r_\mathrm{a}$. {\it Middle panel :} The visibility amplitude for a time without any RFI/instrumental issue is plotted against the {\it uv-}distance. {\it Bottom panel :} The visibility amplitude of a timestamp, 01:25:33.75 UTC, affected by RFI/instrumental issues is plotted against the {\it uv-}distance. This time slice is marked by the red box in the top left panel. In all rows, the left and the right panels represent the XX and the YY polarizations respectively.}
    \label{fig:ds_flag}
\end{figure*}

\subsubsection{Identifying Bad Data from the Dynamic Spectrum for Solar Observations}\label{subsec:bad_data_ds}
Even though the MWA is situated in an exceptionally low radio frequency interference (RFI) environment and is a very stable instrument, occasionally the MWA data does suffer from RFI and/or instrumental issues. It is important to ensure that only healthy data (i.e. data unaffected by RFI and/or instrumental issues) are examined while determining the BPRT and RTF. Sometimes, active solar emissions can mimic bad data in the dynamic spectrum, making it hard to identify bad data based on statistical characteristics in the time and frequency plane alone. 

Hence, I use the fact that, for the MWA, the amplitude distribution with {\it uv-}distance for active/quiet Sun emissions and data affected by RFI/instrumental issues are remarkably different. For the quiet Sun, the visibility distribution represents a disc of about 40 arcmins. It has been found that the compact sources usually associated with active emissions are slightly resolved at MWA resolution \citep{Mohan2019a,Mohan2021a,Mondal2021a,Mohan2021b}. This implies that the visibility distribution for these slightly resolved sources must show a slow drop in visibility amplitudes with increasing baseline length. On the other hand, the small footprint of the MWA and the fact that the RFI sources are mostly far away imply that the entire array tends to see the same RFI environment, and shows a relatively constant visibility amplitudes distribution with {\it uv-}distance. Both solar emission and RFI can vary by multiple orders of magnitude with time and frequency. Hence, a quantity, $r_\mathrm{a}$, is devised which is insensitive to the magnitude of the visibility amplitudes themselves but relies on their distribution as a function of baseline length to identify bad data. $r_\mathrm{a}$ is defined as the ratio of mean visibility amplitudes of long (longest 10~\%) and short baselines ($<20\ \mathrm{m}$).

Figure \ref{fig:ds_flag} shows an example to illustrate the efficacy of this approach. $r_\mathrm{a}$ is calculated independently for each spectral and temporal slice of the observations to produce the dynamic spectrum of $r_\mathrm{a}$. The top panels show the dynamic spectra of $r_\mathrm{a}$ for two parallel hand visibilities (XX and YY). The middle panels show the amplitude distribution for a time and frequency slice for XX (left panel) and YY (right panel) polarizations without any RFI/instrumental issues. The bottom panels show the amplitude distribution for XX (left panel) and YY (right panel) polarizations for a  specific time (01:25:33.75 UTC) and frequency slice (spectral channel number 10) with RFI/instrumental issues. The differences in the visibility distributions for these two panels are self-evident. The data for only the XX polarization are affected and are identified with high contrast in the top left panel. This demonstrates the capability of $r_\mathrm{a}$ dynamic spectra to unambiguously and efficiently identify the data affected by RFI/instrumental issues. Finer levels of identification and flagging of bad data are carried out during later stages of analysis using a custom-developed flagging software \textsf{ankflag} \citep{Kansabanik_paircars_2}, which has been used in several other studies previously \citep[e.g.,][etc.]{Das_2020a,Das_2020,mondal2020_ptf10hgi,Das2022}.

To identify the bad data, median ($r_\mathrm{a,med}$) and median absolute deviation ($r_\mathrm{a,MAD}$) value of $r_\mathrm{a}$ is calculated from the $r_\mathrm{a}$ dynamic spectrum for individual coarse channels separately. Assuming Gaussian distribution of $r_\mathrm{a}$, standard deviation is computed as, $\sigma(r_\mathrm{a})=1.4826\times r_\mathrm{a,MAD}$. Time and frequency slices with $r_\mathrm{a}$ lie outside $r_\mathrm{a,med}\pm5\sigma(r_\mathrm{a})$ are treated as bad time-frequency chunks and are disregarded during the calibration process.

\begin{figure}[!ht]
    \centering
    \includegraphics[trim={0.3cm 0.3cm 0.5cm 0cm},clip,scale=0.7]{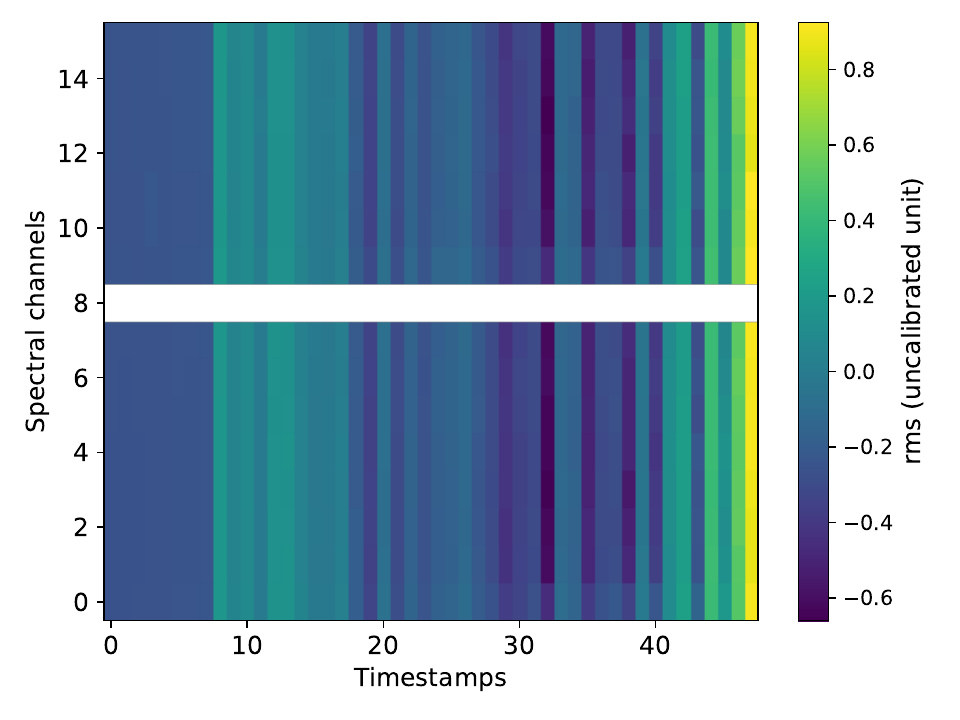}
    \caption[ynamic spectrum of image rms.]{Dynamic spectrum of image rms. The spectral and temporal span of the dynamic spectrum is 1.28 MHz and 240 s respectively. There is significant temporal variation in the image rms, while  variation along the spectral axis is barely evident.}
    \label{fig:rms_ds}
\end{figure}

\subsubsection{Choice of Reference Time and Frequencies}\label{subsec:rtf_choice}
The calibration solutions from the RT are applied to all other timestamps as the initial gain solutions. Hence, it is important to choose a timeslice that enables us to determine gain solutions for each of the antenna tiles with good signal-to-noise. An additional requirement is that the image for this timeslice should also show the quiet Sun disc with sufficient fidelity so that it can be used for alignment of solar images as discussed in Section \ref{subsec:solar_phase_shift}. For the current levels of imaging fidelity achievable with P-AIRCARS using MWA data, these requirements are typically met when a compact source with $T_\mathrm{B}\leq\ 10^7\ \mathrm{K}$ is present on the Sun. 

The MWA is coherent enough to be able to proceed with imaging without any prior gain solutions as demonstrated in Chapter \ref{paircars_principle}. I have found that the rms noise of the dirty images can vary across time due to changes in solar flux density, but it does not vary drastically across frequency. An example dynamic spectrum of the rms measured far away from the Sun is shown in Figure \ref{fig:rms_ds} which illustrates these characteristics. I have examined several datasets and established that the temporal variations of the rms noise are largely independent of the spectral channel over this small bandwidth. At first, time slices that meet the $T_{\mathrm{B}}\leq\ 10^7\ \mathrm{K}$ requirement are identified using the flux density calibrated dynamic spectrum obtained using the method developed by \citet{oberoi2017}. Dirty images are then made for every time slice meeting this requirement, for a single arbitrarily chosen spectral slice. The time slice with the highest imaging DR is chosen to be the RT. The RF channel is identified next by following a similar procedure along the frequency axis for the chosen RT. 

\subsubsection{Choice of Band-pol Reference Time}\label{subsec:bprt_choice}
As the requirements for the bandpass and polarization calibration are different from those for the initial calibration, the criteria for the choice of BPRT are also different from those for the RTF. For reasons discussed in detail in Sections \ref{subsec:bandpass_cal} and \ref{subsec:imaged_based_cor} of Chapter \ref{paircars_algorithm}, bandpass and polarization calibration require data taken under quiet solar conditions. The quiet solar time is identified in the given data, using the flux density calibrated dynamic spectrum arrived at following the method developed by \cite{oberoi2017}. The timestamps with $T_\mathrm{B}$ varying between $10^5-10^6\ \mathrm{K}$ are chosen to represent the quiet sun times. Among these, the timestamp with the maximum DR obtained from frequency-averaged dirty images is selected as the BPRT. 

\subsubsection{Alignment of the Center of Solar Radio Disc}\label{subsec:solar_phase_shift}
\begin{figure*}[!ht]
    \centering
    \includegraphics[trim={1cm 12cm 5cm 0cm},clip,scale=0.6]{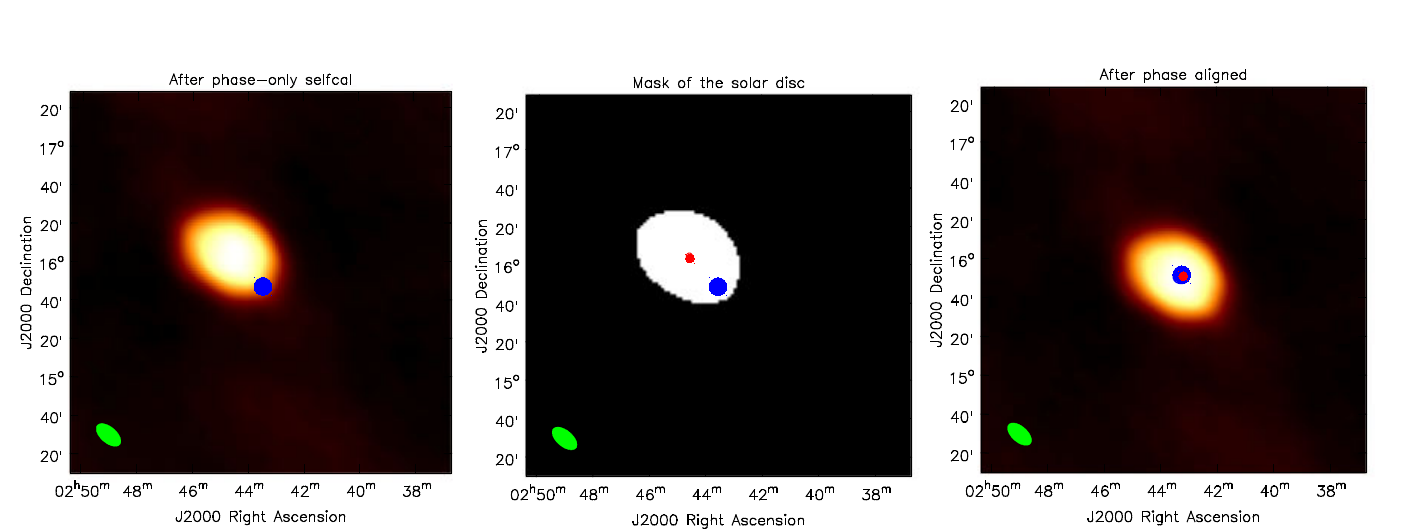}
    \caption[Alignment of the solar radio disc center with the optical solar disc center.]{Alignment of the solar radio disc center with the optical solar disc center. {\it Left panel: }Image after phase-only self-calibration. The center of the optical solar disc shown by the blue dot is not at the center of the radio disc. {\it Middle panel: } It shows the mask of the solar disc and the red dot represents the center of the radio disc. {\it Right panel: }Final image after alignment. The center of the optical and radio disc coincide after the alignment.}
    \label{fig:phase_align}
\end{figure*}

A common problem for any self-calibration-based approach is the loss of information about the absolute position. Hence, the images for the RTFs, which corresponds to a quiet solar time are aligned using an image-plane-based method. P-AIRCARS first performs phase-only intensity self-calibration followed by amplitude-phase self-calibration \citep{Mondal2019,Kansabanik2022_paircarsI}. Once the phase-only intensity self-calibration has converged, an image with the well-demarcated solar disc is available (Left panel of Figure \ref{fig:phase_align}). The blue circle marks the phase center of the radio image, which is set at the center of the optical disc. 

The region with more than $20\sigma$ detection significance is considered to be the solar disc, where $\sigma$ is the rms noise in the image measured close to the Sun. To avoid the intensity weighting, a mask is defined with all the regions more than $20\sigma$ set to unity and the rest of the image set to zero as shown in the middle panel of Figure \ref{fig:phase_align}.  Errors introduced by the presence of either non-uniform disk boundary or weak active emission at the limb are few arcseconds, which is much smaller than the size of the PSF ($\sim50$ arcsec) of the MWA even for its highest observing frequency ($\sim300$ MHz). The center of mass of the masked region is chosen to be the center of the solar radio disc marked by the red circle. The phase center of the source model is shifted to align with the blue circle. Using this aligned source model, a few rounds of phase-only self-calibration are performed. The final set of self-calibration solutions is then applied to the  each 4-minute dataset to bring it to a common phase center.

\subsubsection{Flux Density Calibration}\label{subsec:flux_calib}
Another common limitation of any self-calibration-based approach is the loss of information about the absolute flux density scale. At the MWA, when dedicated calibrator observations are available with the same spectral and attenuation configuration as solar observation, an absolute flux density scale is obtained from the gain solution of the calibrator observations. When no calibrator observation is available with the above-mentioned criteria, P-AIRCARS does flux density calibration using an independent method described in Chapter \ref{fluxcal}.

\subsection{Implementation of Imaging Block}\label{sec:image_block}
\begin{figure*}[!ht]
    \centering
    \includegraphics[trim={1cm 1cm 1cm 1cm},clip,scale=0.5]{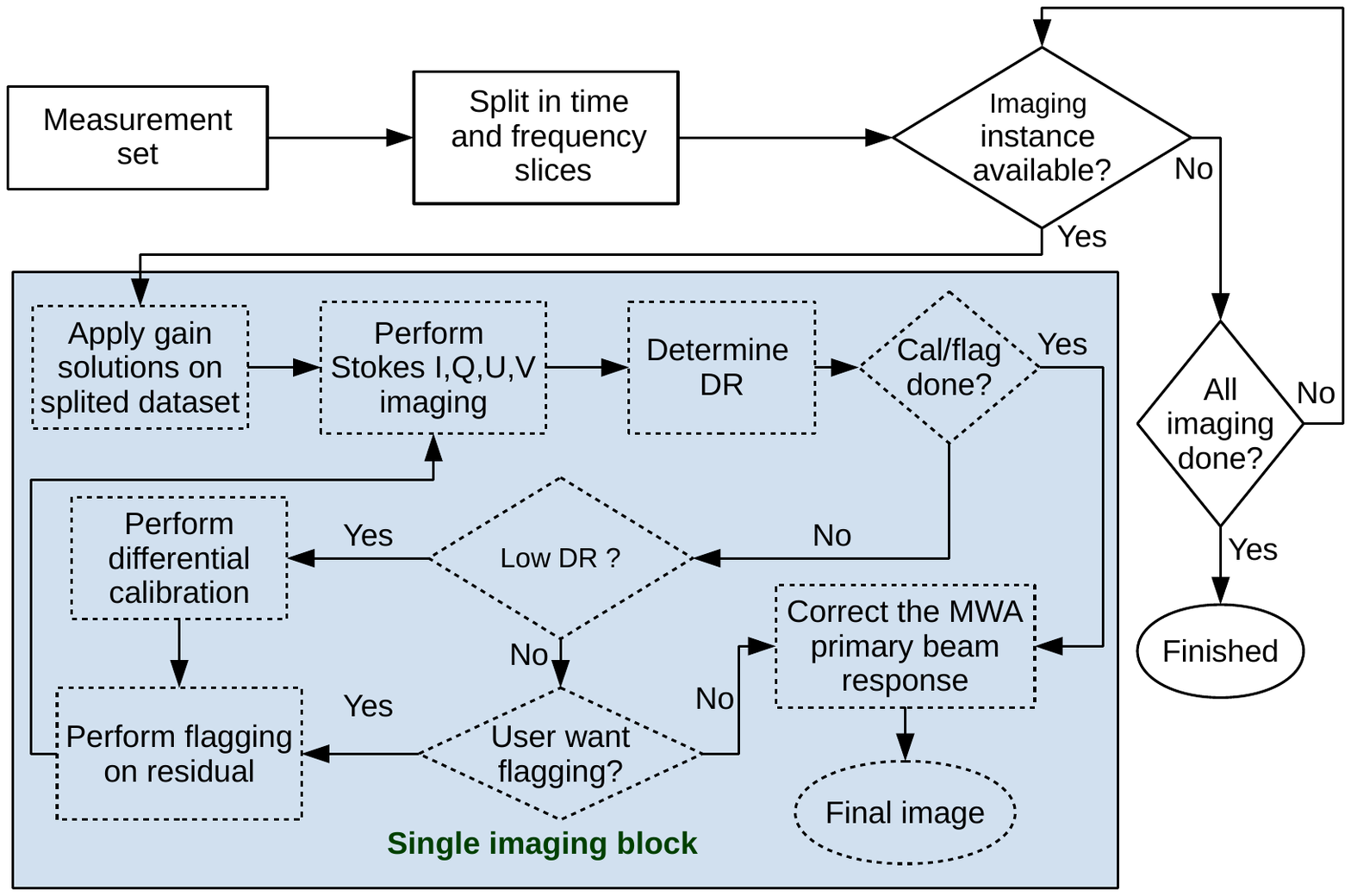}
    \caption[Flowchart describing the imaging block of P-AIRCARS.]{Flowchart describing the imaging block of P-AIRCARS. A single imaging block is shown inside the blue shaded box.}
    \label{fig:imaging_block}
\end{figure*}

Once calibration solutions spanning the time and frequency ranges of interest are available, P-AIRCARS proceeds to perform imaging. In addition to imaging, this block also corrects the images for the instrumental primary beam. The problem is essentially embarrassingly parallel, and hence straightforward to implement in an embarrassingly parallel framework. The key requirement here is to allow the user to allocate a chosen fraction of the compute resources to P-AIRCARS and for P-AIRCARS to make the optimal use of these resources. This is achieved using a custom-developed parallelization algorithm described in Section \ref{subsec:parallel_imaging}. The flowchart of the entire imaging block is shown in Figure \ref{fig:imaging_block}. The functionality in the blue box marked as `single imaging block' is executed in parallel for the different time and frequency slices and is described in Section \ref{subsec:single_imaging}. 

\subsubsection{Parallelization of Imaging Block}\label{subsec:parallel_imaging}
As mentioned in Section \ref{sec:key_principles}, the total number of images to be produced can be as many as $370,000$ for observation with a 30.72 MHz bandwidth and 4 minutes duration.  The number of imaging threads required for this task is much larger than the compute capacity available with most machines. Hence, a scalable mechanism for their efficient parallelization is required. Whenever, the number of imaging jobs, $N_\mathrm{job}$, is smaller than the available CPU threads, $N_\mathrm{thread}$, all jobs are spawned simultaneously. Each job is assigned $n$ numbers of CPU threads, where $n$ is the integer closest to $N_\mathrm{thread}/N_\mathrm{job}$. Otherwise, P-AIRCARS allocates three CPU threads for each single imaging block. The $N_\mathrm{job}$, which can be spawned simultaneously, is given by:
\begin{equation}
    N_\mathrm{job}=\frac{N_\mathrm{thread}}{3}.
\end{equation}
Different imaging jobs may take different run times. To utilize the hardware resources efficiently, as soon as one imaging job is done, a new one is spawned. This process continues until all imaging jobs have been spawned.  

\subsubsection{Single Imaging Block}\label{subsec:single_imaging}
The single imaging block makes the image of a single time and frequency slice and is marked by the blue shaded box in Figure \ref{fig:imaging_block}. Imaging parameters are determined from the data, as discussed in Section \ref{subsec:imaging_params}. Users can choose to either do full polarimetric imaging or only total intensity imaging. 

First, the final calibration solutions are applied to the data. This is followed by a shallow deconvolution (10-$\sigma$ threshold) to ensure that no spurious emission gets included in the source model. Despite the shallowness of this deconvolution, it is sufficient to provide a good check for imaging quality. The DR of these images is compared with the minimum DR ($\mathrm{DR_{min}}$) of the images made during the process of calibration. If the DR of an image after shallow deconvolution is found to be smaller than a pre-defined fraction of $\mathrm{DR_{min}}$, an additional round of calibration is performed to account for the differential antenna gain variations which might have led to the drop in the DR. This pre-defined value is set to $10$\% by default. If the user chooses to perform flagging during the final imaging, independent of whether additional calibration is required or not, a single round of flagging is done on the residual visibilities using a custom-developed flagging software, \textsf{ankflag} \citep{Kansabanik_paircars_2}. Once the flagging is done, a single round of deep deconvolution is performed. These images are then corrected for the instrumental primary beam to arrive at the final images.

\section{Calibration and Imaging Parameters}\label{sec:determine_params}
For reasons discussed in Section \ref{sec:key_principles}, P-AIRCARS is designed to determine the parameters for calibration and imaging in an unsupervised manner. There are only two high-level parameters that the user needs to specify to guide the choices to be made by P-AIRCARS. These are \textsf{quality\_factor} (QF) and \textsf{robustness\_factor} (RF). Both of these parameters take three integer values : 0, 1, and 2. QF relates to the choices of parameters impacting the final image quality, with a higher number corresponding to a better imaging quality. Similarly, RF relates to choices made regarding the convergence criteria and robustness of the self-calibration. The final choice of calibration parameters depends upon the combination of QF and RF chosen, though the final imaging parameters depend only on the choice of QF. In general, larger numbers for QF and RF lead to larger computational loads and hence longer run times.

\subsection{Calibration Parameters}\label{subsec:calib_params}
Multiple different parameters need to be specified for calibration tasks. These include the solution interval along the temporal axis ($t_\mathrm{interval}$), the minimum acceptable signal-to-noise of the antenna gain solutions ($g_\mathrm{min, SNR}$), the shortest baselines to be used, and the changes in DR ($\Delta$ DR) over the past few images defining the convergence of the self-calibration process. The length of the shortest baseline is chosen to avoid contributions from the Galactic diffuse emission as it is hard to model and can dominate the solar signal. By default, P-AIRCARS excludes visibilities below $3\lambda$, which corresponds to $\sim$20$\ \mathrm{degree}$ in angular scale. 

Some additional parameters also need to be specified for the self-calibration process. During intensity self-calibration, deconvolution thresholds are decreased in steps with the self-calibration iterations. The start, stop, and increment values for these thresholds, $th_\mathrm{start}$, $th_\mathrm{stop}$, and $th_\mathrm{step}$ respectively. These are specified in units of image rms measured far away from the Sun, $\sigma$. We define another quantity, the fractional residual flux density, which is the ratio of disc-integrated flux densities obtained from the residual and solar images from the latest self-calibration iteration. Starting from $th_\mathrm{start}$ the deconvolution threshold is lowered by $th_\mathrm{step}$ until it either reaches $th_\mathrm{stop}$ or the fractional residual flux density, $f_\mathrm{res}$, drops below some pre-defined thresholds listed in Table \ref{table:cal_params}.  If the imaging DR exceeds a pre-defined threshold, $DR_\mathrm{max}$, the self-calibration process is stopped even though it might not have converged. The numerical values of all of these parameters chosen based on the combination of QF and RF are listed in Table \ref{table:cal_params}.

\subsection{Imaging Parameters}\label{subsec:imaging_params}
Multiple different parameters need to be specified for imaging. These include -- the size of the image, pixel size, the {\it uv-}taper parameter, visibility weighting scheme, the choice of scales for multiscale deconvolution, the deconvolution threshold, the deconvolution gain, and whether or not to use {\it w-}projection. P-AIRCARS is designed to provide default values of each of these parameters. The expert user always has the flexibility to override the defaults. The default values for the image pixel size and {\it uv-}taper value are estimated from the data. The default values of image size, deconvolution threshold, deconvolution gain, and use of {\it w-}projection are decided based on the choice of the QF. The default values of visibility weighting used during imaging and Gaussian scales used for multiscale deconvolution are chosen independent of the data and QF values. Details of how the default values of imaging parameters are arrived at are available in the documentation available online.  

\section{P-AIRCARS Features}\label{sec:features}
This section briefly highlights some salient features of P-AIRCARS. 
\begin{enumerate}
\item \textbf{Modularity:} As described in Section \ref{sec:pipeline_arch}, P-AIRCARS architecture is highly modular. This not only makes it easy to maintain and upgrade, but it also enables P-AIRCARS to offer the possibility of using multiple different radio interferometric packages. 

\item \textbf{Ease of use:} To facilitate the use by community members with little or no prior experience in radio interferometry, P-AIRCARS provides reasonable defaults for all parameters, which can be overwritten by experienced users. 

\item \textbf{Input validation:} For P-AIRCARS to run successfully, all of the inputs need to be consistent and compatible with the data. To ensure this, P-AIRCARS first checks for this consistency and compatibility before initiating processing. In case some inconsistent or incompatible inputs are found, their values are reset to the default values and a warning is issued to the user.

\item \textbf{Fault-tolerant:} To be able to deal with a wide variety of solar and instrumental conditions in an unsupervised manner, P-AIRCARS has been 
\begin{landscape}
\begin{table}
\centering
\begin{tabular}{|p{0.6cm}|p{0.6cm}|p{1cm}|p{1.0cm}|p{1.5cm}|p{1.5cm}|p{1.2cm}|p{2cm}|p{1.7cm}|p{1.6cm}|p{1.6cm}|}
\hline
   QF & RF & $th_\mathrm{start}$ & $th_\mathrm{step}$ & $th_\mathrm{stop}$ & $g_\mathrm{min, SNR}$ & $\Delta$ DR &  $t_\mathrm{interval}$ (s) & $DR_\mathrm{max}$ & $f_\mathrm{res}$\\ \hline \hline 
   0 & 0 & 9.0 & 1.0 & 6.0 & 2.5 & 25 & 30 & 100 & 0.03\\
   \hline
    0 & 1 & 9.0 & 1.0 & 6.5 & 3.0 & 22 & 20 & 500 & 0.03\\
   \hline
     0 & 2 & 9.0 & 1.0 & 7.9 & 3.5 & 20 & 15 & 1000 & 0.03\\
   \hline
    1 & 0 & 10.0 & 0.5 & 6.0 & 3.5 & 20 & 15 & 1000 & 0.015\\
   \hline
    1 & 1 & 10.0 & 0.5 & 6.5 & 4.0 & 18 & 10 & 5000 & 0.015\\
   \hline
    1 & 2 & 10.0 & 0.5 & 7.0 & 4.0 & 15 & 7 & 10000 & 0.015\\
   \hline
    2 & 0 & 11.0 & 0.25 & 6.5 & 4.0 & 18 & 10 & 10000 & 0.01\\
   \hline
    2 & 1 & 11.0 & 0.25 & 7.0 & 4.5 & 15 & 7 & 50000 & 0.01\\
   \hline
    2 & 2 & 11.0 & 0.25 & 7.0 & 4.5 & 12 & 5 & 100000 & 0.01\\
   \hline
\end{tabular}
\caption[Default self-calibration parameters.]{Self-calibration parameters for different combinations of QF and RF.}
\label{table:cal_params}
\end{table}
\end{landscape}
designed to be fault-tolerant. When it fails to start the intensity self-calibration due to low signal-to-noise, it tries to make data-driven decisions about updating the relevant parameter (e.g., spectral and temporal averaging, choice of reference antenna) values to overcome the source of the problem.   
\item \textbf{Notification over e-mail:} Typical run-time for P-AIRCARS for MWA data can run into days. To make it convenient for the users to stay abreast of its progress, P-AIRCARS can provide regular notifications about its status to a user-specified list of e-mail addresses.
\item \textbf{Graphical User Interface:} P-AIRCARS provides a Graphical User Interface (GUI) for specifying values of input parameters. P-AIRCARS saves a detailed log of the various processing steps and also provides a graphical interface to easily view it.
\end{enumerate}

\section{P-AIRCARS Requirements and Performance}\label{sec:requirement}
This section summarises the hardware and software requirements for P-AIRCARS and provides some information about its run-time for typical MWA data.
\subsection{Hardware Requirements}\label{sec:hardware}
P-AIRCARS is designed to be used on a wide variety of hardware architectures, all the way from laptops and workstations to HPCs. It uses a custom-designed parallelization framework, which also does the scheduling for non-HPC environments. P-AIRCARS has been tested with a minimum configuration of 8 CPU threads and 8 GB RAM, which is increasingly commonplace in commodity laptops. P-AIRCARS has also been tested on workstations with 40$-$70 CPU threads and 256 GB of RAM.  

\subsection{Software Requirements}\label{sec:software}
P-AIRCARS uses multiple radio interferometric software packages (e.g., \textsf{CASA}, \textsf{WSClean}, \textsf{CubiCal/QuartiCal}), each of which has multiple specific software dependencies. P-AIRCARS has been tested successfully on Ubuntu (20.04) and CentOS (7 and 8) Linux environments. P-AIRCARS requires Python 3.7 or higher. To reduce the tedium of dealing with dependency conflicts and make P-AIRCARS deployable out-of-the-box, it has been containerized using {\it Docker} \citep{docker2014}. While P-AIRCARS is under constant development, interested users can download a stable version described here from Zenodo \citep{paircars_zenodo}. 

\subsection{Assessment of Run-time}\label{sec:runtime}
To provide an overall estimate for P-AIRCARS run-time, I list the run times for individual processing blocks:
\begin{enumerate}
    \item Each RTF takes about an hour (marked by blue cells in Figure \ref{fig:parallel_mechanism}).
    \item Bandpass calibration for each coarse channel takes about 15 minutes (marked by green cells with purple borders in Figure \ref{fig:parallel_mechanism}).
    \item Polarization calibration for each coarse channel takes about 45 minutes (marked by green cells with purple borders in Figure \ref{fig:parallel_mechanism}).
    \item Each differential intensity self-calibration takes about 10 to 15 minutes (marked by dark grey and orange cells in Figure \ref{fig:parallel_mechanism}).
\end{enumerate}
The first three steps are done sequentially and add up to a minimum total run time of about 2 hours. Differential intensity self-calibrations are all done in an embarrassingly parallel manner.

\begin{figure}[!ht]
    \centering
    \includegraphics[trim={0.44cm 0.5cm 0.3cm 0cm},clip,scale=0.7]{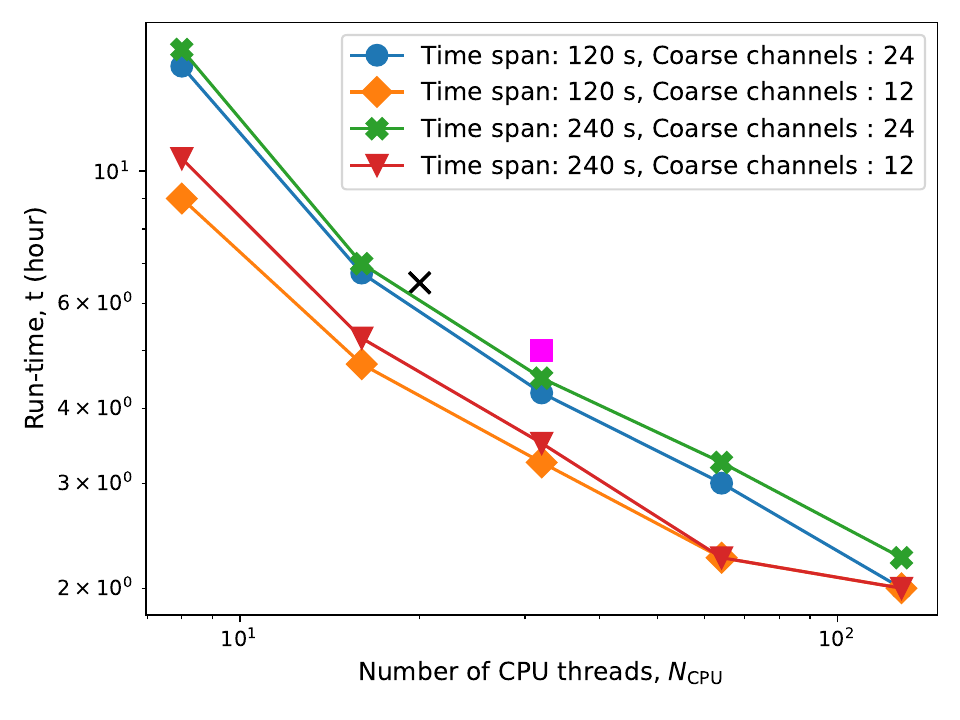}
    \caption[Variation of calibration time with the available number of CPU threads.]{Variation of calibration time with the available number of CPU threads. The green, orange, blue, and red points represent the expected run-time for a combination of temporal and spectral spans. The black cross and magenta square shows the run-time from a real dataset with 20 and 30 CPU threads respectively.}
    \label{fig:core_hours}
\end{figure}

Figure \ref{fig:core_hours} shows the expected variation in run-time, $t$, taken for calibration as a function of the number of CPU threads, $N_\mathrm{CPU}$, for a few different combinations of temporal and spectral spans on a log scale. Orange and red points show the run-time for a dataset with 12 coarse channels with an observing duration of 120 and 240 seconds, respectively. The difference between the two curves is small at the low $N_\mathrm{CPU}$ end and grows even smaller with increasing $N_\mathrm{CPU}$.
At the large $N_\mathrm{CPU}$ end, when there are enough resources available to spawn all of the differential calibration jobs in parallel, there remains no difference in the corresponding $t$s. The blue and green points show the variation of $t$ with $N_\mathrm{CPU}$ for datasets with 24 coarse spectral channels for observing durations of 120 and 240 seconds, respectively, and show similar behavior. Naturally, at the low $N_\mathrm{CPU}$ end, they take significantly longer than the 12 coarse channel datasets and the difference between $t$ for datasets with 24 and 12 coarse channels reduces with increasing $N_\mathrm{CPU}$. These curves have been obtained using a model for P-AIRCARS performance. This model has been benchmarked using measured $t$ for a dataset with 24 coarse channels spanning 240 seconds and processed using 20 and 32 CPU threads, respectively, shown by a black cross and a pink square and lie close in Figure \ref{fig:core_hours}. These values are close to the predicted model values. 

Unlike calibration, imaging jobs are {\it embarrassingly parallel}\footnote{``Embarrassingly parallel" 
is a commonly used term in the field of parallel computing.
It is used to describe a class of problems for which little effort is needed to break down the problem into several parallel tasks. This is usually the case for problems where there is little or no dependency or need for communication between those parallel tasks. This is exactly the case for spectroscopic snapshot imaging.} with $t$ decreasing linearly with increasing $N_\mathrm{CPU}$. A MWA solar observation, with 30.72 $\mathrm{MHz}$ bandwidth and 4 minutes duration, leads to about $50,000$ images at 160 $\mathrm{kHz}$ and 0.5 $\mathrm{s}$ resolution. For such a dataset, P-AIRCARS typically requires about 4 hours for calibration and about 250 hours ($\sim10$ days) for imaging using 32 CPU threads. 

\section{New Results}\label{sec:paircars_discoveries}
High \ DR \ and \ high-fidelity \ spectropolarimetric \ solar \ images \ produced \ by \\P-AIRCARS are already leading to new results. All of these lies in a previously inaccessible part of the phase space, the exploration of which has now been enabled by P-AIRCARS. Here, I briefly highlight these first results next. I note that some of these are preliminary and are intended to provide only glimpses of these interesting and ongoing projects.

\subsection{First Detection of Circular Polarized Thermal Free-free Emission from Undisturbed Solar Corona}\label{subsec:quiet_sun_V}
Measuring the magnetic field at middle and higher coronal heights is an extremely difficult problem. Several observing techniques at different wavelengths have been used to measure coronal magnetic fields \citep{Raja2022}. Observations at radio wavelengths provide several observing tools to estimate coronal magnetic fields \citep[see][for a review]{Alissandrakis2021}. \cite{Dulk1978} reviewed techniques for measuring coronal magnetic fields over active regions using different types of solar radio bursts. There have been many successful reports of coronal magnetic field measurements using polarization measurements of the radio bursts. To the best of my knowledge, no direct measurements of magnetic fields in the middle and higher coronal regions of the ``undisturbed" Sun, where the magnetic fields are much weaker, have been made till now.

In recent work, \cite{Bogod2015} reported polarization of 1.4 – 7\% and magnetic field in the range of 40 – 200 G from the {\it Radio Astronomical Telescope of the Academy of Sciences 600} (RATAN-600) observations in the wavelength range of 2 – 4 cm over the quiet solar region at lower coronal heights. Faraday rotation observations of linearly polarized background radio sources have been used to measure the coronal magnetic fields at several instances \citep[see][for a review]{Kooi2022}. But these measurements are limited to some small numbers of line-of-sights. High-precision polarization measurements of solar emissions are scarce below cm-wavelength \citep{Alissandrakis2021}. This has started to change only recently with the spectro-polarimetric imaging observations with the MWA \citep{Patrick2019}.

Measurement of magnetic field is especially challenging for the {\it undisturbed} solar corona at higher coronal heights, where the magnetic fields are much weaker. Since the work by \cite{Smerd1950}, thermal nature of the {\it undisturbed} solar coronal emission at meter-wavelength has been well established. Theoretically, it should be possible to arrive at an estimate of the {\it undisturbed} coronal magnetic field by measuring the very small level of ($<1\ \%$) induced circular polarization in thermal emission at low radio frequencies \citep{Sastry_2009}. No detection of this weak circular polarization  at meter-wavelength coming from middle and higher coronal heights has been reported yet. 
\begin{figure*}[!ht]
    \centering
    \includegraphics[trim={0cm 0cm 0cm 0cm},clip,scale=0.38]{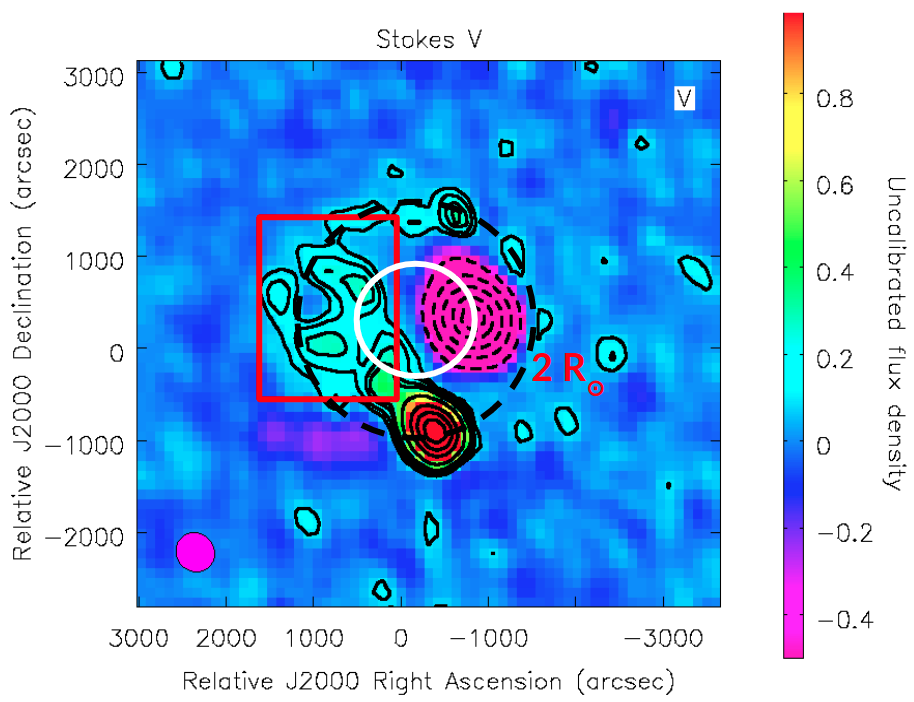}
    \caption[First-ever detection of induced circular polarization from the undisturbed coronal thermal free-free emission.]{First-ever detection of induced circular polarization from the undisturbed coronal thermal free-free emission. Circular polarization image is shown. The inner white circle marks the optical disc of the Sun and the dashed black circle marks $2\ R_{\odot}$. The region with $T_B<10^6$ K is marked by the red box. Two bright compact sources are two active emissions. One shows negative and another shows positive circular polarization. Contour levels are at  -10, -20, -40, -60, -80, 0.1, 0.5, 1, 5, 10, 20, 40, 60, 80\ \% of the peak emission. Negative contours are shown by the dashed lines.}
    \label{fig:quiet_sun_v}
\end{figure*}

High DR and high-fidelity spectropolarimetric images delivered by P-AIRCARS have led to the first robust detection of very low fractional Stokes V emission from undisturbed coronal thermal emission, as shown in Figure \ref{fig:quiet_sun_v}. The image has been made at 96 MHz with a spectro-temporal integration of 160 kHz and 0.5 s. The white circle represents the optical disc of the Sun and the black dashed circle marks 2 $R_\odot$. Regions with  $T_\mathrm{B}\leq10^6\ \mathrm{K}$ are marked by a red box in Figure \ref{fig:quiet_sun_v}. 

Though the average circular polarization fraction detected over the region inside the red box is $\sim0.5\%$, it has been detected over a large region with $>10\sigma$ significance. A robust detection of this very weak signal has been made possible by the ability of P-AIRCARS to reduce the residual instrumental leakage to $<0.07\ \%$.

These imaging Stokes V detections can constrain line-of-sight integrated magnetic field strength of the  undisturbed Sun. The inclusion of these measurements will lead to stronger constraints on different coronal magnetic field models while combined with photospheric magnetic field measurements. Work is currently in progress to use these measurements of circularly polarized emission to constraining coronal magnetic field models, though it is beyond the scope of this thesis. 

\subsection{First-ever Robust Detection of Linearly Polarized Meter-wavelength Solar Emissions}\label{subsec:linear_pol}
Another important discovery made using high-fidelity spectropolarimetric images provided by P-AIRCARS is the first-ever robust imaging detection of linearly polarized emission from meter-wavelength solar radio emission. There have been several studies in the early days \citep[e.g.,][etc.]{Hatanaka1957,Kai1963,Smith1974} which claimed the detection of linearly polarized radio emission from different types of solar radio bursts. However, doubts were cast on these measurements, largely due to limitations of instrumental polarization calibration. On the other hand, \cite{Grognard1973} developed an independent technique to search for linear polarization and concluded that due to the large differential Faraday rotation experienced by the linearly polarized emission while passing through the corona will cause depolarization of the emission and it should not be possible to detect any linear polarization in meter-wavelength solar radio emissions. Since the explanation given in this work seemed very reasonable, this was never questioned later and all other polarimetric observations of the Sun inherently assume that any observed linear polarization must arise exclusively due to instrumental leakages. This assumption has routinely been used for calibrating the instrument for solar polarimetry at meter-wavelength and continues to be in use even in recent studies \citep{mccauley2017, Morosan2022}. Naturally, this approach precludes the possibility of detection of linear polarization in solar radio emission.

\begin{figure*}[!ht]
    \centering
    \includegraphics[scale=0.22]{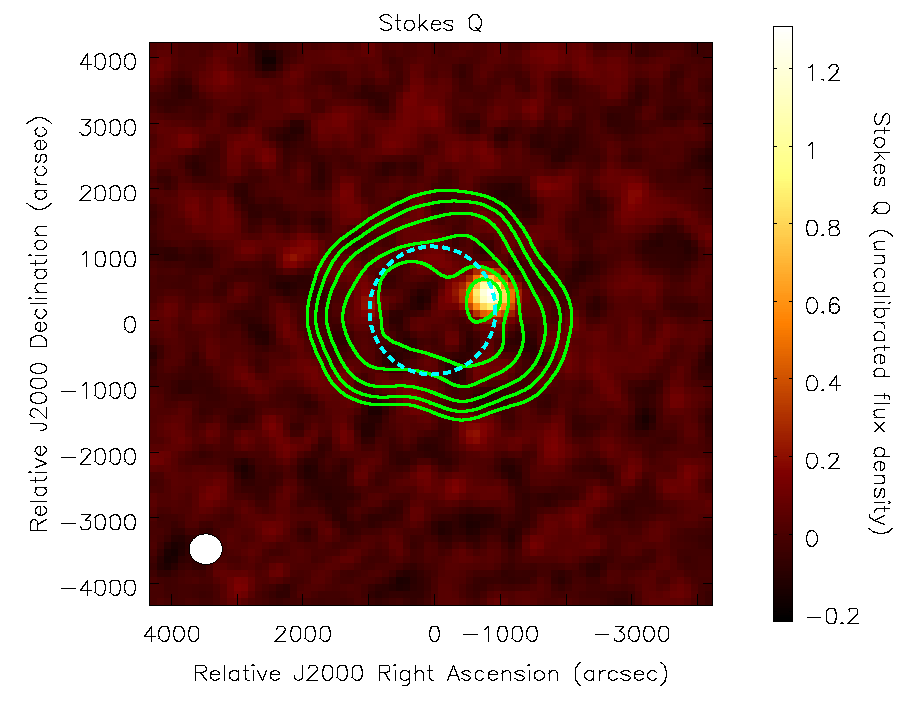}\includegraphics[scale=0.22]{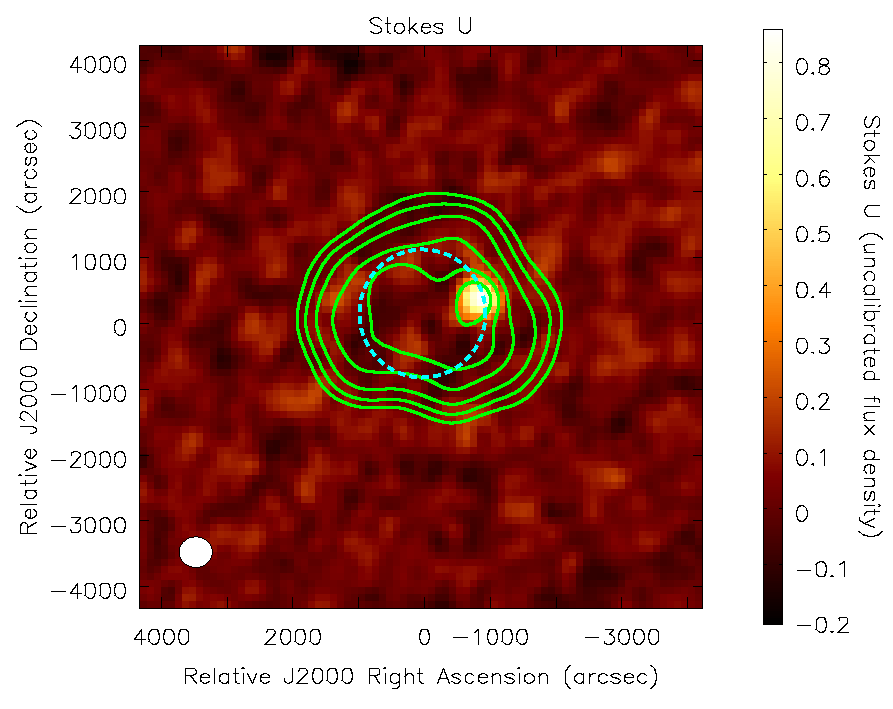}
     \caption[Linearly polarized solar radio emission from a weak type-I solar radio burst.]{Linearly polarized solar radio emission from a weak type-I solar radio burst. {\it Left panel: }Stokes Q image is shown by colormap. {\it Right panel: }Stokes U image is shown by colormap. Green contours represent the Stokes I emission. Contour levels are at 5\%, 10\%, 20\%, 40\%, 60\%, and 80\% of the peak flux density.}
     \label{fig:linear_pol}
\end{figure*}

P-AIRCARS, however, does not rely on this assumption for calibrating instrumental polarization. P-AIRCARS calibrates all known instrumental polarization effects, as has been described in detail in Section \ref{subsec:p_algorithm} of Chapter \ref{paircars_algorithm}, and no systematic instrumental polarization artifacts are present in the final full Stokes images. Due to intrinsic noise associated with any measurement, there is always some uncertainty in the estimation of calibration solutions, which limits the precision of calibration. This is referred to as residual instrumental polarization leakage. P-AIRCARS provides a high DR and high-fidelity imaging capability which in turn improves the robustness of the detection. Using spectropolarimetric images provided by P-AIRCARS, many instances of linearly polarized emissions have been detected from different types of solar radio bursts with widely varying degrees of linear polarization \citep{Dey2022,Majee2022}.

Sample linear polarization (Stokes Q and U) images from a weak type-I solar radio burst are shown in Figure \ref{fig:linear_pol}. There are several arguments that establish that the detected linear polarization is neither of instrumental origin nor could it arise due to processing artifacts. These are -- 
\begin{enumerate}
    \item MWA is a wide FoV instrument and direction-dependent instrumental leakage after modeled primary beam correction is constant over the angular scale of the solar disc. This is demonstrated in Figure \ref{fig:res_stokes_leakage} of Chapter \ref{paircars_algorithm}.
    \item  If there is any systematic polarization artifact, it cannot be different for active and quiet parts of the Sun. So polarization calibration artifacts and/or leakage must give rise to similar polarization fraction for linearly polarized emission both of the active source and quiet Sun regions. This is, however, not observed here.
    \item The observed linearly polarized emission is coincident with the Stokes I peak and is about the size of the point-spread-function (PSF). No linear polarization is expected from the quiet Sun regions and none is detected either. 
    \item As an instrumental artifact cannot be limited in the image plane to the size of a PSF, the observed characteristics of this emission feature imply that it cannot arise due to any sort of instrumental artifact.
\end{enumerate}
The residual instrumental polarization is estimated to be $<$2\%, while the linear polarization of the type-I source is $\sim20\%$. Both small residual leakage compared to the polarization of the source and the above arguments establish that the detection of linearly polarized emission is robust and free of instrumental or calibration artifacts.

These robust detections of linearly polarized emissions at multiple instances force us to question the traditional wisdom that no linear polarization is expected from meter-wavelength coronal emission. We need to come up with feasible scenarios under which the observed degree of linear polarization can be observed. This is actively being pursued by our research group and lies beyond the scope of this thesis. Although, not an apples-to-apples comparison, I note that recently there have been other reports of detection of linearly polarized emission from other stars at frequencies ranging from a few GHz \citep{Bastian2022} down to about 170 MHz \citep{Callingham2021}. 

\subsection{Spectropolarimetric Detection of Radio Emission \\from CMEs at the Largest Heliocentric Distance}\label{subsec:cme}
CMEs produce different types of radio emissions as described in Section \ref{subsec:radio_observation_cme} of Chapter \ref{chapter_intro}. Among them, gyrosynchrotron (GS) emission is produced by the mildly-relativistic electrons trapped in CME plasma. GS emissions are very faint and hard to detect as CMEs proceed to larger coronal heights, except when they appear as type-IV radio bursts. To date, the GS emission has been detected at the largest heliocentric height of 4.7 $R_\odot$ by \cite{Mondal2020a}. High DR images produced by P-AIRCARS have now enabled us to detect the GS emission from a CME out to 8.3 $R_\odot$, the largest heliocentric distance reported yet. This thesis includes a detailed study of these GS emissions for two different CMEs and they are presented in Chapters \ref{cme_gs1} and \ref{cme_gs2}.

\section{Conclusion}\label{sec:conclusion_paircars_implementation}
P-AIRCARS represents the state-of-the-art pipeline for high-fidelity high DR spectropolarimetric snapshot solar imaging at low radio frequencies. This work describes the implementation of the robust polarization calibration and imaging algorithm described in Chapter \ref{paircars_algorithm}. P-AIRCARS benefits from the experience gained and issues encountered during the extensive usage of its predecessor, AIRCARS \citep{Mondal2019}, making it more robust. It is also much more user-friendly than AIRCARS. It delivers solar radio images with residual instrumental polarization leakages comparable to those achieved by high-quality MWA observations of non-solar fields \citep[e.g.][]{lenc2017,lenc2018}. Solar radio imaging has usually been the domain of specialists. Despite the usefulness of solar radio imaging being well established and the increasing availability of large volumes of excellent data in the public domain, the steep learning curve involved has been a hurdle in the large-scale use of these data. Radio interferometric calibration and imaging is a very compute-intensive job. Hence, processing time can only be reduced by devoting large computing resources. By providing a robust tool that dramatically reduces the human tedium involved in making high-quality solar radio images, we hope to help solar radio imaging become more mainstream. P-AIRCARS is optimized for the arrays with a central dense core and dense array configuration. While arrays like the Giant Metrewave Radio Telescope \citep[GMRT,][]{Swarup_1991,Gupta_2017} and to some extent the Jansky Very Large Array \citep[JVLA,][]{VLA2009}, do have a centrally condensed configuration, these arrays are too sparse for application of P-AIRCARS. Nan\c{c}ay Radio Heliograph \citep[NRH; e.g.][]{bonmartin1983,avignon1989} and Gauribidanur Heliograph \citep{ramesh1998} do not have a centrally condensed configuration. But the functionalities in the core module of P-AIRCARS, {\it paircarstools}, can be used efficiently to develop optimized calibration and imaging pipelines for these arrays as well.

The current implementation of P-AIRCARS is optimized for the MWA, however, the underlying algorithm is equally applicable to all centrally condensed arrays, including the upcoming SKAO. The SKAO is expected to be a discovery machine in the field of solar radio and heliospheric physics. P-AIRCARS and its predecessor, AIRCARS, are already leading to explorations of previously inaccessible phase spaces. They have enabled multiple interesting scientific results spanning a large range of solar phenomena using an SKAO precursor, the MWA. It is expected that P-AIRCARS will form the workhorse for solar and heliospheric radio physics with the MWA and the stepping stone for the solar radio imaging pipeline for the SKAO.

\chapter {Deciphering Faint Radio Emissions from CME Plasma}
\label{cme_gs1}

Among the multiple results enabled by P-AIRCARS imaging, one of them is the detection of faint radio emissions from coronal mass ejections (CMEs) at the largest heliocentric distance reported yet. This chapter presents a detailed spectropolarimetric modeling of this radio emission to arrive at a  estimation of the magnetic field and other plasma parameters for one of the observed CMEs. The material presented in this chapter is based on \citet{Kansabanik2023_CME1}, which has been published in the Astrophysical Journal.

\section{Introduction}
Coronal Mass Ejections (CMEs) are large-scale eruptions of magnetized plasma from the solar corona to the heliosphere. CMEs are routinely observed at visible wavelengths using ground and space-based coronagraphs. Observation at visible wavelengths provides several pieces of crucial information about CMEs -- its large-scale three-dimensional structure, velocity, acceleration, electron density \citep[e.g.][]{Webb2012}. There are several models available about the origin and evolution of the CMEs \citep[e.g.,][etc.]{Chen2011,Kilpua2021,Sindhuja_2022}, though the exact mechanisms continue to be debated. Nonetheless, it is well established that CME eruption, evolution, and geo-effectiveness are all primarily driven by their magnetic fields \citep[e.g,][etc.]{aschwanden2004,Vourlidas2020,Temmer2021,Srivastava2021}. Hence measurements of the magnetic fields both inside the CME plasma and at the shock are essential. 

Observations at visible and extreme ultraviolet (EUV) wavelengths have been used to estimate CME magnetic field using some indirect techniques \citep{Savani_etal_2015,Kilpua2021}. Several other indirect techniques have been developed over the last decade or so to measure the average magnetic field strength at the CME shock front \citep[e.g.,][etc.]{Cho2007,Gopalswamy2011,Raja2014,Kumari2017typeII_band,Kumari2017_typeIV,Zhao2019}. Although successful, none of these techniques can be used to measure the magnetic fields entrained in the CME loops.

Radio observations have the potential to estimate the magnetic field entrained in the CME plasma. A recent study by \cite{Ramesh2021} used the induced circular polarization (Stokes V) measurements of thermal emission from CME plasma to estimate the CME magnetic field at heliocentric distance $\sim2\ R_\odot$. Another method that has been used in the past to measure magnetic field entrained in CME plasma is the modeling the spectrum of gyrosynchrotron (GS) emission \citep[e.g.,][etc.]{Boischot1957,Boischot1968,Dulk1973,bastian2001,Maia2007,Tun2013,Bain2014,Carley2017,Mondal2020a,Chhabra_2021}. GS emission is produced by the mildly relativistic electrons trapped in the CME plasma. Spectral modeling the GS emission is regarded as a promising indirect method for estimating magnetic field and other CME plasma parameters remotely using ground-based radio telescopes. Despite the promise it holds and the attention it has commanded, there have been only a handful of successful attempts at the detection of GS emissions from CME plasma at middle and higher coronal heights in the last two decades.

The reason for the limited success of detection of this emission could be that it is challenging to detect the much fainter GS emission from CME plasma (about a few tens to hundreds of $\mathrm{Jy}$) in the vicinity of the much brighter Sun. Even the quiet Sun can be a few SFU (1 SFU = 10$^4 \mathrm{Jy}$), and often, the presence of GS emission overlaps with that of much brighter non-thermal emissions associated with active regions.  An essential requirement for using the modeling of GS emission for the estimation of CME magnetic fields is to first detect it with sufficient significance and spectral sampling. This requires achieving a sufficiently high imaging dynamic range (DR). Using the state-of-the-art calibration and imaging algorithm, AIRCARS \citep{Mondal2019} and its successor P-AIRCARS \citep{Kansabanik2022_paircarsI,Kansabanik_paircars_2} on the MWA data, we have successfully detected GS emission for each of the handful of CMEs studied so far.

There are however additional challenges to overcome beyond total intensity (Stokes I) detection of GS emission from CME plasma. The GS model has ten independent parameters, assuming the non-thermal electron to follow the simplest single power-law distribution \citep{Fleishman_2010,Kuznetsov_2021} and some of them show degeneracies which cannot be broken by Stokes I spectra alone. Hence, it is hard to arrive at firm values for CME magnetic fields only with Stokes I measurements available to constrain the ten parameters of the GS model. The lack of stringent polarization information and the large number of free parameters of the GS models, in comparison to the available constraints, left no choice for the earlier studies but to rely on several assumptions while modeling the observed Stokes I GS spectrum. These assumptions were typically related to the LoS depth, the angle between the magnetic field and LoS, and non-thermal electron density. If additional independent observational information can be provided as a constrain for the modeling, it becomes possible to break the degeneracies in the model and estimate the magnetic field of CME plasma with low uncertainty. \cite{bastian2001} and \cite{Tun2013} did report the presence of circularly polarized emission (Stokes V) from CME GS using the Nan\c{c}ay Radio Heliograph \citep[NRH;][]{bonmartin1983,avignon1989} observations, but did not use the Stokes V information for constraining the GS models. These studies were limited by their spectral sampling and/or the spectral peak not being sampled by the available observations. Observations of Stokes V emission provide independent observational constraints. When used in combination with the Stokes I spectrum, they can break some of the degeneracies of GS models and bring us one step closer toward reliable estimation of CME magnetic field using GS emission. 

Using the high-fidelity spectropolarimetric images from the successor of AIRCARS, P-AIRCARS \citep{Kansabanik2022_paircarsI,Kansabanik_paircars_2}, this work presents spatially resolved estimates of CME GS model parameters using joint constraints from Stokes I spectra and stringent upper limits on Stokes V measurements for a very weak CME. This weak event is chosen to demonstrate the capability of the MWA to detect very faint radio emissions from CME plasma. Though not strictly true, it is not unreasonable to expect stronger CMEs to give rise to stronger emissions. So a  successful detection of a weak CME suggests that detecting emissions from stronger CMEs should usually be well within the capabilities of the MWA. This work also presents the first application of Bayesian analysis to this scenario. Joint constraints from Stokes I and V yield tighter bounds on the distribution of GS model parameters than possible using Stokes I spectra alone.

This chapter is organized as follows -- Section \ref{sec:obs_and_data} describes the observation and the data analysis. The imaging results are presented in Section \ref{sec:result}, along with the arguments for the observed emission arising from the GS mechanism. The impact of variations in the different parameters of the GS model on Stokes I and V spectra are presented in Section \ref{sec:spectrum_sensitivity}. A mathematical framework based on Bayes theorem is presented in Section \ref{subsec:upperlimits_methods}. Sections \ref{sec:spectrum_modeling} and \ref{subsec:magnetic_green} describe the joint Stokes I and V spectral modeling and the estimates of plasma parameters they lead to. Section \ref{sec:discussion_cme1} presents a discussion before presenting the conclusions in Section \ref{sec:conclusion_cme1}.

\section{Observation and Data Analysis}\label{sec:obs_and_data}
The observations presented here were made on 2014 May 04. During this time white-light observations are available from three vantage points in space from three spacecraft -- the Solar and Heliospheric Observatory \citep[SOHO;][]{Domingo1995}, Solar Terrestrial Relations Observatory - Ahead (STEREO-A) and Behind (STEREO-B) \citep{Kaiser2008}. Among different CMEs observed by the MWA Phase-I, this event is one of the weaker events and was observed at the lowest permissible elevation. Among the other weak events, this was among the few that had coronagraph observations available from three vantage points, very useful for independent constraints on the geometrical parameters of the CME.

On this day a total of six active regions were present on the Earth-facing part of the solar disc\footnote{\url{https://www.solarmonitor.org/?date=20140504}}. No large flares (M or X GOES class) were reported. The CME catalog provided by the Coordinated Data Analysis Workshop (CDAW) reported a total of nine CMEs\footnote{\url{https://cdaw.gsfc.nasa.gov/CME_list/UNIVERSAL/2014_05/univ2014_05.html}}, and most of them are reported as ``poor events". Of these, two have overlapping MWA observations -- one is seen to be propagating towards solar north (CME-1) and the other towards southwest (CME-2). In this chapter, I will present a detailed spectropolarimetric imaging analysis of the GS emission from the CME-1 and the results from CME-2 will be presented in Chapter \ref{cme_gs2}.

\subsection{Eruption and Evolution of CME-1}\label{subsec:erup_loc}
\begin{figure}[!ht]
    \centering
     \includegraphics[trim={1.2cm 5cm 1.2cm 4.5cm},clip,scale=0.45]{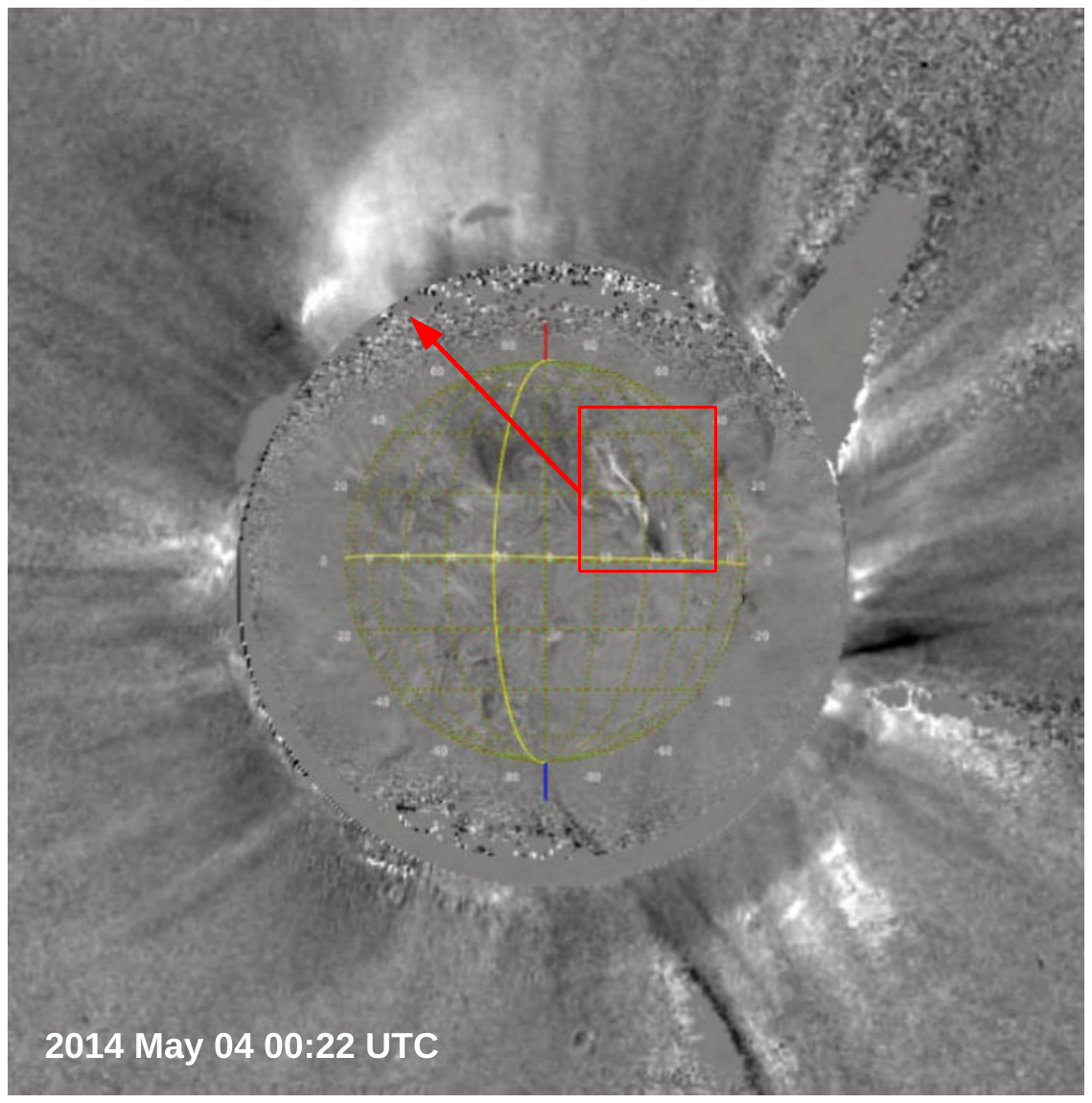}
     \caption[Eruption of northern CME as observed using the STEREO-B spacecraft.]{Eruption of CME-1 as observed using the STEREO-B spacecraft. CME-1 erupted from behind the visible solar disc. A composite base difference image from the Extreme Ultraviolet Imager (EUVI) and COR-1 coronagraph onboard the STEREO-B spacecraft is shown. The red box shows the eruption site and the red arrow shows the propagation direction.}
    \label{fig:north_cme_eruption}
\end{figure}

The CME-1 first appeared in the field-of-view (FoV) of COR1 coronagraph \citep{Thompson2003} onboard STEREO-B spacecraft at 23:52:17 UTC on 2014 May 03. It did not show any eruptive signature in the Extreme Ultra Violet (EUV) images from the Atmospheric Imaging Assembly \citep[AIA;][]{Lemen2012} onboard the Solar Dynamics Observatory \citep[SDO;][]{Pesnell2012}. This suggests that CME-1 has likely erupted from the far side of the Sun. Examining the EUV image from the Extreme Ultraviolet Imager \citep[EUVI;][]{Wuelser2004} onboard STEREO-B, the filament eruption, which gave rise to CME-1, is identified. A composite base difference image from EUVI at 195$\mathrm{\AA}$ and COR1 coronagraph at visible wavelength are shown in Figure \ref{fig:north_cme_eruption}. CME-1 first appeared at 00:12 UTC on 2014 May 04 in the FoV of C2 coronagraph of the Large Angle Spectroscopic Coronagraph \citep[LASCO;][]{Brueckner1995} onboard SOHO and was visible till 02:48 UTC in visible in the C2 FoV. 

\begin{figure}[!ht]
   \centering
    \includegraphics[trim={0.4cm 0.5cm 1cm 0cm},clip,scale=0.5]{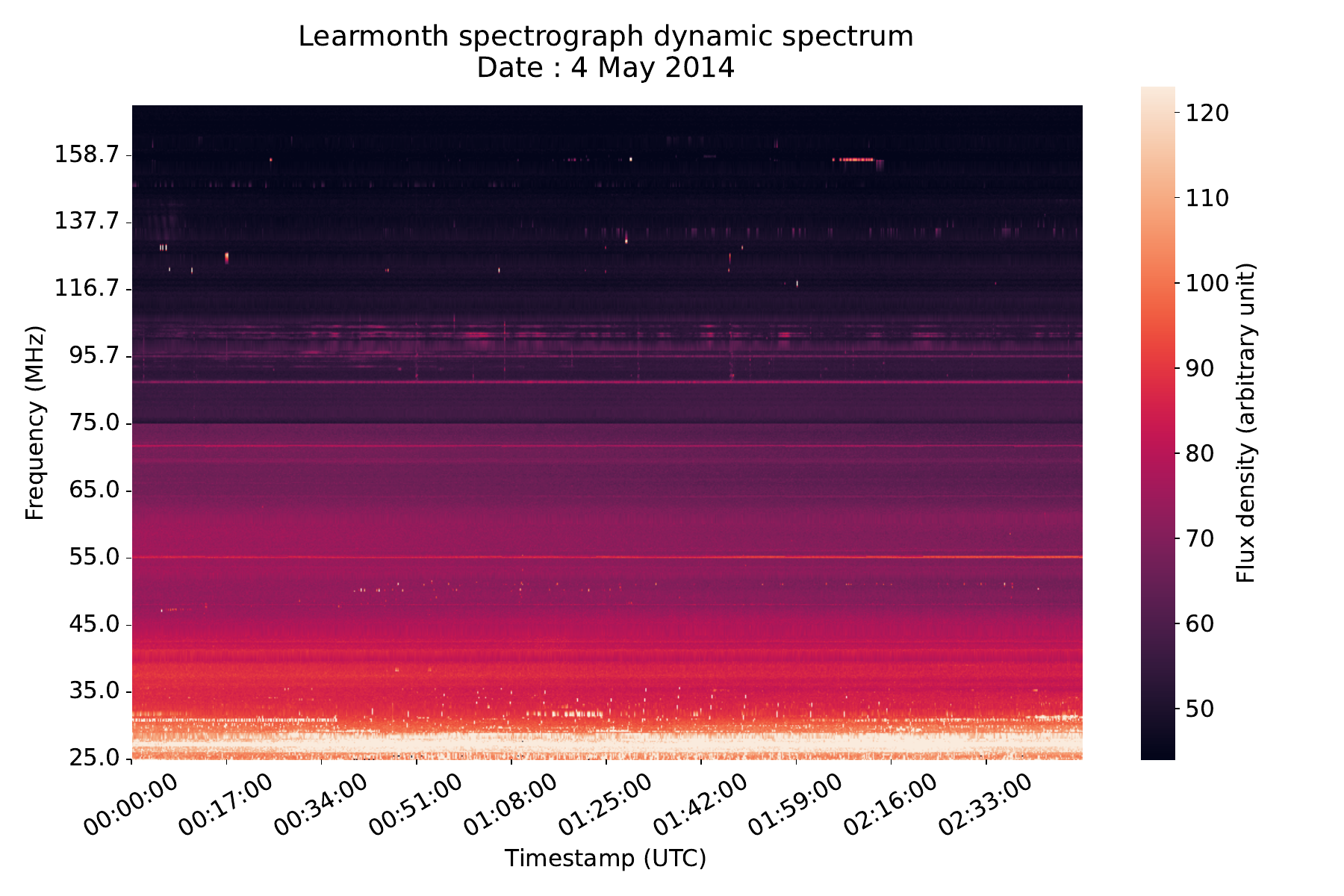}
    \caption[Solar dynamic spectrum from the Learmonth radio spectrograph on 2014 May 04.]{Dynamic spectrum from the Learmonth radio spectrograph. No radio bursts are seen from 00:00-03:00 UTC on 04 May 2014. Several bad channels with persistent radio frequency interference (RFI) have been flagged and interpolated for each time slice independently.}
    \label{fig:learmonth}
\end{figure}

\subsection{Radio Observation and Data Analysis}\label{subsec:radio_data_analysis}
CME-1 was observed at meter-wavelength radio bands using the MWA. On 2014 May 04, the MWA observed the Sun from 00:48 UTC to 07:32 UTC under the project ID  G0002\footnote{\url{http://ws.mwatelescope.org/metadata/find}}. The MWA observations were done in 12 frequency bands, each of width 2.56 MHz, and centered around 80, 89, 98, 108, 120, 132, 145, 161, 179, 196, 217, and 240 $\mathrm{MHz}$. The temporal and spectral resolution of the data were 0.5 $\mathrm{s}$ and 40 $\mathrm{kHz}$, respectively. CMEs are often associated with a variety of active solar emissions -- type-II, -III, and -IV radio bursts \citep{Gopalswamy2011_CME_radio,Carley2020}. The radio dynamic spectrum from the Learmonth Solar Spectrograph, however, does not show any evidence of associated radio emission in the 25--180 MHz band from 00:00 UTC to 03:00 UTC (Figure \ref{fig:learmonth}). No signature of coherent radio emission is seen in the more sensitive data from the MWA either. Observations from S-WAVES radio data \citep{Bougeret1995} onboard the WIND and STEREO-A and B spacecraft were also inspected for any signature of type-II or interplanetary type-II bursts.

\begin{figure*}
    \centering
    \includegraphics[trim={0cm 0cm 0cm 0cm},clip,scale=0.7]{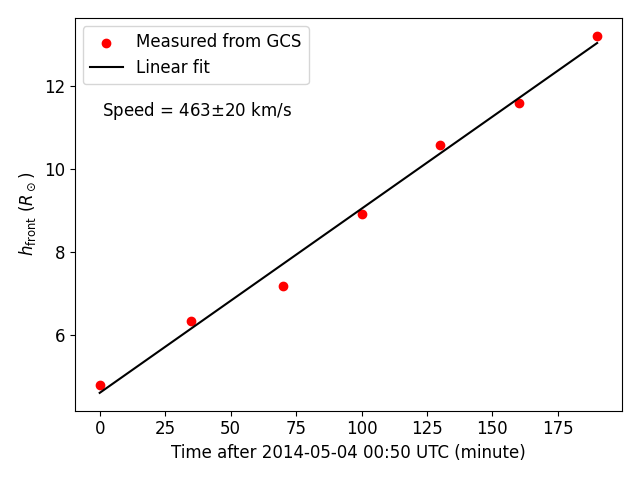}
    \caption[Estimation of three-dimensional speed of CME-1.]{Estimation of three-dimensional speed of CME-1. Red points represent measured values of front height using GCS model and solid black line represent the linear fit to the observed values.}
    \label{fig:gcs_speed}
\end{figure*}

Average plane-of-sky (PoS) speed of CME-1 reported by the Coordinated Data Analysis Workshop (CDAW) catalog is $\sim458$ km/s\footnote{\href{https://cdaw.gsfc.nasa.gov/CME_list/UNIVERSAL_ver1/2014_05/htpng/20140504.001205.p353g.htp.html}{Height time plot and estimated PoS speed from CDAW catalog.}}.  Due to the projection effects involved, the PoS speed is only the lower limit on the true three-dimensional speed. Three-dimensional reconstruction of the CME-1 is done using {\it Python} implementation \citep{gcs_python} of Graduated Cylindrical Shell model \citep[GCS;][]{Thernisien_2006,Thernisien_2011}. GCS modeling is done from about 01:00 UTC to 04:00 UTC. The three-dimensional speed of CME-1 is estimated from a linear fit to the time variation of front height ($h_\mathrm{front}$) obtained from the GCS modeling as shown in Figure \ref{fig:gcs_speed}. The three-dimensional speed of the front height of CME-1 is estimated to be $463\pm20$ km/s.
When a CME is sufficiently faster than the preceding solar wind, a shock wave develops ahead of the CME. CME-1 propagates toward the solar north-polar region, where the solar wind speeds are generally higher. During 2014, the average solar wind speed at higher latitudes, estimated using interplanetary scintillation (IPS) observations, is $\sim600-700$ km/s \citep{Tokumaru_2021}. Since the three-dimensional speed of CME-1 is smaller than the expected background solar wind speed, no shock should be produced by CME-1. In line with this expectation, no evidence of a white-light shock is evident in either COR-1 image from STEREO-B (Figure \ref{fig:north_cme_eruption}) or LASCO-C2 image (Figure \ref{fig:north_cme}), consistent with the absence of a type-II radio burst.

Polarization calibration and full Stokes imaging of the MWA observation are performed using P-AIRCARS. Flux density calibration was done using the technique presented in Chapter \ref{fluxcal}, which is implemented in P-AIRCARS. Integration of 10 s and 2.56 MHz was used for imaging for all 12 frequency bands. All polarimetric images made using P-AIRCARS follow the IAU/IEEE convention of Stokes parameters \citep{IAU_1973,Hamaker1996_3}. 

\section{Results}\label{sec:result}
This section presents the detection of spatially resolved faint radio emission from CME plasma using wideband spectropolarimetric imaging observation from the MWA. 

\subsection{Radio Emission from CME-1}\label{sec:radio_detection}
\begin{figure}[!ht]
    \centering
    \includegraphics[trim={1.7cm 0.3cm 1cm 0cm},clip,scale=0.8]{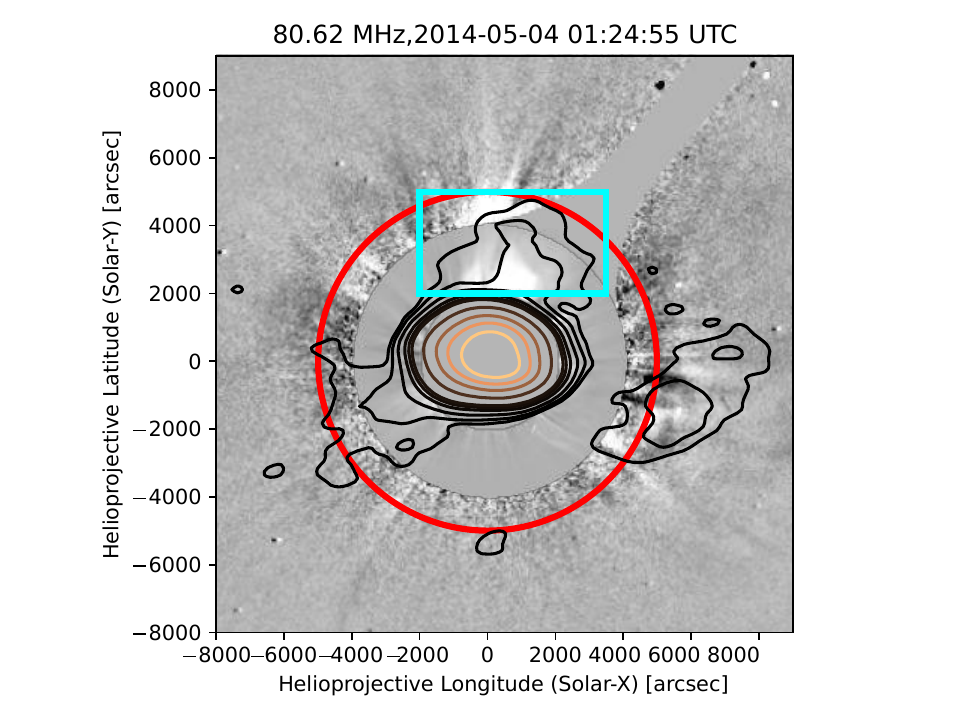}
    \caption[Radio emission from CME-1 at 80 MHz.]{Radio emission from CME-1 at 80 MHz. Stokes I emissions at 80 MHz are shown by the contours overlaid on the base difference coronagraph images. The background shows the LASCO C2 and C3 coronagraph images from the nearest available timestamps. The inner white-light image is from C2 coronagraph and the outer image is from C3 coronagraph. The radio image is at 01:24:55 UTC. Contour levels are at  0.5, 1, 2, 4, 6, 8, 20, 40, 60, and 80 \% of the peak flux density. Radio emission marked by the cyan box is from CME-1, which is detected on the sky plane out to 5.2 $R_\odot$ shown by the red circle.}
    \label{fig:north_cme}
\end{figure}

Figure \ref{fig:north_cme} shows a sample Stokes I radio image at 80.62 MHz overlaid on the closest LASCO C2 and C3 base difference images. This work focuses on the radio emission from CME-I, marked by the cyan box. Other extended radio emissions seen in Figure \ref{fig:north_cme} arise from a different CME (southwest) and a streamer (southeast). A detailed study of the southwestern CME (CME-2) is presented in Chapter \ref{cme_gs2}.

\begin{figure}[!ht]
   \centering
    \includegraphics[trim={0.4cm 0.2cm 0.3cm 0.3cm},clip,scale=0.8]{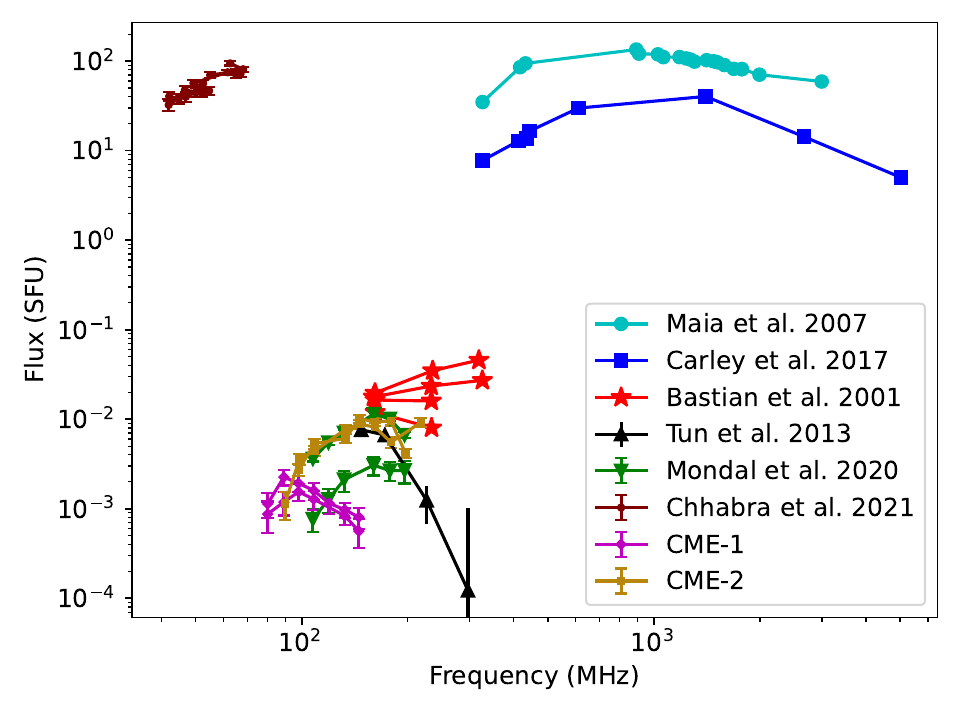}
    \caption[Comparison of gyrosynchrotron emission spectra from CME plasma of some of the previous works with the present works.]{Comparison of gyrosynchrotron emission spectra from CME plasma of some of the previous works with the present works. Magenta and golden yellow points represent sample spectra from CME-1 and CME-2, respectively, which are fainter compared to flux density observed in previous works.}
    \label{fig:past_works}
\end{figure}

\cite{Mondal2020a} (referred to as M20 hereafter) detected spatially resolved radio emission from CME plasma up to 4.73 $R_\odot$. At the time of publication, these detections were at the lowest flux densities and farthest solar distances. Two sample spectra from the CME-1 are shown in Figure \ref{fig:past_works} by magenta points. The flux density of the radio emission from CME-1 is comparable to the weakest flux density detected by M20. The radio emission is detected out to 5.2 $R_\odot$ (Figure \ref{fig:north_cme}), a bit beyond the maximum detection height reported by M20.

\begin{figure}[!htbp]
    \centering
    \includegraphics[trim={1.5cm 0cm 3cm 0cm},clip,scale=0.35]{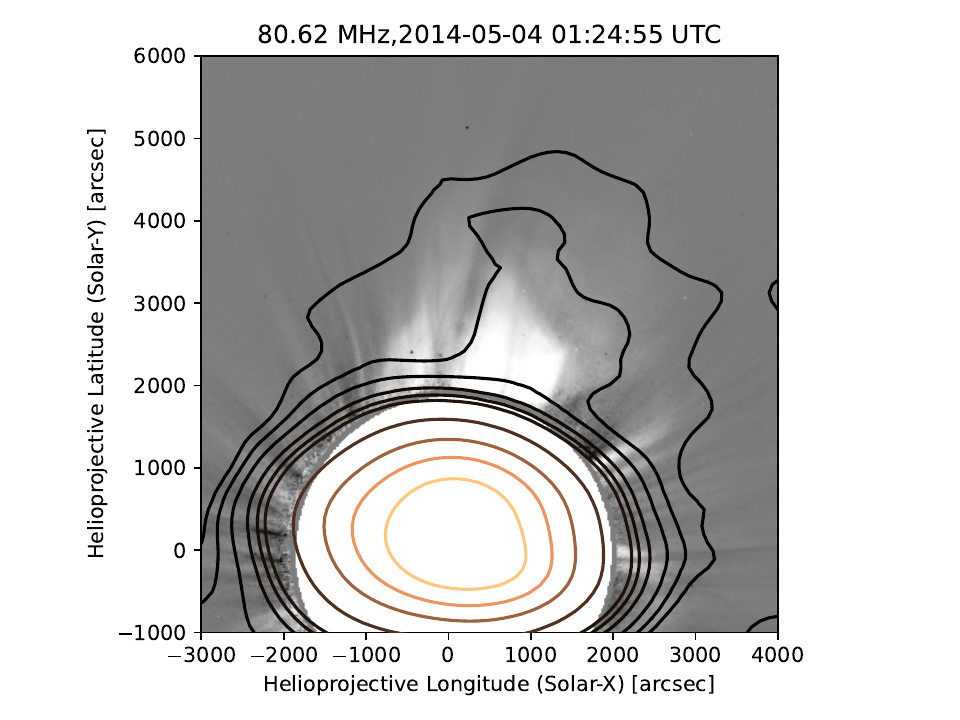}\includegraphics[trim={1.5cm 0cm 3cm 0cm},clip,scale=0.35]{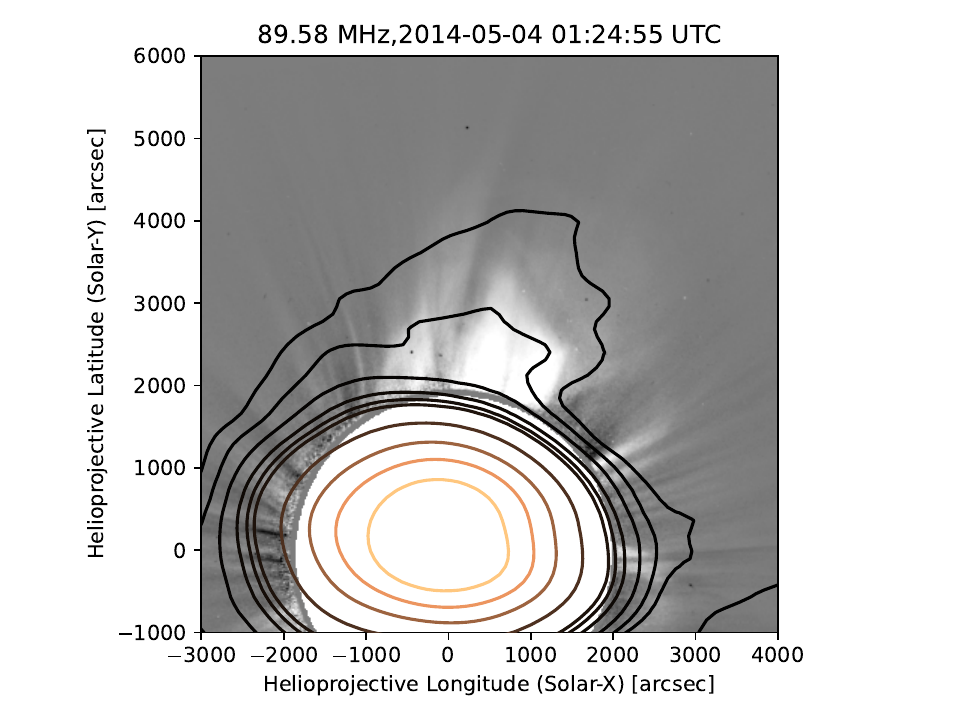}\includegraphics[trim={1.5cm 0cm 3cm 0cm},clip,scale=0.35]{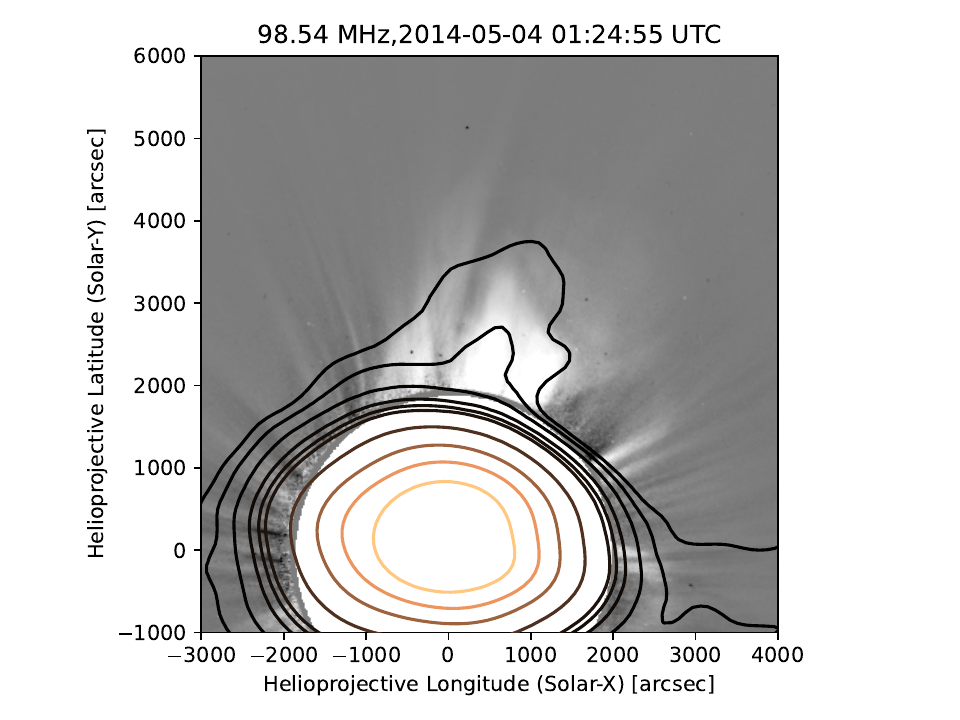}\\
    
    \includegraphics[trim={1.5cm 0cm 3cm 0cm},clip,scale=0.35]{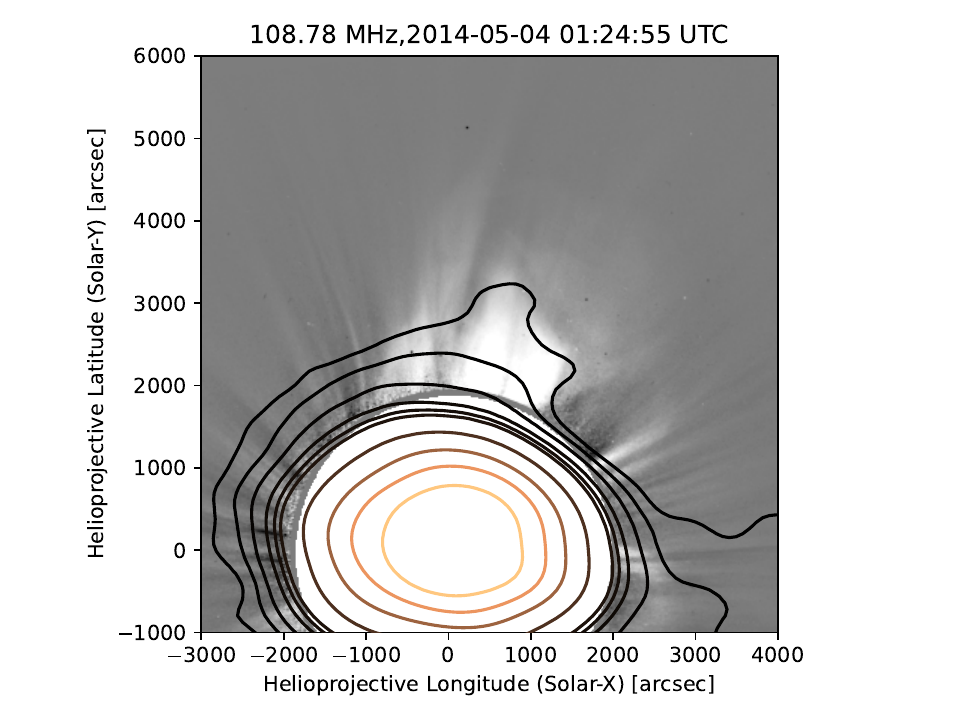}
    \includegraphics[trim={1.5cm 0cm 3cm 0cm},clip,scale=0.35]{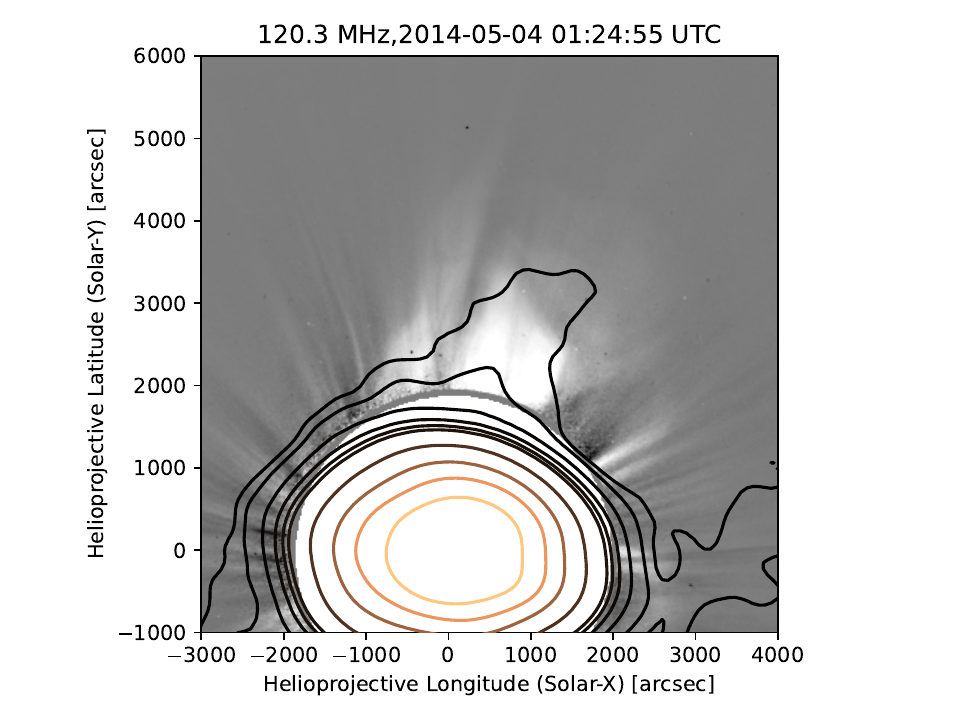}\includegraphics[trim={1.5cm 0cm 3cm 0cm},clip,scale=0.35]{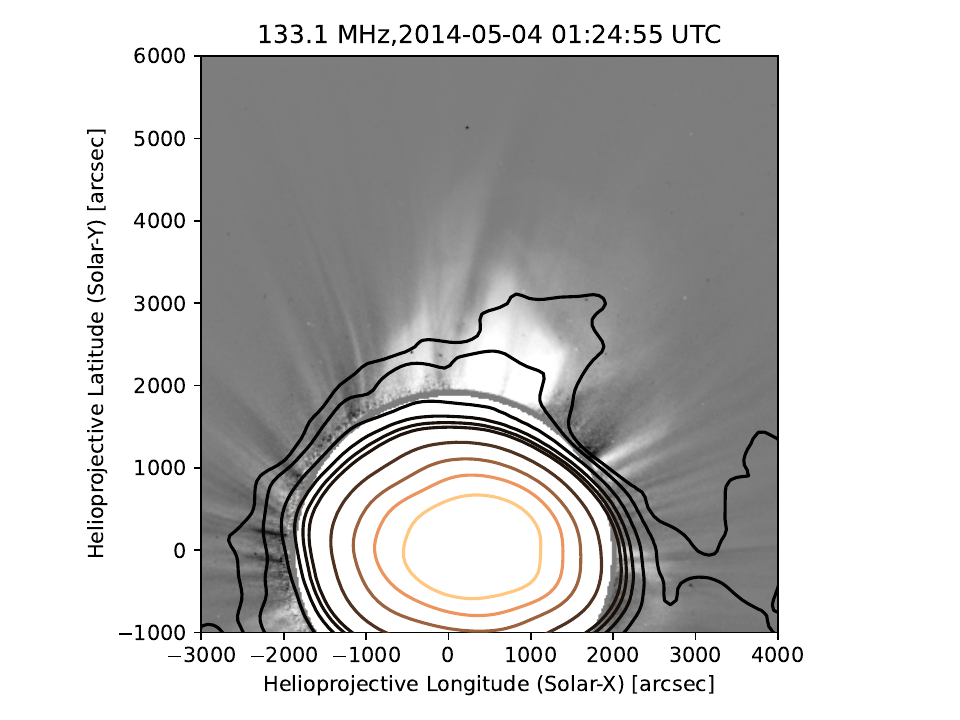}\\
    
    \includegraphics[trim={1.5cm 0cm 3cm 0cm},clip,scale=0.35]{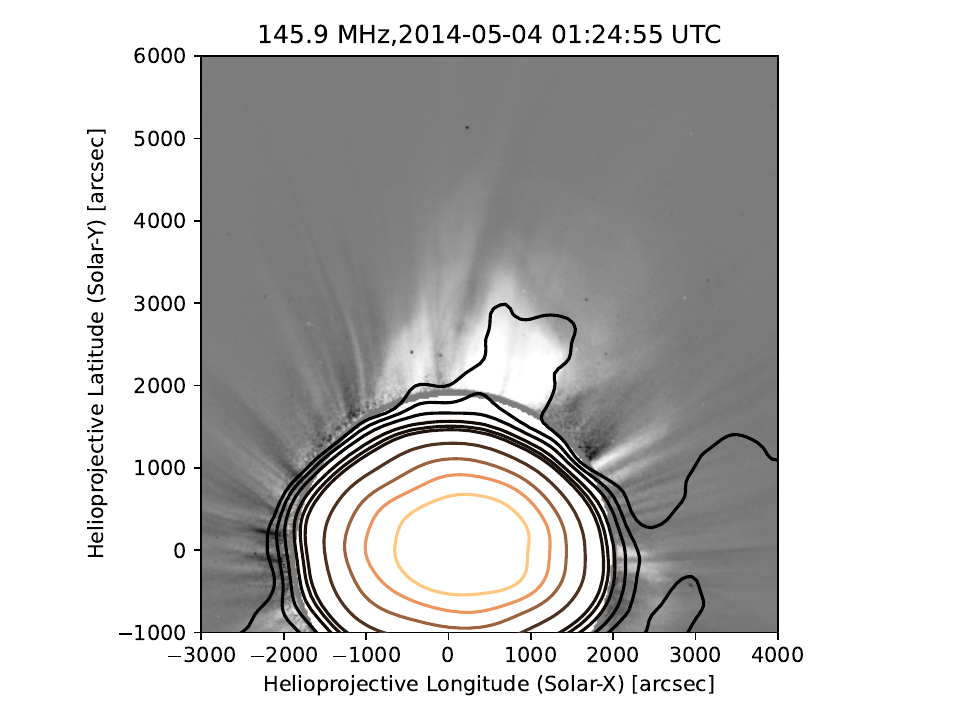}\includegraphics[trim={1.5cm 0cm 3cm 0cm},clip,scale=0.35]{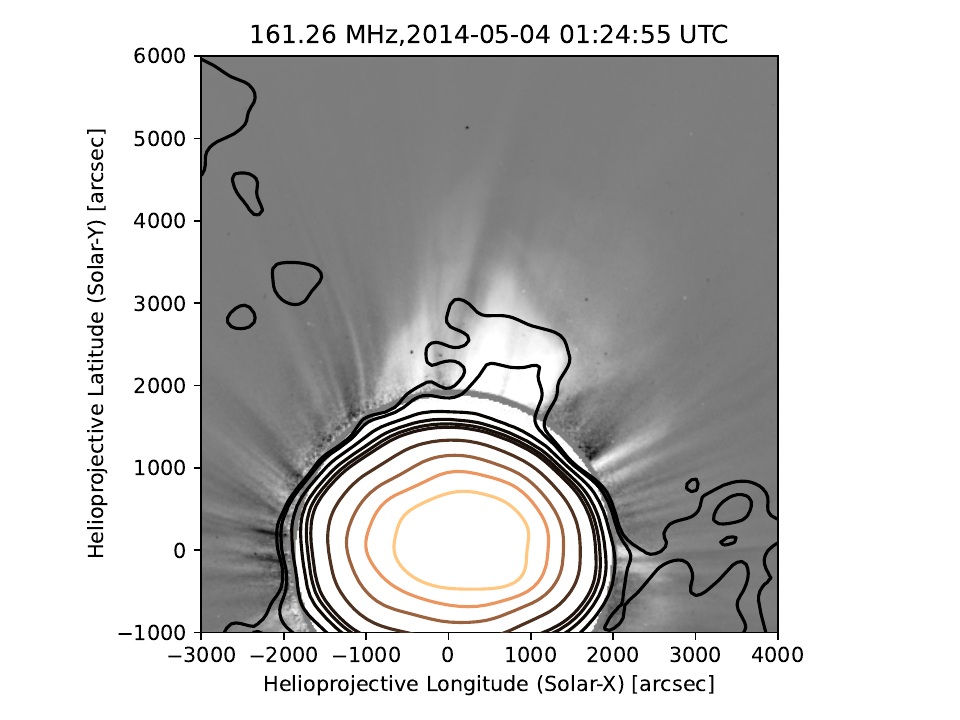}\includegraphics[trim={1.5cm 0cm 3cm 0cm},clip,scale=0.35]{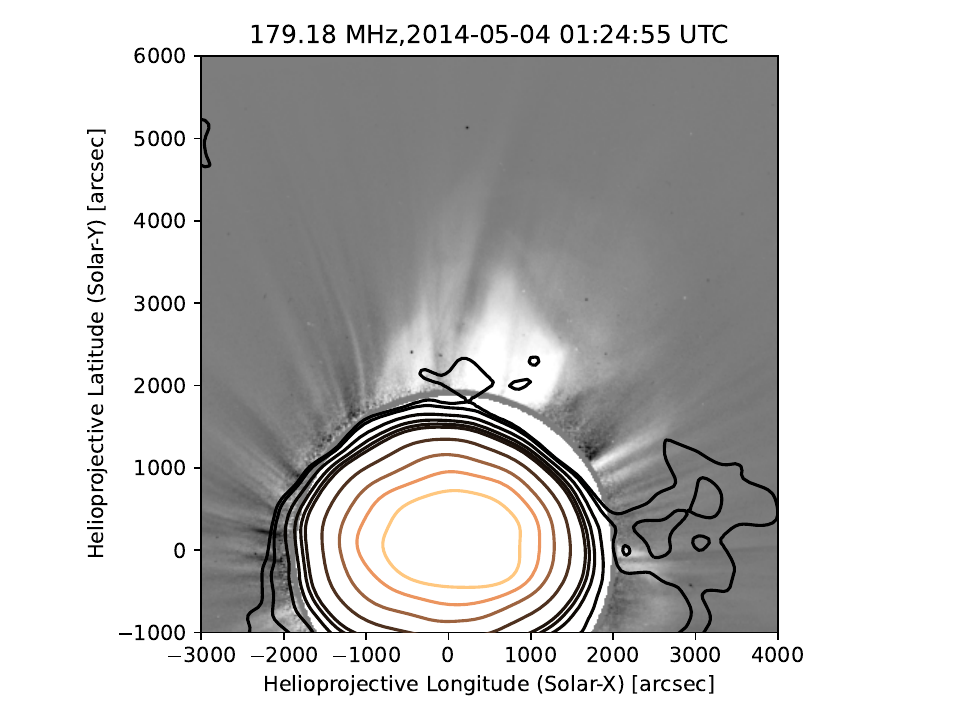}\\
    
    \includegraphics[trim={1.5cm 0cm 3cm 0cm},clip,scale=0.35]{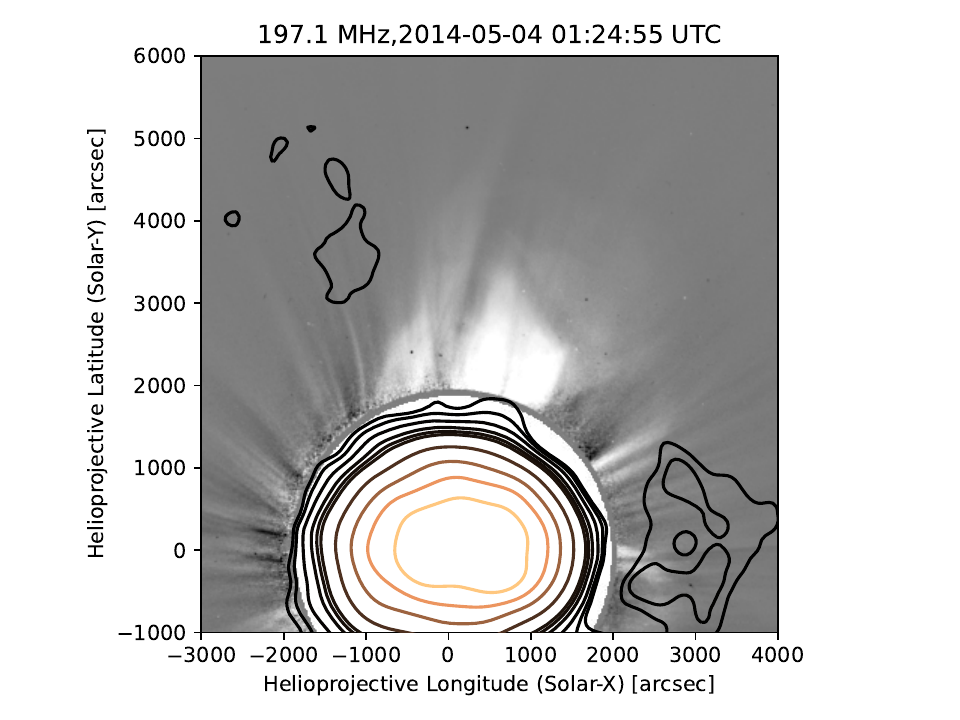}
    \includegraphics[trim={1.5cm 0cm 3cm 0cm},clip,scale=0.35]{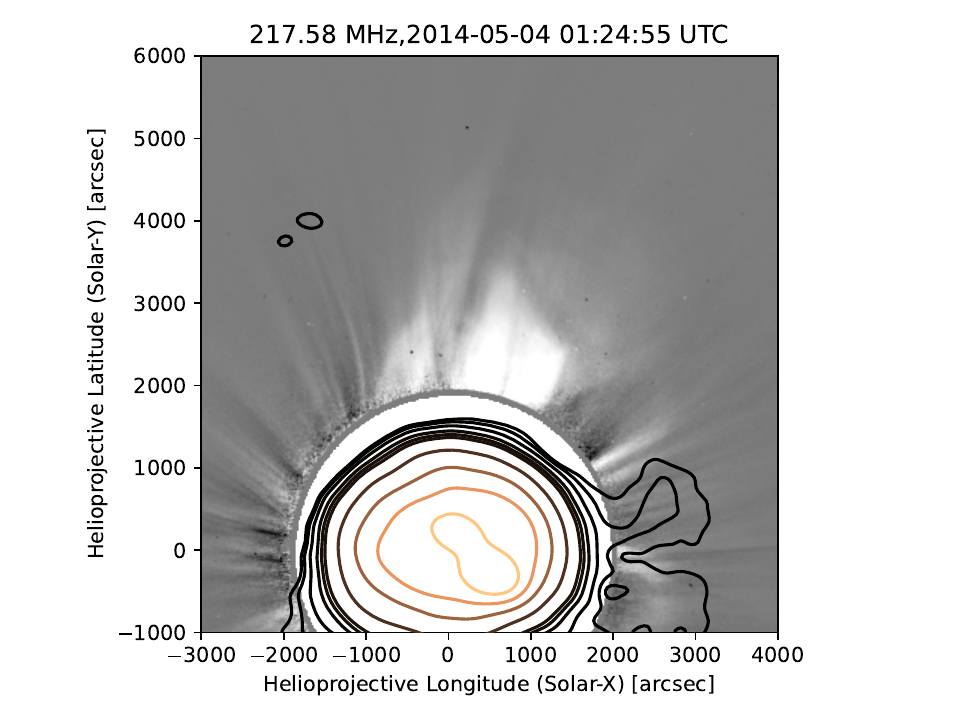}\includegraphics[trim={1.5cm 0cm 3cm 0cm},clip,scale=0.35]{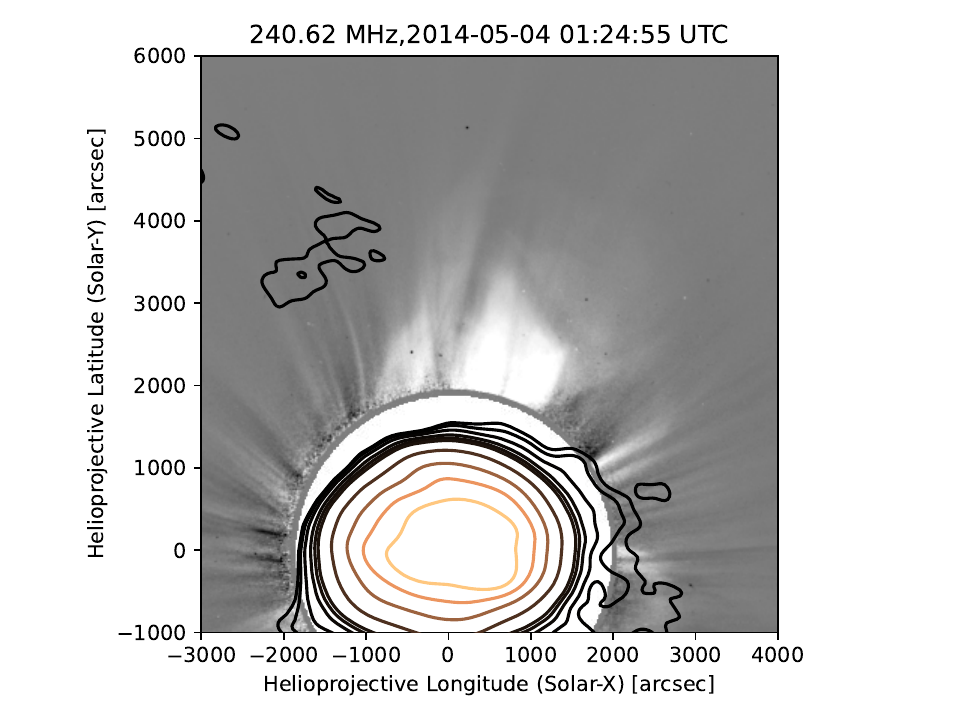}
    \caption[Spectral Stokes I radio images of the northern CME.]{Stokes I radio emission from CME-1 at multiple frequency bands of the MWA. Frequency increases from the top left panel of the image to the bottom right panel. Radio emission from CME-1 is detected up to 161 MHz. Contour levels are at 0.5, 1, 2, 4, 6, 8, 20, 40, 60, and 80 \% of the peak flux density.}
    \label{fig:c2_c3_comp_freq}
\end{figure}

Extended radio emissions are detected at multiple frequencies from the regions co-located with CME-1. The evolution of the radio emission from CME-1 with frequency for a single time slice centered at 01:24:55 UTC is shown in Figure \ref{fig:c2_c3_comp_freq}. Frequency increases from the top left to the bottom right of the figure. Radio emission from the CME-1 is detected upto 161 MHz with more than $5\sigma_\mathrm{I}$ significance, where $\sigma_\mathrm{I}$ is the Stokes I map rms in a region close to the Sun. It is also evident from this figure that the spatial extent of radio emission shrinks with increasing frequency. At the lowest frequency, 80 MHz, the radio emission extends across the entire white-light structure of the CME-1, while at 161 MHz the emission is present only over a small part of it. It is verified that this is not due to any DR limitation.

\subsection{Circularly Polarized Radio Emission from CME-1}\label{subsec:circular_pol}
\begin{figure*}[!ht]
    \centering
    \includegraphics[trim={1.2cm 0.3cm 2cm 0cm},clip,scale=0.9]{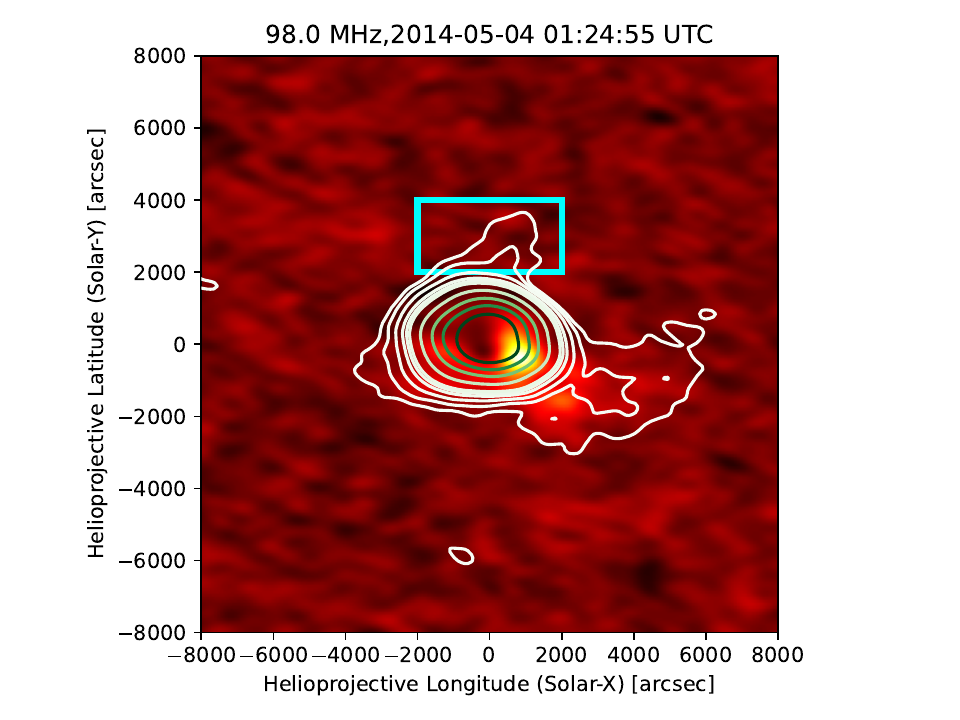}
    \caption[Circular polarization image of CME-1.]{Circular polarization image for CME-1. A sample image at 98 MHz is shown. The background image is Stokes V and Stokes I emission is shown by the contours. Contours at  0.5, 1, 2, 4, 6, 8, 20, 40, 60, 80 \% level of the peak flux density. No Stokes V emission is detected from the CME-1 marked by the cyan box. The background is noise-like and there are no systematic imaging artifacts.}
    \label{fig:circular_pol}
\end{figure*}

Most of the previous studies \citep{Bain2014,Carley2017,Mondal2020a} did not include polarization measurements. \citet{bastian2001} observed a low degree of circular polarization using NRH, but no quantitative information was reported. \citet{Tun2013} reported a high degree of circular polarization but did not quantify the instrumental polarization leakage. In this work, high-fidelity full Stokes images are made using P-AIRCARS. P-AIRCARS allows us to precisely correct all instrumental polarization effects. A quantitative estimation of residual leakage is described in detail in Section \ref{subsec:stress-test} and estimated residual leakage is $<0.1\%$ estimated for this same observation used in this chapter. The background color map shown in Figure \ref{fig:circular_pol} is a sample Stokes V image at 98 MHz and the contours represent the Stokes I emission. 

\subsubsection{Estimating Stringent Upper Limits of Stokes V Emission}\label{subsec:upperlimits_estimation}
Any radio polarization measurement has two primary contributions to its uncertainty -- a fundamental limit imposed by the thermal noise of the measurement and the other arising due to imperfections in correcting for instrumental leakage. Robust polarization calibration provided by P-AIRCARS ensures that the errors introduced due to uncorrected instrumental polarization leakage are extremely small (typically less than 0.1\% for Stokes I to Stokes V, as described in Section \ref{subsec:stress-test} of Chapter \ref{paircars_algorithm}). In addition, there can also be systematic artifacts in the image due to errors incurred during the deconvolution process which radio imaging relies upon \citep{Cornwell1999}. The dense array footprint of the MWA provides an extremely well-behaved point-spread-function (PSF), which is demonstrated in Section \ref{sec:suitability_of_MWA} of Chapter \ref{paircars_principle}. This reduces deconvolution errors to a level below those from other sources \citep{Mondal2019}. This is evident from the Stokes V map shown in Figure \ref{fig:circular_pol}, which clearly shows that the background is noise-like and no systematic artifacts are seen in the image. The measured rms in the Stokes V image ($\sigma_\mathrm{V}$) is only about 1.3 times the expected instrumental thermal noise, further attesting to the high-quality calibration and imaging. The rms values vary with frequency and are listed in Table \ref{table:stokesV_noise}. The Stokes V emission from CME-1 is too weak to be detected at any of the observing bands. Nonetheless, the noise-like nature of these images at the location of CME-1 and the low values of instrumental leakage enables us to place robust upper limits on the absolute value of the Stokes V emission \citep[e.g.,][etc.]{Bastian2000,Lynch2017,lenc2018,Cendes_2022} at each of the frequency bands as discussed further in Section \ref{sec:spectroscopy}.

\begin{table}[!ht]
\centering
    \renewcommand{\arraystretch}{1.4}
    \begin{tabular}{|p{1.8cm}|p{1.5cm}|p{1.8cm}|p{1.5cm}|}
    \hline
       Frequency (MHz) & $\sigma_\mathrm{V}$ (Jy) & Frequency (MHz) & $\sigma_\mathrm{V}$ (Jy)\\ \hline \hline 
        80 & 4.55 & 145 &  1.52\\
        \hline
        89 & 4.38 & 161 &  1.24\\
        \hline
        98 & 3.75 & 179 &  0.71\\
        \hline
        108 & 3.20 & 197 & 0.45\\
       \hline
       120 & 2.18 & 217 & 1.74\\
       \hline
       132 & 1.82 & 240 & 0.95\\
       \hline
    \end{tabular}
    \caption[Measured rms noise from the Stokes V maps at 12 spectral bands.]{Measured rms noise from the Stokes V maps at 12 spectral bands.}
    \label{table:stokesV_noise}
\end{table}

\subsection{Spatially Resolved Spectroscopy}\label{sec:spectroscopy}
\begin{figure*}[!ht]
    \centering
    \includegraphics[trim={1.2cm 0.3cm 2cm 0.2cm},clip,scale=0.9]{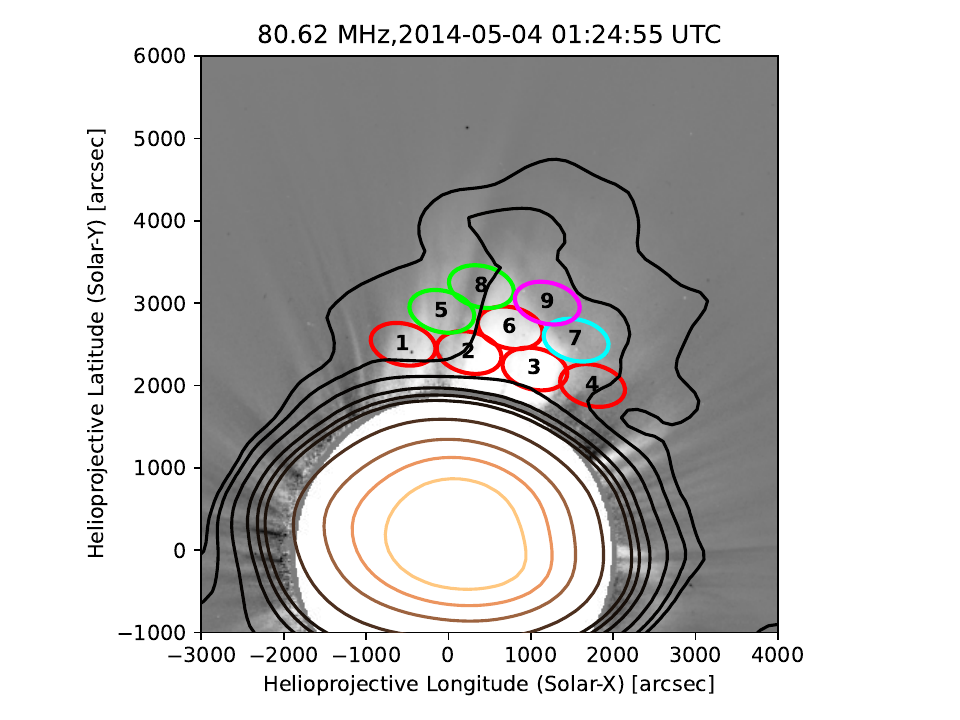}
    \caption[Regions of northern CME where spectra have been extracted.]{Regions where spectra have been extracted. Red regions are those where spectrum fitting is done and spectrum fitting is not done for green regions. Spectrum fitting is also done for region 7 marked by cyan keeping some parameters fixed. Region 9 marked by magenta only has a single spectral point.}
    \label{fig:spectral_regions}
\end{figure*}

Wideband imaging observations allow us to perform spatially resolved spectroscopy of the radio emission from CME-1. Spectra from regions with a size equal to the size of the PSF at the lowest observing frequency of 80 MHz is extracted. These regions are shown in Figure \ref{fig:spectral_regions} and have been chosen to ensure that the Stokes I emission is seen at 0.5\% level or more in at least two spectral bands.

I have calculated rms noise ($\sigma$) and mean ($\mu$) over a comparatively large region close to the Sun. I have also calculated the deepest negative ($n$) over a region close to the CME, and rms noise ($\alpha$) far away from the Sun. The flux density ($f$) for a region at a given frequency is considered to be a reliable detection, if all of the following three criteria are satisfied:
\begin{enumerate}
    \item $f>\mu+5\sigma$
    \item $f>5\alpha$
    \item $f>5|n|$
\end{enumerate}
These stringent selection criteria ensure that any spectral point prone to imaging artifacts is not included. 

\begin{figure}[!ht]
   \centering
    \includegraphics[trim={0.2cm 0.5cm 0cm 0cm},clip,scale=0.7]{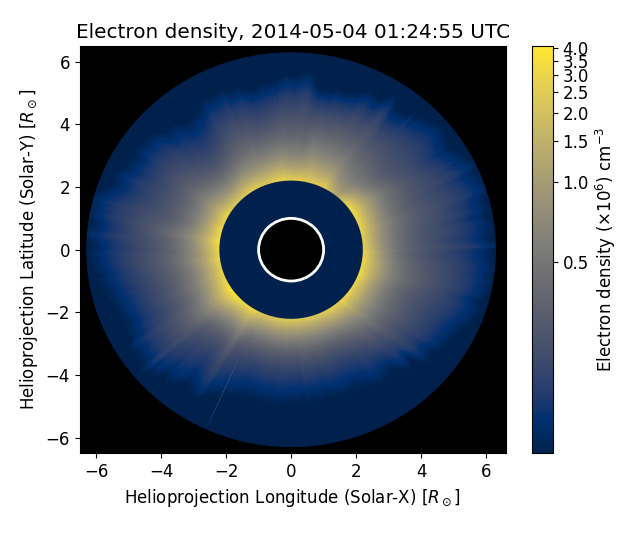}
    \caption[Total coronal electron density estimated from white-light coronagraph images.]{Total coronal electron density at 01:24:55 UTC obtained from LASCO-C2 white-light image. Electron density is estimated using the inversion method developed by \cite{Hayes2001}.}
    \label{fig:electron_density}
\end{figure}

The uncertainty of the Stokes I flux density, $\sigma_\mathrm{I}$, is estimated as,
\begin{equation}
    \sigma_\mathrm{I}=\mathrm{max}(\mu,\sigma).
\end{equation} 
The uncertainty of Stokes V is also estimated similarly. For Stokes V image, $\mu_\mathrm{V}$ is close to zero, and $\alpha$ is comparable to $\sigma$. Hence, I only consider the rms noise calculated from the Stokes V image close to the Sun as the uncertainty on Stokes V, $\sigma_\mathrm{V}$. As there is no Stokes V detection and Stokes V can not be more than Stokes I, $V_\mathrm{u}=\mathrm{min}(5\sigma_\mathrm{V},I)$ is used as the upper limit on the absolute value of Stokes V for each of the frequency bands. 

Spectra are fitted for the red regions which have Stokes I detections at more than five spectral bands (Figure \ref{fig:spectral_regions}). For these regions at least five GS model parameters are fitted as discussed in Section \ref{sec:spectrum_modeling}. Region 7 marked by cyan in the same figure has a peak in the spectrum but is detected only at four spectral bands. Hence spectral fitting for region 7 is performed holding some additional GS model parameters constant. For the regions marked in green, the magnetic field strength is estimated as discussed in Section \ref{subsec:magnetic_green}. Emission from Region 9 is detected at two spectral bands, but one of them falls short of meeting all of the selection criteria.

\subsection{Emission Mechanism}\label{subsec:emission_mechanism}
Possible mechanisms for explaining radio emissions from CMEs are -- plasma emission, free-free emission, and GS emission. All of these mechanisms have a dependence on the local plasma density. 
\begin{figure*}[!ht]
    \centering
    \includegraphics[scale=0.6]{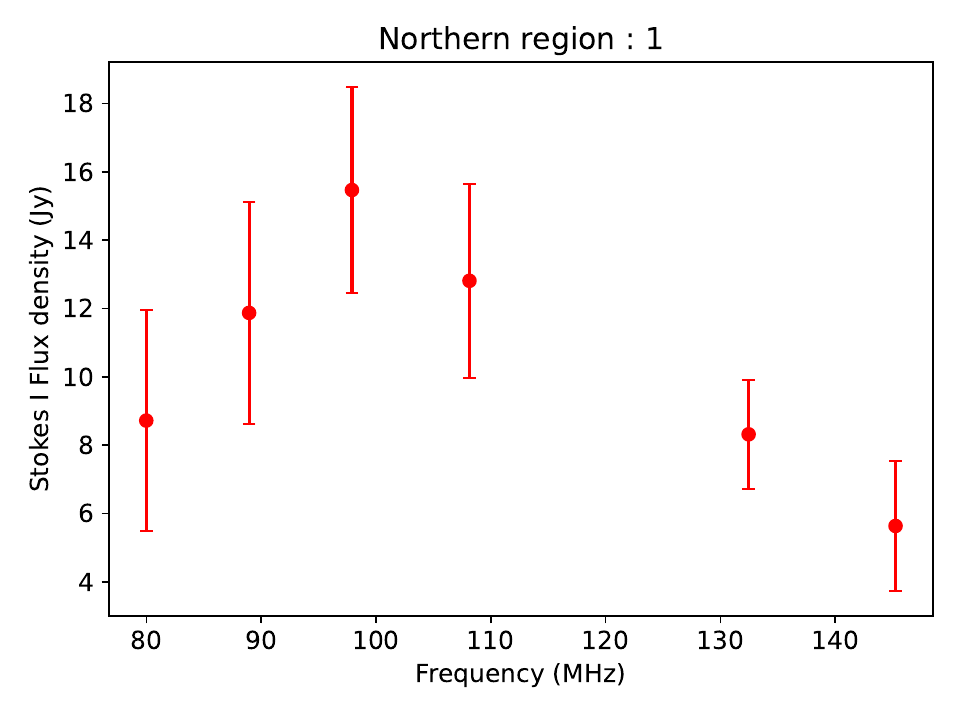}\\
    \includegraphics[scale=0.6]{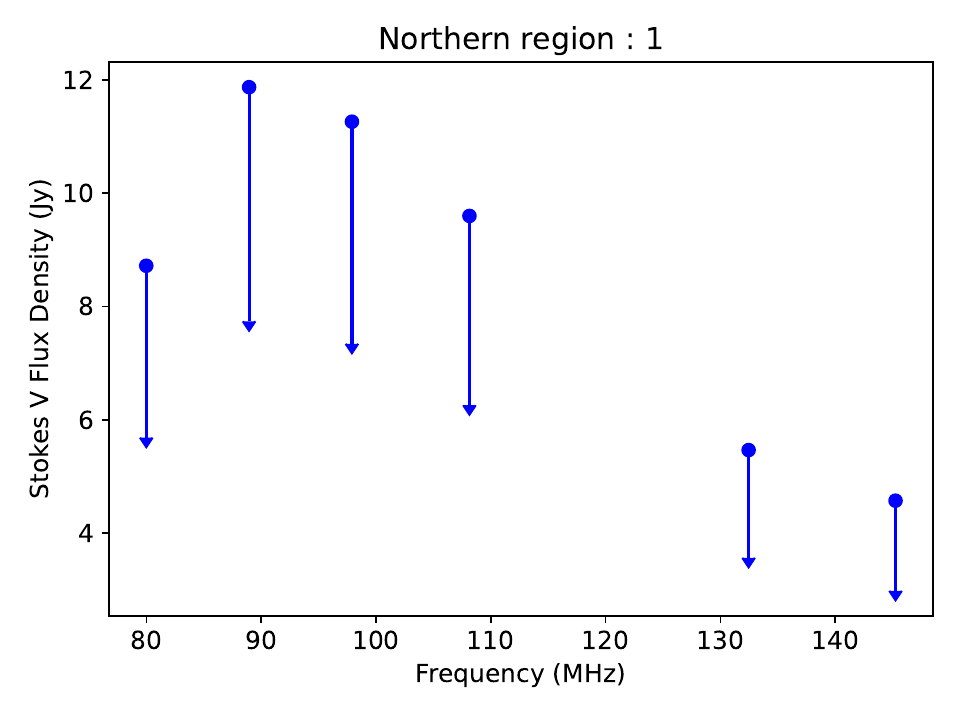}
    \caption[Sample observed Stokes I and V spectra for northern CME.]{Sample observed Stokes I and V spectra for northern CME. {\it Top panel: }Observed Stokes I spectrum for northern region 1. {\it Bottom panel: }Observed Stokes V spectrum for northern region 1.}
    \label{fig:observed_spectra}
\end{figure*}
Coronal electron density is estimated from the LASCO-C2 white light coronagraph images using the inversion method developed by \cite{Hayes2001}. The coronal electron density map is shown in Figure \ref{fig:electron_density}. Average coronal electron density over the region of CME-1 is about $10^6\ \mathrm{cm^{-3}}$ and leads to a corresponding plasma frequency of about $8.5\ \mathrm{MHz}$. However, the radio emission from CME-1 is detected at more than an order of magnitude higher frequency. The observing frequency is much larger than the local plasma frequency convincingly rules out plasma emissions as the possible mechanism. 

The next possibility examined is free-free emission. Considering the coronal plasma temperature $T_\mathrm{e}\approx10^6\ \mathrm{K}$, and neglecting magnetic fields, free-free optical depth is given by \citep{Gary1994},
\begin{equation}
    \tau_\nu\approx0.2\ \frac{\int n_e^2\ dl}{\nu^2\ T_e^{\frac{3}{2}}}.
    \label{eq:free_free_optical_depth}
\end{equation}
Average coronal electron density; $<n>\ \approx 10^6\ \mathrm{cm^{-3}}$ is determined from white-light images. The electron density drops rapidly with increasing solar offset, dropping by more than an order of magnitude between 2 and 5 $R_{\odot}$ \citep{Hayes2001,Patoul_2015}. Hence, we ignore the contributions to $n$ from beyond 5 $R_{\odot}$. For a LoS with a PoS distance of 2 $R_{\odot}$, this leads to a LoS depth of about 9 $R_{\odot}$ within a sphere of 5 $R_{\odot}$. Considering these average values, $\tau_\nu$ becomes unity at $\nu\approx$ 33 MHz. Since the frequency of observation is many times higher than this value, optically thick free-free emission is also ruled out. For optically thin free-free emission, the brightness temperature ($T_\mathrm{B}$) is proportional to $\nu^{-2}$, which implies a flat flux density spectrum. Assuming optically thin free-free emission, $T_\mathrm{B}$ can be written as \citep{Gopalswamy1992},
\begin{equation}
    T_\mathrm{B}=\frac{<n>^2L}{5\ T_e^{0.5}\ \nu^2}
    \label{eq:freefree_Tb}
\end{equation}
Estimated $T_\mathrm{B}$ is $\sim1390$ K. The rms of the image at 100 MHz is $\sim$1100 K. Hence, the contribution from optically thin free-free emission from coronal plasma is below our detection limit. As the observed spectra (a sample spectrum is shown in Figure \ref{fig:observed_spectra}) have well-defined peaks and $T_\mathrm{B}$ is less than our detection limit, optically thin free-free emission can also be ruled out. Hence, the only likely emission mechanism remaining is the GS emission.

\section{Gyrosynchrotron Emission : Parameter Sensitivity}\label{sec:spectrum_sensitivity}
Mildly relativistic electrons gyrating in a magnetic field emit GS emission. GS emission mechanism is well understood theoretically \citep{Melrose1968,Ramaty1969}. However, the exact expressions are computationally very expensive. Over the last decade or so fast GS codes have been developed \citep{Fleishman_2010,Kuznetsov_2021}. These codes are versatile and can produce GS spectra for any given distribution of energy and pitch angles of non-thermal electrons. \cite{Fleishman_2010} quantified the differences between the spectra using exact and approximate expressions and \cite{Kuznetsov_2021} quantified the effects of pitch angle distributions. Building on this and benefiting from the significantly reduced computation time, here I explore the phase space of GS model parameters.

The electron distribution can be described by the expression,
\begin{equation}
    f(E,\mu)=u(E)\ g(\mu),
\end{equation}
where $u(E)$ is electron energy distribution function, $g(\mu)$ is the electron pitch angle distribution function, $E$ is the energy of the electron, $\mu=cos\alpha$, $\alpha$ being the electron pitch angle. The normalization conditions for $u(E)$ and $g(\mu)$ are,
\begin{equation}
    \int_{E_\mathrm{min}}^{E_\mathrm{max}} u(E)\ dE=\frac{n_e}{2\pi};\\
    \int_{-1}^{+1} g(\mu)\ d\mu=1
\end{equation}
where $n_e$ is the non-thermal electron density. For simplicity, an isotropic distribution of electron pitch angle is considered. The simplest form of $u(E)$ is considered described by a single power-law (PLW) as, 
\begin{equation}
    u(E) = NE^{-\delta},\ \mathrm{for}\ E_\mathrm{min}<E<E_\mathrm{max},
    \label{eq:PLW}
\end{equation}
where $E_\mathrm{min}$ is the minimum and $E_\mathrm{max}$ is the maximum energy cutoff and $N$ is a normalization constant. It is also assumed that GS model parameters are following a homogeneous distribution along the relevant part of the LoS.

\begin{figure*}[!ht]
    \centering
    \includegraphics[trim={0.5cm 0.5cm 0.5cm 0.5cm},clip,scale=0.4]{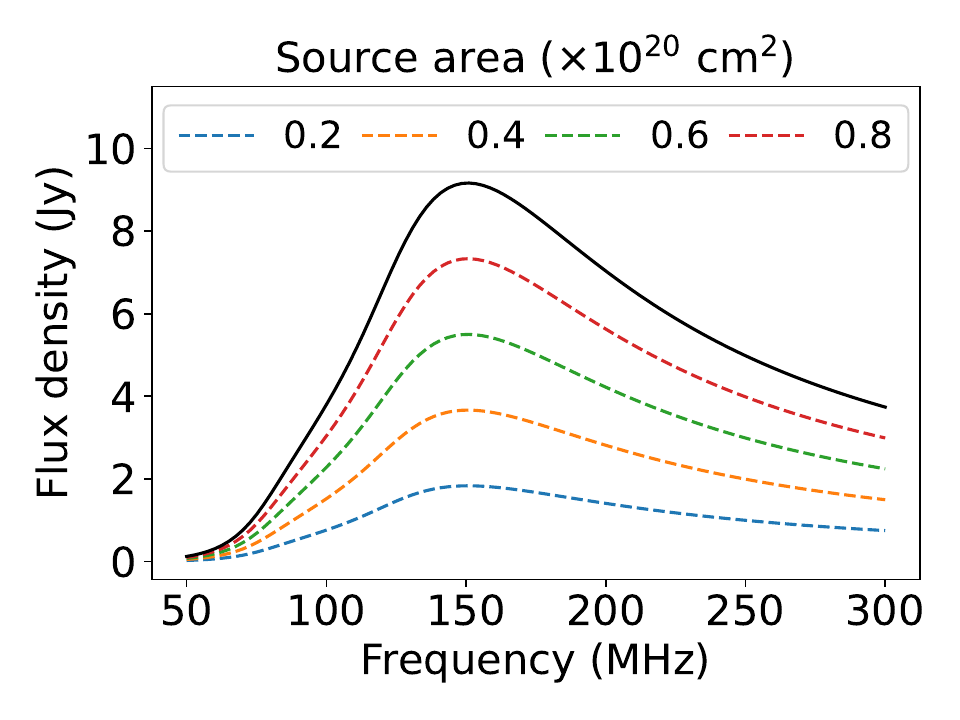}\includegraphics[trim={0.5cm 0.5cm 0.5cm 0.5cm},clip,scale=0.4]{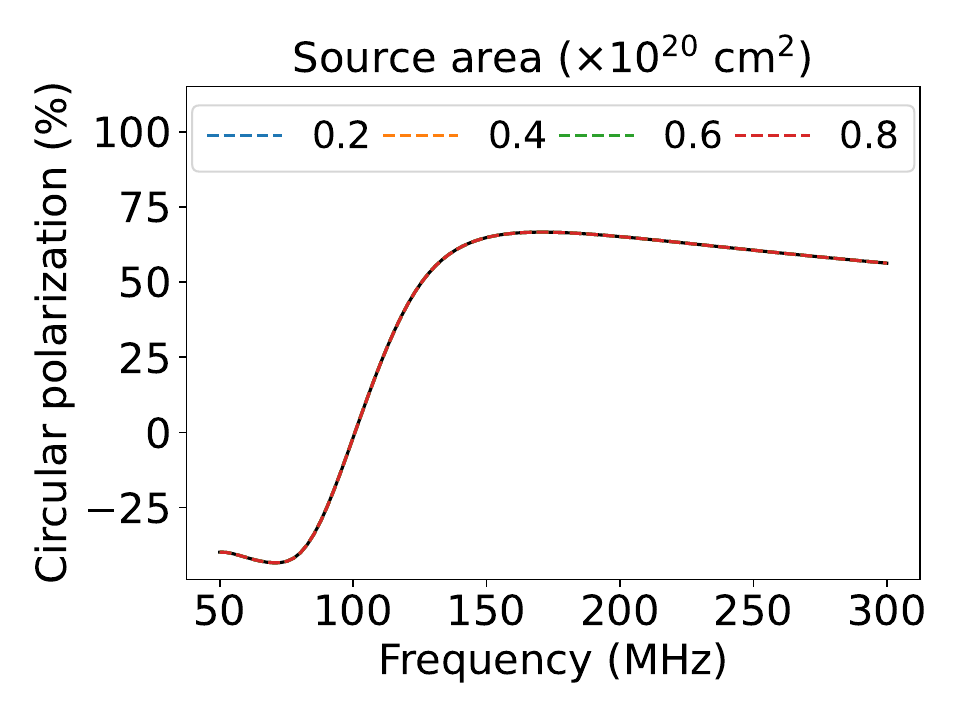}\\
    \includegraphics[trim={0.5cm 0.5cm 0.5cm 0.5cm},clip,scale=0.4]{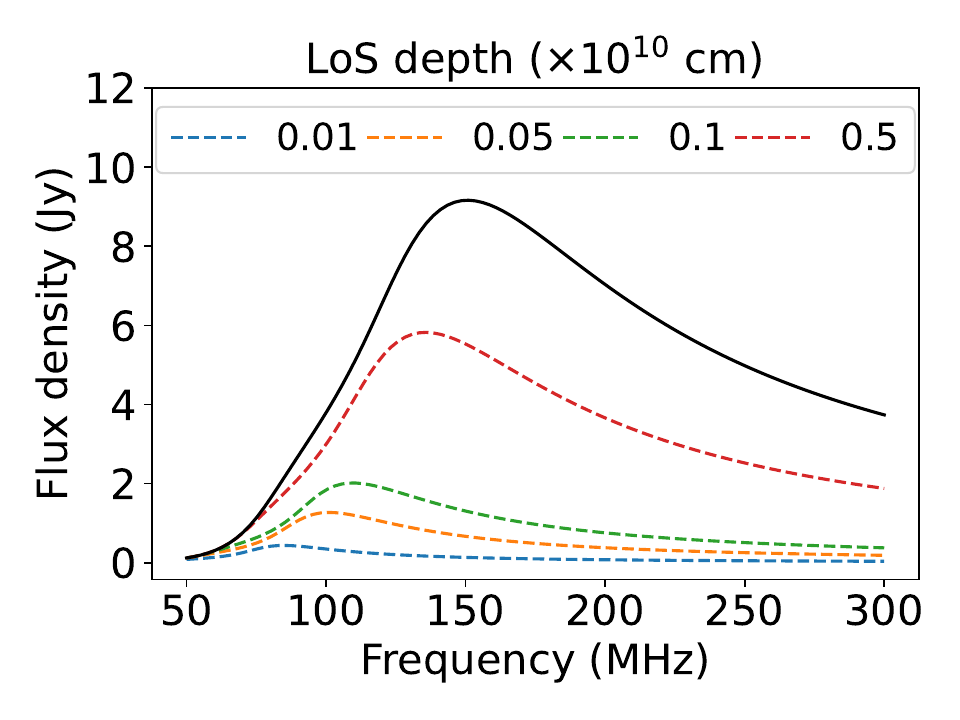}\includegraphics[trim={0.5cm 0.5cm 0.5cm 0.5cm},clip,scale=0.4]{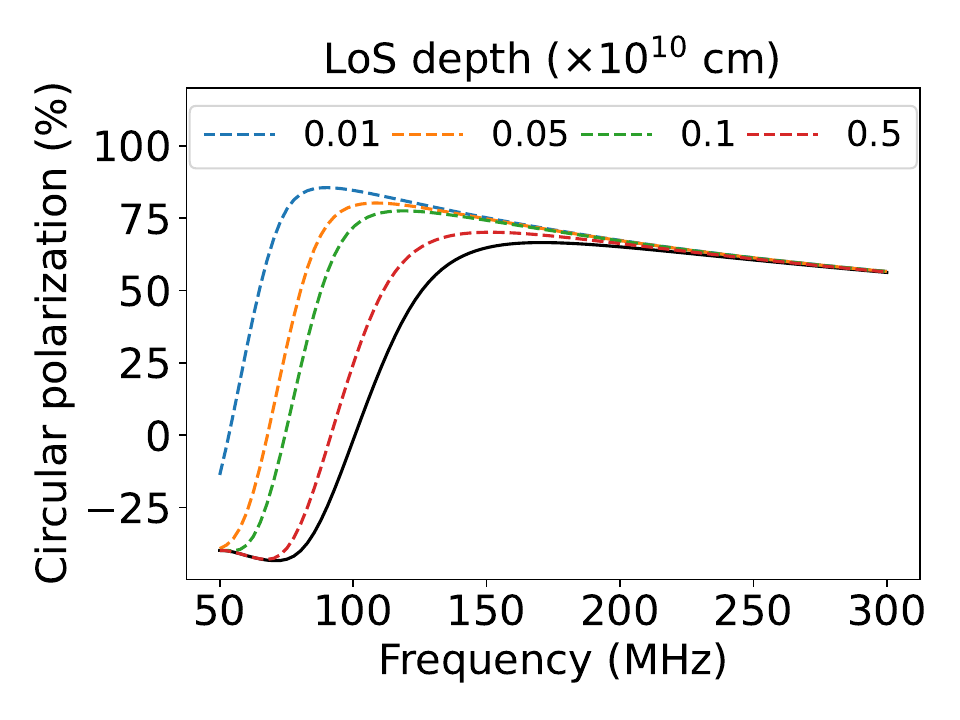}\\
    \caption[Sensitivity of the Stokes I and fractional Stokes V spectra on the geometric parameters of GS models.]{Sensitivity of the Stokes I and fractional Stokes V spectra on the geometric parameters of GS models. First column shows the Stokes I spectra and the second column shows spectra for the Stokes V fraction for two geometrical parameters of the GS model -- source area (top row) and LoS depth (bottom row). The black solid line in different panels represents the GS spectra for the reference parameters.}
    \label{fig:param_sensitivity_geometric}
\end{figure*}

\begin{figure*}[!ht]
    \centering
    \includegraphics[trim={0.5cm 0.5cm 0.5cm 0.5cm},clip,scale=0.4]{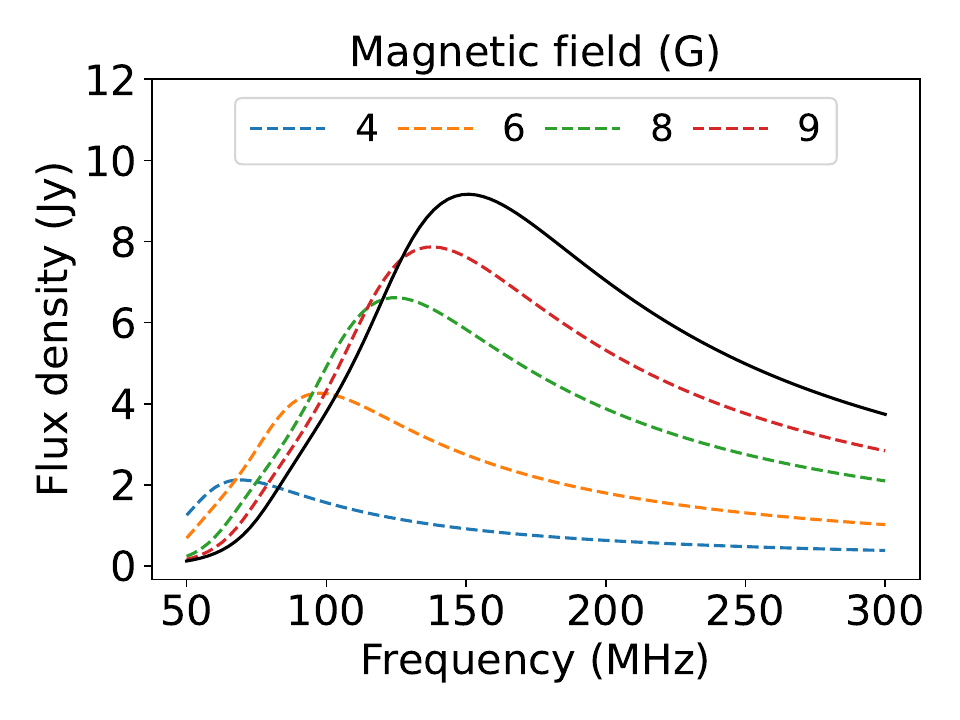}\includegraphics[trim={0.5cm 0.5cm 0.5cm 0.5cm},clip,scale=0.4]{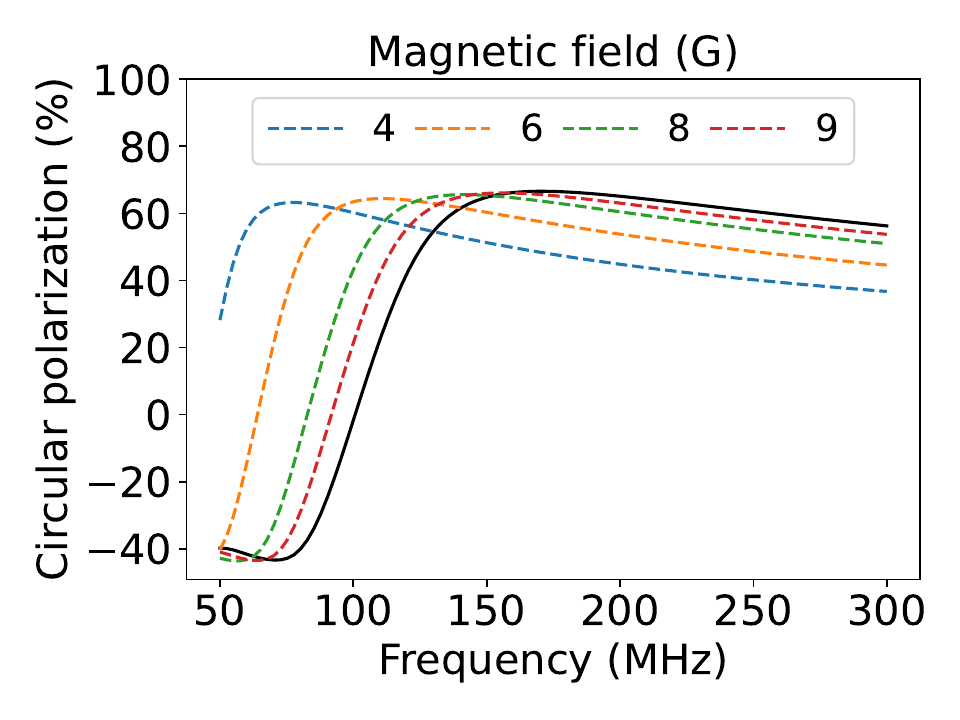}\\
    \includegraphics[trim={0.5cm 0.5cm 0.5cm 0.5cm},clip,scale=0.4]{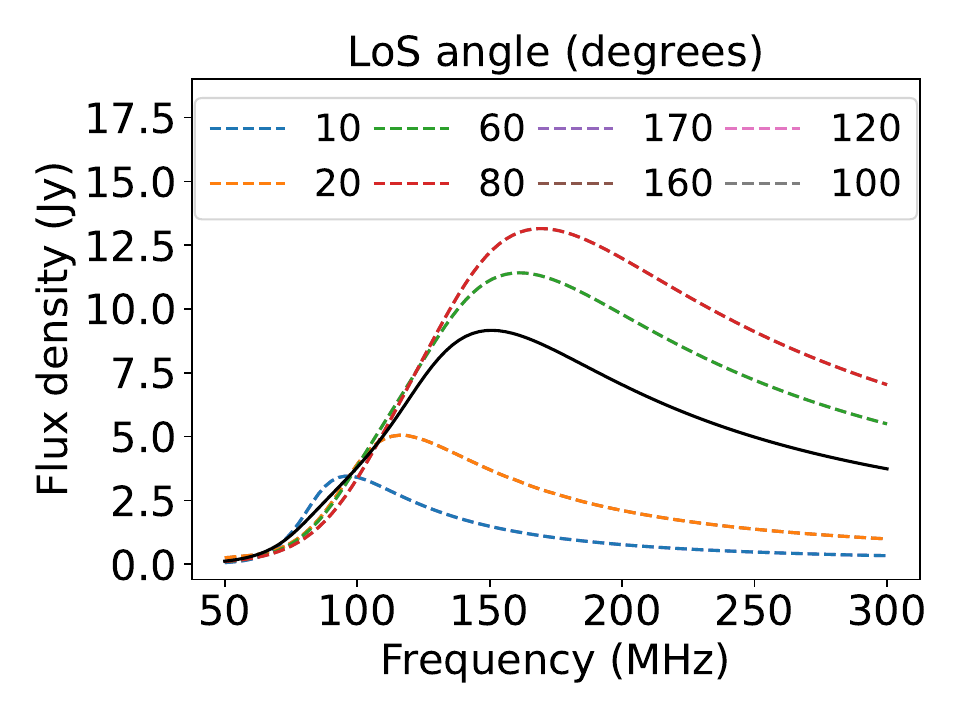}\includegraphics[trim={0.5cm 0.5cm 0.5cm 0.5cm},clip,scale=0.4]{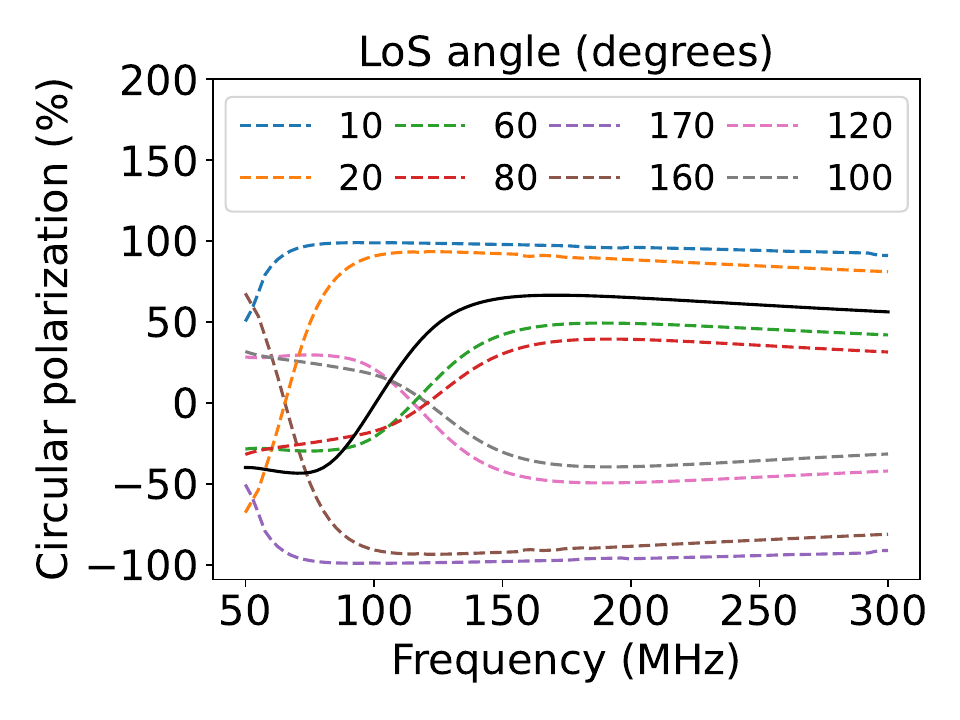}
    \caption[Sensitivity of the Stokes I and fractional Stokes V spectra on the magnetic field parameters of GS models.]{Sensitivity of the Stokes I and fractional Stokes V spectra on the magnetic field parameters of GS models. First column shows the Stokes I spectra and the second column shows spectra for the Stokes V fraction for different magnetic field parameters of the GS model -- magnetic field strength (top row) and the angle between the LoS and the magnetic field (bottom row). The black solid line in different panels represents the GS spectrum for the reference parameters.}
    \label{fig:param_sensitivity_magnetic}
\end{figure*}

\begin{figure*}[!htbp]
    \centering
   \includegraphics[trim={0.5cm 0.5cm 0.5cm 0.5cm},clip,scale=0.38]{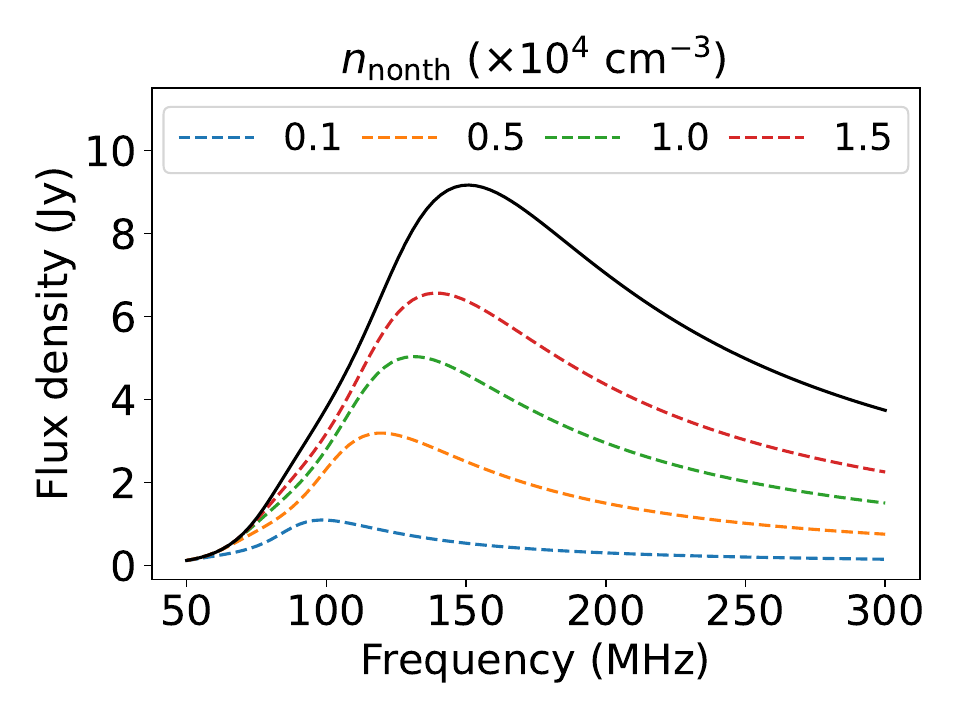}\includegraphics[trim={0.5cm 0.5cm 0.5cm 0.5cm},clip,scale=0.38]{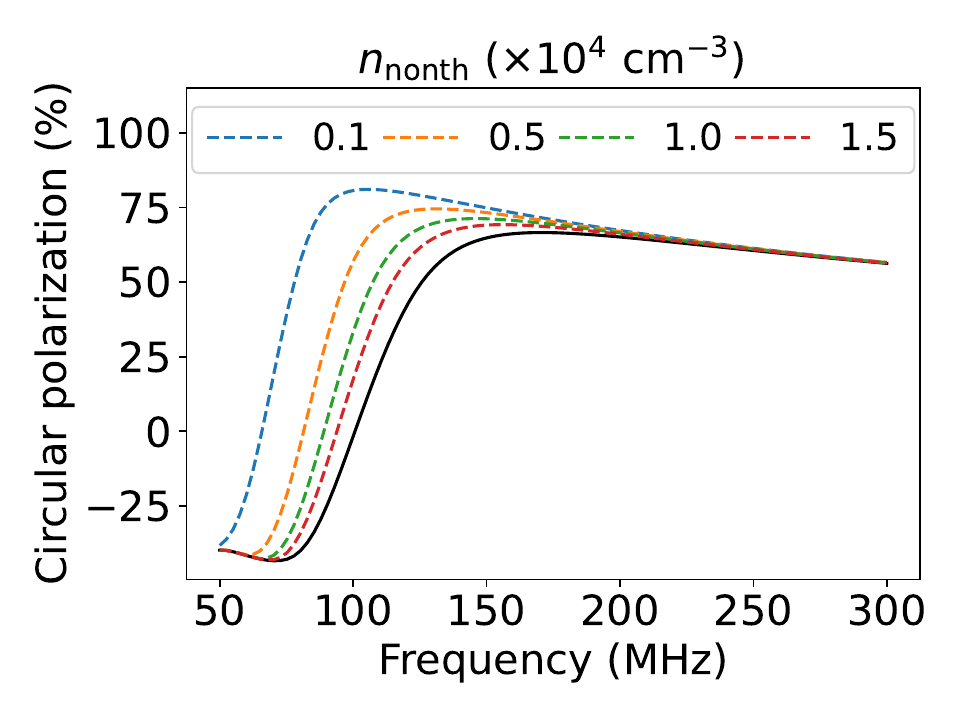}\\
   \includegraphics[trim={0.5cm 0.5cm 0.5cm 0.5cm},clip,scale=0.38]{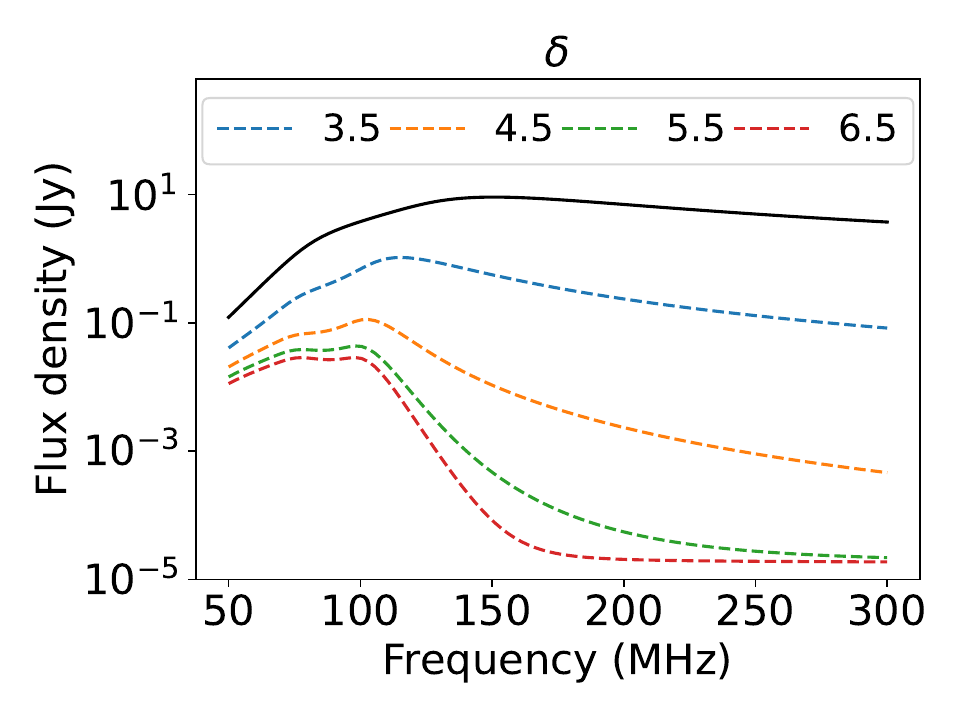}\includegraphics[trim={0.5cm 0.5cm 0.5cm 0.5cm},clip,scale=0.38]{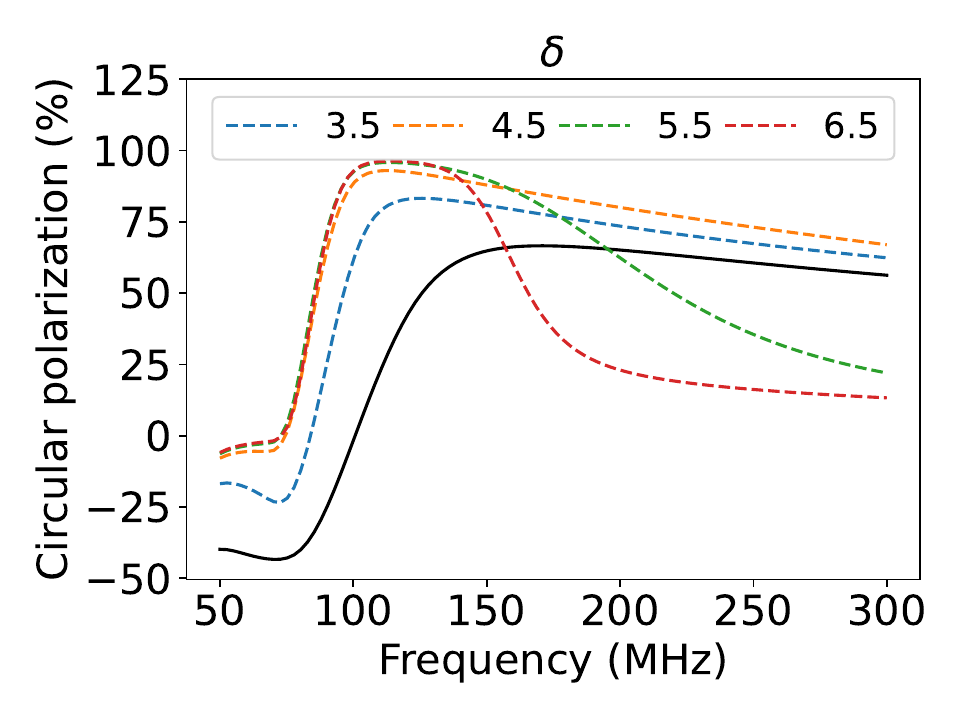}\\
   \includegraphics[trim={0.5cm 0.5cm 0.5cm 0.65cm},clip,scale=0.38]{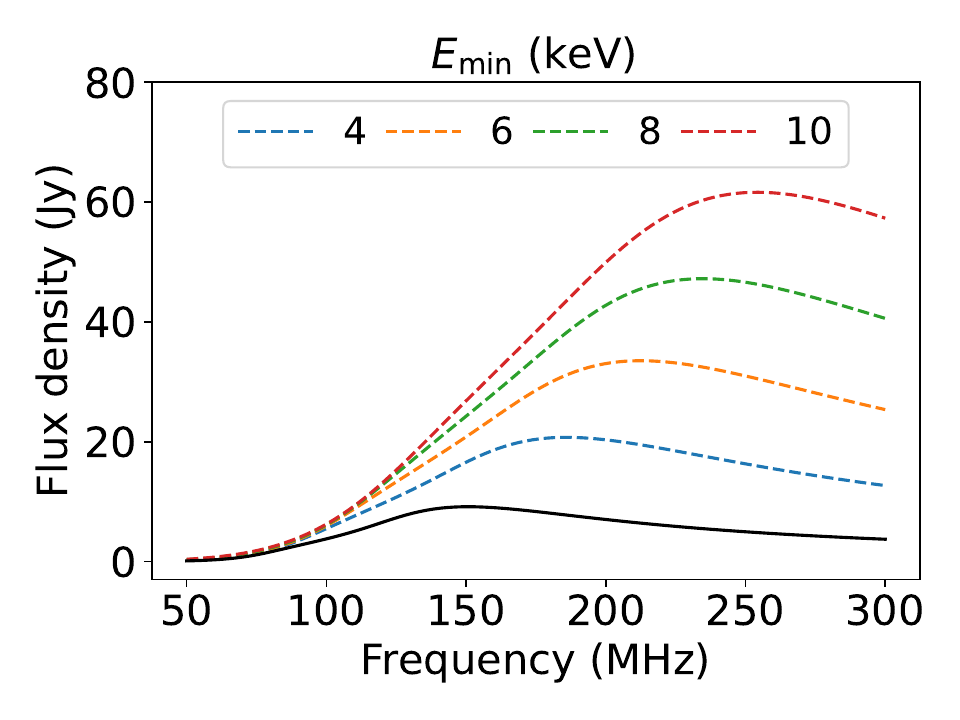}\includegraphics[trim={0.5cm 0.5cm 0.5cm 0.6cm},clip,scale=0.38]{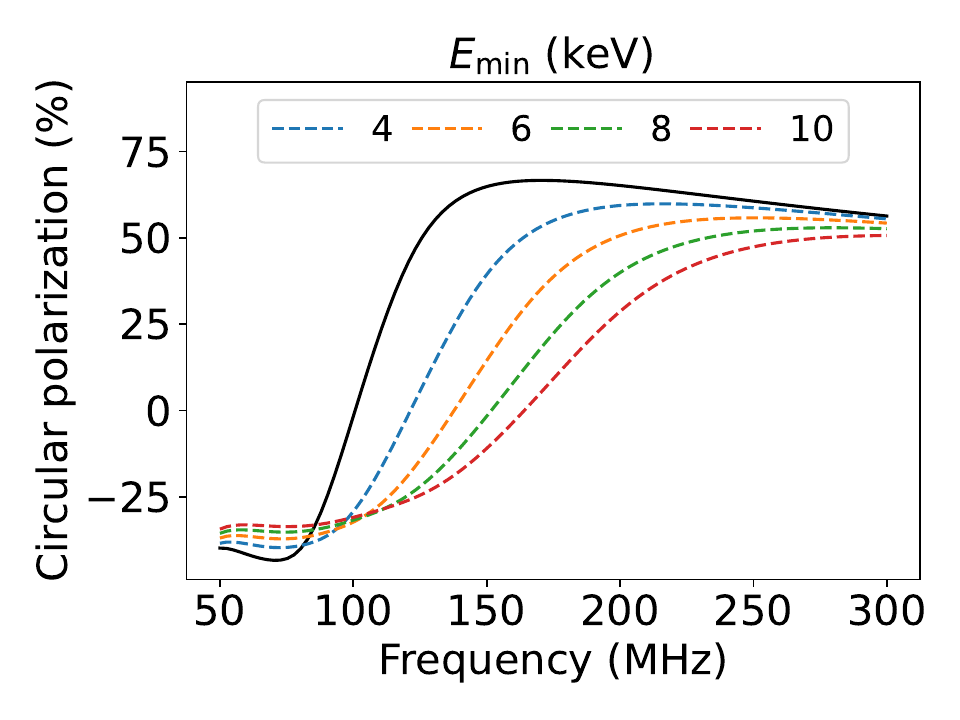}\\
   \includegraphics[trim={0.5cm 0.5cm 0.5cm 0.6cm},clip,scale=0.38]{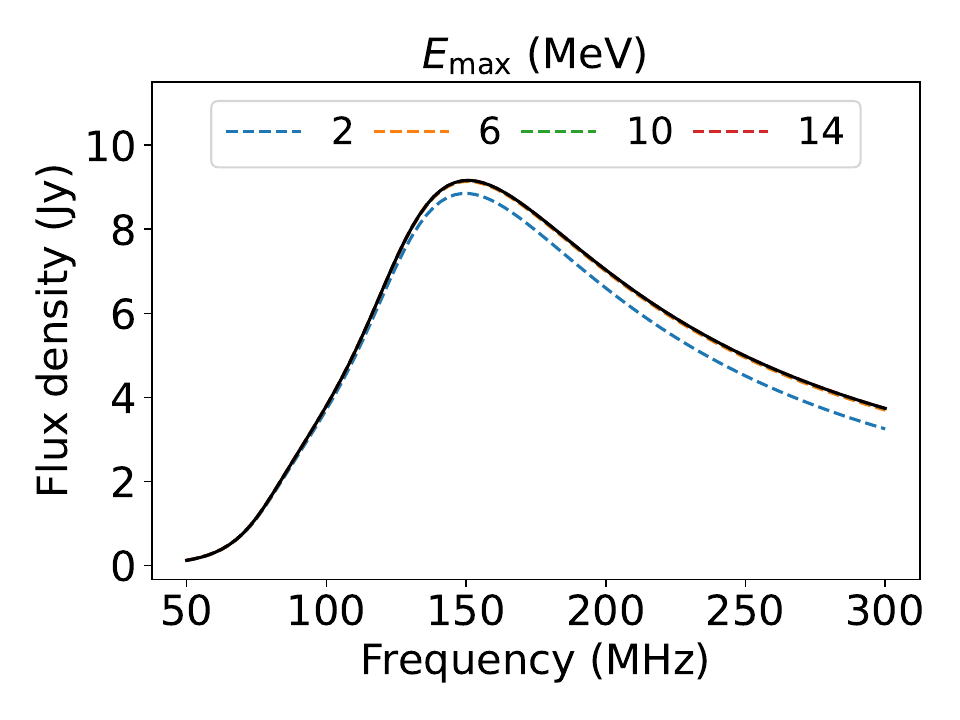}\includegraphics[trim={0.5cm 0.5cm 0.5cm 0.6cm},clip,scale=0.38]{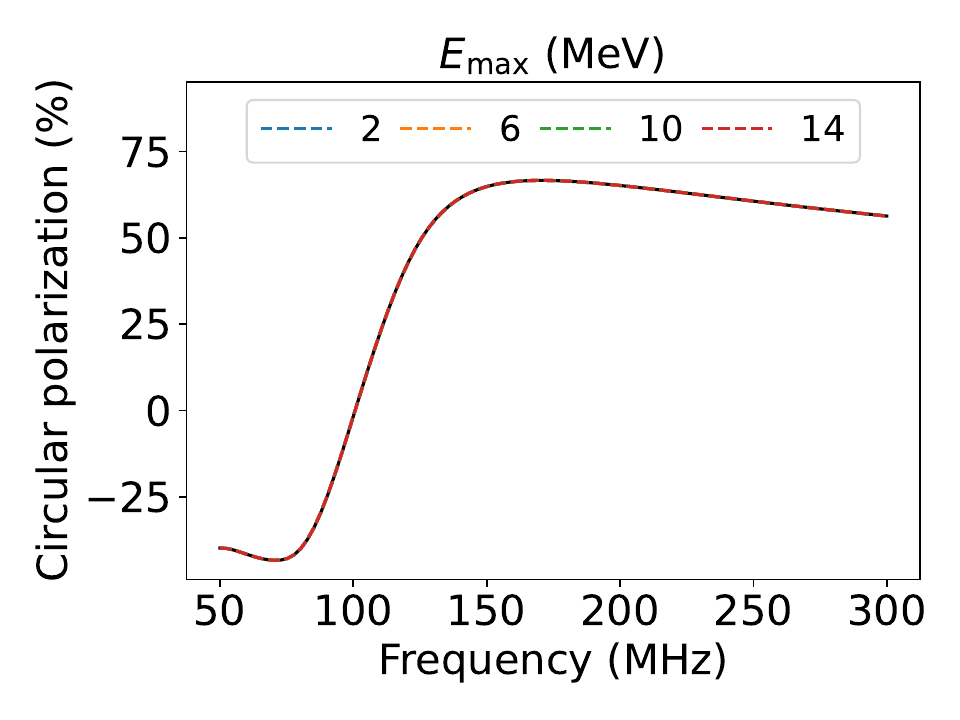}
    \caption[Sensitivity of the Stokes I and fractional Stokes V spectra on the various non-thermal parameters of GS models]{Sensitivity of the Stokes I and fractional Stokes V spectra on the various non-thermal parameters of GS models. First column shows the Stokes I spectra and the second column shows spectra for the Stokes V fraction for the different non-thermal parameters of GS models -- $n_\mathrm{nonth}$, $\delta$, $E_\mathrm{min}$ and $E_\mathrm{max}$ (top to bottom). The black solid line in different panels represents the GS spectrum for the reference parameters.}
    \label{fig:param_sensitivity_non_thermal}
\end{figure*}

\begin{figure*}[!ht]
    \centering
   \includegraphics[trim={0.5cm 0.5cm 0.5cm 0.5cm},clip,scale=0.4]{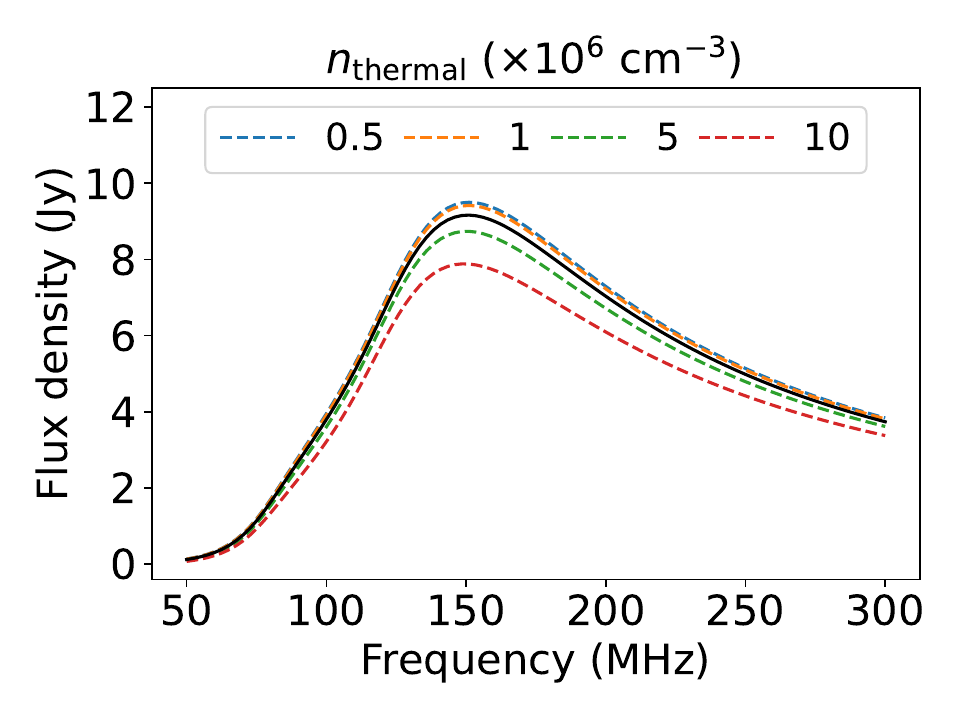}\includegraphics[trim={0.5cm 0.5cm 0.5cm 0.5cm},clip,scale=0.4]{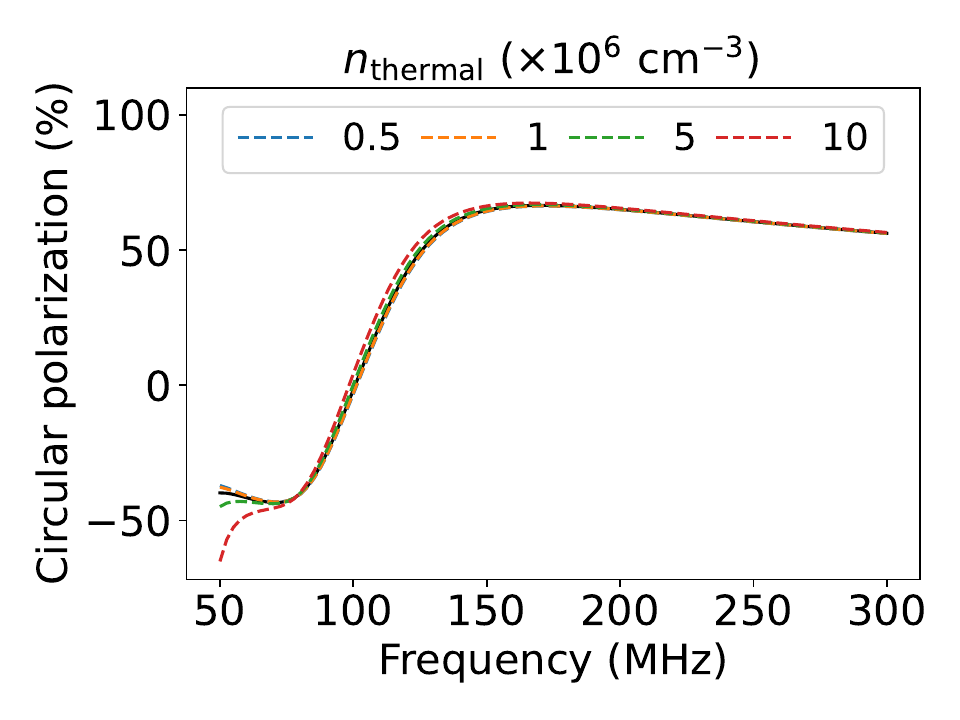}\\
    \includegraphics[trim={0.5cm 0.5cm 0.5cm 0.5cm},clip,scale=0.4]{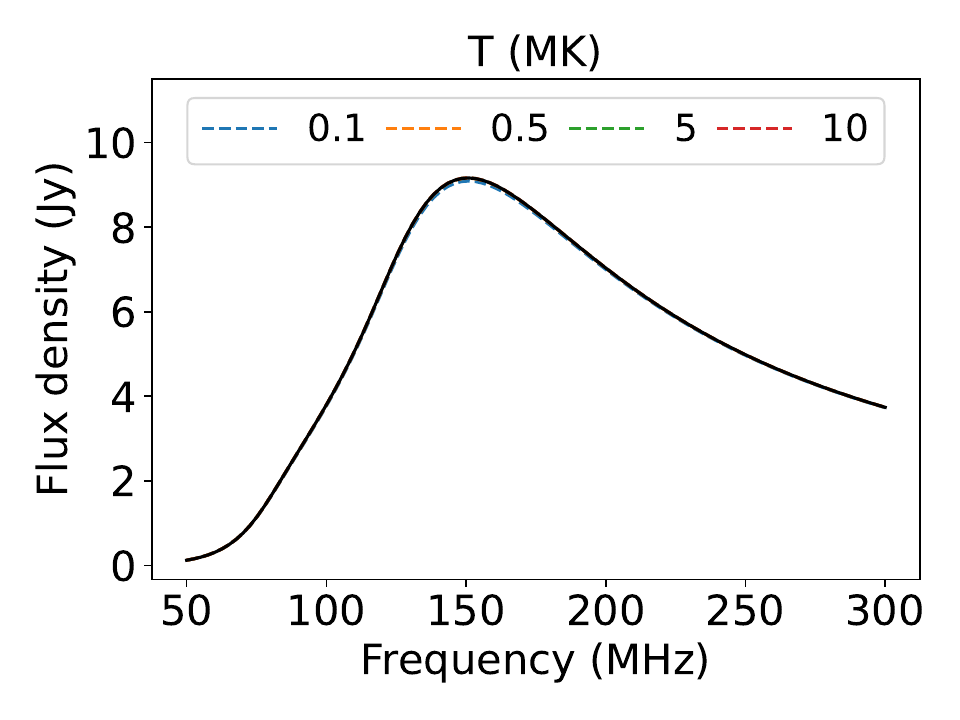}\includegraphics[trim={0.5cm 0.5cm 0.5cm 0.5cm},clip,scale=0.4]{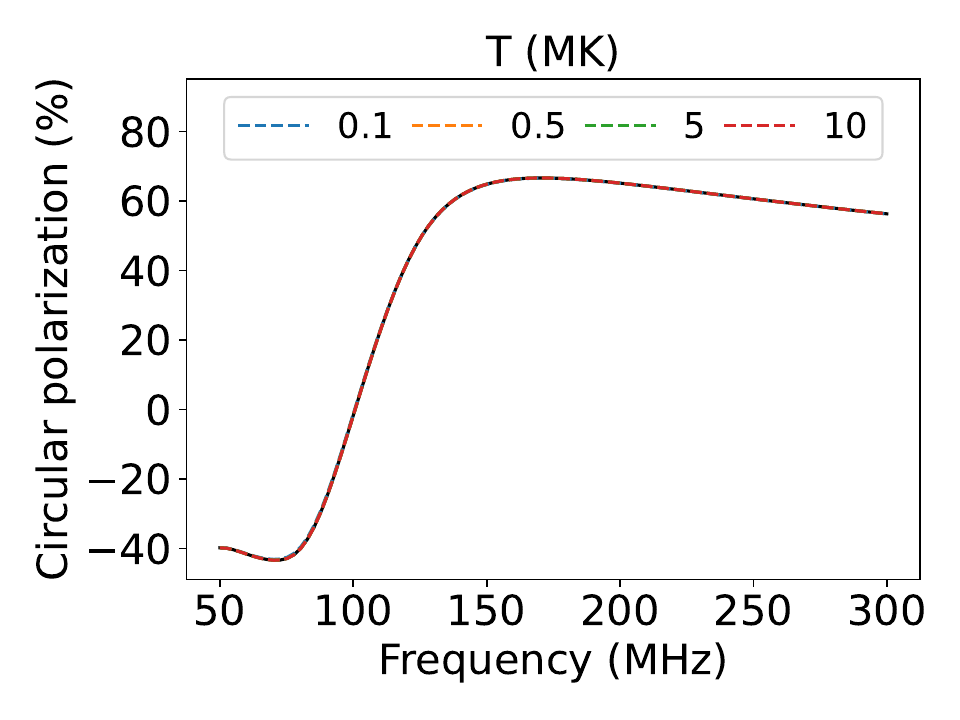}
    \caption[Sensitivity of the Stokes I and fractional Stokes V spectra on the thermal GS model parameters]{\textbf{Sensitivity of the Stokes I and fractional Stokes V 
    spectra on the thermal GS model parameters.} First column shows the Stokes I spectra and the second column shows spectra for the Stokes V fraction for the thermal model parameters -- $n_{termal}$ (top row) and $T$ (bottom row). The black solid line in different panels represents the GS spectrum for the reference parameters.}
    \label{fig:param_sensitivity_thermal}
\end{figure*}

Even for this simplistic case, the GS model requires ten independent parameters -- magnetic field strength ($B$), the angle between the line-of-sight (LoS) and the magnetic field ($\theta$), area of emission ($A$), LoS depth through the GS emitting medium ($L$), temperature ($T$), thermal electron density ($n_\mathrm{thermal}$), non-thermal electron density ($n_\mathrm{nonth}$), power-law index of non-thermal electron distribution ($\delta$),  $E_\mathrm{min}$, and $E_\mathrm{max}$. Varying each of these parameters leads to its specific change in the GS spectra, and the effects on the Stokes I and V spectra can be different. Given the limited number of spectral measurements usually available, it is not feasible to simultaneously constrain all of these model parameters. In addition, there are intrinsic degeneracies in the GS model, which limit the ability to independently constrain the parameter values. This has led the earlier studies to try to constrain some of these parameters using independent measurements (e.g. estimating the thermal electron density from coronagraph observations) and assume reasonable values for some others (e.g. LoS depth, non-thermal electron density, etc.). 

To quantitatively explore the impact of variation of each of these parameters independently, a systematic exploration of the GS model parameters is carried out here, where one parameter is varied over a reasonable range while all others are held constant. The ranges of the parameters are motivated by their values explored and estimated in earlier studies of the GS emission from CME plasma at the meter-wavelengths \citep{bastian2001,Tun2013,Carley2017,Mondal2020a}. Within this range, a fiducial choice of a certain value of each of the parameters is made as the reference value to make comparison convenient. The chosen reference values are -- i) $B=10\ \mathrm{G}$, ii) $\theta=45^{\circ}$, iii) $A = 10^{20}\ \mathrm{cm^{2}}$, iv) $T= 10^6\ \mathrm{K}$, v) $n_\mathrm{thermal}= 2.5\times10^6\ \mathrm{cm^{-3}}$, vi) $n_\mathrm{nonth}= 2.5\times10^4\ \mathrm{cm^{-3}}$, vii) $\delta= 2.8$, viii) $L= 10^{10}\ \mathrm{cm}$, ix) $E_\mathrm{min}=2\ \mathrm{keV}$ and x) $E_\mathrm{max}= 15\ \mathrm{MeV}$. 

Limited exploration of the phase space of GS parameters has been carried out by earlier studies in the context of flare observations at microwave regime \citep{Bastian_2007,Zhou_2006,Wu_2019}. To the best of my knowledge, in the context of the CME plasma, no such explorations have been done. This section presents an exhaustive exploration of the impact of variations in the physical parameters of the GS model on Stokes I and Stokes V spectra.

\subsection{Sensitivity of Stokes I Spectra to GS Model \\Parameters}
\label{subsec:stokesI_sensitivity}
The sensitivity of Stokes I spectra on different GS model parameters are shown in the first columns of Figures \ref{fig:param_sensitivity_geometric}, \ref{fig:param_sensitivity_magnetic}, \ref{fig:param_sensitivity_non_thermal} and \ref{fig:param_sensitivity_thermal}. It is evident from these figures that Stokes I spectra are not sensitive to two of the GS model parameters -- $E_\mathrm{max}$  and $T$. Peak flux density can vary by multiple orders of magnitude as a function of $\delta$, with the peak frequency decreasing with increasing $\delta$. Peak flux density increases with the increase in $A,\ B,\ \theta,\ n_\mathrm{nonth}$ and $E_\mathrm{min}$. On the other hand, peak flux density decreases with the increase in $L$ and $n_\mathrm{thermal}$. Peak frequency is independent of $A$ and $n_\mathrm{thermal}$, while it increases with the increase in $B,\ \theta,\ L,\ n_\mathrm{nonth}$ and $E_\mathrm{min}$. The nature of these variations in the Stokes I spectra implies that there exist degeneracies between values of $B,\ \theta,\  L$, $n_\mathrm{nonth}$ and $E_\mathrm{min}$, in the GS model parameters.  

\subsection{Sensitivity of Stokes V Spectra to GS Model \\Parameters}
\label{subsec:stokesV_sensitivity}
Sensitivities of Stokes V spectra on different GS model parameters are shown in the second columns of Figures \ref{fig:param_sensitivity_geometric}, \ref{fig:param_sensitivity_magnetic}, \ref{fig:param_sensitivity_non_thermal} and \ref{fig:param_sensitivity_thermal}. Some of the GS parameters -- $A$, $n_\mathrm{thermal}$, $E_\mathrm{max}$, and $T$, do not have any noticeable effect on the Stokes V spectra. $B$, $\theta$, and $\delta$ show significant impacts on both the optically thin and thick parts of the Stokes V spectra. $\delta$ has a strong impact on determining the spectral shape of the Stokes V spectra. The polarization fraction in the optically thin part increases with the increase in $B$, while it decreases with the increase in $\theta$. $L,\ E_\mathrm{min}$ and $n_\mathrm{nonth}$ impact only on the optically thick part of the Stokes V spectra, and fractional polarization decreases with the increase in each of these parameters. 

\subsection{Resolving the Degeneracy of GS Model Parameters using Stokes V Spectra}\label{subsec:stokes_V_break_degeneracy}
Different natures of impacts of $B$ and $\theta$ on the optically thin part of the Stokes V spectra break the degeneracy between them observed in Stokes I spectra. The sign of the circular polarization depends on the whether $\theta$ value is less than or greater than 90$^{\circ}$. For both $\theta$ and $180^{\circ}-\theta$, the Stokes I spectra are similar, but the Stokes V spectra are inverted. $L,\ E_\mathrm{min}$ and $n_\mathrm{nonth}$ show similar effects of both Stokes I and V spectra (Figures \ref{fig:param_sensitivity_geometric}, \ref{fig:param_sensitivity_magnetic}, \ref{fig:param_sensitivity_non_thermal} and \ref{fig:param_sensitivity_thermal}). But the availability of multi-vantage point observations allowed us to provide a strong upper limit on the $L$, while no such direct observational constraints are available for $E_\mathrm{min}$ and $n_\mathrm{nonth}$. Hence, geometrical constraints of $L$ allowed us to break the degeneracy. However, the degeneracy between $n_\mathrm{nonth}$ and $E_\mathrm{min}$ can not be resolved even when using both the Stokes I and Stokes V spectra.

\section{Constraining GS Model Parameters Combining Detection and Upper/Lower Limits}\label{subsec:upperlimits_methods}
Most often when fitting a model to the data one makes use of well-measured quantities, each with their corresponding measurement uncertainties, and follows the well-known $\chi^2$ minimization process \citep{Wolberg2006}. However, there are often situations, especially when measuring weak signals, when the quantity being measured lies beyond the detection threshold of the measurement process, but the process can place firm upper/lower limits on the quantity of interest. It seems intuitively reasonable that by constraining the parameters to lie only in the part of the phase space consistent with the limit, the use of such limits should be able to further restrict the allowed parameter space for the model parameters. Though the use of inequality constraints is not common when using $\chi^2$ minimization approaches, well-established techniques for solving such problems exist \citep[see,][for a review]{Borwein2006} and their software implementations are also available in commonly used {\it python} libraries like {\it scipy}\footnote{\href{https://docs.scipy.org/doc/scipy/reference/generated/scipy.optimize.minimize.html}{Link to scipy optimization}} \citep{Scipy2020}. 

In addition to the $\chi^2$ based approaches, there also exist other well-established mathematical frameworks for incorporating the constraints from the availability of limits. A detailed description is available in \citet{Andreon2015_bayesian} along with several examples of applications in physics and astrophysics. Standalone upper limits and combination of measurements and upper limits have often been used with considerable success to constrain physical systems across diverse areas of astrophysics \citep[e.g.,][etc.]{Aditya2015,Kanekar_2015,Kanekar2016,Montmessin2021,Brasseur2022} and cosmology \citep[e.g.,][etc.]{Planck_Col_2016,Ghara2020,Greig2021a,Bevins2022,Maity2022} including solar physics \citep[e.g.,][etc.]{Leer1979,Benz1996,Klein2003}. 

\subsection{Mathematical Framework}\label{subsec:upperlimtit_mathframe}
A few different mathematical approaches can be used to constrain model parameters using limits \citep{Andreon2015_bayesian}. The particular framework suitable for the present needs is described in detail by \cite{Ghara2020} and examples of its applications are available in \cite{Greig2021,Maity2022,Maity2022a}. To place the following analysis in context, this framework is briefly described below. This framework is based on the Bayes theorem \citep{Puga2015,Andreon2015_bayes_thereom}. Bayes theorem states that
\begin{equation}
    \mathcal{P}(\lambda|\mathcal{D})=\frac{\mathcal{L}(\mathcal{D}|\lambda)\ \pi(\lambda)}{\mathcal{P}(\mathcal{D})},
    \label{eq:bayes_theorem}
\end{equation}
where $\mathcal{D}$s are the data points and $\lambda$s the set of free parameters of the model. In this Bayesian framework, the objective is to compute the posterior distribution, $\mathcal{P}(\lambda|\mathcal{D})$, which is the conditional probability of having the set of model parameters $\lambda$ given the data $\mathcal{D}$. $\pi(\lambda)$ is the prior distribution of the model parameters, $\mathcal{P(D)}$ is called the evidence which is the probability distribution of generating observed values given a set of model parameters. Evidence is not relevant from a parameter-finding perspective in general. The standard practice is to set it to unity, implying that a given choice of model parameters leads to a unique set of observed values \citep{brooks2011handbook}. $\mathcal{L}(\mathcal{D}|\lambda)$ is the likelihood function that gives the conditional probability distribution of data given the distribution of the model parameters, $\pi(\lambda)$. In the absence of prior knowledge of the model parameters, the standard practice is to use a uniform prior distribution \citep[e.g.,][etc.]{Kashyap1998,Middleton2015,Li_2019,Ghara2020,Maity2022} of model parameters over a physically meaningful range.

For a well-measured quantity, i.e. when the measurement is above the noise threshold, the likelihood function is defined as,
\begin{equation}
\begin{split}
      \mathcal{L}_\mathrm{1}(\mathcal{D}|\lambda)&=\mathrm{exp}\left(-\frac{1}{2}\sum_{i=1}^N \left[\frac{\mathcal{D}_\mathrm{i}-m_\mathrm{i}(\lambda)}{\sigma_\mathrm{i}}\right]^2\right)\\
      &=\prod_{i=1}^N \mathrm{exp}\left(-\frac{1}{2}\left[\frac{\mathcal{D}_\mathrm{i}-m_\mathrm{i}(\lambda)}{\sigma_\mathrm{i}}\right]^2\right)
\end{split}
\label{eq:likelihood_1}
\end{equation}
where, $N$ is the total number of data points, $\mathcal{D}_\mathrm{i}$, $m_\mathrm{i}(\lambda)$, and $\sigma_\mathrm{i}$ are the observed values, models values and uncertainty on the measurements, respectively. For the case of upper limits, the likelihood function is defined as follows \citep{Ghara2020,Greig2021,Maity2022},
\begin{equation}
\begin{split}
      \mathcal{L}_\mathrm{2}(\mathcal{D}|\lambda)&=\prod_{i=1}^N \frac{1}{2}\left[1-erf\left(\frac{\mathcal{D}_\mathrm{i}-m_\mathrm{i}(\lambda)}{\sqrt{2}\sigma_\mathrm{i}}\right)\right],
\end{split}
\label{eq:likelihood_2}
\end{equation}
where $erf$ refers to the error function. When a mix of detections and upper limits are available, one can define the joint likelihood function as,
\begin{equation}
    \mathcal{L}(\mathcal{D}|\lambda)=\mathcal{L}_\mathrm{1}(\mathcal{D}|\lambda)\ \mathcal{L}_\mathrm{2}(\mathcal{D}|\lambda),
    \label{eq:join_likelihood}
\end{equation}
which allows one to use the constraints from the detections as well as the upper limits. Using this joint likelihood function in the Monte Carlo Markov Chain \citep[MCMC;][]{brooks2011handbook} analysis allows one to use all available information to better infer the model parameters.

Unlike $\chi^2$ minimization, MCMC analysis does not yield a unique set of values for model parameters. Instead, it provides the probability distribution of the parameter values, denoted by the {\it posterior distribution} in MCMC analysis. This allows us to fully understand the degeneracies in the parameter space and thus enables a more robust understanding of the underlying physical system. The {\it true} value of the parameter is close to the value with the highest probability. Thus MCMC based approaches overcome one of the inherent limitations of a $\chi^2$ based approach -- the possibility of converging to one of the many local minima in the $\chi^2$ space, especially when dealing with a large number of free parameters.

\section{Spectrum Modeling}\label{sec:spectrum_modeling}
This section describes the approach to modeling the observed spectra using a GS model to estimate CME plasma parameters. As demonstrated in Section \ref{sec:spectrum_sensitivity}, in the physically motivated range of parameters explored here, the model GS spectra are quite insensitive to variations in $T$ and $E_\mathrm{max}$. Hence, $T$ is kept fixed at an average coronal temperature of 1 MK, and $E_\mathrm{max}$ is kept fixed at a high value, 15 MeV. Thermal electron density is estimated independently from the inversion of the white light images. The value used at any given radial distance is the average over the entire azimuthal range. $n_\mathrm{thermal}$ is kept fixed at this value during GS model fitting. Among the other seven parameters, $B$, $\theta$, $A$, $\delta$ and $E_\mathrm{min}$ are fitted, while setting $n_\mathrm{nonth}$ to 1\% of the $n_\mathrm{thermal}$, similar to what has been assumed in previous works \citep{Carley2017,Mondal2020a}. For spectra from some regions (regions 2 and 3 in Figure \ref{fig:spectral_regions}), $L$ is fitted explicitly, and for the other regions $L$ is kept fixed to a pre-defined value as detailed in Section \ref{subsec:estimate_gcs}.

\begin{figure*}[!htbp]
\centering
    \includegraphics[trim={1.5cm 2.7cm 7cm 1.5cm},clip,scale=0.8]{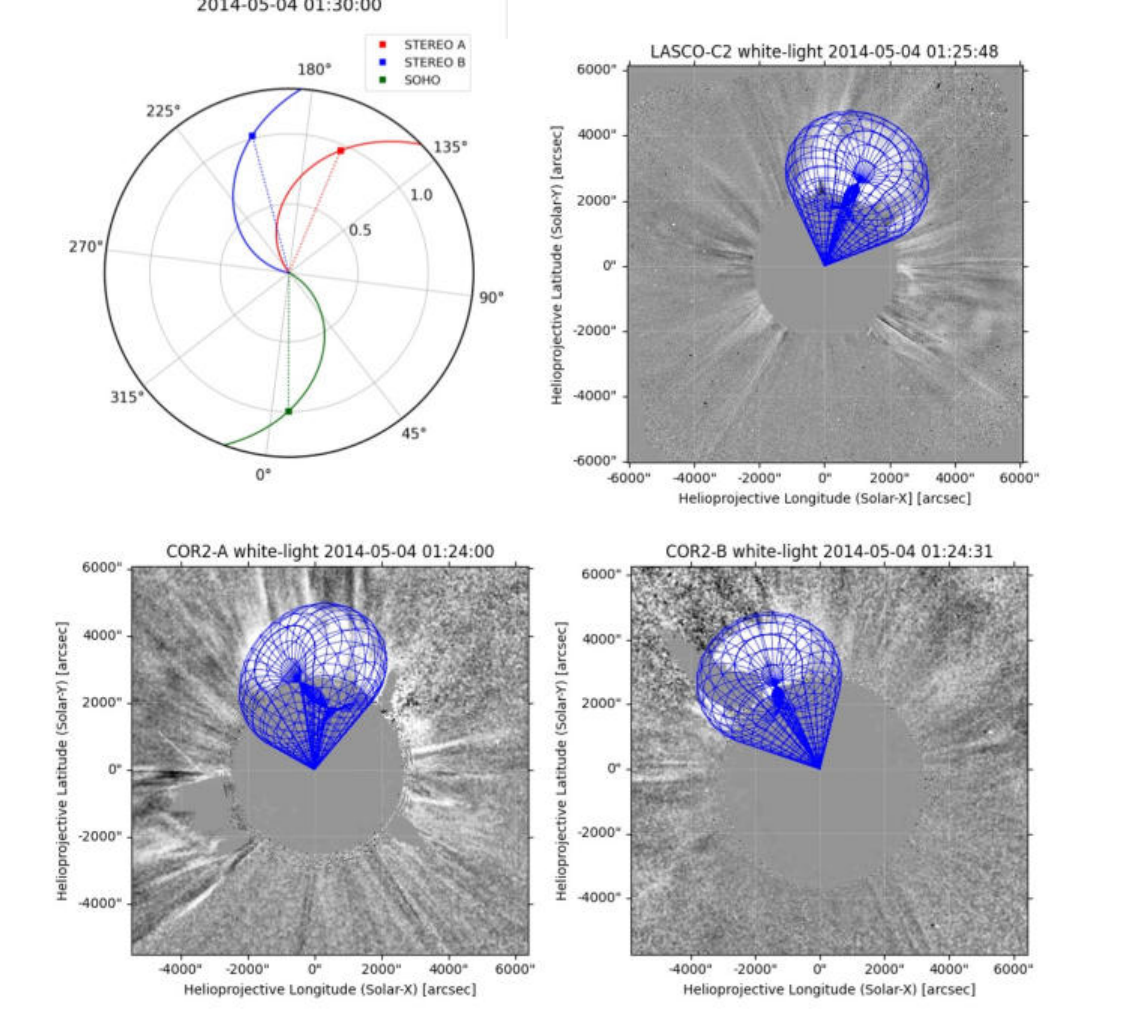}\\
     \caption[Three-dimensional reconstruction of the northern CME.]{Three-dimensional reconstruction of the CME-1 using the Graduated Cylindrical Shell (GCS) model using three vantage point observations. {\it Top left panel: } Position of SOHO, STEREO-A, and STEREO-B spacecraft. STEREO-A and STEREO-B were behind the Sun on 2014 May 04. Positions are marked by squares, and the curved lines represent the Parker spiral connected to each spacecraft. {\it Top right panel:} It shows the GCS model of the CME-1 at about 01:25 UTC using the LASCO-C2. {\it Bottom panel: }GCS model on COR-2 coronagraph images onboard STEREO-A and STEREO-B spacecraft. In LASCO and STEREO-A images the streamer was not bright.}
    \label{fig:gcs}
\end{figure*}

\subsection{Estimation of Geometrical Parameters}\label{subsec:estimate_gcs}
A key reason for choosing this CME for a detailed study was that it has coronagraph observations from multiple vantage points, SOHO, STEREO-A, and STEREO-B, which enable us to build a well-constrained three-dimensional model. The locations of these spacecraft are shown in the top left panel of Figure \ref{fig:gcs} created using Solar-MACH \citep{solar_mach}\footnote{\url{https://solar-mach.github.io/}}. A three-dimensional reconstruction of the CME is done using the Graduated Cylindrical Shell model \citep[GCS;][]{Thernisien_2006,Thernisien_2011} using its {\it python} implementation \citep{gcs_python}. A good visual fit is obtained following the method described by \cite{Thernisien2009}. The GCS model arrived at is shown by blue mesh in Figure \ref{fig:gcs}, where different panels show superposition on LASCO-C2 and COR-2 images from STEREO-A and STEREO-B. The best visual fit GCS model parameters are:
\begin{enumerate}
    \item Front height ($h_\mathrm{front}$) : 6.3 $R_\odot$
    \item Half-angle ($\alpha$) : 21$^{\circ}$
    \item Carrington Longitude ($\Phi$) : 153$^{\circ}$ 
    \item Heliospheric Latitude ($\Theta$) : 48$^{\circ}$ 
    \item Aspect Ratio ($\kappa$) : 0.34
    \item Tilt Angle ($\gamma$) : 40$^{\circ}$
\end{enumerate}
At 80 MHz, the radio emission is detected up to the leading edge observed in LASCO-C2 white-light image (Figure \ref{fig:north_cme}). The projected distance of the radio emission in the sky plane is 5.2 $R_\odot$. The corresponding three-dimensional distance computed from the GCS model based on the multi-vantage point observations puts this at $\sim6.3\ R_\odot$. 

\begin{figure}[!ht]
\centering
    \includegraphics[trim={0cm 0cm 0cm 0cm},clip,scale=0.47]{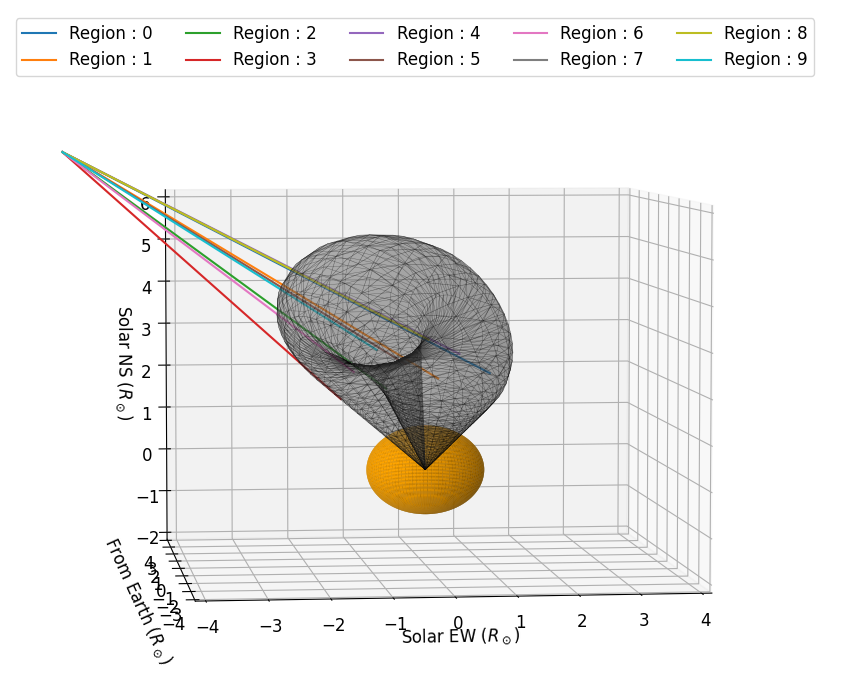}
    \caption[Ray-tracing through the reconstructed CME for different PSF-sized regions.]{Ray-tracing through the Graduated Cylindrical Shell (GCS) model flux rope for different PSF-sized regions. Origin of the coordinate system is chosen to lie at the center of the Sun. Note that for ease of representation, the distance along the Sun-Earth direction is shown in units of 50 $R_\odot$. Different rays originated from the Earth at 214 $R_\odot$ (0, 0, 214 $R_\odot$) are traced through the GCS flux rope to find out the geometrical LoS depth of a certain PSF-sized region. The orange sphere represents the Sun, and the GCS flux rope is shown by grey mesh. Rays are shown by colored lines.}
    \label{fig:gcs_ray}
\end{figure}

\begin{table}
\centering
    \renewcommand{\arraystretch}{1.4}
    \begin{tabular}{|p{1.5cm}|p{1.5cm}|p{1.5cm}|p{1.5cm}|p{1.5cm}|p{1.5cm}|}
    \hline
       Region & $L_\mathrm{geo} \newline{(R_\odot)}$ & $\sigma(L_\mathrm{geo}) \newline{(R_\odot)}$ & Region & $L_\mathrm{geo} \newline{(R_\odot)}$  & $\sigma(L_\mathrm{geo}) \newline{(R_\odot)}$\\ \hline \hline 
        1 & 2.4 & 0.8 &  5 & 2.9 & 1.1\\
        \hline
        2 & 3.2 & 1.8 &  6 & 3.1 & 1.1\\
        \hline
        3 & 2.9 & 1.2 &  7 & 2.3 & 1.3\\
        \hline
        4 & 2.8 & 1.0 & 8 & 3.5 & 1.8\\
       \hline
    \end{tabular}
    \caption[Estimated geometric LoS depth from GCS modeling.]{Estimated geometric LoS depth from GCS modeling. The geometric LoS depths are obtained for different PSF-sized regions using ray tracing from Earth through that region. Geometric LoS depths are given in units of the solar radius.}
    \label{table:los_depth}
\end{table}

For LoS originating from the Earth, ray-tracing is done through the GCS model and computed the geometrical path length through the CME ($L_\mathrm{geo}$) for each PSF-sized region using {\it python}-based ray-tracing code {\it trimesh} \citep{trimesh}. The ray paths for different regions are shown in Figure \ref{fig:gcs_ray}. Estimated $L_\mathrm{geo}$ for each of these regions are listed in Table \ref{table:los_depth}. 

\begin{figure*}[!ht]
    \centering
    \includegraphics[trim={0cm 0cm 0cm 0cm},clip,scale=0.55]{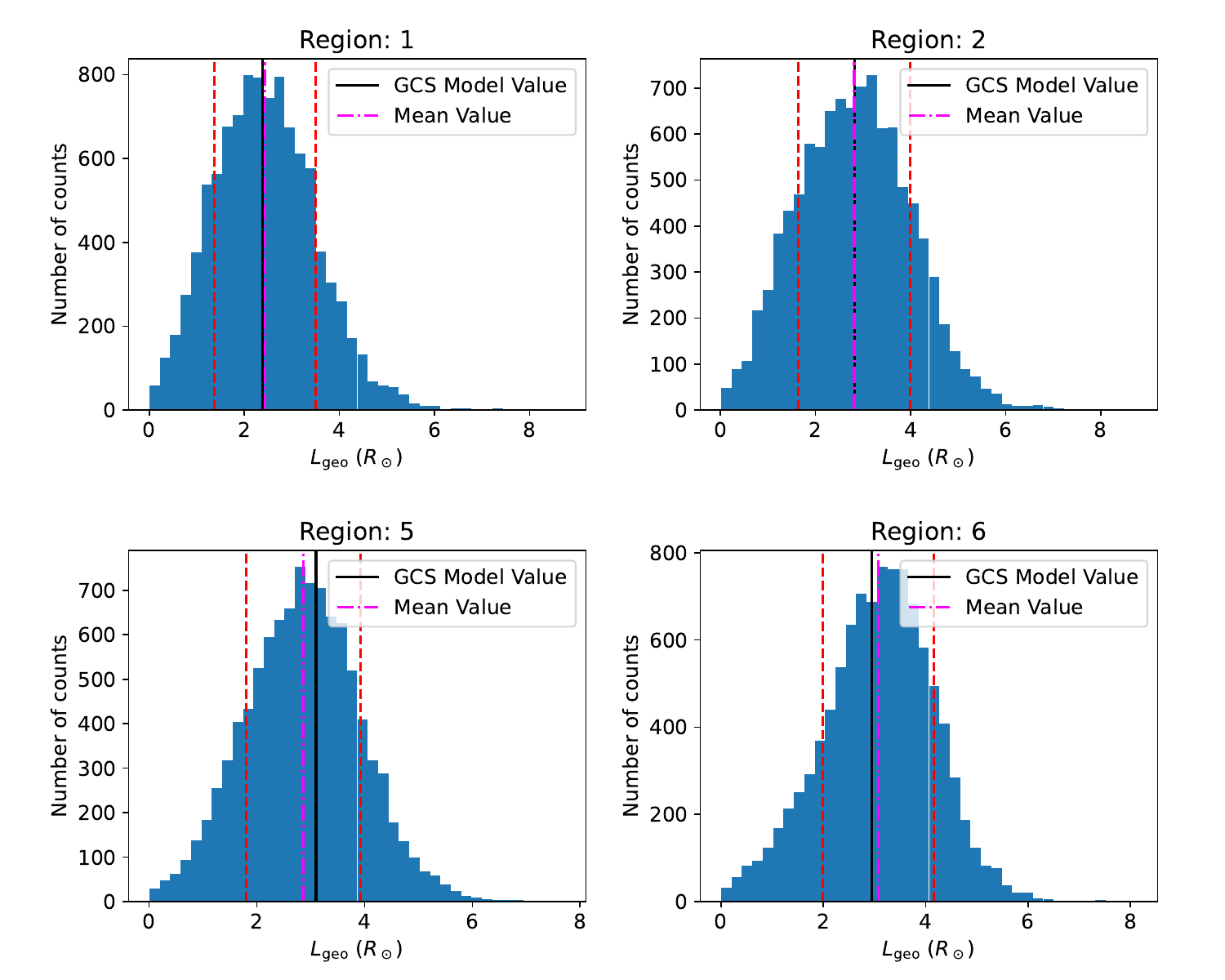}
    \caption[Distributions of $L_\mathrm{geo}$ for northern CME.]{Distributions of $L_\mathrm{geo}$ for some sample PSF-sized regions for the northern CME. Solid black lines represent $L_\mathrm{geo}$ for the GCS model parameters mentioned above. The dot-dashed magenta lines represent the mean and the red-dashed lines represent the standard deviation around the mean. The mean and corresponding standard deviations are mentioned in Table \ref{table:los_depth}.}
    \label{fig:cme1_los_hist}
\end{figure*}

Determining the best fit GCS model is not the result of a formal optimization procedure. In addition to the limitations imposed by the sensitivity of measurements, it is prone to errors for reasons ranging from human subjectiveness to relative locations of the vantage points. To quantify these errors, \cite{Thernisien2009,VERBEKE2022} examined a large number of synthetic or forward-modeled CMEs of different kinds observed using different numbers and configurations of spacecraft. As there is no analytic relationship between GCS model parameters and $L_\mathrm{geo}$, usual error propagation cannot be used to estimate the uncertainty on $L_\mathrm{geo}$ ($\sigma(L_\mathrm{geo})$). To overcome this limitation, 10,000 realizations of GCS model parameters are generated from independent Gaussian distributions for each of the parameters. The mean of these distributions was set to the fitted values and the standard deviation to the uncertainty was reported in \cite{VERBEKE2022}. $L_\mathrm{geo}$ was computed for each of these realizations. This was done independently for each of the PSF-sized regions. Some sample histograms of distributions of $L_\mathrm{geo}$ for some example regions are shown in Figure \ref{fig:cme1_los_hist}. Histograms of distributions of $L_\mathrm{geo}$ for other regions are similar. The mean and standard deviation of the distribution of $L_\mathrm{geo}$ values so obtained are given in Table \ref{table:los_depth}. 

It is important to note that the geometrical value of LoS angle and depth can differ from those for the best-fit GS model. This is because $\theta$ and $L$ describe the GS source along a given LoS, while the geometric parameters are derived from the white-light CME morphology. Also note that while the angle with the sky plane can not provide any constraint on $\theta$, $L$ on the other hand, is tightly constrained to be smaller than $L_\mathrm{geo}$. This constraint on $L$ has not been used in earlier studies. It is evident from Figure \ref{fig:param_sensitivity_geometric} that the peak flux density and peak frequency of the Stokes I spectrum and the Stokes V fraction in the optically thick part are all sensitive to $L$. Hence to constrain $L$ using GS models, it is important that the spectral peak be included in the observed spectrum and it has at least seven measurements. For this reason, $L$ is used as a free parameter for regions 2 and 3, but not for other regions. The maximum value of $L$ for a given region is chosen to be $L_\mathrm{max}=L_\mathrm{geo}+\sigma(L_\mathrm{geo})$ as listed in Table \ref{table:los_depth}. An average fraction is calculated to be, $f=L_\mathrm{fit}/L_\mathrm{max}$ for these two regions, where $L_\mathrm{fit}$ is the estimated value of $L$ from GS modeling. For regions 2 and 3 the values of $f$ are 0.29 and 0.23, respectively, and have a mean of $\sim$0.26. Assuming the filling fraction of the GS sources from different regions to lie in the same ballpark, $L$ is kept fixed at 0.26 times $L_\mathrm{geo}+\sigma(L_\mathrm{geo})$ for all other regions. 
\begin{figure*}[!htbp]
    \centering
     \includegraphics[trim={0.3cm 0.5cm 0.0cm 0.3cm},clip,scale=0.4]{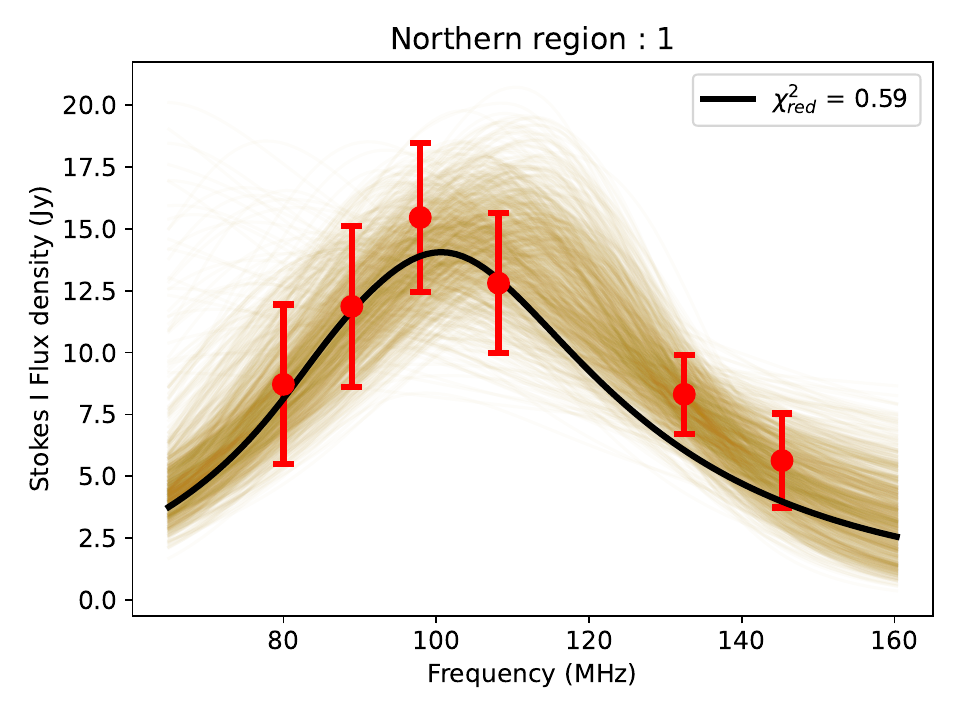} \includegraphics[trim={0.3cm 0.5cm 0.0cm 0.3cm},clip,scale=0.4]{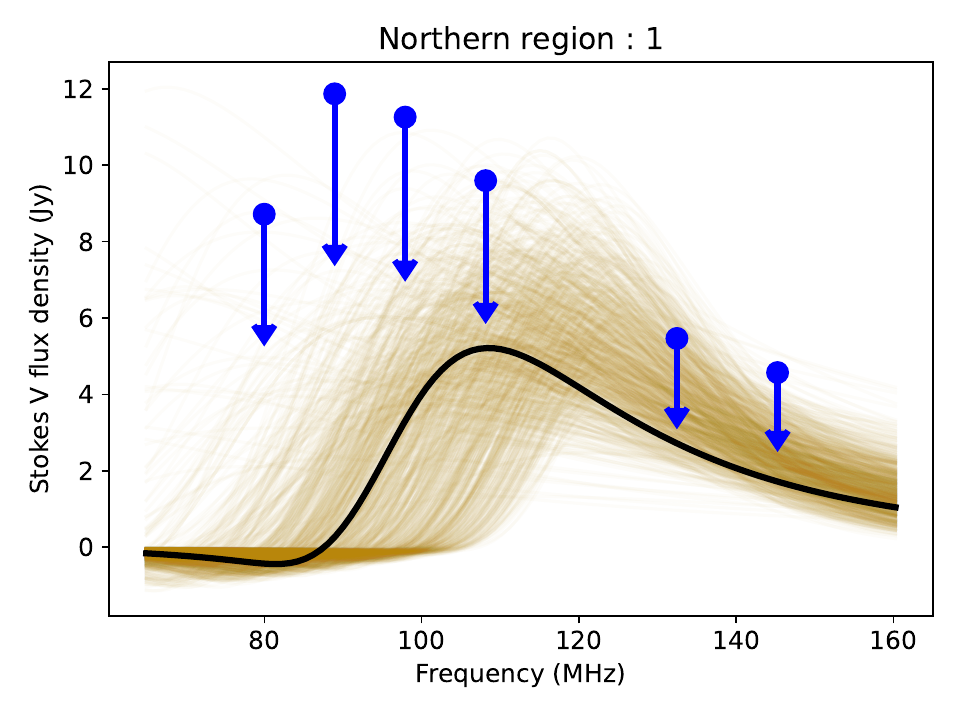}\\
     
     \includegraphics[trim={0.3cm 0.5cm 0.0cm 0.3cm},clip,scale=0.4]{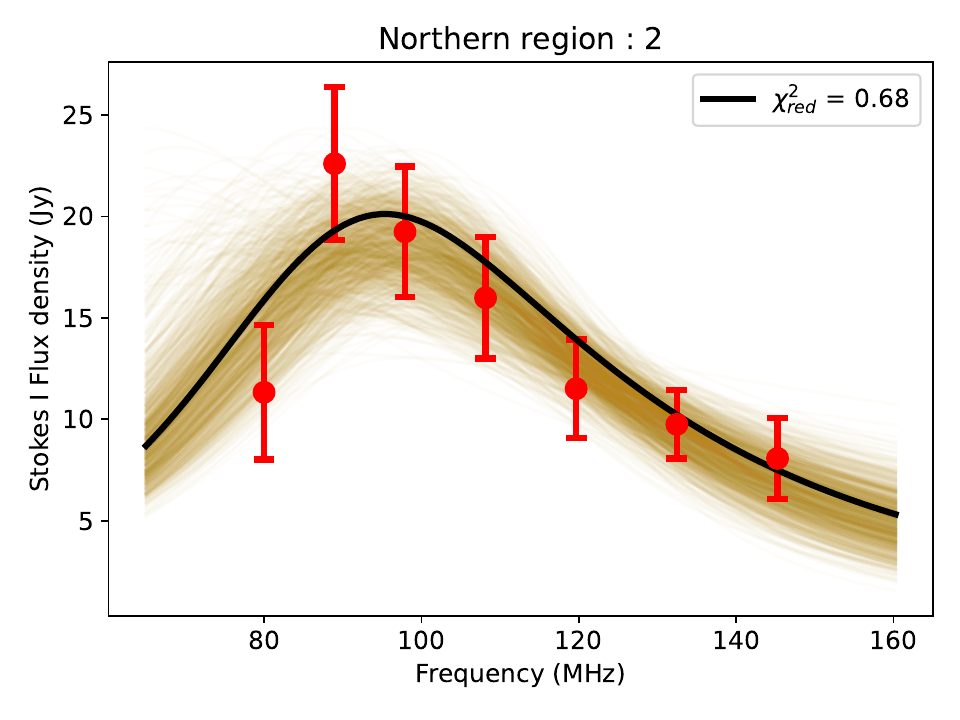}\includegraphics[trim={0.3cm 0.5cm 0.0cm 0.3cm},clip,scale=0.4]{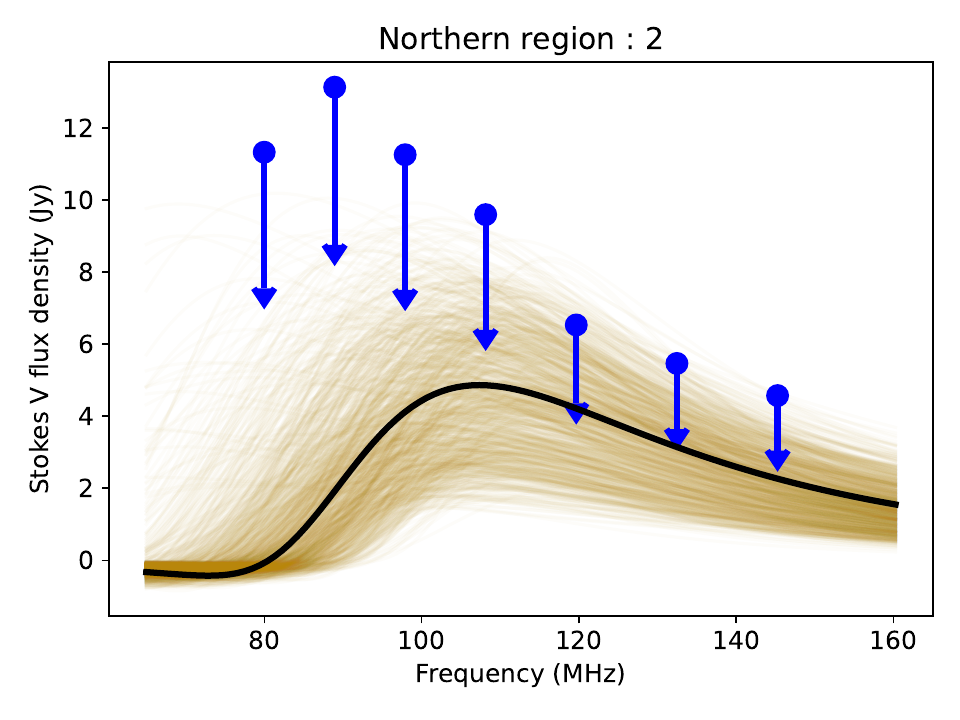}\\
     
     \includegraphics[trim={0.3cm 0.5cm 0.0cm 0.3cm},clip,scale=0.4]{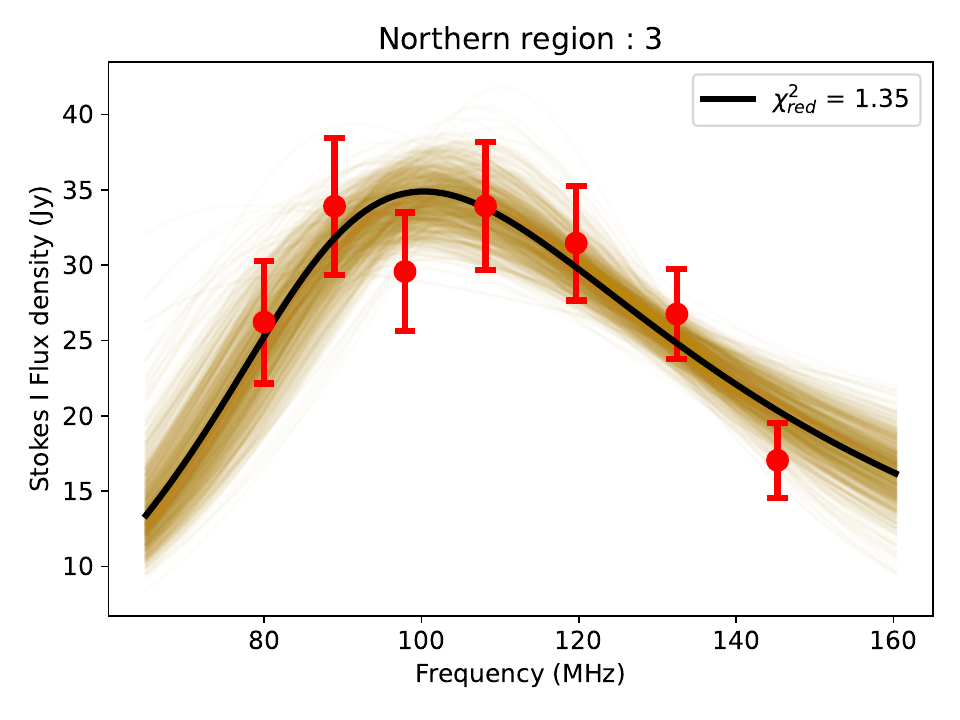}\includegraphics[trim={0.3cm 0.5cm 0.0cm 0.3cm},clip,scale=0.4]{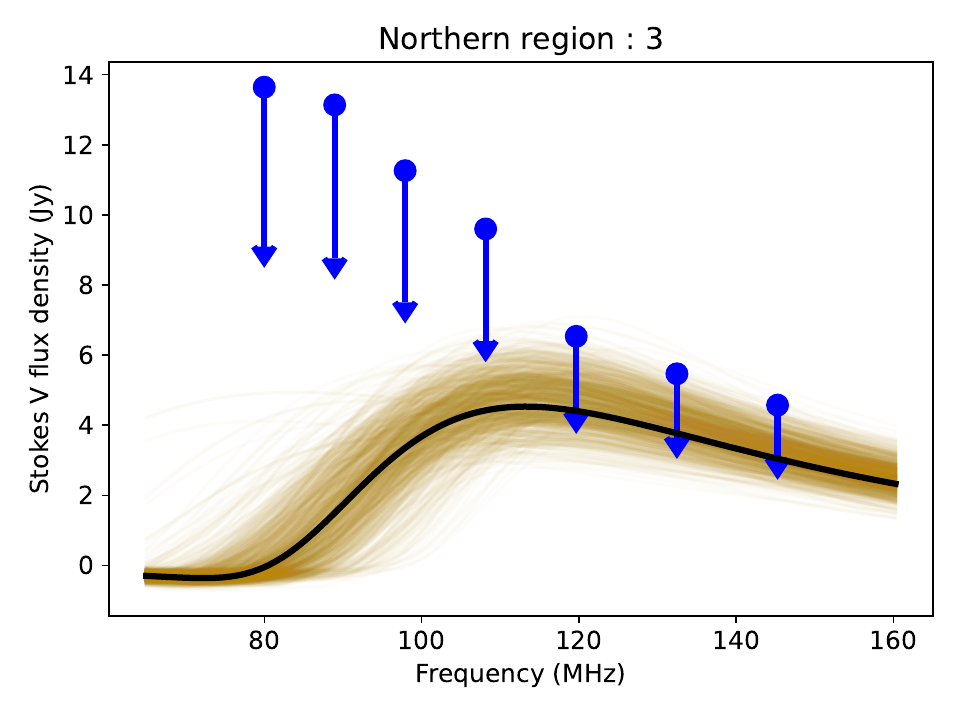}
    \caption[Observed and fitted spectra for regions 1, 2, and 3 of northern CME.]{Observed and fitted spectra for regions 1, 2, and 3 of CME-1. {\it First column: }Stokes I spectra are shown. Red points represent the observed flux densities. {\it Second column: }Stokes V spectra are shown. Blue points represent the upper limits at each of the frequencies. The black lines represent the Stokes I and V GS spectra corresponding to GS parameters reported in Table \ref{table:north_params}. Light yellow lines show the GS spectra for 1000 realizations chosen randomly from the posterior distributions of the GS model parameters. Sample posterior distributions for region 3 are shown in Figure \ref{fig:corner_northern}.}
    \label{fig:spectra1}
\end{figure*}

\subsection{Joint Spectral Fitting of Stokes I and V}\label{subsec:joint_fitting}
A joint spectral fit is performed using the Stokes I and V spectra for the red and cyan regions marked in the right panel of Figure \ref{fig:spectral_regions}. I followed the mathematical framework described in Section \ref{subsec:upperlimtit_mathframe}. Combing the Stokes I detection and Stokes V upper limits, following Equation \ref{eq:join_likelihood} the joint likelihood function is defined as
\begin{equation}
    \mathcal{L}(\mathcal{D}|\lambda)=\mathcal{L}_\mathrm{I}(\mathcal{D}_\mathrm{I}|\lambda)\ \mathcal{L}_\mathrm{V}(\mathcal{D}_\mathrm{V}|\lambda),
\end{equation}
where $\mathcal{D}_\mathrm{V,i}$, $m_\mathrm{V,i}(\lambda)$, and $\sigma_\mathrm{V,i}$ are the upper limits of absolute Stokes V flux density, GS model Stokes V flux density and uncertainties in the Stokes V. $\mathcal{L}_\mathrm{I}(\mathcal{D}_\mathrm{I}|\lambda)$ is the likelihood function for the Stokes I detection following the Equation \ref{eq:likelihood_1} and $\mathcal{L}_\mathrm{V}(\mathcal{D}_\mathrm{V}|\lambda)$ is the likelihood function for the Stokes V upper limits defined in Equation \ref{eq:likelihood_2}. 

The posterior distribution is sampled using the Metropolis-Hastings algorithm \citep{Metropolis1953} of the MCMC method. I use publicly available {\it python} package {\it lmfit} \citep{lmfit} for this purpose, which runs the MCMC chains using another {\it python} package {\it emcee} \citep{emcee2013}. A total of 1,000,000 MCMC chains are executed per spectrum. 

I have used uniform priors, $\pi(\lambda)$, for the model parameters as follows,
\begin{enumerate}
    \item $B\ (\mathrm{G})$ : $(0,\ 20]$
    \item $\theta\ (\mathrm{degree})$ : $(0,\ 90)$
    \item $\delta$ : $(1,\ 10]$
    \item $A \times10^{20}\ (\mathrm{cm^{2}})$ : $[0.0001,\ 100]$
    \item $E_\mathrm{min}\ (\mathrm{keV})$ : $(0.1,\ 100]$
    \item $L\ (R_\odot)$ : $(0.01,L_\mathrm{max}]$ 
\end{enumerate}

The range of $B$ is guided by the choices made in previous works \citep{Vourlidas2020}. $\delta$ is also chosen based on previous studies and direct X-ray imaging observations \citep{Carley2017}. In principle, $\theta$ can take values ranging from 0$^{\circ}$ to 180$^{\circ}$. The value of $\theta$ and $180^{\circ}-\theta$ produce similar Stokes I spectra and their Stokes V spectra are inverted with respect to each other, as shown in the bottom right panel of Figure \ref{fig:param_sensitivity_magnetic}. Since only upper limits on the absolute Stokes V are available in the present case, this degeneracy between $\theta$ and $180^{\circ}-\theta$ can not be broken. However, that does not impact the estimated value of $B$. Hence, I have chosen the $\theta$ to lie in the range 0$^{\circ}$--90$^{\circ}$. The minimum value of $A$ is chosen at a similar order of magnitude to that reported in M20, and the maximum value is chosen to be equal to the PSF area.  Given that there are no direct measurements of non-thermal electron distributions at these heights, the minimum value of $E_\mathrm{min}$ is chosen to be slightly higher than the energy of thermal electrons at $10^6\ \mathrm{K}$, close to the minimum value of $E_\mathrm{min}$ found by M20. 
\begin{figure*}[!ht]
    \centering
    \includegraphics[trim={0cm 0.6cm 0cm 0cm},clip,scale=0.4]{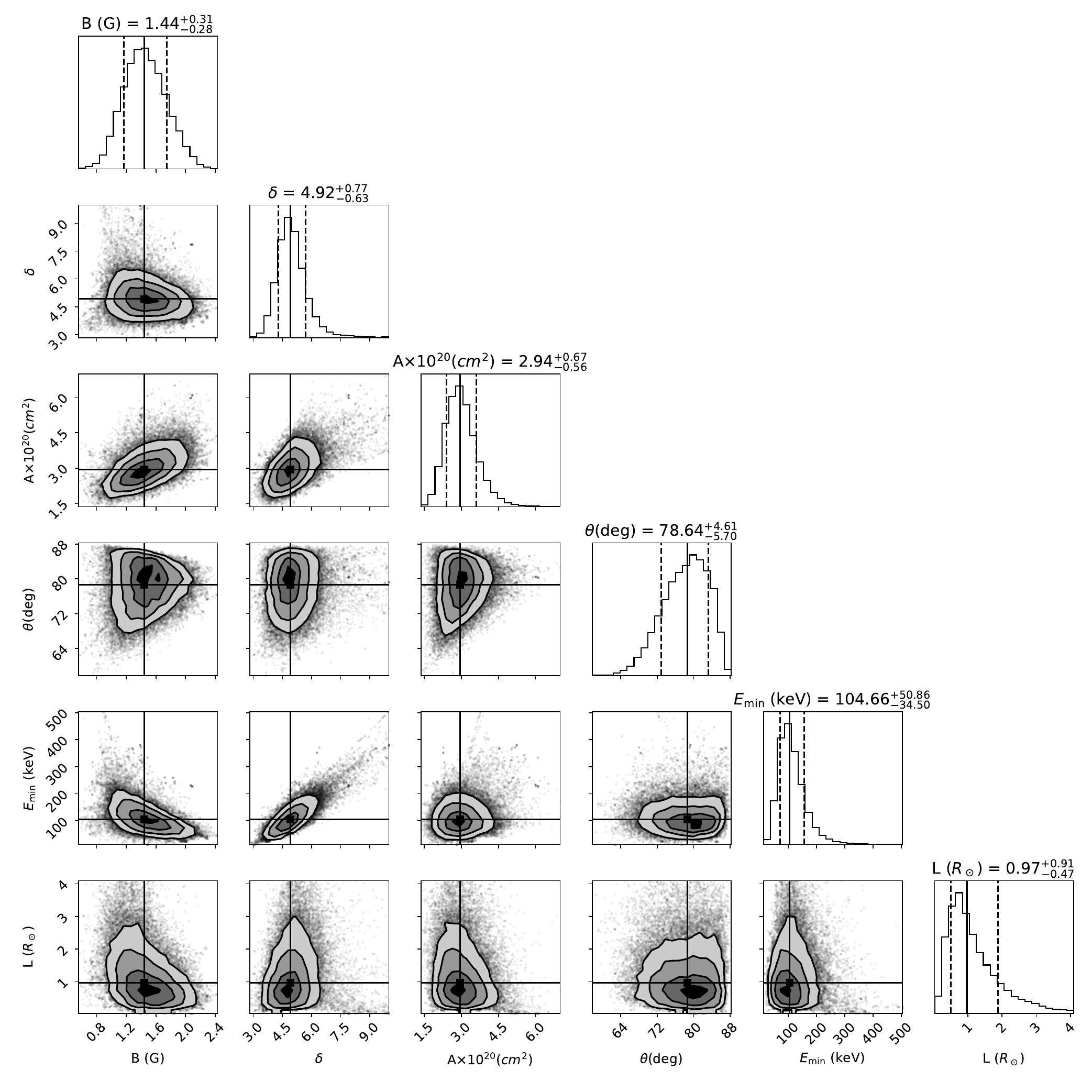}
    \caption[Correlation of posterior distribution of GS model parameters for region 3.]{Correlation of posterior distribution of GS model parameters for region 3. 2-dimensional plots show the joint probability distribution of any two parameters. The contours are at 0.5, 1, 2, and 3$\sigma$. The solid lines in the 1-dimensional histogram of posterior distributions mark the median values, and the vertical dashed lines mark the 16$^\mathrm{th}$ and 84$^\mathrm{th}$ percentiles. The median values are also marked in the panels showing the joint probability distribution.}
    \label{fig:corner_northern}
\end{figure*}

\subsection{Estimation of GS Model Parameters}\label{sec:plasma_parameters}
\begin{figure}
    \centering
      \includegraphics[trim={0.3cm 0.5cm 0.0cm 0.3cm},clip,scale=0.4]{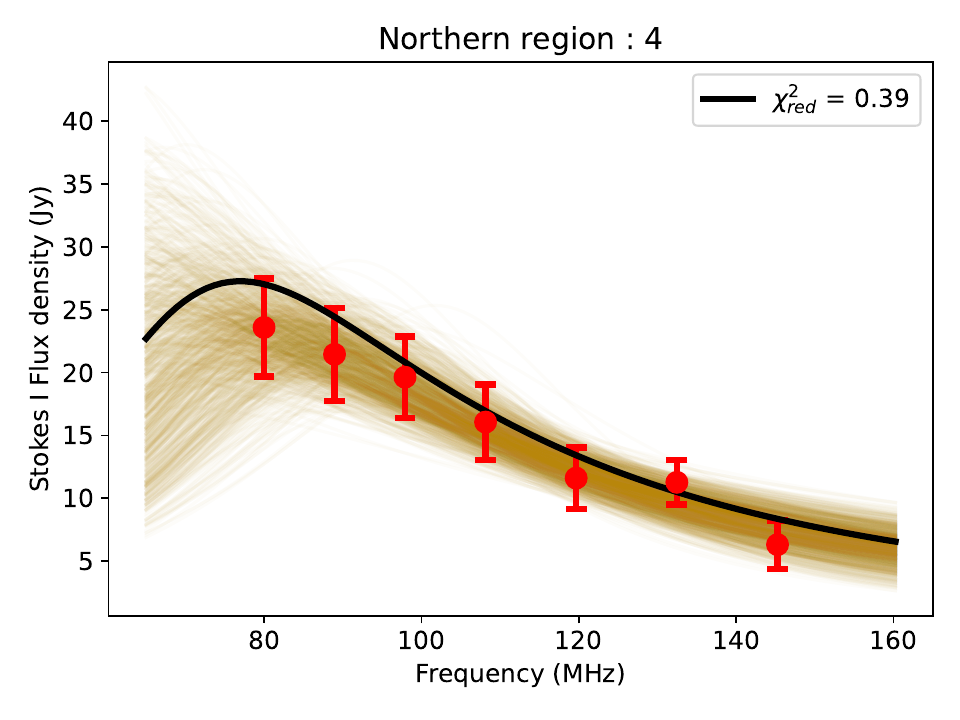} \includegraphics[trim={0.3cm 0.5cm 0.0cm 0.3cm},clip,scale=0.4]{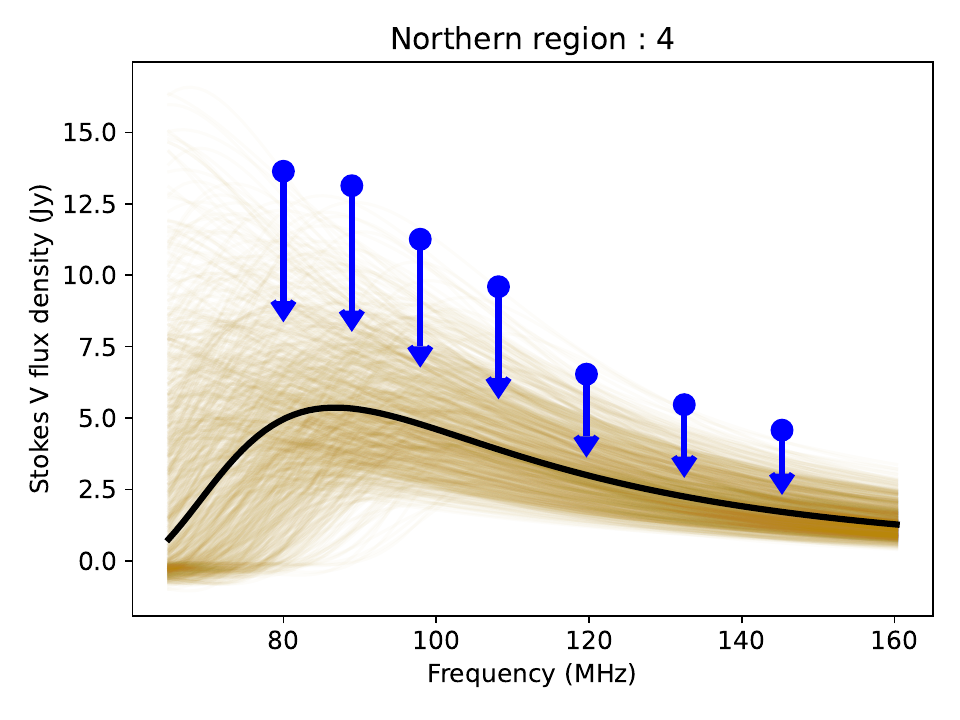}\\
      
      \includegraphics[trim={0.3cm 0.5cm 0.0cm 0.3cm},clip,scale=0.4]{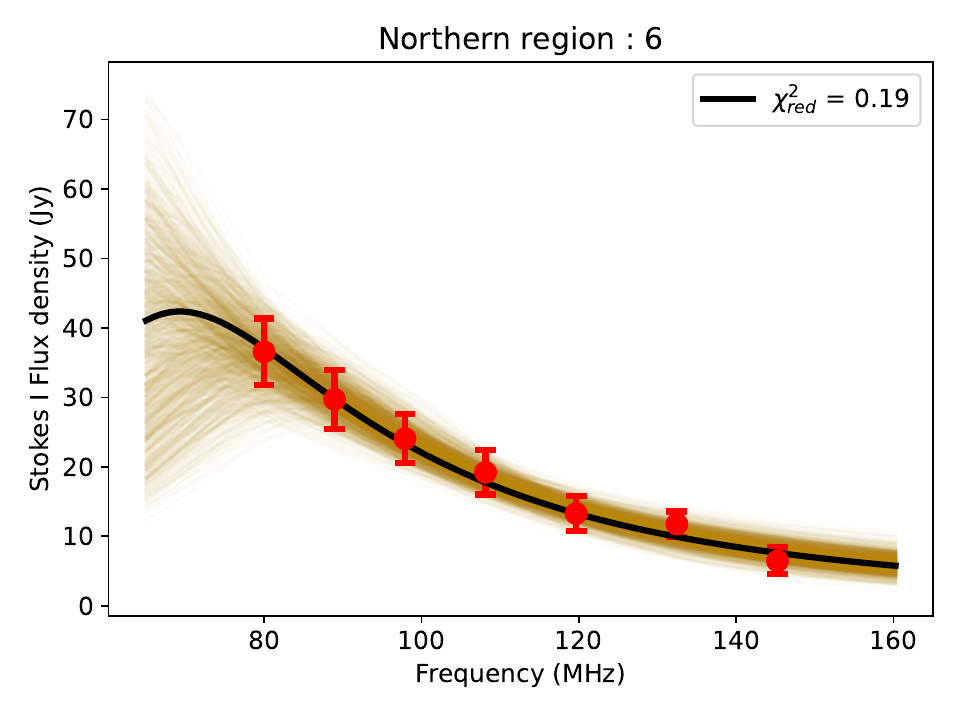}\includegraphics[trim={0.3cm 0.5cm 0.0cm 0.3cm},clip,scale=0.4]{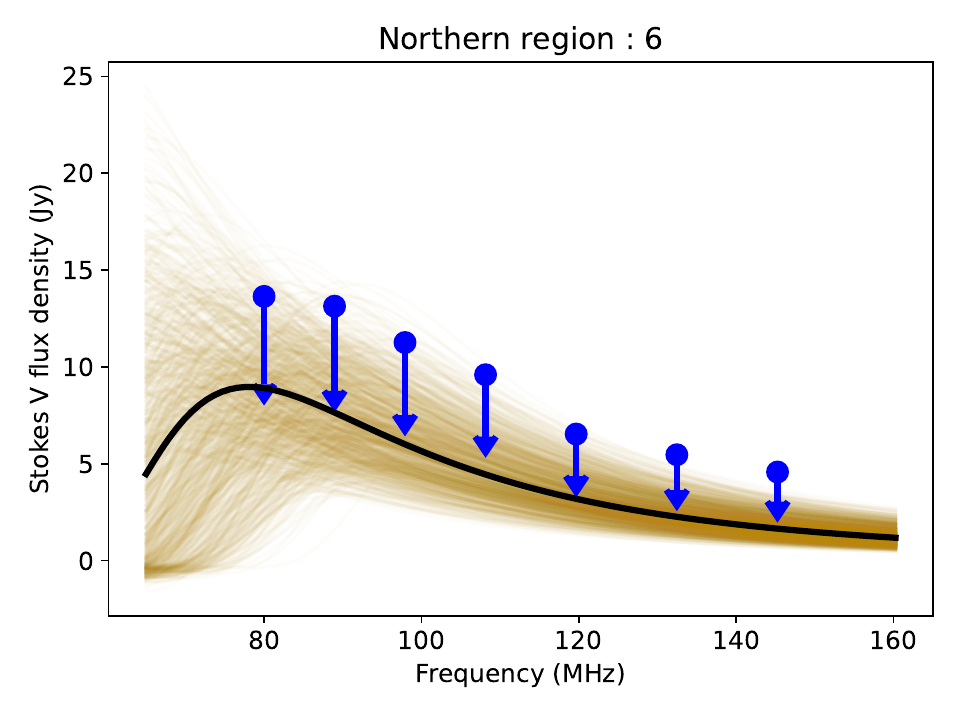}\\
      
      \includegraphics[trim={0.3cm 0.5cm 0.0cm 0.3cm},clip,scale=0.4]{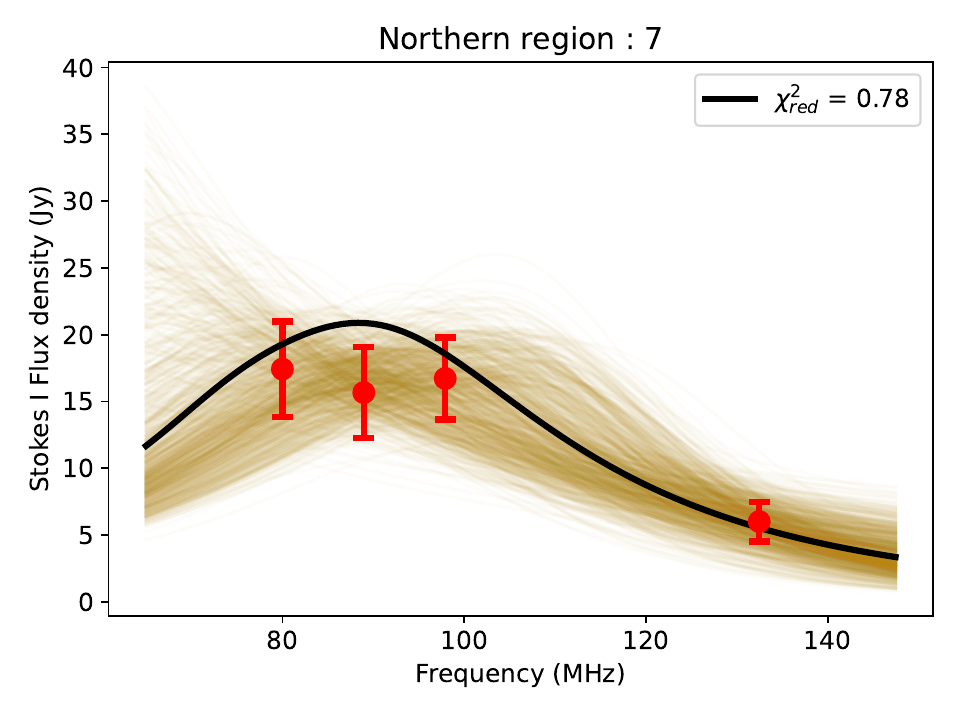}\includegraphics[trim={0.3cm 0.5cm 0.0cm 0.3cm},clip,scale=0.4]{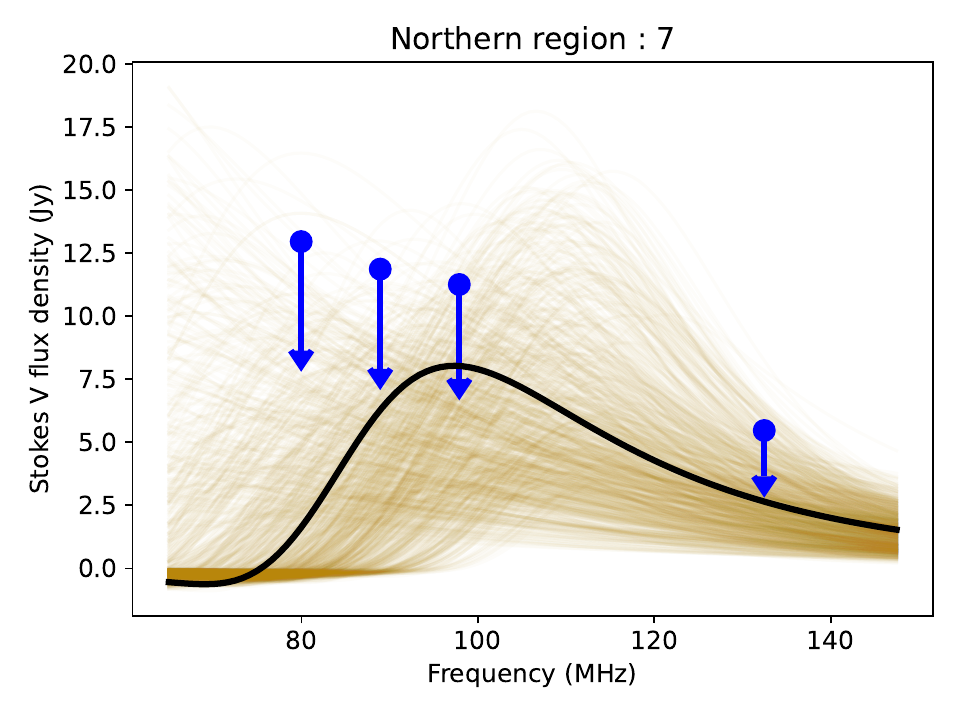}
      
    \caption[Observed and fitted spectra for regions 4, 6, and 7 of northern CME.]{Observed and fitted spectra for regions 4, 6, and 7 of CME-1. {\it First column: }Stokes I spectra are shown. Red points represent the observed flux densities. {\it Second column: }Stokes V spectra are shown. Blue points represent the upper limits at each of the frequencies. The black lines represent the Stokes I and V GS spectra corresponding to GS parameters reported in Table \ref{table:north_params}. Light yellow lines show the GS spectra for 1000 realizations chosen randomly from the posterior distributions of the GS model parameters. Sample posterior distributions for region 4 are shown in Figure \ref{fig:corner_northern_4}.}
    \label{fig:spectra2}
\end{figure}

The regions for which good spectral sampling was obtained are marked in red and cyan in the right panel of Figure \ref{fig:spectral_regions} their spectra are shown in Figures \ref{fig:spectra1} and \ref{fig:spectra2}. The first columns in these figures show the Stokes I spectra and the second columns show the Stokes V spectra. The black lines represent the GS spectra corresponding to the median values of the posterior distributions of GS model
\begin{landscape}
\begin{table}
\centering
    \renewcommand{\arraystretch}{1.5}
    \begin{tabular}{|p{1.2cm}|p{2cm}|p{1.5cm}|p{1.5cm}|p{1.5cm}|p{2cm}|p{1.8cm}|p{1.5cm}|p{1.4cm}|p{1.4cm}|}
    \hline
       Region No. & Heliocentric \newline{Distance} & B (G) & $\delta$ & $ A \times 10^{20}$\newline{$(cm^{2})$} & $E_\mathrm{min}$ (keV) & $\theta$ \newline{(degrees)} & $L\ (R_\odot)$ & $n_\mathrm{thermal}$ \newline{$\times 10^6$}\newline{$(cm^{-3})^*$}  & $n_\mathrm{nonth}$\newline{$\times 10^4$} \newline{$(cm^{-3})^*$} \\ \hline \hline 
        1 & 2.5 & $4.55_{-0.99}^{+1.14}$ & $5.34_{-1.22}^{+2.28}$ & $8.01_{-3.09}^{+6.80}$ & $19.97_{-10.78}^{+22.31}$ &  $71.47_{-11.97}^{+10.92}$  & $0.78^*$ &1.5 & 1.5\\
        \hline
        2 & 2.5 & $1.28_{-0.34}^{+0.40}$  & $6.40_{-1.13}^{+1.31}$  & $2.43_{-0.61}^{+0.87}$  & $178.72_{-72.49}^{+10.94}$  & $59.23_{-12.60}^{+15.81}$ & $1.52_{-0.93}^{+1.70}$ & 1.5 & 1.5\\
        \hline
        3 & 2.5 & $1.44_{-0.28}^{+0.30}$ & $4.92_{-0.63}^{+0.77}$ & $2.94_{-0.56}^{+0.67}$ & $104.66_{-34.50}^{+50.86}$ & $78.64_{-5.69}^{+4.61}$ & $0.97_{-0.47}^{+0.91}$ & 1.5 & 1.5\\
        \hline
        4 & 2.5 & $1.39_{-0.38}^{+0.50}$ & $4.86_{-0.62}^{+0.85}$ & $4.87_{-1.52}^{+3.06}$ & $75.67_{-34.29}^{+53.35}$ & $69.82_{-15.10}^{+13.15}$ & $0.98^*$ &1.5 & 1.5\\
       \hline
       5 & 3.0 & $<1.28$ & $1.68^*$ & $9.57^*$ & $122.45^*$ & $65.94^*$ & $1.01^*$ & 0.7 & 0.7\\
       \hline
        6 & 3.0 & $1.27_{-0.34}^{+0.45}$ & $5.21_{-0.53}^{+0.65}$ & $10.03_{-3.55}^{+8.43}$ & $95.13_{-36.98}^{+54.87}$ & $67.19_{-10.12}^{+12.19}$ & $1.09^*$ & 0.7 & 0.7\\
       \hline
       7 & 3.0 & $1.99_{-0.87}^{+0.71}$ & $6.76_{-1.82}^{+2.16}$ & $5.80_{-2.39}^{+3.24}$ & $122.45^*$ & $54.03_{-15.03}^{+22.35}$ & $0.95^*$ & 0.7 & 0.7\\
       \hline
        8 & 3.0 & $<1.42$ & $2.14^*$ & $9.57^*$ & $122.45^*$ & $65.94^*$ & $1.40^*$ & 0.7 & 0.7\\
       \hline
    \end{tabular}
    \caption[Estimated plasma and GS source parameters of northern CME.]{\textbf{Estimated plasma and GS source parameters of CME-1.} These parameters are estimated for 01:24:55 UTC. Parameters marked by $^*$ are kept fixed during the fitting.}
    \label{table:north_params}
\end{table}
\end{landscape}
\noindent parameters presented in Table \ref{table:north_params}. The reduced $\chi^2$ ($\chi^2_\mathrm{red}$) for each spectrum is listed in the corresponding Stokes I panels. Stokes V model spectra always lie below the upper limits. 

Estimated plasma parameters are listed in Table \ref{table:north_params}. The parameters marked by stars are kept fixed to the mentioned values. As an example, posterior distributions of fitted parameters for region 3 are shown in Figure \ref{fig:corner_northern}. Distributions of all of the physical parameters show unimodal and sharply peaked clusters. Those for Region 2 also show similar behavior.

\begin{figure}
    \centering
    \includegraphics[trim={0.5cm 0.6cm 0cm 0cm},clip,scale=0.46]{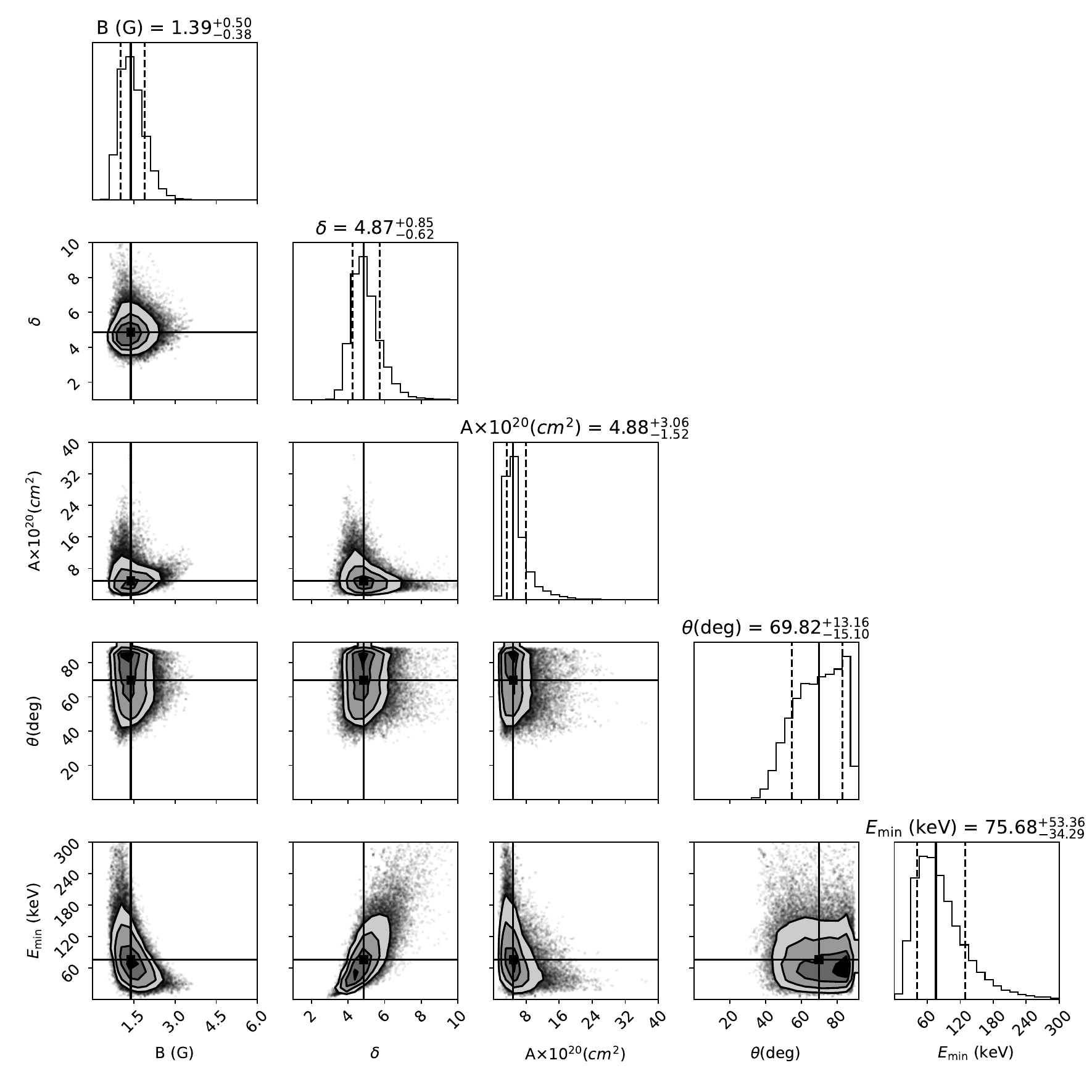}
    \caption[Correlation of posterior distributions of GS model parameters for region 4.]{Correlation of posterior distributions of GS model parameters for region 4. 2-dimensional plots show the joint probability distribution of any two parameters. The contours are at 0.5, 1, 2, and 3$\sigma$. The solid lines in the 1-dimensional histogram of posterior distributions mark the median values, and the vertical dashed lines mark the 16$^\mathrm{th}$ and 84$^\mathrm{th}$ percentiles. The median values are also marked in the panels showing the joint probability distribution.}
    \label{fig:corner_northern_4}
\end{figure}

For regions 4 and 6, the observations do not sample the spectral peak. As discussed in Section \ref{sec:spectrum_sensitivity}, the variation in $A$ impacts only the Stokes I peak flux density, while other free parameters impact the location of the spectral peak as well as the spectral shape for both Stokes I and V. The spectral peak depends on several parameters; $B,\ \theta,\ A, E_\mathrm{min}$ and $n_\mathrm{nonth}$ as evident from Figures \ref{fig:param_sensitivity_magnetic}, \ref{fig:param_sensitivity_geometric} and \ref{fig:param_sensitivity_non_thermal}. Among these, $A,\ E_\mathrm{min}$ and $n_\mathrm{nonth}$ do not have any significant impact on the optically thin part of the Stokes V spectra, while $B$ and $\theta$ have significant impacts on the optically thin part of the Stokes V spectra. Hence, the upper limits of Stokes V provide constraints on the $B$ and $\theta$. As the peak flux density is not known for these two regions, $A$ remains poorly constrained. This is evident from the posterior distribution of the parameters for region 4 shown in Figure \ref{fig:corner_northern_4}, though the ability to constrain the other parameters is \ not \ compromised much. Similar is the case for Region 6. For region 7 marked by cyan in the right panel of Figure \ref{fig:spectral_regions}, although the spectrum samples the peak, it only has four spectral points. It is, hence, not reasonable to fit five free parameters to this spectrum. For this reason, we have kept the $E_\mathrm{min}$ fixed to a value close to that obtained for the adjacent region 6 and only fitted the other four parameters, $B$, $\delta$, $A$, and $\theta$. 

\section{Estimation of Plasma Parameters for Poorly Sampled Spectra}\label{subsec:magnetic_green}
\begin{figure}[!ht]
    \centering
    \includegraphics[trim={0.3cm 0.4cm 0cm 0cm},clip,scale=0.6]{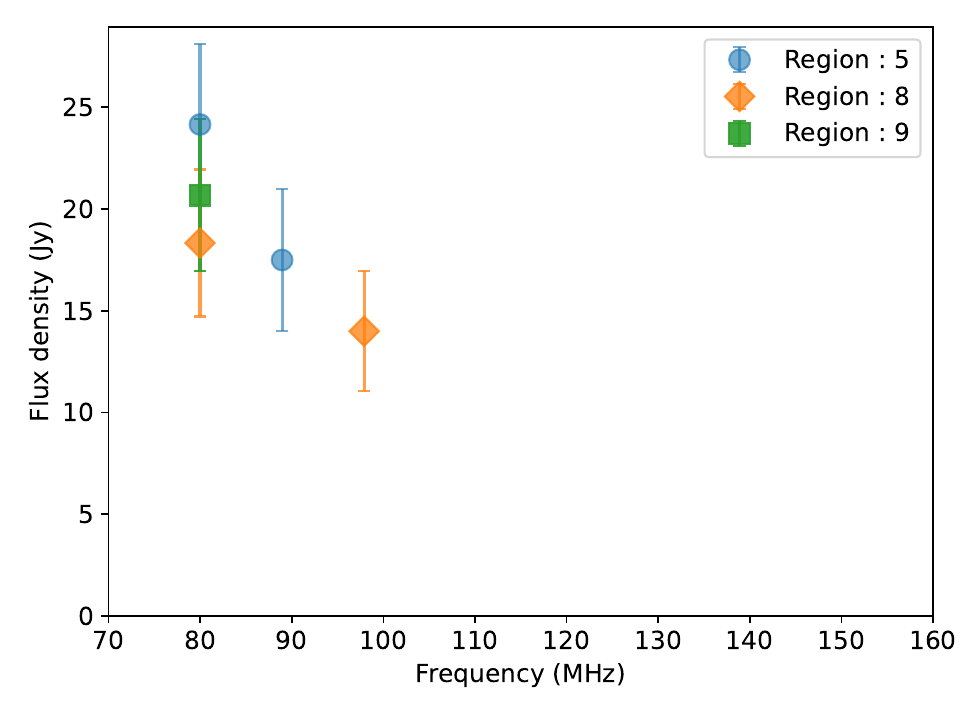}
    \caption[Observed spectra of the green PSF-sized regions for the northern CME.]{Observed spectra of the green PSF-sized regions for the CME-1. We have not done the fitting to these spectra because the number of spectral points is less than four.}
    \label{fig:green_region_spectra}
\end{figure}

The spectra for regions 5, 8, and 9 are shown in Figure \ref{fig:green_region_spectra}. The emission from Region 9 is detected only at one frequency, 80 MHz, and cannot be modeled. Regions 5 and 8 have only two spectral points, too few for the GS spectral modeling approach adopted in the earlier section. For these regions, I follow a different approach to estimating GS parameters. 

As for other regions, the thermal electron densities are available independently from the inversion of LASCO-C2 coronagraph image (Figure \ref{fig:electron_density}). For the non-thermal electron distributions, I use values estimated from the adjacent red regions, which lie at the same heliocentric height. Values of $n_\mathrm{nonth}$ are set to 1\% of $n_\mathrm{thermal}$, while $E_\mathrm{min}$, and $\theta$ are set to those determined for the nearby region 6. From the two spectral points for regions 5 and 8 shown in Figure \ref{fig:green_region_spectra}, it is evident that the spectra are in an optically thin part. Following \citet{Dulk1982,Carley2017}, $\delta$ for these regions is estimated using the spectral index ($\alpha_\mathrm{thin}$) of the optically thin part of spectrum as,
\begin{equation}
    \delta=|-1.1(\alpha_\mathrm{thin}-1.2)|,
    \label{eq:delta_from_alpha}
\end{equation}
where the optically thin part of the spectrum is given by $S(\nu)=\ S_\mathrm{peak}(\nu/\nu_\mathrm{peak})^{\alpha_\mathrm{thin}}$ and $S_\mathrm{peak}$ is the peak flux density. Thus estimated values of $\delta$ for regions 5 and 8 are $1.68$ and $2.14$, respectively. 

The simplified expression for $\nu_\mathrm{peak}$ \citep{Dulk1982} is valid under the assumptions I have already been making -- a power-law distribution of non-thermal electron and a homogeneous GS source. This expression is accurate for limited ranges of GS model parameters -- $\theta$ between $\sim$20$^\circ$ and $\sim$80$^\circ$, $\delta$ between $\sim$2 and $\sim$7, and $E_\mathrm{min}$ between $\sim10$ keV to $\sim1$ MeV. The estimated values of $\theta$, $E_\mathrm{min}$, and $\delta$ from regions 1 through 6 lies in the permissible ranges. Assuming identical values for regions 5 and 8 suggests that $B$ can be estimated using the simplified expression for $\nu_\mathrm{peak}$ given as \citet{Dulk1982},
\begin{equation}
\begin{split}
    \nu_\mathrm{peak}=&2.73\times10^3\ 10^{0.27\delta}\ (sin\theta)^{0.41+0.03\delta}\times\\&(n_\mathrm{nonth}L)^{0.32-0.03\delta}\ B^{0.68+0.03\delta},
\end{split}
\label{eq:dulk_and_marsh}
\end{equation}
To determine $B$ using this analytical expression, one needs to provide the values of $n_\mathrm{nonth}$, $L$, and $\theta$. These parameters are set to their values estimated for adjacent regions. Since the peak of the spectra is not sampled for these regions, only an upper limit on $B$ can be estimated, which is listed in Table \ref{table:north_params}.

\section{Discussion}\label{sec:discussion_cme1}
This work presents a spatially resolved spectropolarimetric modeling of faint GS emission from a CME. As mentioned in Section \ref{sec:challenges} of Chapter \ref{chapter_intro}, there are two challenges in using GS emission to estimate magnetic field entrained in CME plasma -- observational challenge and modeling challenge. In this work, a challenging CME data set was chosen, which comes from 2014 May 04 when the MWA was pointed to its lowest permissible elevation, where the sensitivity of the MWA is the poorest. Even for this challenging dataset, Stokes I GS emission is detected covering the entire white-light structure of CME-1. This demonstrates the capability of the MWA to overcome observational challenges in detecting GS emissions from CME plasma. Hence, it is possible to detect GS emission, whenever it is present, from strong events observed at more favorable observing conditions.

As discussed in Section \ref{subsec:joint_fitting}, the phase space of GS model parameters explored here has been motivated by physical arguments and earlier studies. This detailed exploration of the GS model parameter space allowed us to identify the underlying degeneracies between parameters. Earlier studies only used the Stokes I spectrum to estimate CME plasma parameters under certain assumptions. Here, the focus was to go one step ahead to address the challenges in spectrum modeling of GS emission for the estimation of CME plasma parameters.

Although the event was analyzed with the aim of detecting the Stokes V emission, the current sensitivity did not allow us to detect Stokes V emission from CME-1. While jointly used with Stokes I spectrum, non-detection of Stokes V does not make the observations useless, but improves the credibility of the estimated parameters of a physics-based GS emission model compared to the case when no Stokes V constraints are available at all. The robust polarization calibration, high-fidelity and high DR imaging capabilities of P-AIRCARS, and excellent snapshot PSF allow us to provide a sensitive upper limit on the Stokes V emission. Including these upper limits along with Stokes I spectrum significantly reduces the spread in the distribution function of the model parameters and breaks some of the degeneracies in the GS model, compared to the situation when only Stokes I spectrum has been used. This section quantifies this improvement and also the benefits of the improved methodology used here as well as the limitation of the current study which can be improved further.
\begin{figure*}[!htbp]
    \centering
    \includegraphics[trim={0.5cm 0.6cm 0cm 0cm},clip,scale=0.46]{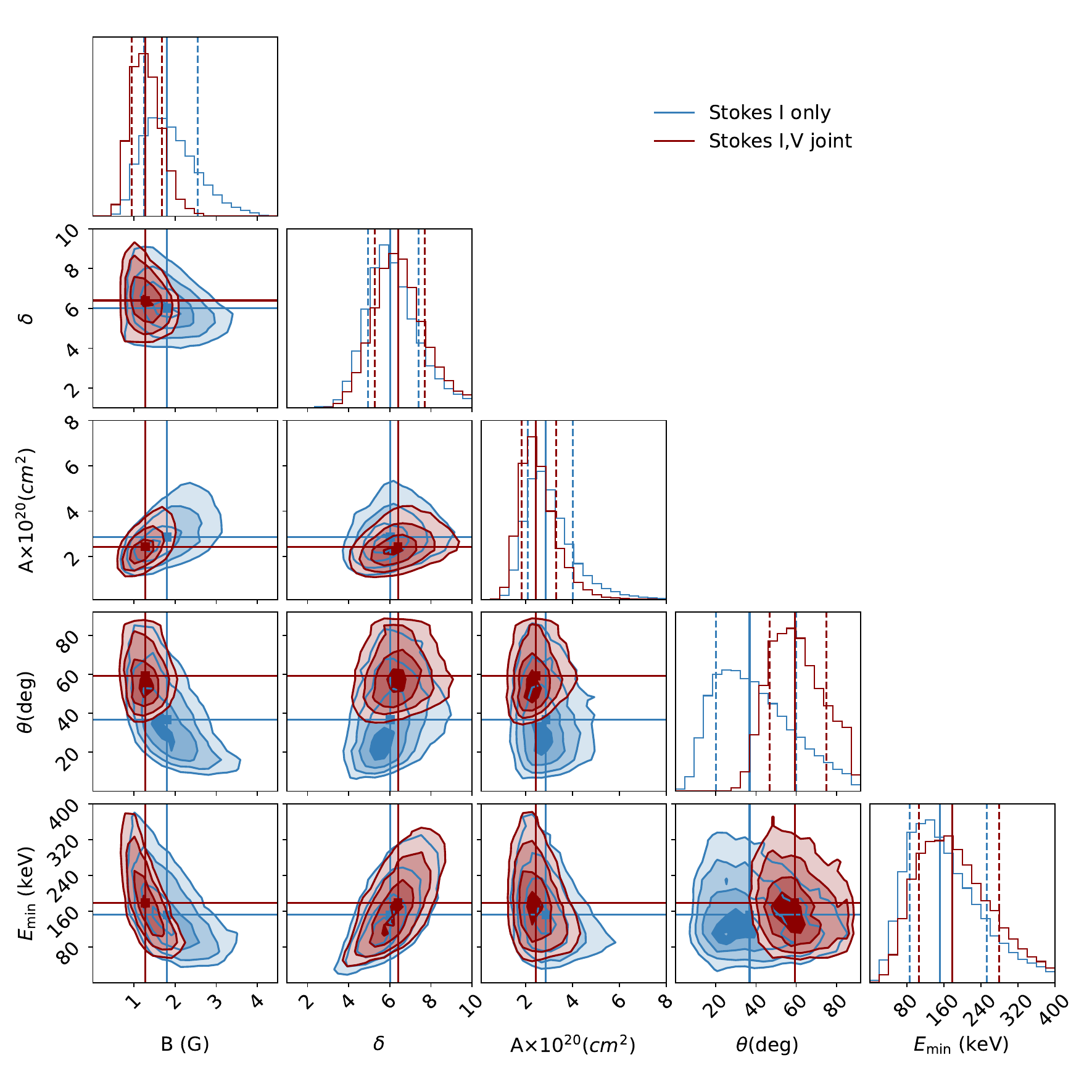}
    \caption[Comparison between Stokes I only and Stokes I, V joint modeling for region 2.]{Comparison between Stokes I only and Stokes I, V joint modeling for region 2. 2-dimensional plots show the joint probability distribution of any two parameters. The contours are at 0.5, 1, 2, and 3$\sigma$. Blue contours represent the posterior distribution of parameters using only the Stokes I spectrum. Maroon contours represent the posterior distribution using the joint Stokes I and V spectrum. $B$ and $\theta$ are better constrained for Stokes I and V joint fitting. The vertical dashed lines mark the 16$^{th}$ and 84$^{th}$ percentiles. The median values are also marked in the panels showing the joint probability distribution.}
    \label{fig:combine_corner}
\end{figure*}

\subsection{Advantages of Using Stokes V Spectra Jointly With Stokes I Spectra}\label{subsec:importance_stoks_V}
The sensitivity of the Stokes V spectra to the physical parameters of the GS model has already been demonstrated in Section \ref{subsec:stokesV_sensitivity}. Even though the present work only uses upper limits on Stokes V emission, it already leads to a better-constrained of GS model parameters. The use of stringent Stokes V upper limits jointly with Stokes I spectra enables us to exclude the part of the parameter space of GS models, which is consistent with the Stokes I spectra but not with the Stokes V upper limits. To substantiate this, I compare the posterior distribution of parameters obtained using only Stokes I constraints (shown in blue in Figure \ref{fig:combine_corner}) with those obtained using joint constraints from Stokes I and V measurements (shown in maroon in Figure \ref{fig:combine_corner}). To keep the number of free parameters below the number of constraints available for Stokes I only modeled and do an apples-to-apples comparison, $L$ was fixed to the value mentioned in Table \ref{table:north_params}. The significant improvement in the ability to constrain $\theta$, $B$, and $E_\mathrm{min}$ is self-evident in Figure \ref{fig:combine_corner}. Examining the ranges spanned by the vertical dashed lines marking the 16$^{th}$ and 84$^{th}$ percentiles shows that the uncertainties in the estimates of $\theta$ have reduced by $\sim$44\% each and that in $B$ by $\sim$30\% on using joint Stokes I and V modeling.

\subsection{Importance of Sampling the Spectral Peak}\label{subsec:importance_peak}
A crucial feature of the spectrum is its peak, an accurate determination of which robustly constraints 
several GS model parameters. The spectral peak depends on several GS model parameters; $B,\ \theta,\ A, E_\mathrm{min},\ L$ and $n_\mathrm{nonth}$. As discussed in Section \ref{sec:spectrum_sensitivity}, changes in $A$ only impact the peak flux density and leave the fractional Stokes V spectra unchanged (Figure \ref{fig:param_sensitivity_geometric}). The observed spectra for regions 4 and 6 do not sample the spectral peak, and this leads $A$ to be poorly constrained (Figure \ref{fig:corner_northern_4}). By contrast in the cases where the spectral peak has been sampled (regions 2 and 3), the uncertainty in $A$ is lower by about an order of magnitude. Not only $A$, but the unavailability of the spectral peak affects the uncertainty of other parameters as well.   

\subsection{Prediction of the Presence or Absence of Stokes V Emission}\label{subsec:stokesV_presence}
\begin{figure*}
    \centering
    \includegraphics[trim={0.3cm 0.5cm 0.0cm 0.3cm},clip,scale=0.42]{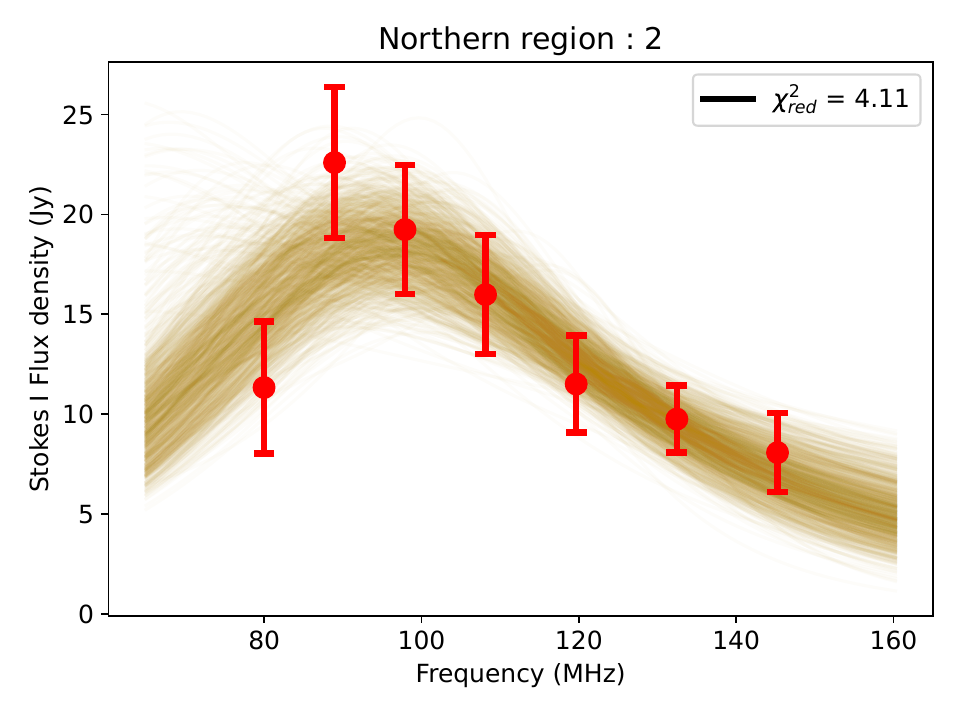}\includegraphics[trim={0.3cm 0.5cm 0.0cm 0.3cm},clip,scale=0.42]{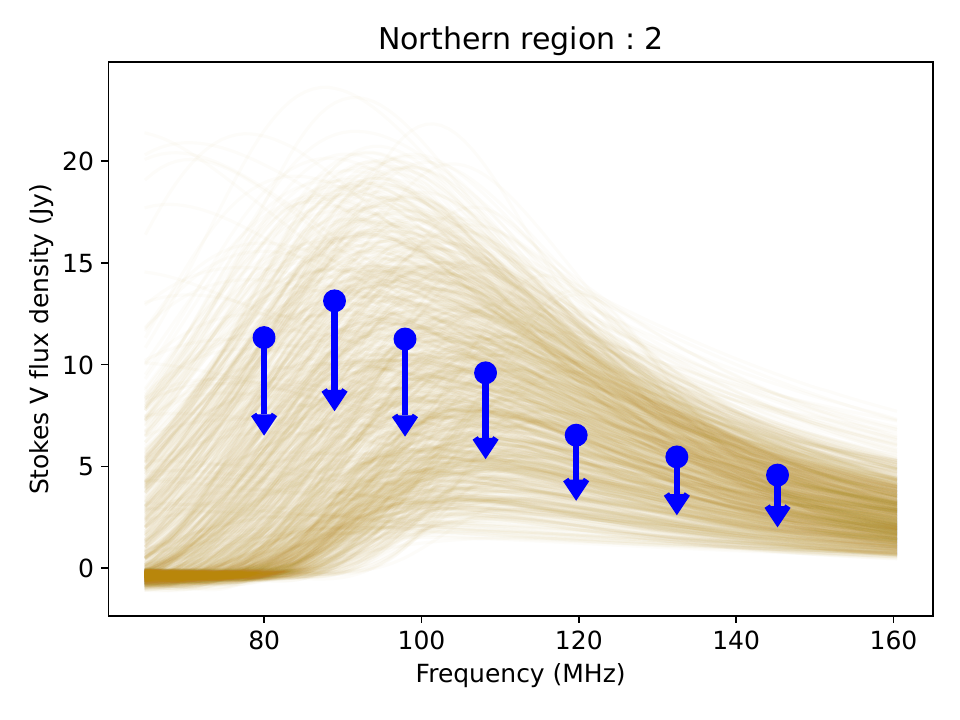}
     \caption[Observed and fitted spectra for region 2 only using Stokes I spectrum.]{Observed and fitted spectra for region 2 only using Stokes I spectrum. {\it Left panel: }Stokes I spectra are shown. Red points represent the observed flux densities. {\it Right panel: }Stokes V spectra, computed based on constraints from only Stokes I observations, are shown. Blue points represent the upper limits at each of the frequencies. Light yellow lines show the GS spectra for 1000 realizations chosen randomly from the posterior distributions shown by blue contours in Figure \ref{fig:combine_corner}.}
    \label{fig:stokesI_only}
\end{figure*}

The only firm statement that can be made when using the stringent Stokes V upper limits is that the Stokes V emission cannot exceed the upper limits. It is, however, not possible to comment on the presence or absence of Stokes V emission based only on the upper limit. Irrespective of the availability of Stokes V observations, the parameter space of a GS model can be constrained only using Stokes I observations and can predict the  Stokes V spectra corresponding to the constrained parameter space. These GS spectra predict Stokes V emission to be non-zero across most of the spectral bands being modeled. This is illustrated in Figure \ref{fig:stokesI_only}. As evident, there is not a single Stokes V spectrum that is uniformly zero at all frequencies both in optically thick and thin parts of the spectrum.

\subsection{Filling Factor of GS Source}\label{subsec:filling_factor}
Earlier studies did not use any physically motivated constraints on $L$. For lack of a better estimate, typically $L$ was fixed to the value of the PSF diameter \citep{Mondal2020a, Vourlidas2020} assuming spherical symmetry. The validity of this assumption was never tested. The CME studied here was chosen specifically to have coronagraph observations from multiple vantage points. This enabled us to build a detailed and well-constrained three-dimensional model for it and use it to estimate an upper limit on $L$ (Section \ref{subsec:estimate_gcs}). The major axis of the PSF for the current observation is $\sim9\times10^{10}$ cm, while the estimated values of $L$ vary between $\sim3\times10^{10}$ and $10\times10^{10}$ cm (Table \ref{table:north_params}). These values are close to that taken by M20, but smaller or equivalent to the PSF size. 

Estimated values of $A$ are of order $10^{20}-10^{21}\ \mathrm{cm^{2}}$. The area of PSF at the lowest observing frequency is $\sim10^{22}\ \mathrm{cm^{2}}$. This leads to an areal filling fraction of 0.01$-$0.1. Assuming that the filling fractions in the sky plane and along the LoS are similar, M20 concluded that either the non-thermal electrons have a small filling fraction and/or the emission comes from regions of the concentrated magnetic field. The presence of such regions has been suggested under the names of magnetic knots in the literature \citep{Karpen2012}. Having an independent estimate of $L$ from GS modeling and the $L_\mathrm{geo}$ from geometric modeling of the CME enables us to compute the volumetric filling factor, $f$, without relying on the assumption of the filling factor in the plane of the sky and along LoS being the same. $f$ is defined as,
\begin{equation}
    f=\frac{AL}{A_\mathrm{PSF}L_\mathrm{geo}}
    \label{eq:filling_frac}
\end{equation}
where, $A_\mathrm{PSF}$ is the area of the PSF. The average volumetric filling factor of the GS source for the CME under study turns out to be $\sim$0.1$-$1\%. The low value of $f$ obtained here is consistent with the one that arrived at by M20.

\subsection{The Path Forward}\label{sec:limitation_cme1}
While this work represents an advancement from earlier works, it is limited by the fact that it relies on Stokes V upper limits rather than actual measurements. Naturally Stokes V measurements would lead to even tighter constraints on the GS models. Additionally, as shown earlier, sampling the peak of the GS spectrum is essential for constraining the GS model parameters. This spectral peak can lie over a large frequency range depending on the plasma parameters of the CME at different heights. The new generation instruments span a large frequency range across them and also offer sufficient sensitivity and imaging quality to be able to pursue this science better.

\section{Conclusion}\label{sec:conclusion_cme1}
Since the first detection and attempt of modeling of the GS emission from CME by \cite{bastian2001}, radio emissions from CME plasma have been detected only for a handful of fast CMEs. M20 presented the first detection of GS radio emissions from a slow CME. The flux densities of radio emission reported by M20 and the present work are among the lowest reported. These works furnish further evidence that the earlier non-detections of GS emission from slow CMEs can be attributed to the limited DR achieved in those attempts and that these limitations can now be overcome with the high DR imaging yielded by the combination of data from instruments like the MWA and imaging pipelines like P-AIRCARS \citep{Kansabanik_principle_AIRCARS,Kansabanik2022_paircarsI, Kansabanik_paircars_2}. 

Even with detection of CME GS emissions whenever it is present, the limited number of spectral points at which measurements are typically available, in contrast with the large numbers of GS model parameters and the degeneracies between some of them pose significant complications. These issues force one to seek independent estimates for some of the model parameters and assume physically motivated values for others. This has, in the past, limited the robustness of the GS model parameter estimates and, hence, the usefulness of this approach to arrive at conclusive estimation of CME magnetic fields. 

This work uses a homogeneous source model and the GS model parameter phase space explored here has been motivated by physical arguments and earlier studies (Section \ref{subsec:joint_fitting}). Under these assumptions, it presents a detailed quantitative analysis of the sensitivity of the observed Stokes I GS spectra to the various model parameters and the degeneracies present. It also demonstrates that Stokes V spectra have a different dependence on GS model parameters than Stokes I spectra and can be used effectively to break many of these degeneracies.

This work uses both Stokes I and V spectra for constraining the GS model parameters. Even though only sensitive upper limits on Stokes V spectra are available, they are still useful to reduce the uncertainty in the model parameters of most interest ($B$ and $\theta$) by as much as $\sim$40\% while jointly used with Stokes I spectrum. It has also been found that for the GS model parameters to be well constrained, the peak of the GS spectrum must be included in the observed part of the spectrum. Another aspect of this work is a demonstration of the usefulness of a good geometric model of the CME for determining the volume filling factor of GS emission and estimates it to be $\sim$0.1$-$1\%. Constraining the geometric model parameters requires coronagraph observations from multiple vantage points. 

This work marks the next step beyond the earlier attempts of estimating CME plasma parameters using GS emission. It also demonstrates the usefulness of upper limits when used appropriately in conjunction with other available constraints. Based on the results from present-day instruments like the MWA, there is no doubt that the even more sensitive and wider bandwidth spectropolarimetric imaging from the upcoming instruments, like the Square Kilometre Array Observatory \citep[SKAO;][]{SKAO2021}, the Next Generation Very Large Array \citep[ngVLA:][]{ngVLA2019}, and the Frequency Agile Solar Radiotelescope \citep[FASR:][]{Gary2003,Bastian2005,Bastian2019,Gary2022_FASR}; aided by the multi-vantage point coronagraph observations will provide a routine and a robust remote sensing technique for estimating CME plasma parameters spanning a large range of coronal heights.

\chapter {Indication of Insufficiency of Homogeneous Gyrosynchrotron \\Model}
\label{cme_gs2}

Coronal mass ejections (CMEs) are large-scale magnetized structures showing inhomogeneity in terms of density and magnetic field \citep{Mishra2015,Owens2017,Song2021}. Despite their obvious limitations, all of the earlier studies have assumed homogeneous and isotropic distributions of plasma parameters when modeling the CME gyrosynchrotron (GS) spectra \citep{Boischot1957,Dulk1973,bastian2001,Maia2007,Tun2013,Carley2017,Mondal2020a}. These assumptions are made out of necessity, as the number of observational constraints available is grossly insufficient for constraining the much larger number of model parameters required to describe an inhomogeneous GS model. Additionally, these simple models have been able to fit the GS spectra observed in all of the attempts so far, and hence deemed sufficient. In this chapter, I present the GS modeling study associated with a different CME, which is marked as CME-2 in Chapter \ref{cme_gs1}. This study presents robust detection of Stokes V CME GS emission.  When modeling the Stokes V detection simultaneously with Stokes V upper limits and Stokes I spectrum, this study finds that there is no reasonable GS model consistent with the data. We explore the possibility of this situation arising because of a violation of the various assumptions made by this model and commonly employed in all prior works. 

\section{Introduction}
Observations at radio wavelengths provide a few different useful methods to measure the magnetic fields of the CMEs at the coronal and heliospheric heights. At the coronal heights GS emission from mildly relativistic electrons gyrating in the CME magnetic field \citep[e.g.,][etc.]{bastian2001,Tun2013,Mondal2020a} is one of the few methods which can measure the CME-entrained magnetic fields. Since the first imaging detection and modeling by \cite{bastian2001}, there have been only a handful of studies that have successfully managed to detect the GS emission from CME loops \citep{Maia2007,Mondal2020a}. This scenario has changed over the past few years with the availability of high dynamic range (DR) imaging observations from the MWA. Though the sample is statistically small, these data have shown the presence of Stokes I GS emission from all of the CMEs that have been examined thus far \citep{Mondal2019,Kansabanik2023_CME1}.

Even with the routine and reliable detection of GS emission from CMEs, estimating the plasma parameters from the observed GS spectrum remains challenging. The GS model requires ten free parameters even for the simplest homogeneous and isotropic plasma distributions with a single power-law energy distribution of non-thermal electrons \citep{Fleishman_2010,Kuznetsov_2021}. Constraining all of these GS model parameters only using the total intensity (Stokes I) spectrum is not possible and requires several assumptions to be made. Using the high-fidelity and high DR spectropolarimetric imaging with the MWA, \cite{Kansabanik2023_CME1} demonstrated that even the availability of strong upper limits of the Stokes V measurements along with the Stokes I spectrum can significantly improve the constraints on the GS model parameters and lift some of the degeneracies in the model parameters.

As additional observational constraints become available, using high-fidelity spectropolarimetric imaging provided by P-AIRCARS, one expects to be able to better constrain the GS models. Contrary to this, one finds that no model in the reasonable part of the solution phase space is able to meet all of the observational constraints imposed by the data. This situation, where less constraining data leads to a good model fit and more constraining data does not, strongly suggests that one should examine the model and the assumptions it makes. It is well known from magnetic flux-rope models \citep[e.g.,][etc.]{Isavnin_2016,Mostl2018} that CME plasma parameters are not expected to be homogeneous along the LoS. Nonetheless, they have always been assumed to be so in all prior works studying CME GS emission. In fact, these simplifying assumptions have been essential because the constraints available from the observations are not enough to constrain the more detailed and physically meaningful GS source models which require many more free parameters. Among other things, this chapter presents a systematic, though limited, study of the impact of the violation of the assumption homogeneity on the observed GS spectra.

This chapter is organized as follows -- Section \ref{sec:obs} describes the observation and the data analysis. The imaging results are presented in Section \ref{sec:results}, along with the discussion about the origin of the radio emission. Section \ref{sec:spectra_model} describes spectrum modeling using a homogeneous GS model. Validity of the homogeneous and isotropic assumptions of GS emission to model the observed spectra is discussed in Section \ref{subsec:homo_insuff} followed by simulations to explore the effects of inhomogeneity of plasma parameters along the LoS in Section \ref{sec:nonuniform}. Section \ref{sec:discussion_cme2} presents a discussion followed by the conclusions in Section \ref{sec:conclusion_cme2}.

\begin{figure}[!ht]
    \centering
     \includegraphics[trim={1.2cm 6cm 1.2cm 6.5cm},clip,scale=0.5]{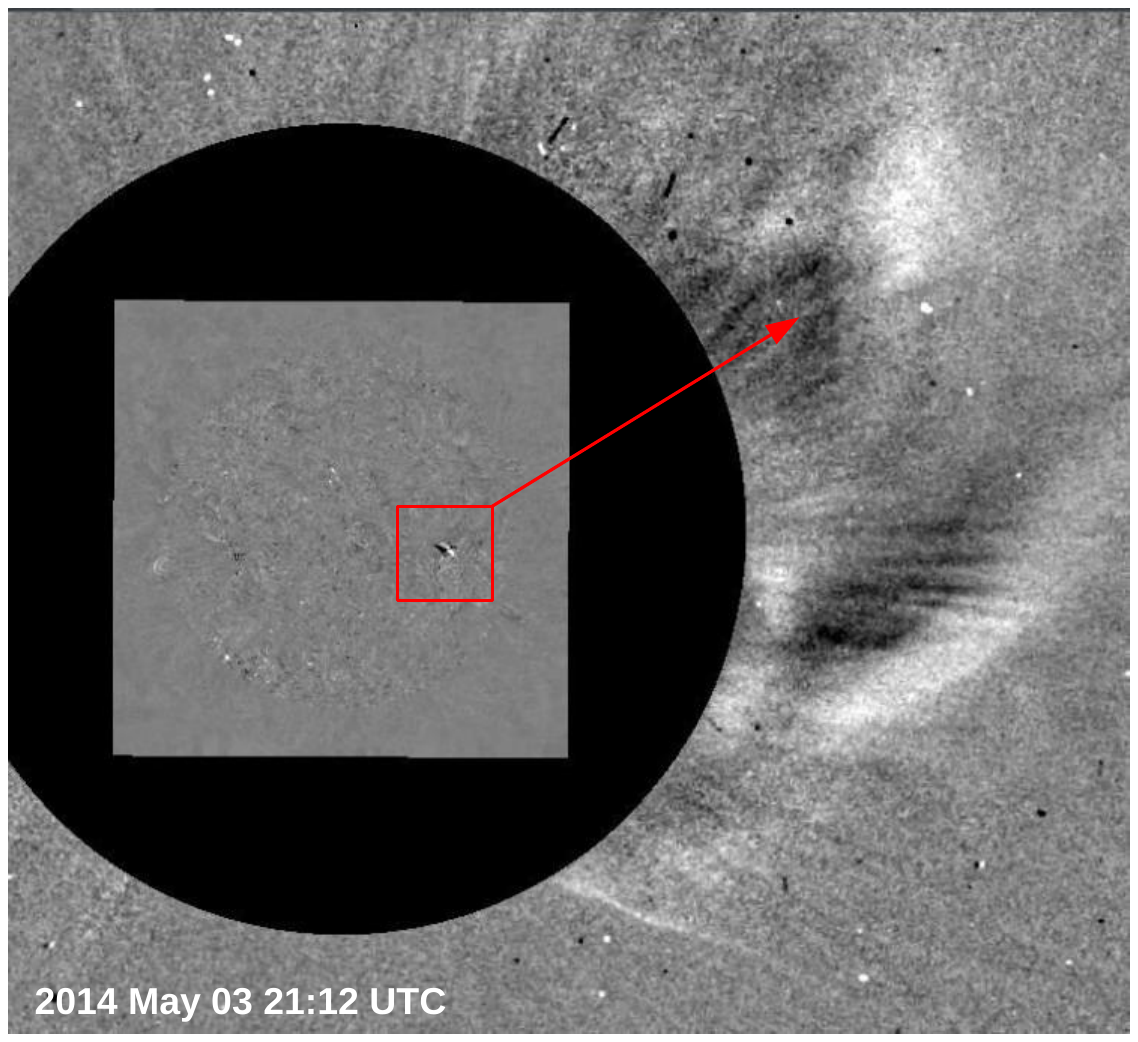} 
     \caption[Eruption of CME-2 observed using SDO/AIA spacecraft.]{Eruption of CME-2 observed using SDO/AIA spacecraft. CME-2 erupted from the visible part of the solar disc. A composite base difference image from the Atmospheric Imaging Assembly (AIA) onboard SDO at 94 \AA\ and LASCO C2 coronagraph image onboard the SOHO spacecraft is shown. The red box shows the active region 12047, which is the eruption site for CME-2, and the red arrow shows the propagation direction.}
    \label{fig:south_cme_eruption}
\end{figure}

\section{Observation and Data Analysis}\label{sec:obs}
The observation presented here were made on 2014 May 04 using the MWA and a detailed description is provided in Section \ref{sec:obs_and_data} of Chapter \ref{cme_gs1}. Among the two CMEs seen to be overlapping in time in these data, in this chapter I focus on the CME propagating towards the south-western direction (CME-2). A detailed spectropolarimetric imaging and modeling study of the CME propagating towards solar north (CME-1) was presented in Chapter \ref{cme_gs1}. 

\subsection{Eruption and Evolution of CME-2}\label{subsec:evolution}
The CME-2 erupted from active region 12047 present on the visible part of the Sun. The eruption site is marked by a red box in Figure \ref{fig:south_cme_eruption}. CME-2 first appeared in the field-of-view (FoV) of the C2 coronagraph of the Large Angle Spectroscopic Coronagraph \citep[LASCO;][]{Brueckner1995} onboard the Solar and Heliospheric Observatory \citep[SOHO;][]{Domingo1995} at 20:48 UTC on 2014 May 03. It was visible in LASCO C3 coronagraphs until 02:06 UTC on 2014 May 04 up to about 17 $R_\odot$. The CDAW catalogue\footnote{\url{https://cdaw.gsfc.nasa.gov/CME_list/UNIVERSAL/2014_05/univ2014_05.html}} reported CME-2 as a partial halo CME.

\subsection{Radio Observation and Data Analysis}\label{subsec:radio_analysis}
CME-2 was observed at meter-wavelength radio bands using the MWA on 2014 May 04 from 00:48 UTC to 07:32 UTC under the project ID  G0002\footnote{\url{http://ws.mwatelescope.org/metadata/find}}. The MWA observations were done in 12 frequency bands, each of width 2.56 MHz, and centered around 80, 89, 98, 108, 120, 132, 145, 161, 179, 196, 217, and 240 $\mathrm{MHz}$. The temporal and spectral resolution of the data were 0.5 $\mathrm{s}$ and 40 $\mathrm{kHz}$, respectively. CMEs are often associated with a variety of active solar emissions -- type-II, -III, and/or -IV radio bursts \citep{Gopalswamy2011_CME_radio,Carley2020}. As discussed in Section \ref{subsec:radio_data_analysis} of Chapter \ref{cme_gs1}, no solar radio bursts were reported on this day. Polarization calibration and full Stokes imaging of the MWA observations were done using P-AIRCARS. Integration of 10 s and 2.56 MHz was used for imaging for all 12 frequency bands. All polarization images follow the IAU/IEEE convention of Stokes parameters \citep{IAU_1973,Hamaker1996_3}. 

\section{Results}\label{sec:results}
This section presents the results from the wideband spectropolarimetric imaging observation of CME-2 using the MWA and the possible mechanisms which can give rise to it.

\subsection{Radio Emission from CME-2}\label{subsec:radio_emission_CME2}
\begin{figure}[!ht]
    \centering
    \includegraphics[trim={1.5cm 0.3cm 2cm 0cm},clip,scale=0.8]{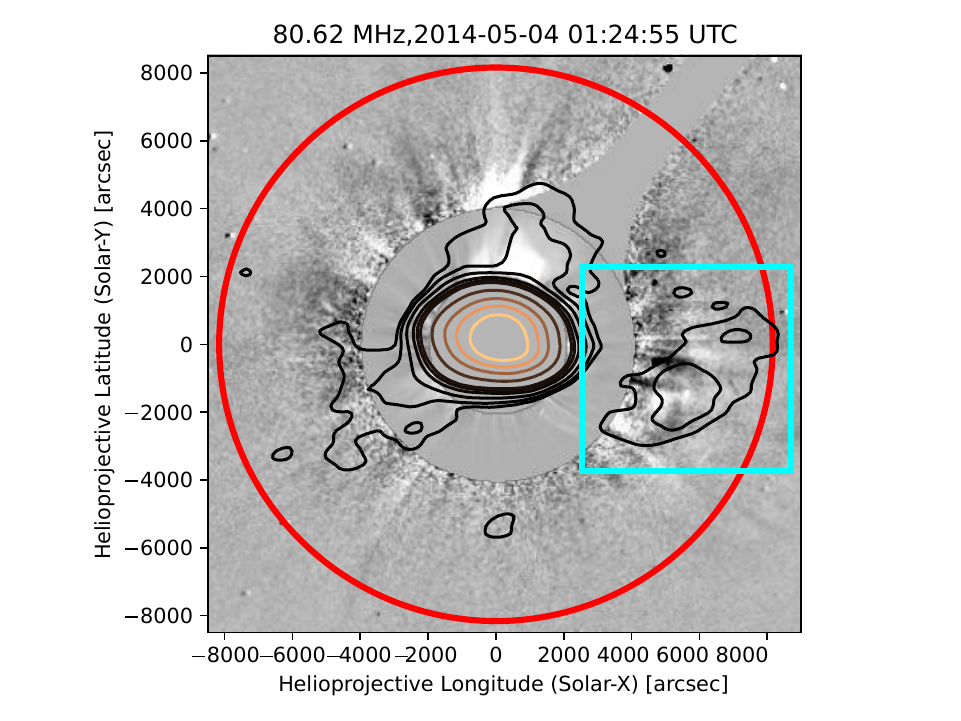}
    \caption[Stokes I radio emission from CME-2 at 80 MHz.]{Radio emission from CME-2 at 80 MHz. Stokes I emissions at 80 MHz are shown by the contours overlaid on the LASCO C2 base difference image at 01:24:54 UTC. The radio image is at 01:24:55 UTC. Contour levels are at  0.5, 1, 2, 4, 6, 8, 20, 40, 60, and 80 \% of the peak flux density. Radio emission marked by the cyan box is from CME-2 and is the focus of this work. The emission is detected up to 8.3 $R_\odot$, shown by the red circle.}
    \label{fig:south_cme}
\end{figure}

Figure \ref{fig:south_cme} shows a sample Stokes I image at 80.62 MHz using the contours overlaid on LASCO C2 and C3 base difference images. This work focuses on the radio emission from CME-2 marked by the cyan box. The study of another extended radio emission feature seen in Figure \ref{fig:south_cme}, which arises from a streamer (south-east), is beyond the scope of this work. The GS emission from CME-2 is detected up to a heliocentric distance of 8.3 $R_\odot$. This is the largest heliocentric distance to which GS radio emission from CME plasma has been detected to date. 

\begin{figure*}[!htbp]
    \centering
    \includegraphics[trim={1.5cm 0cm 2cm 0cm},clip,scale=0.35]{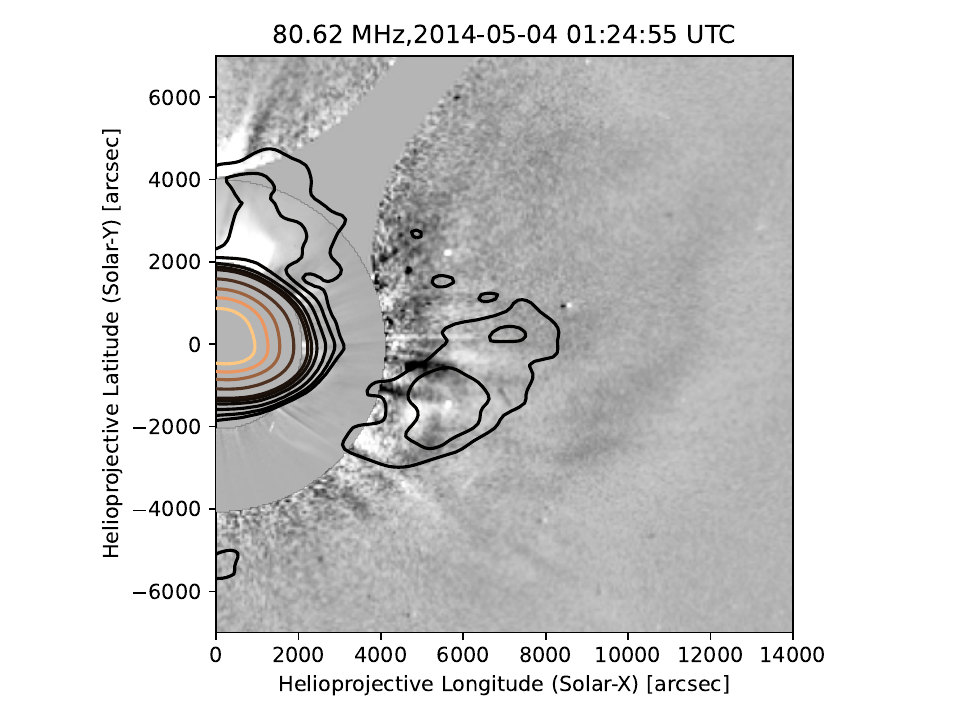}\includegraphics[trim={1.5cm 0cm 2cm 0cm},clip,scale=0.35]{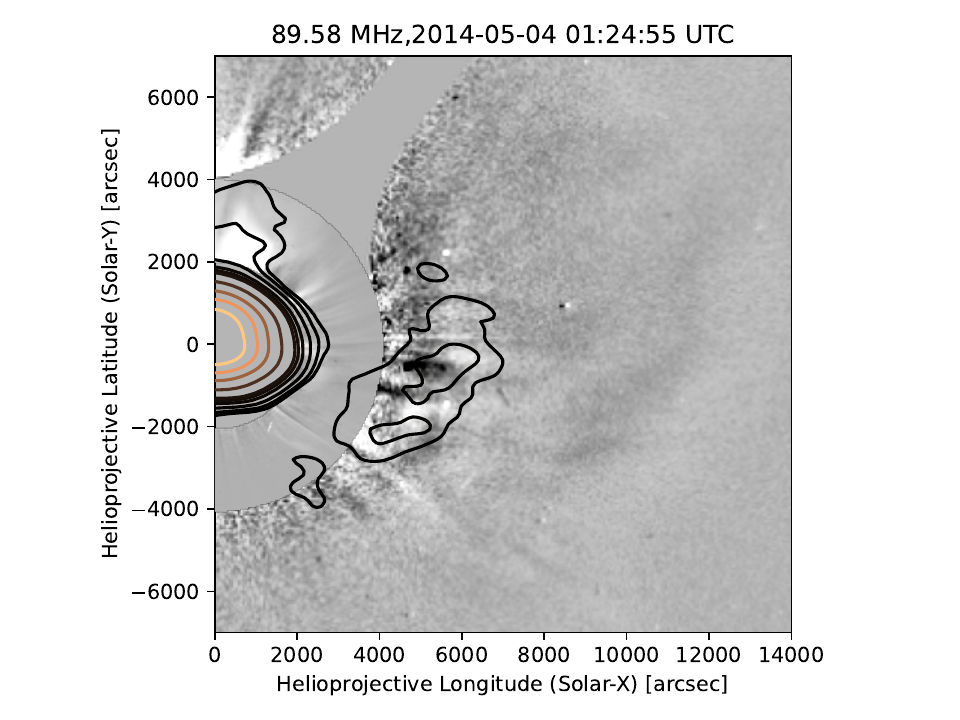}\includegraphics[trim={1.5cm 0cm 2cm 0cm},clip,scale=0.35]{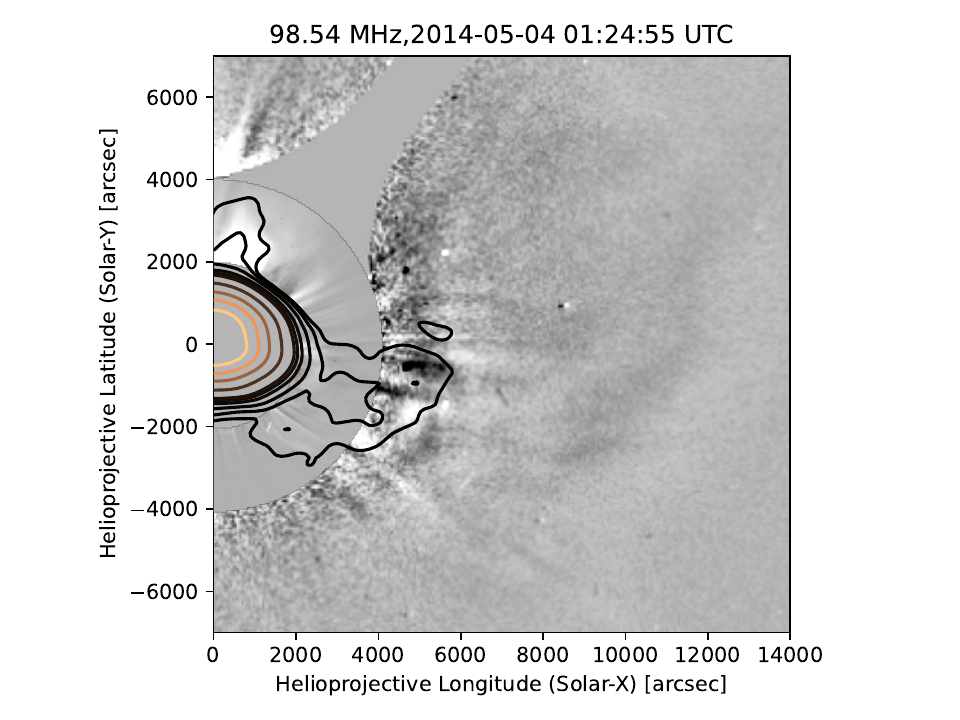}\\
    
    \includegraphics[trim={1.5cm 0cm 2cm 0cm},clip,scale=0.35]{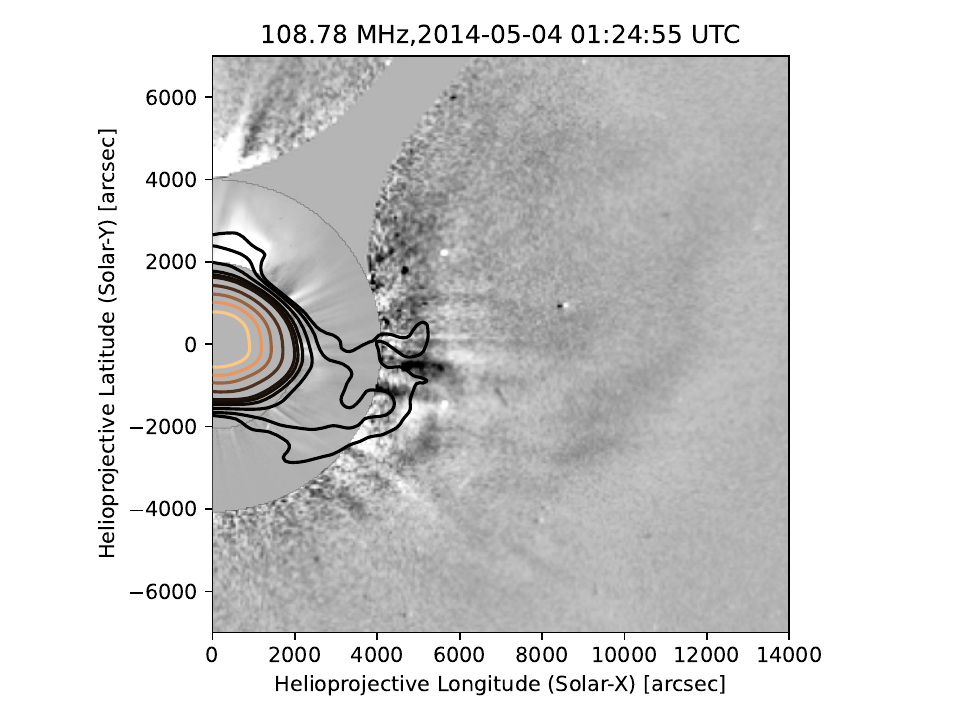}\includegraphics[trim={1.5cm 0cm 2cm 0cm},clip,scale=0.35]{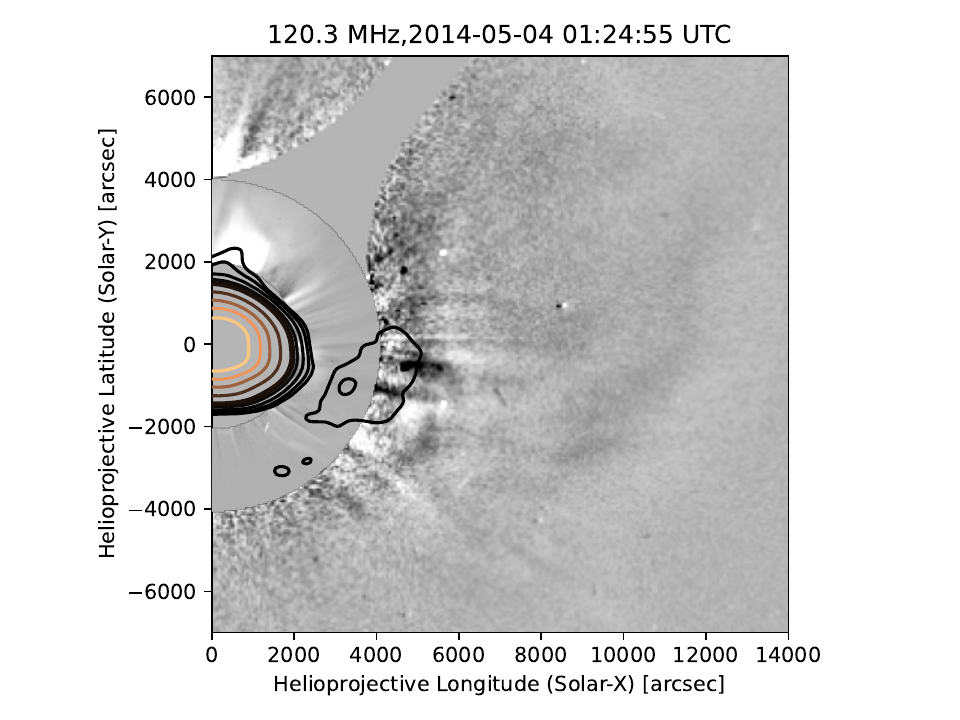}\includegraphics[trim={1.5cm 0cm 2cm 0cm},clip,scale=0.35]{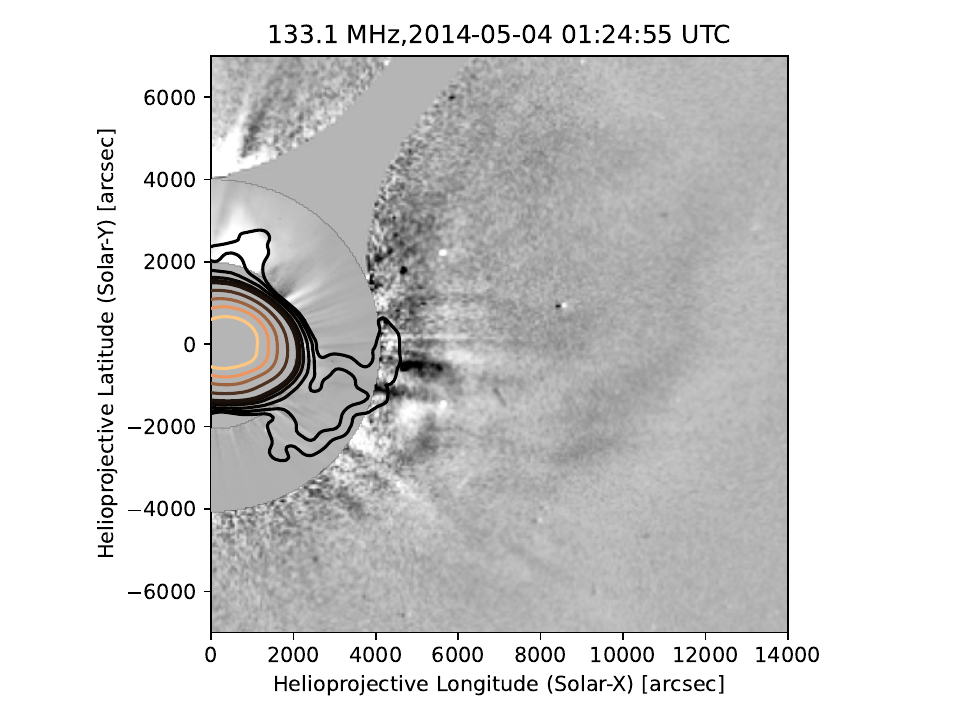}\\

    \includegraphics[trim={1.5cm 0cm 2cm 0cm},clip,scale=0.35]{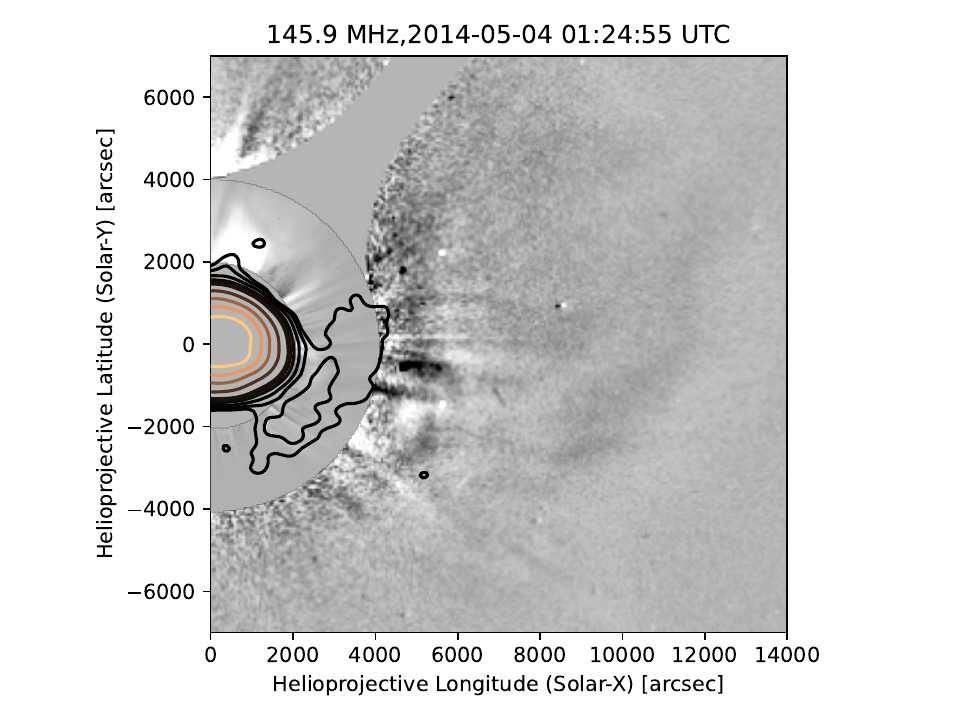}\includegraphics[trim={1.5cm 0cm 2cm 0cm},clip,scale=0.35]{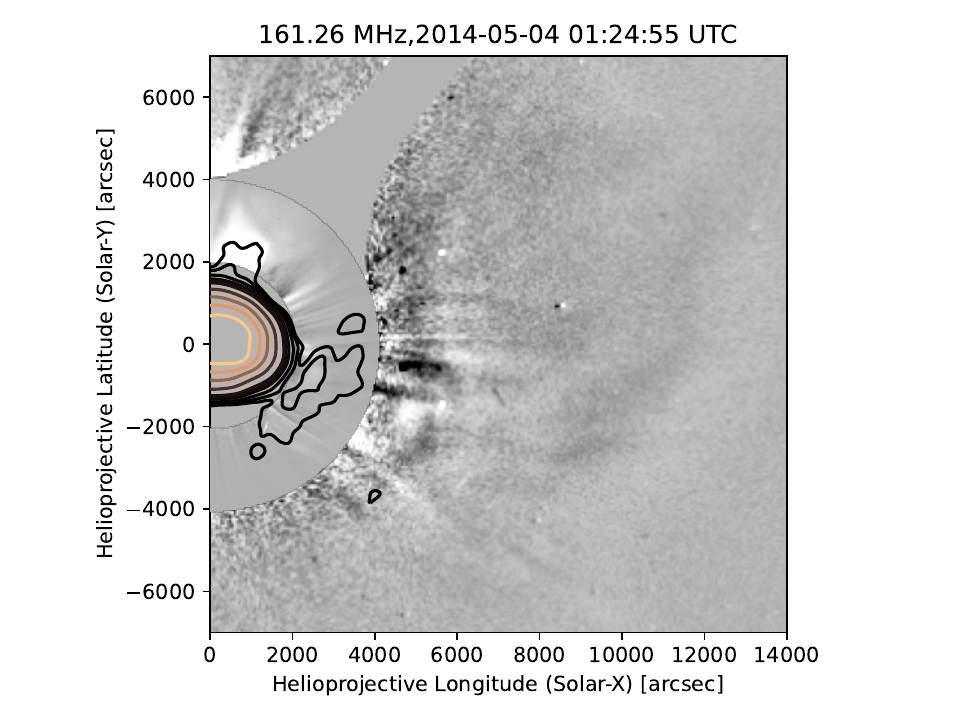}\includegraphics[trim={1.5cm 0cm 2cm 0cm},clip,scale=0.35]{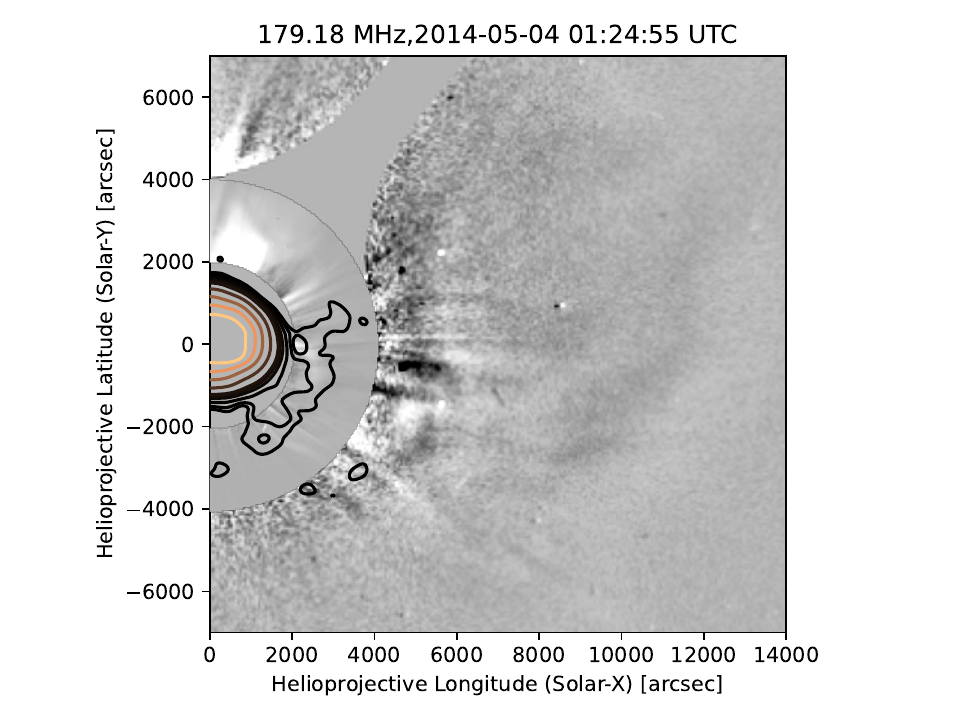}\\
    
    \includegraphics[trim={1.5cm 0cm 2cm 0cm},clip,scale=0.35]{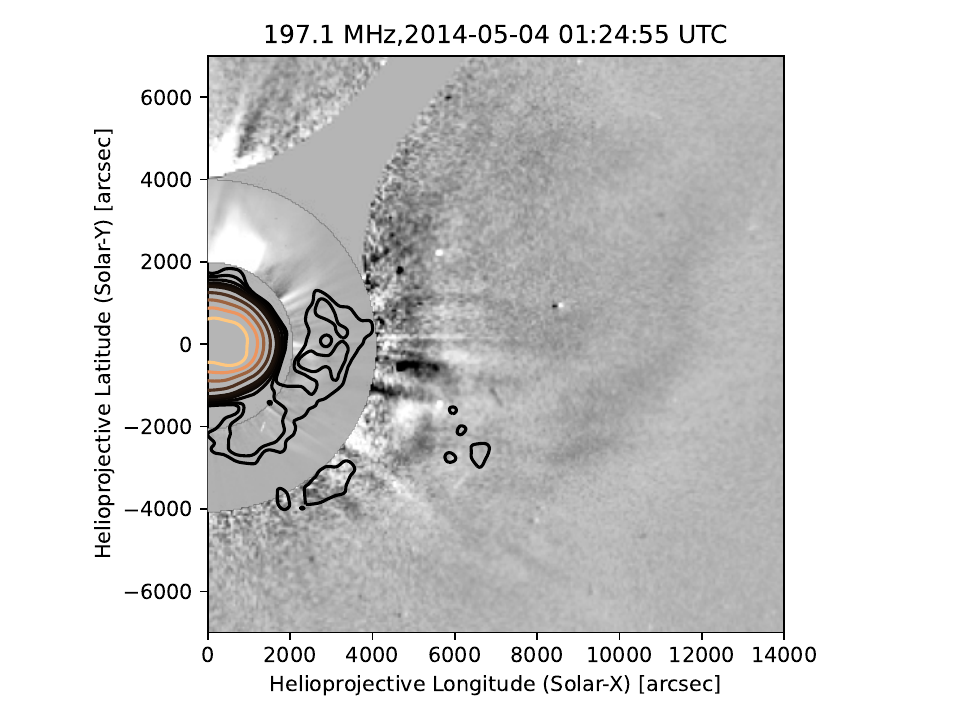}
    \includegraphics[trim={1.5cm 0cm 2cm 0cm},clip,scale=0.35]{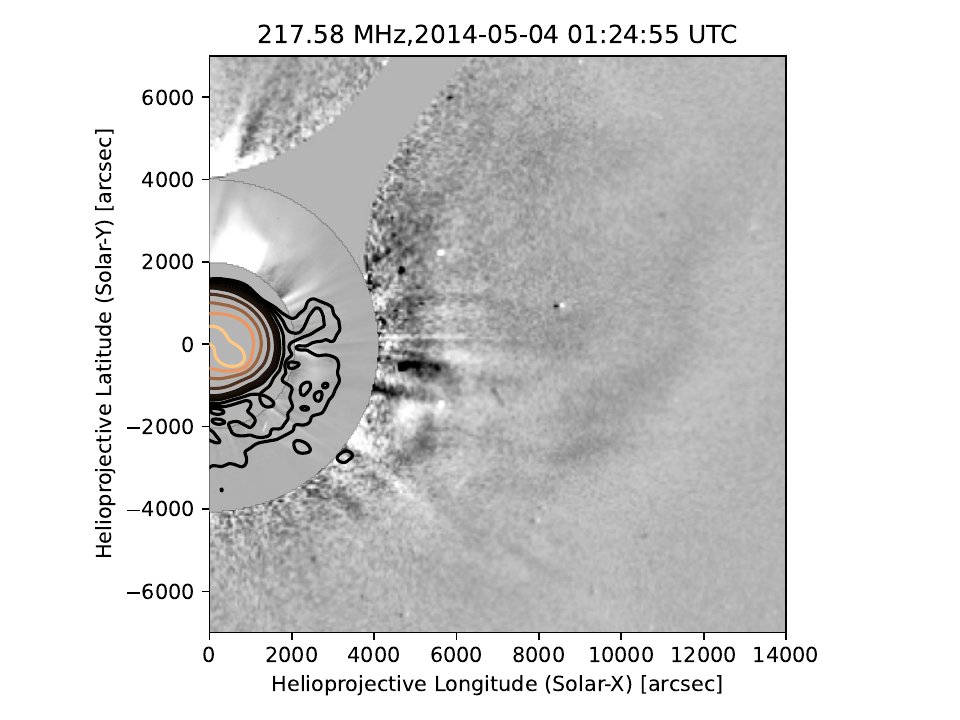}\includegraphics[trim={1.5cm 0cm 2cm 0cm},clip,scale=0.35]{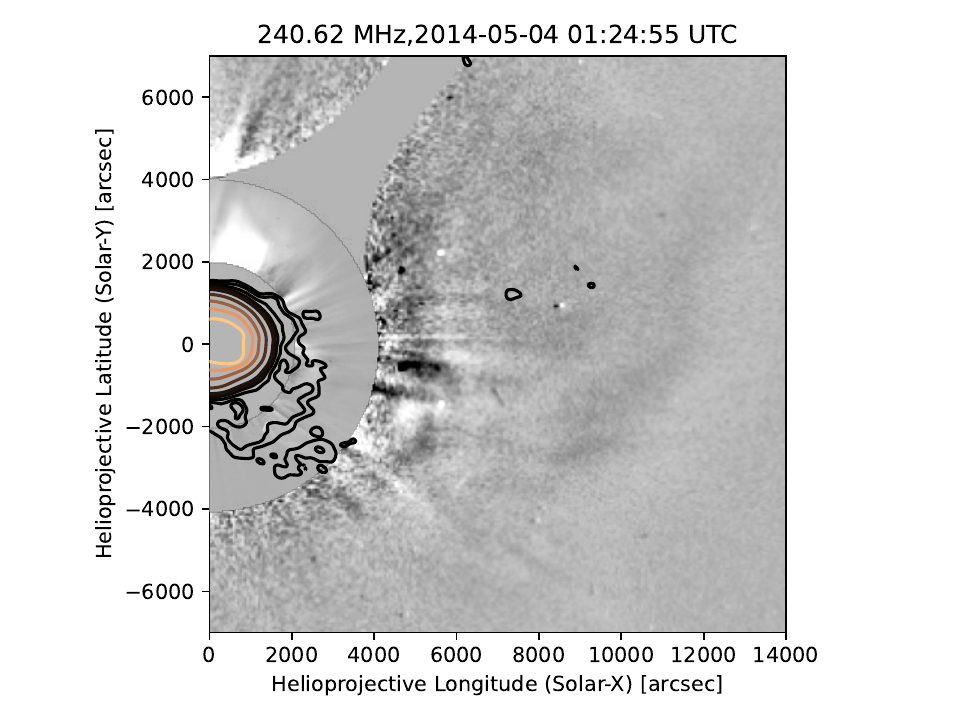}
    \caption[Stokes I radio emission from CME-2 at MWA  frequency bands.]{Stokes I radio emission from CME-2 at MWA frequency bands. Stokes I emissions are shown by contours overlaid on LASCO C2 and C3 base difference images. Frequency increases from the top left panel of the image to the bottom right panel. Contour levels are at 0.5, 1, 2, 4, 6, 8, 20, 40, 60, and 80 \% of the peak flux density.}
    \label{fig:c2_c3_comp_freq_cme2}
\end{figure*}

The extended radio emission is detected in all 12 coarse bands of the MWA spanning 80 to 240 MHz. The evolution of the radio emission with frequency for a single time slice centered at 01:24:55 UTC is shown in Figure \ref{fig:c2_c3_comp_freq_cme2}. Frequency increases from the top left to the bottom right of the figure. The spatial extent of radio emission shrinks towards a lower heliocentric height with increasing frequency. At the lowest frequency, 80 MHz, the radio emission extends up to 8.3 $R_\odot$, while at 240 MHz the emission is seen only out to $\sim2\ R_\odot$. 

\begin{figure}[!ht]
    \centering
    \includegraphics[scale=0.6]{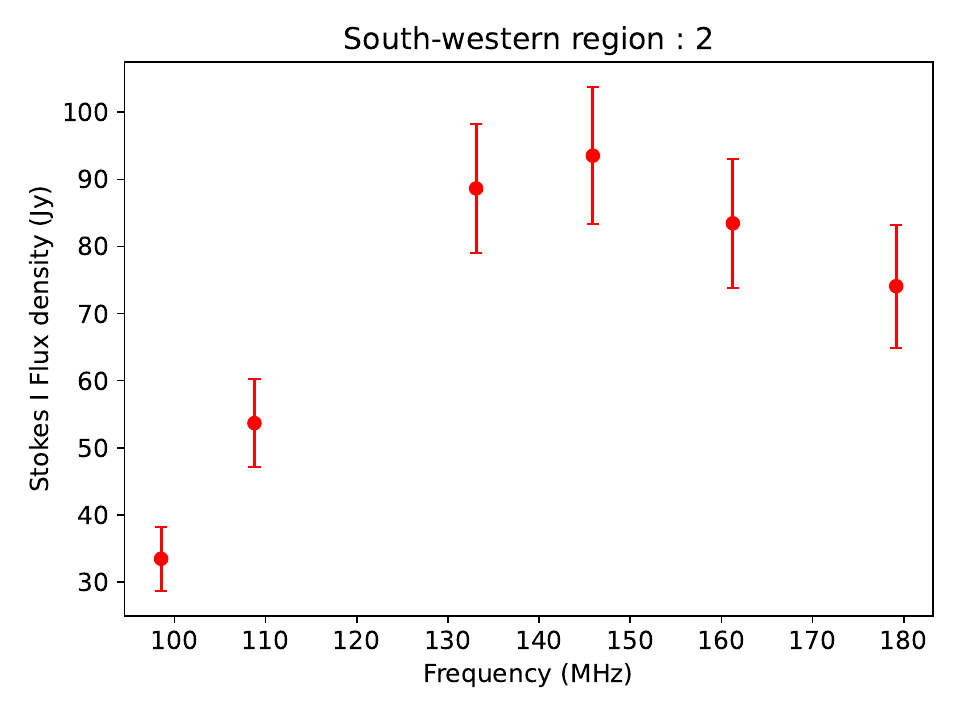}\\
    \includegraphics[scale=0.6]{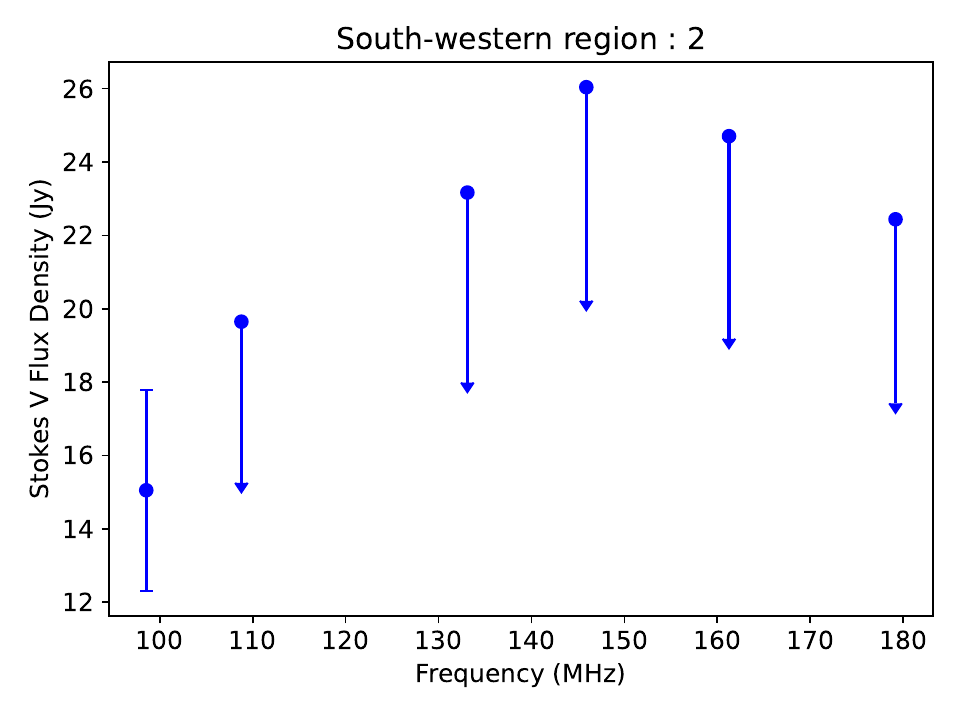}
    \caption[Sample observed Stokes I and V spectra for south-western CME.]{Sample observed Stokes I and V spectra for south-western CME. {\it Top panel: }Observed Stokes I spectrum for northern region 2 marked in Figure \ref{fig:cme2_regions}. {\it Bottom panel: }Observed Stokes V spectrum for south-western region 2.}
    \label{fig:observed_spectra_cme2}
\end{figure}

\subsection{Possible Source of the Radio Emission}
\label{subsec:origin_of_emission}
\begin{figure}[!ht]
    \centering
    \includegraphics[trim={0cm 0cm 0cm 0cm},clip,scale=0.06]{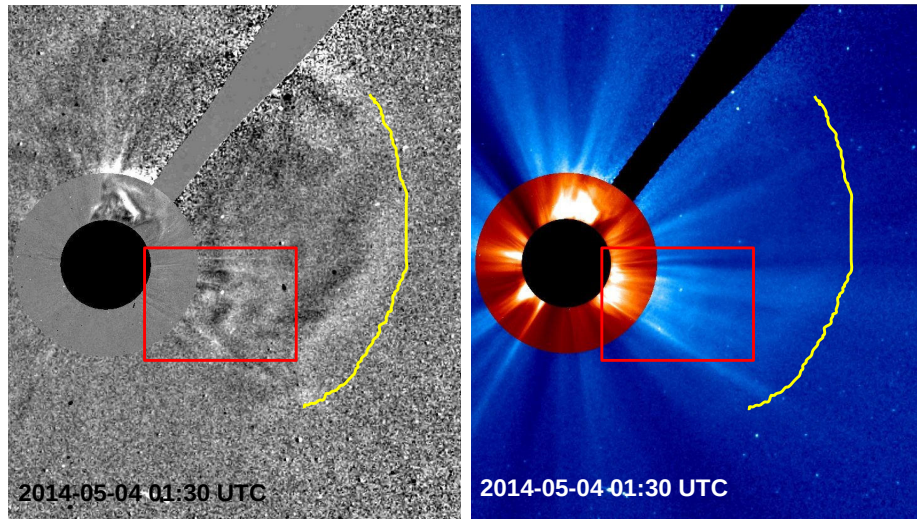}
    \caption[Location of CME-2 and pre-existing streamer.]{CME-2 and a pre-existing streamer are seen in LASCO C2 and C3 images. Left panel shows the LASCO C2 base difference image of CME-2 at 01:30 UTC on 2014 May 04 when radio analysis is done. In the right panel, bright streamers are seen in LASCO C2 continuum image. The red box marks the region from where the radio emission is coming. The yellow curved line shows the leading edge of the CME-2.}
    \label{fig:c2c3_combine}
\end{figure}
There are several possible emission mechanisms -- plasma emission, thermal free-free emission, and GS emission -- which can give rise to meter-wavelength radio emission. We have estimated the coronal electron density from the LASCO-C2 white light coronagraph images using the inversion method developed by \cite{Hayes2001}, which is about $\sim(0.7-1.25)\times10^6\ \mathrm{cm^{-3}}$ at the location of CME-2. This leads to a corresponding plasma frequency of about $7.1\ \mathrm{MHz}$, which is more than an order of magnitude lower than the frequencies where radio emission from CME-2 is detected. This rules out plasma emissions as a possible mechanism. Free-free emission is also ruled out due to reasons along the same lines as those discussed in Section \ref{subsec:emission_mechanism} of Chapter \ref{cme_gs1}. As the observed spectra (a sample spectrum is shown in Figure \ref{fig:observed_spectra_cme2}) have a spectral peak, the only feasible emission mechanism is the GS emission.

It is found that radio emission is not coming from the leading edge of the CME-2. The CME leading edge at 01:30 UTC is marked by the yellow curved line in Figure \ref{fig:c2c3_combine} and the radio emission is coming from the region marked by the red box in the same figure. The radio emission is coming from behind the CME leading edge.  Based on an examination of LASCO-C2 base difference images it is seen that as the CME passes, the pre-existing streamer structures are disturbed, their density is enhanced, but they are not completely disrupted even after the CME has moved out\footnote{\href{https://www.youtube.com/shorts/KZMkSo5d_50}{A movie of LASCO C2 base difference images}}. Similar density enhancements and disturbances in streamer structures are also seen in STEREO-A COR2 base difference images.

We conjecture that the disturbed magnetic field from this region of interaction between the CME and the pre-existing streamer may led to some magnetic reconnection activity, which, in turn, produces mildly relativistic electrons. These electrons in the presence of the magnetic fields of that region give rise to the observed GS radio emission. The available observations are not sufficient to establish this conjecture.

\begin{figure}[!ht]
    \centering
    \includegraphics[trim={13cm 0.5cm 8cm 1cm},clip,scale=0.4]{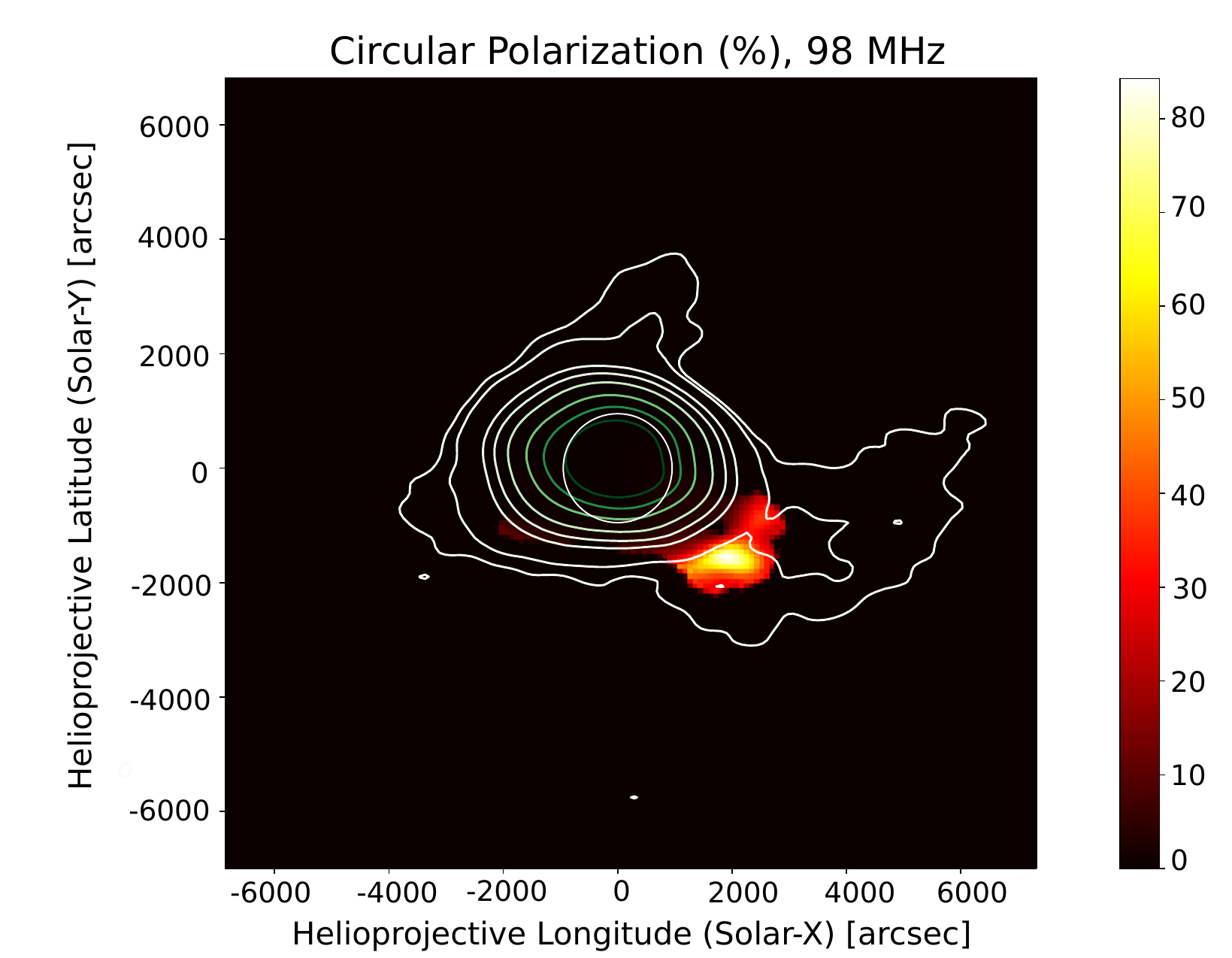}
    \caption[Circular polarization image at 80 MHz.]{Circular polarization image at 98 MHz. Background colormap shows percentage circular polarization and the contours represent the Stokes I emission. Contours at 0.5, 1, 2, 4, 6, 8, 20, 40, 60, 80 \% level of the peak Stokes I flux density.}
    \label{fig:circular_pol_regions}
\end{figure}
\begin{figure}[!ht]
    \centering
    \includegraphics[scale=0.1]{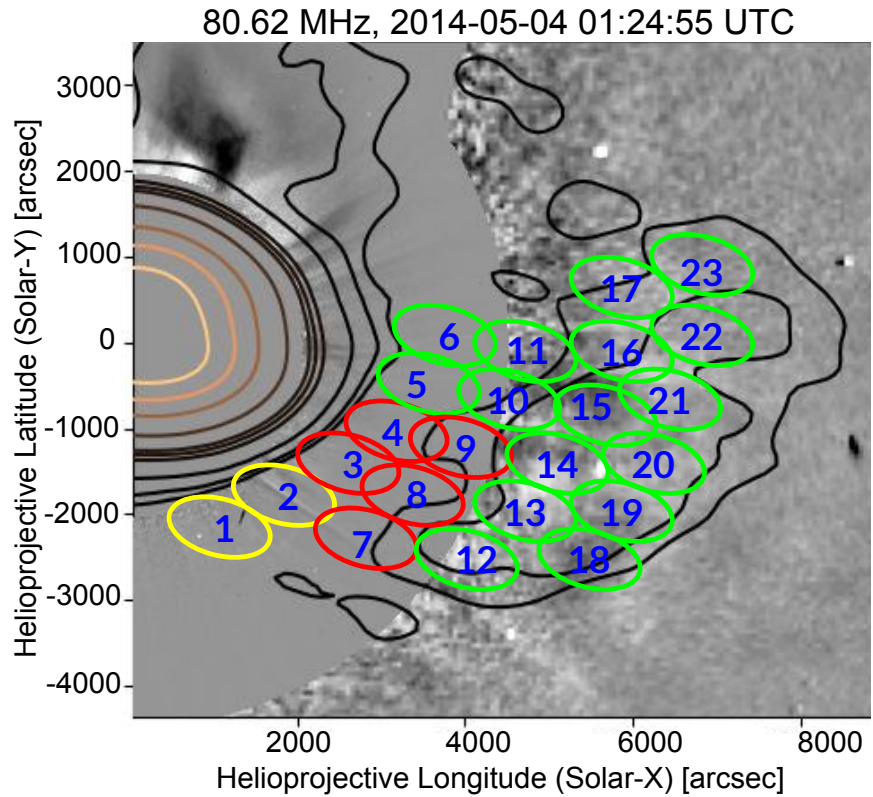}
    \caption[Regions of south-western CME where spectra have been extracted.]{Regions of northern CME where spectra have been extracted. Contours are at 80 MHz overlaid on LASCO coronagraph images. Red regions are those where spectrum fitting is done. Spectrum modeling is not done for green regions. Spectrum fitting is also done for yellow regions, which also have Stokes V detection at 98 MHz.}
    \label{fig:cme2_regions}
\end{figure}

\subsection{Robust Detection of Circularly Polarized Radio Emission Associated With CME-2}\label{subsec:circular_pol_cme2}
Unlike most of the previous studies \citep{Bain2014,Carley2017,Mondal2020a}, this study presents high-fidelity full Stokes imaging of GS emission from a CME.  It has already been established the quality of polarimetric calibration and imaging earlier in this thesis (Sections \ref{subsec:linear_pol}  and  \ref{subsec:upperlimits_estimation}).

Previous Chapter \ref{cme_gs1} presented a study of CME-1 which did not have a Stokes V detection from the CME plasma, but had stringent upper limits on it. In the same dataset, Stokes V emission is detected with high significance over a small part of Stokes I emission associated with the present event. Percentage Stokes V image at 98 MHz is shown as the background color map of the left panel of Figure \ref{fig:circular_pol_regions} and corresponding Stokes I emission is shown by contours. Stokes V emission is only detected at 98 MHz and over the two PSF-sized regions marked by yellow ellipses in Figure \ref{fig:cme2_regions}. For other frequencies and other regions, I have used the stringent Stokes V upper limits estimated following the same method as described in Section \ref{subsec:upperlimits_estimation} in Chapter \ref{cme_gs1}. Residual instrumental polarization leakage is estimated following the prescription described in Section \ref{subsec:res_leakage_estimation} of Chapter \ref{paircars_algorithm}. Residual instrumental leakage for Stokes V is $<|1\%|$. The average polarization fraction detected over the regions marked by yellow ellipses in Figure \ref{fig:cme2_regions} is $\sim50\%$, which is more than an order of magnitude larger than the residual instrumental polarization leakage. This further establishes the robustness of the Stokes V detection.

\subsection{Spatially Resolved Spectroscopy}\label{subsec:spectroscopy_cme2}
Wideband spectropolarimetric imaging observations enable one to perform spatially resolved spectroscopy of the radio emission from CME-2. Spectra are extracted from regions with a size equal to the size of the PSF at the lowest observing frequency of 80 MHz. These PSF-sized regions are marked in Figure \ref{fig:cme2_regions}. For the detection at a given frequency to be considered reliable, I have followed the same criteria as used earlier and described in Section \ref{sec:spectroscopy} of Chapter \ref{cme_gs1}. These are that all of the following must be satisfied:
\begin{enumerate}
    \item $f>\mu+5\sigma$
    \item $f>5\alpha$
    \item $f>5|n|$
\end{enumerate}
where, $f$ is the flux density, $n$ is the deepest negative close to the Sun, $\sigma$ and $\mu$ are the rms noise and mean, respectively, calculated over a region close to the Sun, and $\alpha$ is the rms noise estimated far away from the Sun. These stringent criteria ensure that flux densities are not severely contaminated by any imaging artifacts. Uncertainty on Stokes I flux density is denoted by $\sigma_I$ and of Stokes V is denoted by $\sigma_V$. Spectra are fitted for the red regions which have Stokes I detections in more than five spectral bands (marked by red and golden yellow ellipses in Figure \ref{fig:cme2_regions}) satisfying the stringent criteria mentioned above. Although radio emission is detected up to 240 MHz, emissions at the high-frequency bands do not cover the entire region having an area of the PSF at 80 MHz, and hence estimated flux densities will be wrong. These spectral points are not included in further analysis even though they satisfy the above three criteria. 

\section{Spectral Modeling Using Homogeneous \\Source Model}\label{sec:spectra_model}
To date, all modeling of the GS radio emissions from CMEs has been done assuming a homogeneous source model along the LoS \citep[e.g.,][etc.]{bastian2001,Bain2014,Tun2013,Mondal2020a,Kansabanik2023_CME1}. Such a homogeneous GS source is considered with mildly-relativistic electrons following a single power-law energy distribution. This simple GS model has ten independent parameters -- magnetic field strength ($|B|$), angle between the line-of-sight (LoS) and the magnetic field ($\theta$), area of emission ($A$), LoS depth through the GS emitting medium ($L$), temperature ($T$), thermal electron density ($n_{thermal}$), non-thermal electron density ($n_{nonth}$), power-law index of non-thermal electron distribution ($\delta$),  $E_\mathrm{min}$, and $E_\mathrm{max}$. I have used fast GS code developed by \cite{Fleishman_2010} and \cite{Kuznetsov_2021} for GS modeling.

As presented in Section \ref{sec:spectrum_sensitivity} of Chapter \ref{cme_gs1}, the model GS spectra are quite insensitive to variations in $T$ and $E_\mathrm{max}$. Hence, $T$ and $E_\mathrm{max}$ are kept fixed at 1 MK and 15 MeV respectively. $n_\mathrm{thermal}$ is estimated from inversion of the white light coronagraph images \citep{Hayes2001} and used a fixed azimuthally averaged value at a certain radius. $B$, $\theta$, $A$, $\delta$ and $E_\mathrm{min}$ are fitted, while setting $n_\mathrm{nonth}$ to 1\% of the $n_\mathrm{thermal}$, similar to what has been assumed in earlier works \citep{Carley2017,Mondal2020a,Kansabanik2023_CME1}. $L$ is explicitly fitted only for region 3 which has at least seven spectral points, with its upper limit set to $L_\mathrm{max}$ obtained from 3D reconstruction from multi-vantage point white-light observations as described in Section \ref{subsec:geometrical_params}.

The mathematical framework used for joint spectral fitting using Stokes I detection and Stokes V upper limits has been discussed in detail in Section \ref{subsec:upperlimits_methods} of Chapter \ref{cme_gs1}. Bayes theorem is used \citep{Puga2015,Andreon2015_bayes_thereom} to estimate the posterior distribution, $\mathcal{P}(\lambda|\mathcal{D})$ of model parameters, $\lambda$; given the data, $\mathcal{D}$ and a likelihood function, $\mathcal{L}(\mathcal{D}|\lambda)$. When either Stokes I or Stokes V detections are used, the likelihood function is defined as
\begin{equation}
\begin{split}
      \mathcal{L}_\mathrm{1}(\mathcal{D}|\lambda)&=\mathrm{exp}\left(-\frac{1}{2}\sum_{i=1}^N \left[\frac{\mathcal{D}_\mathrm{i}-m_\mathrm{i}(\lambda)}{\sigma_\mathrm{i}}\right]^2\right)\\
      &=\prod_{i=1}^N \mathrm{exp}\left(-\frac{1}{2}\left[\frac{\mathcal{D}_\mathrm{i}-m_\mathrm{i}(\lambda)}{\sigma_\mathrm{i}}\right]^2\right),
\end{split}
\label{eq:likelihood_1_cme2}
\end{equation}
where $N$ is the total number of data points, $\mathcal{D}_\mathrm{i}$, $m_\mathrm{i}(\lambda)$, and $\sigma_\mathrm{i}$ are the observed values, models values and uncertainty on the measurements respectively. For the case of upper limits, the likelihood function is defined as follows \citep{Ghara2020,Greig2021,Maity2022},
\begin{equation}
\begin{split}
      \mathcal{L}_\mathrm{2}(\mathcal{D}|\lambda)&=\prod_{i=1}^N \frac{1}{2}\left[1-erf\left(\frac{\mathcal{D}_\mathrm{i}-m_\mathrm{i}(\lambda)}{\sqrt{2}\sigma_\mathrm{i}}\right)\right],
\end{split}
\label{eq:likelihood_2_cme2}
\end{equation}
where $erf$ refers to the error function. When both detections and upper limits are available, one can define the joint likelihood function as,
\begin{equation}
    \mathcal{L}(\mathcal{D}|\lambda)=\mathcal{L}_\mathrm{1}(\mathcal{D}|\lambda)\ \mathcal{L}_\mathrm{2}(\mathcal{D}|\lambda),
    \label{eq:join_likelihood_cme2}
\end{equation}
which allows one to use the constraints from the detections as well as the upper limits. We use the Monte Carlo Markov Chain \citep[MCMC;][]{brooks2011handbook} analysis to estimate the posterior distribution of parameters using the joint likelihood function.

\subsection{Estimation of Geometrical Parameters}\label{subsec:geometrical_params}
\begin{figure*}[!ht]
\centering 
    \includegraphics[trim={0.5cm 5.5cm 0cm 5cm},clip,scale=0.35]{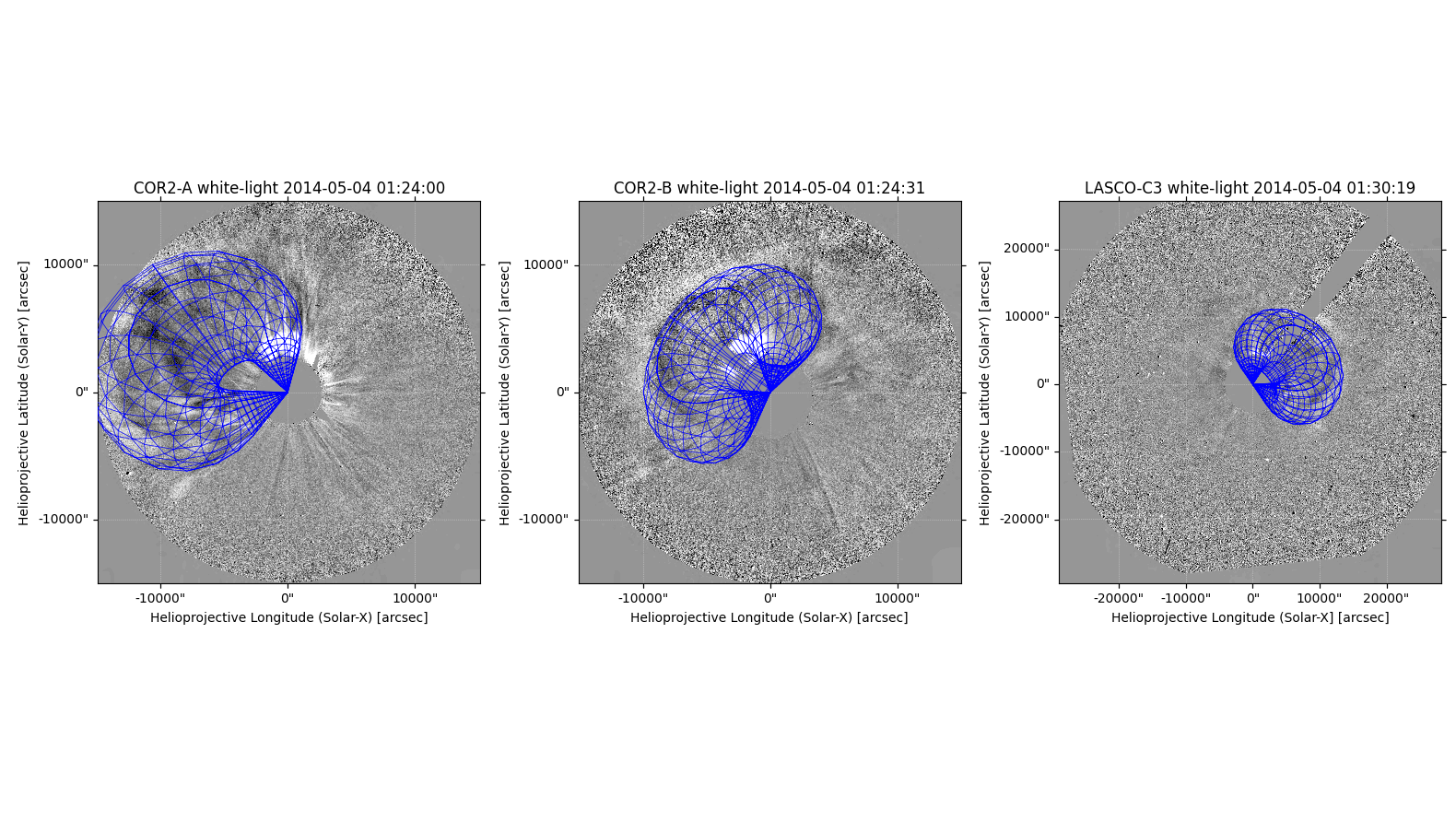}   
    \caption[Three-dimensional reconstruction of the CME-2 using Graduated Cylindrical Shell (GCS) model.]{Three-dimensional reconstruction of the CME-2 using Graduated Cylindrical Shell (GCS) model. GCS fitting is constrained using three vantage point observations from the LASCO-C2, C3 and COR-2 coronagraphs onboard STEREO-A and STEREO-B spacecraft. Blue meshes show different views of the GCS model of the CME-2 at 01:30 UTC on 2014 May 04.}
    \label{fig:gcs_cme2}
\end{figure*}
\begin{figure*}[!ht]
    \centering
    \includegraphics[trim={0cm 0cm 0cm 0cm},clip,scale=0.55]{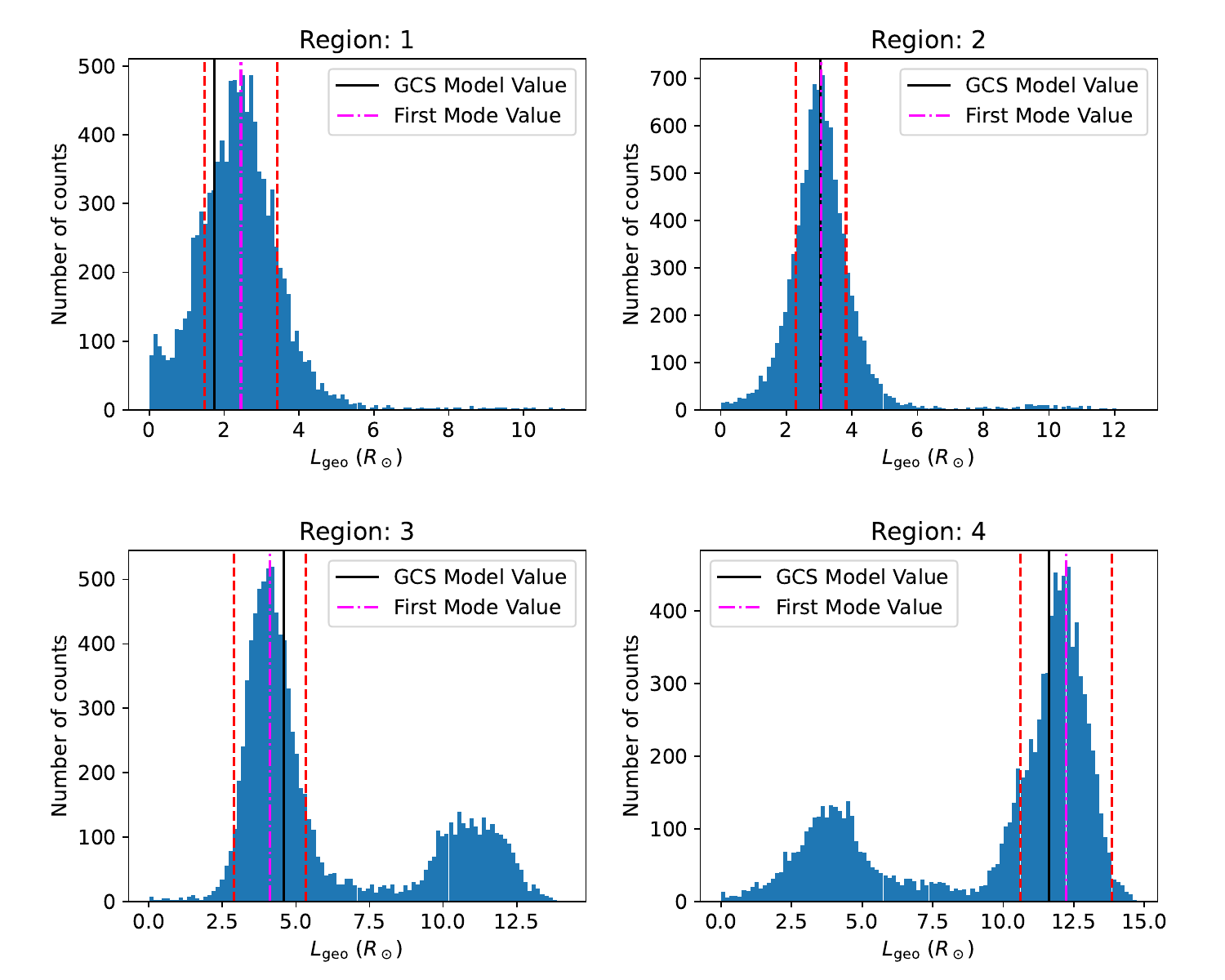}
    \caption[Distributions of $L_\mathrm{geo}$ for south-western CME.]{Distributions of $L_\mathrm{geo}$ for some sample PSF-sized regions of CME-2. Solid black lines represent $L_\mathrm{geo}$ for the GCS model parameters mentioned above. Dash-dot magenta lines represent the mode and red dashed lines represent the median absolute deviation around the mode. The mode and corresponding standard deviations are mentioned in Table \ref{table:los_depth_cme2}.}
    \label{fig:cme2_los_hist}
\end{figure*}
\begin{table}[!ht]
\centering
    \renewcommand{\arraystretch}{1.4}
    \begin{tabular}{|p{1.4cm}|p{1.8cm}|p{2.2cm}|p{1.4cm}|p{1.8cm}|p{2.2cm}|}
    \hline
       Region & $L_\mathrm{geo} (R_\odot)$ & $\sigma(L_\mathrm{geo})\ (R_\odot)$ & Region & $L_\mathrm{geo}\ (R_\odot)$  & $\sigma(L_\mathrm{geo})\ (R_\odot)$\\ \hline \hline 
        1 & 2.45 & 0.98 &  7 & 3.57 & 1.25\\
        \hline
        2 & 3.05 & 0.77 &  8 & 4.68 & 3.52\\
        \hline
        3 & 4.12 & 1.23 &  9 & 11.94 & 1.27\\
        \hline
        4 & 12.23 & 1.62 & -- & -- & --\\
       \hline
    \end{tabular}
    \caption[Estimated geometric LoS depth from GCS modeling of south-western CME.]{\textbf{Estimated geometric LoS depth from GCS modeling of south-western CME.} The geometric LoS depths are obtained for red regions using ray tracing from Earth through that region. Geometric LoS depths are given in units of the solar radius.}
    \label{table:los_depth_cme2}
\end{table}
Multiple vantage point observations using SOHO, STEREO-A, and STEREO-B spacecraft allow us to perform a 3D reconstruction of the CME-2. 3D reconstruction is performed using the Graduated Cylindrical Shell \citep[GCS;][]{Thernisien_2006,Thernisien_2011} model using its {\it python} implementation \citep{gcs_python}. A good visual fit is obtained following the method described by \cite{Thernisien2009}. The GCS model at 01:30 UTC is shown by blue mesh in Figure \ref{fig:gcs_cme2}, where different panels show superposition on COR-2 images from STEREO-A and STEREO-B and C3 coronagraph images from LASCO. The best visual fit GCS model parameters at 01:30 UTC are:
\begin{enumerate}
    \item Front height ($h_{front}$) : 17.9 $R_\odot$
    \item Half-angle ($\alpha$) : 46.8$^{\circ}$
    \item Carrington Longitude ($\Phi$) : 36.5$^{\circ}$ 
    \item Heliospheric Latitude ($\Theta$) : 14.1$^{\circ}$ 
    \item Aspect Ratio ($\kappa$) : 0.45
    \item Tilt Angle ($\gamma$) : -56.7$^{\circ}$
\end{enumerate}
\subsubsection{Estimating $L_\mathrm{geo}$ from GCS Model}\label{subsubsec:los_depth_estimate_south}
I have performed ray-tracing through the GCS model and computed the geometrical path length ($L_\mathrm{geo}$) through the CME-2 for each PSF-sized region using {\it python}-based ray-tracing code {\it trimesh} \citep{trimesh}. All rays originate from the Earth. 
Error on $L_\mathrm{geo}$, $\sigma(L_\mathrm{geo})$ is  calculated following the same method as described in Section \ref{subsec:estimate_gcs} of Chapter \ref{cme_gs1}. 
\begin{figure*}[!ht]
    \centering
     \includegraphics[trim={0.3cm 0.5cm 0.0cm 0.3cm},clip,scale=0.35]{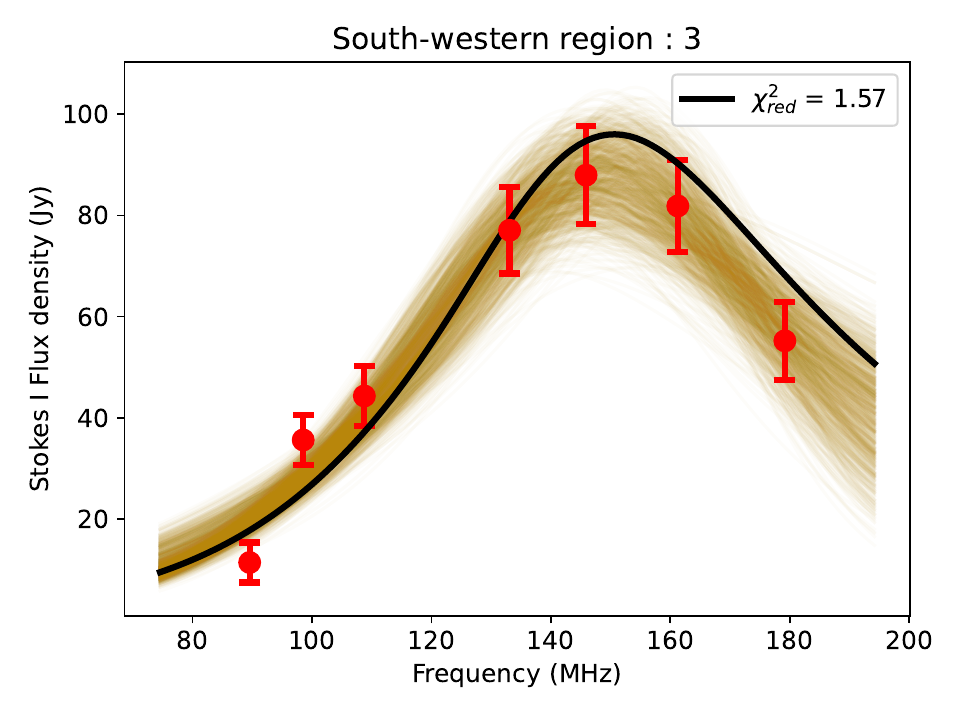} \includegraphics[trim={0.3cm 0.5cm 0.0cm 0.3cm},clip,scale=0.35]{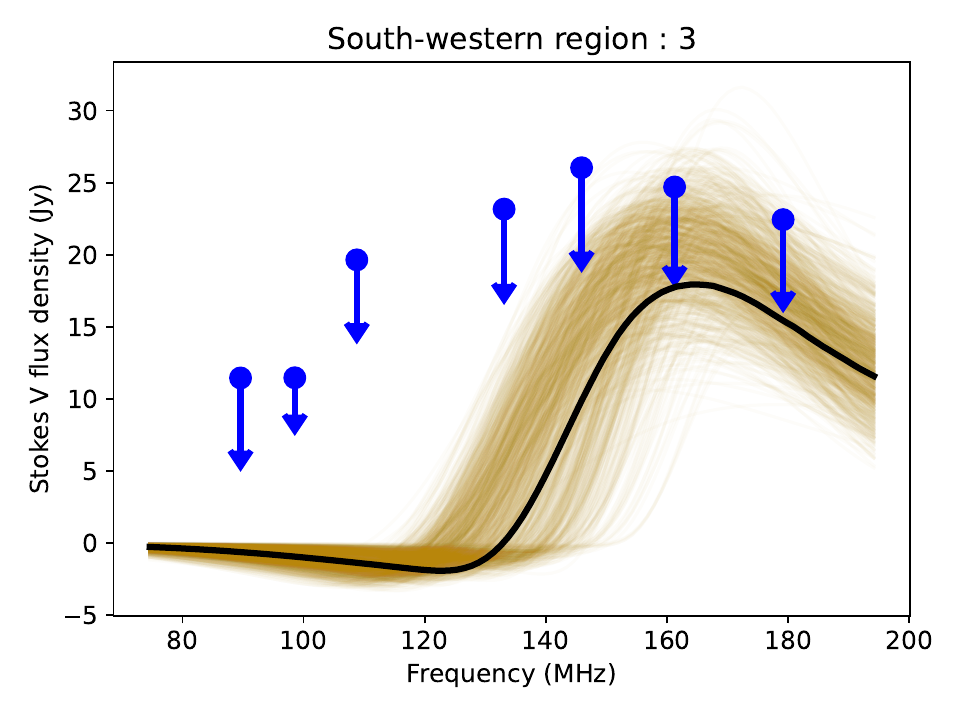}\\
     
     \includegraphics[trim={0.3cm 0.5cm 0.0cm 0.3cm},clip,scale=0.35]{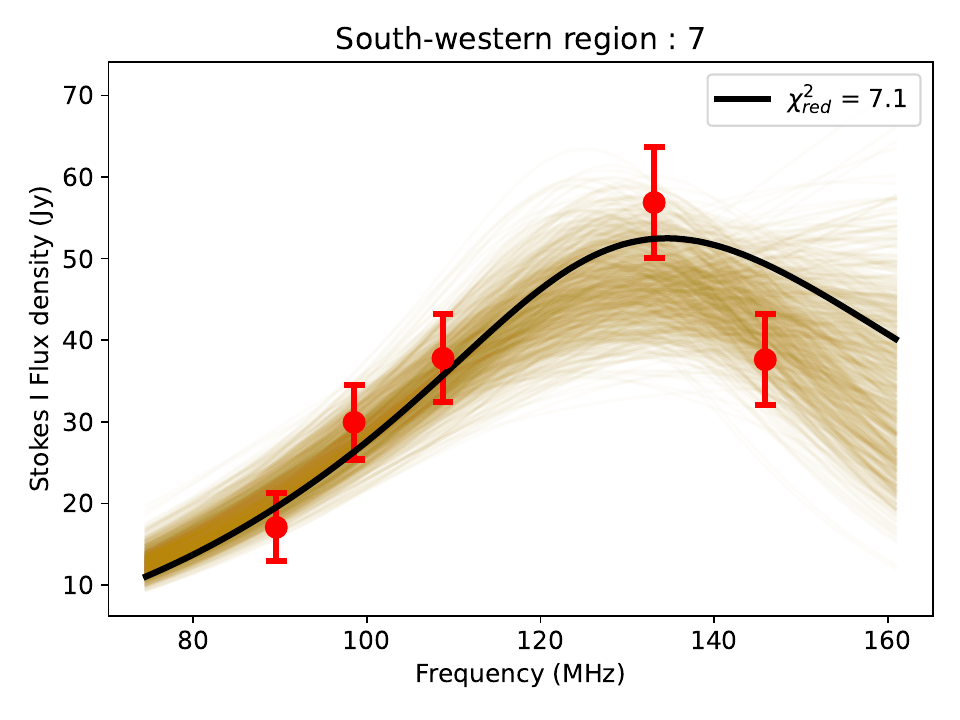}\includegraphics[trim={0.3cm 0.5cm 0.0cm 0.3cm},clip,scale=0.35]{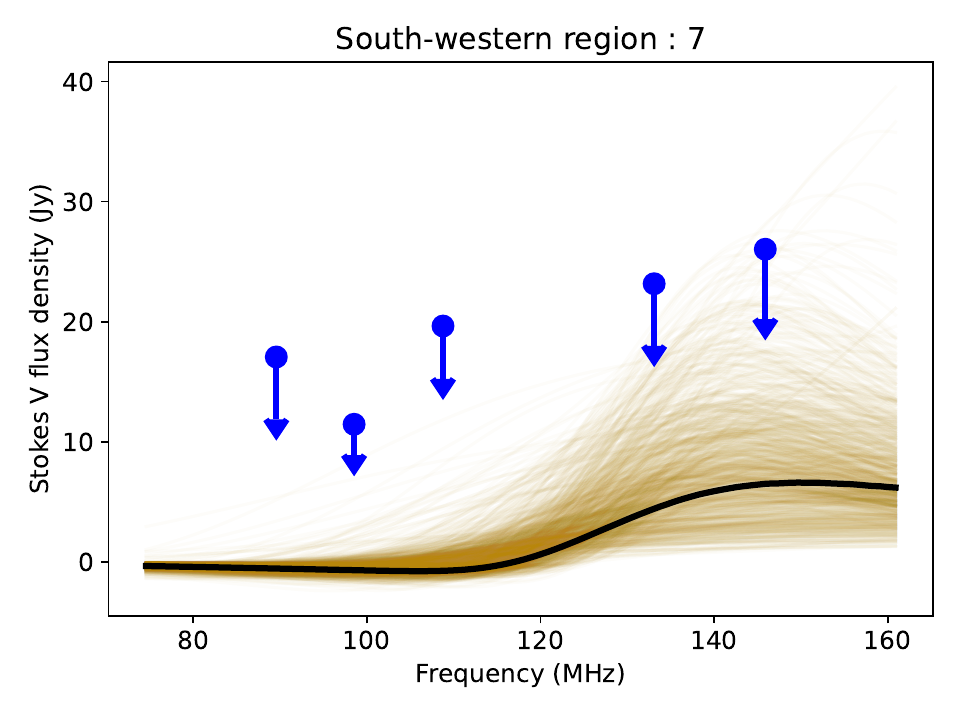}

    \includegraphics[trim={0.3cm 0.5cm 0.0cm 0.3cm},clip,scale=0.35]{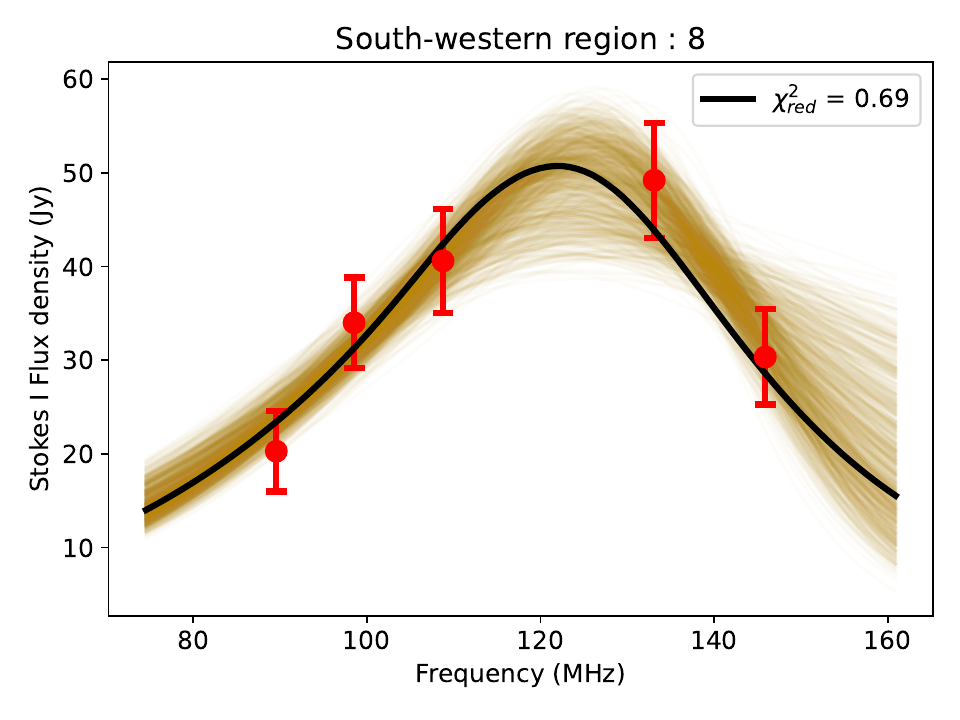} \includegraphics[trim={0.3cm 0.5cm 0.0cm 0.3cm},clip,scale=0.35]{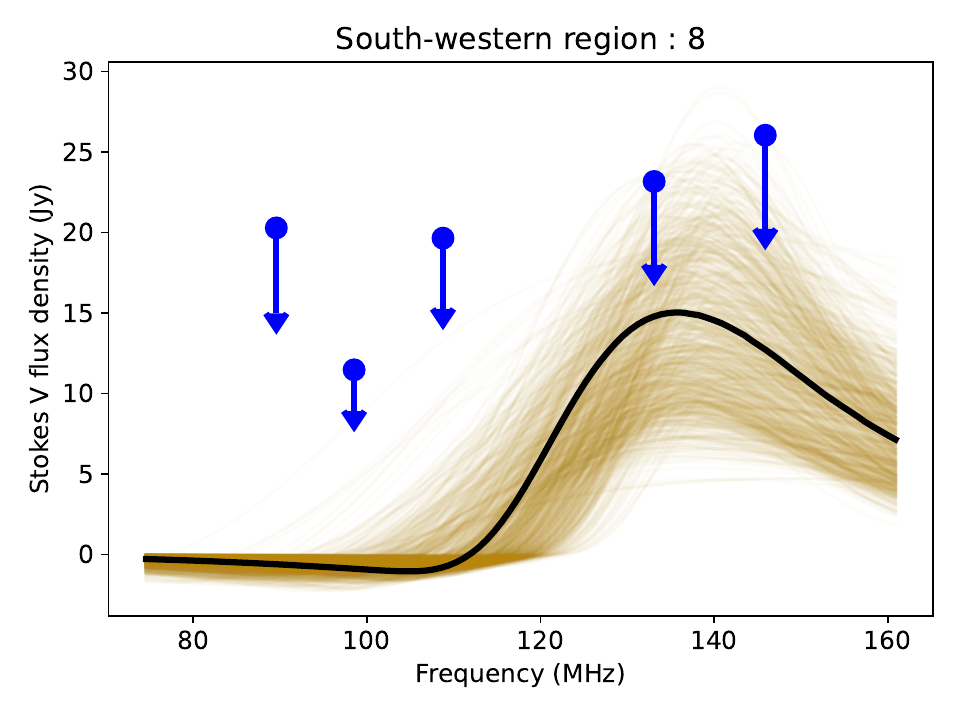}\\
    
     \includegraphics[trim={0.3cm 0.5cm 0.0cm 0.3cm},clip,scale=0.35]{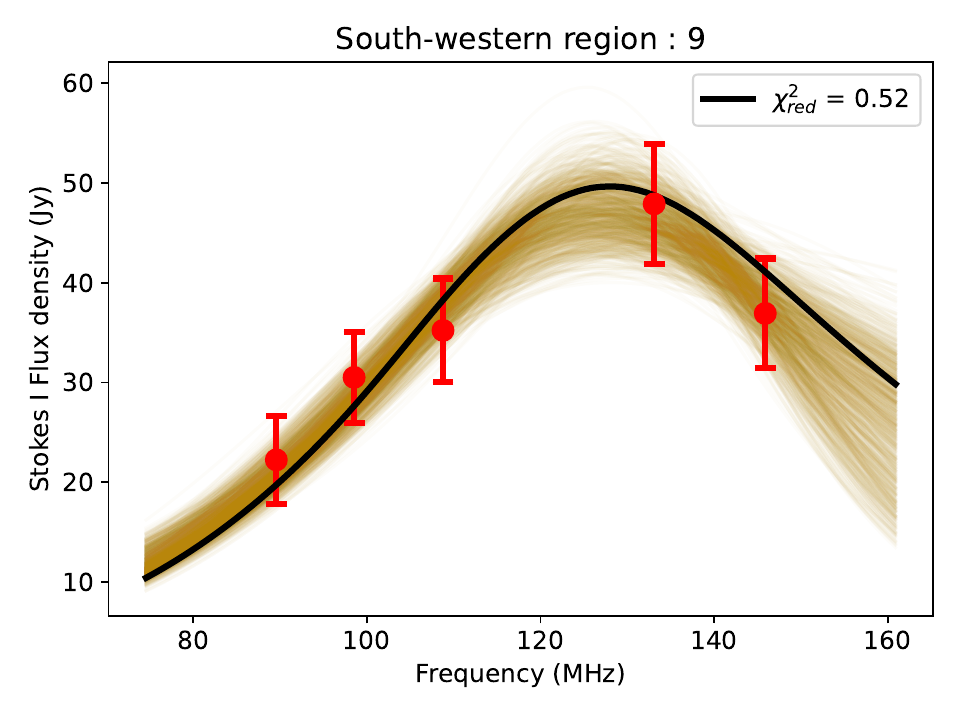}\includegraphics[trim={0.3cm 0.5cm 0.0cm 0.3cm},clip,scale=0.35]{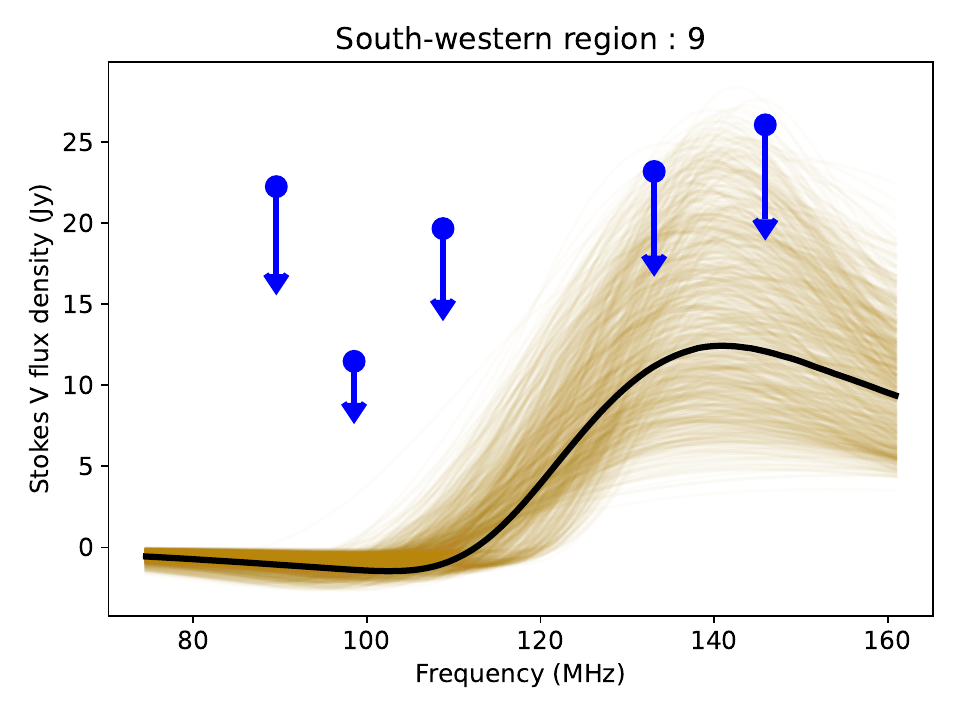}
    \caption[Observed and fitted spectra for regions 3, 7, 8, and 9 of the southwestern CME.]{Observed and fitted spectra for regions 3, 7, 8, and 9 of CME-2. {\it Left column: }Stokes I spectra are shown. Red points represent the observed flux densities. {\it Right column: }Stokes V spectra are shown. Blue points represent the observed upper limits. The black lines represent the GS spectra corresponding to GS parameters reported in Table \ref{table:south_params}. Light yellow lines show the GS spectra for 1000 randomly chosen realizations from the posterior distributions of the GS model parameters resulting from a total of 1,000,000 MCMC chains.}
    \label{fig:spectra_homo1}
\end{figure*}
While the histograms of distributions of $L_\mathrm{geo}$ for most of the PSF-sized regions show a unimodal distribution, those for regions 3 and 4 show a bimodal distribution with a secondary peak of much lower amplitude, as shown in Figure \ref{fig:cme2_los_hist}. It has been found that the $L_\mathrm{geo}$ (marked by solid black lines in Figure \ref{fig:cme2_los_hist}) corresponding to the GCS model parameters mentioned above lie close to the mode of the distribution (marked by a dash-dot magenta line in Figure \ref{fig:cme2_los_hist}). Hence, instead of using the mean or the median, the mode value is used as $L_\mathrm{geo}$ and $\sigma(L_\mathrm{geo})$ is estimated as $1.4826\times$ MAD, where MAD is the ``median absolute deviation" with respect to mode value (assuming the distribution is quasi-Gaussian around the mode). $L_\mathrm{geo}\pm\sigma(L_\mathrm{geo})$ are shown by dashed red lines in Figure \ref{fig:cme2_los_hist}. It is important to note that LoS depth in the GS model ($L$) could be different from $L_\mathrm{geo}$. $L_\mathrm{geo}$ and $\sigma(L_\mathrm{geo})$ are tabulated in Table \ref{table:los_depth_cme2}. 
The maximum value of $L$ for a given region is chosen to be $L_\mathrm{max}=L_\mathrm{geo}+\sigma(L_\mathrm{geo})$.

\begin{landscape}
\begin{table*}[!htpb]
\centering
    \renewcommand{\arraystretch}{1.5}
    \begin{tabular}{|p{1.2cm}|p{2cm}|p{1.8cm}|p{1.5cm}|p{1.6cm}|p{2.2cm}|p{1.8cm}|p{1.5cm}|p{1.2cm}|p{1.2cm}|}
    \hline
       Region No. & Heliocentric \newline{Distance} & $|B|$ (G) & $\delta$ & $ A \times 10^{20}$\newline{$(cm^{2})$} & $E_\mathrm{min}$ (keV) & $\theta$ \newline{(degrees)} & $L\ (R_\odot)$ & $n_\mathrm{thermal}$ \newline{$\times 10^6$}\newline{$(cm^{-3})^*$}  & $n_\mathrm{nonth}$\newline{$\times 10^4$} \newline{$(cm^{-3})^*$} \\ \hline \hline 
        1 & 2.65 & $0.20_{-0.02}^{+0.02}$ & $2.59_{-0.37}^{+0.53}$ & $2.06_{-0.93}^{+1.08}$ & $278.74_{-148.64}^{+293.77}$ & $38.41_{-2.86}^{+4.31}$  & $1.42^*$ &1.25 & 1.25\\
        \hline
        2 & 2.65 & $1.72_{-0.52}^{+0.64}$  & $6.82_{-1.56}^{+1.88}$  & $4.76_{-1.84}^{+2.94}$  & $139.65_{-67.09}^{+107.86}$  & $79.34_{-10.22}^{+5.96}$ & $1.41^*$ & 1.25 & 1.25\\
        \hline
        3 & 2.65 & $3.99_{-0.76}^{+0.71}$ & $5.94_{-1.10}^{+1.89}$ & $10.24_{-3.39}^{+7.72}$ & $35.69_{-16.70}^{+26.07}$ & $77.67_{-5.69}^{+4.58}$ & $1.98_{-1.15}^{+2.04}$ & 1.25 & 1.25\\
        \hline
        4 & 2.65 & $1.44_{-0.44}^{+0.64}$ & $4.02_{-1.05}^{+2.34}$ & $0.94_{-0.47}^{+1.15}$ & $67.77_{-40.94}^{+133.62}$ & $53.01_{-18.38}^{+23.68}$ & $5.12^*$ &1.25 & 1.25\\
       \hline
       7 & 3.2 & $1.61_{-0.86}^{+0.51}$ & $6.42_{-2.34}^{+2.52}$ & $2.88_{-2.17}^{+4.81}$ & $139^*$ & $71.12_{-17.83}^{+13.33}$ & $1.78^*$ & 0.7 & 0.7\\
       \hline
        8 & 3.2 & $2.36_{-0.56}^{+0.39}$ & $8.57_{-2.42}^{+2.81}$ & $10.27_{-6.93}^{+9.67}$ & $139^*$ & $64.73_{-12.45}^{+15.70}$ & $3.03^*$ & 0.7 & 0.7\\
       \hline
        9 & 3.2 & $2.49_{-0.49}^{+0.66}$ & $6.51_{-0.86}^{+1.98}$ & $5.30_{-2.27}^{+8.68}$ & $67^*$ & $63.06_{-12.19}^{+17.10}$ & $4.88^*$ & 0.7 & 0.7\\
       \hline
    \end{tabular}
    \caption[Estimated GS model parameters of CME-2 considering homogeneous GS source model.]{\textbf{Estimated GS model parameters of CME-2 considering homogeneous GS source model.} These parameters are estimated for 01:24:55 UTC. Parameters marked by $^*$ are kept fixed during the fitting.}
    \label{table:south_params}
\end{table*}
\end{landscape}
\subsection{Joint Spectral Fitting of Stokes I and V Spectra Using Homogeneous GS Model}\label{subsec:homo_modeling}
A joint spectral fitting is done using Stokes I and V spectra for all red and yellow regions marked in Figure \ref{fig:cme2_regions}. Uniform priors, $\pi(\lambda)$, used for the GS model parameters are as follows,
\begin{enumerate}
    \item $B\ (\mathrm{G})$ : $(0,\ 10]$
    \item $\theta\ (\mathrm{degree})$ : $(0,\ 90)$
    \item $\delta$ : $(1,\ 10]$
    \item $A \times10^{20}\ (\mathrm{cm^{2}})$ : $[0.0001,\ 100]$
    \item $E_\mathrm{min}\ (\mathrm{keV})$ : $(0.1,\ 100]$
    \item $L\ (R_\odot)$ : $(0.01,L_\mathrm{max}]$.
\end{enumerate}

\begin{figure*}[!ht]
    \centering
    \includegraphics[trim={0cm 0.6cm 0cm 0cm},clip,scale=0.4]{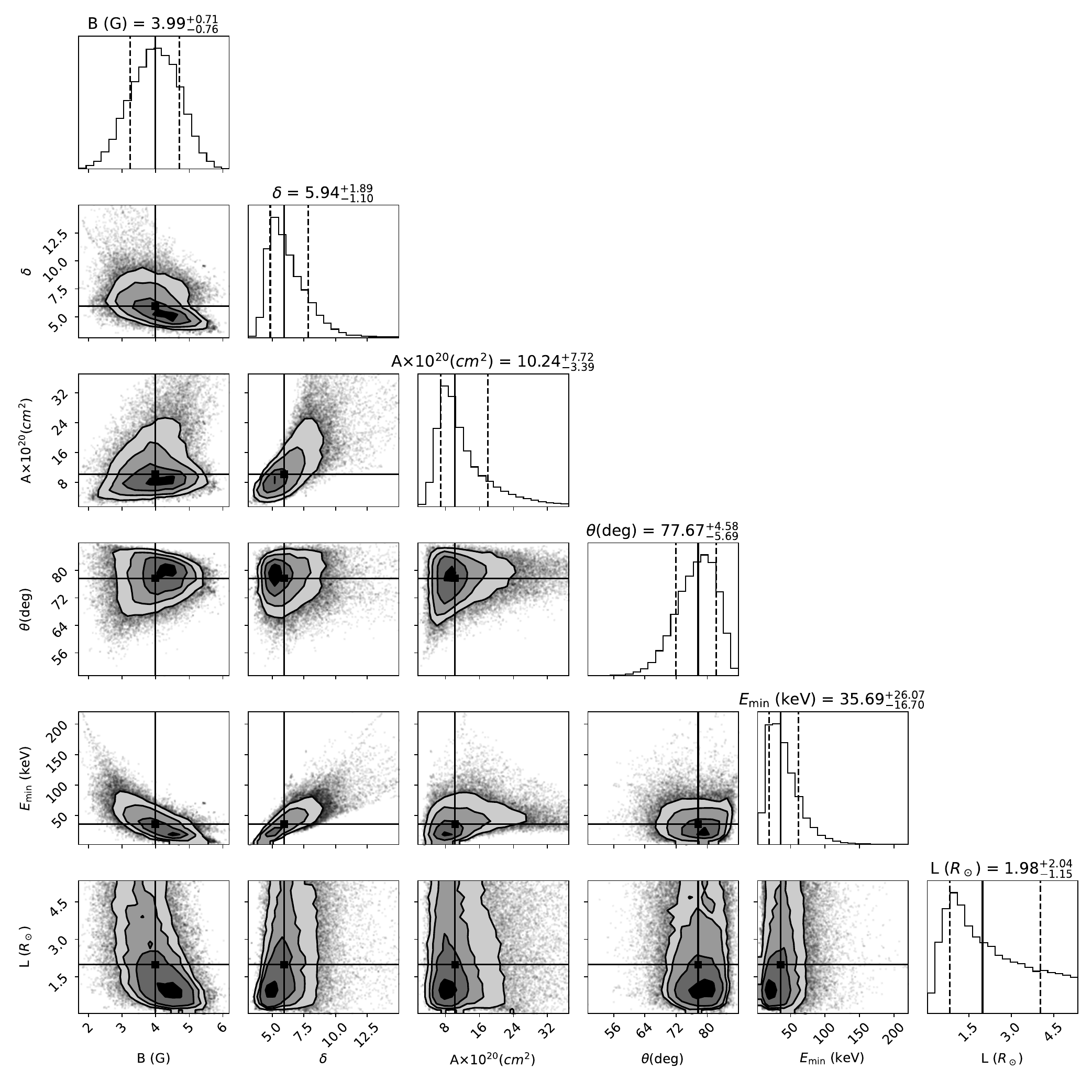}
    \caption[Correlation of posterior distributions of GS model parameters for region 3.]{Correlation of posterior distributions of GS model parameters for region 3. These panels show the joint probability distribution of any two parameters. The contours are at 0.5, 1, 2, and 3$\sigma$. The solid lines in the 1-dimensional histogram of posterior distributions mark the median values, and the vertical dashed lines mark the 16$^\mathrm{th}$ and 84$^\mathrm{th}$ percentiles. The median values are also marked in the panels showing the joint probability distribution.}
    \label{fig:corner_reg3}
\end{figure*}

Modeled and observed Stokes I and V spectra are shown in Figures \ref{fig:spectra_homo1}, \ref{fig:spectra_homo2}, and \ref{fig:spectra_homo3}. Modeled spectra are consistent with the observed Stokes I spectra and Stokes V upper limits for regions 3, 4, 7, 8, and 9 as evident from Figures \ref{fig:spectra_homo1} and \ref{fig:spectra_homo2}. Most GS model parameters are well-constrained as evident from the posterior distribution of GS model parameters for region 3 shown in Figure \ref{fig:corner_reg3}. 

The spectrum for Region 3 has seven Stokes I spectral points. Hence, for region 3, $L$ is kept as a free parameter. Fractional GS source depth, $f=L/L_\mathrm{max}$, is $\sim0.37$. For other regions, there are less than seven spectral points. Hence, to keep the number of free parameters in check, for other regions $f$ is assumed to be similar to that for region 3, and $L$ is kept fixed at $L=f\times L_\mathrm{max}$ and mentioned in Table \ref{table:south_params}. For regions 7, 8, and 9, there are 5 spectral points. For these regions, $E_\mathrm{min}$ is also kept fixed at the values arrived at for nearby regions. GS model parameters for all these regions are presented in Table \ref{table:south_params}. 

\subsection{Insufficiency of Optically Thick GS Spectrum}\label{subsec:reg1_optically_thick}
\begin{figure*}[!ht]
    \centering
     \includegraphics[trim={0.3cm 0.5cm 0.0cm 0.3cm},clip,scale=0.4]{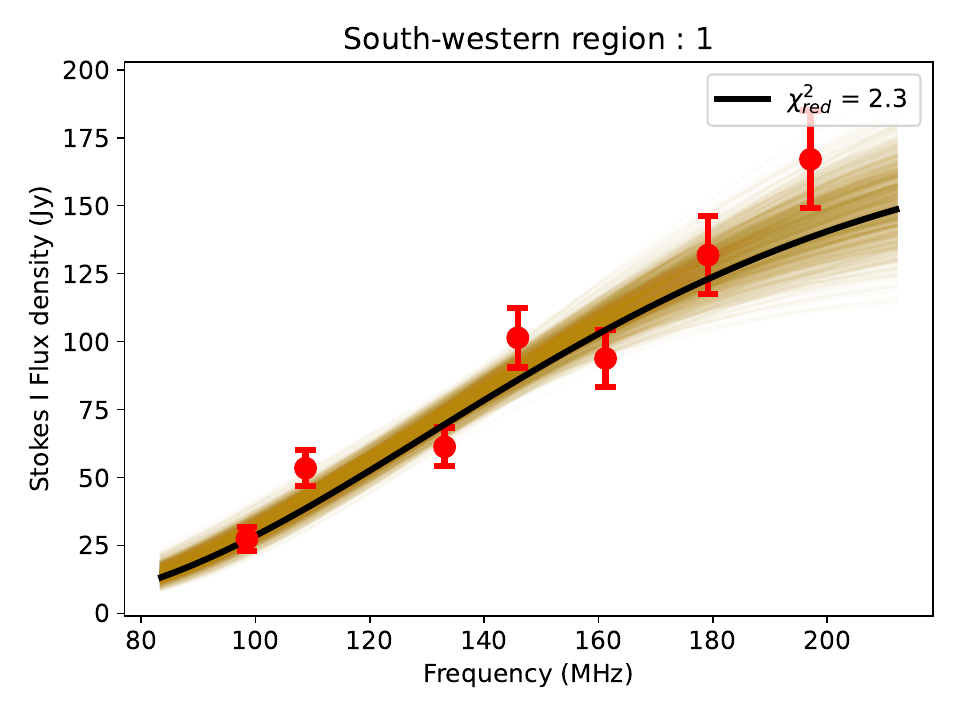} \includegraphics[trim={0.3cm 0.5cm 0.0cm 0.3cm},clip,scale=0.4]{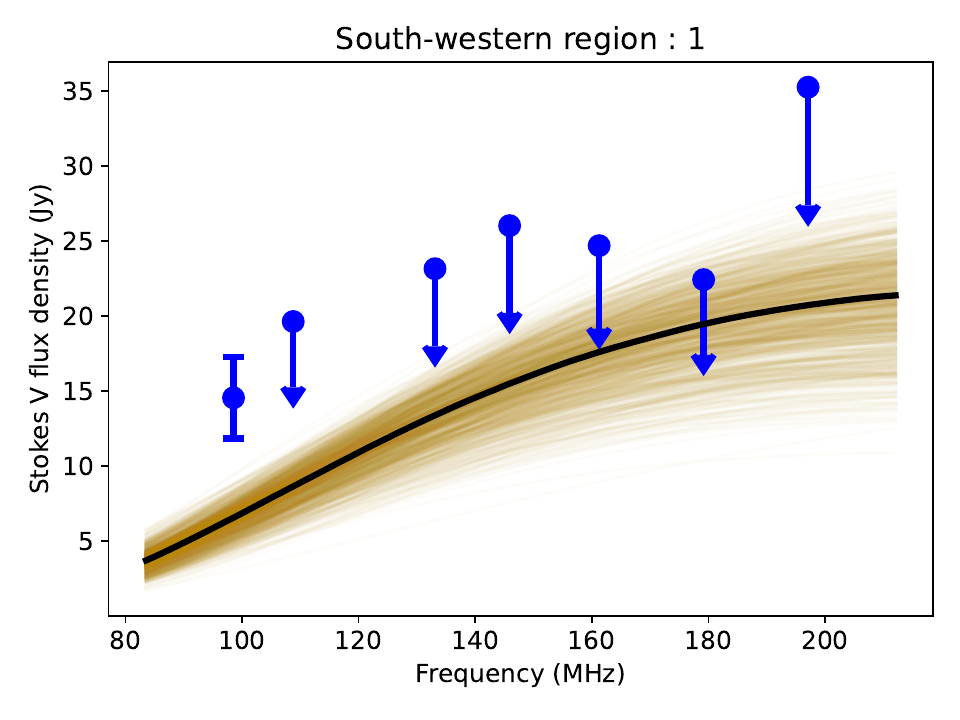}\\

     \includegraphics[trim={0.3cm 0.5cm 0.0cm 0.3cm},clip,scale=0.4]{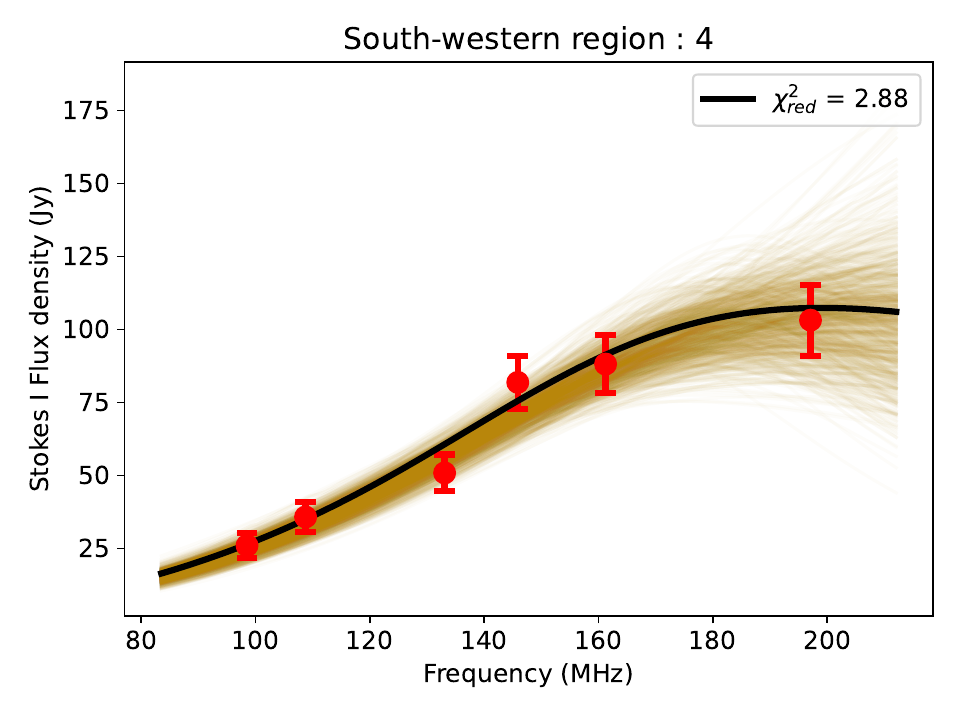}\includegraphics[trim={0.3cm 0.5cm 0.0cm 0.3cm},clip,scale=0.4]{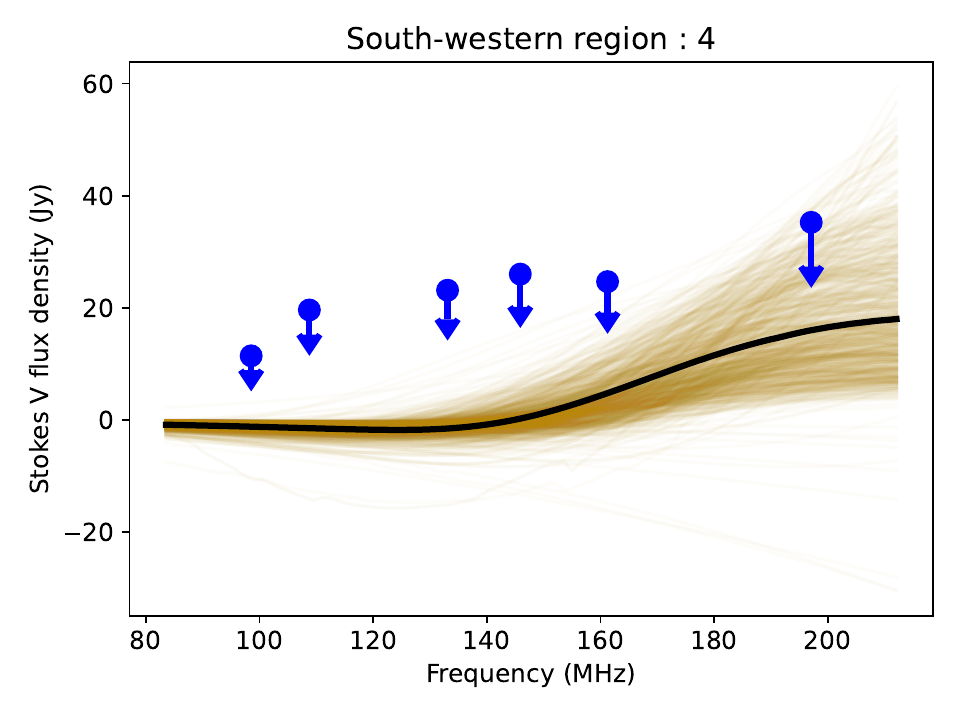}\\
     
    \caption[Observed and fitted spectra for regions 1 and 4 of south-western CME.]{Observed and fitted spectra for regions 1 and 4 of south-western CME. {\it Left column: }Stokes I spectrum is shown. Red points represent the observed flux densities. {\it Right column: }Stokes V spectrum is shown. Blue points represent the upper limits at each of the frequencies. The black lines represent the Stokes I and V GS spectra corresponding to GS parameters reported in Table \ref{table:south_params}. Light yellow lines show the GS spectra for 1000 realizations chosen randomly from the posterior distributions of the GS model parameters.}
    \label{fig:spectra_homo2}
\end{figure*}

\begin{figure*}[!ht]
    \centering
    \includegraphics[trim={0cm 0.6cm 0cm 0cm},clip,scale=0.55]{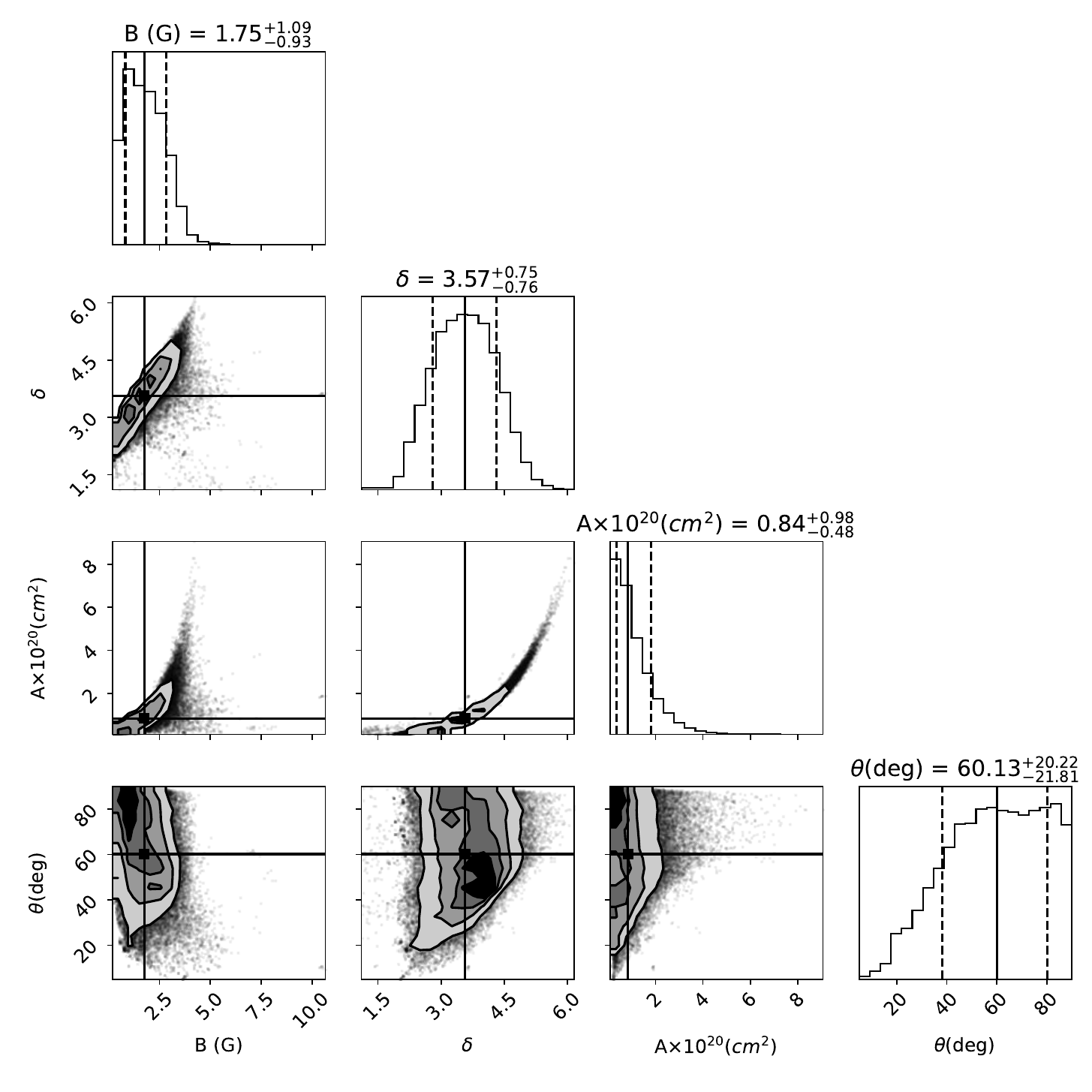}
    \caption[Correlation of posterior distributions of GS model parameters for region 4.]{Correlation of posterior distributions of GS model parameters for region 4. These panels show the joint probability distribution of any two parameters. The contours are at 0.5, 1, 2, and 3$\sigma$. The solid lines in the 1-dimensional histogram of posterior distributions mark the median values, and the vertical dashed lines mark the 16$^\mathrm{th}$ and 84$^\mathrm{th}$ percentiles. The median values are also marked in the panels showing the joint probability distribution.}
    \label{fig:corner_reg4}
\end{figure*}

The importance of sampling the spectral peak of GS spectra has been discussed in Section \ref{subsec:importance_peak} of Chapter \ref{cme_gs1}. Examples shown in Section \ref{subsec:importance_peak} of Chapter \ref{cme_gs1} are for regions 4 and 6 of CME-1. For those regions, although the spectral peak is not sampled, optically thin parts of the spectra are sampled. It is evident from Figures \ref{fig:param_sensitivity_magnetic}, \ref{fig:param_sensitivity_non_thermal} and \ref{fig:param_sensitivity_geometric} of Chapter \ref{cme_gs1}, optically thin part of the GS spectrum is more sensitive to GS model parameters, as compared to optically thick part. Hence, estimates of GS model parameters using only the optically thick part of the GS spectrum may not be that tightly constrained. 
I note that for regions 1 and 4 of CME-2, only the optically thick part of the spectra have been sampled, as shown in Figure \ref{fig:spectra_homo2}. For these regions, GS model parameters are not well constrained, which is evident from the posterior distribution of region 4 shown in Figure \ref{fig:corner_reg4}.

\section{Validity of Homogeneous and Isotropic \\Assumptions of GS Model}\label{subsec:homo_insuff}
\begin{figure*}[!ht]
    \centering
     \includegraphics[trim={0.3cm 0.5cm 0.0cm 0.3cm},clip,scale=0.4]{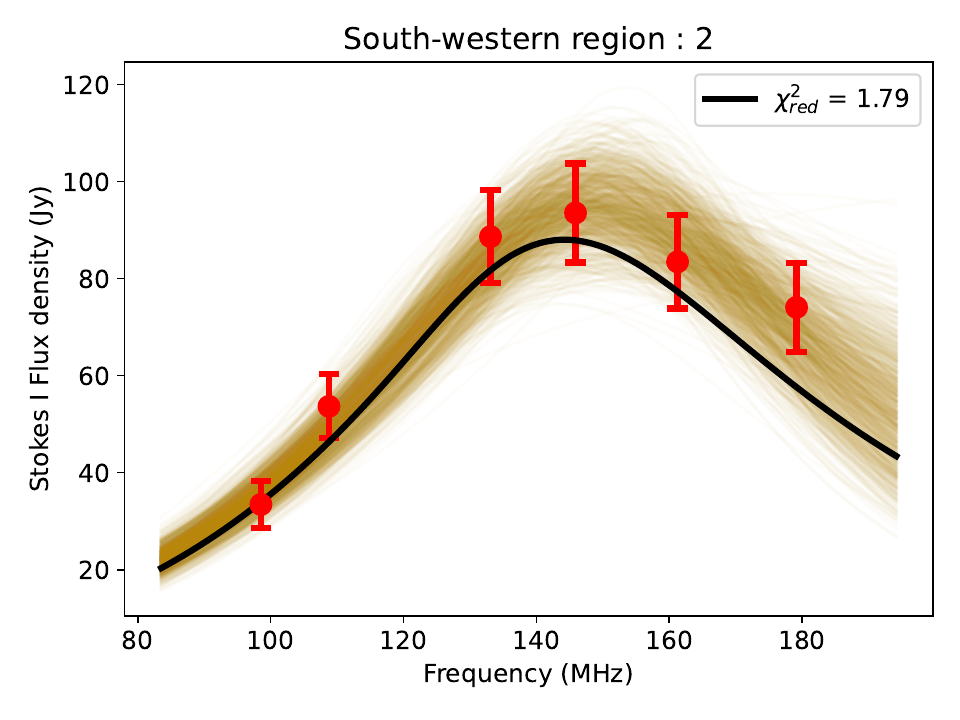}\includegraphics[trim={0.3cm 0.5cm 0.0cm 0.3cm},clip,scale=0.4]{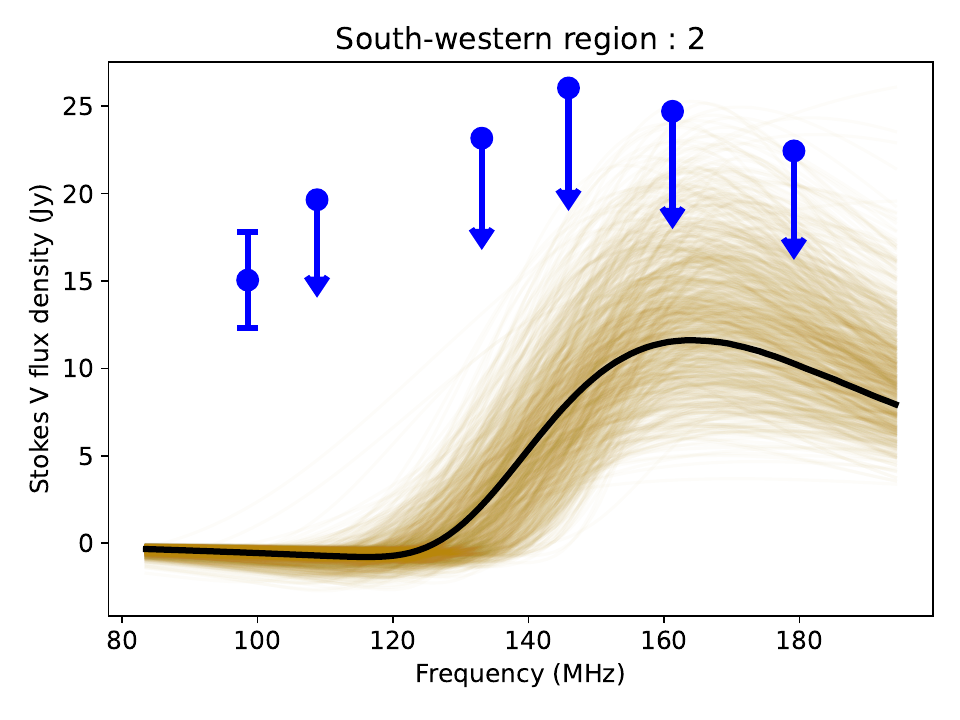}
    \caption[Observed and fitted spectrum for region 2 of south-western CME.]{Observed and fitted spectrum for region 2 of south-western CME. {\it Left column: }Stokes I spectrum is shown. Red points represent the observed flux densities. {\it Right column: }Stokes V spectrum is shown. Blue points represent the upper limits at each of the frequencies. The black lines represent the Stokes I and V GS spectra corresponding to GS parameters reported in Table \ref{table:south_params}. Light yellow lines show the GS spectra for 1000 realizations chosen randomly from the posterior distributions of the GS model parameters.}
    \label{fig:spectra_homo3}
\end{figure*}
\begin{figure*}[!ht]
    \centering
    \includegraphics[trim={0cm 0.6cm 0cm 0cm},clip,scale=0.45]{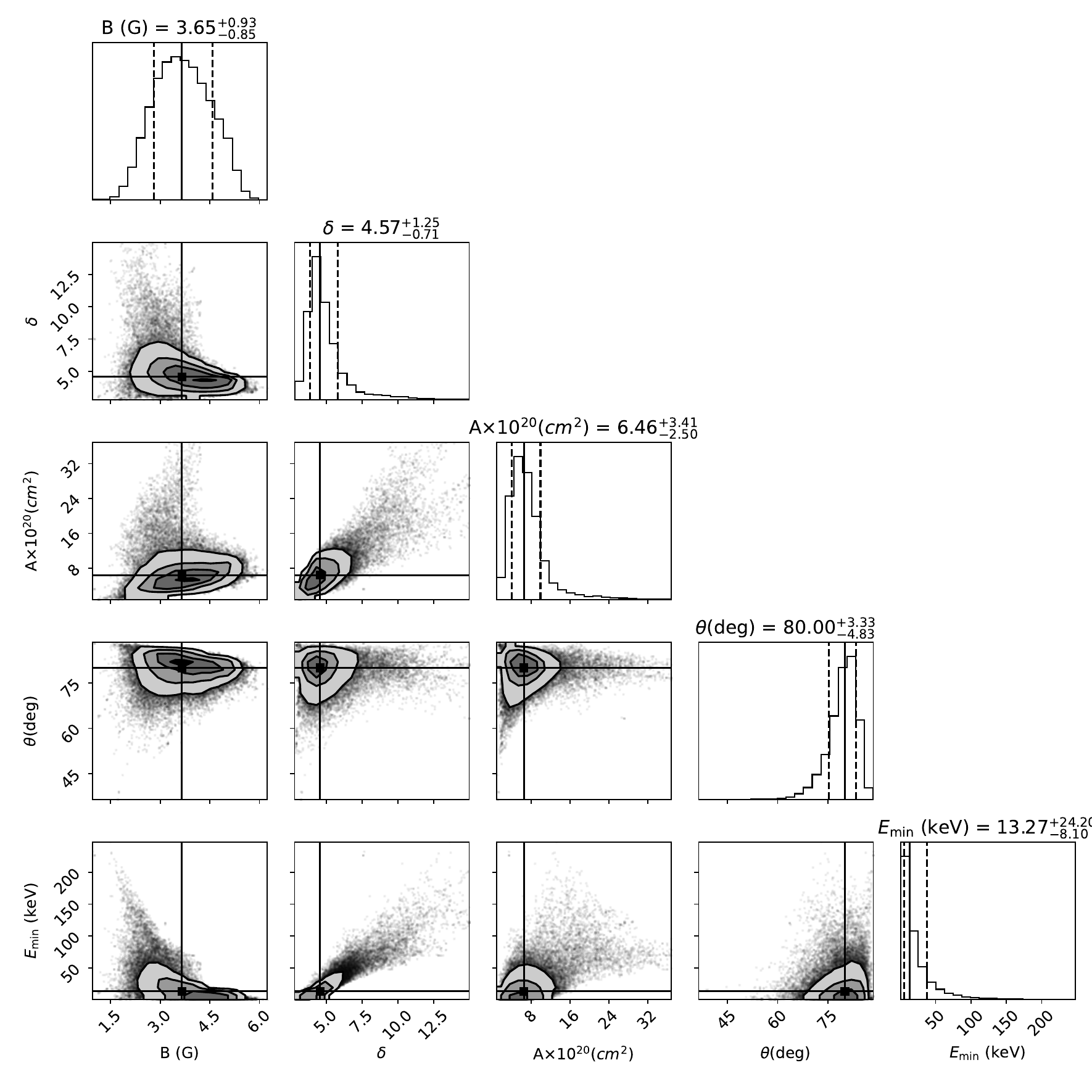}
    \caption[Correlation of posterior distributions of GS model parameters for region 2.]{Correlation of posterior distributions of GS model parameters for region 2. 2-dimensional plots show the joint probability distribution of any two parameters. The contours are at 0.5, 1, 2, and 3$\sigma$. The solid lines in the 1-dimensional histogram of posterior distributions mark the median values, and the vertical dashed lines mark the 16$^\mathrm{th}$ and 84$^\mathrm{th}$ percentiles. The median values are also marked in the panels showing the joint probability distribution.}
    \label{fig:corner_reg2}
\end{figure*}

Since the first attempt to model GS emission from CME loops to estimate plasma parameters and magnetic field by \cite{bastian2001}, all studies have assumed a homogeneous and isotropic GS source model. At the same time, there is also not been any observational evidence to claim that these assumptions are not valid. The modeling of the observed spectra of CME-2 (regions 3, 7, 8, and 9) and CME-1 (presented in Chapter \ref{cme_gs1}) including Stokes I spectra and stringent Stokes V upper limits also use homogeneous and isotropic GS models and can find well-constrained model parameters consistent with the observations.   

This is, however, not the case for regions 1 and 2 of CME-2. These two regions are different from others in that Stokes V emission has been detected from these regions at 98 MHz. Observed and modeled spectra for regions 1 and 2 are shown in the top panel of Figure \ref{fig:spectra_homo2} and Figure \ref{fig:spectra_homo3}, respectively. While the peak of the spectrum lies beyond the MWA frequency range for Region 1, the spectral peak has been sampled well for Region 2. As discussed in Section \ref{subsec:reg1_optically_thick}, sampling the spectral peak is essential for placing tight constraints on the GS model parameters. It is evident from Figure \ref{fig:corner_reg2} that this is indeed the case for Region 2. More importantly perhaps, while it is possible to find GS models consistent with the Stokes I measurements and the Stokes V upper limits, there is no GS model in the entire phase space explored which is simultaneously also consistent with the lone Stokes V measurement at 98 MHz. The ranges of the physical parameters explored here are sufficiently wide and it would be hard to justify expanding them beyond their present values. The inability or ability of the GS model used to find plasma parameters consistent with the data is likely a consequence of one or more of the assumptions made by the model being violated. To examine this possibility and attempt to identify the specific assumption being violated in the data, I systematically examine these assumptions, one at a time in the remainder of this Section. The key assumptions examined are -- restricting the electron energy distribution to a single power law (Section \ref{subsec:dpl_tnt_modeling}), ignoring any anisotropy in the nonthermal electron pitch-angle distribution (Section \ref{subsec:anisotropy_effect}), and the assumption of homogeneity in the plasma present in the volume being modeled by the GS model (Sections \ref{subsec:obs_evidence_inhom} and \ref{sec:nonuniform}).

\subsection{Modeling Observed Spectrum Using Different Electron Distributions}\label{subsec:dpl_tnt_modeling}
The fast GS code developed by \cite{Kuznetsov_2021} provides some analytical electron energy distribution functions as listed below:
\begin{enumerate}
    \item {\bf Single power-law distribution (PLW): }This is the simplest electron energy distribution which has typically been used in modeling of CME GS spectra and is given by Equation \ref{eq:PLW} of Chapter \ref{cme_gs1}.
    \item {\bf Double power-law distribution (DPL): }In this case, the electron energy distribution consists of two parts; high-energy and low-energy, where both the high-energy and low-energy parts are individually described by power-laws with different power-law indices. This double power-law distribution can be described as,
    \begin{equation}
    \begin{split}
        u_\mathrm{DPL}(E)dE=A_1E^{-\delta_1}dE, E_\mathrm{min}\leq E<E_\mathrm{break};\\
        =A_2E^{-\delta_2}dE, E_\mathrm{break}\leq E\leq E_\mathrm{max};
    \end{split}        
        \label{eq:dpl}
    \end{equation}
    $\delta_1$ and $\delta_2$ are power-law indices for lower and higher energy parts of the double power-laws. 
    In the above expression, $E_\mathrm{break}$ is referred to as the break energy and $A_1E_\mathrm{break}^{-\delta_1}=A_2E_\mathrm{break}^{-\delta_2}$ is imposed to make $u_\mathrm{DPL}(E)$ continuous. 
    The normalization factor is given by the following expression:
    \begin{equation}
        A_1^{-1}=\frac{2\pi}{n_\mathrm{nonth}}\left(\frac{E_\mathrm{min}^{1-\delta_1}-E_\mathrm{break}^{1-\delta_2}}{\delta_1-1}+E_\mathrm{break}^{\delta_2-\delta_1}\frac{E_\mathrm{break}^{1-\delta_2}-E_\mathrm{max}^{1-\delta_2}}{\delta_2-1}\right)
        \label{eq:dpl_norm}
    \end{equation}
    and $A_2$ is determined from the continuity condition. 
    
    \item {\bf Thermal/nonthermal distribution over energy (TNT):} This distribution behaves like a thermal distribution (THM) at low energies and a single power-law nonthermal distribution at high energies, with a continuous transition at some energy, $E_\mathrm{cr}$. The distribution is given as 
    \begin{equation}
    \begin{split}
         u_\mathrm{TNT}(E)dE= u_\mathrm{THM}(E)dE, E<E_\mathrm{cr};\\
        =AE^{-\delta}dE, E_\mathrm{cr}\leq E \leq E_\mathrm{max},
    \end{split}
    \label{eq:tnt}
    \end{equation}
     where, $u_\mathrm{THM}(E)$ is relativistic thermal distribution and $A=u_\mathrm{THM}(E_\mathrm{cr})E_\mathrm{cr}^{-\delta}$ to make the function continuous. 
     For 1 MK plasma, thermal energy is $\sim100$ eV, while $E_\mathrm{min}$ estimated in the present context many orders of magnitude larger, lying in the range of a few tens to hundreds of keV (Table \ref{table:south_params}). Hence, this distribution is not considered for this exercise.
    \item {\bf Isotropic thermal and power-law over energy (TPL): } In this case the electron distribution function represents a sum of an isotropic thermal distribution and a single power-law distribution given by
    \begin{equation}
        u_\mathrm{TPL}(E)dE=u_\mathrm{THM}(E)dE+u_\mathrm{PLW}(E)dE,
    \end{equation}
    where, $u_\mathrm{TPL}(E)$ is thermal electron distribution function and $u_\mathrm{PLW}(E)$ is single power-law distribution presented in Equation \ref{eq:PLW} of Chapter \ref{cme_gs1}.
    
    \item {\bf Isotropic thermal and double power-law over energy (TPD): } In this case, electron distribution is similar to TPL distribution, except that the nonthermal component has the double power-law energy dependence, as given in Equation \ref{eq:dpl}. The electron distribution is given as,
     \begin{equation}
        u_\mathrm{TPD}(E)dE=u_\mathrm{THM}(E)dE+u_\mathrm{DPL}(E)dE,
    \end{equation} 
\end{enumerate}
\begin{figure*}[!htbp]
    \centering
     \includegraphics[trim={0.3cm 0.5cm 0.0cm 0.3cm},clip,scale=0.4]{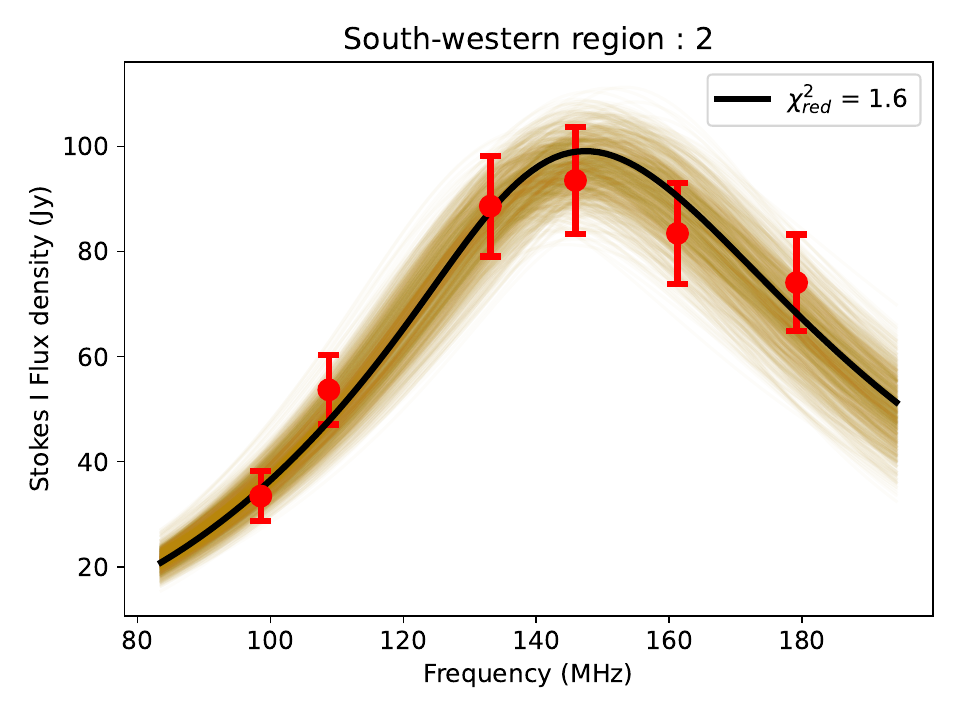} \includegraphics[trim={0.3cm 0.5cm 0.0cm 0.3cm},clip,scale=0.4]{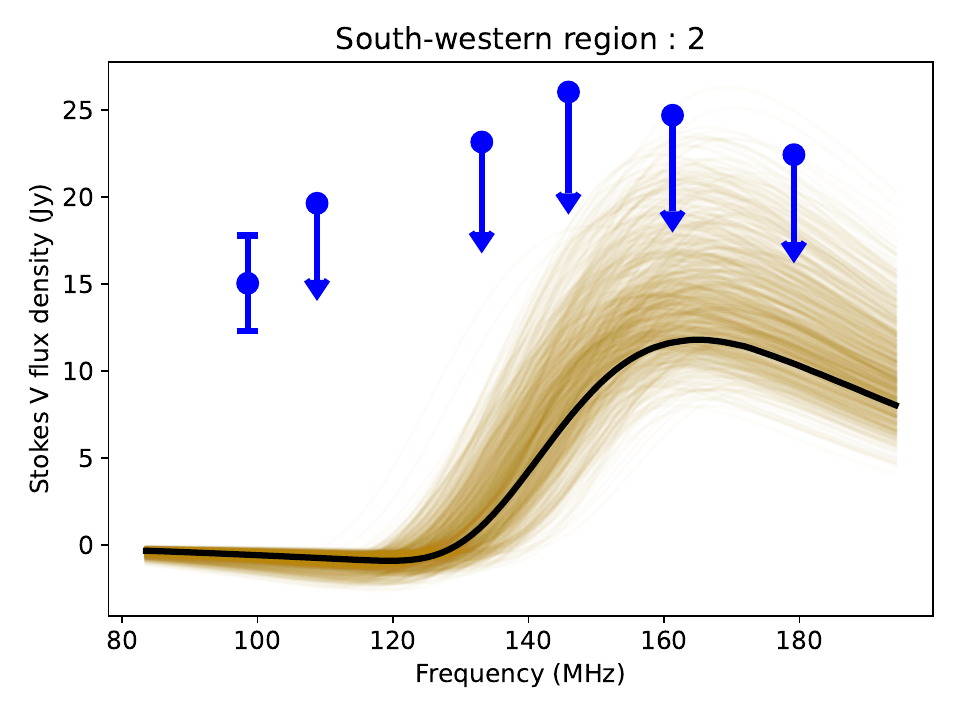}\\

     \includegraphics[trim={0.3cm 0.5cm 0.0cm 0.3cm},clip,scale=0.4]{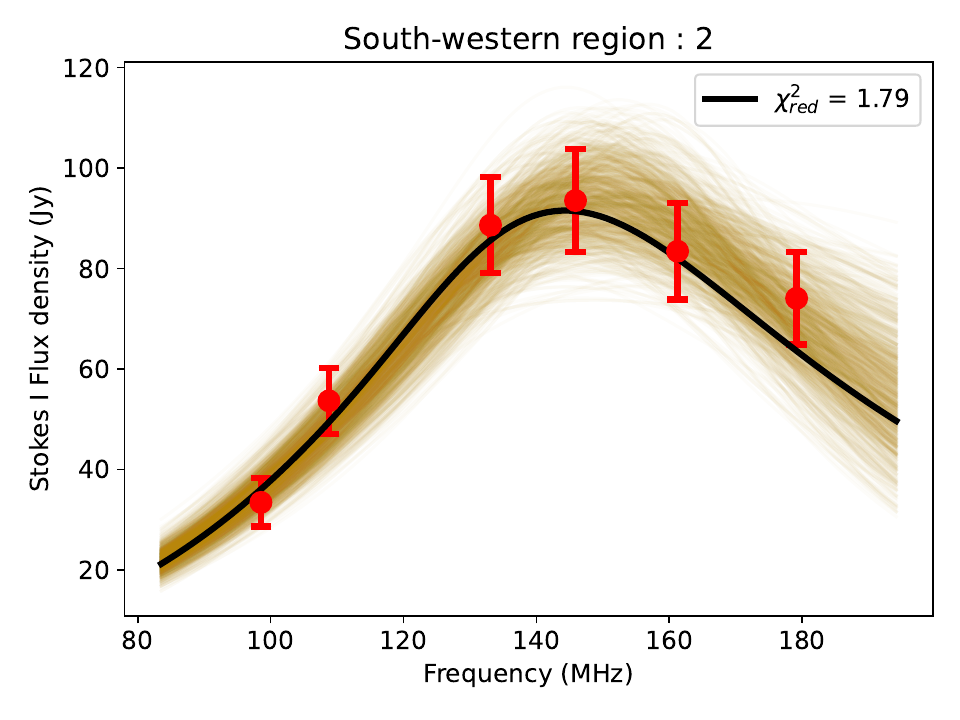}\includegraphics[trim={0.3cm 0.5cm 0.0cm 0.3cm},clip,scale=0.4]{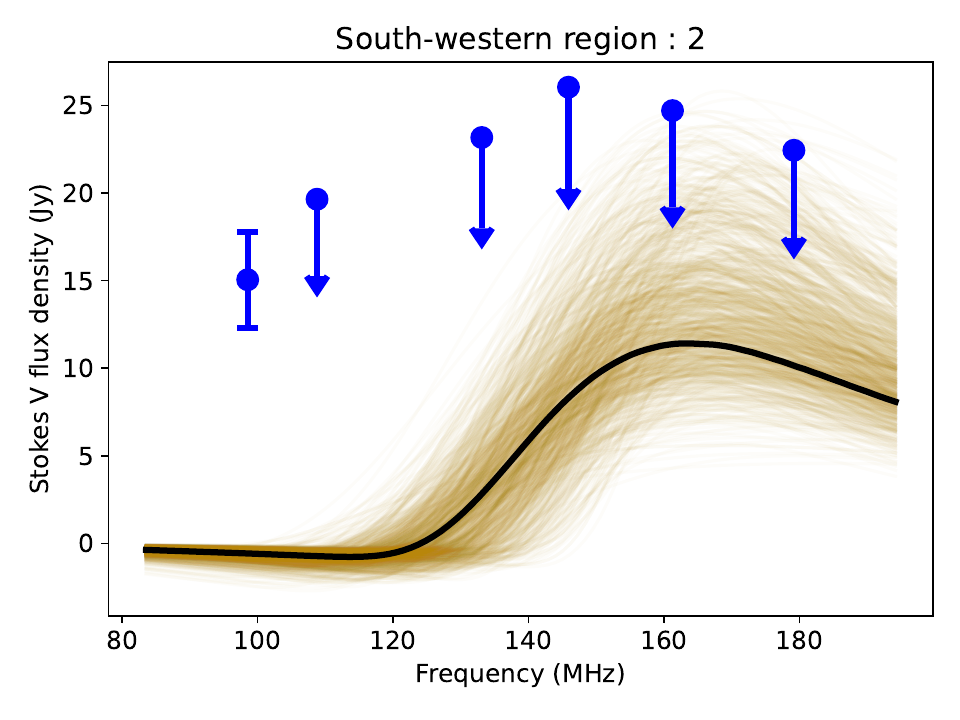}\\

      \includegraphics[trim={0.3cm 0.5cm 0.0cm 0.3cm},clip,scale=0.4]{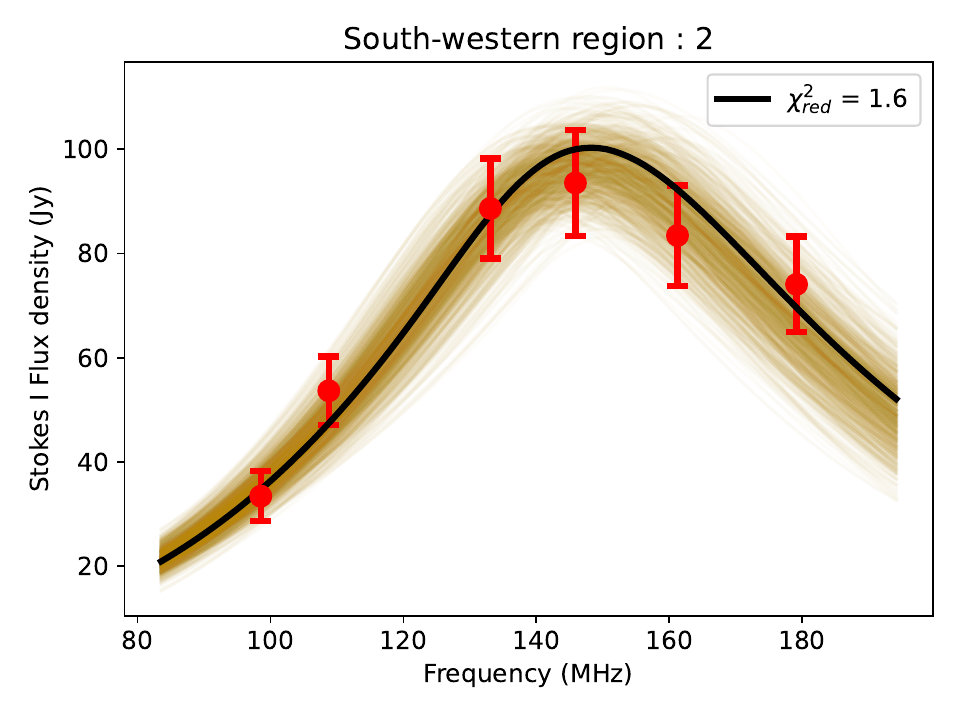}\includegraphics[trim={0.3cm 0.5cm 0.0cm 0.3cm},clip,scale=0.4]{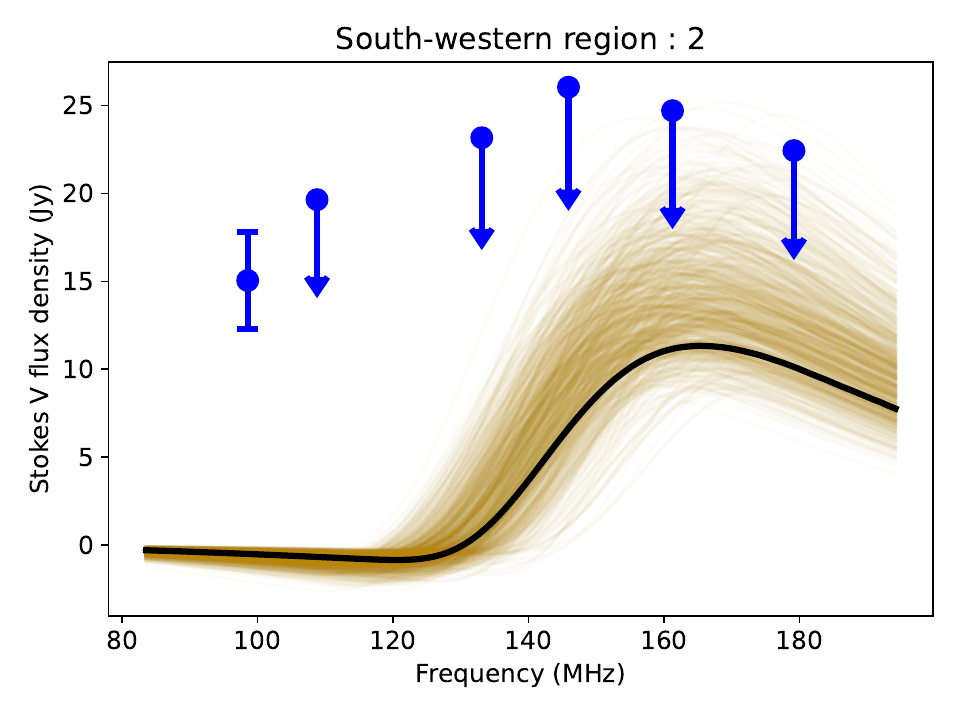}\\
     
    \caption[Observed and fitted spectra for region 2 of south-western CME using different electron distribution.]{Observed and fitted spectra for region 2 of south-western CME using different electron distribution. Spectra for DPL, TPL, and TPD electron distributions are shown in the top, middle, and bottom panels, respectively. {\it Left column: }Stokes I spectrum is shown. Red points represent the observed flux densities. {\it Right column: }Stokes V spectrum is shown. Blue points represent the upper limits at each of the frequencies. The black lines represent the Stokes I and V GS spectra corresponding to GS parameters reported in Table \ref{table:south_params}. Light yellow lines show the GS spectra for 1000 realizations chosen randomly from the posterior distributions of the GS model parameters.}
    \label{fig:spectra_dpl_tnt}
\end{figure*}

\begin{figure*}[!htbp]
    \centering
    \includegraphics[trim={0cm 0.6cm 0cm 0cm},clip,scale=0.4]{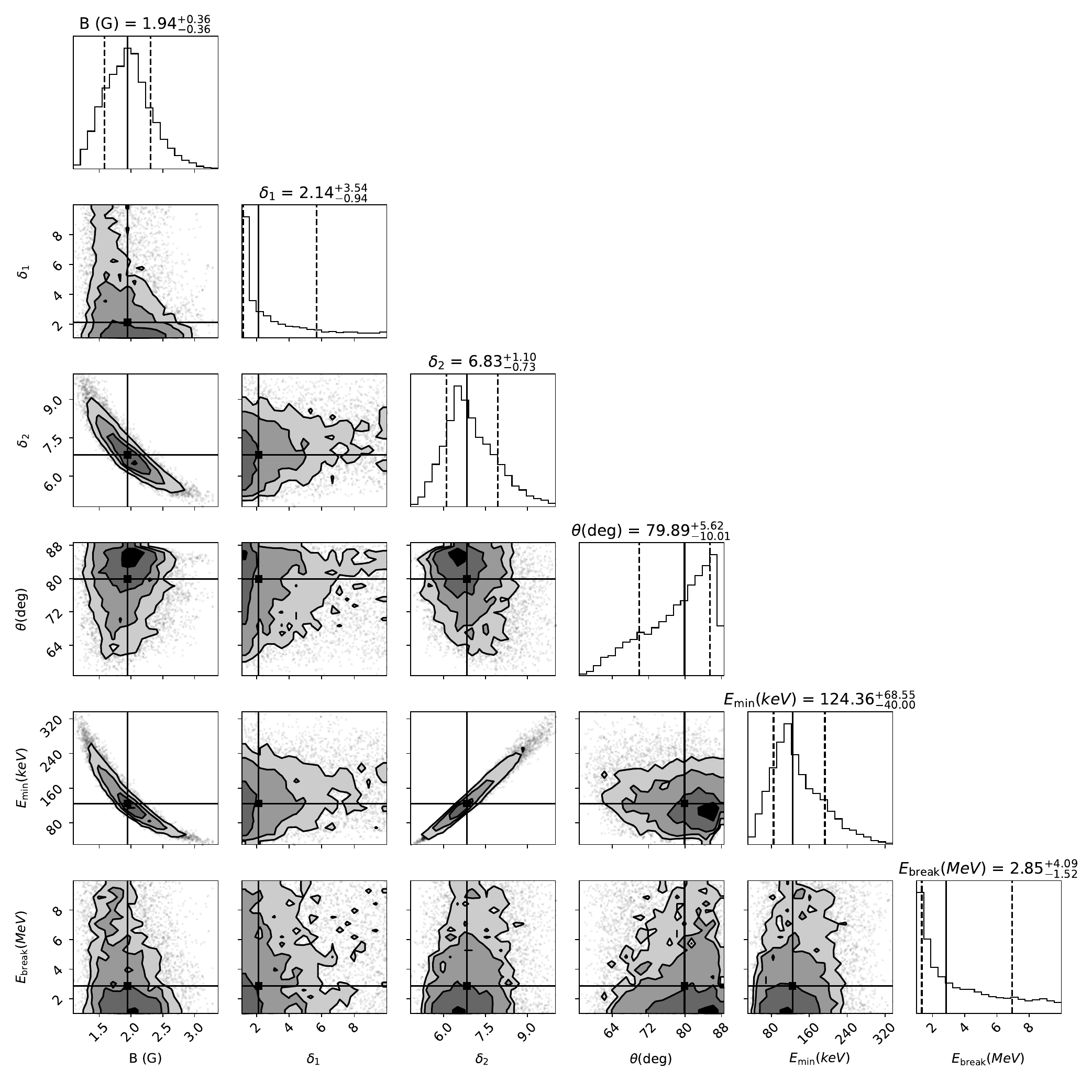}
    \caption[Correlation of posterior distributions of GS model parameters for region 2 considering double power-law electron energy distribution.]{Correlation of posterior distributions of GS model parameters for region 2 considering double power-law electron energy distribution. These panels show the joint probability distribution of any two parameters. The contours are at 0.5, 1, 2, and 3$\sigma$. The solid lines in the 1-dimensional histogram of posterior distributions mark the median values, and the vertical dashed lines mark the 16$^\mathrm{th}$ and 84$^\mathrm{th}$ percentiles. The median values are also marked in the panels showing the joint probability distribution.}
    \label{fig:dep_corner_reg2}
\end{figure*}
\begin{table}[!ht]
\centering
    \renewcommand{\arraystretch}{1.4}
    \begin{tabular}{|p{2cm}|p{2.2cm}|p{2.2cm}|p{2.2cm}|p{2.2cm}|p{2.2cm}|}
    \hline 
       Electron \newline{distribution} & PLW & DPL & TPL & TPD\\ \hline \hline
        $|B|$ (G) & $1.72_{-0.52}^{+0.64}$ &  $1.94_{-0.36}^{+0.36}$ &  $1.82_{-0.42}^{+0.49}$ & $1.89_{-0.35}^{+0.39}$\\
        \hline
        $\delta$ & $6.82_{-1.56}^{+1.88}$ &  -- &  $6.27_{-1.26}^{+2.03}$ & -- \\
        \hline
        $\delta_1$ & -- & $2.14_{-0.94}^{+3.54}$ &  -- & $2.87_{-1.47}^{+4.07}$ \\
        \hline
        $\delta_2$ & -- & $6.83_{-0.73}^{+1.1}$ &  -- & $6.97_{-0.81}^{+1.09}$ \\
       \hline
        $E_\mathrm{min}$ (keV) & $139.65_{-67.09}^{+107.86}$ & $124.36_{-40.00}^{+68.55}$ & $115.19_{-49.31}^{+85.44}$ & $133.09_{-45.37}^{+68.89}$ \\
       \hline
        $E_\mathrm{break}$ (MeV) & -- &  $2.85_{-1.52}^{+4.09}$ & -- & $2.68_{-1.43}^{+4.11}$ \\
       \hline
        $\theta$ & $79.34_{-10.22}^{+5.96}$ &  $79.89_{-10.01}^{+5.62}$ & $80.38_{-9.92}^{+5.36}$ & $80.47_{-10.10}^{+5.27}$ \\
       \hline
    \end{tabular}
    \caption[Estimated GS model parameters for different electron energy density distributions.]{\textbf{Estimated GS model parameters for different electron energy density distributions.}}
    \label{table:reg2_params_diff_elecdis}
\end{table}

I have already been using the PLW distribution for GS modeling. 
Of the remaining 4 analytic models just mentioned, for the reasons already pointed out, the TNT model is not appropriate for the present application. As it was discussed in Section \ref{subsec:emission_mechanism} of Chapter \ref{cme_gs1}, contributions due to free-free emission from thermal plasma are so weak that they lie below the rms noise of the present measurements. For this reason, for the data available, the TPL and TPD models should be indistinguishable from PLW and DPL models, respectively. However, to verify this expectation, I also consider TPL and TPD distributions for the present exercise.

I have considered three different electron distributions -- DPL, TPL, and TPD. The observed Stokes I and V spectra for region 2 are fitted jointly considering homogeneous and isotropic GS model with each of these electron distributions. DPL and TPD models require a larger number of free parameters. To keep the problem well constrained, the geometric parameters $A$ and $L$ are kept fixed at the value mentioned in Table \ref{table:south_params} for all three models considered. 

Modeled spectra for DPL, TPL, and TPD electron distributions are shown in the top, middle, and bottom panels of Figure \ref{fig:spectra_dpl_tnt}, respectively. Posterior distributions of GS modeled parameters for DPL electron distribution, as well as $\theta$ and $B$, are shown in Figure \ref{fig:dep_corner_reg2}. GS model parameters for TPL and TPD distributions are essentially the same as obtained for PLW and DPL models, as is evident from the parameter values listed in Table \ref{table:reg2_params_diff_elecdis}. It is indeed interesting to note that the estimated values of the three common GS parameters in these models, $|B|$, $E_\mathrm{min}$, and $\theta$, are consistent within their uncertainties. 

For all of these distributions, the models are consistent with the Stokes I flux densities and Stokes V upper limits, but the lone observed Stokes V detection is not consistent with any of the models. This exercise establishes that none of the prevalent homogeneous and isotropic electron density models can reproduce the observed Stokes I and V spectra simultaneously.

\subsection{Possible Effects of Anisotropic Electron Distribution}\label{subsec:anisotropy_effect}
In general, mildly-relativistic electrons injected in CME plasma either due to magnetic reconnection \citep{DuBois2017,agudelo_2021} or shock acceleration or both could well be anisotropic during the initial phases and the anisotropy can sustain till later times \citep{Simnett_2002,Giacalone_2021}. This anisotropic distribution becomes isotropic over time due to collisional or turbulent scattering \citep{Kuznetsov_2021}. Hence, it is interesting to consider the impact of an anisotropic pitch-angle distribution for modeling the observed GS spectra for region 2. In this case, electron distribution is given by
\begin{equation}
    f(E,\mu)=u(E)\ g(\mu),
\end{equation}
where, $u(E)$ is the electron energy density distribution and $g(\mu)$ is the pitch angle distribution, $\mu=cos\ \alpha$, with $\alpha$ being the pitch angle.
\begin{figure*}[!ht]
    \centering
     \includegraphics[trim={0.3cm 0.5cm 0.0cm 0.3cm},clip,scale=0.4]{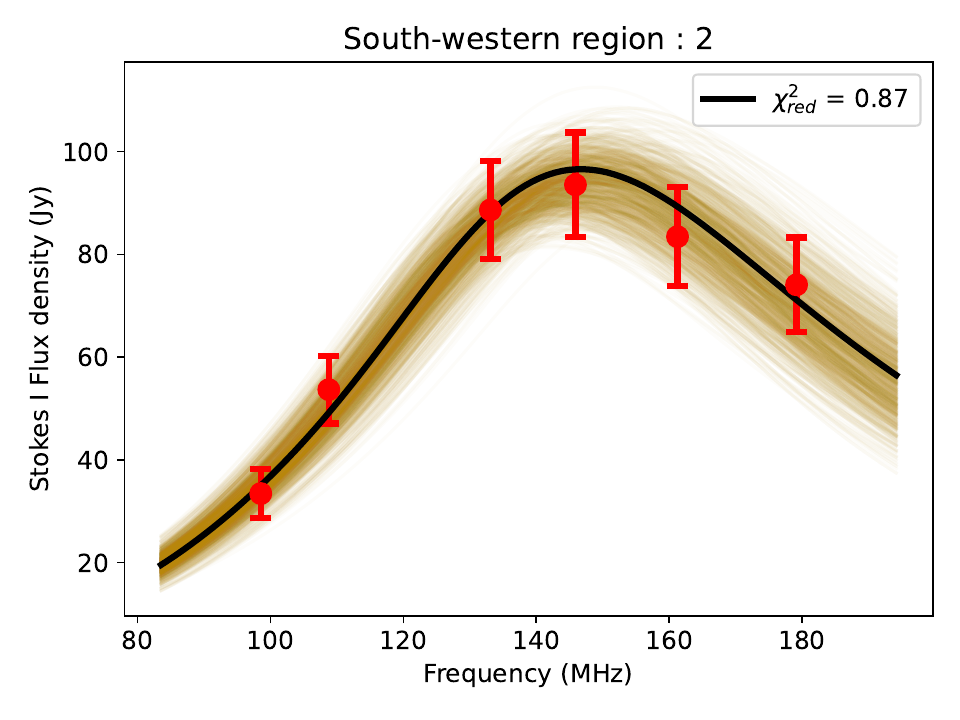} \includegraphics[trim={0.3cm 0.5cm 0.0cm 0.3cm},clip,scale=0.4]{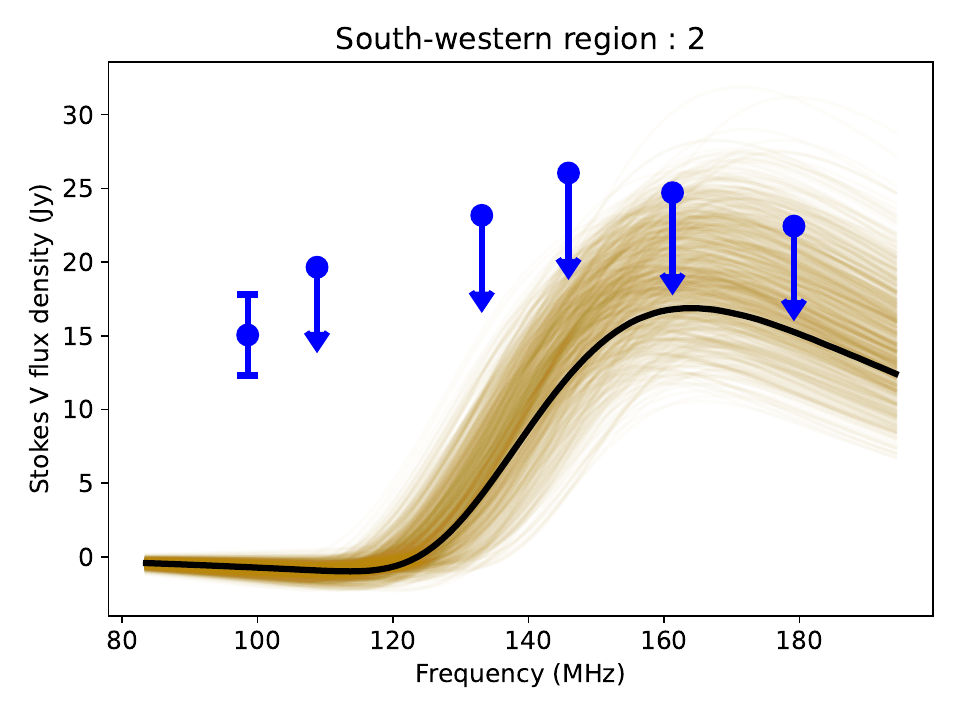}\\

     \includegraphics[trim={0.3cm 0.5cm 0.0cm 0.3cm},clip,scale=0.4]{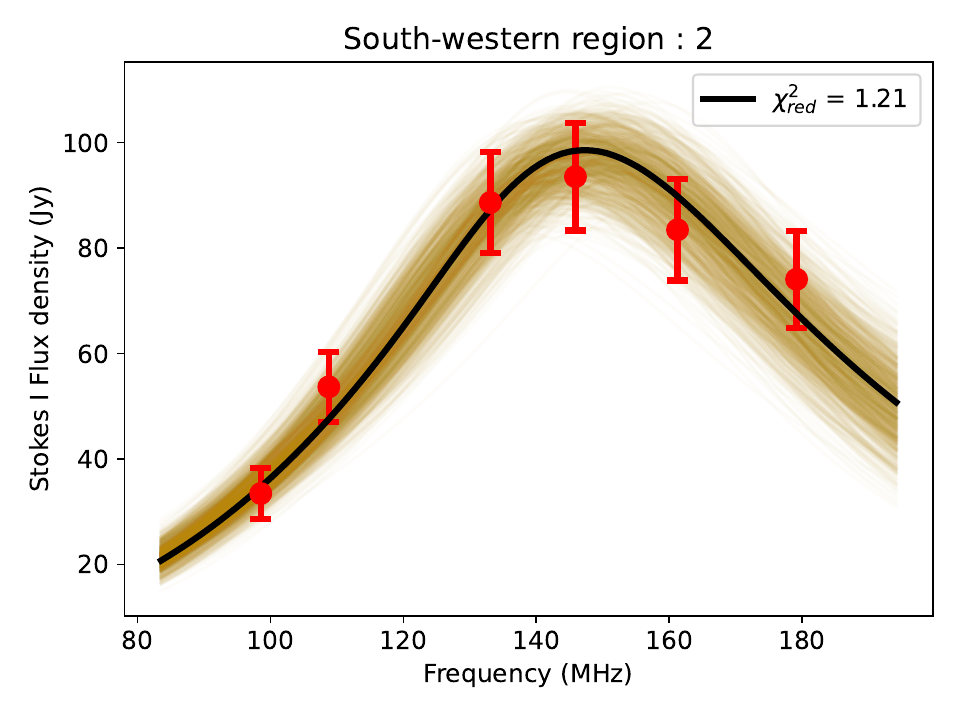}\includegraphics[trim={0.3cm 0.5cm 0.0cm 0.3cm},clip,scale=0.4]{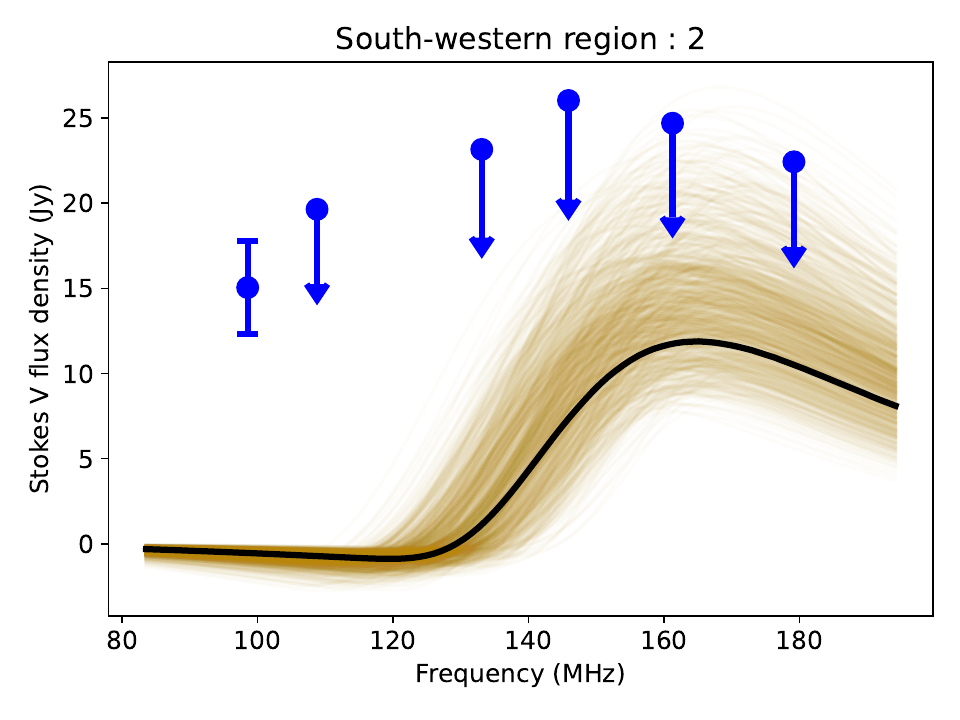}\\

    \caption[Observed and fitted spectra for region 2 of south-western CME using different electron pitch-angle distributions.]{Observed and fitted spectra for region 2 of south-western CME using different electron pitch-angle distributions. Spectra for GAU and GLC pitch-angle distributions are shown in the top and bottom panels, respectively. {\it Left column: }Stokes I spectrum is shown. Red points represent the observed flux densities. {\it Right column: }Stokes V spectrum is shown. Blue points represent the upper limits at each of the frequencies. The black lines represent the Stokes I and V GS spectra corresponding to GS parameters reported in Table \ref{table:south_params}. Light yellow lines show the GS spectra for 1000 realizations chosen randomly from the posterior distributions of the GS model parameters.}
    \label{fig:glc_gau_spectra}
\end{figure*}
\begin{figure*}[!htbp]
    \centering
    \includegraphics[trim={0cm 0.6cm 0cm 0cm},clip,scale=0.45]{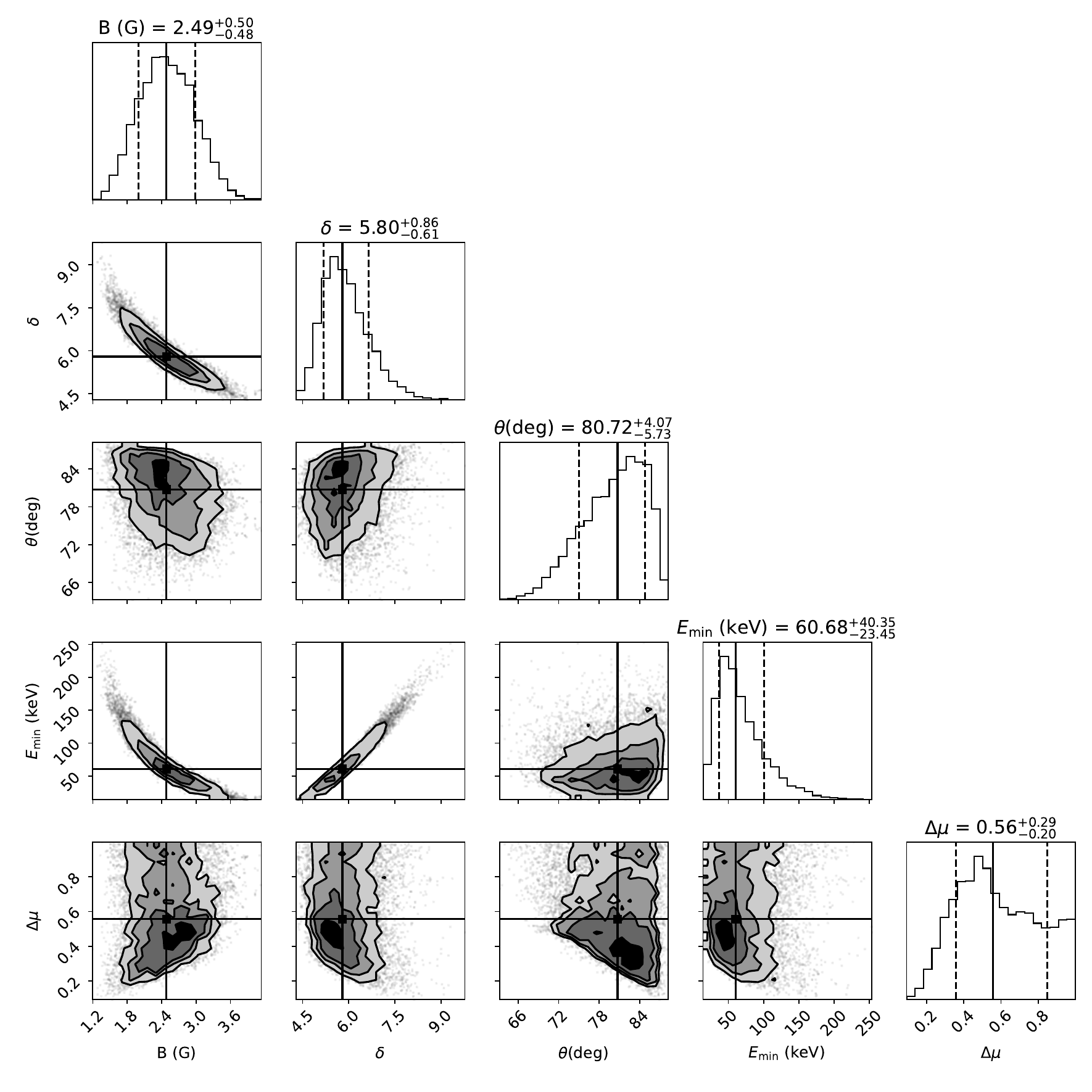}
    \caption[Correlation of posterior distributions of GS model parameters for region 2 considering GAU electron pitch-angle distribution.]{Correlation of posterior distributions of GS model parameters for region 2 considering GAU electron pitch-angle distribution. These panels show the joint probability distribution of any two parameters. The contours are at 0.5, 1, 2, and 3$\sigma$. The solid lines in the 1-dimensional histogram of posterior distributions mark the median values, and the vertical dashed lines mark the 16$^\mathrm{th}$ and 84$^\mathrm{th}$ percentiles. The median values are also marked in the panels showing the joint probability distribution.}
    \label{fig:gau_corner}
\end{figure*}
\begin{figure*}[!ht]
    \centering
    \includegraphics[trim={0cm 0.6cm 0cm 0cm},clip,scale=0.45]{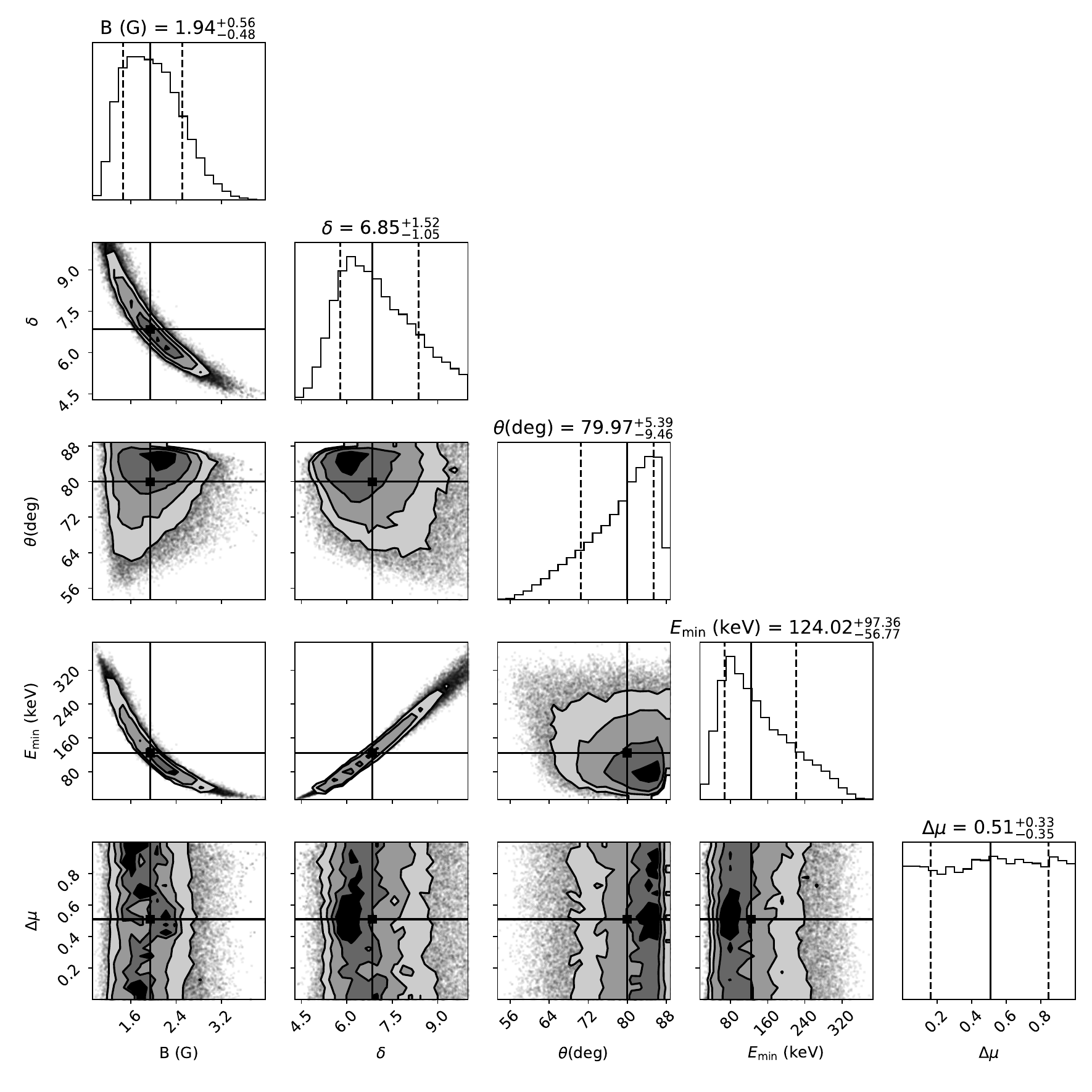}
    \caption[Correlation of posterior distributions of GS model parameters for region 2 considering GLC electron pitch-angle distribution.]{Correlation of posterior distributions of GS model parameters for region 2 considering GLC electron pitch-angle distribution. These panels show the joint probability distribution of any two parameters. The contours are at 0.5, 1, 2, and 3$\sigma$. The solid lines in the 1-dimensional histogram of posterior distributions mark the median values, and the vertical dashed lines mark the 16$^\mathrm{th}$ and 84$^\mathrm{th}$ percentiles. The median values are also marked in the panels showing the joint probability distribution.}
    \label{fig:glc_corner}
\end{figure*}

I have considered two types of analytical pitch-angle distribution available in fast GS code \citep{Fleishman_2010,Kuznetsov_2021}. These are:
\begin{enumerate}
     \item {\bf Gaussian beam distribution (GAU): }In this case, $g(\mu)$ is given as,
    \begin{equation}
        g(\mu)=A\ \mathrm{exp} \left[-\frac{(\mu-\mu_0)^2}{\Delta\mu^2}\right],
    \end{equation}
    where, $\mu_0=cos\ \alpha_0$; $\alpha_0$ is the beam direction, and $\Delta\mu$ depends on beam angular width. $\alpha_0$ is kept fixed at 90 degrees. $A$ is the normalization constant.
    \item {\bf Gaussian loss-cone distribution (GLC): } In this case, $g(\mu)$ is given as
    \begin{equation}
    \begin{split}
        g_\mathrm{GLC}(\mu)&=A, |\mu|<\mu_c;\\
        &=A\ \mathrm{exp}\left(-\frac{(|\mu|-\mu_c)^2}{\Delta\mu^2}\right), |\mu|\geq\mu_c;
    \end{split}
    \end{equation}
    \label{eq:GLC}
    where, $\mu_c=cos\ \alpha_c\ > 0$ is the loss-cone boundary and $\Delta\mu$ determines the sharpness of loss-cone boundary. $\alpha_c$ is kept fixed at 45 degrees. Normalization factor $A$ is given as,
    \begin{equation}
        A^{-1}=2\left[\mu_c+\frac{\sqrt{\pi}}{2}\Delta\mu\ erf(\frac{1-\mu_c}{\Delta\mu})\right],
    \end{equation}
    where $erf$ is the error function.
\end{enumerate}
\begin{table}[!ht]
\centering
    \renewcommand{\arraystretch}{1.4}
    \begin{tabular}{|p{3cm}|p{2.2cm}|p{2.2cm}|p{2.2cm}|}
    \hline 
       Pitch-angle \newline{distribution} & Isotropic & GAU & GLC \\ \hline \hline
        $|B|$ (G) & $1.72_{-0.52}^{+0.64}$ &  $2.49_{-0.48}^{+0.50}$ &  $1.94_{-0.48}^{+0.56}$\\
        \hline
        $\delta$ & $6.82_{-1.56}^{+1.88}$ &  $5.80_{-0.61}^{+0.86}$ &  $6.85_{-1.05}^{+1.52}$\\
        \hline
        $E_\mathrm{min}$ (keV) & $139.65_{-67.09}^{+107.86}$ & $60.68_{-23.45}^{+40.35}$ & $124.02_{-56.77}^{+97.35}$\\
       \hline
        $\theta$ & $79.34_{-10.22}^{+5.96}$ &  $80.72_{-5.73}^{+4.07}$ & $79.97_{-9.46}^{+5.39}$ \\
       \hline
       $\Delta\mu$ & -- & $0.56_{-0.20}^{+0.29}$ & $0.51_{-0.35}^{+0.33}$\\
       \hline
    \end{tabular}
    \caption[Estimated GS model parameters for different electron pitch-angle distributions.]{\textbf{Estimated GS model parameters for different electron pitch-angle distributions.}}
    \label{table:reg2_params_diff_pitch}
\end{table}

GS spectrum modeling is performed for region 2 of CME-2 considering homogeneous PLW electron distribution with both GLC and GAU pitch-angle distributions. The modeled spectra for GAU and GLC pitch-angle distributions are shown in the top and the bottom panels of Figure \ref{fig:glc_gau_spectra}. To avoid increasing the number of free parameters to be constrained during the modeling process, I have kept the geometrical parameters fixed at the values listed in Table \ref{table:south_params}. 

It is evident from Figure \ref{fig:glc_gau_spectra} that introducing anisotropy of electron pitch-angle distributions (using a GAU or GLC model) in the GS model is unable to reproduce the observed Stokes I and V simultaneously. The model values estimated for isotropic, GAU, and GLC pitch-angle distributions are listed in Table \ref{table:reg2_params_diff_pitch}. Other plasma parameters have similar estimated values that have been estimated considering an isotropic pitch-angle distribution. The posterior distribution of model parameters for GAU and GLC pitch-angle distributions are shown in Figures \ref{fig:gau_corner} and \ref{fig:glc_corner}, respectively. The pitch-angle distribution parameter, $\Delta\mu$ is poorly constrained when using the GAU model and even more so when using the GLC model.

\subsection{Insufficiency of Homogeneous GS Model}\label{subsec:obs_evidence_inhom}
Earlier in this thesis, I demonstrated that including even the sensitive upper limits from Stokes V observations along with the Stokes I spectrum can significantly improve the ability to constrain the GS model parameters (Chapter \ref{cme_gs1}). The preceding discussion shows that despite the availability of stringent Stokes V upper limits and Stokes I spectrum, a Stokes V detection even at a single spectral point can provide significant additional information. However, the routinely used GS modeling approach is unable to find a model consistent with Stokes V detection, Stokes V upper limits, and Stokes I spectrum simultaneously. This led us to critically examine the assumptions made during the GS modeling process. I have just demonstrated that considering more general electron energy distributions and accounting for the impact of their pitch-angle distributions, while holding the assumption of homogeneity, are not sufficient to meet these constraints. The remaining assumption is that of homogeneity in the CME plasma along the LoS being modeled. Given that it is already well known that the plasma and magnetic field comprising a CME are inhomogeneous, the need for exploring such models is hardly surprising. As evident, the Stokes I spectrum is always consistent with homogeneous and isotropic GS model. Since all earlier studies only used the Stokes I spectrum to model the GS emission, there was never a pressing need to consider an inhomogeneous GS model.

Given Stokes V detection only at a single frequency, it is not possible to constrain an inhomogeneous GS model using the current observation. But, in light of the demonstration of the inability of the homogeneous models to explain both Stokes I and V observations simultaneously, I have taken the first steps toward understanding the impact of inhomogeneity on the GS spectra using simple toy-simulations. The results from this toy-simulation will help assess the sensitivity of different GS model parameters to inhomogenity. As and when Stokes V measurements become available across the band this information will serve to guide the modeling process. These toy-simulations are described next.

\section{Effects of Inhomogeneity on GS Spectrum}\label{sec:nonuniform}
As discussed in Chapter \ref{cme_gs1}, a homogeneous and isotropic GS model with single power-law distribution of nonthermal electrons has ten free parameters. If an inhomogeneous GS model is considered, the number of free parameters will increase substantially. Modeling these large numbers of free parameters with a limited number of spectral measurements reduces the ability to effectively constrain them. Hence, I start by trying to isolate and quantify the impact of inhomogeneity in individual parameters of the GS models on the Stokes I and V GS spectra using toy models, such that the parameter with higher sensitivities on inhomogeneity can be identified.
 
\subsection{Description of the Simulation}\label{subsec:simulation}
I have performed different kinds of simulations to examine and understand the effects of inhomogeneities on the GS spectra. The GS emission model for mildly-relativistic electrons following a single power-law distribution has the following ten independent parameters -- $|B|$, $\theta$, $A$, $L$, $T$, $n_\mathrm{thermal}$, $n_\mathrm{nonth}$, $\delta$, $E_\mathrm{min}$, and $E_\mathrm{max}$. Two of the ten GS model parameters are related to the geometry of the CME ($A$ and $L$) and the concept of inhomogeneity does not apply to them. For all other parameters, except $\theta$, I have simulated the Stokes I and V spectra for a Gaussian distribution of these parameters along the LoS.

\begin{figure}[!htbp]
    \centering
    \includegraphics[trim={1cm 0cm 0cm 0cm},clip,scale=0.35]{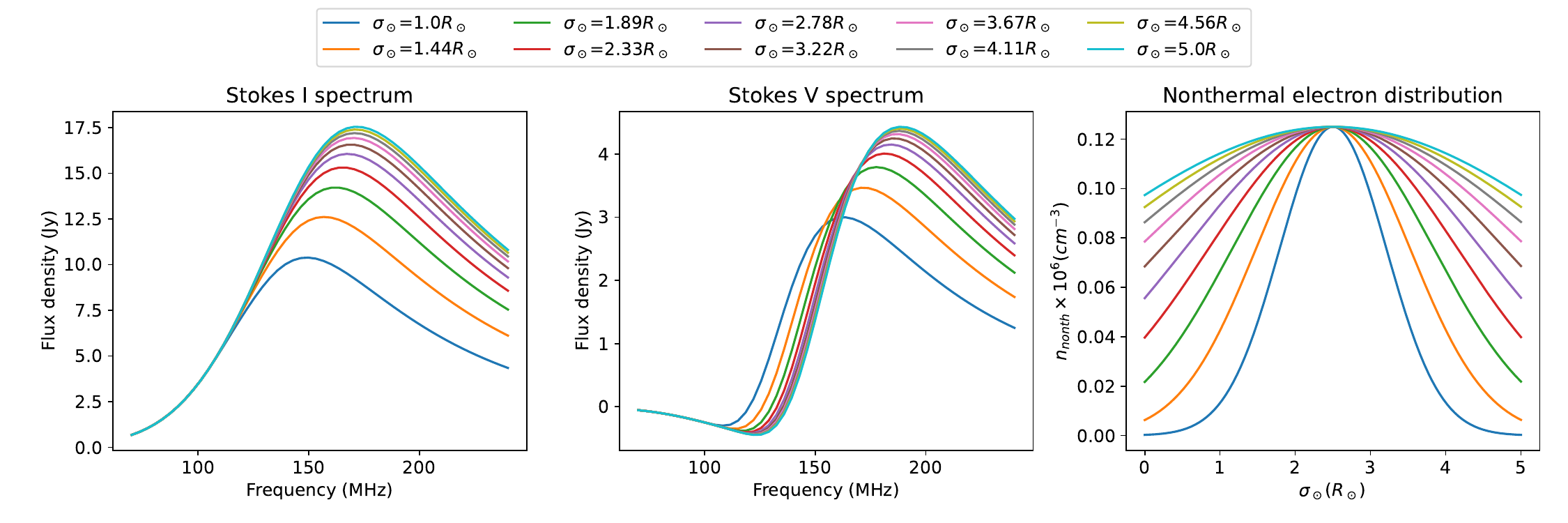}\\
    \includegraphics[trim={1cm 0cm 0cm 2cm},clip,scale=0.35]{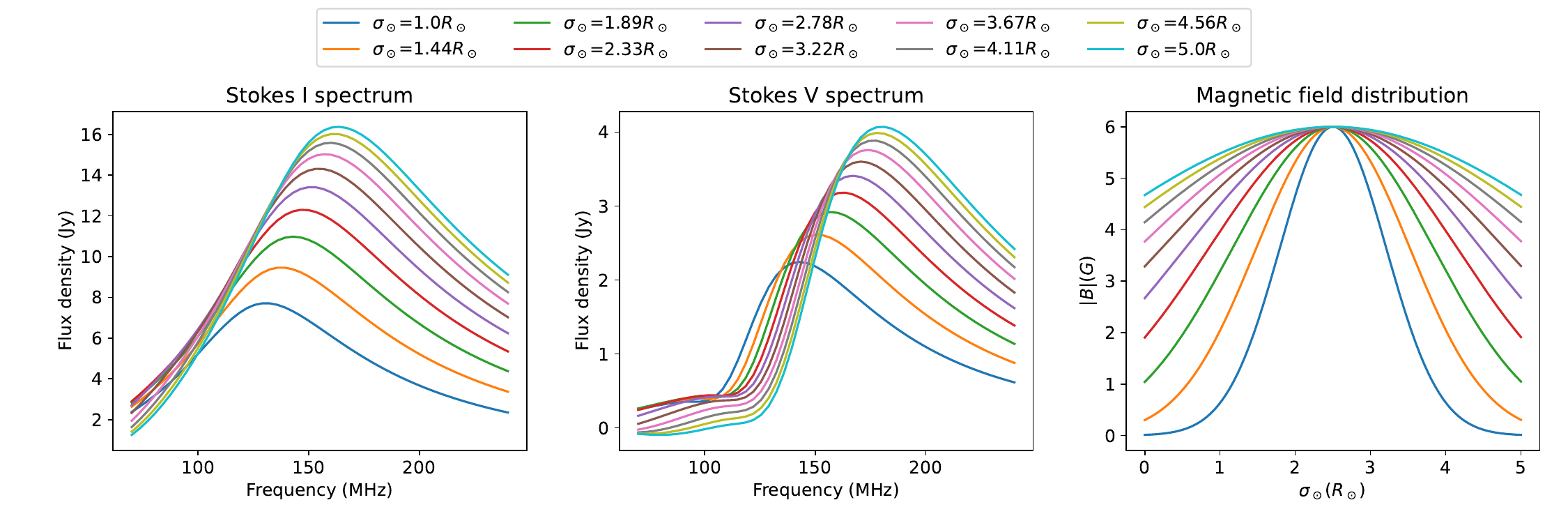}\\
    \includegraphics[trim={1cm 0cm 0cm 2cm},clip,scale=0.35]{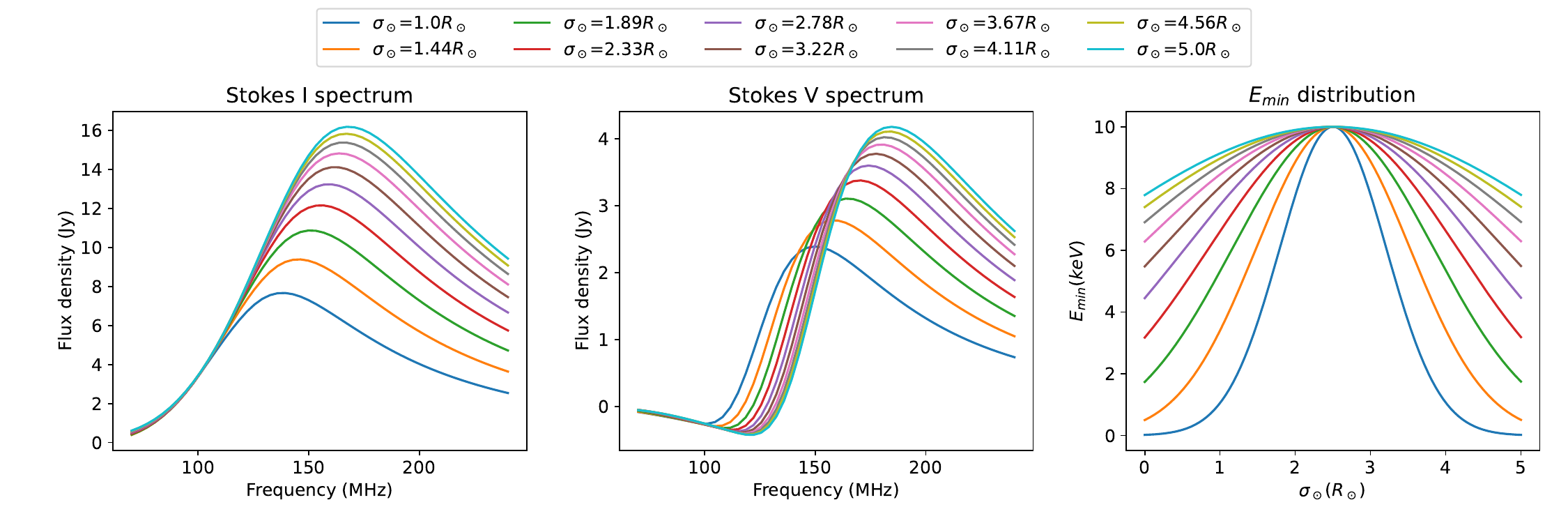}\\
    \includegraphics[trim={1cm 0cm 0cm 0cm},clip,scale=0.35]{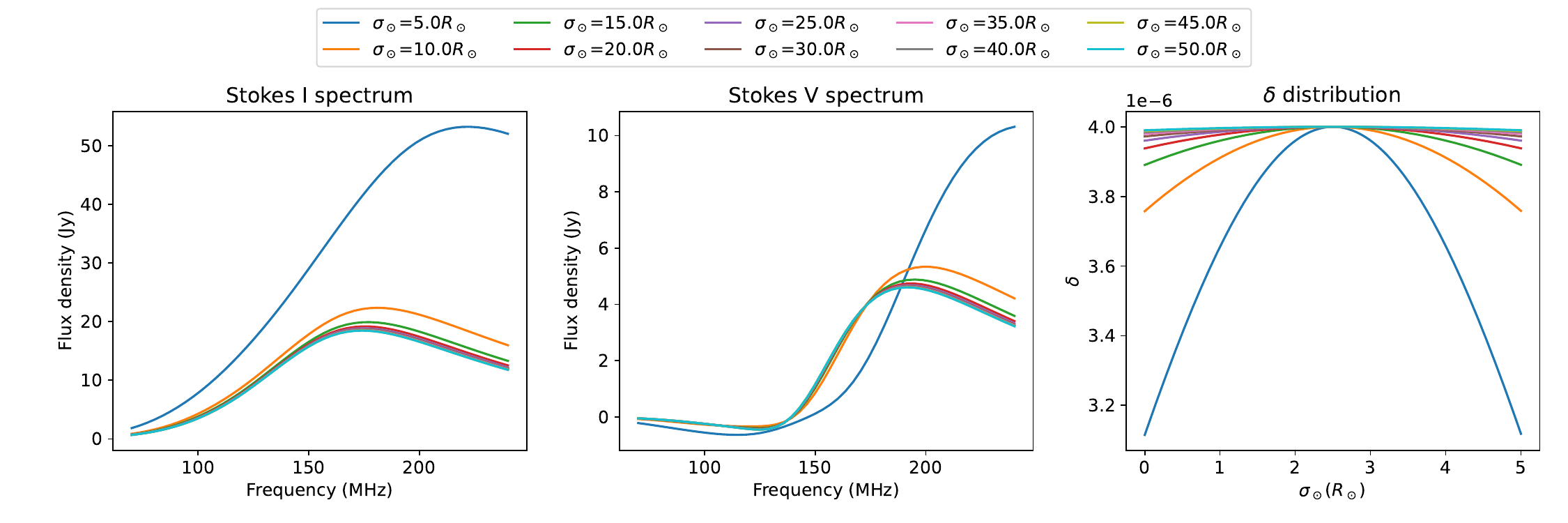}\\
    \caption[Effects of inhomogeneous distributions of magnetic field and nonthermal parameters on simulated GS spectra.]{Effects of inhomogeneous distributions of magnetic field and nonthermal parameters on simulated GS spectra. {\it First column: }Simulated Stokes I spectra. {\it Second column:} Simulated Stokes V spectra. {\it Third column: }Distribution of the plasma parameter along the LoS which is considered to be inhomogeneous. Different colors represent Gaussian distribution with different widths.}
    \label{fig:simulation_1_1}
\end{figure}

\begin{figure*}[!ht]
    \centering
    \includegraphics[trim={1cm 0cm 0cm 0cm},clip,scale=0.35]{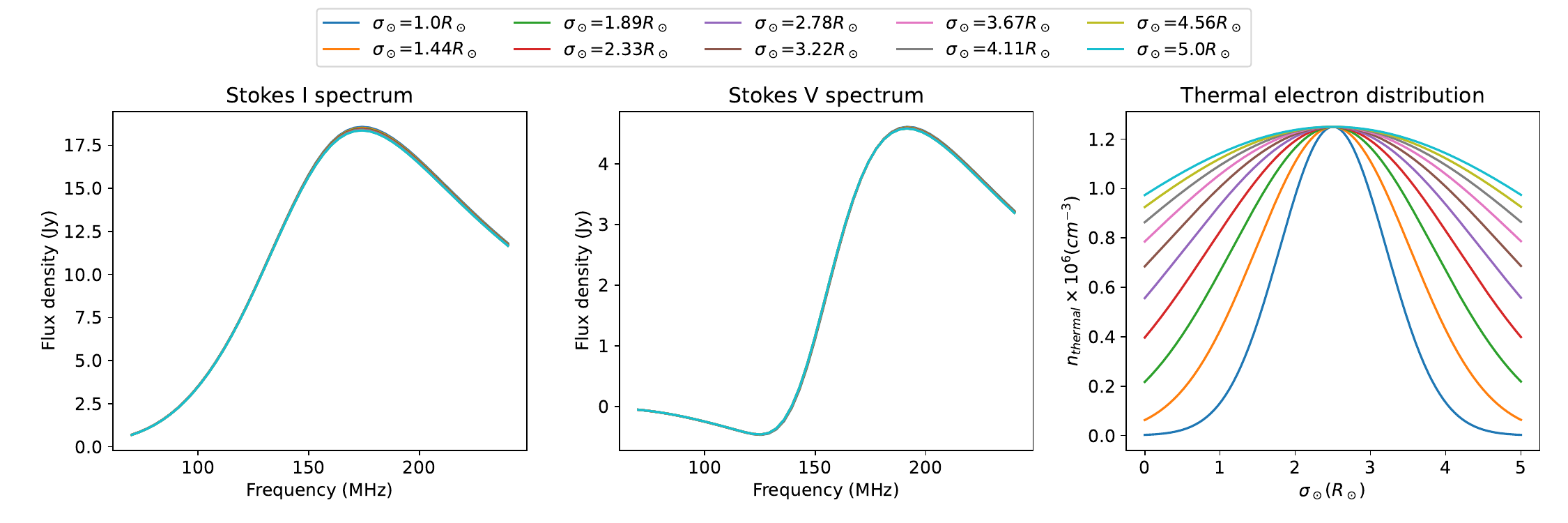}\\
    \includegraphics[trim={1cm 0cm 0cm 2cm},clip,scale=0.35]{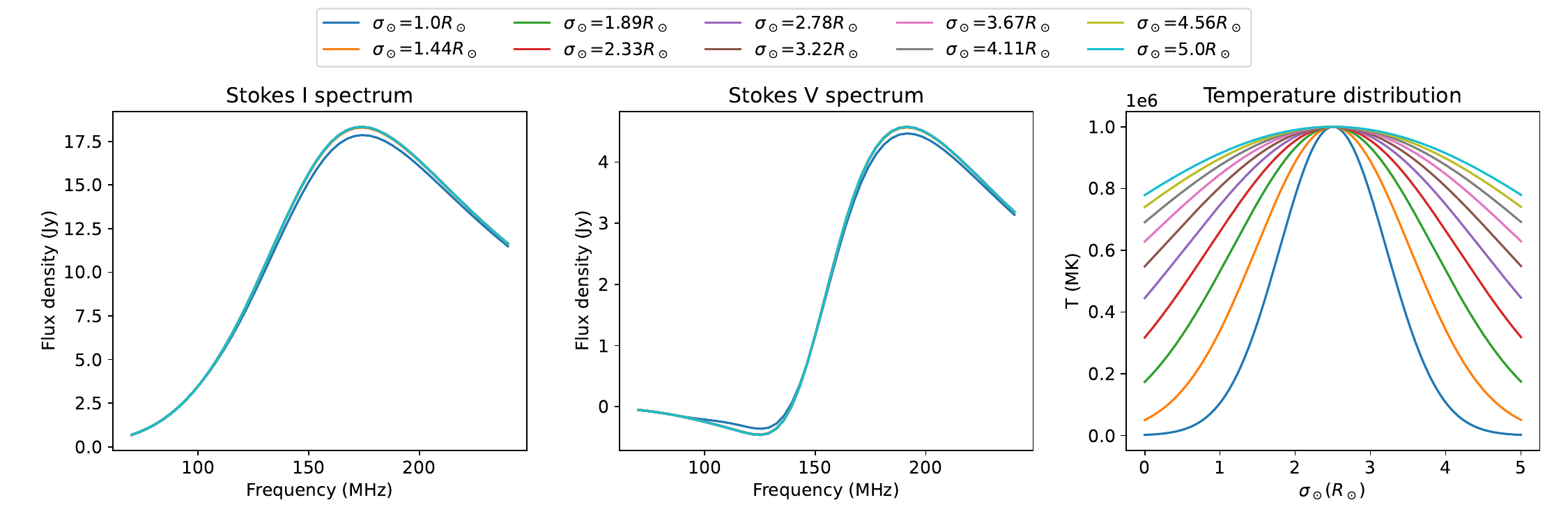}\\
    \caption[Effects of inhomogeneous distributions of thermal parameters on simulated GS spectra.]{Effects of inhomogeneous distributions of thermal parameters on simulated GS spectra. {\it First column: }Simulated Stokes I spectra. {\it Second column:} Simulated Stokes V spectra. \textbf{Third column: }Distribution of thermal parameters along the LoS. Different colors represent Gaussian distribution with different widths.}
    \label{fig:simulation_1_2}
\end{figure*}

\subsubsection{Simulation 1: Effects of Gaussian Distribution of Plasma Parameters}\label{subsec:inhomo_smooth}
Stokes I and V spectra are simulated for different plasma parameters following a Gaussian distribution along the LoS having a form,
\begin{equation}
    p(l)=p_0\ exp\left[-\left(\frac{l-\frac{l_\mathrm{max}}{2}}{\sigma}\right)^2\right],
    \label{eq:inhomo_dist}
\end{equation}
where $p$ is the value of the plasma parameter at an LoS segment, $l$, $p_0$ represents the maximum value of a given plasma parameter, $l_\mathrm{max}$ is the maximum number of LoS segments and $\sigma$ is the width of distribution in units of LoS segments. $\sigma$ values are presented in $R_\odot$ in Figures \ref{fig:simulation_1_1} and \ref{fig:simulation_1_2}, and denoted by $\sigma_\odot$. Within a physically motivated range of parameter values, a fiducial choice of a certain value of each of the parameters is used as the reference or maximum value ($p_0$) for the simulation. The chosen values are -- i) $|B|_0=6\ \mathrm{G}$, ii) $\theta=80^{\circ}$, iii) $A = 10^{20}\ \mathrm{cm^{2}}$, iv) $T_0= 10^6\ \mathrm{K}$, v) $n_\mathrm{thermal,0}= 1.25\times10^6\ \mathrm{cm^{-3}}$, vi) $n_\mathrm{nonth,0}= 1.25\times10^4\ \mathrm{cm^{-3}}$, vii) $\delta= 4$, viii) $L= 5\times10^{10}\ \mathrm{cm}$, ix) $E_\mathrm{min}=10\ \mathrm{keV}$ and x) $E_\mathrm{max}= 15\ \mathrm{MeV}$. I divided the GS source into 1000 LoS segments for this simulation each of length 0.005 $R_\odot$. $\sigma$ is varied between 1 $R_\odot$ (shown by blue lines in third column of Figures \ref{fig:simulation_1_1} and \ref{fig:simulation_1_2}) to 5 $R_\odot$ (shown by cyan lines in third column of Figures \ref{fig:simulation_1_1} and \ref{fig:simulation_1_2}). These limits are chosen in such a way that at the lowest $\sigma$ the Gaussian distribution is highly peaked and essentially vanishes beyond a few solar radii and at the highest $\sigma$ the distribution comes much closer to the usual homogeneous distribution considered in earlier simulations. These choices have been made to allow me to examine the impact of homogeneity on the spectra.

In this toy simulation, only one GS plasma parameter is considered to be inhomogeneous at any time. All other model parameters are regarded to be homogeneous and set to their respective reference values. Simulated Stokes I and V spectra are shown in Figures \ref{fig:simulation_1_1} and \ref{fig:simulation_1_2}. It is evident from Figure \ref{fig:simulation_1_2} that the inhomogeneity in $n_\mathrm{thermal}$ and $T$ does not have any effect on the observed Stokes I or V spectra. Other plasma parameters, $|B|,\ \delta,\ n_\mathrm{nonth}$ and $E_\mathrm{min}$ show significant effects on both Stokes I and V spectra (Figure \ref{fig:simulation_1_1}). With increasing inhomogeneity (i.e., decreasing $\sigma$), the peak frequency and peak flux density of the Stokes I and V spectra decreases for $n_\mathrm{nonth}$, $E_\mathrm{min}$ and $|B|$. It is also found that the effect of inhomogeneity on the Stokes I spectra is seen only in the optically thin part of the spectra (i.e. at frequencies above the peak of the spectrum), while it is visible in both optically thick and thin parts of the Stokes V spectra. Flux density increases with the increase in inhomogeneity in the optically thick part, while it decreases in the optically thin part. Unlike these three parameters, a small amount of inhomogeneity in $\delta$ shows significant changes in both Stokes I and V spectra, as shown in the bottom panel of Figure \ref{fig:simulation_1_1}.
\subsubsection{Simulation 2: Effects on Modeling an Inhomogeneous GS Source with Homogeneous Model}
\label{subsec:simulation_2}
\begin{figure*}[!htbp]
    \centering
    \includegraphics[trim={1cm 0cm 0cm 0cm},clip,scale=0.35]{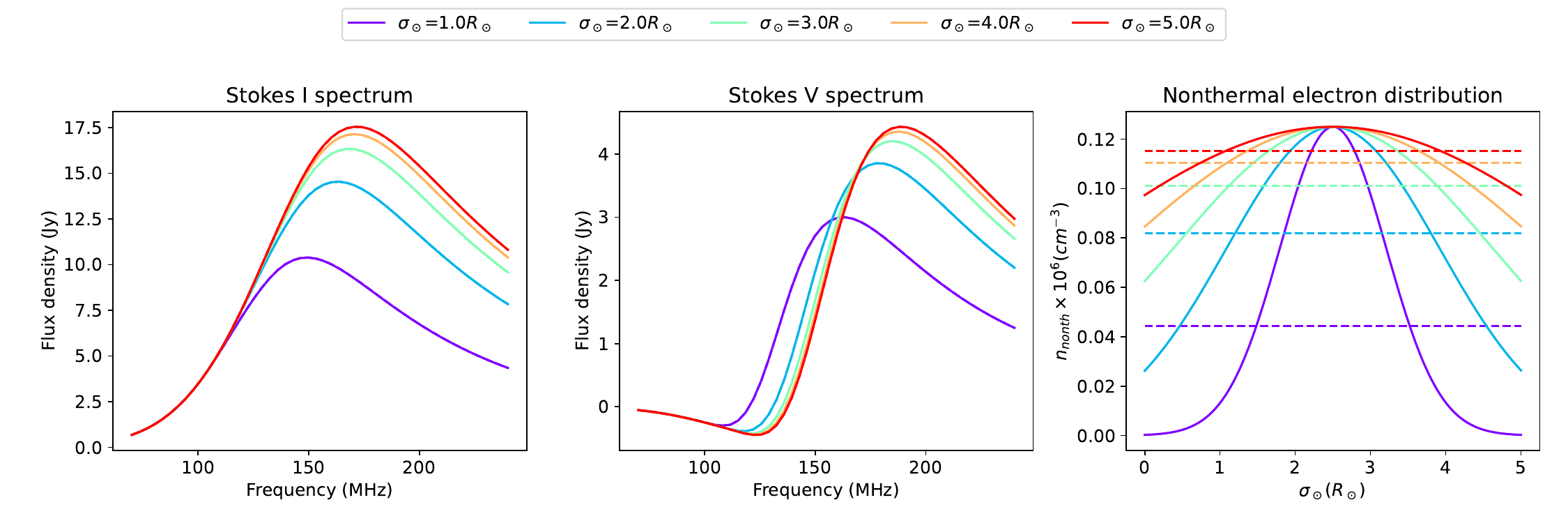}\\
    \includegraphics[trim={1cm 0cm 0cm 2cm},clip,scale=0.35]{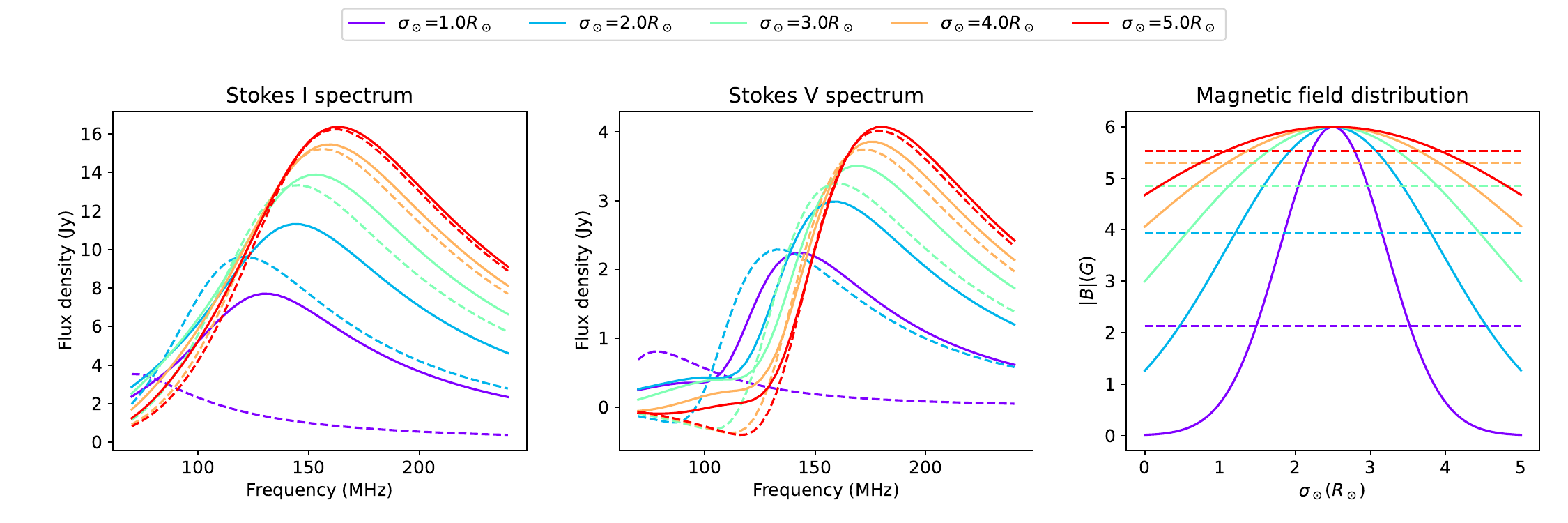}\\
  \includegraphics[trim={1cm 0cm 0cm 2cm},clip,scale=0.35]{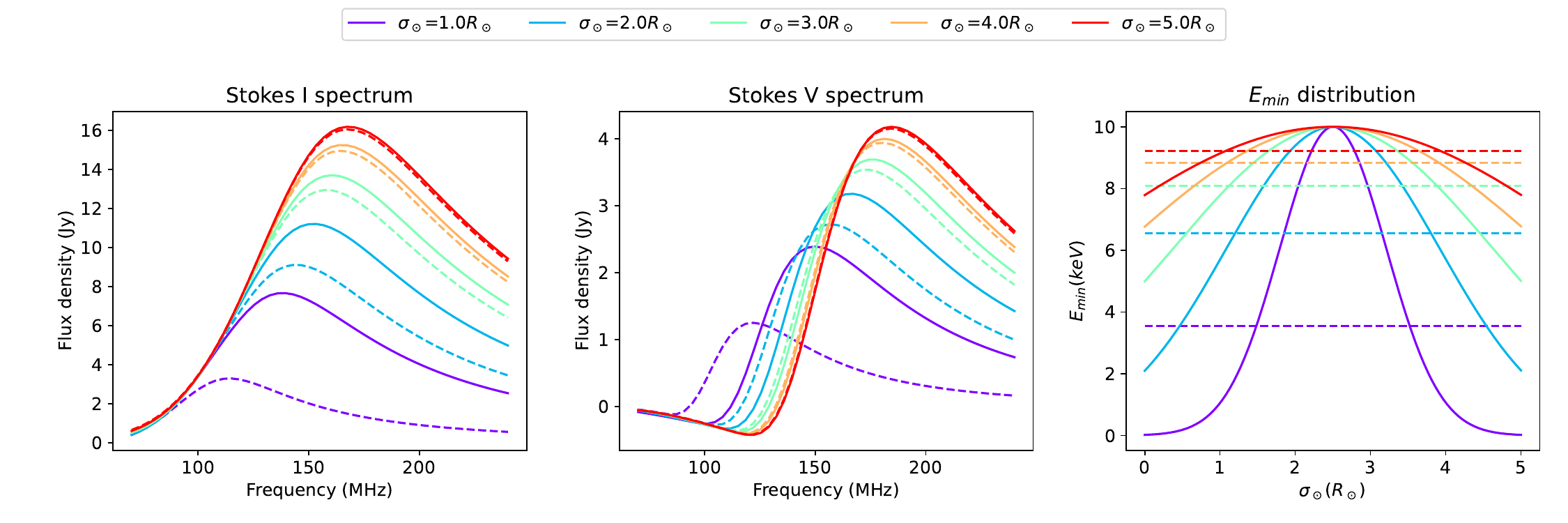}\\
  \includegraphics[trim={0cm 0cm 0cm 0cm},clip,scale=0.35]{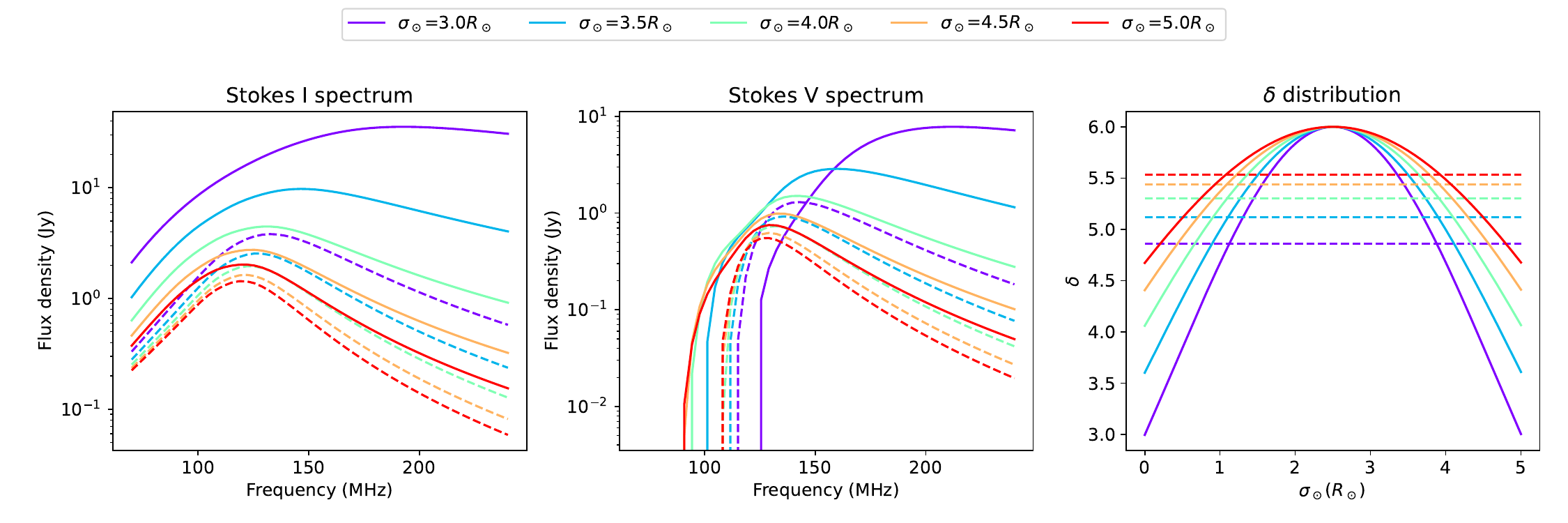}\\
    \caption[Simulated Stokes I and V spectra for inhomogeneous and mean homogeneous distribution of CME plasma parameters.]{Simulated GS spectra for inhomogeneous and mean homogeneous distribution of CME plasma parameters. {\it First column: }Simulated Stokes I spectra. {\it Second column:} Simulated Stokes V spectra. {\it Third column: }Distribution of plasma parameters along the LoS. Simulated spectra for inhomogeneous and mean homogeneous distributions are shown by solid and dashed lines, respectively.}
    \label{fig:simulation_2_1}
\end{figure*}
Since, to date, homogeneous GS models have been used routinely, it is important to understand whether homogeneous models can be used to at least estimate the mean value of the distribution of the relevant plasma parameters. 
To understand this, I have simulated Stokes I and V GS spectra considering a Gaussian distribution of the test parameter along the LoS -- $|B|$, $n_\mathrm{nonth}$, $E_\mathrm{min}$ and $\delta$, with different widths of the Gaussians, while keeping other parameters fixed at the reference values mentioned in Section \ref{subsec:inhomo_smooth}. These are shown by solid lines in Figure \ref{fig:simulation_2_1}. 

I have then taken the mean of the inhomogeneous distribution of the test parameter and built a homogeneous model with the value of the test parameter set to this mean value. The mean values are marked by dashed lines in Figure \ref{fig:simulation_2_1} and the corresponding simulated Stokes I and V spectra are also shown by dotted lines in the same figure. 

It is evident from the first row of Figure \ref{fig:simulation_2_1} that while different Gaussian distributions of $n_\mathrm{nonth}$ produce different Stokes I and V spectra, the corresponding homogeneous GS model set to the mean value of $n_\mathrm{nonth}$ lead to essentially an identical Stokes I and V spectra. This implies that a homogeneous GS model can provide an accurate estimation of the mean $n_\mathrm{nonth}$ along the LoS. For the other three parameters -- $|B|$, $E_\mathrm{min}$ and $\delta$, there are significant differences between spectra resulting from inhomogeneous and homogeneous distributions set to the corresponding mean value of the parameter, inhomogeneity in which is being explored. These differences grow larger as the width of Gaussian distribution grows smaller, i.e. with an increase in the degree of inhomogeneity along LoS. Conversely, as should be expected, the two spectra come closer to each other as $\sigma$ increases and the degree of inhomogeneity decreases.

This toy simulation shows that the ability of a homogeneous model to represent the mean of the true inhomogeneous distribution for $|B|$, $E_\mathrm{min}$ and $\delta$, even for the simplest of inhomogeneous models, is dependent on the level of inhomogeneity present in the medium and grows poorer with increasing degree of inhomogeneity.
If the level of inhomogeneity is large enough, the GS model spectra will differ significantly from the ones corresponding to the mean values of the distributions.

\section{Discussion}\label{sec:discussion_cme2}
\subsection{Learnings From the Current Work}\label{subsec:cme2_importance}
Since Stokes V is only a fraction of Stokes I flux density, it is harder to detect than Stokes I emission. The dataset chosen for this work was particularly challenging, and even for that extended Stokes I emission is detected across multiple frequencies. This work also presents robust detection of Stokes V emission over a small region from the CME-2 at a single spectral point as discussed in Section \ref{subsec:circular_pol_cme2}. 

This work gives the first possible indication of the insufficiency of a homogeneous GS model for modeling both the observed Stokes I and V spectra simultaneously. It is found that while the homogeneous GS model with PLW electron energy distribution and isotropic pitch-angle distribution can explain the observed Stokes I spectra and stringent Stokes V upper limits, it cannot simultaneously fit the lone Stokes V detection. To explore the reasons for the discrepancy between observed and model spectra, I have systematically explored different possibilities -- effects of different feasible electron energy distributions, electron pitch-angle distribution, and inhomogeneity along the LoS. In the limited but illuminating exploration done here, it is found that a homogeneous GS model with different electron energy distribution or different pitch-angle distribution cannot reproduce the observed Stokes I and V simultaneously. 

To understand the effects of inhomogeneity, toy simulations have been performed keeping a single test parameter inhomogeneous, while the others are kept homogeneous. These toy simulations allow us to identify the effects of inhomogeneity of different plasma parameters on Stokes I and V GS spectra. The first set of simulations demonstrates that even if there could be inhomogeneity in thermal electron distribution and temperature, their effects on the GS spectrum are negligible. Inhomogeneity in magnetic field and nonthermal electrons, on the other hand, have noticeable effects on both the Stokes I and V GS spectrum. The second set of simulations examines the outcome of modeling a GS source with an inhomogeneous magnetic field and nonthermal electron distribution as a homogeneous source. The results from this exercise indicate that the inhomogeneities in $|B|$, $E_\mathrm{min}$, and $\delta$ can lead to changes in GS spectra which cannot be correctly modeled assuming homogeneous GS models.

\subsection{Limitations and Path Forward}\label{subsec:cme2_limitation}
It is reasonable to expect the reality to only be more complex, with the distributions of multiple, perhaps all, parameters being inhomogeneous and more complicated than the simplest possibility of a Gaussian distribution considered for the toy simulations. Despite the indications of insufficiency of the homogeneous GS model, the currently available data does not have the ability to constrain the much larger number of free parameters of an inhomogeneous model.

However, one aspect that has not yet been explored in the GS modeling of CMEs is to regard the entire extent of the spatially resolved emission as a single structure that can be modeled. As our ability to detect spatially resolved Stokes I and V emission over extended regions improves, this approach will be very interesting to explore. By moving away from constraining the parameters of each PSF-sized region independently, this approach has the potential to significantly reduce the number of free parameters in the problem. 

Although, not an apples-to-apples comparison, similar physics-based three-dimensional modeling of GS emission from flare loops have already been performed \citep[e.g.][]{Kuznetsov_2011,Reznikova_2014,Doorsselaere2016}. Using spectroscopic imaging observations of flares at microwave frequencies, these models have been used to constrain flare parameters. Following a similar approach for constraining GS emission from CMEs can be the next step in this area of research.

\section{Conclusion}\label{sec:conclusion_cme2}
Since the first attempt at spatially resolved modeling of the GS emission from CME plasma by \cite{bastian2001}, all previous studies have assumed homogeneous and isotropic GS models to estimate the CME plasma parameters. The same assumptions were used to model the GS spectrum of CME-1 (Chapter \ref{cme_gs1}) for which stringent Stokes V upper limits were used for the first time, in conjunction with Stokes I spectra. The same approach was also used for the CME-2 for regions that only had Stokes V upper limits, but fell short for the only region for which a Stokes V detection was available at one spectral point and the spectral peak was also sampled by the observations.

This led me to examine the various assumptions made by the GS modeling procedure, quantify their impacts and attempt to identify the one(s) which might not be satisfied. I find that relaxing the assumptions of a single power law distribution for electron energies and allowing a distribution of electron pitch angles is not sufficient for reproducing the observed spectra. My simulations suggest that the assumption which is most likely cause to be violated is the one about the homogeneity of plasma properties in the volume being modeled. The inability to model these inhomogeneities appropriately is also the likely cause for the mismatch between the model and the observed spectra.

Making substantial progress in modeling of CME GS spectra will require progress on two separate fronts -- gathering a larger number of stringent constraints, which will come from more sensitive and wideband observations, and will provide Stokes V measurements at multiple spectral points; and making progress on physics-based 3D forward modeling approaches for CME GS modeling. These are discussed in more detail in Section \ref{sec:future_work_final} of Chapter \ref{conclusion}.

\chapter {Conclusions and Future Works}
\label{conclusion}

The magnetic field of CMEs is one of the crucial parameters that determines the eruption, evolution, propagation, and geo-effectiveness of CMEs. Over the decades, CMEs have routinely been observed using white-light coronagraphs. Although white-light coronagraph observations provide several crucial pieces of information about the CME geometry and dynamics, they cannot provide a direct estimation of magnetic fields entrained in different parts of the CME. Hence, one has to rely on models and several assumptions to estimate the magnetic fields of CMEs from white-light observations \citep[e.g.,][etc.]{Savani_etal_2015,Gopalswamy2017,Kilpua2021}. On the other hand, there are several observational tools available at radio wavelengths to remotely measure the magnetic fields of CMEs, as described in Section \ref{subsec:radio_observation_cme} of Chapter \ref{chapter_intro}. Although these methods have been known for several decades, their applications remain limited due to instrumental and modeling challenges. The main goal of this thesis is to address these challenges so that ground-based radio observations can be used to provide measurements of magnetic fields and other plasma parameters of CMEs at coronal heights. This thesis addresses both the observational and modeling aspects and the underlying challenges in estimating CME plasma parameters using GS emission from CME plasma. It presents significant advances towards this and also acknowledges the limitations of current observational capabilities and modeling approaches. I present some concluding remarks on the thesis in Section \ref{sec:conclusion_final} followed by a discussion about the limitations in Section \ref{sec:limitation_of_current_work} and the path forward in the future in Section \ref{sec:future_work_final}. 

\section{Conclusions}\label{sec:conclusion_final}
This thesis presents a detailed exploration of spatially resolved spectropolarimetric modeling of GS emission from CMEs using high-fidelity and high DR spectropolarimetric radio imaging using the MWA. This involves overcoming both the observational and modeling challenges discussed in Section \ref{sec:challenges} of Chapter \ref{chapter_intro}. 

\subsection{State-of-the-art Spectropolarimetric Solar Imaging Algorithm}\label{subsec:conclusion_paircars}
Both the flux density and polarization fraction of metric solar emission varies by several orders of magnitudes (Figure \ref{fig:different_emissions} in Chapter \ref{paircars_principle}). These emissions also show high spectro-temporal variability. Hence, to study these emissions and use them to estimate other physical properties of the solar corona and coronal eruptive phenomena, one needs high DR and high-fidelity spectropolarimetric snapshot solar imaging capability. 
Although the technical requirements to achieve this is met rather well by the MWA, producing large numbers of spectropolarimetric snapshot solar images of high quality involves several non-trivial steps. To streamline the process of producing science-ready images from the MWA solar observations, I have developed a state-of-the-art spectropolarimetric calibration and imaging algorithm and software pipeline -- ``Polarimetry using Automated Imaging Routine for the Compact Arrays for the Radio Sun (P-AIRCARS)" \citep{Kansabanik2022_paircarsI,Kansabanik_paircars_2}. This algorithm is based on the {\it Measurement Equation} framework \citep{Hamaker1996_1,Hamaker2000}, which forms the basis of all modern radio interferometric calibration and imaging and takes advantage of the compact and dense array configuration of the MWA \citep{Kansabanik_principle_AIRCARS}.

P-AIRCARS can provide spectroscopic snapshot solar images with DR between $>300$ -- $10^5$. These are possibly the highest quality spectroscopic snapshot metric solar images, both during the active and quiet periods, available to date. Not only does P-AIRCARS provide high DR, but it also provides high-fidelity in flux density and polarization calibration. An independent technique developed by \cite{Kansabanik2022} and implemented in P-AIRCARS provides absolute solar flux density with an uncertainty of $\sim10\%$. Polarization purity of the full Stokes images provided by P-AIRCARS is similar to that obtained for high-quality astronomical observations -- residual Stokes I to Q leakage is $\leq1\%$ and Stokes I to U, V leakages are $\leq0.1\%$. Although P-AIRCARS has been developed and tested on the MWA solar observations, its algorithm is general and flexible. It can be adapted for any radio interferometric array with a compact, dense core. Essentially all of the future radio interferometric arrays, currently in varying phases of design and construction, such as the Square Kilometre Array Observatory \cite[SKAO;][]{ska_concept,SKAO2021}, the Next Generation Very Large Array \citep[ngVLA,][]{ngVLA2019}, and the Frequency Agile Solar Radiotelescope \citep[FASR,][]{Gary2003,Bastian2005,Bastian2019,Gary2022_FASR}, will have a dense central core. It should, hence, be straightforward to adapt P-AIRCARS for these arrays, and I hope and anticipate that P-AIRCARS will indeed provide a very good starting point for the dedicated solar imaging pipelines which will eventually be developed for these instruments. 

P-AIRCARS has been successful in achieving its desired goals. It is already leading to discoveries and interesting results. All of these lie in a previously inaccessible part of the phase space, the exploration of which has been enabled by P-AIRCARS. They include -- the first detection of a very low-level induced circular polarization in the quiet Sun thermal emission, the first robust imaging detection of linearly polarized metric solar emission, and the detection of faint GS emission from CME at the largest heliocentric distance reported yet. I have used spectropolarimetric images provided by P-AIRCARS for a detailed exploration of GS emission from two CMEs. P-AIRCARS is not only providing high-quality solar radio images, but it is also user-friendly. It has been designed in a manner that does not require the non-specialist user to have prior radio interferometry expertise or experience for generating science-ready images. Despite the usefulness of solar radio imaging being well established and the increasing availability of large volumes of excellent data in the public domain, the steep learning curve involved in radio interferometric has been the key hurdle limiting the large-scale use of these data by the larger solar community. I hope that the availability of a robust tool like P-AIRCARS will help make solar radio imaging more mainstream. I also note that for the well-informed advanced user, the P-AIRCARS design provides all of the flexibility and features of the underlying software packages. So it is well-placed to be the analysis tool of choice for both early and advanced users.

\subsection{Improvement in Robustness of Modeling GS Emission from CMEs using Spectropolarimetry}\label{subsec:conclusion_cme1}
From the first attempt at modeling spatially resolved GS spectra from a CME by \cite{bastian2001} to the latest study by \cite{Mondal2020a}, all of them use only Stokes I spectra to estimate the GS model parameters. Even the simplest GS model has ten free parameters, and if using only Stokes I spectra, some of these parameters are degenerate. Hence, one needs to make several assumptions to estimate magnetic field strength and other plasma parameters of the CME plasma from the observed GS spectra. The robustness of the estimated plasma parameters crucially depends on the degree to which these assumptions hold. It is known that some of these degeneracies can be broken by including Stokes V measurements with the Stokes I measurements. Some of the earlier studies on GS modeling have reported the detection of Stokes V emission \citep{bastian2001,Tun2013}, but they used only Stokes I measurements for GS spectral modeling. 

High-fidelity spectropolarimetric images provided by P-AIRCARS allow us to use Stokes V measurements or even sensitive upper limits to constrain GS model parameters. The effort of inclusion of Stokes V measurement in a joint spectropolarimetric modeling framework is described in Section \ref{subsec:upperlimtit_mathframe} of Chapter \ref{cme_gs1}. In that section, I have performed a detailed spatially resolved spectropolarimetric modeling of the CME propagating towards solar north, which was referred to in Chapter \ref{cme_gs1} as CME-1. Although no Stokes V emission is detected from CME-1, the stringent upper limits on it provided by robust calibration by P-AIRCARS allows us to provide stronger constraints on GS model parameters and break some of the degeneracies between GS model parameters (Section \ref{subsec:importance_stoks_V} of Chapter \ref{cme_gs1}). It also should be noted that multi-vantage point white-light observations are available from LASCO/SOHO. STEREO-A and STEREO-B provided strong independent constraints on the geometrical parameters of the CME. This study also demonstrated the importance of sampling the spectral peak of the GS spectrum. If the spectral peak is not sampled, even the inclusion of Stokes V measurements does not improve the robustness of the GS model parameter estimates. Based on the findings from work presented in Chapter \ref{cme_gs1}, I conclude that for robust modeling of the GS emission, one needs observations such that the spectral peak is sampled, along with Stokes I and V detections. Achieving these will require wide bandwidth high sensitivity measurements, which the upcoming radio interferometers will be able to provide.

\subsection{Possible Observational Indication for Insufficiency of Homogeneous GS Model}\label{subsec:conclusion_cme2}
In Chapter \ref{cme_gs2}, I have studied GS emission from a region of CME-streamer interaction (CME-2 as defined in Chapter \ref{cme_gs2}), which is possibly only the second such example, the earlier one was reported by \cite{Mondal2020a}. In all earlier modeling studies of GS emissions from CME plasma, the distribution of plasma parameters has been assumed to be homogeneous along the line-of-sight (LoS). Although it is well-known that CMEs are large-scale inhomogeneous plasma structures, the homogeneity assumption is made because the available data is barely sufficient to constrain even the simplest homogeneous GS models. Besides that, there was never any indication that the homogeneous GS model is insufficient to explain the observation. In Chapter \ref{cme_gs2}, I have demonstrated possibly the first observational indication for the insufficiency of a homogeneous GS model that can not fit the both Stokes I and V measurements simultaneously.

Unlike the CME-1, I have detection of Stokes V emission over a small part of CME-2 at a single spectral slice. For other regions and other spectral slices, stringent Stokes V upper limits are available. While the homogeneous GS models consistent with the observed Stokes I spectra as well as Stokes V upper limits could be found, no model spectra consistent with the observed Stokes V detection at the single spectral slice were found. This suggests that one or more of the three assumptions of the GS model is not valid. These three assumptions are -- homogeneity of distributions of plasma parameters along the LoS, isotropy of pitch-angle distribution of electrons, and the electron energy distribution following a single power law. A systematic exploration of these assumptions is presented in Section \ref{subsec:homo_insuff} of Chapter \ref{cme_gs2}. It suggests that observed Stokes V emission can not be reproduced by any homogeneous GS models with different pitch-angle distributions and a few other electron energy distributions considered. This indicates that the inhomogeneity of plasma parameters along the LoS is the most probable cause behind the discrepancy between these models and observations.  

The inclusion of inhomogeneity will increase the number of free parameters substantially, and hence it will no longer be possible to constrain them with the current observations. Hence, instead of modeling the observed spectrum using an inhomogeneous GS model, I have performed some toy simulations to understand the effects of inhomogeneity on the estimated GS model parameters. It turns out from the simplistic toy simulations (Section \ref{sec:nonuniform} of Chapter \ref{cme_gs2}) that homogeneous source model assumption may not always provide robust estimates of magnetic field and some of the nonthermal electron parameters depending on the strength of the inhomogeneity. If the strength of the inhomogeneity is not sufficiently strong, the homogeneous models can provide reasonable estimates of GS model parameters. Although this work does not constrain an inhomogeneous model from the observations, it provides possibly the first strong indication of the effects of inhomogeneity on the GS emission spectra. This work also demonstrates that the insufficiency of the homogeneous model cannot be observationally verified when only Stokes I and Stokes V upper limits are available. With the availability of Stokes V spectra, along with Stokes I spectra, using more sensitive observations either for stronger events or from more sensitive future instruments, it should become possible to test the insufficiency of homogeneous CME GS models in greater detail and explore ways to constrain inhomogeneous GS models as discussed in the next section.

\section{Limitations}\label{sec:limitation_of_current_work}
One of the major challenges in detecting spatially resolved CME GS emission is its very low flux density, as compared to most other solar radio emissions. Robust and precise instrumental calibration and imaging provided by P-AIRCARS deliver a performance close to the thermal sensitivity of the MWA (Section \ref{subsec:compare_with_gleam}). This implies that we are already operating close to the instrumental sensitivity limit and there is little more to be gained by pushing the calibration and imaging front. The current sensitivity of the MWA permitted the detection of Stokes I emission from both the events presented in Chapters \ref{cme_gs1} and \ref{cme_gs2}. However, it could only provide stringent upper limits on Stokes V emission for CME-1 and detection of Stokes V emission over a small spatial patch at a single frequency for CME-2. The ability to make robust Stokes V detection over a larger spectral and spatial span will require instrumentation which can provide higher sensitivity and excellent imaging quality.

To use GS emission to study the CME plasma properties from close to its eruption at lower coronal heights out to about 10$R_\odot$, one needs to cover a much larger spectral band than provided by the MWA. Frequency coverage of the MWA allows the detection of CME GS emission in the heliocentric height range from $\sim2-6\ R_\odot$.  Instruments capable of producing high DR solar radio images have already started operations both at much lower frequencies, like the Owens Valley Long Wavelength Array \citep[OVRO-LWA;][]{Hallinan2023} and NenuFAR \citep{Zarka2018}, and at higher frequencies like MeerKAT \citep{Chen_meerkat2021}. In coming years, SKA-Low and SKA-Mid will provide much higher sensitivities over a frequency range from $\sim50$ MHz -- $15$ GHz. With these capable instruments, it is expected that it will become possible to detect Stokes I and Stokes V GS emissions from CME over a large heliocentric height range.

As discussed in detail in Chapters \ref{cme_gs1} and \ref{cme_gs2}, although high-quality observations have started becoming available, modeling the observed GS spectrum to estimate the CME magnetic field robustly remains challenging due to the large numbers of free parameters required by the models. It had been suggested that the inclusion of Stokes V measurements should improve the robustness of the estimated CME plasma parameters \citep{Mondal2020a}. This has been demonstrated in Chapter \ref{cme_gs1} using stringent Stokes V upper limits. In Chapter \ref{cme_gs2}, when Stokes V detection is available at a single spectral point, it became evident that a homogeneous GS model is insufficient to explain both the Stokes I and V spectra simultaneously. We seem to be at a juncture where the observations have grown detailed enough to rule out the most simplistic models. However, they don't have enough information to constrain the more sophisticated models which require a much larger number of free parameters to describe them. As Stokes V detections become available across the spectrum, a possible new approach to handle these large numbers of free parameters is worth considering and is discussed in Section \ref{subsec:cme_modeling_future}.

\section{Future Works}\label{sec:future_work_final}
This work represents significant progress that has been made both in producing high-quality solar radio images and establishing the methods to robustly constrain CME plasma parameters at middle and higher coronal heights using GS emission from CME plasma. This work pushes the MWA  observations close to its limits and reaches very close to its formal thermal sensitivity as discussed in Section \ref{subsec:compare_with_gleam} of Chapter \ref{fluxcal}. At the same time, it also establishes the limitations of the current observational and modeling capabilities. Successful detection of Stokes I emission from one of the poor CME events observed with the MWA indicates that it is reasonable to expect to detect CME GS emission from stronger events with the MWA Phase-I and -II. It is also anticipated that future instruments will be able to  provide improved sensitivity and  allow us to observe fainter and fainter GS emissions from the CME plasma. This will come up against modeling challenges. In the following sections, I will discuss the path forward for the future to overcome these challenges. 

\subsection{Future Developments and Upgrades of P-AIRCARS}\label{subsec:paircars_future}
Although P-AIRCARS produces high DR and high-fidelity spectropolarimetric snapshot solar imaging routinely, several implementation and algorithm-related improvements can still be made. The most important ones among these are:
\begin{enumerate}
    \item {\bf Relaxing the assumption of a homogeneous array} -- At present, P-AIRCARS assumes a homogeneous array (i.e., all antenna tiles/dishes in the array have a similar response) and performs a direction-independent polarization self-calibration towards the direction of the Sun. The homogeneous array assumptions may not hold for the MWA and the future SKAO when multiple dipoles in antenna tiles or stations are not working. 
    \item {\bf Wide field-of-view (FoV) polarimetric calibration and imaging} -- High DR imaging provided by P-AIRCARS already demonstrates the ability to detect multiple faint background galactic/extra-galactic radio sources in the presence of the Sun in the FoV. Some of these background sources, especially the Galactic background, would be linearly polarized and can potentially be used to measure the Faraday rotation while a CME is passing across the LoS to these sources. To be able to do this, one needs wide FoV polarimetric calibration and imaging. A possible approach to enable wide FoV polarimetry even using a heterogeneous array has been briefly discussed in Section \ref{Conclusion_paircars} of Chapter \ref{paircars_algorithm}. Combined with wide FoV polarimetric imaging, the ability to see multiple faint background sources can be used to perform image-based polarization leakage correction. This is also an essential step toward developing the ability to measure the Faraday rotation of linearly polarized radiation from background sources due to the magnetized CME plasma, a long-term goal that we are pursuing.
    \item {\bf Keeping P-AIRCARS abreast of ongoing developments} -- The modular design of P-AIRCARS allows it to benefit easily from the developments and improvements being continually made in the underlying software packages it uses. We plan to incorporate the recent developments of these software packages and data structures in P-AIRCARS. The next generation of {\it measurement set} format (MS-v.3) has recently been released\footnote{\url{https://casacore.github.io/casacore-notes/264.pdf}}, as a part of the Next Generation CASA infrastructure \textsf{(ngCASA)}\footnote{\url{https://cngi-prototype.readthedocs.io}} effort. The MS-v.3 offers a major advantage by significantly reducing the input-output (IO) overheads incurred during the calibration and imaging process. We expect this to lead to significant benefits. Incorporating MS-v.3 in P-AIRCARS, however, needs calibration, and imaging software is used to be compatible with the MS-v.3 data structures. This requirement is already met by \textsf{QuartiCal}, which P-AIRCARS relies upon for calibration.
    \item {\bf Building a database of P-AIRCARS calibration solutions} -- While they have not been activated yet, P-AIRCARS has internal mechanisms for each run of P-AIRCARS by any user anywhere in the world to contribute calibration solutions to a common database. This has been done with a vision to build a central repository of all available calibration solutions for MWA solar data accessible to all P-AIRCARS users. It will benefit the individual users by providing them with pre-existing calibration solutions when available and reducing their run-time. Over time, as the usage of P-AIRCARS grows, we expect this to become a useful resource for the community. It will also enable us to examine the long-term evolution of calibration solutions and array performance etc.
    \item {\bf Containerization of P-AIRCARS} -- To ease the installation of P-AIRCARS across a diverse set of Linux-based operating platforms, efforts are already underway to containerize P-AIRCARS.
    \item {\bf An HPC/Cloud implementation of P-AIRCARS } -- While the P-AIRCARS architecture has been designed to be compatible with HPC deployment, it has not been deployed on one yet, primarily due to a lack of a suitable opportunity. Currently, P-AIRCARS takes on the tasks of both parallelizations as well as scheduling. In an HPC environment, the scheduling is usually done by a dedicated job scheduler, like Portable Batch System \citep[PBS,][]{pbs} or Slurm \citep{SLURM}.  Work is in progress to adapt P-AIRCARS for a cluster environment by incorporating an interface to a job scheduler. In parallel, we are also exploring the possibility of adapting P-AIRCARS for cloud computing platforms like Amazon Web Services (AWS) and Google Cloud Platform (GCP). HPC and cloud implementations are necessary to analyze significant fractions of more than 3,000 hours of existing MWA solar observations and the ongoing and planned observations with the MWA, which is steadily progressing toward its phase-III. 
    \item {\bf Automating ways to extract information from large numbers of  high-quality spectropolarimetric snapshot} solar radio images -- With the problem of making high-quality solar radio images largely addressed, it is now possible to make tens or even hundreds of thousands of images from just a few minutes of MWA data. With the new-generation radio instruments, the quality of spectroscopic snapshot solar radio images at meter-wavelength improved by a lot. At higher frequencies, new generation instruments like MeerKAT \citep{meerkat2016,Chen_meerkat2021} have already started producing spectroscopic snapshot solar images with unprecedented quality \citep{Kansabanik2023_meerkat}. Comparing radio images with solar images from other wavelengths is not really an apples-to-apples comparison. It should be noted in passing that these new generation radio instruments can provide imaging dynamic range ($\sim10^5$) and spectro-temporal resolution much higher than possible with current generation visible, EUV, or X-ray instruments. These images enable studies of the temporal and spectral evolution of solar radio emissions in unprecedented detail. The need for spectroscopic snapshot solar radio imaging is now well established by several studies \citep{Mohan2017spreads,Mondal2021a,Mondal2023,Oberoi2023}. This is also evident from Figure \ref{fig:different_emissions} following the discussion in Section \ref{sec:intro_chapter_paircars_principle} of Chapter \ref{paircars_principle}. However, one quickly runs into the next challenge -- it is infeasible to examine these very large numbers of images and process them to extract quantitative information about features of interest manually. One will necessarily need to come up with automated ways to identify emission features of interest; track them across the image plane, time, frequency, and polarization; and build robust automated approaches to quantify them. The demands of this problem seem to be well matched to the strengths of Artificial Intelligence and Machine Learning (AI/ML) approaches. So  AI/ML techniques are expected to be a fertile ground for exploration in this context and efforts towards this are already  leading to interesting results \citep{Bawaji2023}.
\end{enumerate}

\subsection{Constraining CME parameters using Physics-Based Forward Models of GS Emission}\label{subsec:cme_modeling_future}
The work presented in this thesis has demonstrated that modeling CME GS emission is indeed a promising approach for estimating the physical parameters of CMEs. The work presented here is more detailed and rigorous than earlier attempts. It is also better constrained as it makes use of both Stokes I and V measurements. This work has also highlighted the limitations of the methodology currently in use for modeling CME GS emission.

The most promising path to explore in the future would be to relax some of the assumptions made by the GS models currently in use. While this is expected to lead us to a more realistic model for the CME, it also rapidly leads to a situation where the number of degrees of freedom becomes too large for the available constraints. There is, however, a possible approach for keeping the number of degrees of freedom in check while still relaxing the most constraining assumptions. The idea is that rather than modeling each PSF-sized region independently, one should use a ``forward modeling" approach. Essentially build a single model for the CME plasma parameters and, from this model, generate the synthetic spectropolarimetric GS emission maps. The minimization process is then designed to minimize the differences between the synthetic and the observed image cubes. The CME forward model parameters are modified iteratively to lead to a synthetic GS image cube which is closer to the observations. This differs from the current approach, where the emission from each PSF-sized region is modeled independently, and the continuity of the plasma parameters in nearby regions is not made use of. The forward modeling approach suggested here has been used successfully for modeling GS emission from flares.

\section{The End}\label{sec:end}
This thesis presents a detailed exploration of spectropolarimetric modeling of GS emission from CME plasma using high-fidelity and high DR spectropolarimetric solar imaging provided by the P-AIRCARS from MWA solar observations. A key aspect of this thesis is to showcase the improvement in the robustness of GS model parameters due to the inclusion of circular polarization measurements. The other key highlight of this thesis is that it presents possibly the first observational evidence of the insufficiency of homogeneous GS source models. I anticipate that the more sensitive observations from the upcoming ground-based radio interferometers will routinely provide sensitive Stokes I and V measurements across broad observing bands. The spectropolarimetric image cubes from these measurements, along with multi-vantage point white-light observations, will form the constraints for the physics-based forward modeling of CME GS emissions. I envisage that this approach will establish itself as a unique and robust remote-sensing observational probe for CME plasma at coronal heights.

\backmatter
\pagestyle{plain}
\begin{center}

\vspace{3.0cm}

\begin{Large}
\textbf{LIST OF PUBLICATIONS} \\
\end{Large}
\end{center}

\section*{Publications relevant to this thesis}
\hrule
\vskip 0.5cm
\subsection*{Refereed Publications}
\hrule
\vskip 0.5cm
\begin{enumerate}
\item \textbf{Robust absolute solar flux density calibration for the Murchison Widefield Array} \\
\textit{\textbf{D. Kansabanik}, S. Mondal, D. Oberoi, A. Biswas and S. Bhunia} \\ 
\href{https://doi.org/10.3847/1538-4357/ac4bba}{The Astrophysical Journal 2022, Volume 927, Issue 1, 17}.

\item \textbf{Tackling the unique challenges of low-frequency solar-polarimetry with the Square Kilometre Array Low precursor : The Algorithm} \\
\textit{\textbf{D. Kansabanik}, S. Mondal and D. Oberoi} \\ 
\href{https://doi.org/10.3847/1538-4357/ac6758}{The Astrophysical Journal 2022, Volume 932, Issue 2, 110}.

\item \textbf{Working Principle of the Calibration Algorithm for High dynamic-range Solar Imaging with Square Kilometre Array Precursor} \\
\textit{\textbf{D. Kansabanik}} \\ 
\href{https://doi.org/10.1007/s11207-022-02053-x}{Solar Physics, 2022, Volume 297, Issue 9, 122}.

\item \textbf{Tackling the unique challenges of low-frequency solar-polarimetry with the Square Kilometre Array Low precursor : Pipeline Implementation} \\
\textit{\textbf{D. Kansabanik}, A. Bera, D. Oberoi and S. Mondal} \\ 
\href{https://doi.org/10.3847/1538-4365/acac79}{The Astrophysical Journal Supplement Series, 2023, Volume 264, Issue 2, 47}.

\item	\textbf{Deciphering the Faint Gyrosynchrotron Emission from Solar Coronal Mass Ejections with Spectro-polarimetric Radio Observation} \\
\textit{\textbf{D. Kansabanik}, S. Mondal and D. Oberoi} \\ 
\href{https://doi.org/10.3847/1538-4357/acc385}{The Astrophysical Journal, 2023, Volume 950, Issue 2, 164}.

\end{enumerate}

\pagebreak
\subsection*{Conference Proceedings}
\hrule
\vskip 0.3cm
\begin{enumerate}
    \item \textbf{A novel algorithm for high fidelity spectro-polarimetric snapshot imaging of the low-frequency radio Sun using SKA-low precursor} \\
    \textit{\textbf{D. Kansabanik}, D. Oberoi and S. Mondal} \\ 
    \href{https://doi.org/10.48550/arXiv.2207.11924}{URSI AT-AP-RASC 2022}
\end{enumerate}

\bibliography{thesis}{}
\bibliographystyle{aas} 

\end{document}